\documentclass[11pt, twoside, openright, dvipsnames]{scrbook}

\usepackage{layout}
\usepackage[german, ngerman, english]{babel}
\usepackage[english]{babel}
\usepackage[a4paper,textwidth=450pt,textheight=680pt,bindingoffset=13.9mm]{geometry}

\usepackage{slashed}
\usepackage{latexsym}
\usepackage{amssymb} 
\usepackage{amsmath}
\usepackage{array}
\usepackage{eucal}
\usepackage{mathrsfs}
\usepackage{makeidx}
\usepackage{graphicx}
\usepackage{float}
\usepackage{framed}
\usepackage{booktabs}
\usepackage{multirow}
\usepackage{lineno}
\usepackage[table]{xcolor}
\usepackage{placeins}
\usepackage{arydshln}
\usepackage{textcomp}
\usepackage{tikz}
\usetikzlibrary{shapes.geometric}
\usepackage{url}
\usepackage{comment}
\usepackage[figuresright]{rotating}
\usepackage{booktabs}
\usepackage{multirow}
\usepackage{float}
\usepackage{cite}
\usepackage{dsfont}
\usepackage{braket}
\usepackage{pdfpages}
\usepackage[titletoc]{appendix}
\usepackage{wrapfig}
\usepackage{tablefootnote}

\usepackage[Sonny]{fncychap} 
\usepackage{dcolumn}
\usepackage[nottoc]{tocbibind}
\usepackage{epigraph}

\usepackage{overpic}
\usepackage{subfig}

\usepackage[squaren]{SIunits}
\usepackage{Style/atlasphysics}

\usepackage[colorlinks=true, pdfstartview=FitV, linkcolor=blue, 
           citecolor=blue, urlcolor=blue,
            pdftex,
            pdfauthor={Joshua Wyatt Smith},
            pdftitle={Top quarks and photons},
            pdfsubject={Dissertation},
            pdfkeywords={top quarks, photons, ttgamma, ATLAS}
           ]{hyperref}

\newcommand{\HRule}{\hrulefill}

\newcommand{\TITLE}{Fiducial cross-section measurements of the production of a prompt photon in association with a top-quark pair at $\sqrt{s}=13$~\TeV\ with the ATLAS detector at the LHC}
\newcommand{\AUTHOR}{Joshua Wyatt Smith}
\newcommand{\COUNTRY}{Cape Town, South Africa}

\ChNameVar{\Large}
\ChNumVar{\Huge}
\ChTitleVar{\Large}
\ChRuleWidth{0.5pt}
\ChNameUpperCase

\makeatletter
\g@addto@macro\bfseries{\boldmath}
\makeatother

\title{\TITLE}
\author{\AUTHOR}

\makeindex

\begin{document}

\frontmatter
\begin{titlepage}
\begin{center}
\begin{otherlanguage}{ngerman}



\HRule \\[0.4cm]
{\Large \bfseries \TITLE \\[0.4cm] }

\HRule \\[1.5cm]
{\large
Dissertation\\[1.0cm]
for the award of the degree\\
``Doctor of Philosophy" PhD Division of Mathematics and Natural Sciences\\
of the Georg-August-Universit\"at G\"ottingen\\[1.0cm]
within the ProPhys doctoral program \\
of the Georg-August University School of Science (GAUSS)\\[2.0cm]
submitted by\\[1.0cm]
\AUTHOR \\[0.5cm]
from \COUNTRY \\[2.0cm]


\vfill
G\"ottingen, 2018
}
\end{otherlanguage}
\end{center}

\end{titlepage}

\begin{otherlanguage}{ngerman}
\thispagestyle{empty}
{\small
\vfill
\noindent
\underline{Thesis Committee:}\\[0.2cm]
Prof. Dr. Arnulf Quadt \\
Prof. Dr. Stan Lai\\[1.0cm]

\noindent
\underline{Members of the Examination Board:}\\[0.2cm]
\begin{tabular}{ll}
Reviewer: & Prof. Dr. Arnulf Quadt \\
{}&{\footnotesize II. Physikalisches Institut, Georg-August-Universit\"at G\"ottingen}\\[0.2cm]
Second Reviewer: &Prof. Dr. Stan Lai \\
{}&{\footnotesize II. Physikalisches Institut, Georg-August-Universit\"at G\"ottingen}\\[0.2cm]
\end{tabular}\\[1.0cm]

\noindent
\underline{Further Members of the Examination Board:}\\[0.2cm]
Prof. Laura Covi, PhD \\
{\footnotesize Institut f\"ur Theoretische Physik, Georg-August-Universit\"at G\"ottingen}\\[0.2cm]
Prof. Dr. Ariane Frey \\
{\footnotesize II. Physikalisches Institut, Georg-August-Universit\"at G\"ottingen}\\[0.2cm]
Prof. Dr. Wolfram Kollatschny \\
{\footnotesize Institut f\"ur Astrophysik, Georg-August-Universit\"at G\"ottingen}\\[0.2cm]
Prof. Dr. Steffen Schumann \\
{\footnotesize Institut f\"ur Theoretische Physik, Georg-August-Universit\"at G\"ottingen}\\[1.5cm]
}
Date of the Oral Examination: 6$^\text{th}$ November 2018\\[1.5cm] 

\noindent
Reference: II.Physik-UniG\"o-Diss-2018/01

\clearpage
\end{otherlanguage}
\cleardoublepage
\thispagestyle{empty}
\begin{center}

\includegraphics[width=0.6\textwidth]{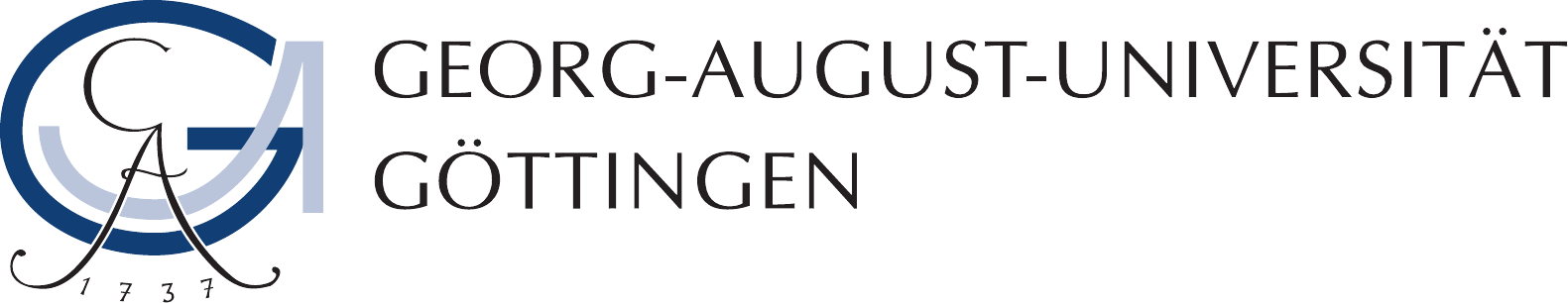}~\\[1cm]



\HRule \\[0.2cm]
{\Large \bfseries \TITLE \\[0.2cm] }

\HRule \\[1.0cm]

\textbf {Joshua Wyatt Smith}\\[1.0cm]
\end{center}

{
\noindent
The cross sections for top-quark pair production in association with a photon are measured in a fiducial volume with the \ATLAS detector at a centre-of-mass energy of 13~\TeV. Results are presented using proton-proton collision data collected by the LHC during 2015 and 2016, amounting to a total of $36.1~\fb^{-1}$. This also presents the first  \ttgamma \xsecModifyNoun measurements performed in the \chll channels.
Exactly one photon is required to have $\pt > 20$~\GeV and be isolated based on track and calorimeter information. At least two (four) jets are required in the \chll (\chljets) channels, with at least one jet originating from a $b$-quark.
Two separate neural network algorithms are used to help reduce the impact backgrounds play in the final measurements. 
The Prompt Photon Tagger is trained on information from energy deposits in the calorimeters to distinguish prompt photons from hadronic fake photons. The output of this neural network is fed into the Event-level Discriminator that uses event information to classify signal from the sum of all backgrounds.
A maximum likelihood fit is performed on the output of the Event-level Discriminator to determine the fiducial \xsec of the signal process. The fiducial \xsec for the \chljets and \chll channel are measured to be $521 \pm 9 \text{(stat.)} \pm 41 \text{(sys.)}$~\fb and $69 \pm 3 \text{(stat.)} \pm 4\text{(sys.)}$~\fb, respectively.
In total, eight \xsecModifyNoun measurements are performed and all agree with theoretical next-to-leading-order predictions.
}

\cleardoublepage


\chapter*{Acknowledgements}
\thispagestyle{empty}
{
\noindent
I would first like to thank Arnulf Quadt, my PhD supervisor, for the opportunity to be a part of the II. Physikalisches Institut in G\"ottingen. I learnt a tremendous amount during my time in G\"ottingen, CERN, and the various conferences around the world that I was lucky enough to participate in. I picked up a very broad range of skills that will certainly help me in my next endeavours, whatever those may be.

\noindent
Thank you to Elizaveta Shabalina who has been incredibly helpful in the day-to-day supervision of this analysis. Her experience and willingness to try innovative techniques ensured that this measurement was interesting and unique.

\noindent
A special thank you to Dan Guest for many fruitful discussions (often over pasta) on machine learning topics, and just general help and advice on many subjects during the course of my PhD. Also, thank you for not letting us fall off or get crushed by mountains. That is important.

\noindent
Thanks to Mar\'ia Moreno Ll\'acer who helped determine the strategy the analysis took, as well as provide assistance on the fitting procedures. Thank you to Yichen Li and Julien Caudron. As analysis contacts they moved the analysis forward and provided essential tips, comments and studies. Yichen was one of the main driving forces in making it possible to show these results at a public conference.
Likewise, to the group members of the \ttgamma analysis, thank you for making this happen!

\noindent
Thank you to Thomas Peiffer and Royer Edson Ticse Torres for interesting discussions and for proofreading this thesis. Many helpful comments were provided.

\noindent
To my friends and colleagues in G\"ottingen, around the world, and those constantly scattered throughout the French and Swiss Alps (you should be working!), thank you for the general help and companionship that is needed when pursuing a PhD.

\noindent
Thank you to Marie-Pier for the encouragement over the last year. You understand the demands of finishing a PhD thesis and have been nothing but supportive.

\noindent
Finally, thank you to my family for the continued love and support throughout the years. It goes without saying that none of this would have been possible without you. For that I am grateful.
}
\cleardoublepage

\pagenumbering{Roman}
\tableofcontents

  
\chapter{Prologue}
\pagenumbering{arabic}
\setcounter{page}{1}
The study of high energy particle physics revolves around understanding our universe at the most fundamental level.
At this level, the constituents of matter and the forces they feel (the electromagnetic, weak, and strong forces) are described by the Standard Model of particle physics. 
The Standard Model has been confirmed to incredibly high precision over countless experiments for more than 50 years. And yet, we know it is not the full picture. Perhaps the most exciting and profound part is that it can only explain around 4\% of \emph{everything} we know in the entire universe.
The Large Hadron Collider at CERN was built to explore and probe the Standard Model.
By colliding (mainly) protons at high energies, quarks, gluons, leptons, and bosons are created in abundance. The study of these particles and their interactions provide physicists with the data to continue to expand our knowledge of the universe.

The top quark is especially interesting for several reasons. Its very large mass and its strong coupling to the recently discovered Higgs~boson seem to hint that this particle is special. The top quark is also the only quark to decay before it has a chance to hadronise. It presents us with a unique opportunity to study the direct fundamental properties of a quark through the decay products.
By probing the top-photon coupling at increasing centre-of-mass energies we are continuously putting the Standard Model to the test. 
While making precision measurements useful for other analyses, we are essentially looking for a scale at which our model does not describe the data anymore. It must exist, we are just not sure where and how this new physics will manifest. 

The analysis presented in this thesis measures the \xsec for photons radiated from top-quark pairs, the so-called \ttgamma process.
The data was collected in 2015 and 2016, and amounts to $36.1$~\fb{}$^{-1}$ of proton-proton collisions at $\sqrt{s} = 13$~\TeV.
Similar measurements of the $t\gamma$ coupling have been performed by the CDF, \ATLAS, and \CMS experiments, all with less data and lower centre-of-mass energies. For the first time the \ttgamma \xsecModifyNoun measurements are performed in the \chll channels, as well as the \chljets channels. Modern machine learning algorithms play an important role in the strategies chosen for this analysis.

This thesis is laid out as follows: the introduction (Chapter~\ref{sec:intro}) provides brief theoretical background material in the form of the Standard Model and machine learning in high energy physics.
Chapters~\ref{sec:atlasexp}~-~\ref{sec:objectID} lay the groundwork for the experimental setup as well as the experiment-wide algorithms essential for object reconstruction and identification.
Chapter~\ref{sec:ttgammaprocess} describes the analysis strategy, signal, and background processes for the $\sqrt{s} = 13$~\TeV analysis.
Two unique neural networks using object- and event-level information are presented in Chapter~\ref{sec:NNs}.
An overview of the systematic uncertainties used in the analysis is provided in Chapter~\ref{sec:systematics}, and the final results with in-depth discussions are presented in Chapter~\ref{sec:results}. This is followed by conclusions in Chapter~\ref{sec:conclusions}.

In general the appendix is home to supporting material for the above studies. However, Appendix~\ref{sec:arm} presents an overview of work done during this PhD in porting the \ATLAS software stack (Intel x86 based) to the ARM architecture. 
Appendices~\ref{sec:analysisstrategyappendix}~-~\ref{sec:resultsappendix} contain support material for the analysis strategy, the object- and event-level neural networks, the systematic uncertainties, and the final results.

The analysis presented in this thesis~\cite{ATLAS-CONF-2018-048} was first presented at the $11^{\text{th}}$ International Workshop on Top Quark Physics (TOP 2018). 
\mainmatter
\setcounter{page}{3}
\pagestyle{headings}


\chapter[Introduction]{Introduction}
\label{sec:intro}

\section{Brief summary of the Standard Model}
\label{sec:theSM}

The Standard Model (SM) describes the behaviour of three out of the four fundamental forces we can observe. 
These are electromagnetism, the weak force, and the strong force.  
Many physicists have contributed to the SM over the last century. While the theory itself was formalised in the 1970's, the building blocks such as Quantum Electrodynamics (QED)~\cite{Dirac:1927dy,RevModPhys.4.87,Feynman:1948ur,Schwinger:1948iu,Schwinger:1948yk} were created well before that.

The SM is the combination of three renormalisable, local gauge invariant quantum field theories, two of which (electromagnetism and the weak force) are unified into the electroweak force~\cite{Glashow:1961tr,Salam:1964ry,Weinberg:1967tq,Glashow:1970gm}.
The vector fields of the local symmetry $SU(3)_{C} \times SU(2)_{L} \times U(1)_{Y}$ give rise to these three fundamental interactions. 
The first term ($SU(3)$) describes the strong interaction in the framework of Quantum Chromodynamics (QCD)~\cite{Politzer:1974fr,Politzer:1973fx,Gross:1973ju}. The second two terms ($SU(2)_{L} \times U(1)_Y$) describe the electroweak interaction.

Particles can be described as excitations of scalar and vector fields.
There are 12 spin-$\frac{1}{2}$ fermions and five force mediators or gauge bosons.
Interactions between fermions are mediated by the exchange of gauge bosons.
Photons ($\gamma$) mediate the electromagnetic force, $W^{\pm}$- and $Z^{0}$-bosons mediate the weak force, and gluons ($g$) carry the strong force.

The masses of the $W^{\pm}$- and $Z^{0}$-bosons, as well as the prediction of a new heavy scalar boson, arise through spontaneous symmetry breaking of the  $SU(2)_{L} \times U(1)_{Y}$ gauge symmetry. This is achieved through the Brout-Englert-Higgs mechanism~\cite{Englert:1964et,Higgs:1964pj}, first predicted in 1964. By interacting with the heavy scaler field, fermions also acquire mass.
In 2012 the Higgs-boson particle was discovered by the \ATLAS and CMS collaborations at the LHC~\cite{HIGG-2012-27,CMSHiggs}.

The interaction between fermions and the photon can be described by the electromagnetic Lagrangian in the framework of QED:
\begin{align}\label{eq:eqdL}
\mathcal{L}_{\text{QED}} &= \mathcal{L}_{\text{fermion}} + \mathcal{L}_{\text{photon}}+ \mathcal{L}_{\text{interaction}} \notag \\
 &= \bar{\psi} (i \slashed{\partial} + m)\psi - \frac{1}{4}F_{\mu\nu}F^{\mu\nu} - q \bar{\psi}\gamma^{\mu}A_{\mu}\psi.
\end{align}
The first term describes the kinematics of the spin-$\frac{1}{2}$ fermion field ($\psi/\bar{\psi}$). The second term describes the kinematics of the photon field $A_{\mu}$, where $F_{\mu\nu}$ is the electromagnetic field tensor defined as $F_{\mu\nu} = \partial_{\mu}A_{\nu} - \partial_{\nu}A_{\mu}$.

The last term describes the interaction between the photon and fermion field, and is directly proportional to the charge ($q$) of the fermion. Thus, probing this vertex probes the structure of the coupling as well as the coupling strength of photons to fermions.
It is the particles which we can measure in collider experiments that enable us to infer properties about the associated interacting fields. 

Of the fermions, six are called quarks and six are called leptons. They are ordered in three generations (essentially by mass), where each quark has an ``up-type"  (up ($u$), charm ($c$), top ($t$)) and ``down-type" (down ($d$), strange ($s$), bottom ($b$)) particle.
The leptons consist of charged particles ($e$, $\mu$, $\tau$) and their neutral counterparts, the neutrinos ($\nu_{e}$, $\nu_{\mu}$, $\nu_{\tau}$).
Each of the up-type quarks carries a fractional electric charge of $\frac{2}{3}e$,\footnote{Throughout this thesis natural units are used. This involves setting $c=\hbar=1$. Masses, energies and momenta are expressed in~\GeV. Additionally, charges are expressed as absolute values of the electron, $e$.} while down-type quarks carry $-\frac{1}{3}e$. 
Charged leptons carry integer charge ($-1e$), while the neutrinos are neutral and only interact via the weak force.
All quarks also carry colour charge labelled (somewhat arbitrarily) as red, green or blue. Due to colour confinement~\cite{Wilson:1974sk}, quarks and gluons can not be observed as free particles but rather hadronise to form colour neutral composite particles called hadrons. Two examples of hadrons are baryons, which have three quarks ($uud$ in the case of protons), and mesons ($u\bar{d}$ in the case of a charged pion)\footnote{Tetraquarks (two quarks and two antiquark-quarks) and pentaquarks (four quarks and one antiquark) have also been observed at the Large Hadron Collider~\cite{Aaij:2015tga,D0:2016mwd}.}.
All of the observable matter in the universe is comprised of three particles from the first generation: $u$-quarks, $d$-quarks  and electrons.  

For each fermion there exists an antiparticle, essentially identical, except that additive quantum numbers (such as charge and weak isospin) have opposite values.
For quarks and neutrinos these particles are denoted with bars (i.e. $\bar{u}$, $\bar{t}$, $\bar{\nu}_e$), while for charged leptons the charge is simply reversed ($e^{+}$, $\mu^{+}$, $\tau^{+}$).

Leptons, both charged and neutral, can interact via the weak interaction and so are organised into left-handed doublets and right-handed singlets, defined as:
\begin{align*}
\binom{\nu_{l}} { l^{-}}_{L}, \hspace{0.5cm}
l^{-}_{R}.\hspace{0.5cm}
\end{align*}
Of the left-handed doublets, the charged leptons have a corresponding third-component of weak isospin\footnote{This quantum number must be conserved in all weak interactions.}, $I_3 = -\frac{1}{2}$, while the neutral leptons have $I_3 = \frac{1}{2}$. Right-handed singlets have $I_3=0$.
The SM does not accommodate right-handed neutrinos, and all neutrinos in the SM are assumed to be massless. However, small oscillations between neutrino flavours have been observed, meaning their masses are actually non-zero~\cite{Fukuda:1998mi}.

Quarks are sensitive to the weak and strong forces and therefore also transform as doublets and singlets. However, there exists cross-generational coupling between the quarks. The representations can be written as:
\begin{align*}
\binom{q_{\text{u-type}}}{q'_{\text{d-type}}}_{L}, \hspace{0.5cm}
q_{\text{u-type,}R}, \hspace{0.5cm}
q_{\text{d-type,}R},\hspace{0.5cm}
\end{align*}
where for the left-handed doublets the up-type quarks have $I_3 = \frac{1}{2}$ and the down-type quarks have $I_3 = -\frac{1}{2}$. For the right-handed singlets, $I_3=0$. 
The weak eigenstates ($q'$) are translated to the mass eigenstates ($q$) of the down-type quarks via the CKM matrix~\cite{Cabibbo:1963yz,Kobayashi:1973fv}.

\begin{figure}[!h]
\centering
\includegraphics[width=0.9\linewidth]{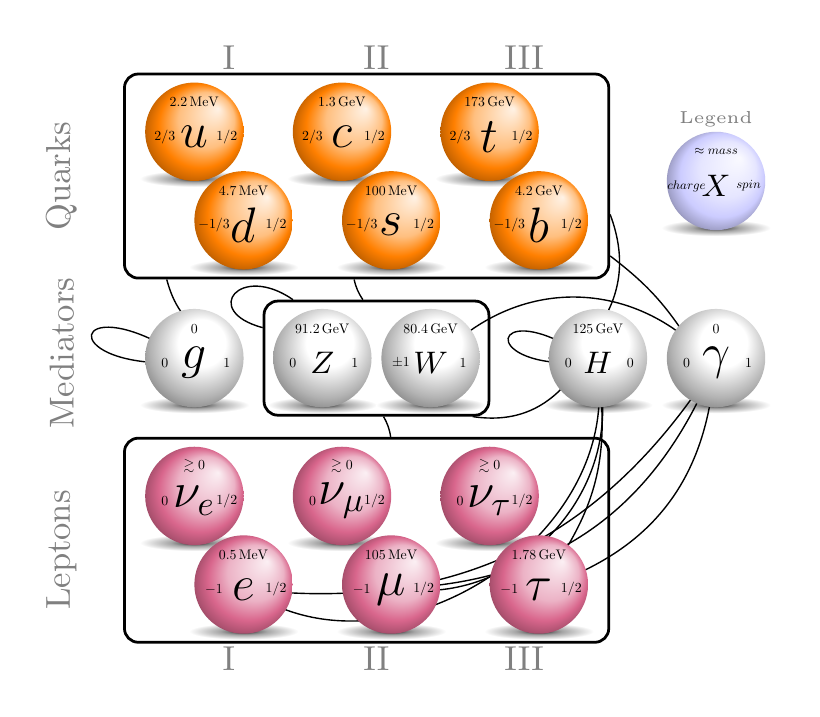}
\caption {The Standard Model of particle physics ordered by generation and particle type; quarks, bosons (force mediators) and leptons. Also included are the possible interactions for each boson represented by connecting lines.}\label{fig:SM}
\end{figure}

The SM of particle physics as described above is summarised in Figure~\ref{fig:SM}. 
For each particle the charge, mass and spin is depicted and grouped according to generation number (and implicitly by $I_3$).
Also included are the interaction terms represented by lines connecting various particles. For example, gluons have colour charge, as do quarks. Thus, gluons can couple to themselves as well as quarks. Photons can interact with any charged particle; the electron, muon, tau, all six quarks and the $W$-boson.
The $W$- and $Z$-vector bosons\footnote{In this thesis $V$-boson refers to the $W$- and $Z$-boson, and not the photon.} couple to themselves and the Higgs-boson ($H$). Since they lack colour charge, they do not couple to gluons.

The predictions of the SM have been measured to high degrees of accuracy~\cite{PDG2018} and have yet to be conclusively proven wrong. However, there are gaps missing in the formalism of the SM.
A few such examples are: 
\begin{itemize}
\item Gravity is not included in the SM.

\item The masses of particles are not predicted by the SM but rather must be experimentally determined (along with other parameters).

\item Neutrinos do not have mass in the SM, yet experimental observation shows their ability to oscillate between different flavours. This depends on the mass squared differences, and thus requires them to have some mass~\cite{Fukuda:1998mi}.

\item The SM describes only about 4\% of the entire universe. The rest, 74\% dark energy and 22\% dark matter, are believed to exist through indirect experimental evidence~\cite{Rubin:1970zza,Bartelmann:1999yn,Clowe:2006eq}. 
We know that dark energy and dark matter exist, however, we know very little about them.

\end{itemize}
By interpreting our measurements in the framework of the SM, these gaps could manifest themselves as new particles or unexplained interactions.

\FloatBarrier

\subsection{The top quark}
This section gives a brief overview of the properties of the top quark. A more thorough treatment can be seen in~\cite{Quadt:2007jk}.

After the discovery of the $b$-quark at Fermilab in 1977~\cite{Herb:1977ek}, a weak isospin partner was needed to fulfil the prediction of the SM, just as the previous two particle generations have weak isospin partners.
The top quark was discovered in 1995 at the Tevatron at Fermilab in proton-antiproton collisions~\cite{D0:1995jca, Abe:1995hr}.


\subsubsection{Production}
Top quarks are predominantly created in pairs (top-antitop) through the processes of quark-antiquark annihilation ($q\bar{q} \to \ttbar$) or gluon-gluon fusion ($gg\to \ttbar$) in proton-proton collisions. At $\sqrt{s}=13~\TeV$, approximately 90\% of top-quark pairs are created through gluon-gluon fusion.

In 2009, 14 years after the discovery of the top quark, electroweak production of single-top quarks was observed in the $s$- and $t$-channels\footnote{Here, $s$-, $t$-, $u$-channels refer to the Mandelstam variables~\cite{Mandelstam:1958xc}.} at the Tevatron~\cite{Abazov:2009ii,Aaltonen:2009jj}.
Single-top quarks are also created at the LHC and have a cross section around two to three times lower than that of pair production. The dominant process (about 70\%) is from the $t$-channel production, $ub \to dt$ or $\bar{d}b \to \bar{u}t$ via a virtual $W$-boson. Other processes include the $s$-channel, $u\bar{d} \to t\bar{b}$ (also via a virtual $W$-boson) as well as production of a top quark with a real $W$-boson, $gb \to Wt$.

In hadron colliders, top-quark pair production can be factorised into two ``pieces" and convoluted with one another, known as the factorisation theorem~\cite{Collins:1989gx}. As a function of the centre-of-mass energy (CME, $\sqrt{s}$) and the top-quark mass, the cross section of \ttbar production from proton-proton collisions is described as  
\begin{align}
\sigma^{pp\to \ttbar}(\sqrt{s},m_{t}) = \sum_{i,j = q, \bar{q},g} & \int dx_{i}dx_{j}f_{i}(x_{i},\mu_{f}^2)f_{i}(x_{j},\mu_{f}^2) \notag
\\
 &\cdot \hat{\sigma}^{ij\to \ttbar}(m_t,\sqrt{\hat{s}},x_i,x_j,\alpha_s(\mu_R^2),\mu_R^2).
\end{align}
The term on the first line is governed by non-perturbative QCD and describes the modelling of the protons, where $f_i(x_i,\mu_f^2)$ is the parton distribution function (PDF) for parton $i$. The PDFs describe the fraction of momenta ($x_i$) the partons possess from the parent hadron. The PDFs are experimentally determined~\cite{Altarelli:1977zs,Gribov:1972ri,Dokshitzer:1977sg} and are evaluated as a function of the factorisation scale $\mu_{f}$, which separates perturbative from non-perturbative QCD.
The bottom term describes the interaction using perturbative QCD where the cross section is a function of the top-quark mass, the CME in the parton-parton rest frame ($\sqrt{\hat{s}}$), the strong coupling constant $\alpha_s$, and the renormalisation scale $\mu_R$. 
This term includes the matrix element calculation which can occur at various levels of precision.

\subsubsection{Decay and backgrounds}

The SM predicts that top quarks decay essentially 100\% of the time to a $W$-boson and $b$-quark. Decays to the $s$- and $d$-quarks are strongly suppressed by the CKM matrix.
The $W^{+}$-boson decays to a pair of up-type ($u$- or $c$-) and down-type ($\bar{d}$-, $\bar{s}$-, $\bar{b}$-) quarks 67\% of the time, or a charged lepton ($l^{+}$) and the corresponding neutrino ($\nu_{l}$) around 33\% of the time (summed total for the three charged leptons). The decay of the $W$-boson categorises the channels in which we choose to make top-quark measurements. For instance $t\to W^{+}b \to l^{+}\nu b$ is categorised as a leptonic decay, whereas  $t\to W^{+}b \to q\bar{q} b$ is classified as a ``hadronic" decay.

\begin{figure}[!h]
\centering
\includegraphics[width=0.7\linewidth]{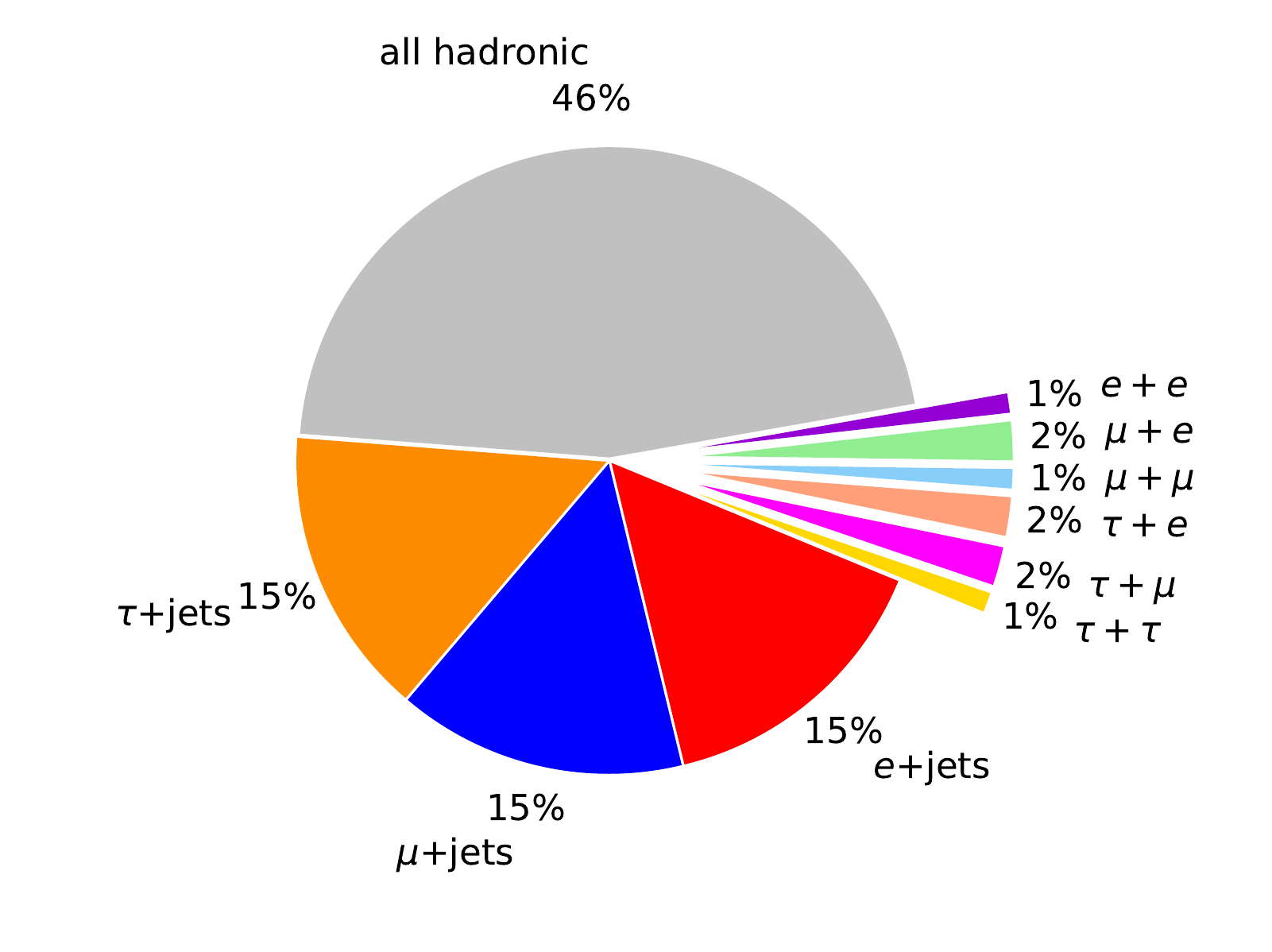}
\caption {Branching ratios for \ttbar decay~\cite{PDG2018}.}\label{fig:ttbardecay}
\end{figure}

\begin{figure}[!h]
\centering
\subfloat[\chljets]{
\includegraphics[width=0.37\linewidth]{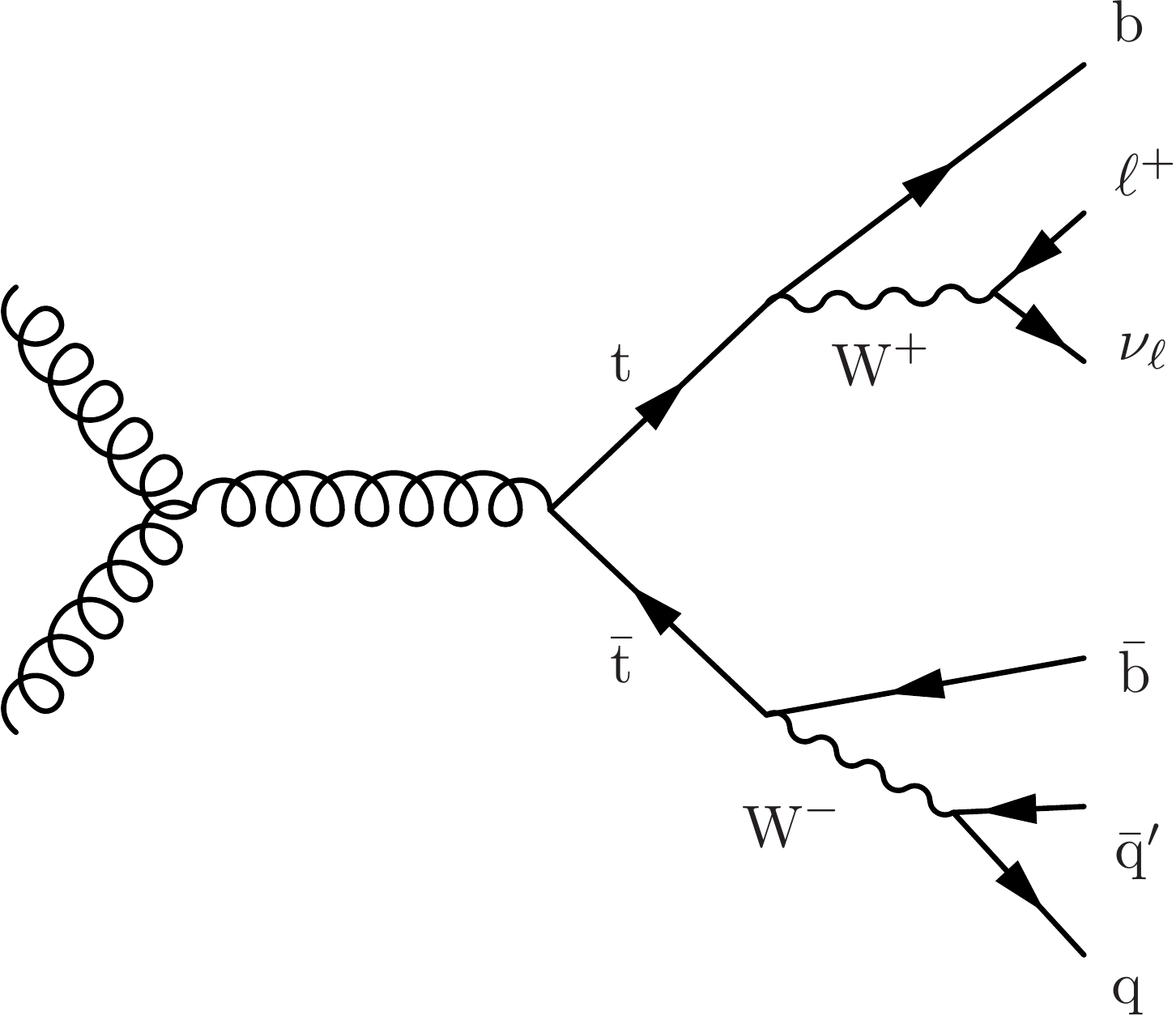}
}
\subfloat[\chll]{
\includegraphics[width=0.37\linewidth]{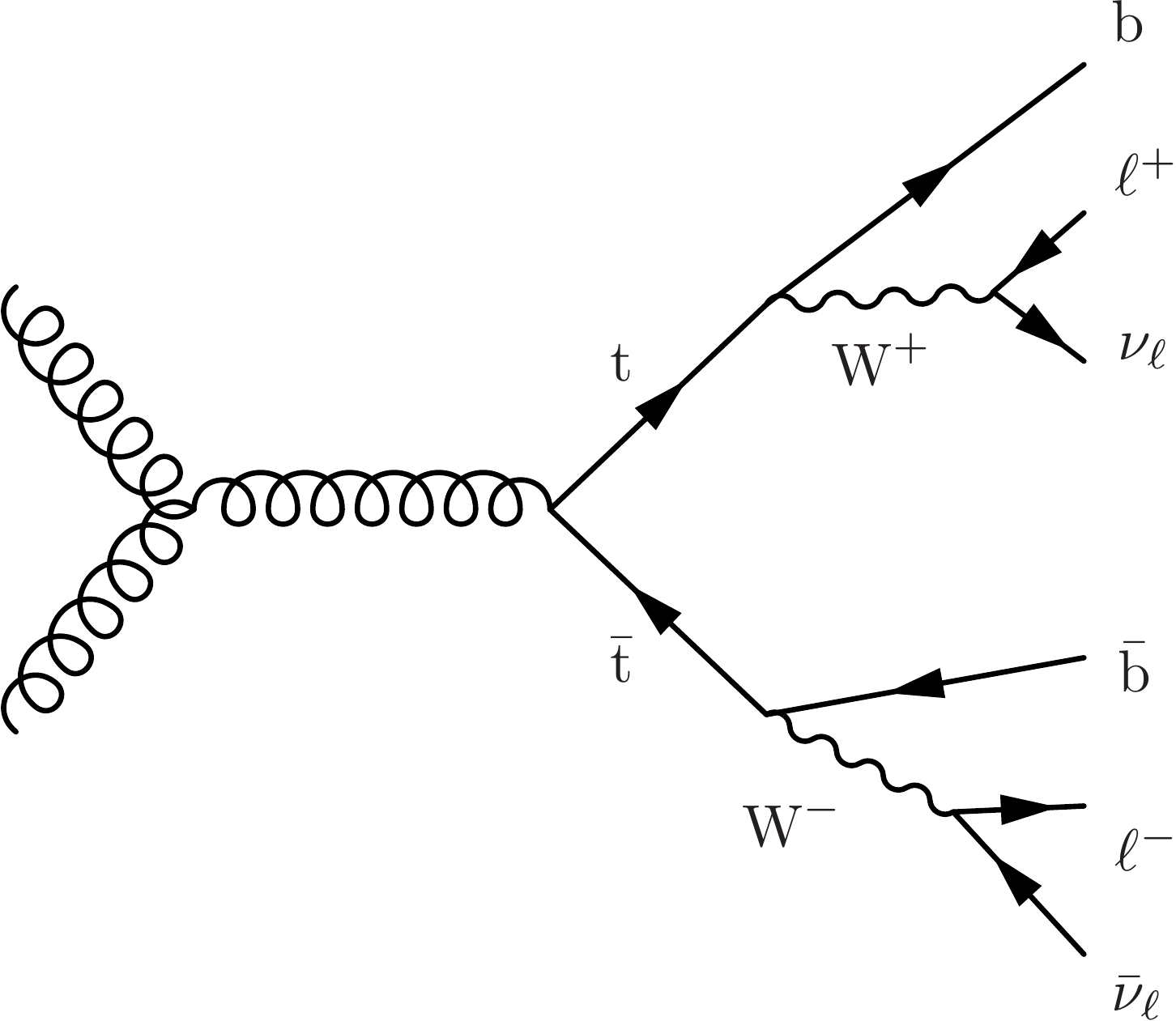}
}
\caption {Examples of the \ttbar process for gluon-gluon production and \chljets and \chll decay channels.}\label{fig:channelOverview}
\end{figure}

The branching ratios for \ttbar decay are shown in Figure~\ref{fig:ttbardecay}. 
When both $W$-bosons decay to a quark-antiquark pair and a $b$-quark, this is classified as ``all hadronic" and occurs 46\% of the time. When one $W$-boson decays leptonically ($l=e,\mu,\tau$) this is classified as the \chljets channel and occurs 45\% of the time. 
If both $W$-bosons decay leptonically then this is called the \chll channel and occurs 9\% of the time. 
In the context of this thesis, an ``inclusive" measurement includes all but the ``all hadronic" channels.
Figure~\ref{fig:channelOverview} shows the diagrams for gluon-gluon initiated \ttbar production for the \chljets and \chll decay channels.

The main irreducible\footnote{Irreducible backgrounds have the same final state as the signal} backgrounds in \ttbar measurements arise from the production of a leptonically decaying $W$- or $Z$-boson with jets in the \chljets and \chll channels, respectively. 
The electroweak single-top processes can also significantly contribute to the \ttbar background with extra jets from background QCD processes.
However, the flavour of the jets is a powerful discriminant to help distinguish these backgrounds from the signal.
As the CME increases the \ttbar production \xsec increases faster than that of the respective backgrounds, essentially turning the LHC into a top-quark factory. Thus, the signal to background ratio becomes larger.

\begin{figure}[!h]
\centering
\includegraphics[width=0.75\linewidth]{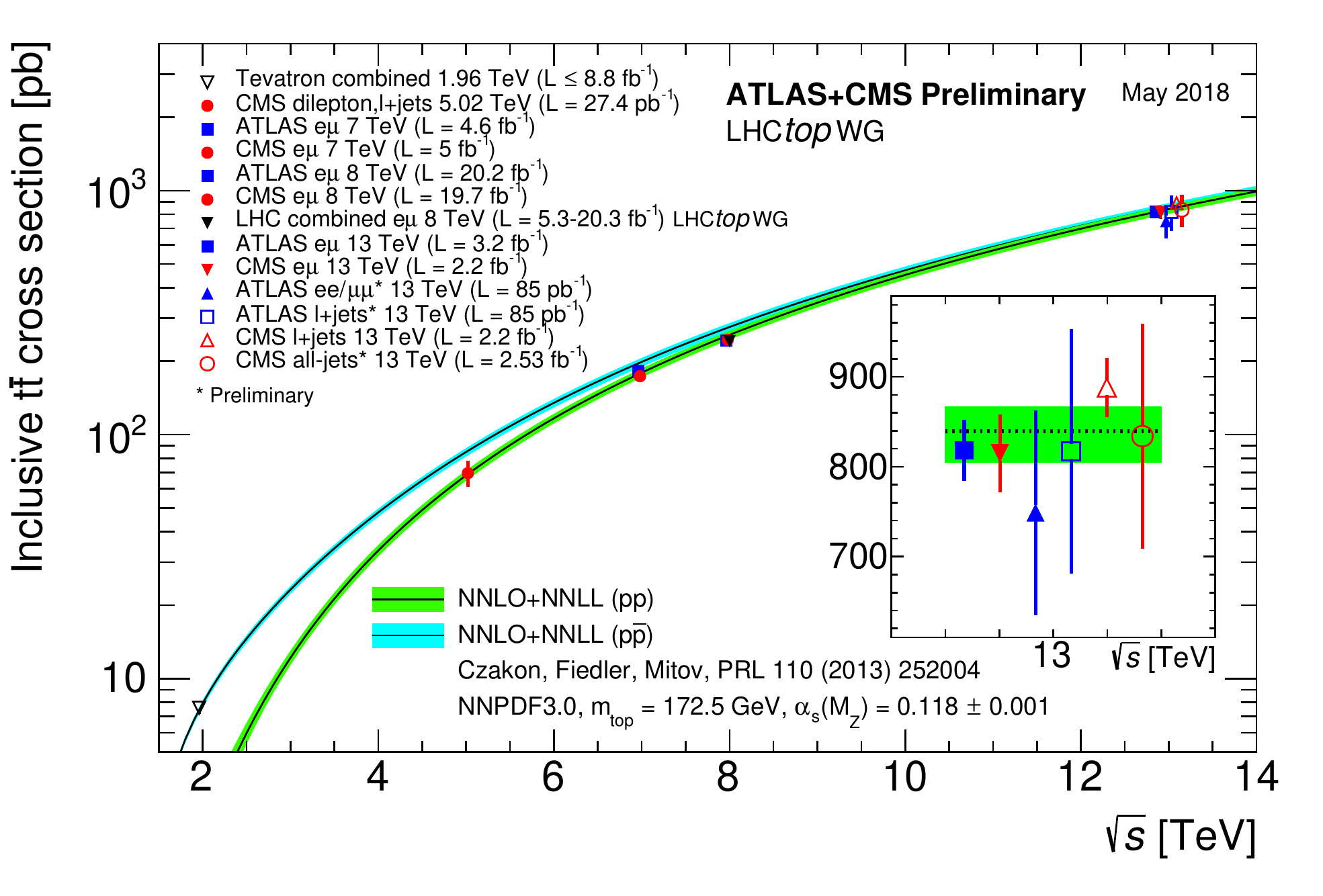}
\caption {A summary of all \ttbar \xsec measurements performed at the Tevatron and LHC as a function of the CME, compared to the NNLO QCD calculation with NNLL resummation. Measurements made at the same CME are slightly offset for clarity~\cite{TopSummaryPlots}.}
\label{fig:ttbarmeasurements}
\end{figure}

Figure~\ref{fig:ttbarmeasurements} shows all \ttbar \xsec measurements performed at the Tevatron and LHC as a function of the CME. Theoretical calculations to which the measurements are compared are performed at next-to-next-to-leading order (NNLO) with next-to-next-to-leading logarithm (NNLL) order calculations for soft gluon resummation~\cite{Cacciari:2011hy}. All measurements agree with the theoretical prediction.

\subsubsection{Properties}
Due to the very large mass of the top quark it is expected that it plays an important role in electroweak symmetry breaking.
The lifetime is approximately $0.5\times10^{-24}$s, which is about an order of magnitude smaller than the timescale for the hadronisation process of quarks. Thus, it decays before it has a chance to hadronise. This means that the properties are transferred directly to the decay products. This makes for interesting measurements in which the underlying dynamics of quarks can be studied. 
A select few properties and results are highlighted, while a full overview can be found in~\cite{PDG2018}.

Direct mass measurements of the top quark have been performed at the Tevatron~\cite{TevatronElectroweakWorkingGroup:2016lid} and at ATLAS~\cite{TOPQ-2016-03} and CMS~\cite{Khachatryan:2015hba,Sirunyan:2017huu} at the LHC. A global average results in $173.0\pm0.4$~\GeV~\cite{PDG2018}.
Indirect measurements of the top-quark mass from cross sections have also been performed at D$\emptyset$~\cite{Abazov:2016ekt} at the Tevatron, as well as at ATLAS~\cite{TOPQ-2013-04,TOPQ-2014-06,TOPQ-2015-02} and CMS~\cite{Khachatryan:2016mqs,Sirunyan:2017uhy}. The global average results in $173.1\pm0.9$~\GeV~\cite{PDG2018}, consistent with the direct mass measurements.

After the discovery of what was believed to be the top quark, one hypothesis was that it could be an exotic quark with charge $-\frac{4}{3} e$~\cite{Chang:1998pt}. Measurements made at the Tevatron, and more recently at the LHC, strongly reject this hypothesis~\cite{Abazov:2006vd,TOPQ-2011-13}.

The top-quark width is expected to be fairly large given that the lifetime is very small. Deviations from the expected decay width could hint at new physics. A global average results in $\Gamma_t = 1.41^{+0.19}_{-0.15}$~\GeV~\cite{PDG2018}, which agrees with theoretical predictions.

An interesting measurement of top-quark pairs is that of the spin correlation~\cite{Abazov:2011gi}. 
The SM predicts that the angles of the decay products from the top quark are correlated with the top polarisation. In \ttbar pairs, the spin correlations between each of the top quarks can be inferred from angles between the respective decay products.
At $\sqrt{s} = 7$~\TeV the measurements made by \ATLAS were consistent with the SM~\cite{TOPQ-2011-11,TOPQ-2013-01}. However, in a recent measurement at $\sqrt{s} = 13$~\TeV with the \ATLAS detector, a $3.2\sigma$ deviation from the SM is observed~\cite{ATLAS:2018rgl}. 
It should be mentioned that the same mis-modelling is observed in the analysis presented in this thesis at a lower significance of $1.5\sigma$. However, it will not be presented and differential distributions can be seen in~\cite{ATLAS-CONF-2018-048}.

In $q\bar{q}\to t\bar{t}$ events at leading order (LO), the top quark (antitop quark) moves in the direction of the incoming quark (antiquark). However, at next-to-leading order (NLO) interference terms from additional Feynman diagrams cause an asymmetry between this symmetrical angular production. The asymmetry can be exaggerated by Beyond SM (BSM) physics.
Tension with the SM was first seen at measurements made at the Tevatron~\cite{Aaltonen:2012it}.
However recent measurements done at $\sqrt{s}=8$~\TeV in \ATLAS and \CMS~\cite{TOPQ-2014-16,TOPQ-2015-08,TOPQ-2015-10,TOPQ-2016-16,Khachatryan:2015oga,Khachatryan:2015mna,Khachatryan:2016ysn} all agree with theoretical SM predictions.
Interestingly, the charge asymmetry between top-quark pairs produced at the LHC is expected to be magnified in \ttgamma production~\cite{Aguilar-Saavedra:2014vta}, however this measurement has yet to be performed.

Lastly, a measurement of how the top quark couples to photons probes the electromagnetic coupling. This is the central theme in this thesis and is explained in the next section.

\FloatBarrier

\subsection{Top-quark pair production in association with a photon}

A \xsecModifyNoun measurement of a prompt photon in association with a top-quark pair (\ttgamma) can be interpreted as a direct probe of the electromagnetic coupling of the top quark as explained above.
Anomalous top-quark couplings could manifest as shape discrepancies in various kinematic distributions or in \xsecModifyNoun measurements~\cite{Baur:2004uw,Bouzas:2012av}.
The results can also be interpreted in the framework of an effective field theory in the search for new physics~\cite{Schulze:2016qas,eft_1601.08193}.

In addition to being radiated from a top quark, photons can be radiated from any of the charged particles resulting from top quark decay, as well as the incoming partons in proton-proton collisions. The processes that contribute to the background of \ttbar are essentially the same for \ttgamma, albeit with smaller \xsec{}s. 
However, overall it is fake photons and leptons that contribute the most to the \ttgamma backgrounds. These are discussed further in Chapter~\ref{sec:ttgammaprocess}.

\begin{figure}[h!]
\centering
\subfloat[\label{fig:t1channelttgamma}]{
\includegraphics[width=0.37\linewidth]{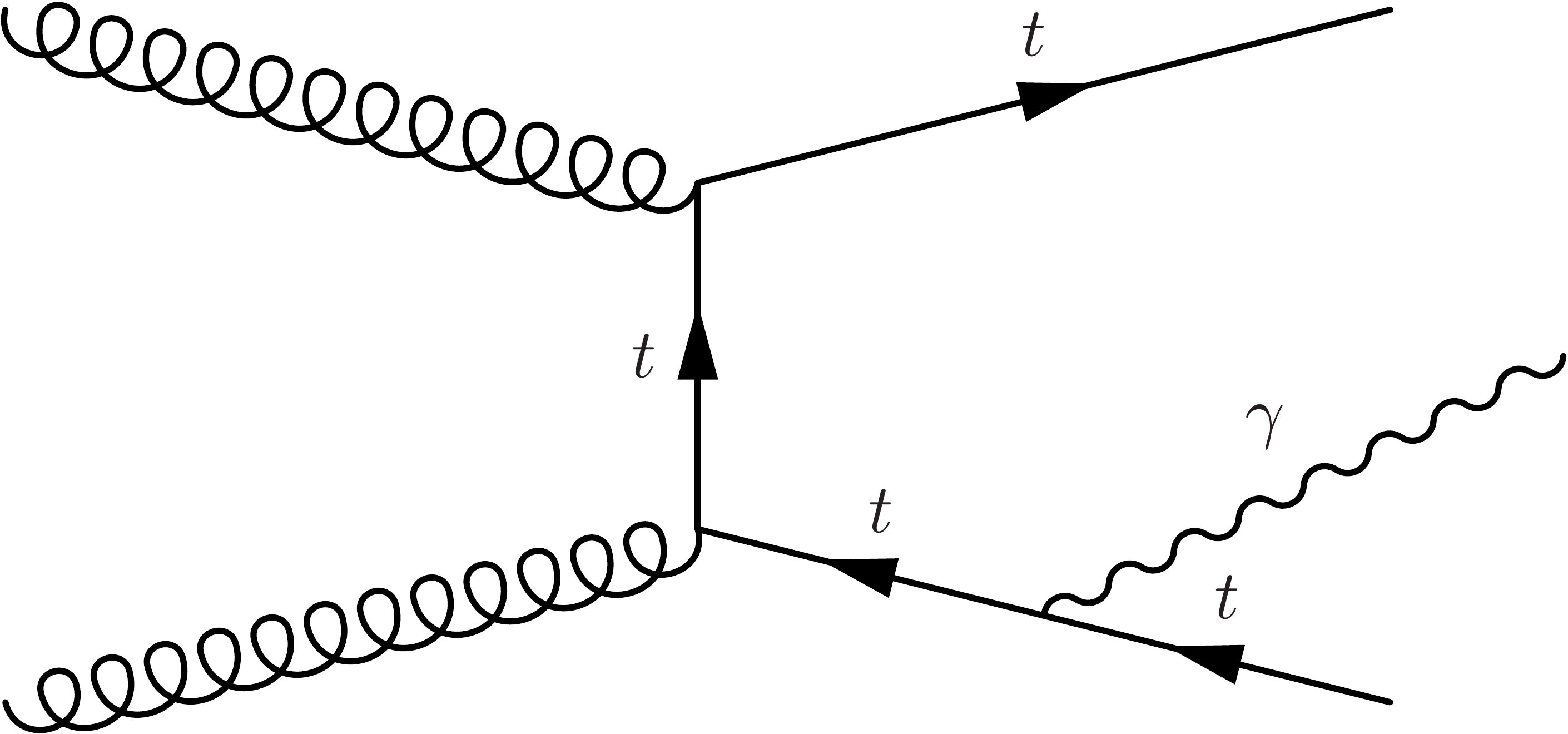}
}\vspace{1cm}
\subfloat[\label{fig:t2channelttgamma}]{
\includegraphics[width=0.37\linewidth]{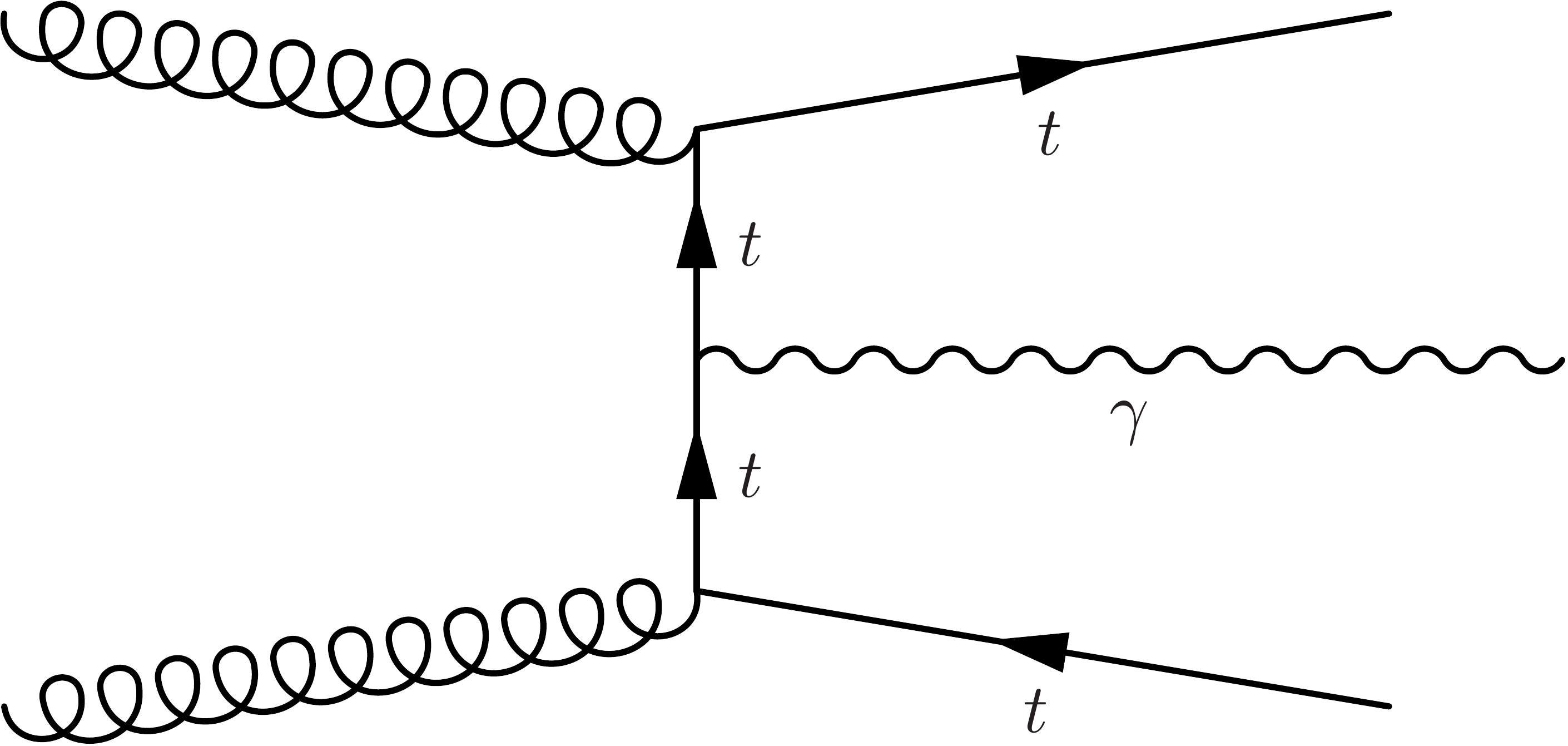}
}

\subfloat[\label{fig:schannelttgamma}]{
\includegraphics[width=0.37\linewidth]{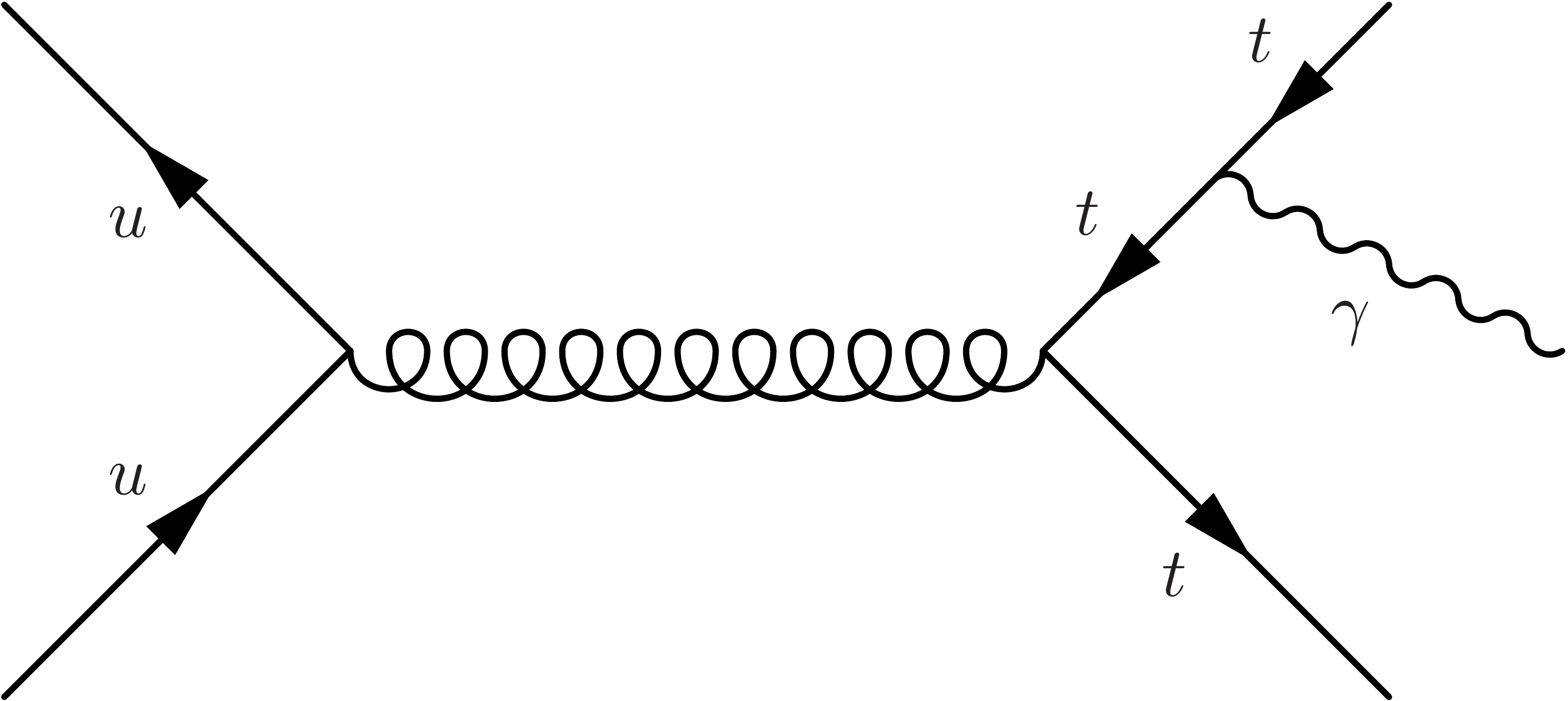}
}
\subfloat[\label{fig:isr}]{
\includegraphics[width=0.37\linewidth]{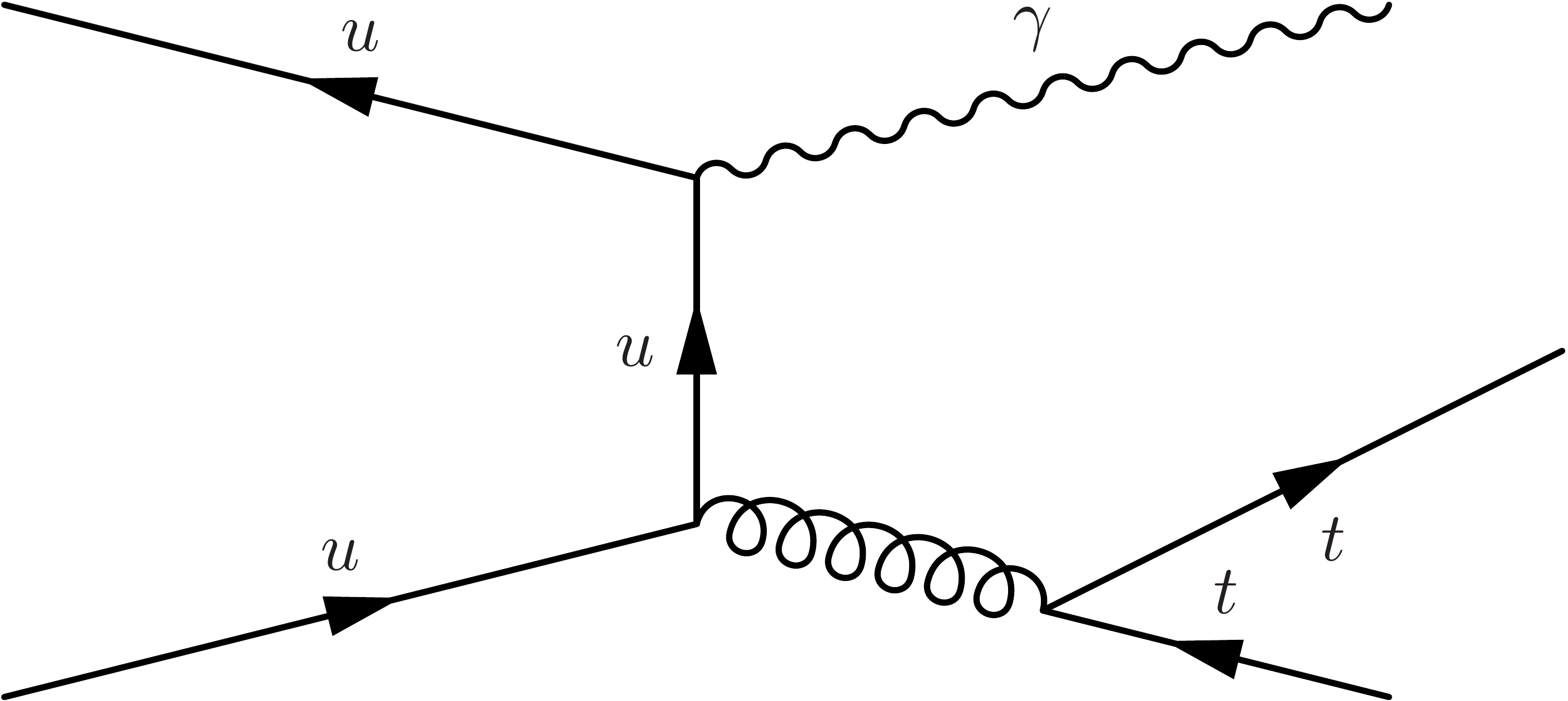}
}
\caption {Examples of the \ttgamma process at LO from quark-antiquark annihilation and gluon-gluon fusion.  Photons are radiated during top quark production or from initial partons. All diagrams are considered as signal.}\label{fig:rp}
\end{figure}

\begin{figure}[!h]
\centering
\subfloat[\label{fig:rdbquark}]{
\includegraphics[width=0.37\linewidth]{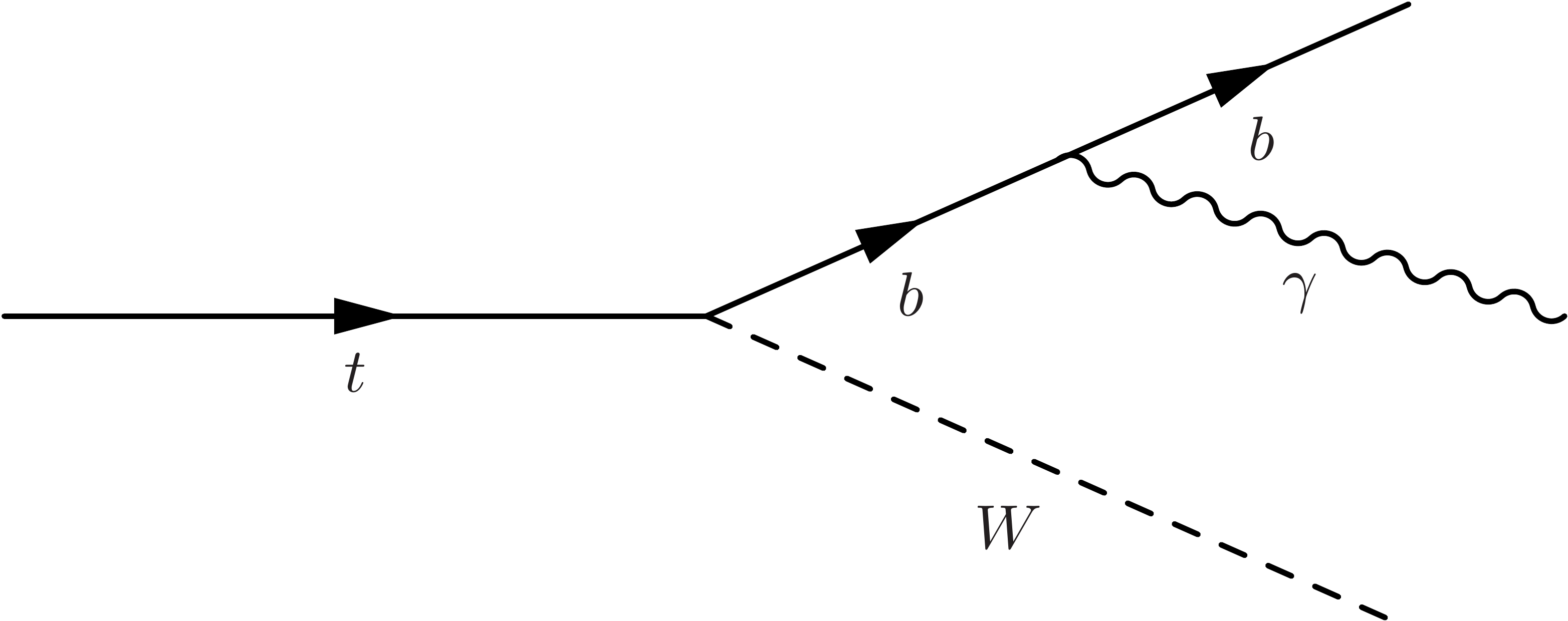}
}
\subfloat[\label{fig:rdwboson}]{
\includegraphics[width=0.37\linewidth]{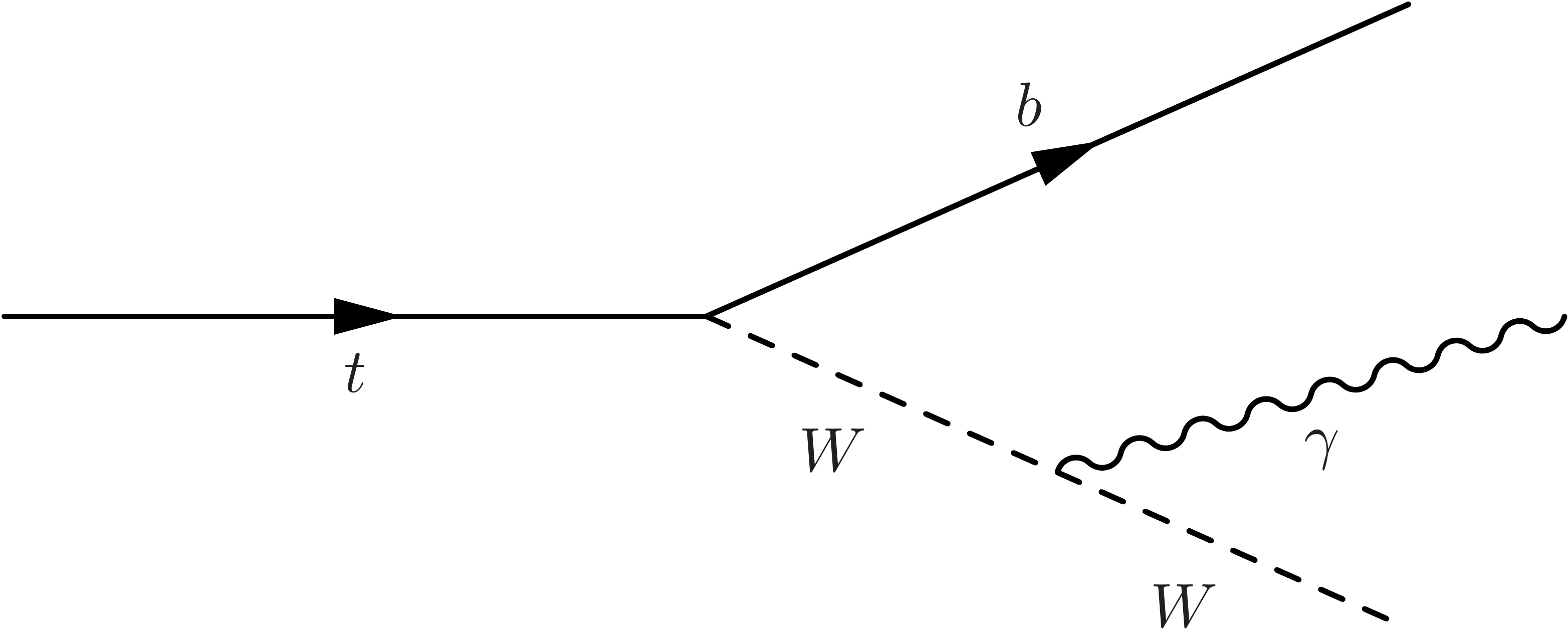}
}

\subfloat[\label{fig:rdlepton}]{
\includegraphics[width=0.37\linewidth]{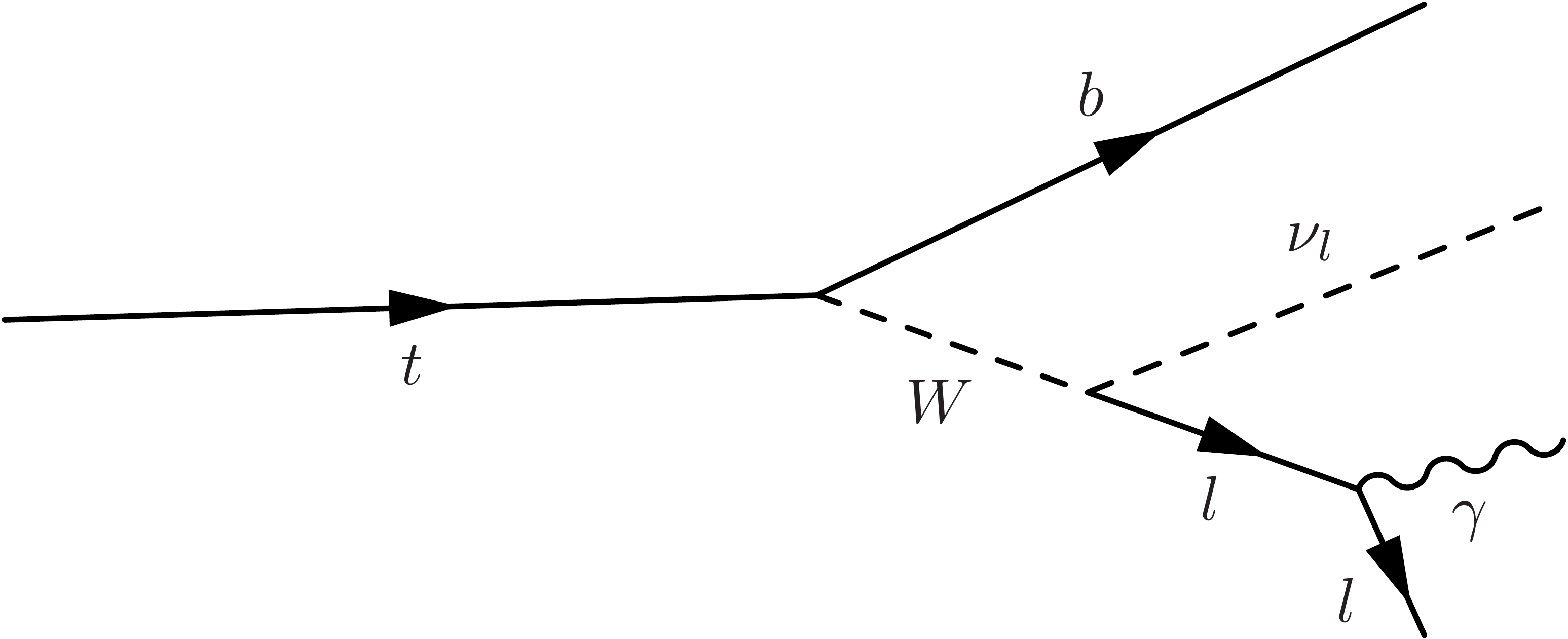}
}
\caption {Example diagrams of top quark decays where photons are radiated from charged particles. All diagrams are considered as signal.}\label{fig:rd}
\end{figure}

We can artificially construct two ``processes" in which to classify photons considered as signal. 
\begin{itemize}
\item \emph{Radiative top production}: This is the sought after signal process, specifically when photons are radiated from off-shell top quarks (Figure~\ref{fig:t1channelttgamma}, \ref{fig:t2channelttgamma}, \ref{fig:schannelttgamma}).
An interference term can also enter in the form of initial state radiation (ISR) from the incoming quarks (Figure~\ref{fig:isr}).
As mentioned previously, as the CME increases, so too does the \xsec of \ttbar production from gluon-gluon fusion. Thus, the contribution of photons radiated from the incoming quarks is only $\mathcal{O}(10\%)$.

\item \emph{Radiative top decay}: Photons are radiated from any of the charged particles from the decay of the top quark. This includes photons from $b$-quarks (Figure~\ref{fig:rdbquark}), $W$-bosons (Figure~\ref{fig:rdwboson}), and leptons (Figure~\ref{fig:rdlepton}).
Also included is the radiation from the on-shell top quark.

\end{itemize}
In all diagrams, the coupling of a quark to the photon is described by the last term in Equation~\ref{eq:eqdL}.
Distinguishing these processes is difficult (and in the case of ISR from top-quark radiation, an ill posed question) as all result in prompt photons. However, for radiative top-decay contributions we can exploit the kinematic properties and constituents of each event.
Placing cuts between leptons and photons in the $\eta$-$\phi$ plane helps reduce this contribution. This will be explained in Chapter~\ref{sec:cuts}.

\FloatBarrier

\subsubsection{Status of \ttgamma analyses}

Evidence of the \ttgamma process occurred at the CDF experiment at the Tevatron~\cite{Aaltonen:2011sp} at a CME of 1.96~\TeV. A total dataset of 6.0~fb$^{-1}$ from $p\bar{p}$ collisions was used to measure the \xsec resulting in $\sigma_{\text{\ttgamma}}=180\pm 80$~\fb. 
A ratio of the \ttbar to \ttgamma \xsec was also measured resulting in $\sigma_{t\bar{t}}/\sigma_{t\bar{t}\gamma} = 0.024\pm0.009$. Measuring the ratio allows for smaller uncertainties due to various experimental effects either cancelling out or being reduced.
Subsequent measurements have been carried out at the LHC in both the \ATLAS and \CMS experiments.

Observation at 5.3$\sigma$ was seen at the LHC by \ATLAS at $\sqrt{s} = 7 $~\TeV  with a total dataset size of 4.59~fb${^{-1}}$~\cite{Aad:2015uwa}. The measurement was carried out in the \chljets channel resulting in a fiducial \xsec, $\sigma_{\text{\ttgamma}} \times \text{BR} = 64 \pm 8 \text{(stat)}^{\ +17}_{\ -13} \text{(syst)} \pm 1 \text{(lumi)}$~\fb per lepton flavour.

\CMS performed an analysis at $\sqrt{s} = 8 $~\TeV with an integrated luminosity\footnote{The integrated luminosity is the total amount of data collected over a certain period.
The instantaneous luminosity corresponds to the incoming number of particles per second and is determined by the accelerator properties.} of 19.7~fb$^{-1}$. The fiducial measurement was done in the \chljets channel where they measured the ratio to the \ttbar \xsec,  $\sigma_{t\bar{t}}/\sigma_{t\bar{t}\gamma} = (5.2\pm1.1)\times 10^{-4}$~\cite{Sirunyan:2017iyh}. The \ttgamma \xsec is extrapolated to all \chljets final states resulting in  $\sigma_{\text{\ttgamma}}\times \text{BR} =515\pm 108 \text{(stat+syst)}$~\fb per \chljets flavour.

\ATLAS carried out another measurement at  $\sqrt{s} = 8$~\TeV in the \chljets channel with an integrated luminosity of 20.2~fb$^{-1}$~\cite{TOPQ-2015-21}. The \xsec yielded $\sigma_{\text{\ttgamma}}=139 \pm 7 \text{(stat.)} \pm 17 \text{(syst.)}$~\fb. For the first time this analysis included differential measurements of the transverse momentum and pseudorapidity of the photon. 

\begin{figure}[!h]
\centering
\includegraphics[width=0.74\linewidth]{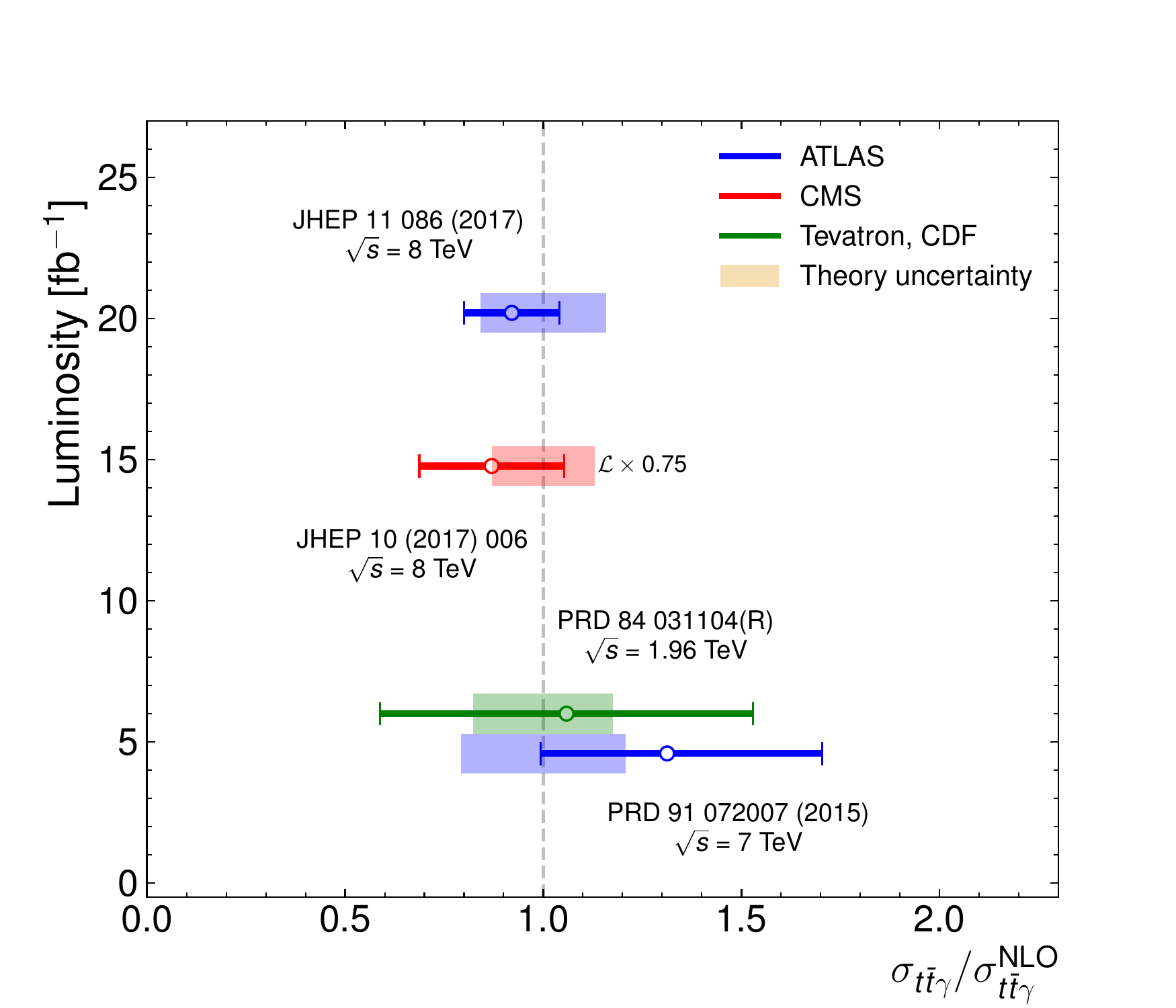}
\caption {Summary of the previous \ttgamma \xsecModifyNoun measurements performed at the Tevatron and LHC, scaled and compared to the respective theoretical predictions (represented by the dashed vertical line at one). The luminosity for the \CMS measurement is scaled by 0.75 for better readability.
The shaded bands represent the theoretical errors obtained from dedicated calculations for each analysis.
}\label{fig:ttgammaSummary}
\end{figure}

In summary, all the above measurements have been carried out in some form of the \chljets channel, and all agree with SM predictions.
The \xsecModifyNoun results scaled to the respective theoretical NLO predictions (represented by the dashed vertical line) are summarised in Figure~\ref{fig:ttgammaSummary}. 

\FloatBarrier

\section{Machine learning in high energy physics}
\label{sec:ML}

The goal of all detector components and algorithms used in high energy physics (HEP) is essentially to identify particles more efficiently with smaller uncertainties. The goal of all analyses is to examine and extract meaningful information from the data. 

Multivariate analyses (MVAs) in HEP take many inputs (called variables or features) and reduce this complexity to find new features in data. Some examples that CERN has conventionally used in the past are Likelihoods, Kalman fitters~\cite{Fruhwirth:1987fm}, Boosted Decision Trees (BDTs, which the Higgs~boson discovery of 2012 used) and neural networks (\NN{}s). The latter two belong to a class of algorithms called Machine Learning (ML).
When ML algorithms are used two steps are essential: training a model which is time and central processing unit (CPU) intensive, and inference of the model. Inference refers to the process of applying the trained algorithm to unseen data. This is a CPU in-expensive  task.
ML techniques for training (and inference) have progressed significantly in the last few years with specific emphasis on NNs. A new class, \emph{Deep neural networks} (DNNs), has large implications for the workflows in HEP\footnote{The term ``Deep" is very subjective. Essentially, it can mean many trainable parameters from hundreds of thousands to tens of millions or even larger. For this reason DNN and \NN are often used interchangeably.}. 
With the huge amount of data collected (and still to be collected) at the LHC, deeper \NN{}s are possible, thus further exploring the fundamentals of physics interactions.

Some topics in which ML can help HEP are highlighted in dedicated reports~\cite{Albertsson:2018maf,Guest:2018yhq} as well as detailed studies and research papers. A selection of some general themes (with examples where available) are explained below.

\begin{itemize}

\item (Fast) Simulation~\cite{Paganini:2017hrr}: Simulation plays a crucial role as it enables us to compare our collected data to a theory (Chapter~\ref{sec:MCsim}). The events in the simulated data must have gone through the same interactions as the real data do through the detectors. That means each particle and its interaction with matter needs to be simulated accurately. This is an incredibly CPU intensive task, especially in calorimeters where particle showers occur. An area of research is to learn these ``shower shape variables" through DNNs, or more specifically Generative Adversarial Networks (GANs)~\cite{Goodfellow:2014upx}. 
Inference of already simulated datasets could save significant amounts of CPU time.

\item Triggering: The decision to accept or reject events is made by the online and offline components of a triggering system in a detector (Chapter~\ref{sec:triggers}). Key to the success is the speed and accuracy in which decisions are made to keep only a tiny fraction of interesting events. Also key is how to define what constitutes an interesting event.
DNNs essentially have the capability of increasing the efficiency of various triggers by using all reconstructed particle information and kinematics. This could mitigate the possibility of throwing away interesting events based on a more fuller description of the constituents of the event.

\item Object (and event) reconstruction and identification~\cite{Shimmin:2017mfk,Louppe:2017ipp,Komiske:2016rsd,Kagan:2016wnu,Barnard:2016qma,deOliveira:2015xxd,Stoye:DLPS2017,PhysRevD.94.112002,ATL-PHYS-PUB-2017-004}: 
A large proportion of DNN studies focus around jet physics. This involves studying the quark/gluon jet substructure that leaves different signatures in calorimeters, an ideal problem for Recurrent Neural Networks (RNN) with Long Short Term Memory (LSTM) units or Convolutional Neural Networks (CNN). These algorithms have originally been developed for the field of natural language processing and image recognition.
This thesis makes use of novel techniques for object identification (Chapter~\ref{sec:PPT}) and event classification (Chapter~\ref{sec:ELD}).

\item Uncertainty mitigation~\cite{Louppe:2016ylz}: An interesting area of research involving NNs it to mitigate the systematic uncertainties that arise in any HEP analyses. 
The key lies in ``pivoting" between sensitivity or robustness in a classification problem. This introduces a novel technique using GANs.

\item Anomaly detection~\cite{Wielgosz:2017yrb}: 
From the nominal operation of each component of a detector, to storing massive amounts of data on hardware, everything needs constant monitoring.
In the case of failure an appropriate response needs to be carried out. Conventionally this is done to some degree by electronic sensors, but ultimately always ends up at an ``expert" taking action. By detecting or even forecasting anomalies based on enormous amounts of data, algorithms could detect and fix, or even just notify the correct expert significantly faster than humans.
Another type of anomaly detection applies to physics processes. By training a \NN on simulation (which is a theory such as the SM) we create output distributions. These outputs are non-linear transformations and so can represent non-trivial properties that are otherwise hidden. By inferring on unseen collected data, assumptions can be made on the validity of the theory.
 
\end{itemize}
The above points generally rely on some well simulated datasets that provide the labels that are used in training. In cases where labels are not available (for example training on actual data) or when the labels are not implicitly trusted, novel methods have been developed~\cite{greedywise1,greedywise2,Metodiev:2017vrx,Dery:2017fap}.
While some of these techniques are still in early development, we can imagine the advantages that DNNs without labels may provide. An example would be not telling the network what to look for, but rather looking for patterns that occur naturally in the data, which can be interpreted as re-discovering fundamental physics with very little or no bias to a given theory. 

\subsection{Core concepts}

A brief introduction to some core concepts, common algorithms, and architectures used in this thesis will be presented here.

The typical goal of an NN algorithm is to search for a function which maps a given space or set $\mathbf{X}$ (which are some observed data points with associated features) to lower dimensional space $\mathbf{Y}$ (which are the target labels). In doing so, a loss function or cost function is optimised.  

The transformation of an input variable ($\boldsymbol{x}$) into a hidden state ($\mathbf{h}_{i}$) is expressed as 
\begin{equation}
\boldsymbol{h}_{i+1} = g_i(W_i \boldsymbol{h}_i + \boldsymbol{b}_i),
\end{equation}
where $i$ represents the $i$th transformation. For the first transformation, $\boldsymbol{h}_{0}=\boldsymbol{x}$.
$W_i$ is called a weight matrix and $\mathbf{b}_i$ is the bias vector. These are the parameters that are updated in the process of optimising a loss function, the so-called training of the NN.
For this thesis, weight matrices and bias vectors are initialised uniformly in the range $[0,1]$.
 $g_i$ is called the activation function. This is often a non-linear function with the only requirement being that it needs to be differentiable. This thesis makes use of three different activation functions;
\begin{itemize}
\item ReLu: $g(z) = max(0,z)$
\item Sigmoid: $g(z) = \frac{1}{1+e^{-z}}$
\item Softmax: $g(z) =\frac{e^{z_{j}}}{\sum_{k=1}{e^{z_{k}}}} $, for $j$ number of inputs to final layer, with $k$ final outputs.
\end{itemize}
For all given weights and biases we want an algorithm that describes $\boldsymbol{y}$ for all inputs $\boldsymbol{x}$. In this thesis, ideally, signal and background events are classified such that
\begin{equation}
  y=\left\{
  \begin{array}{@{}ll@{}}
    0, & \text{if}\ \text{background} \\
    1, & \text{signal.}
  \end{array}\right.
  \end{equation}
Thus, \emph{binary cross entropy} is used as the loss function and defined as:
\begin{equation}
\label{eq:lossfunction}
\text{Loss}(\boldsymbol{y},\boldsymbol{p}) = - {\boldsymbol{y}} \text{log}( \boldsymbol{p}) + (1-\boldsymbol{y}) \text{log}(1-\boldsymbol{p}), 
\end{equation}
where $\boldsymbol{y}$ are the given labels and $\boldsymbol{p}$ are the predictions made by the network.
The loss becomes small (close to 0) when $\boldsymbol{y}$ is roughly equal to the output, $\boldsymbol{p}$. 
The training stage of a \NN involves minimising the loss function using gradient descent. 
The significant advancements in ML that we have seen since the 1990's is largely due to the invention of \emph{back-propagation}~\cite{1986Natur}. 
By taking partial derivatives, weights and biases in all hidden layers are able to be updated with respect to the loss function in significantly less time. This allows for the training of deeper networks.

Overtraining occurs when the \NN has learned too much about the training dataset. Generalisation to unseen data could result in artificial trends emerging based on the learned fluctuations in the training set.
Two regularisation methods in the form of layers are used to reduce any overtraining that may occur (in addition to the general $k$-fold cross validation checks presented later).
\begin{itemize}
\item Batch Normalisation~\cite{BN}: During training, the distribution of inputs at each layer changes as the parameters of the previous layer change. This leads to slower training as the overall learning rate is reduced due to parameter initialisation. It is also difficult to train models with saturating non-linearities (such as for a sigmoid or $tanh$ activation layers). This is referred to internal covariate shift. The solution is surprisingly simple and effective: normalise each batch so that it has a mean of 0 and a standard deviation of 1.
\item Dropout~\cite{dropout}: A dropout layer sacrifices some performance for the generalisation of the \NN. During training, a percentage (supplied as a hyper-parameter in which to optimise) of neurons are evicted from the training epoch\footnote{An epoch refers to one full cycle of training on the training dataset.}. This ensures the \NN does not become too dependent on specific neurons, which can lead to overtraining. Essentially what this technique is doing is training many slightly different \NN{}s  simultaneously.
\end{itemize}

\subsection{Tools and libraries}

An underlying theme in this thesis is the use of industry standard, open-source machine learning libraries in \ATLAS. While this may not sound profound, it is. It represents a large change to the workflow of how ML is used within HEP at CERN, specifically within \ATLAS. By using modern libraries such as TensorFlow~\cite{tensorflow2015-whitepaper}, Theano~\cite{2016arXiv160502688short} and  Keras~\cite{chollet2015keras}, new approaches to ML can be achieved making the ideas presented above possible.

This thesis presents work on an analysis that was one of the first studies to use modern \NN libraries on \ATLAS data and not on phenomenological datasets that most of the studies presented above use. 
It presents a challenge because the \ATLAS software stack is large (about 6.5 million lines) with the vast majority written in C++. It was built at a time when machine learning in HEP was not foremost in people's minds. 
Modern ML libraries generally have python (and only much more recently, C++) interfaces, which would need significant restructuring of our code to be able to use.
For this reason ROOT~\cite{Brun:1997pa} (the standard ``swiss army knife" of tools in HEP) and the TMVA library~\cite{Hocker:2007ht} have been popular in HEP, simply because the code was designed around ROOT data structures. 

ROOT and TMVA were developed at CERN and are maintained by developers based at CERN. 
However, there are good reasons to use industry standard (yet open-source) libraries.
For instance, at time of writing, TensorFlow and Keras have around 1600 and 700 contributors to their respective codebases since their ``founding" in 2015.\footnote{Taken from their respective Github code repositories.} In comparison, ROOT has just over 200 contributors spanning over about 18 years\footnote{Taken from \url{www.openhub.net}.}.
A larger user-base ensures access to cutting edge algorithms, better documentation and more helpful forums.

A strategy that has propelled the use of ML in HEP is based on the premise that training and inference of a \NN are two steps that are completely independent of each other. The training stage can take place in any environment, often making use of graphical processing units (GPUs), which considerably speeds up the training of complex networks.
Inference has more restrictions, and in the case of \ATLAS needs to have a low CPU footprint and be written in C++. 
The analysis presented in this thesis makes extensive use of Lightweight Trained Neural Network (\lwtnn)~\cite{lwtnn}, which was created to bridge the gap between training using modern ML libraries, and inference of the \NN within \ATLAS (and CERN in general). It has enabled rich ML studies using novel techniques that are coming to fruition.


\chapter{The \LHC and the \ATLAS experiment}
\label{sec:atlasexp}

The European Organisation for Nuclear Research (CERN) is currently the largest scientific laboratory in the world. CERN is a high energy particle physics laboratory founded in 1954, where many important discoveries have been made. These include (among others) the discovery of neutral currents in 1973~\cite{Gargamelle}, $W$- and $Z$- bosons~\cite{Arnison:1983rp,Arnison:1983mk}, direct charge-parity (CP) violation~\cite{Christenson:1964fg} and the Higgs boson in 2012~\cite{CMSHiggs,HIGG-2012-27}.
Various accelerators have been commissioned at CERN with the latest being the Large Hadron Collider (LHC). This chapter discusses the LHC and describes the ATLAS detector which sits on one of the four main interaction points (IP) on the LHC ring. 

\section{The Large Hadron Collider}

\begin{figure}[!htbp]
\centering
\includegraphics[width=0.85\linewidth]{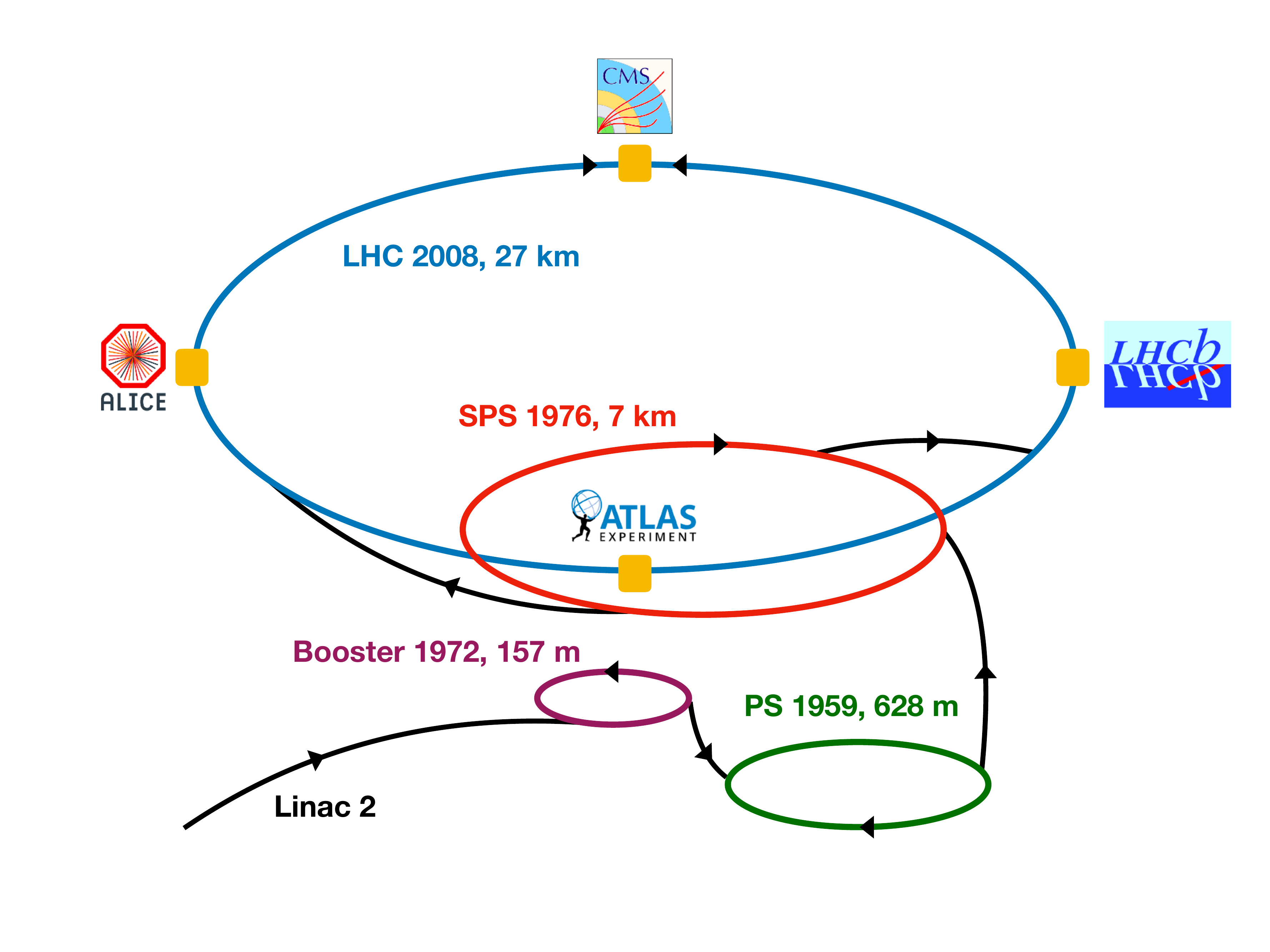}
\caption {A simplified view of the CERN accelerator complex. Interaction points are shown as yellow squares. Each element in the chain serves to accelerate particles (black arrows) to higher energies, culminating at the LHC. This schematic applies to protons being injected into the LHC. For heavy ions, a slightly different initial injector scheme is used, which is not shown.}
\label{fig:CERNComplex}
\end{figure}

Figure~\ref{fig:CERNComplex} shows the schematics for how particles are injected into the LHC.
The protons originate from a bottle of hydrogen where the electrons are stripped off and injected into the Linear accelerator 2 (Linac 2). They are accelerated to an energy of 50~MeV where they then enter the Booster and are accelerated up to 1.4~GeV. From there the protons enter the Proton Synchrotron (PS) where they are accelerated up to 25~GeV within a ring with a circumference of 628~m. The protons then enter the Super Proton Synchrotron (SPS), which is the second largest accelerator at CERN. It has a circumference of 7~km and was first switched on in 1976. It was at this accelerator that the W and Z bosons were discovered. Protons are accelerated up to 450~\GeV before they are injected into the final accelerator: the LHC.

The LHC~\cite{LHC} is 27~km in circumference and crosses the border of Switzerland and France in multiple places. It is the final stage in the chain of accelerating particles to the design centre of mass energy ($\sqrt{s}$) of 14~TeV. It is made up of eight straight segments and eight arcs which contain dipole magnets to keep the particles in a circular orbit. There are 1232 dipole magnets and 392 quadrupole magnets that steer the beam along its 27~km trajectory. Special focusing magnets squeeze the beam at four IPs to induce collisions. The LHC was designed to accommodate 2808 proton bunches per beam (with $10^{11}$ protons in a bunch) and with a spacing of 25~ns per bunch. This means a peak instantaneous luminosity of around $1.2 \times10^{34} ~\text{cm}^{-2}\text{s}^{-1}$, which translates to just under 1 billion collisions per second.

A detector sits at each of these IPs. There are the two general purpose detectors: 

\begin{itemize}

\item \ATLAS (A large Toroidal LHC ApparatuS)~\cite{PERF-2007-01} is a general purpose detector designed to probe many aspects of particle physics from ``bump hunting" (detection of new particles) to SM precision measurements, to Beyond the SM (BSM) and supersymmetry (SUSY) searches. It is the largest HEP detector in the world. \ATLAS will be described in detail in Section~\ref{sec:ATLAS}.
\item \CMS (Compact Muon Solenoid)~\cite{CMS} is a second general purpose detector with the same physics goals as \ATLAS. The redundancy is needed to ensure that discoveries are verifiable. While the physics goals are the same, the detector itself and the software and algorithms are different to \ATLAS.
\end{itemize}

There are two other detectors with more specific aims\footnote{There are multiple smaller experiments that branch off from the LHC, but these four detectors are the only ones that sit at beam IPs.}: 

\begin{itemize}
\item ALICE (A Large Ion Collider Experiment)~\cite{ALICE} studies the properties of the universe an infinitesimal time after the Big Bang. Specifically, it studies the quark-gluon plasma that forms under conditions of extremely high temperatures and densities. They require heavy nuclei to be collided such as lead-lead, proton-lead or even Xenon atoms. Special runs of the LHC are requested where protons are switched to heavy ions.

\item LHCb (LHC beauty)~\cite{LHCB} focuses on $b$-physics (hadrons that contain $b$~quarks) which play an important role in probing the asymmetries of nature. It could answer questions like the imbalance of matter versus anti-matter in our universe. LHCb takes advantage of the production asymmetry of the signatures of interest and only has detectors in a more forward region of one of the beam directions.

\end{itemize}

\section{The \ATLAS detector}
\label{sec:ATLAS}

\begin{figure}[!htbp]
\centering
  \begin{minipage}{0.9\textwidth}
      \begin{overpic}[width=\linewidth]{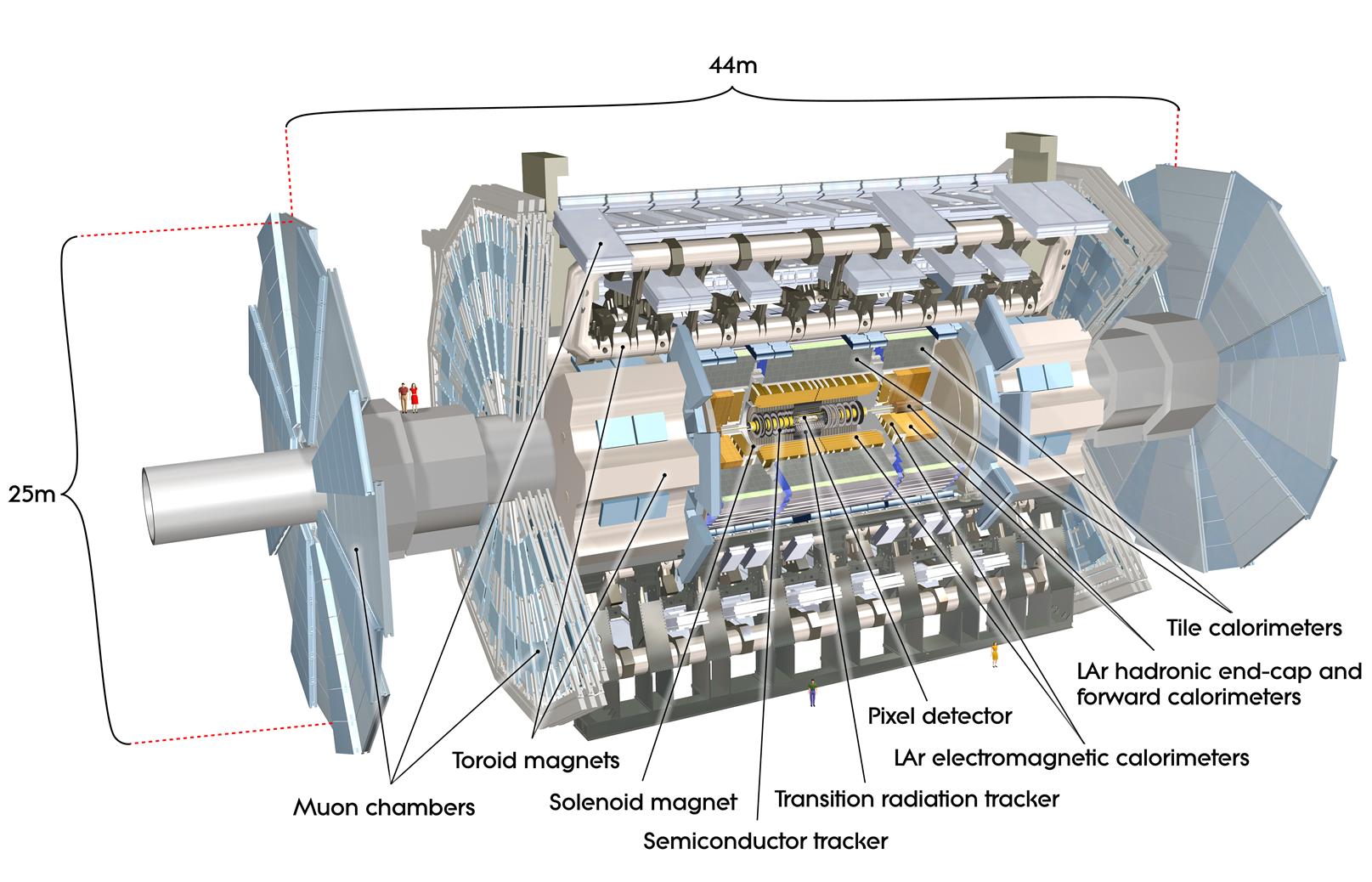}
      \put(75,1){\tiny\fontfamily{phv}\selectfont ATLAS Experiment \textcopyright~2018 CERN }
    \end{overpic}
  \end{minipage}
\caption {An overview of the \ATLAS detector (including humans for scale) with some of the sub-detectors labelled.}
\label{fig:atlas}
\end{figure}

The \ATLAS detector, shown in Figure~\ref{fig:atlas}, is 25~m tall and 44~m long, weighing about 7000~tons. It is symmetric and is made up of many sub-detectors. 
The IP is located at the centre of the detector, which defines the origin of the coordinate system used henceforth. The beam axis is defined by the $z$-axis with the $x$-$y$ plane transverse to the beam axis. The polar angle, $\theta$ is measured from the beam axis and is often expressed in terms of the pseudorapidity\footnote{This quantity is preferred over $\theta$ since the differences in pseudorapidity are Lorentz invariant under boosts along the beam axis. The relationship can be shown as: 

\begin{tikzpicture}[scale=0.4]
    \draw [<->,thick] (0,2) node (yaxis) [above] {$\theta = 90^{\circ}$, $\eta = 0$}
        |- (2,0) node (xaxis) [right] {$\theta = 0^{\circ}$, $\eta = \infty$};
    \draw (0,0) coordinate (a_1) -- (1.5,1.5) coordinate (a_2) node (middle) [right] {$\theta = 45^{\circ}$, $\eta = 0.88$};
\end{tikzpicture}
}, $\eta = -\text{ln tan}(\theta/2)$, while the azimuthal angle, $\phi$ is measured around the beam axis. A cone $\Delta R = \sqrt{\Delta \eta^2 + \Delta \phi^2} $, can be expressed as the angular distance in the $\eta-\phi$ plane. Quantities such as the transverse momentum (\pt), transverse energy (\et) and missing transverse energy (\met) are expressed in the $x$-$y$ plane.

\begin{figure}[!htbp]
\centering
  \begin{minipage}{0.67\textwidth}
      \begin{overpic}[trim={2.8cm 1.1cm 1.2cm 0.5cm},clip,width=\linewidth]{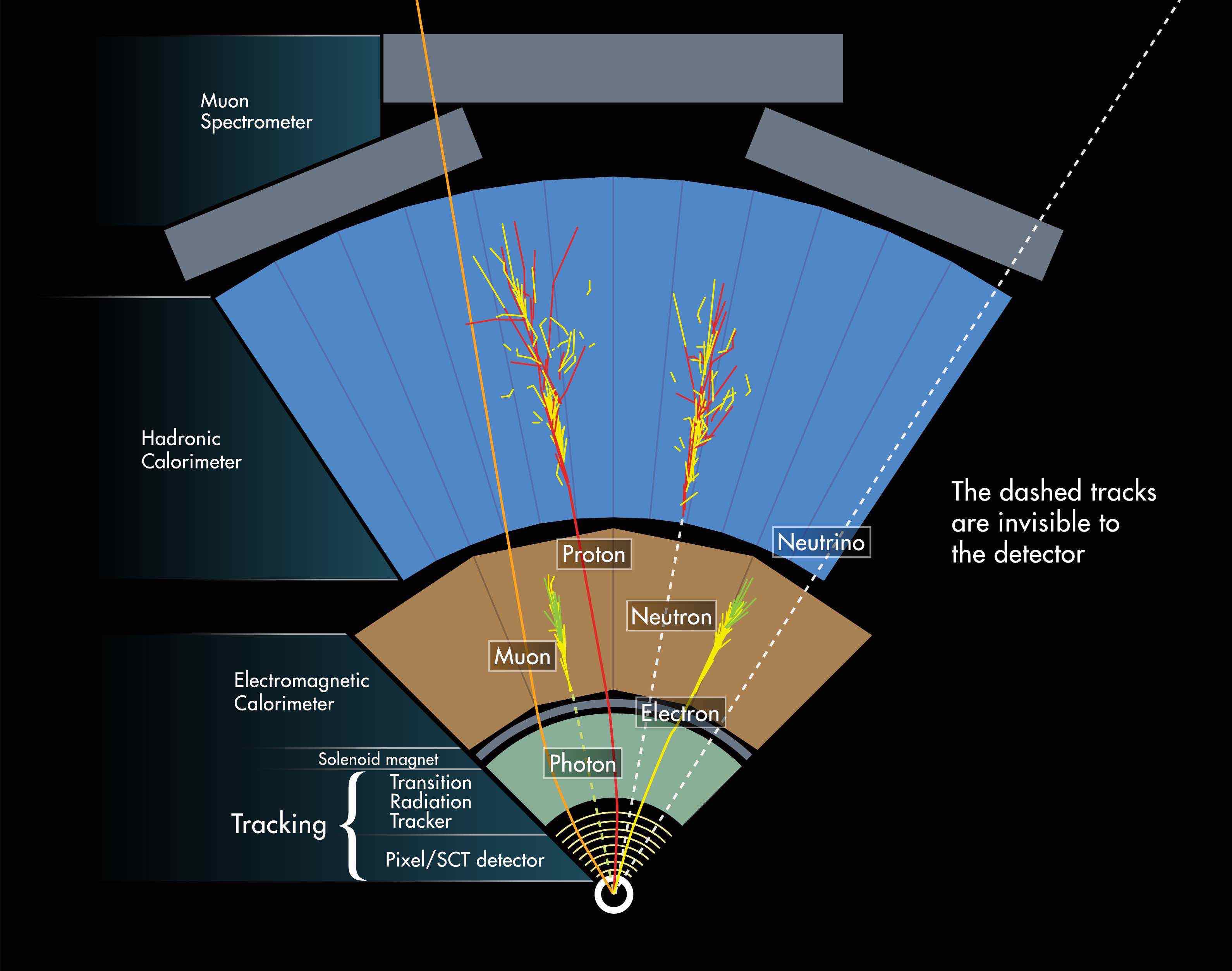}
      \put(68,1){\tiny\fontfamily{phv}\selectfont \textcolor{white}{ATLAS Experiment \textcopyright~2018 CERN} }
    \end{overpic}
  \end{minipage}
\caption {A cross section of the \ATLAS detector showing the main detector components and the different points at which particles interact with the detector. Deflection or curvature of particles due to the solenoid magnet are also shown. Deflection or curvature due to the toroid magnet are in and out of the page and so can not be seen.}
\label{fig:atlas_cross_section}
\end{figure}

Figure~\ref{fig:atlas_cross_section} shows a cross section of the \ATLAS detector and how it can be split into four main components (each of which have further sub-components). The inner detector (ID) is largely responsible for tracking. With millions of read-out channels and thus high resolution, this detector provides a snapshot of tracks for charged particles. Ionising gases or semiconductor material is used in these detectors. \ATLAS has two critical magnet systems.  The first surrounds the inner detector with a homogeneous 2~T magnetic field to detect charge type and particle momentum.
The second critical magnet is the large toroidal magnet for which \ATLAS is named. This toroidal magnet provides an in-homogeneous field of between 0.5 and 4~Tesla.
The electromagnetic (EM) calorimeters measure energy from photons and electrons due to their interactions with matter. Pair production and bremsstrahlung are such examples. The hadronic calorimeters measure the energy deposits of heavier particles (such as protons and neutrons) due to their interaction with matter.
Muons are minimal ionising particles and so are able to traverse the entire detector. Thus, the muon detectors encapsulate the other detectors and form the outer layer. 
The interaction of neutrinos with matter is negligible. For this reason neutrinos pass through the detector and are reconstructed as ``missing transverse energy" (\met), which is possible due to energy conservation.
These components are described further in the next sections.

\subsection{Magnets}
\label{sec:magnets}

\begin{figure}[!htbp]
\centering
  \begin{minipage}{0.6\textwidth}
    \begin{overpic}[width=\linewidth]{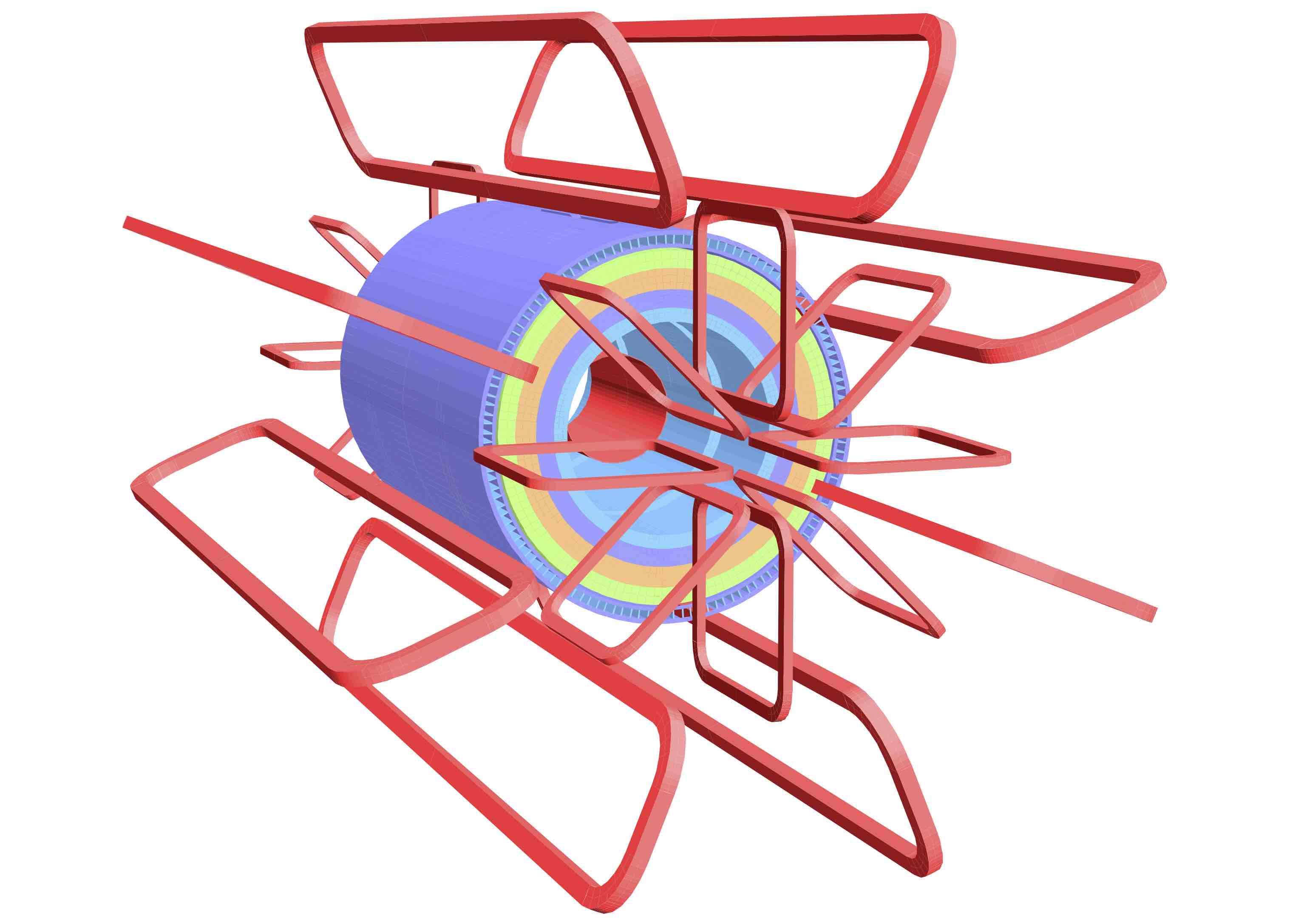}
      \put(75,1){\tiny\fontfamily{phv}\selectfont ATLAS Experiment \textcopyright~2018 CERN }
    \end{overpic}
  \end{minipage}
\caption {The solenoid magnet (the last layer on the cylinder) surrounds the ID. The eight air-coils for the toroid magnets as well as the end-caps are also shown.}
\label{fig:magnets}
\end{figure}

Charged particles feel a force due to an electric and magnetic field (the Lorentz Force). Their bending radius is proportional to their momentum. Thus, magnets are essential to determine the momentum and charge type of various particles. \ATLAS has two magnet systems which influenced much of the design for the rest of the experiment. These are the solenoid and toroid (with separate end-caps) magnets. Figure~\ref{fig:magnets} shows the geometry of the magnet systems. At the centre lies the ID (the first four layers) which is surrounded by the solenoid magnet. The eight toroid coils as well as the end-cap coils are also shown. 

The solenoid magnet has an outer diameter of 2.56~m and a length of 5.8~m and has its conductive windings between the inner and outer layer of width 12~mm. Care had to be taken to reduce the amount of material before the calorimeter, yet remain steady and lightweight.
The conductive coils are made from Niobium-titanium (NbTi), which is reinforced with an aluminium alloy. 
The solenoid fully encompasses the ID and provides a homogeneous 2~T field, while only contributing a total of 0.66 radiation lengths, $X_{0}$\footnote{The radiation length is a property of a material defined as the length at which a charged particle loses 0.368 (or $1/e$) of its energy when traversing the material.}. 
The toroid magnet system consists of a barrel and end-cap segments. The barrel components consists of eight coils encased in circular stainless-steel vacuum tubes. The overall barrel toroid length is 25.3~m while the outer diameter is 20.1~m. 
The net Lorentz force acting on each coil is approximately 1400~tonnes, directed inwards. The structural deflection of the barrel toroid had to be taken into account in the construction of magnet infrastructure. 
The end-cap toroids consists of eight flat, square coil units and eight wedges. These are joined to form a structure that can withstand a Lorentz force of 250~tonnes. Both the barrel and the end-caps toroids consist of pure Al-stabilised Nb/Ti/Cu conductor.

\subsection{The Inner Detector}
\label{sec:ID}

\begin{figure}[!htbp]
\centering
\subfloat[\label{fig:ID1}]{
\includegraphics[width=0.7\linewidth]{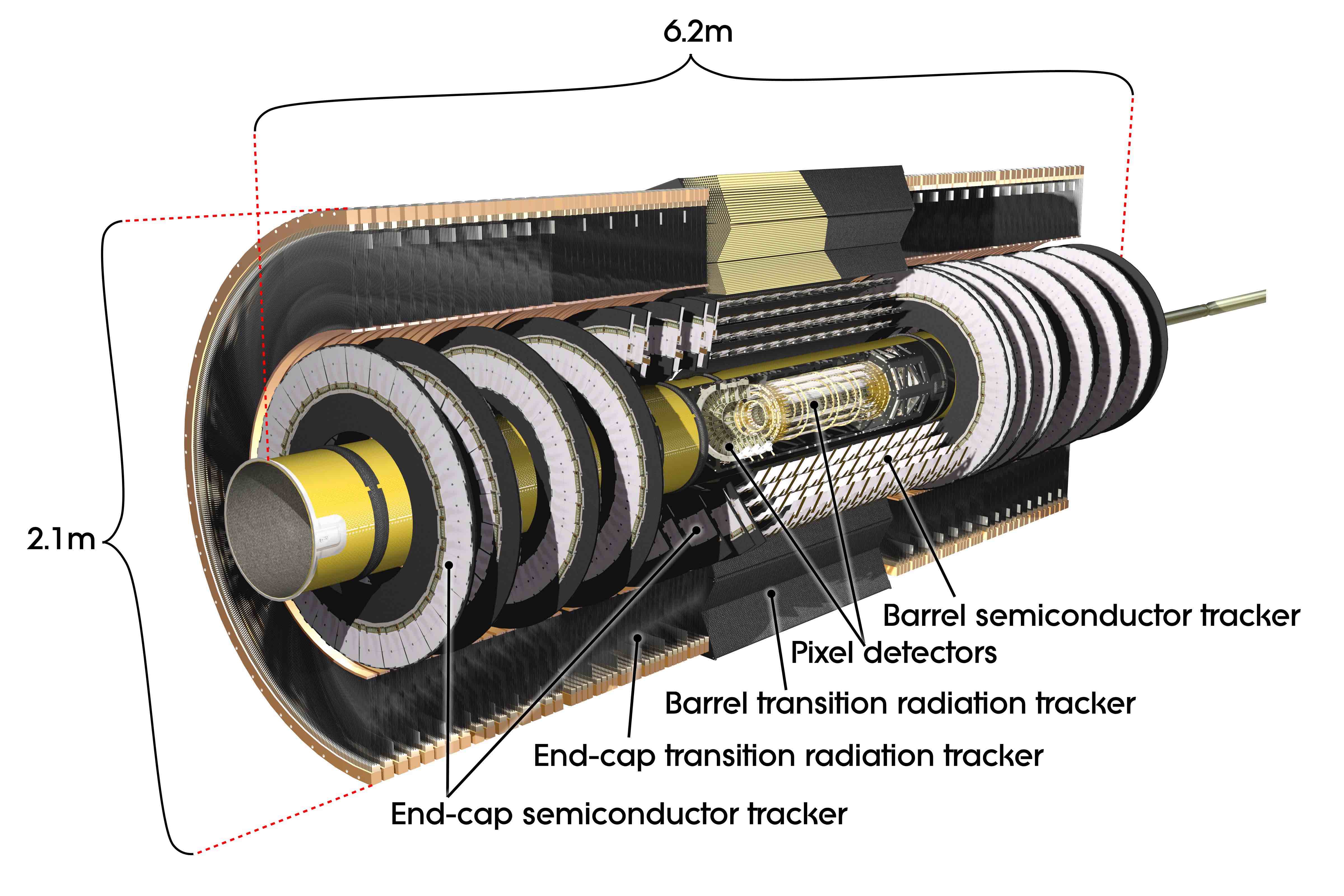}
}

\subfloat[\label{fig:ID2}]{
\includegraphics[width=0.9\linewidth]{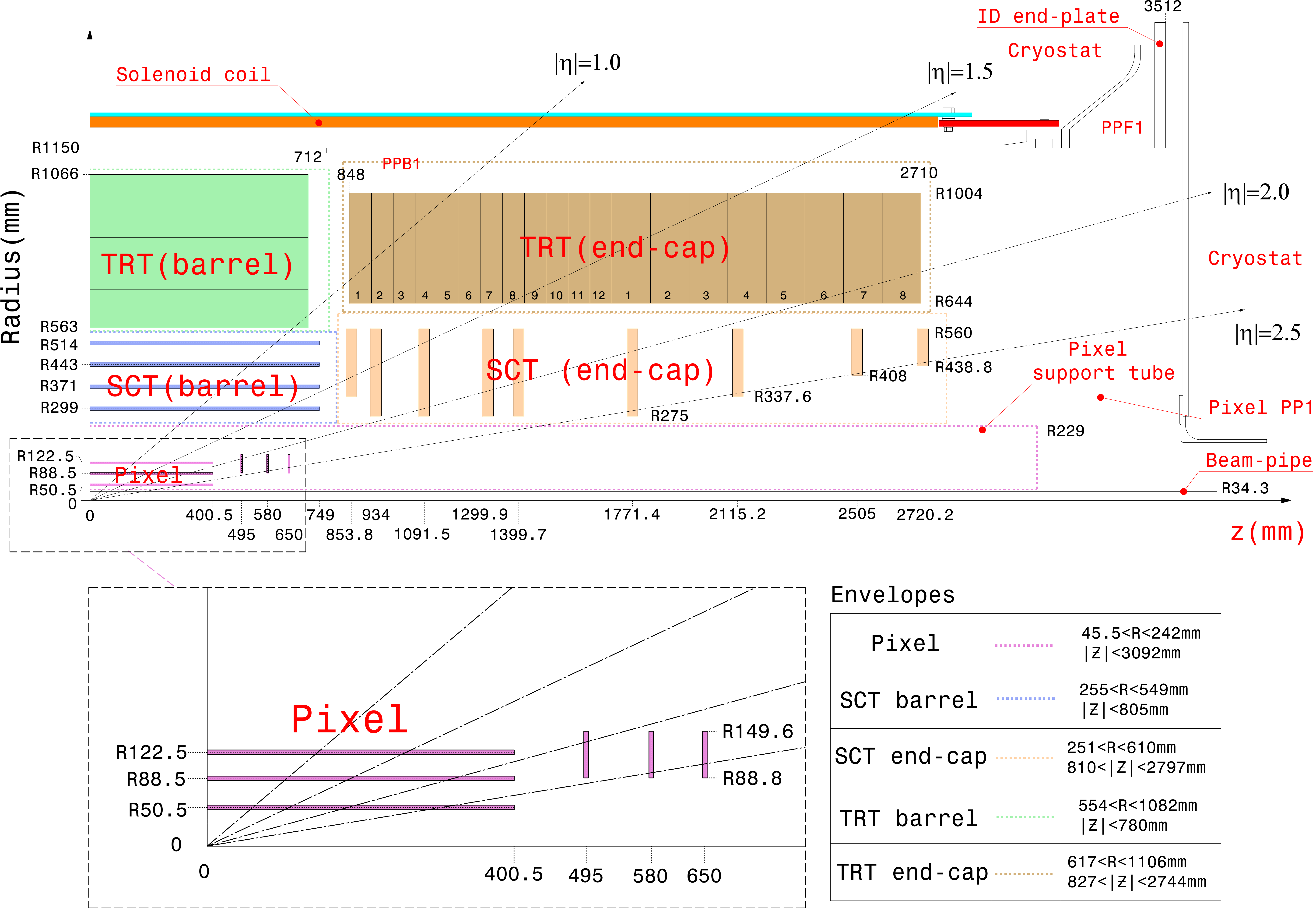}
}

\caption {a) An overview of the \ATLAS ID with the different components. b) Quarter-slice of the ID showing the main detectors and dimensions. Also shown is the solenoid magnet. 
Not shown is the IBL which was added during the first long shutdown of the LHC in 2013. It would be shown in b) at a radius of $R=33.2$~mm from the beam-pipe~\cite{PERF-2007-01}.}
\end{figure}

Three sub-components make up the Inner Detector (ID), all placed within a homogeneous 2~T magnetic field provided by the solenoid. This allows for accurate determination of charged particles' momenta.
By order of distance from the beam-pipe these are the Pixel Detector, the Semiconductor Tracker (SCT) and the Transition Radiation Tracker (TRT). Figure~\ref{fig:ID1} shows the location of each component. All components have a barrel and end-cap elements. Not shown is the Insertable B-Layer (IBL), which is part of the pixel layer and was added after Run~1.
A crucial consideration in designing the ID was to minimise the material used in the construction, and thus ensure particles reach the calorimeters. Depending on the value of $|\eta|$, the radiation length varies from about 0.5~$X_{0}$ (for around $|\eta|<0.7$) to approximately 2.5~$X_{0}$, the densest material being due to the barrel and end-cap infrastructure plates at approximately $|\eta| = 1.5$ and $ |\eta | = 2.7$.
Figure~\ref{fig:ID2} shows a more detailed schematic of the ID, which includes these end-plates. Also shown are radii for each component from the beam-pipe, as well as the lengths of the detectors in the $z$-plane.
The resolution for the ID as a function of the transverse momenta is $\sigma(\pt)/\pt\approx 0.05\% \pt [\text{GeV}]\oplus 1\%$.

\subsubsection{Pixel detectors}
The Pixel detectors (based on silicon semiconductor technology) provide high resolution vertex measurements for charged particles within $|\eta | < 2.5$ and consist of more than 80~million pixels. Originally it consisted of three layers but after Run~1, \ATLAS installed an Insertable B-Layer (IBL)~\cite{Capeans:1291633}, which contains smaller pixels and thus enhances the capabilities of detecting $b$-tagged jets.
The IBL is placed 33.2~mm away from the beam-pipe, with the three original layers placed at 50.2~mm, 88.5~mm and 122.5~mm from the beam-pipe, as shown in Figure~\ref{fig:ID2}.
The older three layers and disks of the Pixel Detector modules have an accuracy of $10~\micro$m in the $R$-$\phi$ plane and $115~\micro$m in the $z$ or radial plane for single hits. The IBL is accurate to $8~\micro$m in the $R$-$\phi$ plane and $40~\micro$m in the $z$ direction.

\subsubsection{Semiconductor Tracker}
The SCT contributes to high precision measurements of track impact parameters and has a coverage of $|\eta | < 2.5$. It functions similar to the Pixel detectors in that it is based on silicon semiconductor technology. It is composed of four cylindrical barrel layers and 18 planar end-cap discs.
A slight tilt and overlaying sensors allows for 2D measurements of particle hits.
The SCT barrel and disk modules have an accuracy of $17~\micro$m in the $R$-$\phi$ plane and $580~\micro$m in the $z$ or radial plane.

\subsubsection{Transition Radiation Tracker}
The TRT is made up of 4~mm diameter tubes of polyimide, cut to 144~cm for the barrel and 37~cm for the end-caps. Around 73 layers in the barrel and 160 straw planes in the end-cap provide high track curvature measurements through transition radiation up to a coverage of $|\eta | < 2.0$. The TRT is also able to discriminate between pions and electrons due to high thresholds on the signal energy.
The barrel straws are placed parallel to the beam axis, while the end-cap disks place the straws perpendicular to the beam-axis.
The straws are filled with a Xe/CO$_{2}$/O$_{2}$ gas mixture and the anodes consist of tungsten wires plated with gold.
The TRT modules have an electron collection time of approximately 48~ns and a drift-time accuracy of around $130~\micro$m.

\subsection{Calorimeters}
\label{sec:calorimeters}

\begin{figure}[!htbp]
\centering
\includegraphics[width=0.85\linewidth]{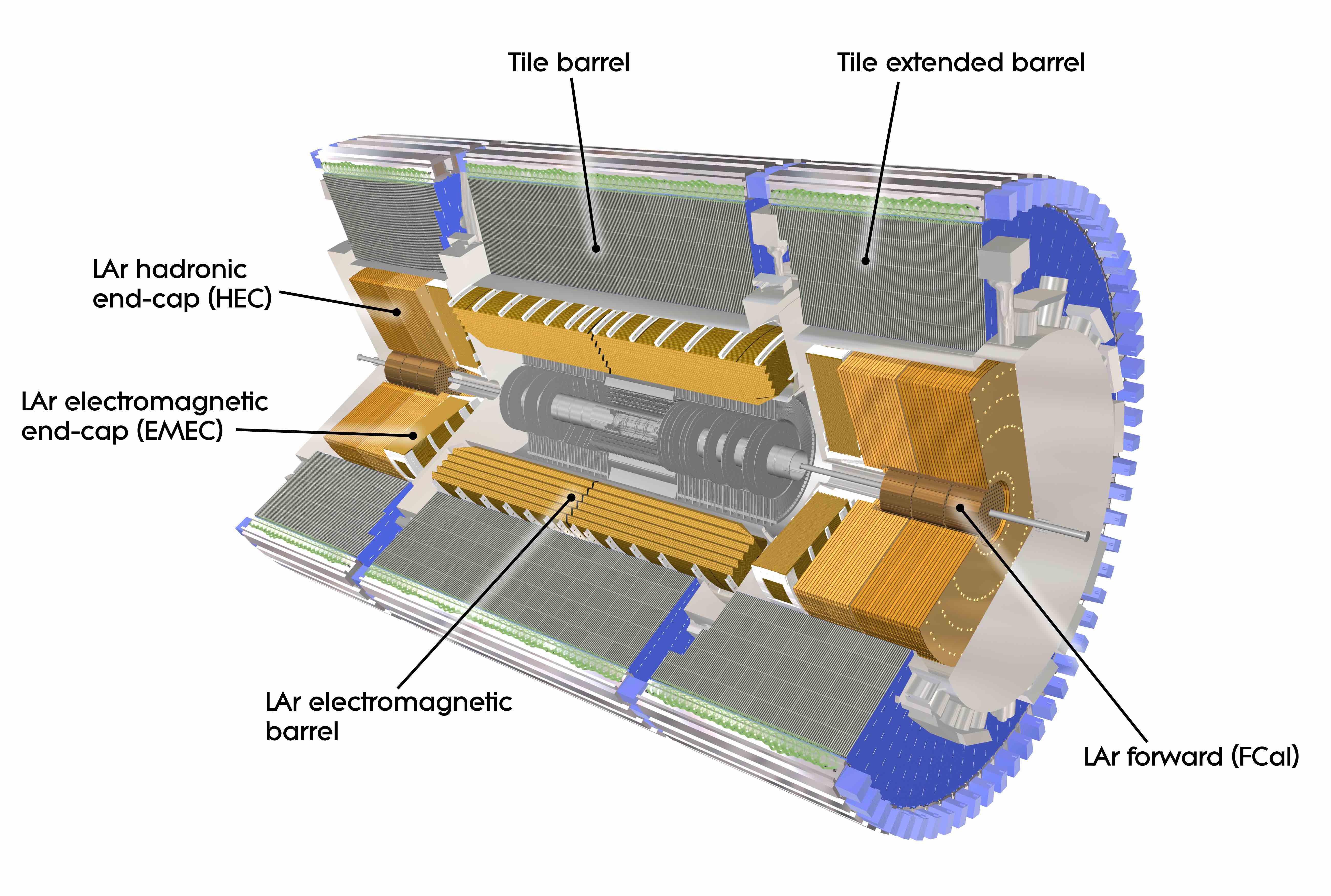}
\caption {The layout of the electromagnetic and hadronic calorimeters in the \ATLAS detector~\cite{PERF-2007-01}.}
\label{fig:calorimeters}
\end{figure}

Figure~\ref{fig:calorimeters} shows the layout of the EM and hadronic calorimeters in \ATLAS. Both have full $\phi$ coverage around the beam axis with a total pseudorapidity coverage of $|\eta| < 4.9$. The EM calorimeters consist of a barrel (EMB) and end-cap (EMEC) components. 

\begin{figure}[!htbp]
\centering
\includegraphics[width=0.68\linewidth]{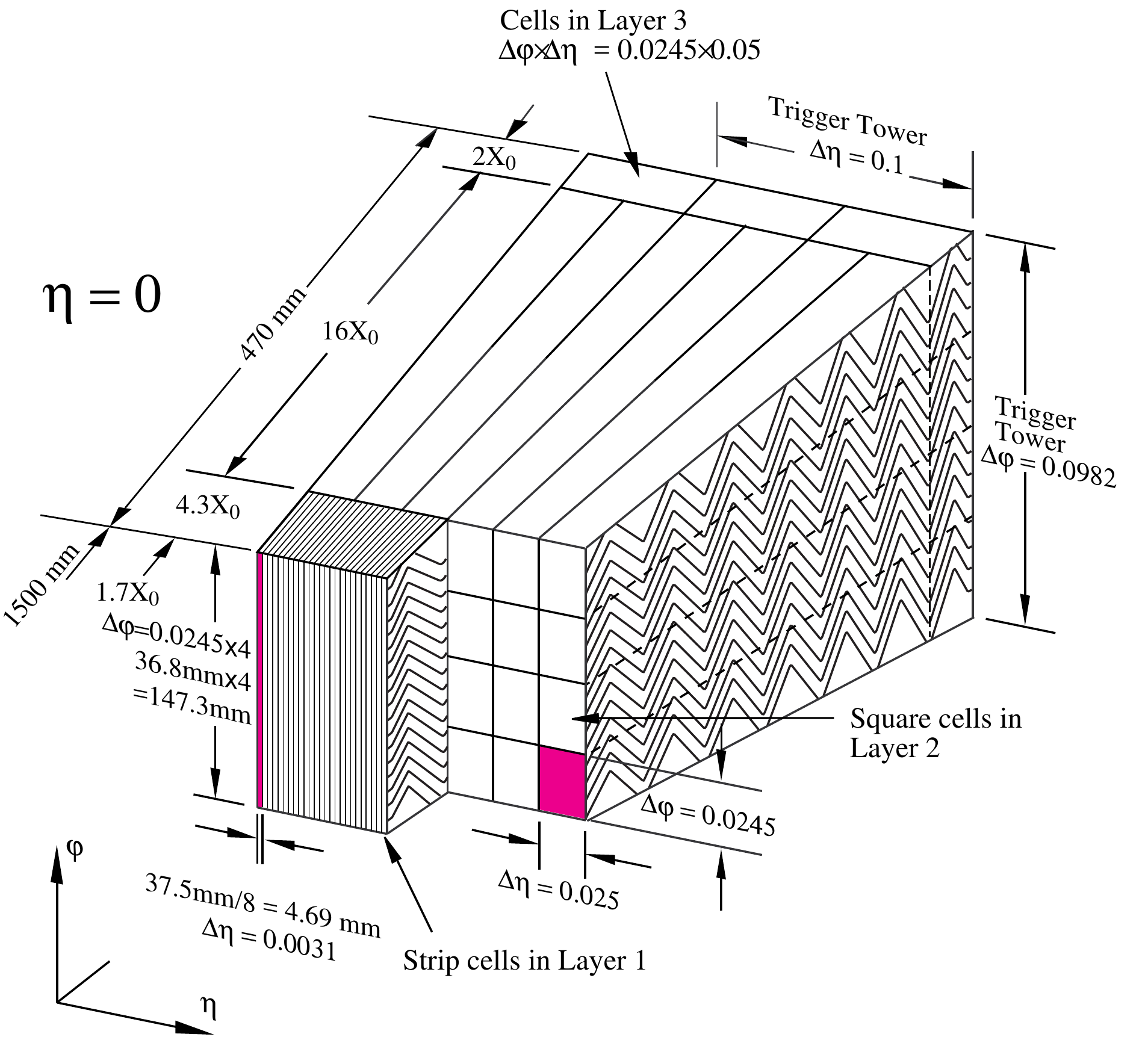}
\caption {A sketch of a barrel module for the electromagnetic calorimeter. Shown are the three layers with the $\eta$-$\phi$ resolution and radiation lengths~\cite{LAR_TDR}.}
\label{fig:LAR}
\end{figure}

The hadronic calorimeter sits behind the EM calorimeters and includes the central barrel part and extended barrel part (the Tile calorimeter). Liquid Argon (LAr) hadronic end-cap calorimeters (HEC) and LAr forward calorimeters (FCal) also form part of the overall hadronic calorimeter, which is most suited to \met measurements and jet reconstruction.
The EMB has a total coverage of $|\eta| < 1.475$, while the EMECs have a coverage of $1.375< |\eta| < 3.2$. The EM calorimeter is a lead-liquid argon detector where lead plays the role of the absorber and LAr the active material. The accordion shape is motivated to avoid cracks in coverage in the azimuthal direction. The EMEC is split into two separate wheels that cover the range $1.375 < |\eta| < 2.5$ and $2.5 < |\eta| < 3.2$.
The first and most finely grained layer of the EMB (and the first wheel of the EMECs) is the LAr strip layer which gives high precision measurements of EM energy clusters in the $\eta$-direction. The second layer collects the majority of the energy deposition of the EM shower and so has a coarser granularity. The third layer collects the tail of the energy distribution and so is the least granulated layer.
Figure~\ref{fig:LAR} shows a slice of the EM calorimeter at $\eta=0$ (i.e. perpendicular to the beam axis and therefore a barrel component where high precision physics is carried out), specifically the three layers and the accordion-like geometry. The resolution in the $\eta$-$\phi$ plane is also given for each layer, as well as the radiation lengths.

The Tile calorimeter has coverage of $|\eta| < 1.7$ and is made of steel (the absorbing material) and silicon as the scintillating material.
The HEC has coverage of $1.5 < |\eta| < 3.2$  and is a copper/LAr sampling calorimeter.
The FCal has coverage of $3.1 < |\eta| < 4.9$ and is split into three modules. The first, an EM calorimeter makes use of a copper absorber and a LAr active material. The second two modules cater to hadronic needs and use mainly a tungsten absorber (to cater for high absorption lengths) with LAr active material.

The energy resolution for the calorimeters is parameterised by $\sigma(E)/E = (a / \sqrt{E[\text{GeV}]} ) \oplus b$. For the EM calorimeters $a\approx10\%$ and $b\approx0.7\%$, while for the hadronic calorimeters $a\approx50\%$ and $b\approx3\%$ in the barrel and end-cap, and $a\approx100\%$ and $b\approx10\%$ in the forward region.

\begin{figure}
\centering
 \captionsetup[subfigure]{width=0.35\textwidth}

  \subfloat[Energy deposit signatures from a photon candidate.]{
      \begin{overpic}[width=0.3\linewidth]{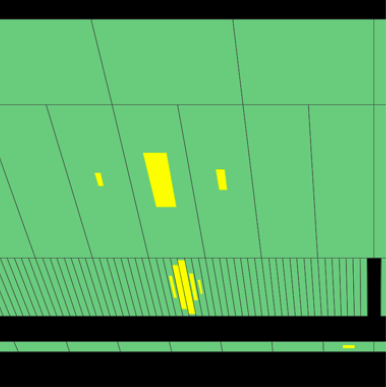}
      \put(15,85){\tiny\fontfamily{phv}\selectfont \textcolor{black}{ATLAS Experiment \textcopyright~2018 CERN} }
          \end{overpic}
    }\hspace{0.08\linewidth}
  \subfloat[Energy deposit signatures from a $\pi^{0}$ candidate.]{
      \begin{overpic}[width=0.3\linewidth]{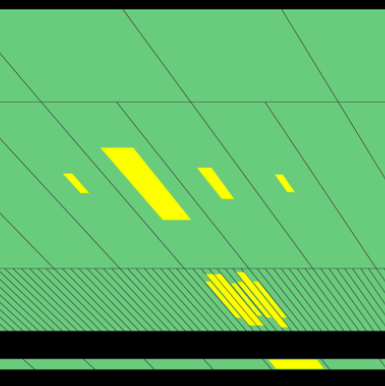}
      \put(15,85){\tiny\fontfamily{phv}\selectfont \textcolor{black}{ATLAS Experiment \textcopyright~2018 CERN} }
    \end{overpic}
    }

\caption {Two event displays showing a a) photon candidate and a b) $\pi^{0}$ candidate decaying to two photons in the presampler and three different layers of the EM calorimeter. Lead separates the presampler from the three layers of the EM calorimeter.\label{fig:event_displays}}
\end{figure}

A relevant example detailing the effectiveness of the different layers in the EM calorimeter for this thesis is shown in Figure~\ref{fig:event_displays}. Two event displays depict the presampler\footnote{The presampler is a thin layer of LAr that sits in front of the EM calorimeter. It is designed to correct for energy loss from particles in the ID and solenoid.}, then a layer of lead, followed by the finely grained LAr strip layer and the two coarser layers.
A prompt photon candidate (left image) and a $\pi^{0}$ candidate (right image) in the three layers of the EM calorimeter are shown. A $\pi^{0}$ decays into two photons and so for many analyses looking at prompt photons, these are considered fake photons from hadron decays and thus background.  The prompt photon candidate shows a more collimated energy deposition in the finely grained first layer of the EM calorimeter, whereas the $\pi^{0}$ decay shows a more spread out energy deposit due to the presence of two photons. In the second layer of the calorimeter the two energy deposits look similar. Thus, the LAr strips are essential for prompt versus fake photon identification.

\subsection{Muon spectrometers}
\label{sec:MS}

\begin{figure}[!htbp]
\centering
\includegraphics[width=0.7\linewidth]{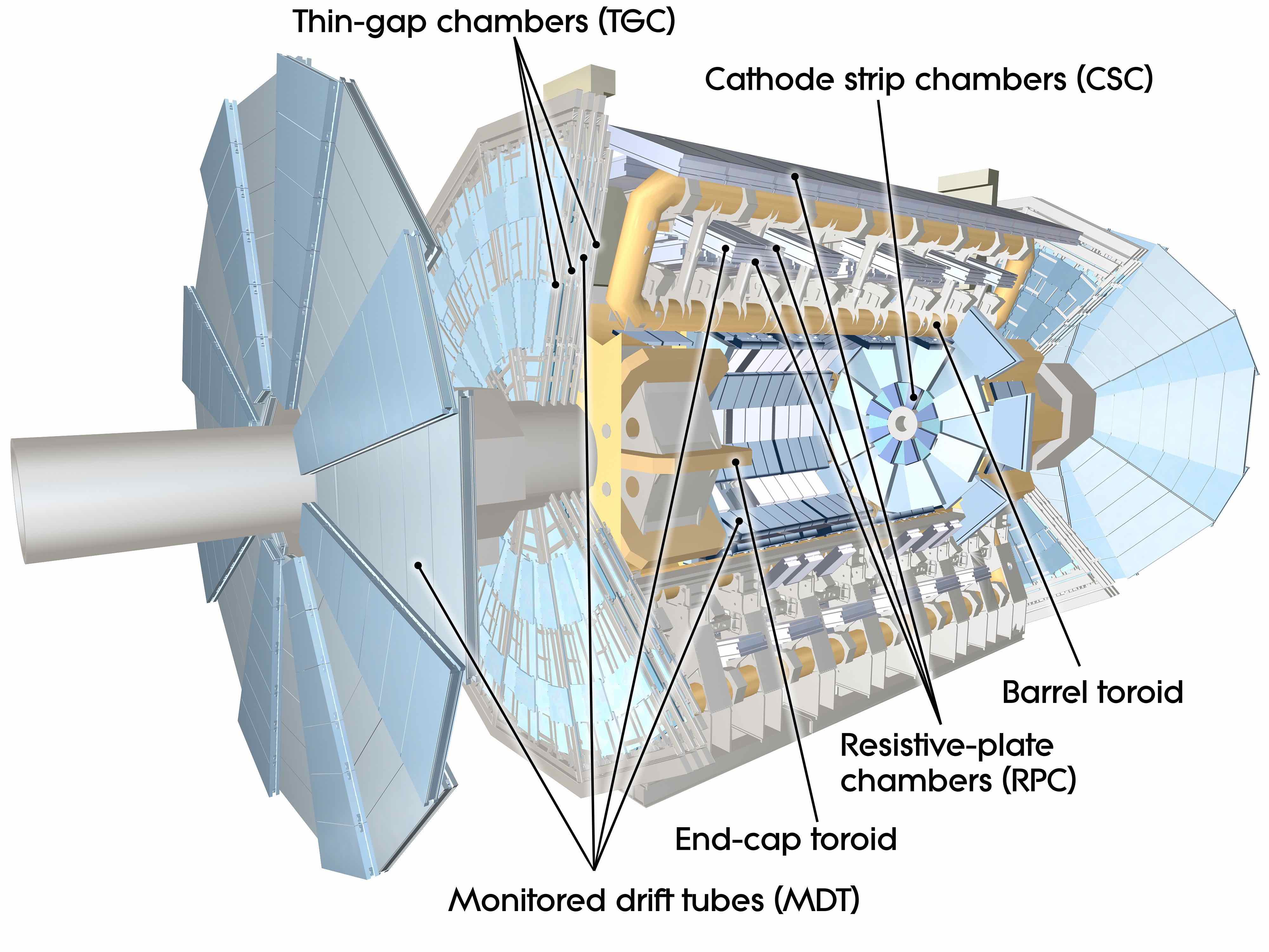}
\caption {The layout of the four main muon detector components for the \ATLAS detector. Also shown are the two toroid magnet components~\cite{PERF-2007-01}.}
\label{fig:muons}
\end{figure}

The components of the muon spectrometers have a coverage of $|\eta| < 2.7$ (but trigger on $|\eta|<2.4$) and work together with the in-homogeneous field provided by the air-coil toroidal magnet (for $|\eta| < 1.4$) and the two end-cap magnets (for $ 1.6 < |\eta| < 2.7$). The region $1.4 < |\eta| < 2.6$ receives magnetic field contributions from both magnets.  
Thus, muons are curved due to the magnetic field lines that are mostly perpendicular to the particles, and the detectors can make accurate measurements of the particles' momenta.
The momentum resolution for muons with \pt greater than 1~\GeV is approximately $\sigma(\pt)/\pt = 10\%$.
The layout of the muon detector systems is shown in Figure~\ref{fig:muons}. 
It consists of four major types of sub-detectors. Two serve as tracking: monitored drift tubes (MDT) and cathode strip chambers (CSC); and two serve as muon triggers: resistive plate chambers (RPC) and thin gap chambers (TGC). 

\subsubsection{MDTs and CSCs}
The MDTs are made up of drift tubes filled with an Ar/CO$_{2}$ gas mixture with a diameter of 30~mm. A gold-plated tungsten-rhenium wire of diameter 50~$\micro$m at a potential of 3080~V collects charges due to the ionisation of gas when a particle traverses the chamber. This allows for an average resolution of 80~$\micro$m per tube. They cover up to $|\eta| < 2.7$, except in the innermost end-cap layer where $|\eta| < 2.0$.
The CSCs cover the innermost layer and have a coverage of $2.0 < |\eta| < 2.7$. To account for higher than expected backgrounds in this forward region, multi-wire proportional chambers with cathodes segmented into strips are used creating a higher granularity. A potential of 1900~V is applied. The ionising gas used is a mixture of CO$_2$ and CF$_4$. The resolution of a CSC chamber is $40~\micro$m in the $z$-direction and 5~mm in the $\phi$-direction.
The locations of the MDT wires and CSC strips are aligned with an optical alignment system and are actively monitored. This ensures that the resolution of a muon trajectory is known to less than $30~\micro$m.

\subsubsection{RPCs and TGCs}
The RPC and TGC form a muon triggering system in the pseudorapidity region of $|\eta| \leq 2.4$. They provide fast information of muons traversing the \ATLAS detector. Some examples that they help with are muon transverse momentum measurements, bunch crossing identification and tracking information.
RPCs are used in the barrel region. They consist of a parallel electrode-plate filled with a CO$_{2}$/C$_{5}$H$_{10}$ mixture at a distance of 2~mm from each other. The electric field in the gap is 4.9~kV/mm. Three concentric cylindrical layers around the beam axis cover a pseudorapidity region of $|\eta| \leq 1.05$ in the barrel and provide good time and spatial resolution.
The TGCs are used in the end-cap regions, $1.05 < |\eta| < 2.4$ and make use of multi-wire proportional chambers. The wire-to-cathode distance is always smaller than the wire-to-wire distance, and thus they provide a high rate capability and good time resolution. In addition to functioning as a trigger, they also provide the second $\phi$ coordinate which compliments the MDT measurements. The ionising gas is a mixture of CO$_2$ and n-C$_{5}$H$_{12}$ and has an operation potential of 200~V.

\subsection{Triggers}
\label{sec:triggers}
With a bunch crossing rate of 40~MHz at the \ATLAS IP, a system in which only interesting events are saved is crucial.
In Run~1 \ATLAS used a three level trigger system (one level was hardware based and two were software) to reduce the number of data saved  to the order of a few hundred Hertz. 
For Run~2, \ATLAS revisited its trigger and data acquisition pipeline to be able to cope with the increased data and more challenging run conditions. 

An upgraded custom-made electronic hardware trigger is used as the first trigger (L1). It uses information from a subset of detectors such as the triggering calorimeters and muon detectors to highlight candidate objects in regions of interest (RoI).  Candidate objects are muons, EM clusters, jets, taus, missing transverse energy, and total energy. This reduces the event rate down to about 100~kHz. These RoIs then seed the second trigger.
This is called the High Level Trigger (HLT) and replaces the older level 2 (L2) and event filter layers. A dedicated HLT computing farm makes use of shared algorithms to merge RoIs to allow for more efficient CPU processing, and thus a more efficient trigger. The HLT reduces the event rate from the L1 trigger of 100~kHz to a more manageable 1~kHz, which is then written to disk.

\chapter{Analysed data and Monte Carlo}

\section{Data}
\label{sec:data}

\begin{figure}[!htbp]
\centering
\subfloat[2015]{
\includegraphics[width=0.5\linewidth]{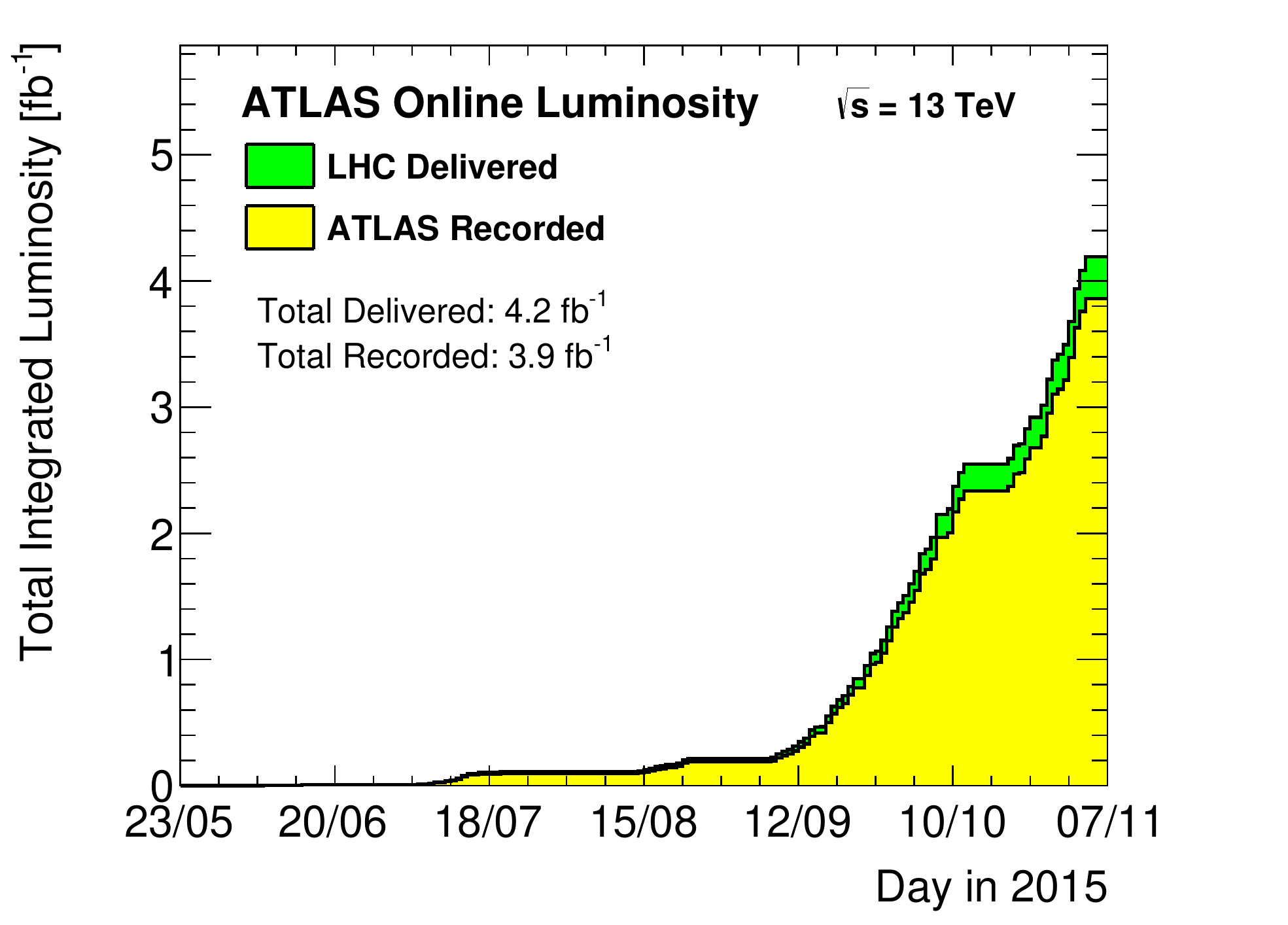}
}
\subfloat[2016]{
\includegraphics[width=0.5\linewidth]{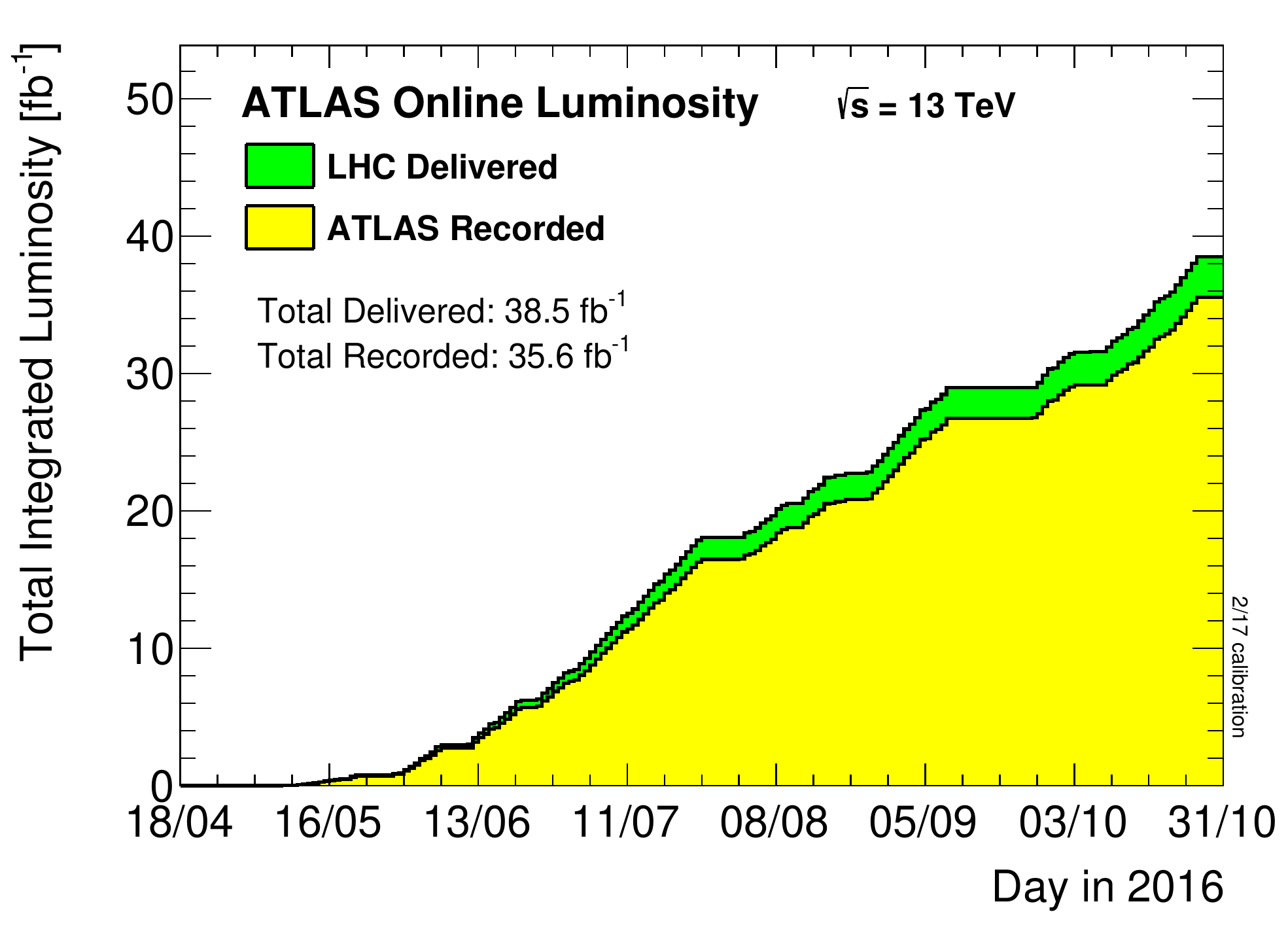}
}
\caption {The total integrated luminosity delivered by the LHC and recorded by the \ATLAS detector in 2015 and 2016~\cite{lumiplots}.}
\label{fig:Lumi}
\end{figure}

This thesis makes use of data collected with the \ATLAS detector in 2015 and 2016 at $\sqrt{s}=13$~TeV. This makes up only part of the LHC's Run 2 dataset\footnote{A total of around 120~fb$^{-1}$ of data is expected to be collected by the end of Run 2, i.e. the end of 2018.}. Strict requirements are placed on the quality of the data, requiring only good luminosity blocks\footnote{A run in which \ATLAS is collecting data is made up of luminosity blocks corresponding to around a minute of data-taking each. A good luminosity block defines whether those data are good for physics analyses. This requirement can be different if, for example, an analysis does not make use of a certain phase space of the detector that was not operating nominally.}. This corresponds to 3213~pb$^{-1}$ and 32885~pb$^{-1}$ of data collected in 2015 and 2016, respectively.  
Figure~\ref{fig:Lumi} shows the total integrated luminosity that the LHC delivered and what \ATLAS recorded for the years 2015 and 2016. The difference in delivered and recorded luminosity can be due to a number of reasons ranging from problems in \ATLAS data acquisition software and hardware, to the process of the \ATLAS detector reaching data taking conditions once stable beams have been announced by the LHC.

\begin{figure}[!htbp]
\centering
\includegraphics[width=0.5\linewidth]{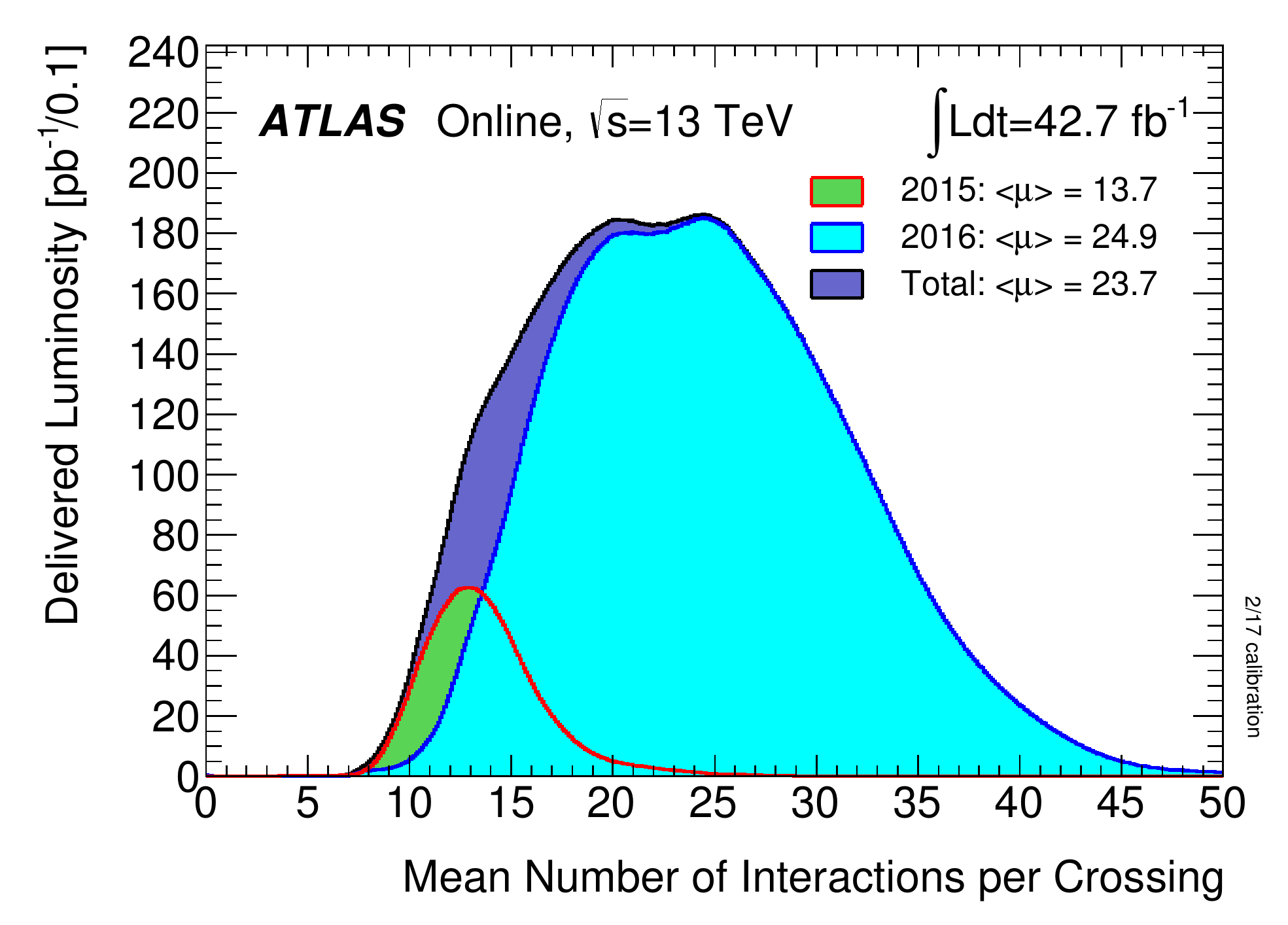}
\caption {The pileup profile for 2015 and 2016 data recorded with the \ATLAS detector~\cite{lumiplots}.}
\label{fig:mu_2015_2016}
\end{figure}

Figure~\ref{fig:mu_2015_2016} shows the mean number of interactions per bunch crossing (\emph{pileup}) for the years of 2015 and 2016. This corresponds to the mean of the Poisson distribution for the number of interactions calculated per bunch crossing. 
Higher pileup makes object identification more difficult and thus increases the systematic contributions that have to be taken into account. As the LHC increases its energy and instantaneous luminosity, the problem of larger pileup profiles becomes more prolific and needs to be addressed.

\section{Monte Carlo datasets}
\label{sec:MCsim}

Theoretical models are needed for which collected data can be compared against. The process of going from a theoretical prediction to a Monte Carlo (MC) dataset that is in the same format and has the same detector biases as \ATLAS data is a complex process requiring three main steps (and many sub-steps). These are: event generation, detector simulation and object reconstruction.

\subsubsection{Event Generation}
Event generation is the first step, which essentially has four ``ingredients".
Protons are composite particles made up of quarks and gluons, together forming partons.
When protons collide Parton Distribution Functions (PDF) are needed to determine the fraction of momenta that these partons have. These PDFs are experimentally determined and are dependent on the CME. The PDF sets that are used in this analysis are: \NNPDFLO~\cite{Pumplin:2002vw}, \NNPDFNLO~\cite{Ball:2014uwa} and \CTten~\cite{CT10}.
PDFs are used as parameterisations to the next step, the hard scattering process, which is the Matrix Element (ME) calculation. The ME is the perturbative approach to modelling the process of interest, which can be represented by Feynman Diagrams. Different orders of precision can be modelled, with higher precisions being computationally more expensive and sometimes not practical.
Amplitudes or probabilities for certain processes are a result of the ME calculation. Once the ME has been calculated and final state particles produced accordingly, ME particles are showered. 
Partons emit photons or gluons via the QCD or QED mechanisms due to bremsstrahlung, which can cause multiple cascades of particles. These then need to be taken into account when calculating the final kinematic distributions for an event. The process of adding QCD and QED bremsstrahlung is the showering of the ME.
The last stage, called hadronisation, is when hadrons are formed from partons and are allowed to decay until stable particles are formed. This is due to QCD confinement which says quarks and gluons can not exist as individual particles, but need to form hadrons.

The steps above can in full or in part be carried out by what is called a MC generator. They can be interfaced to different PDFs or showering algorithms if necessary.
The main MC generators used in this analysis are: \texttt{MadGraph5\_aMC@NLO} (\madgraph)~\cite{madgraph}, \sherpaAll~\cite{Gleisberg:2008ta}, \powheg{v1} and \powheg{v2}~\cite{powhegST,powheg1v2,powheg2v2,powheg3v2,powheg4v2}, and \pythia{8}~\cite{Sjostrand:2014zea}.
\sherpaAll provides its own showering and hadronisation model, while other generators need to be interfaced to a generator like \pythia{8} or \herwig{7}~\cite{Bahr:2008pv,Bellm:2015jjp}.

There are many parameters that are not necessarily well modelled in the above steps. MC tuning is needed to account for unforeseen circumstances such as multiple parton interactions or an increase in the strong coupling constant at low energies.
This is done using tunes provided through empirical studies to data. The tunes used in this analysis are \Afourteen~\cite{ATL-PHYS-PUB-2014-021} and \perugia~\cite{perugia}.

\subsubsection{Detector Simulation}

The simulation of particles through the \ATLAS detector is done using \textsc{Geant}~4~\cite{Agostinelli:2002hh}. This step is the most CPU intensive as the interaction of each particle and its decay products needs to be simulated with the respective parts of the detector. The simulation of a single event can take anywhere from around 2 minutes to about 8 minutes depending on the complexity of the event. Millions of events for a given process need to be simulated. 
As previously mentioned, pileup plays an important role in the performance of object reconstruction and needs to be simulated accurately. \pythia{8} is responsible for simulating the pileup due to QCD processes. Tuneable parameters called \Atwo~\cite{ATL-PHYS-PUB-2012-003} and the \MSTWLO~\cite{Martin:2009iq} parton distribution functions (PDF) set are used.
Techniques such as using pre-simulated electromagnetic showers can be used with a small penalty on accuracy, and as mentioned in Chapter~\ref{sec:ML}, further research using ML techniques is being done to significantly speed up the simulation process by several orders of magnitude.

\subsubsection{Object reconstruction}

Detector properties combined with software algorithms are able to reconstruct objects for real and simulated data that pass through the detector (as shown in Figure~\ref{fig:atlas_cross_section}). In general this is done using the \ATLAS software infrastructure~\cite{SOFT-2010-01}. Object reconstruction is explained in more detail in Chapter~\ref{sec:objectID}.

\subsection{Nominal signal and background MC samples}

The \ttgamma signal sample is modelled at LO using \madgraph and the \NNPDFLO pdf set. Parton showering is simulated with \pythia{8} using the \Afourteen tune. The five flavour scheme is used where all quark masses except the top mass are set to 0. The top quark mass is set to 172.5~\GeV with a decay width of 1.320~\GeV and the fine structure constant is set to 1/137.
A set of kinematic cuts is needed to avoid infrared and collinear divergences. The photon and lepton transverse momenta are required to be larger than 15~\GeV, while the absolute value of the pseudorapidity is required to be less than 5.0. A minimum $\Delta R$ distance of 0.2 is required between the photon and any charged particle in the final state. 
Higher order predictions of the \ttgamma process are incredibly CPU intensive and are not realistic to compute. NLO \emph{k-factors} are used as a correction factor to account for higher orders in the QCD calculation (Section~\ref{sec:kfactor}).

The nominal \ttbar MC sample is generated with \powheg{v2} with \NNPDFNLO, and interfaced to \pythia{8} using the \Afourteen tune and \NNPDFLO set.

The production of $W/Z$+jets is done with \sherpa{2.2.1} while the dedicated $W/Z$+jets+$\gamma$ ($W/Z\gamma$) samples are generated with \sherpa{2.2.2}. Both use the \NNPDFNLO pdf set.

The single top-quark processes ($t$-, $s$- and $tW$-channels) are produced with \powheg{v1}. 
Each channel is generated using the \CTten PDF set and interfaced to \pythia{6}, with the $t$- channel also using the \perugia tune.

The diboson ($WW$, $WZ$ and $ZZ$) samples are simulated using \sherpa{2.1} with the \CTten pdf set.
The \ttbar{}$+Z/W$ (\ttV) samples are produced with \madgraph, interfaced to \NNPDFNLO PDF set and showered with \pythia{8}.



\subsection{Signal sample NLO normalisation}
\label{sec:kfactor}
The nominal \ttgamma MC sample at reconstruction level is LO as explained above. Thus, so called $k$-factors are used to correct the prediction to NLO precision. 
This calculation has been done in the approximation of a stable top quark~\cite{duan1}, and extended to allow the decay of top quarks at $\sqrt{s} = 14$~\TeV~\cite{melnikov}. The authors from~\cite{melnikov} performed dedicated calculations for $\sqrt{s}=13$~\TeV in the \chljets and \chll channels, which are used in this analysis. The NLO calculation was performed in the fiducial regions defined in Chapter~\ref{sec:fiducial} at parton level, separately for the \chljets and \chll channels. The LO calculation was performed at particle level\footnote{Particle level is defined as events generated in the hard-process before detector simulation and reconstruction occurs.} and care is taken so that objects at particle level correspond to those at parton level. For example, photons and leptons should be produced from the  matrix element and not the parton shower.
The $k$-factor is calculated as 
\begin{equation}
k\text{-factor} = \frac{\sigma^{\text{parton}}_{\text{NLO}}}{\sigma^{\text{particle}}_{\text{LO}}}.
\end{equation}

Systematic uncertainties in this theoretical calculation enter from sources such as the renormalisation and factorisation scales, PDFs, jet cone sizes (to evaluate the impact of additional radiation) and multiple parton interactions. This yields different systematic uncertainties in the \chljets and \chll channels, with the total uncertainty equating to 20\% and 15\%, respectively. Future versions of this analysis should place high priority on reducing these theoretical uncertainties.
For more information on the $k$-factors and NLO cross-section calculation, refer to~\cite{ATLAS-CONF-2018-048}.


\chapter{Object reconstruction and identification}
\label{sec:objectID}

Crucial to any successful analysis in HEP is the localisation, reconstruction, and identification of the physics objects within the ATLAS detector. A combination of the detector's geometry and material composition as well as advanced algorithms allow physicists to determine the object type, be it a photon, electron,  muon, jet, or missing transverse energy (\MET). The different elements in the detector that play crucial roles in determining each object have been shown in Chapter~\ref{sec:ATLAS}. 
This Chapter presents an overview of the different algorithms used in \ATLAS. 

\section{Electron and photon objects}
\label{sec:egamma}

Due to the similarity between electrons and photons and the comparable signatures they leave in the calorimeters, the reconstruction of these objects happens simultaneously. Information from the ID, presented in Chapter~\ref{sec:ID}, and both the electromagnetic calorimeter (EMC) and the hadronic calorimeter (HCAL), presented in Chapter~\ref{sec:calorimeters}, are used.
An overview of the reconstruction of electrons and photons follows.
\begin{itemize}
\item First, a \emph{seed cluster} in the EMC is formed. A grid of $\Delta \eta \times \Delta \phi = 0.025\times0.025$ (corresponding to the granularity in the barrel region of the EMC middle layer) is formed in the $\eta$-$\phi$ plane of the EMC.
A \emph{tower} for each element in the grid is formed by integrating the total energy over the presampler, first, second and third layers of the EMC.
Then, for every element in the defined EMC space, a sliding window algorithm~\cite{ATL-LARG-PUB-2008-002} is used to form EM energy cluster candidates. A defined region of $3\times5$ cells\footnote{This cluster size changes slightly for barrel versus endcap parts of the EMC.} in the $\eta$-$\phi$ plane is moved 0.025 steps in either the $\eta$ or $\phi$ direction and searches for energy deposits whose summed transverse energy is greater than 2.5~\GeV. These energy deposits form a seed cluster. Often candidate seeds can overlap and so an overlap removal is defined. If two candidate seeds overlap within an area of $\eta\times\phi = 5\times9$ cells, the candidate with the larger transverse energy is taken. If the transverse energy of the clusters is within 10\% of each other, then the cluster with the largest individual tower is taken. 

\item Track reconstruction in the ID proceeds in two steps. The first is a pattern recognition technique which is based on energy loss due to interactions with detector material~\cite{Cornelissen:1020106}.
The second step is a track fit. The \ATLAS Global $\chi^{2}$ Track Fitter~\cite{Cornelissen:2008zza}  is used with either a pion or an electron hypothesis.
These tracks are then loosely matched to the seed clusters in the EMC.

\item These loosely-matched tracks can have hits in different parts of the detectors. From these hits, vertices are used for conversion reconstruction. For example, a converted photon can be associated with two-track conversion vertices. An unconverted photon would have no associated tracks.

\item Finally, based on track information, conversion vertices and the seed cluster, an algorithm decides whether an object will be reconstructed as an electron, photon or both.
\end{itemize}

Central to electrons and photons is the use of ``shower shape" variables. These are various properties of the electromagnetic showers that are exploited to allow for effective object differentiation. A list with explanations of these variables is shown in Table~\ref{tab:showershapes}. References to these shower shape variables will be made in this, and later chapters.

\begin{table}[htb]
	\centering
	\begin{tabular}{c p{12cm}}
	\toprule
	\textbf{Name} & \textbf{Description} \\
	\hline
		\multicolumn{2}{l}{\textbf{Hadronic leakage}}\\
		\hline
		$R_\mathrm{had}$ or $R_\mathrm{had1}$ & Transverse energy leakage in the HCAL normalised to transverse energy of the photon candidate in the EMC. In the region $0.8\leq |\eta|\leq 1.37$, the entire energy of the photon candidate in the HCAL is used ($R_\mathrm{had}$), while in the region $|\eta| < 0.8$ and $|\eta| > 1.37$ the energy of the first layer of the HCAL is used ($R_\mathrm{had1}$)\\
		\hline
		\multicolumn{2}{l}{\textbf{Energy ratios and width in the second layer of EMC}}\\
		\hline
		$R_\eta$ & Energy ratio of $3\times 7$ to $7\times 7$ cells in the $\eta\times\phi$ plane\\
		$R_\phi$ & Energy ratio of $3\times 3$ to $3\times 7$ cells in the $\eta\times\phi$ plane\\
		$w_{\eta 2}$ & Lateral width of cluster in $\eta\times\phi = 3\times 5$: $\sqrt{\frac{\sum_i E_i\eta_i^2 }{\sum_i E_i} - \left(\frac{\sum_i E_i\eta_i }{\sum_i E_i}\right)^2}$\\
		
		\hline		
		\multicolumn{2}{l}{\textbf{Energy ratios and widths in the first (strip) layer of EMC}}\\
		\hline
		$w_{\eta 1}(w_{s3})$ & Energy weighted width using 3 strips around the maximum: $\sqrt{\frac{\sum_i E_i(i - i_\mathrm{max})^2}{ \sum_i E_i}}$\\
		$w_\mathrm{tot,s1}(w_s)$ & Energy weighted width using 20 strips around the maximum, see $w_{\eta,1}$\\
		$f_\mathrm{side}$ & Energy within 7 strips without 3 central strips normalised to energy in 3 central strips\\
		$E_\mathrm{ratio}$ & Ratio between difference of first $2$ energy maxima divided by their sum ($E_\mathrm{ratio}=1$ if there is no second maximum)\\
		$\Delta E$ & Difference between the second energy maximum and the minimum between first and second maximum ($\Delta E = 1$ if there is no second maximum)\\
	\bottomrule
	\end{tabular}
	\caption{Summary of the shower shape variables used for electron and photon reconstruction and identification. Summations over $i$ run over the cluster constituents.}
	\label{tab:showershapes}
\end{table}

From here, different requirements on photon and electron identification and isolation determine the purities and efficiencies in which these objects are selected.

\subsection{Photons}
\label{sec:photonDefs}

Photon identification~\cite{ATL-PHYS-PUB-2016-014} is performed by using rectangular cuts on the shower shape variables shown in Table~\ref{tab:showershapes}.
As shown in Figure~\ref{fig:event_displays}, photons originating from different processes are characterised by the shapes of the electromagnetic shower developments through the EMC and HCAL. The variables in Table~\ref{tab:showershapes} characterise this. Two identification reference points are calibrated within \ATLAS; \loose and \tight.
The \loose selection is based on information provided by the second layer of the EMC as well as the HCAL. This corresponds to the variables $R_{\text{had}}$, $R_{\text{had1}}$, $R_{\eta}$ and $\omega_{\eta_{2}}$.
This analysis mostly uses the \tight selection, which makes use of all shower shape variables in Table~\ref{tab:showershapes}\footnote{In some studies, such as for the \hfake background in Chapter~\ref{sec:hfake} this requirement is reversed.}.
Information from the strip layer in the EMC is included, thus \tight events are a subset of the \loose events. To account for the slightly different shower shape energy deposits between converted and unconverted photons, the working points are optimised separately. To account for the calorimeter geometry and the effect of slightly different materials, the working points are also calibrated as a function of $\eta$.

To obtain a purer sample of prompt photons, isolation plays a crucial role. It significantly reduces the contribution from photons both from hadronic decays or hadrons misidentified as photons.
Isolation for photons is based on the transverse energy in a $\Delta R$ size cone around the photon (excluding the photon's transverse energy). Transverse energy can originate either from tracks or from calorimeter deposits.

\begin{itemize}

\item $E_{\text{T}}^{\text{topoetcone}x}$ is defined as the calorimeter isolation and is obtained from summing the transverse energy of the topological clusters around a cone of $\Delta R < \frac{x}{100}$. Contributions from the photon candidate itself and pileup are subtracted.

\item $p_{\text{T}}^{\text{cone20}}$ is defined as the track isolation and is obtained by summing the transverse momenta of all tracks (excluding those from photon conversions) with $\pt > 1$~\GeV around of cone of $\Delta R < 0.2$. A  distance requirement to the primary vertex along the beam axis, $|z_{0}\text{sin}\theta| < 3$~mm, must also be fulfilled.
\end{itemize}
There are three provided working points within \ATLAS, which are shown along with their definitions in Table~\ref{tab:photonIsolation}. This analysis generally makes use of the \FCT isolation working point.

\begin{table}[!htbp]
	\centering
	\begin{tabular}{lcc}
	\toprule
	\textbf{Working point} & \textbf{Calorimeter isolation} & \textbf{Track isolation} \\
	\hline
         \FCTCO & $E_{\text{T}}^{\text{topoetcone40}} < 0.022 \pt +2.45$~[\GeV{}] & none\\
         \FCT & $E_{\text{T}}^{\text{topoetcone40}} < 0.022 \pt +2.45$~[\GeV{}] & $p_{\text{T}}^{\text{cone20}}/\pt < 0.05$ \\
         \FCL & $E_{\text{T}}^{\text{topoetcone20}} < 0.065 \pt$ & $p_{\text{T}}^{\text{cone20}}/\pt < 0.05$ \\
	\bottomrule
	\end{tabular}
	\caption{The 3 defined isolation working points for photons.}
	\label{tab:photonIsolation}
\end{table}

\begin{figure}[!htbp]
\centering
\subfloat[Converted photons]{
\includegraphics[width=0.45\linewidth]{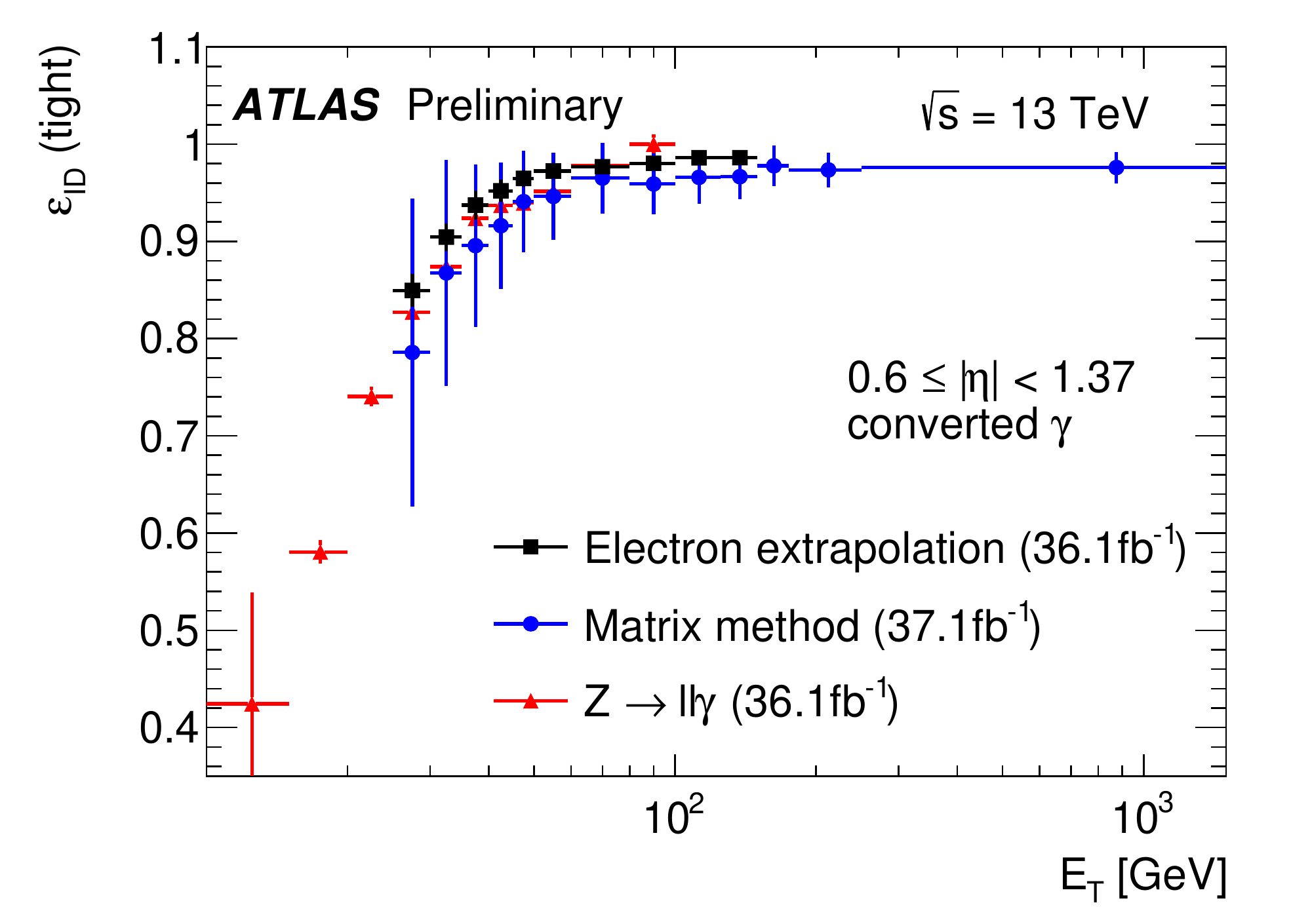}
}\hspace{-0.038\linewidth}
\subfloat[Unconverted photons]{
\includegraphics[width=0.45\linewidth]{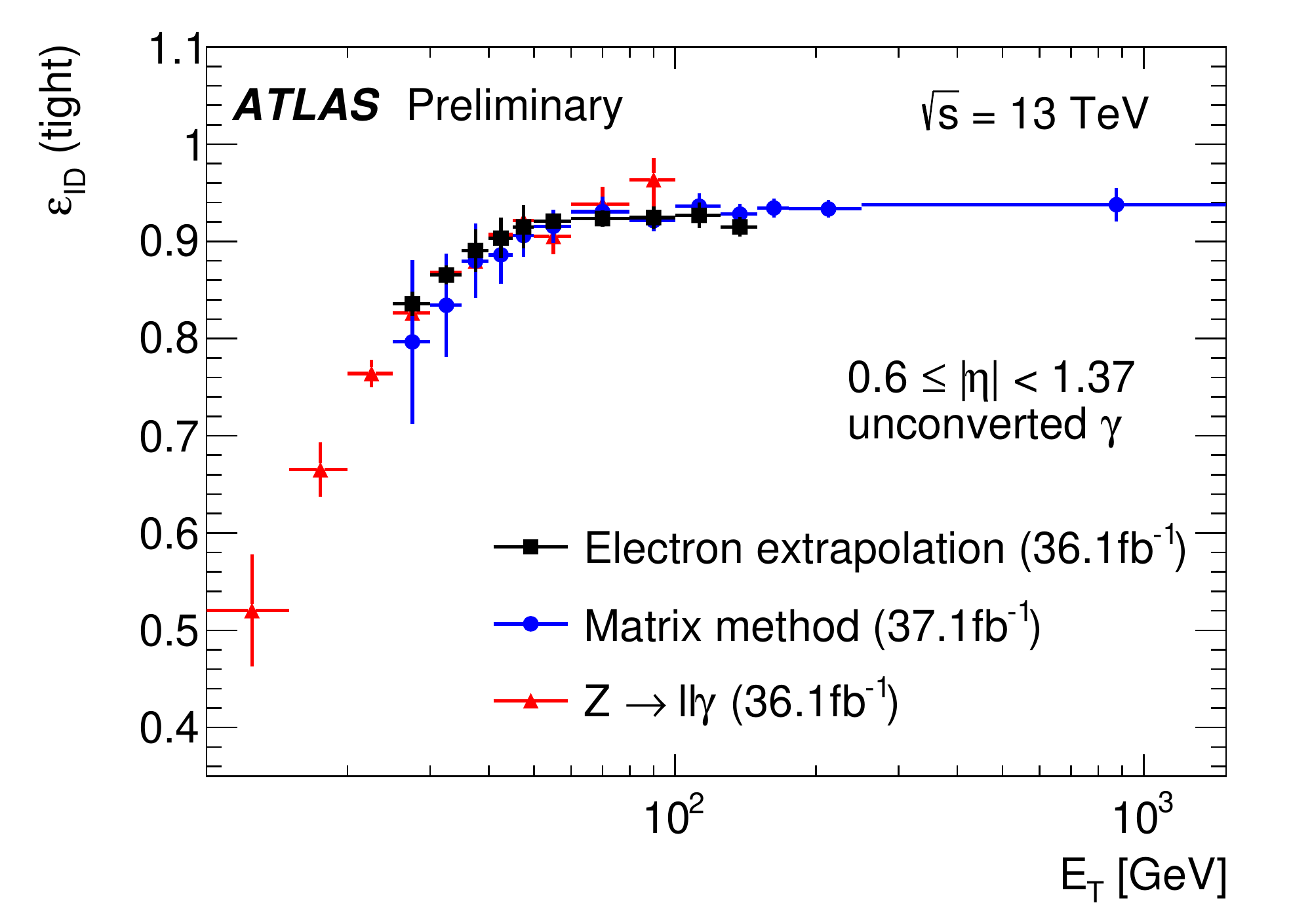}
}\hspace{-0.038\linewidth}
\caption {The efficiencies for \tight photons shown for the three complementary methods~\cite{egammapublic}.}\label{fig:photonIDeff}
\end{figure}

The efficiencies of \tight photons are measured in data using three methods. This enables coverage of a wide \ET that is not possible otherwise.
The cleanest source of prompt photons is measured in $Z\to ll \gamma$ decays for the efficiencies of $\ET = 10~\GeV$ up to approximately $100$~\GeV. Event yields after this are too low.
For photons of about $25 < \ET < 150$~\GeV the \emph{electron extrapolation} method is used from $Z\to ee$ decays. Here, the showers that electrons make in the calorimeters are slightly altered to reflect those of photons.
The last method in which efficiencies are measured applies to photons in the range 25~\GeV $ < \ET < 1.5$~\TeV. A photon sample is collected using single-photon triggers. The efficiency is essentially derived by comparing those events that pass the \tight track-based isolation criteria to the full \tight photon sample.
In all the above methods various isolation criteria for the photons are applied.
Figure~\ref{fig:photonIDeff} shows the efficiencies for the range $0.6 \le |\eta| < 1.37$ for converted and unconverted \tight photons. The three methods used to obtain the efficiencies complement each other at different energy scales. The error bars include the statistical and systematic uncertainties associated with each method. The large spikes are from fewer statistics.
These efficiency measurements are combined to provide scale factors to correct for data/MC mis-modelling. 

\begin{figure}[!htbp]
\centering
\includegraphics[width=0.67\linewidth]{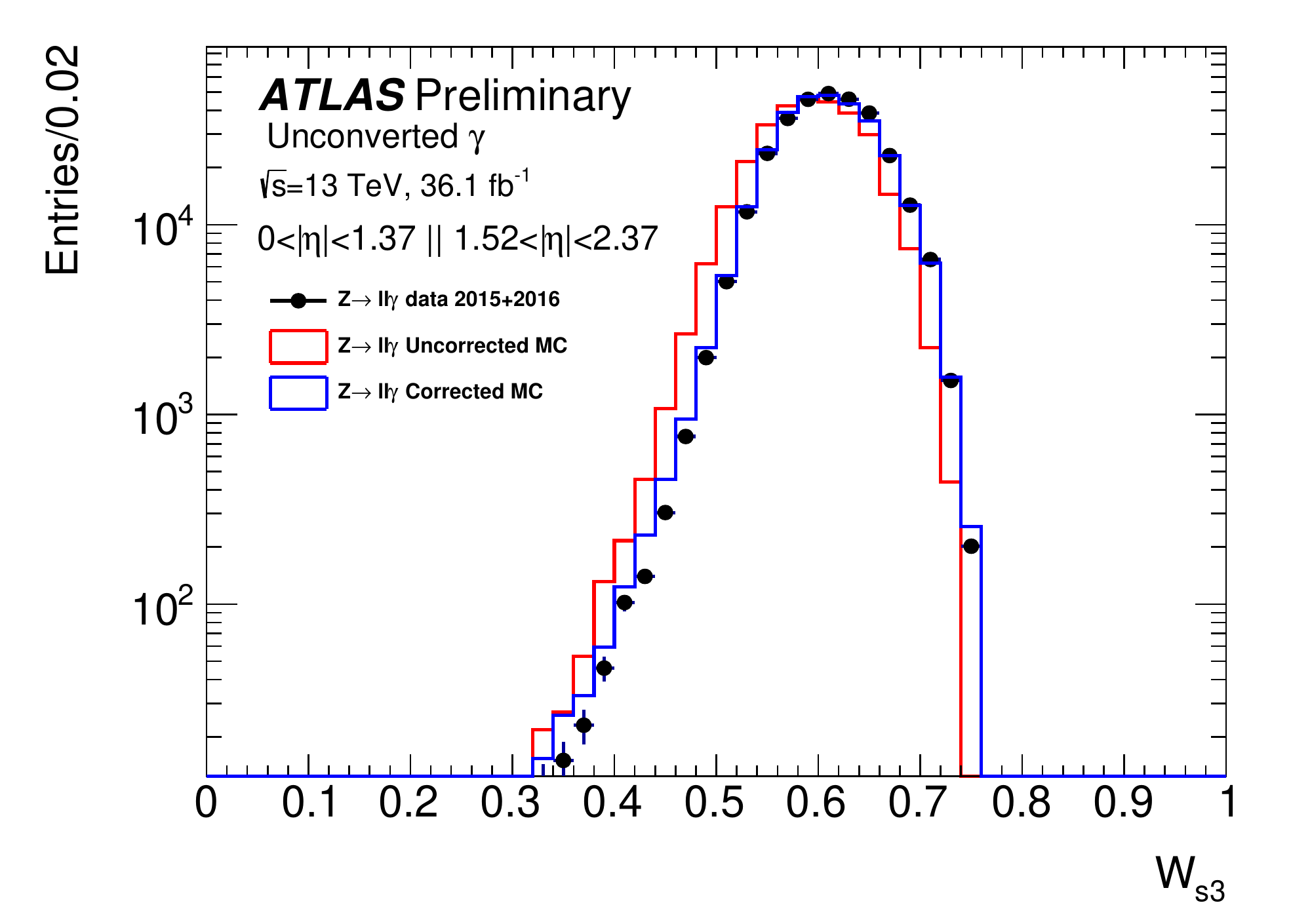}
\caption {The $\omega_{\text{s3}}$ shower shape variable. This compares the original simulated events to real data, as well as the final corrected simulated events~\cite{egammapublic}.}\label{fig:showershape}
\end{figure}

An important aspect to the shower shape variables is that of the correction factors that are applied to simulated samples. 
Samples used in photon identification are simulated with \textsc{Geant4} and the \ATLAS reconstruction pipeline. However, \textsc{Geant4} reproduces the shower shapes of photons imperfectly and so an offset occurs when comparing these MC samples to real data. 
Corrections are derived from $Z\to ll \gamma$ and $\gamma$+jet MC samples which take into account systematic and statistical errors.
An example is shown in Figure~\ref{fig:showershape}. This shows the MC simulation before and after corrections have been applied, compared to the measured data for a given bin of $\eta$. Corrections to shower shape variables are essential for photon identification in \ATLAS.
All above efficiencies are performed using the corrected shapes, and from hereon, this analysis only makes use of the corrected variables.

\subsection{Electrons}
\label{sec:electronDef}

Electron identification~\cite{ATLAS-CONF-2016-024} follows an MVA likelihood based approach in which variables from Table~\ref{tab:showershapes} are used as inputs. An overall probability that the object is a signal electron is calculated. These probabilities form a discriminant, which at different values or ``working points" define different purities of electrons. Three working points are provided in \ATLAS, which in increasing order of background rejection are \loose, \medium and \tight. This analysis makes use of the \tight working point. This equates to efficiencies (measured in $Z\to ee$ and $J/\psi \to ee$ events) of around 80\% for \ET$=30$~\GeV and increases with higher \ET.

Electrons are often required to fulfil isolation requirements.  
Two variables are defined; $E_{\text{T}}^{\text{cone0.2}}$ and $p_{\text{T}}^{\text{varcone0.2}}$.
\begin{itemize}
\item $E_{\text{T}}^{\text{cone0.2}}$ is the sum of the transverse energy of clusters formed in the calorimeters, within a cone of $\Delta R = 0.2$ around the electron cluster.

\item $p_{\text{T}}^{\text{varcone0.2}}$ is the sum of the transverse momenta of all tracks within a cone of \\ $\Delta R = \text{min}(0.2,10 \GeV/\ET)$ around the candidate electron track.
\end{itemize}
This analysis makes use of the \gradient isolation working points, which uses cuts on the two above variables to target specific efficiencies.
For \gradient isolation, this targets a total isolation efficiency of electrons (between the combination of calorimeter and track based isolation) of 90\% (99\%) at 25~\GeV (60~\GeV).

\section{Muons}
\label{sec:muons}
Muon reconstruction~\cite{PERF-2015-10} in \ATLAS relies on information from the ID (Chapter~\ref{sec:ID}), the muon spectrometer (MS, Chapter~\ref{sec:MS}) and calorimeters (Chapter~\ref{sec:calorimeters}). 
The ID provides track measurements and information related to distances from the IP. The MS provides accurate momentum measurements with high resolutions for muons. The calorimeters supplement the MS and ID with information regarding energy loss.
From combining information from the sub-detectors, four main types of muons are identified: 
\begin{itemize}
\item Combined muons: Tracks are present in both the MS and the ID for which reconstruction is performed separately.
A combined track is formed with a global refit to the hits of both sub-detectors. 

\item Segment-tagged muons: A track in the ID can be extrapolated to the MS, but is only associated with one track segment in the MDT or CSC chambers. These muons are often low \pt tracks, or muons that fall into the reduced MS acceptance regions.

\item Calorimeter-tagged muons: Tracks in the ID and a minimum-ionisation energy deposit in the calorimeter can be identified as a muon. The purity of this type of muon is very low, but it recovers acceptance in areas of the \ATLAS detector in which the MS has poor coverage.

\item Extrapolated muons: Tracks in the MS that are not matched to tracks in the ID are extrapolated back to the IP taking into account the energy loss in the calorimeter. These muons are predominantly used in the region $2.5< |\eta| < 2.7$, for which the ID does not cover.

\end{itemize}
This thesis makes use of \medium identification criteria. This default selection in \ATLAS is one that minimises the systematic uncertainties during reconstruction and calibration.
Combined muon tracks are required to have at least three hits in at least two MDT layers. Extrapolated muon tracks are required to have at least three hits in MDR/CSC layers and are used only in the region $2.5<|\eta| < 2.7$. Contamination from hadrons misidentified as muons is reduced by a requirement on the ratio of the muon's \pt in the MS and ID.
Efficiencies are measured using a tag-and-probe method in almost pure $J/\psi \to \mu \mu$ or $Z \to \mu \mu$ events with at least one \medium muon. 
In the region $|\eta| < 2.5$, the efficiency is measured to be close to 99\% for $\pt > 5~\GeV$.

Prompt muons from $W$- and $Z$-boson decays are more isolated then muons originating from semileptonic decays in jets. Thus, isolated muons are important when selecting final channels as well as suppressing background. 
Isolation makes use of track and calorimeter based isolation variables.
The track based variable, $p_{\text{T}}^{\text{varcone30}}$, is defined as the scalar sum of all transverse momenta of the tracks with $\pt > 1~\GeV$ (excluding the muon track itself) in a cone of $\Delta R = \text{min}(10~\GeV / p_{\text{T}}^{\mu},0.3)$ around the transverse momentum of the muon, $p_{\text{T}}^{\mu}$.
The calorimeter based isolation variable, $E_{\text{T}}^{\text{topoetcone20}}$, is the sum of the transverse energy from topological clusters in a cone of $\Delta R = 0.2$ around the muon (excluding the muon itself).

Relative isolation variables are used to define various working points that are used in different analyses.
The \gradient based isolation working point is used in this thesis, described as cuts placed on ($p_{\text{T}}^{\text{varcone30}}/p_{\text{T}}^{\mu},E_{\text{T}}^{\text{topocone20}}/p_{\text{T}}^{\mu}$). 
The efficiency for this working point (measured in $Z\to \mu \mu$ events) achieves greater than 90\% (99\%) efficiency for muons at 25~\GeV (60~\GeV).

\section{Jets}

Due to colour confinement and the requirement for particles to have zero net colour charge, quarks and gluons can not exist by themselves, but rather form hadrons through the process of hadronisation (Chapter~\ref{sec:intro}). These hadrons then decay into a collimated spray of particles termed \emph{jets}. 
The reconstruction and identification of jets~\cite{ATL-PHYS-PUB-2015-036} at the LHC only increases in importance as the pileup due to higher luminosities increases. They can form dominant backgrounds, or be flags for topological signatures in various processes.
Jets are reconstructed using the anti-$k_{t}$ algorithm~\cite{Cacciari:2008gp} using a radius of $R=0.4$.
The anti-$k_{t}$ algorithm is a sequential clustering algorithm based on a distance parameter for topological calorimeter clusters (topo-clusters)~\cite{PERF-2014-07} $i$ and $j$, and another distance parameter between  topo-cluster $i$ and the beam ($B$):
\begin{align}
d_{ij} &= \text{min}(k_{ti}^{-2},k_{tj}^{-2}) \frac{\Delta^{2}_{ij}}{R^2} \\
\text{and} \notag \\
d_{iB} &= k_{ti}^{-2}.
\end{align}
Here, $R$ is the radius and $\Delta^{2}_{ij} = (y_{i}-y_{j})^2 + (\phi_{i}-\phi_{j})^2$, where $k_{ti}$, $y_{i}$ and $\phi_{i}$ are the transverse momentum, rapidity and azimuthal angle of particle $i$. 
The minimum between $d_{ij}$ and $d_{iB}$ is taken. If the minimum distance is $d_{ij}$, the two topo-cluster entities are combined. If the minimum distance is $d_{iB}$ then $i$ is called a jet and removed from the list of entities. The process iterates over all topo-clusters until no entities are left.
From the $1/k^2_{ti,j}$ terms, this tells us that harder particles are clustered with a higher priority.  If no other hard jets are within a distance $2R$, it accumulates the softer particles resulting in a conical shape. If two hard particles are present, depending on the distribution of momentum, one may be more conical than the other. Essentially, soft particles do not alter the shape of the jet.

Corrections are applied to the jet energy scale (JES)~\cite{PERF-2016-04} and the jet energy resolution (JER)~\cite{ATL-PHYS-PUB-2015-015} in the form of \pt and \eta dependent scale factors. The JES corrects the jet energy to that of the truth jet energy at the particle level. This essentially corrects the reconstructed jet energy and the orientation in the calorimeter.

Essential to the reconstruction of jets is the ability to reduce the contribution from pileup and enhance the jets that result from the hard process. For this, a multivariate tool called the Jet Vertex Tagger (JVT) is used~\cite{ATLAS-CONF-2014-018}. 
The JVT defines the fraction of total tracks in the jet that are associated to the primary vertex.

\subsection{$b$-jets}
\label{sec:btagging}
In top quark analyses $b$-jets are crucial since the branching ratio of $t \to W + b$ is completely dominant.
However, tagging $b$-jets ($b$-tagging) is difficult due to busy environments. Specifically, distinguishing $light$-jets ($u,d,s$ or gluons) and $c$-jets from $b$-jets uses MVA techniques using inputs containing many different sources of information.

$b$-jets originate from hadrons containing $b$-quarks. Compared to $c$- and $light$-quarks, $b$-quarks have unique properties that can be exploited.
The $b$-hadron lifetime of around 1.5~ps allows the particle to travel further from the beam pipe before decaying, thus creating a secondary vertex.
A second useful property of the $b$-quark is its large mass. Thus, when it decays it creates more charged particles.

A BDT algorithm called \mvtwo{}~\cite{ATL-PHYS-PUB-2015-022} is used for this purpose.
 The most important inputs are the tracks obtained from the ID. Additionally, there are outputs from three algorithms~\cite{Aad:2009wy,ATLAS-CONF-2011-102} that get fed into \mvtwo.
\begin{itemize}
\item Impact Parameter based algorithms (IP2D and IP3D): These algorithms make use of impact parameter information in the longitudinal ($z_{0}$) and transverse ($d_0$) planes.
A likelihood is built using the signed significance of each ($z_{0}/\sigma_{z_{0}}$, $d_{0}/\sigma_{d_{0}}$).
IP3D makes use of transverse and longitudinal impact parameters (thus making it 3D), while IP2D only uses the transverse impact parameters.
\item Secondary Vertex finding algorithm (SV): This algorithm exploits the secondary vertex due to the longer decay lifetimes of $b$-quarks.
\item Decay chain multi-vertex algorithm (JetFitter): This algorithm attempts to reconstruct the full $b$- and $c$- hadron decay chains. A Kalman filter algorithm tries to find the different vertices from which the $b$- and $c$-hadrons originate.
\end{itemize}
The jet \pt and \eta are also included in the training of \mvtwo.

Several working points are defined in which analyses can specify the required $b$-jet efficiency. This is defined as $\epsilon_b = N_{b\text{-tagged}}/N_{b\text{-true}}$. In this analysis, the working point $\epsilon_b = 77\%$ is used in the main event selections.

Another way to use information from $b$-tagging is in the form of the \emph{pseudo-continuous weights}. For each jet in an event, a $b$-tagging score from the \mvtwo algorithm is assigned. This creates a histogram of fixed bins corresponding to different efficiencies. 
\emph{Pseudo-continuous} refers to the fact that there are only five bins that are calibrated and the distribution is not truly continuous.
 These are shown in Table~\ref{tab:btaggin_bins}. These distributions for leading, sub-leading and subsub-leading jets are used extensively in the Event-level Discriminator (Chapter~\ref{sec:ELD}).

\begin{table}[!htbp]
	\centering
	\begin{tabular}{cc}
	\hline
	Bin \# & $\epsilon_b$ \\
	\hline
	\hline
	1 & $>85\%$ \\
	2 & $77\%-85\%$ \\
	3 & $70\%-77\%$ \\
	4 & $60\%-70\%$ \\
	5 & $<60\%$\\
	\hline
	\end{tabular}
	\caption{The definition of the pseudo-continuous $b$-tagging weights by bin number.}
	\label{tab:btaggin_bins}
\end{table}

\section{Missing transverse energy}
The \ATLAS detector allows for some energy in collisions to go undetected. This can be due to technological imperfections or limited detector coverage, but also the nature of particles such as neutrinos. Often perceived as massless (a good approximation), neutrinos are not charged and do not leave any tracks in the ID or energy deposits in the calorimeters. They pass undetected through the detector. 
However, due to momentum conservation in the plane transverse to the beam axis of collider experiments, the overall transverse momentum of a collision should sum to 0. The residual energy is known as \emph{missing transverse energy} (\MET)~\cite{ATL-PHYS-PUB-2015-023}.
The approximation of \MET is given by the negative vectorial sum of the momenta for each of the calibrated objects; electrons, photons, taus, muons, jets, and a soft term. The soft term enters as the sum of all transverse momenta for tracks that have not already been matched to a physics object, but are still associated to the primary vertex.
Thus, 

\begin{align}
\label{eq:met}
 E^{\text{miss}}_{x(y)} &= -1 \cdot (p_{x(y)}^{e} + p_{x(y)}^{\gamma} + p_{x(y)}^{\tau} + p_{x(y)}^{\mu} + p_{x(y)}^{\text{jets}}  + p_{x(y)}
^{\text{soft}}) \\
\MET &= \sqrt{(E^{\text{miss}}_x)^2 + (E^{\text{miss}}_y)^2}.
\end{align}

The \MET can essentially be thought of as a particle originating from the hard scattering. This is useful for some BSM searches, which predict other minimally-interacting particles.

\section{Object overlap removal procedure}

Particle tracks and energy deposits can in some cases be used to reconstruct multiple objects. To prevent double counting a standard overlap removal procedure is performed.
First, electrons that share tracks with any other electrons are removed.
Any electron found sharing a track with a muon is then removed.
Any jet found within $\Delta R < 0.2$ of an electron is removed. 
Then, if any electron is found within $\Delta R < 0.4$ of a jet it is removed.
Any jet with less than 3 tracks associated to it found within $\Delta R < 0.2$ of a muon is removed.
Then, for any muon subsequently found within $\Delta R < 0.4$ of a jet is removed.
Any photon found within a $\Delta R < 0.4$ of an electron or a muon is removed.
Any jet found within  $\Delta R < 0.4$ of a \loose photon is removed.
%

\section{Event overlap removal and categorisation}

Double counting of prompt photons may occur due to the overlap of MC samples (Chapter~\ref{sec:MCsim}), for example, in \ttgamma and \ttbar, and $W/Z\gamma$ and $W/Z$+jets processes.
Thus, a truth based matching scheme is used to remove those types of events.
The \emph{truth origin} defines the mother type of the particle of interest while \emph{truth type} indicates the actual particle type.
It is possible to classify the type of photon into four general categories based on these truth labels. These will be used extensively and are called: \emph{signal}, \emph{\efakes}, \emph{\hfakes} and \emph{\Other}.

\subsubsection{\efakes}
Photons are classified as ``\efakes{}" when a reconstructed photon is truth matched to an electron or if there is a truth electron in a cone of radius 0.2 around the reconstructed photon.

\subsubsection{Hadronic fakes}
Photons are classified as ``\hfakes{}" when the particle's truth origin corresponds to a hadron. The truth type demands that this photon originates from a boson or a lepton and does not originate from initial or final state radiation. A further requirement is that the reconstructed photon is not an \efake based on the above mentioned criteria.

\subsubsection{Signal}
Signal events are categorised by reversing the requirements placed on the \efake and \hfake categories. Specifically, the reconstructed photon should not be classified as originating from a hadron, should not be a ``background photon" and should not be truth matched as or to an electron. These requirements are met by the nominal signal MC.

\subsubsection{\Other}

Photons classified as ``\Other{}" are defined in the same way as those in signal. The difference being that this category does not include signal MC but rather diboson, single-top and \ttbar{}$V$ MC samples. Depending on the channel, this category can also contain \Zgamma for \chljets and \Wgamma for \chll channels. This is to account for when the \Zgamma or \Wgamma backgrounds are negligible or dominant in the respective channel.

The signal and background processes introduced above will be presented in more detail in Chapter~\ref{sec:ttgammaprocess}.

\FloatBarrier


\chapter[Top-quark pairs and photons at \texorpdfstring{$\sqrt{s}}{sqrt(s)} = 13$~\TeV]{Top-quark pairs and photons at $\sqrt{s} = 13$~\TeV}
\label{sec:ttgammaprocess}

This chapter focuses on the general analysis strategy and the signal and backgrounds for the \ttgamma process at $\sqrt{s} = 13$~\TeV. 
Section~\ref{sec:analysisstrategy} presents the method of extracting the cross-sections using a maximum likelihood fit.
Section~\ref{sec:eventselections} presents the triggers and event selections used to obtain a signal region.
This is followed by the \efake (Section~\ref{sec:efake}) and \hfake (Section~\ref{sec:hfake}) background data-driven scale factor extractions using the tag-and-probe and ABCD method, respectively.
Section~\ref{sec:fakelepton} shows the data-driven estimate of the \QCD background using the matrix method.
Section~\ref{sec:promptbackgrounds} presents the prompt photon background contributions.
Finally, Section~\ref{sec:preFitDists} presents kinematic distributions in the signal region with all scale factors derived in the previous sections applied.



\section[Analysis strategy]{Analysis strategy}
\label{sec:analysisstrategy}

This section discusses the strategy and methods to obtain fiducial \xsecModifyNoun results. 
The fiducial region and theoretical NLO predictions are presented in Section~\ref{sec:fiducial}.
Then, the method to extract the \xsecModifyNoun via a maximum likelihood fit is discussed in Section~\ref{sec:MLF} 
Lastly, the different types of fitting scenarios are presented in Section~\ref{sec:fitscenarious}.

Important to note is that in addition to fiducial \xsecModifyNoun measurements, the analysis team also carried out \emph{differential} \xsecModifyNoun measurements. These are not presented in this thesis, and more information can be seen in~\cite{ATLAS-CONF-2018-048}.

\subsection{Fiducial cross-section measurements}
\label{sec:fiducial}

The final \xsecModifyNoun measurements are carried out in a subset of the phase space defined by the event selections presented further in this chapter. This is labelled the \emph{fiducial region}. This new region is defined at particle level and is designed to be as close as possible to the region defined at reconstruction level (when detector effects and sensitivities come into play), thus, it is used to reduce the impact that detector effects may have on the measurement.

The object requirements at particle level are defined as follows:
Leptons do not originate from hadron decays and have $\pt > 25$~\GeV and $|\eta| < 2.5$.
Photons have $\ET > 20$~\GeV and $|\eta| < 2.37$ and must not be from hadron decays or dressed to a lepton. 
The photon is dressed to the lepton if it does not originate from a hadron and is within $\Delta R < 0.1$ around the lepton.
They must also be isolated in that $\pt^{\text{cone}} < 0.1$.
Jets are re-clustered with the anti-$k_t$ algorithm with a radius of $\Delta R=0.4$ and have $\pt > 25$~\GeV and $|\eta| < 2.5$.
A ghost matching scheme~\cite{Cacciari:2008gn} is used to determine the flavour of the jets.

The event requirements are as follows:
The \chljets (\chll) channels require at least four (two) jets with at least one of them being tagged as coming from a $b$-hadron. Exactly one (two) leptons are required. Events are removed if $\Delta R(\gamma, jet) < 0.4$ and $\Delta R (\gamma,lep) < 1.0$.

We can define two quantities; the acceptance ($A$), 
\begin{align}
A &= \frac{N^{\text{fid}}_{\text{gen}}}{N^{\text{all}}_{\text{gen}}} \label{eq:acceptance},
\end{align}
and the correction ($C$),
\begin{align}
C &= \frac{N_{\text{reco}}}{N^{\text{fid}}_{\text{gen}}} \label{eq:correction}.
\end{align}
The acceptance factor ($A$) gives the ratio of the generated signal particle level events that pass the fiducial selections ($N^{\text{fid}}_{\text{gen}}$) to the total number of generated signal events ($N^{\text{all}}_{\text{gen}}$). It is used to extrapolate the theoretical cross section to a smaller phase space.
The correction factor gives the ratio of the signal events that pass the full event selection at reconstruction level ($N_{\text{reco}}$) to the generated particle level signal events that pass the fiducial selections. It essentially turns the final \xsec into a fiducial \xsec. 

Experimentally, a fiducial \xsec for a given channel is determined from the equation
\begin{align}
\label{eq:cross_section}
\sigma = \frac{N_{\text{data}} - \sum_{i} N_{\text{bkg}, i}}{\mathcal{L}  \times C},
\end{align}
where $N_{\text{data}}$ is the number of data, $N_{\text{bkg}, i}$ is the sum of all background contributions, $\mathcal{L}$ is the luminosity, and $C$ is the correction factor from Equation~\ref{eq:correction}.

In practice, a result often quoted is the scaling of the measured cross section to the (NLO) theoretical prediction. The so called \emph{signal strength}, $\mu$, is defined as 
\begin{align}
\label{eq:signalstrength}
\mu = \frac{\sigma_{\text{fid}}}{\sigma^{\text{NLO}}_{\text{fid}}}, 
\end{align}
where the fiducial NLO theory prediction is 
\begin{align}
\label{eq:nlo}
\sigma^{\text{NLO}}_{\text{fid}} = \sigma^{\text{LO}}_{\text{total}} \times A \times k.
\end{align}
From the $k$-factors and the acceptances defined earlier, we can calculate the theoretically predicted NLO cross sections for each channel with their total uncertainties. These values are summarised in Table~\ref{tab:fiduxsec}.
It is important to note that tau leptons decay hadronically ($\tau^{-} \to \nu_{\tau} d \bar{u}$) and leptonically ($\tau^{-} \to \nu_{\tau} l^{-} \bar{\nu}_{l}$, $l=e,\mu$) and so a dedicated measurement in the ``$\tau$-channel" is difficult. In this analysis, in addition to the \chljets and \chll channels, five further channels are defined (\chejets, \chmujets, \chee, \chmumu and \chemu) in which \xsecModifyNoun measurements are performed. These channels do not distinguish electrons, muons and jets from tau leptons. The theoretical \xsec{} calculations are consistent and follow this same strategy. Thus, the \chejets, \chmujets, \chee, \chmumu and \chemu measurements should \emph{not} be thought of as measurements made in the \ttbar channels presented in Figure~\ref{fig:ttbardecay}.

 \begin{table}[h!]
 \centering
\begin{tabular}{c|c|c|c|c|cc}
\hline
 &\multicolumn{2}{c|}{\chljets}&\multicolumn{3}{c}{\chll}  \\\hline
 \hline
Cross section [fb]   & \multicolumn{2}{c|}{495 $\pm$   99}&\multicolumn{3}{c}{63 $\pm$    9}   \\

\hline
 & \chejets&\chmujets&\chee&\chemu&\chmumu\\\hline
Cross section [fb]  & 247 $\pm$   49& 248 $\pm$   50&  16 $\pm$    2&  31 $\pm$    5&  16 $\pm$    2\\
\hline

 \end{tabular}
 \caption{The theoretical fiducial NLO cross sections for each channel with their total uncertainties.}
 \label{tab:fiduxsec}
 \end{table}

\FloatBarrier

\subsection{Maximum likelihood fit}\label{sec:MLF}

To extract the signal strength described in Equation~\ref{eq:signalstrength} a binned maximum likelihood fit is performed on a distribution believed to be able to effectively separate signal and background processes. The Event-level Discriminator (\ELD) serves this purpose and will be presented in Chapter~\ref{sec:ELD}.
A general likelihood can be described as 
\begin{equation}
\mathcal{L} = \mathcal{L}(\mu; \ELD, \theta) = P \times G.
\end{equation}
This gives the probability of observing a certain signal signal strength for the \ELD distribution given a set of systematic uncertainties. The first term on the right refers to a Poisson distribution modelling the signal region. The second term refers to a Gaussian distribution which constrains the nuisance parameters (NPs).
This can be expanded: 

\begin{equation}
\label{eq:likelihood}
\mathcal{L} = \prod_{j \in \text{bins}} P(N_{j}^{\text{data}} | \mu \cdot N_{j}^{\text{sig.}}(\vec{\theta})+\sum_{b\in \text{bkgs.}}N_{j}^{b}(\vec{\theta})) \times  \prod_{t} G(0 | \theta_{t},1).
\end{equation}
The first Poisson distribution gives the probability of observing a number of data events ($N^{\text{data}}_{j}$) for a given number of signal ($\mu \cdot N^{\text{sig.}}_{j}$) and total background ({$N^{b}_{j}$}) events in bin $j$ of the \ELD.
The number of signal and background events in each bin are parameterised by all NPs ($\vec{\theta}$), where each NP ($\theta_t$) is constrained by a Gaussian distribution. Each NP is constrained with an ``up" and ``down" variation\footnote{The up and down choice of names is arbitrary and do not necessarily reflect a positive and negative variation. At times both variations can be positive or negative.}, where each variation is described as $\pm 1 \sigma$ away from the nominal value. 
 
A NP can also enter as an un-constrained or free floating parameter where no prior is assumed. A typical example for a free floating NP is a dominant background's normalisation.
The NPs give the signal and background the freedom to adjust their predictions to observed data.
The likelihood function is then maximised where the global maximum is the point at which the parameters are estimated most accurately. 
In practice the negative log-likelihood is minimised, with $\pm 1\sigma$ taken as the uncertainties. 
The likelihood and fits were built and performed using custom wrappers on top of the RooFit framework~\cite{Verkerke:2003ir}.

\subsection{Fit scenarios}
\label{sec:fitscenarious}

\begin{figure}[!htbp]
\centering
    \includegraphics[width=0.88\linewidth]{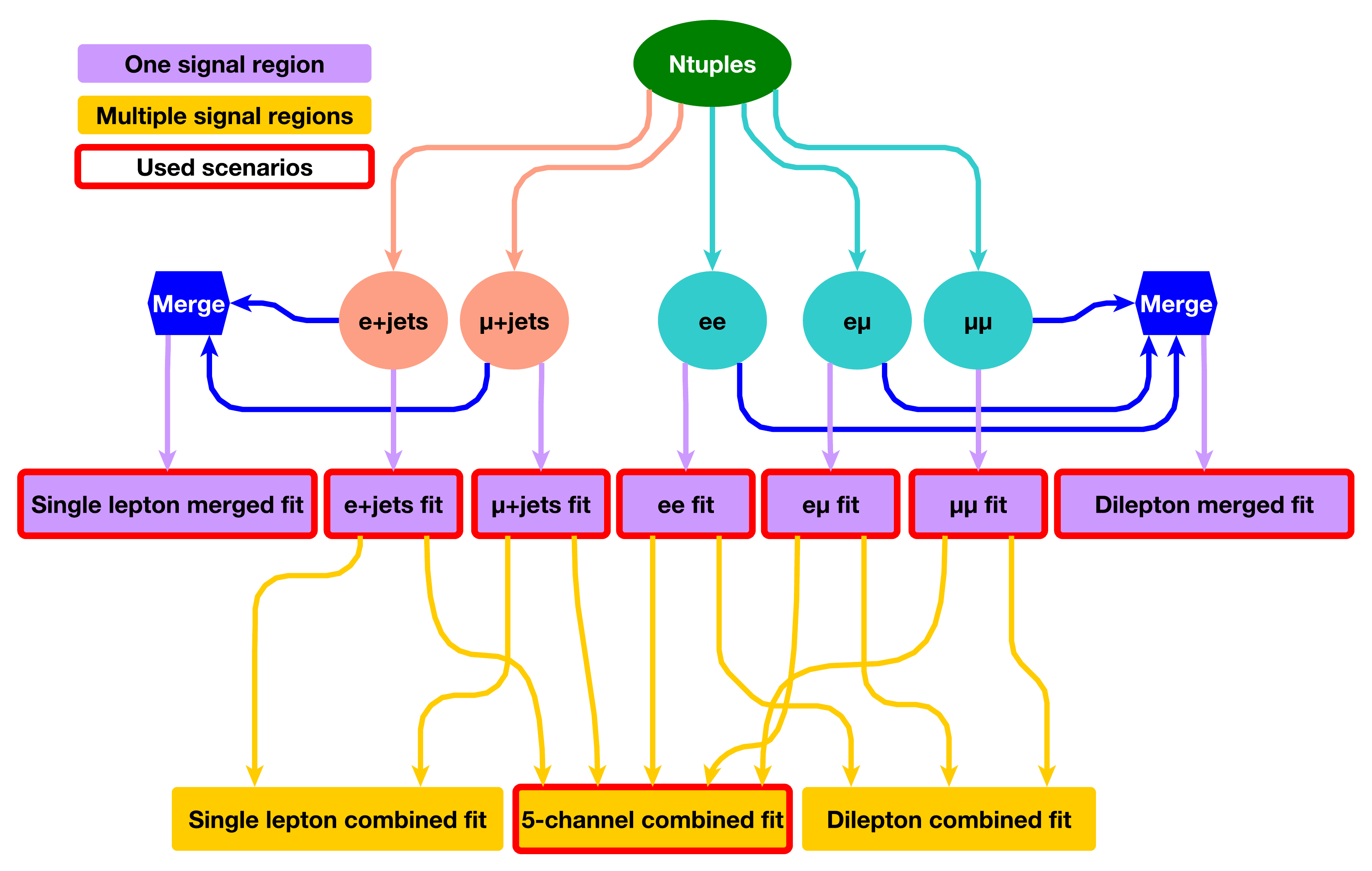}
\caption [] {An overview of the different scenarios for how fits can be carried out.

Purple (second to last row) indicates a merged or single region fit, yellow (last row) indicates a combined or multiple region fit. The boxes with borders indicate the final chosen strategy.}\label{fig:fitscenarios}
\end{figure}

There are multiple ways to define the channel that enters the maximum likelihood fit.
An overview of the scenarios is shown in Figure~\ref{fig:fitscenarios} and explained below.

For the individual channels this is straightforward and separate likelihood fits in each of the five channels (\chejets, \chmujets, \chee, \chemu and \chmumu) are performed. 
When it comes to carrying out \chljets and \chll channel fits there are two options considered. 

\begin{itemize}
\item Merged: This is when histograms from the \chejets and \chmujets channels are essentially added together to form the \chljets channel, and histograms from \chee, \chemu and \chmumu are added to form the \chll channel. 
The advantage (and ultimately deciding factor) to using this method is that only one signal region is used in the fit, meaning there is only one final signal strength and set of histograms for each \chljets and \chll channel. This makes differential measurements more straightforward. 
For some background studies and the training of the \ELD, the merged channels were used to take advantage of the increased statistics.

\item Combined: This is when multiple signal regions are fit simultaneously, which allows the possibility for nuisance parameters to be more effectively constrained. Essentially, more terms are added to the likelihood in Equation~\ref{eq:likelihood}.  For example, a ``combined" fit could be carried out on the \chejets, \chmujets, \chee, \chemu and \chmumu histograms, which outputs one signal strength but five sets of new histograms. 
This strategy is used only for a 5-channel combined fit to obtain an ``inclusive" cross-section.

\end{itemize}

Figure~\ref{fig:fitscenarios} shows the chosen scenarios used for the final fits, which have red borders. This includes the five individual channels, the \chljets and \chll merged fits, and the ``5-channel combined fit".

\section{Signal region event selections}
\label{sec:eventselections}

Selections are placed on the objects defined in Chapter~\ref{sec:objectID} to suppress background contributions and thus enrich the \ttgamma purity. The strategy is to follow the event selections required for \ttbar, but with the additional requirements on the photon. The triggers used are presented first, followed by the selection cuts.

\subsubsection{Triggers}
Different HLT triggers are used for data taking periods in 2015 and 2016~\cite{TRIG-2016-01,ATL-DAQ-PUB-2017-001,Jones:2290123}.
For this analysis single lepton triggers for electrons and muons are used to define parameter spaces of interest. These are described below.

For the 2015 electron candidates, the first trigger requires a minimum \ET threshold of 24~\GeV and a \medium identification requirement. It is seeded by a L1 trigger requiring an \ET of 20~\GeV and applies a veto against any energy deposited in the hadronic calorimeter (L1EM20VH). At $\ET > 60$~\GeV, this is superseded by a trigger without the L1 seed, and at $\ET>120$~\GeV a trigger with \loose identification is used. 
For 2016 electron candidates of threshold $\ET > 26$~\GeV, a \tight identification, and a \loose isolation is required. At $\ET > 60$~\GeV the isolation requirement is dropped and the identification relaxed to \medium. At $\ET > 140$~\GeV the identification is further relaxed to a \loose identification requirement. In all cases the 2016 triggers no longer require transverse impact parameter cuts.

For the 2015 muon candidates, the trigger requires a minimum of $\pt \ge 20$~\GeV with a \loose isolation criterion. This trigger is seeded by the L1 trigger where the muon candidate has a minimum \pt of 15~\GeV (L1MU15).
Once the \pt is greater than 50~\GeV, the trigger is superseded by another trigger that does not require isolation. This compensates for a decrease in efficiency due to higher \pt muons.
The 2016 muon trigger requires candidates to have transverse momentum greater than 26~\GeV. A \medium isolation is also required. Again, at \pt greater than 50~\GeV the trigger is superseded by another trigger that does not require isolation.
A summary of all triggers used is shown in Table~\ref{tab:triggers}.


\begin{table}[!htb]
	\centering
    \scalebox{0.90}{
    \begin{tabular}{|c|c|c|c|c|}
    \hline
    Year & \multicolumn{4}{c|}{electron triggers} \\ \hline
    \hline
    
    \multirow{4}{*}{2015} & \pt threshold [\GeV] & Identification menu & Isolation menu & L1 seed  \\ \cline{2-5}
    				     & $\ge$24 & \medium & None & L1EM20VH \\
    				     & $\ge$60 & \medium & None & - \\
    				     & $\ge$120 & \loose & None & - \\		     
				      \hline 
  \multirow{3}{*}{2016}   & $\ge$26 & \tight & Gradient(\loose) & - \\
    				     & $\ge$60 & \medium & None & - \\
    				     & $\ge$140 & \loose & None & - \\
				      \hline

    Year & \multicolumn{4}{c|}{muon triggers} \\ \hline
    \hline

    \multirow{3}{*}{2015} & \ET threshold [\GeV] & Identification menu & Isolation menu & L1 seed  \\ \cline{2-5}
    				     & $\ge$20 & None & Gradient(\loose) & L1MU15 \\
    				     & $\ge$50 & None & None & - \\
				      \hline 
  \multirow{2}{*}{2016}   & $\ge$26 & None & Gradient(\medium) & - \\
    				     & $\ge$50 & None & None & - \\

    \hline
    \end{tabular}
    }
	\caption{The requirements of the lepton triggers considered in the event selections.}
	\label{tab:triggers}
\end{table}

\subsubsection{Signal region cuts}
\label{sec:cuts}

After potentially interesting data has been selected using single lepton triggers, using objects reconstruction methods as described in Chapter~\ref{sec:objectID}, further selection cuts are made to enhance the purity of the \ttgamma signal.

The presence of electrons and muons in an event is characterised by the triggers mentioned above. Additional cuts are applied.
Electrons are required to have $\pt > 25$~\GeV ($\pt > 27.5$~\GeV) for the 2015 (2016) data, while 
muons are required to have $\pt > 27.5$~\GeV for the 2015 and 2016 data. 
Only one reconstructed lepton is required for the \chljets channels, while exactly two leptons of opposite sign are required for the \chll channels. 
All electrons are required to have \tight identification, while muons require \medium identification. Both are required to be isolated based on calorimeter and track based information to obtain high efficiencies. 

At least four (two) jets are required in the \chejets and \chmujets (\chee, \chemu and \chmumu) channels. At least one of these reconstructed jets should be tagged as a $b$-jet with a working point efficiency of 77\%.

In all the \chll channels a minimum invariant mass between the two leptons is such that $m(l,l) > 15$~\GeV. This suppresses low mass Drell-Yan events where a quark and an anti-quark form a lepton pair through the propagation of a virtual photon. Low mass resonances such as the $J/\psi$ ($c\bar{c}$) and the $\Upsilon$ ($b\bar{b}$) are also suppressed.
In the \chee and \chmumu channels an invariant mass veto of the two leptons is placed around the mass of the $Z$-boson, such that $m(l,l) \notin$ [85,95]~\GeV.
Similarly, an invariant mass veto between the two leptons and the photon ($m(l,l,\gamma)$) is placed in the same way around the mass of the $Z$-boson for the same channels.

In the \chee and \chmumu channels a cut on the \MET is used such that $\MET > 30$~\GeV.

In the \chejets channel a veto is placed on the invariant mass of the photon and the electron  such that $m(\gamma,e) \notin$ [85,95]~\GeV. This suppresses \efakes from $Z$-boson decays.

To enhance the \ttgamma signal region, cuts are made on photons. In all channels, exactly one photon of $\pt>20$~\GeV is required. This photon needs to be \tight and isolated.
A final cut is placed on the $\Delta R$ distance between the photon and the lepton, $\Delta R(\gamma,l) > 1.0$. This cut reduces the contribution of photons from radiative decay, as shown in Figure~\ref{fig:rd}.

\begin{figure}[!htbp]
\centering
\subfloat[\chljets]{
\includegraphics[width=0.4\linewidth]{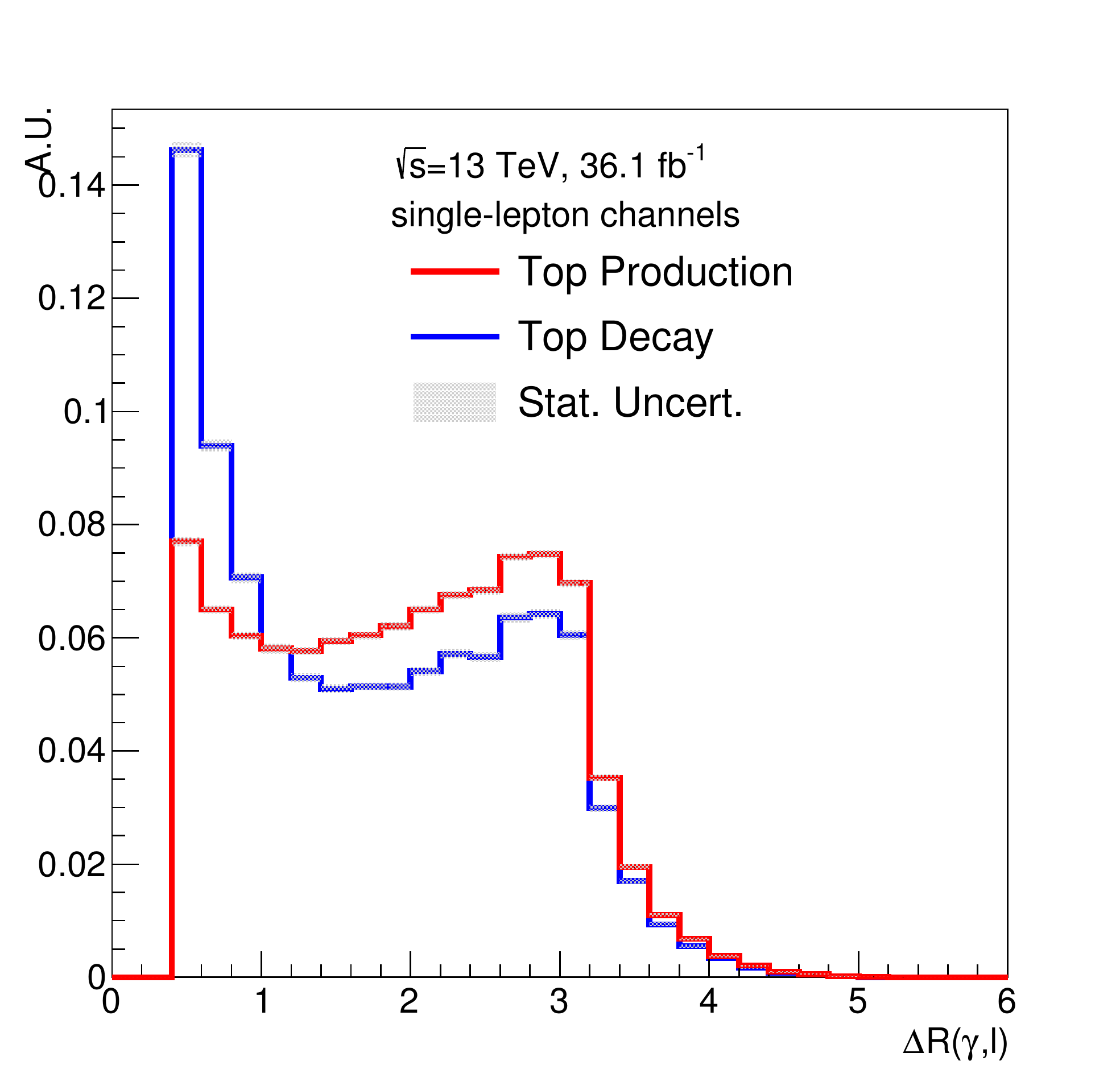}
}
\subfloat[\chll]{
\includegraphics[width=0.4\linewidth]{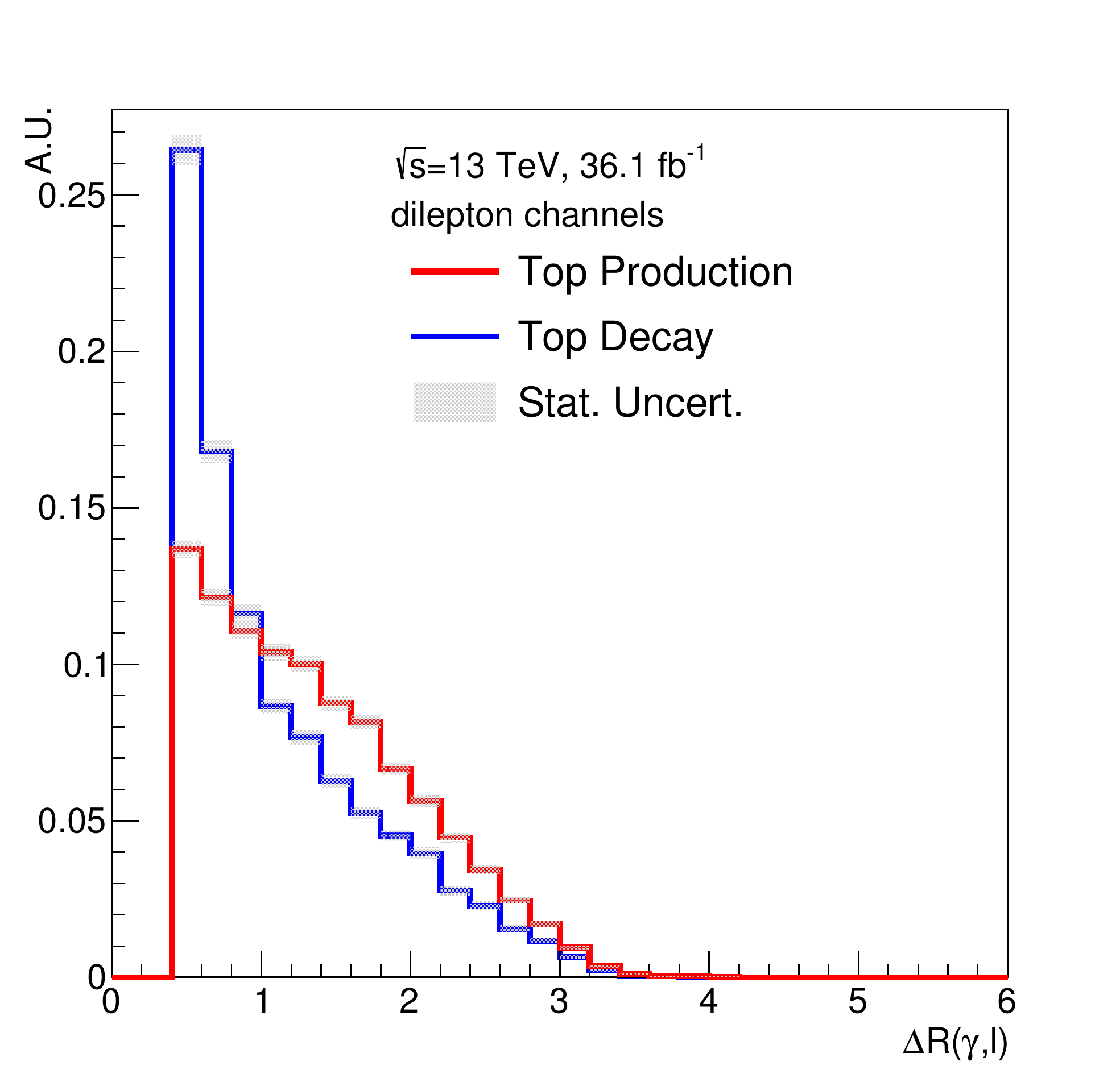}
}
\caption [] {The $\Delta R$ distance between the lepton and the photon. A cut of $\Delta R > 1.0$ was chosen as the ideal value.}
\label{fig:dRcuts}
\end{figure}

The $\Delta R(\gamma,l)$ distribution with all cuts mentioned above applied (except the final $\Delta R$ cut) is shown in Figure~\ref{fig:dRcuts}.
Top production includes all types of processes shown in Figure~\ref{fig:rp}, while top decay ideally contains events only from top quark decay products. There is some ambiguity with truth records due to the matrix element calculation and recording off-shell particles.
 Even after this cut there is still a considerable contribution of photons from radiative decay. The second peak at $\Delta R(\gamma,l)\approx \pi$ for the \chljets channels occurs when the photon is radiated from the top quark or top quark decay products which lies opposite (in the $\eta$-$\phi$ plane) to the $W$-boson decaying into a lepton and neutrino. 
Future analyses can benefit from MVA techniques to further reduce this background contribution.

Validation plots for the region defined above but before photon selections can be seen in Appendix~\ref{sec:validation_plots}, where good agreement is seen between data and MC. This is essentially a \ttbar validation region.

A summary of all the cuts mentioned above is shown in Table~\ref{tab:cuts}. Table~\ref{tab:prefiteventYields} shows the yields for each process before any data-driven scale factors have been applied. Only statistical uncertainties are included.
In the \chejets  (\chmujets) channel the \efake contributes is the largest background and makes up around 14\%  (12\%) of the signal + background MC. The \Wgamma background contribution is second largest with approximately 10\% (11\%).
This is followed by the \hfake background of around 8\%  (9\%). The biggest difference between the two channels is the \QCD contribution which makes up approximately 5\% (1\%).
For the \chee  (\chmumu) channel the largest background contribution comes from the \Zgamma background at around 12\% (23\%), with the other background processes essentially negligible. Because of the much smaller contribution of \Zgamma in the \chemu channel, this is grouped together as ``\Other{}". For the \chemu channel, all backgrounds combined make up about 6\% of the signal+background contribution making it the purest channel.

\begin{figure}[!htbp]
\centering
\subfloat[Merged \chljets channels]{
\includegraphics[width=0.45\linewidth]{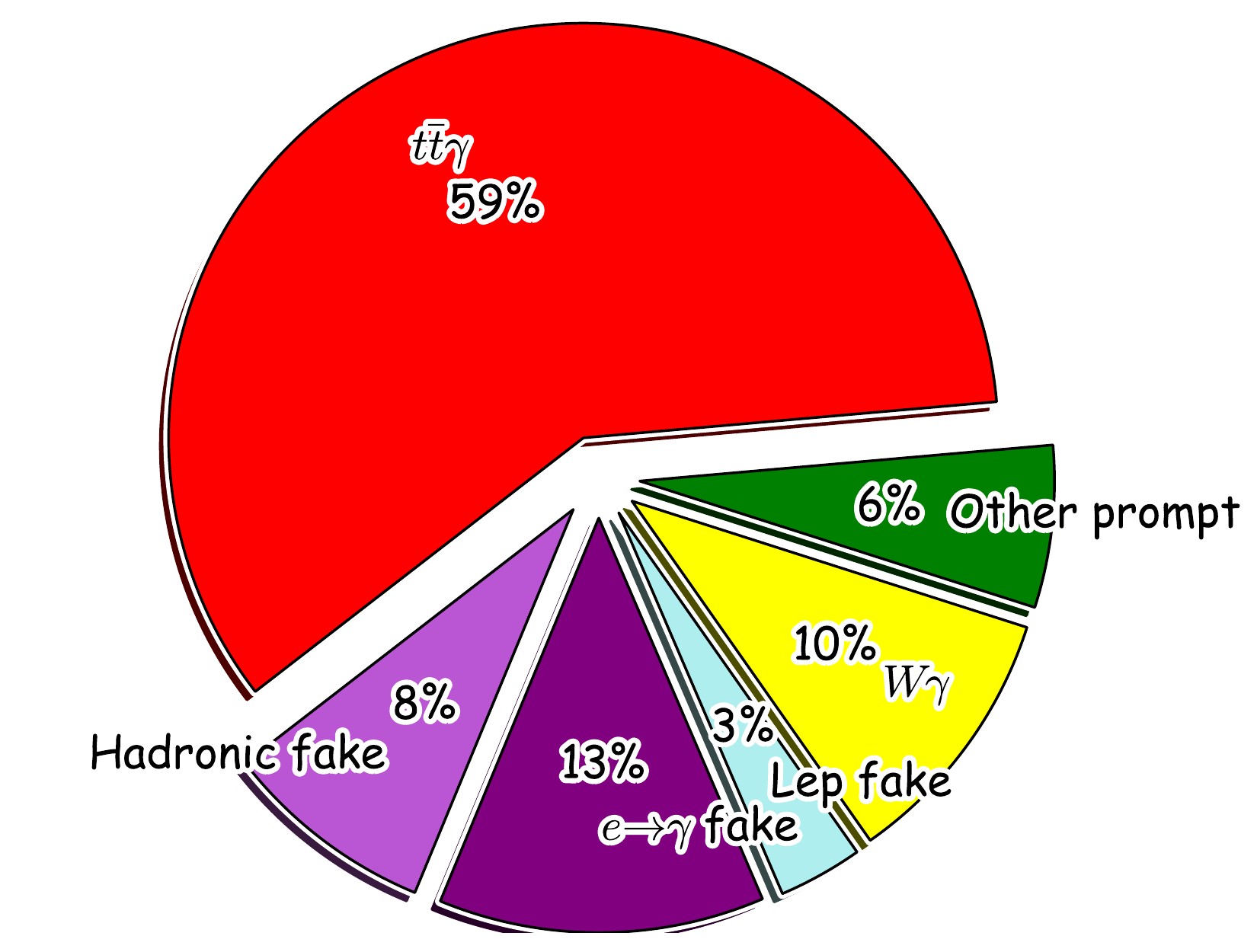}
}
\subfloat[Merged \chll channels]{
\includegraphics[width=0.45\linewidth]{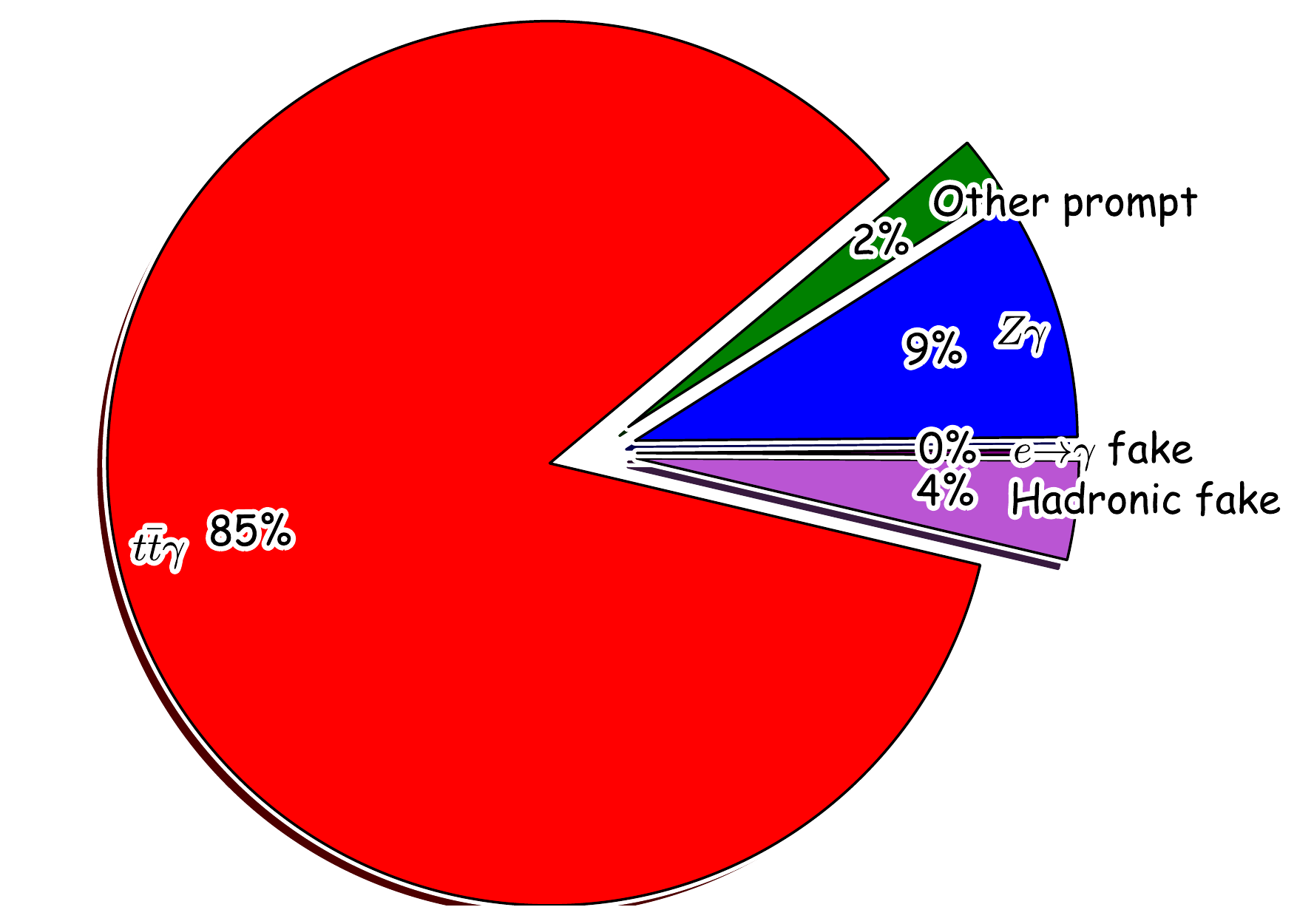}
}
\caption{Relative contributions of each process for the \chljets and \chll channels after merging. No data-driven scale factors have been applied.}
\label{fig:pieplots}
\end{figure}

Thus, it is deemed that the composition of backgrounds for the \chejets and \chmujets channels are similar enough that these channels are often merged to form the \chljets channel. Similarly, for the \chee, \chmumu and \chemu channels, these are combined to form the \chll channel. 
This is often due to simplicity but also because the merged channels benefit from the increased statistics in many studies. 
Figure~\ref{fig:pieplots} summarises the processes that make up the combined \chljets and \chll channels by percentage, before data-driven scale factors have been applied.
The different sources of the background processes as well as how scale factors are derived for various backgrounds will be discussed in upcoming sections.

\begin{table}[!htb]
	\centering
        \begin{tabular}{|c|c|c|c|c|c|}
        \hline
        Channel & \chejets & \chmujets & \chee & \chmumu & \chemu \\
        \hline
        \hline
        \multirow{2}{*}{Common}  & \multicolumn{5}{c|} {Primary vertex, Event cleaning, Run number, etc.} \\
        \cline{2-6}
          & 1 $e$ & 1 $\mu$ & 2 $e$, OS & 2 $\mu$, OS & 1 $e$ + 1 $\mu$, OS \\
        \cline{2-6}
        \hline
        Photon  & \multicolumn{5}{c|} {1 $\gamma$ with \pt $>$ 20 GeV} \\
        \hline
        Jets	& \multicolumn{2}{c|} {$\geq$ 4} & \multicolumn{3}{c|} {$\geq$ 2} \\
        \hline
        $b$-jets	& \multicolumn{5}{c|} {$\geq$ 1} \\
        \hline
        \multirow{2}{*}{$m(\ell, \ell)$}  & \multicolumn{2}{c|} {-} & \multicolumn{2}{c|} {$\notin [85,95]$~\GeV} & - \\
        \cline{2-6}
                 & \multicolumn{2}{c|} {-} & \multicolumn{3}{c|} {$> 15$~GeV} \\
	
        \hline
        $m(\ell, \ell, \gamma)$  & \multicolumn{2}{c|} {-} & \multicolumn{2}{c|} {$\notin [85,95]$~\GeV} & -\\
        \hline
        \MET  & \multicolumn{2}{c|} {-} & \multicolumn{2}{c|} {$>$ 30 GeV} & -\\
        \hline
        $m(\gamma, e)$ & $\notin [85,95]$~\GeV & \multicolumn{4}{c|} {-} \\
        \hline
        $\Delta R(\gamma, \ell)$ & \multicolumn{5}{c|} {$>$ 1.0} \\ 
        \hline
        \end{tabular}
	\caption{A summary of the event selections for each channel.}
	\label{tab:cuts}
\end{table}

\begin{table}[!htbp]
\begin{center}
\scalebox{0.85}{
\begin{tabular}{|c|r|r|r|r|r|}
\hline

 & \chejets  & \chmujets & \chee & \chmumu & \chemu  \\ \hline \hline

\ttgamma & 3207 $\pm$ 14 &  3200 $\pm$ 13 &  145 $\pm$ 3.0 &  174 $\pm$ 3.1 &  401 $\pm$ 5.2 \\ \hline  

\hfake & 441 $\pm$ 13 &  457 $\pm$ 14 &  7 $\pm$ 2 &  6 $\pm$ 2 &  18$\pm$ 3 \\ \hline  

\efake & 774 $\pm$ 21 &  604 $\pm$ 16 &  0.5 $\pm$ 0.3 &  0.2 $\pm$ 0.5 &  1 $\pm$ 1 \\ \hline  

\QCD & 293 $\pm$ 35 &  63 $\pm$ 22  & - & - & - \\ \hline

\Wgamma & 541 $\pm$ 29 &  579 $\pm$ 34  & - & - & - \\ \hline  

\Zgamma & - & - & 21 $\pm$ 4 &  54 $\pm$ 14 & - \\ \hline

\Other & 411 $\pm$ 47 &  275 $\pm$ 15 &  5 $\pm$ 1 &  4 $\pm$ 1 &  8 $\pm$ 1 \\ \hline  

Total & 5665 $\pm$ 71 &  5178 $\pm$ 50 &  179 $\pm$ 5 &  238 $\pm$ 14 &  428 $\pm$ 5.9 \\ \hline  

Data & 6002 &  5660 &  196 &  233 &  473 \\ \hline 

\end{tabular}
}
\caption{Yields for signal and backgrounds for each channel. Only statistical uncertainties are included. Scale factors for the different backgrounds are not included.}
\label{tab:prefiteventYields}
\end{center}
\end{table}

\FloatBarrier
\section[\texorpdfstring{\efake}{Electron fake} background]{\efake background}
\label{sec:efake}

An important background to the \ttgamma \chljets channels arises when electrons are misidentified as photons. This makes up around 13\% of total events in the signal region, and thus is the largest background contribution.
This occurs mainly from the $\ttbar$ dileptonic decays, specifically from the \chemu and \chee channels. Another source is from $Z\to$ \chee decays, hence the invariant mass window cut in the \chejets channel. For this reason the \efake background is negligible in the \chll channels.
The approach to estimating the \efake background follows a data-driven \emph{tag-and-probe} method. The derived scale factors are used to correct the number of fake photons predicted by the MC samples.

Two control regions are needed to carry out this study. The first is a region enriched with fake photons called the $e\gamma$ control region. Here, one electron is trigger matched and at least one photon is required. The azimuthal angle between the electron and the leading \pt photon is required to be larger than 150 degrees implying that they are back-to-back.
An invariant mass requirement between the electron and the photon is placed so that $m(l,\gamma) \in [40,140]$~\GeV.
In this CR the electron is called the \emph{tag electron} while the photon is called the \emph{probe photon}.
This CR implies that an electron has either been missed in the reconstruction, or that the photon is fake.
The second control region, called the $ee$ CR has the same definitions as the $e\gamma$ CR. The only key difference is that the requirement on the photon is replaced with a requirement to be an electron of opposite charge to the initial tag electron. This second electron now acts as the probe with the implication that this CR only contains real electrons.

Truth studies can be performed on the probe photon in the $e\gamma$ CR to determine the source of the \efake{}s. The study is done by looking at the seed cluster in the EM calorimeter and calculating the angular distance to the truth particle before simulation. 
In addition the probe photon's kinematics can also be compared to those of the probe electron (not shown in this thesis). From these studies, four cases have been identified:
\begin{itemize}
\item Approximately 60\% of the photons are simply mis-reconstructed as an electron. 

\item Approximately 25\% of the photons are mis-matched to a true photon. The reconstructed photon's \pt is larger than the truth photon's \pt by more than 10\%. In addition there is a truth electron which satisfies $\Delta R < 0.05$ to the seed cluster. 

\item Approximately 2\% of the photons are truth matched to a reconstructed photon with no electrons in the nearby vicinity. This indicates that these are actual prompt photons possibly from the decay $Z\to ee + \gamma$. These photons are not counted as fake.

\item Approximately 3\% of photons are matched to a true photon, have a \pt difference of less than 10\%, and true electron is within $\Delta R < 0.05$ of the photon. This indicates a real non-prompt photon.
\end{itemize}

We can infer a fake rate from MC, which is defined as

\begin{equation}
\text{FR}^{\text{e-fake}}_{\text{MC}} = \frac{N_{e,\gamma}}{N_{e,e}},
\label{eq:efakeSF}
\end{equation}
where $N_{e,\gamma}$ ($N_{e,e}$) is the number of events observed in the $Z\to e \gamma$ ($Z\to ee$) CR. The MC sub index means that this estimate is derived using truth information and the number of events described by the MC.
To define the data-driven fake rate, in both CRs the backgrounds that do not contribute to the $Z$-boson peak are subtracted using a sideband fit of the $m(l,\gamma)$ distribution. 
Equation~\ref{eq:efakeSF} can be adjusted so that

\begin{equation}
\text{FR}^{\text{e-fake}}_{\text{d.d.}} = \frac{N_{e,\gamma}^{\text{data}} - N_{e,\gamma}^{\text{non-Z}}}{N_{e,e}^{\text{data}} - N_{e,e}^{\text{non-Z}}},
\label{eq:efakeSFDD}
\end{equation}
where $N_{e,\gamma/e}^{\text{non-Z}}$ indicates the tails of the $Z$-boson peak in the respective CRs.

Finally, the ratio of the data-driven fake rate to the MC fake rates gives scale factors that can be applied as corrections to the \efake MC samples, 

\begin{equation}
\text{SF}_\text{FR}^{\text{e-fake}} = \frac{\text{FR}_\text{d.d.}^{\text{e-fake}}}{\text{FR}_\text{MC}^{\text{e-fake}}}.
\end{equation}
To give a sense of the size, an overall SF is derived to be $0.97\pm0.02 \text{(stat)}$. However, in practice 2D fake rates and therefore SFs are derived for bins of \pt and \eta.

A number of variations to the above procedure are considered to account for systematic uncertainties. For the data-driven fake rate calculation the signal (background) modelling is varied from a Crystal-ball\footnote{A convolution of a Gaussian function as the main contribution and below some threshold in the low end part of the tail, a power law function.} (Bernstein 4th order polynomial) function to the MC predicted template (a Gaussian function). The MC model considered to subtract the prompt photon contribution from the $Z\to ee$ sample is switched to a dedicated $Z\to ee\gamma$ sample. Finally, the range of the fit in the low and high ends of the tail are shrunk between 5 and 10\%. Each of these contributions is summed in quadrature. To give a sense of the size, the overall relative systematic uncertainty on the \efake scale factor is 22\%. As with the scale factors, in practice, the systematic uncertainties are calculated in \pt and \eta bins.
Further theoretical uncertainties taken into consideration are explained in Chapter~\ref{sec:systematics}.

The final component of the \efake background is a closure test showing the extrapolation from the two CRs used above to essentially the signal region defined in Section~\ref{sec:cuts}. The difference being that the required photon is replaced by an electron. The \chejets channel is now dominated by \ttbar and $Z$+jets events. A ratio of data over prediction is calculated, which yields $0.98 \pm 0.01 \text{(stat.)}$. This is applied on top of the 3D scale factors in the \chljets channels. For the \chll channels, due to lack of \efake events, an extra 50\% normalisation uncertainty is assumed instead.

\FloatBarrier
\section{Hadronic fake background}
\label{sec:hfake}

The \hfake background plays an important role in many analyses when photons are required. It is for this reason that the Prompt Photon Tagger (Chapter~\ref{sec:PPT}) was developed. The majority of \hfake photons in the \chljets channels comes from \ttbar events where a final state jet radiates a photon. Small contributions also arise from $W$+jets and single top processes. In the \chll channels there are small contributions from the $Z$+jets events.
Before requiring an isolated photon this background is the dominant process in the \chljets channels. The new strategy for this analysis (compared to the 8~\TeV result) significantly reduces the contribution of \hfake photons by requiring isolated photons. In the \chljets (\chll) channel it contributes around 8\% (4\%) to the overall number of events. This is a non-negligible contribution (specifically in the \chljets channel) and so correct modelling needs to be ensured.
A more accurate estimation of this background that does not rely on MC modelling follows a data driven ABCD method.
Scale factors to match MC prediction to data are derived in control regions where the isolation and identification of the photon have been reversed, thus enriching the \hfake contribution.

\begin{figure}[!htbp]
\centering
    \begin{tikzpicture}[square/.style={minimum width=2.5cm,regular polygon,regular polygon sides=4}]
        \node at (0,0) [square,draw] (B) {B};
        \node at (2,0) [square,draw] (C) {C};
        \node at (0,2) [square,draw] (A) {A};
        \node at (2,2) [square,draw,fill=gray!40] (D) {D};
        
    \draw [-,thick] (-1.3,3) node (yaxis) [above] {}  |- (3,-1.3) node (xaxis) [right] {};

\node[below,yshift=-1.5cm] at (C) {Tight};
\node[below,yshift=-1.5cm] at (B) {!Tight};
\node[left,xshift=-1.5cm] at (A) {Isolated};
\node[left,xshift=-1.5cm] at (B) {!Isolated};

    \end{tikzpicture}
\caption{The control regions with the orthogonal photon definitions defined for the \hfake ABCD method. Region D is the signal region. An exclamation mark indicates the reverse of the cut.}
\label{fig:abcd}
\end{figure}
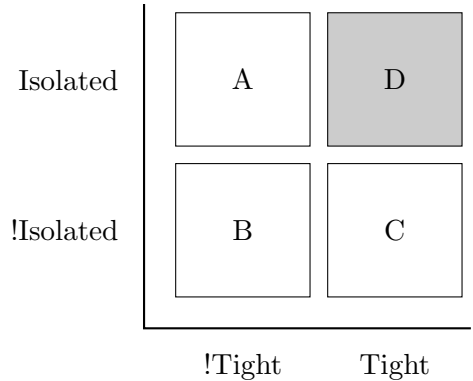

 The ABCD method relies on having four orthogonal and non-correlated regions. 
Using the same event selections defined for the \chljets channels in Section~\ref{sec:cuts}, three new regions are defined by tweaking or inverting a photon related requirement. Region A requires that the photon fail at least two out of four shower shape variables used for tight photon identification ($\omega_\text{s3}, F_{\text{side}}, \Delta E, E_{\text{ratio}}$)\footnote{The choice of failing two shower shape variables is based on studies that maximise statistics while keeping prompt photon contamination low.}. Region B requires photons to fail the same ID as A and additionally fail the isolation requirement. Region C requires that photons fail the same isolation requirement as B, but pass the tight ID. Region D is the signal region. 
An extra cut of $\pt^{\text{cone20}} > 3$~\GeV is also required in region B and C to reduce the prompt photon contamination.
This is illustrated by the schematic in Figure~\ref{fig:abcd}.

If the photon isolation and ID are uncorrelated, then the assumption holds that the ratio of tight vs non-tight photons should not change depending on if the photons are isolated or non-isolated. 
Similarly, the ratio of isolated vs non-isolated photons should not change if the photon is tight or non-tight. Thus, 
\begin{align}
\frac{N^{\text{h-fake}}_{\text{D}}}{N^{\text{h-fake}}_{\text{C}}}  &= \frac{N^{\text{h-fake}}_{\text{A}}}{N^{\text{h-fake}}_{\text{B}}} \\
&\text{and}  \notag \\ 
\frac{N^{\text{h-fake}}_{\text{D}}}{N^{\text{h-fake}}_{\text{A}}}  &= \frac{N^{\text{h-fake}}_{\text{C}}}{N^{\text{h-fake}}_{\text{B}}}.
\label{eq:abcd}
\end{align}

Apart from the two variables needing to be uncorrelated, the ABCD method requires that enough statistics are in each region (which is why this method can not be used in the \chll channels), and that there is only a small signal contamination in region A, B and C. 
However, in practice photon isolation and identification are not uncorrelated, and so this needs to be taken into account. A double ratio, $\theta_{\text{MC}}$ is defined as

\begin{equation}
\theta_{\text{MC}}=\frac{{}^{N_\text{D, MC}^{\text{h-fake}}}/_{N_\text{C, MC}^{\text{h-fake}}}}{{}^{N_\text{A, MC}^{\text{h-fake}}}/_{N_\text{B, MC}^{\text{h-fake}}}}.
\label{eq:theta_mc}
\end{equation}
This quantifies the correlation between isolation and identification. Assuming 0 correlation the term will be 1, which often is not the case. The $\text{MC}$ sub index indicates that these are the predicted number of events for the \hfake contribution according to MC\footnote{A data-driven approach to the double ratio term can be done but it requires a more complicated region definition since $N_\text{D, data}^{\text{h-fake}}$ is the quantity sought after.}.

From Equation~\ref{eq:abcd} and Equation~\ref{eq:theta_mc}, we can define the estimated \hfake contribution in the signal region:

\begin{eqnarray}
N_{\text{D, est.}}^{\text{h-fake}}&=&\frac{N_{\text{A, data}}^{\text{h-fake}}~.~N_{\text{C, data}}^{\text{h-fake}}}{N_{\text{B, data}}^{\text{h-fake}}}~\times~
\frac{{}^{N_\text{D, MC}^{\text{h-fake}}}/_{N_\text{C, MC}^{\text{h-fake}}}}{{}^{N_\text{A, MC}^{\text{h-fake}}}/_{N_\text{B, MC}^{\text{h-fake}}}},
\label{eq:ABCD_full}
\end{eqnarray}
where $N_{\text{A/B/C, data}}^{\text{h-fake}}$ indicates the number of \hfake events estimated by subtracting the sum of all background events (including the \hfake MC) and signal events from the total data events in the respective region. 
Finally, the \hfake scale factor is defined as

\begin{equation}
\text{SF}^{\text{h-fake}} = \frac{N_{\text{D, est.}}^{\text{h-fake}}}{N_{\text{D, MC}}^{\text{h-fake}}}.
\end{equation}

The scale factors are extracted in bins of \pt, $\eta$ and converted/unconverted photon type. They range from 0.8 to 3.2 and are applied to the \hfake MC samples in both the \chljets and \chll channels. 
Several sources of systematic uncertainties are considered as a result of the ABCD method. The statistical uncertainty of the number of \hfake events estimated from data as well as from the $\theta_{\text{MC}}$ factor are considered. Uncertainties due to \ttgamma and the background subtraction is also included, where the \ttgamma signal  is varied by 100\%, the other prompt sources and the \efake background are varied by 50\%, and the \QCD background is varied accordingly as explained in Section~\ref{sec:fakelepton}.
Further theoretical background modelling uncertainties taken into consideration are explained in Chapter~\ref{sec:systematics}.

\FloatBarrier

\section{Non-prompt and fake lepton background}
\label{sec:fakelepton}

Leptons from $W$-boson decays categorise the channels in which this analysis is performed. These are called prompt leptons. However, non-prompt and fake leptons may satisfy the event selection criteria. From hereon, both of these types of events are categorised as ``\QCD{}" background and are predominantly in the \chljets channels. The main contribution comes from the QCD driven multi-jet processes in association with a photon. 
Non-prompt or fake electrons and muons can arise from the semi-leptonic decay of $b$ and $c$ quarks. For electrons there can be additional contributions from photon conversions and jets in the electromagnetic calorimeter. Muons can be non-prompt or faked from energetic showers in the hadronic calorimeter, or from energetic hadrons that punch-through to the hadronic calorimeter.
In the \chll channels a fake and a real lepton are required, or two fake leptons. This contribution is negligible and so the \QCD background for the \chll channels is neglected.
The estimation of the \QCD background in the \chljets channels follows the fully data-driven Matrix Method~\cite{ATLAS-CONF-2014-058} approach explained in the following.
 
An additional loose sample is created that uses looser lepton identification and isolation requirements than those outlined in Section~\ref{sec:eventselections}. For the loose sample the electrons are required to have \medium identification with no isolation requirement, while muons keep the same identification (\medium) but the requirement on isolation is dropped. At this point, no requirement on a photon is made on either the tight or the loose samples.

An assumption can be made that the tight sample will contain mostly real leptons, while the loose sample is enriched with events that contain fake leptons.
Thus, we can write

\begin{align}
N^{\text{loose}} &= N^{\text{loose}}_{\text{real}} + N^{\text{loose}}_{\text{fake}} 
\\
N^{\text{tight}} &= N^{\text{tight}}_{\text{real}} + N^{\text{tight}}_{\text{fake}},
\end{align}

which says that the total number of loose (tight) leptons is a linear combination of the real and fake leptons in the loose (tight) sample.
We define $\epsilon_{\text{real}}$ as the probability of a real lepton in the loose sample to pass the tighter selection. Similarly, $\epsilon_{\text{fake}}$ is the probability of a fake lepton in the loose sample to pass the tighter selection.  The total leptons in the tight sample can be written as 

\begin{align}
N^{\text{tight}} = \epsilon_{\text{real}} N^{\text{loose}}_{\text{real}} + \epsilon_{\text{fake}} N^{\text{loose}}_{\text{fake}}.
\end{align}

If the real and fake efficiencies are known, then from the equations above, the number of events with fake leptons in the tight sample can be estimated using 

\begin{align*}
N^{\text{tight}}_{\text{fake}} = \frac{\epsilon_{\text{fake}}}{\epsilon_{\text{real}}-\epsilon_{\text{fake}}}(\epsilon_{\text{real}} N^{\text{loose}} - N^{\text{tight}}).
\end{align*}

The real efficiencies are estimated using the tag-and-probe method in the $Z\to ee$ and $Z\to \mu\mu$ control regions. In the region of 80 to 100~\GeV a pure sample is obtainable.
The fake efficiencies are estimated in data samples dominated by non-prompt and fake leptons. Among other requirements, one \loose lepton, at least one jet and low \MET are required.
These measurements can be done with respect to a number of variables that depend on the lepton or event kinematics, for example the lepton \pt and \eta, number of $b$-jets, and jet \pt. Thus, they are parametrised as a function of these variables. The efficiency measurements are provided centrally in \ATLAS and are explained in more detail in \cite{ATLAS:2014ffa}.

\begin{figure}[!htpb]
\centering
\subfloat[\chejets]{
\includegraphics[trim={0 1cm 0 0.5cm },clip,width=0.7\linewidth]{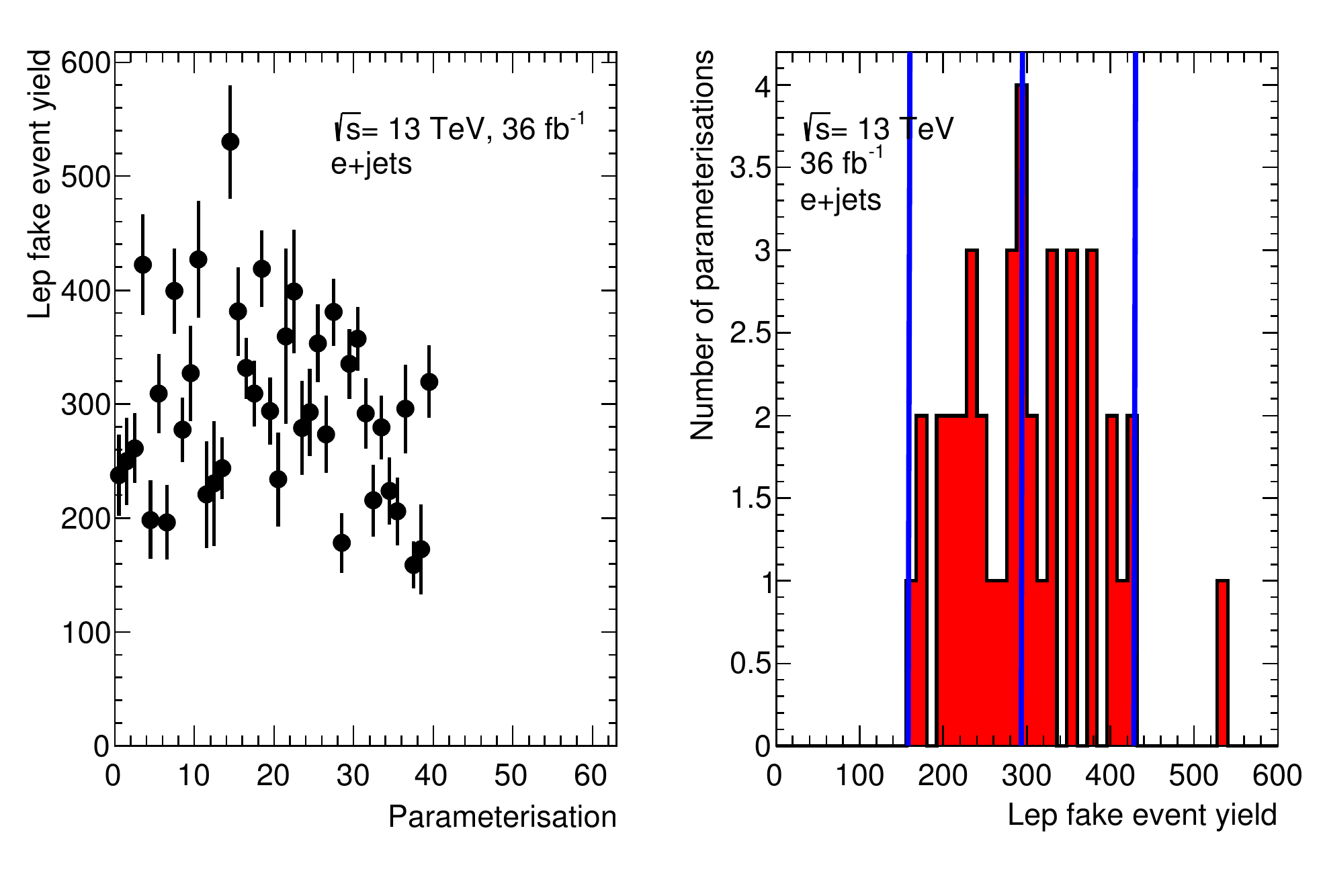}
}

\subfloat[\chmujets]{
\includegraphics[trim={0 1cm 0 0.5cm },clip,width=0.7\linewidth]{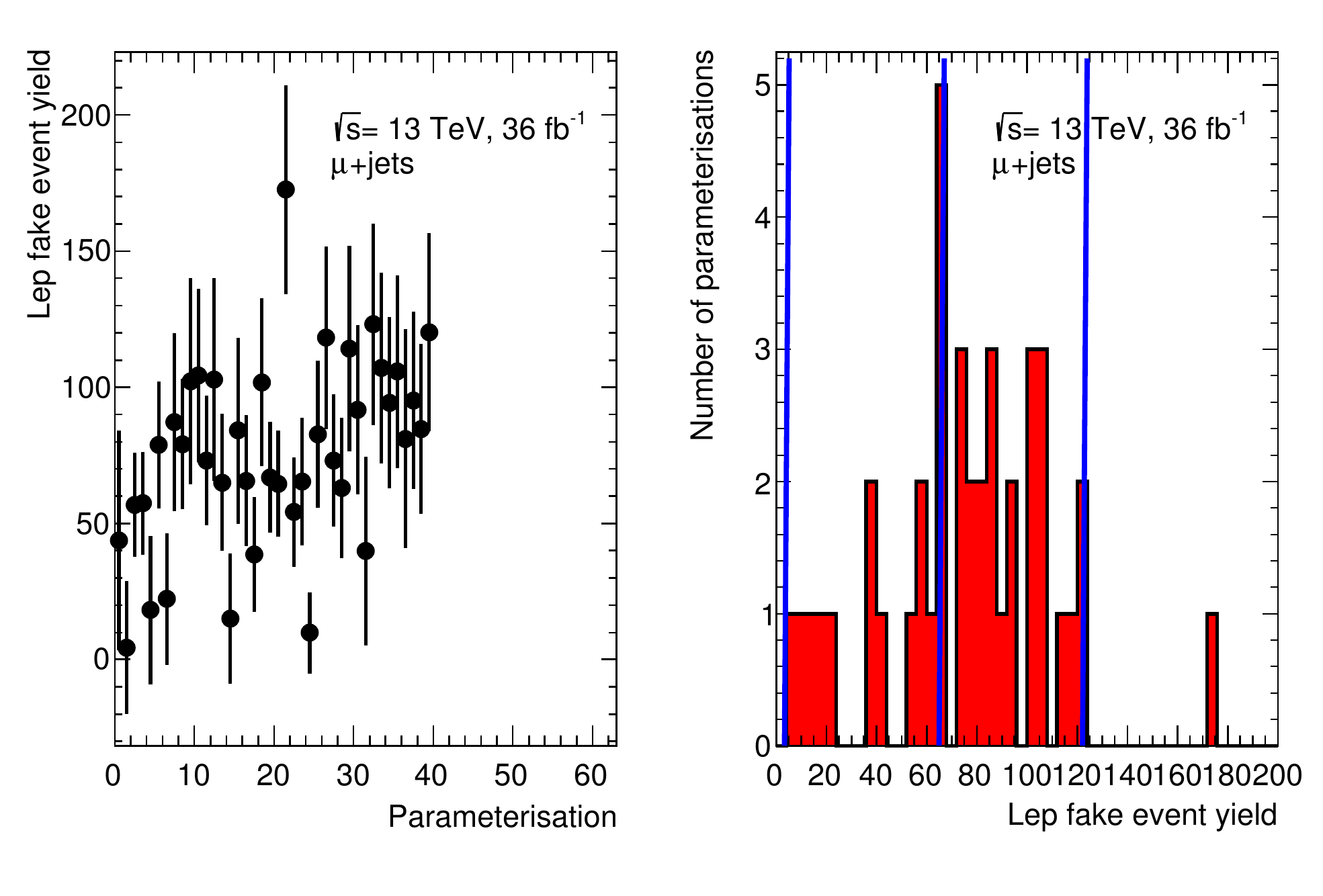}
}
\caption [The \QCD yields for each parametrisation with the $y$-axis projection displayed on the right.] {The \QCD yields for each parametrisation with the $y$-axis projection displayed on the right. The blue lines indicate the mean (the nominal weight) and the up/down systematic variations which form the extremes of the histogram.}
\label{fig:qcd_parametrisations}
\end{figure}

Individual event weights are calculated as functions of the efficiencies and take the form
\begin{equation}\label{eq:qcdweights}
w_i = \frac{\epsilon_{\text{fake}}}{\epsilon_{\text{real}} - \epsilon_{\text{fake}}}(\epsilon_{\text{real}} - \delta_i),
\end{equation}
where $\delta_{i}$ is 1 if the event $i$ passes the tight selection, otherwise 0. These event weights are applied to the loose data sample.
Lastly, by adding the photon related cuts a \QCD estimation in the full signal region is obtained.\\
\indent
Figure~\ref{fig:qcd_parametrisations} shows the spread of different event yields for the \chejets and \chmujets channels. The left plots show the event yield as a function of each parameterised set of weights, while the right plots show the projection of the $y$-axis. The closest parametrisation to the mean of the new histogram is taken as the nominal weight. The chosen up and down systematic variations are chosen as the extreme edges of the histogram (neglecting the obvious outliers). Both the mean and up/down systematic variations are indicated by the blue vertical lines.
Further parameterisations are neglected due to being outliers and are not included in the plot. For the \chejets channel these are \pt related, while for \chmujets these are $\Delta \phi$ related.

\FloatBarrier

\section{Prompt photon backgrounds}
\label{sec:promptbackgrounds}

Events in which a real prompt photon is radiated from anything except top quarks (or incoming quarks) make up about 16\% (11\%) of total events in the combined \chljets (\chll) channels.
For the \chljets channel this is largely due to the \Wgamma process, whereas for the \chll channel the main contribution is from the \Zgamma process. Other processes such as single-top, diboson and \ttV, where each have an additional prompt photon in the final state have small or even negligible contributions. Table~\ref{tab:promptBkg} shows the prompt background composition for each of the five channels after event selections discussed in Section~\ref{sec:cuts} have been applied.

In the \chljets (\chll) channel the \Wgamma (\Zgamma) contribution is largest and is therefore separated from the rest of the prompt photon backgrounds. These are then grouped together as ``\Other{}".

A validation region (VR) is defined for the \Zgamma and \Wgamma backgrounds primarily to ensure that the modelling of these processes is adequate. 
A maximum likelihood fit to the jet $\pt$ distribution is performed, of which the mechanics of the fit are explained in Chapter~\ref{sec:analysisstrategy}\footnote{The maximum likelihood fit to the signal region was repurposed for this validation region, and so will only be presented later. Essentially, a value of $\mu=1$ indicates the theoretical SM prediction.}. All systematics explained in Chapter~\ref{sec:systematics} have been included. 
The \QCD data-driven background estimate from the previous section as well as all previously derived scale factors for various backgrounds are also applied. While these should be derived specifically for the region in question, given the function of these VRs this approximation is sufficient.
It is important to emphasise that these are validation regions and so none of the results in this section actually enter the final result. They are purely used as a cross-check and closure test.

\begin{table}[h]
\small
\centering
\vspace*{0.5cm}
 \begin{tabular}{|c|r|r|r|r|r|}
    \hline 
   \centering
    Process       & \chemu           & \chmumu                    & \chee                                  & \chejets                         & \chmujets        \\ \hline
    \hline
    \Wgamma   &    -         		 &    -        			     &    -          				 & 541 $\pm$ 29 	& 579 $\pm$ 34 \\ \hline
    \Zgamma    & -			 & 54 $\pm$ 14     & 21 $\pm$ 4	& 295 $\pm$ 46		& 152 $\pm$ 13 \\ \hline
    Single-top   & 6 $\pm$ 1  & 3 $\pm$ 1         & 34 $\pm$ 1	& 85 $\pm$ 5   		& 92 $\pm$ 6   \\ \hline
    Diboson     &    -         		 & -          & -  		& 4 $\pm$ 1    		& 6 $\pm$ 1    \\ \hline
    \ttV	    & 3.0 $\pm$ 0.2 &1.5 $\pm$ 0.1        & 1.3 $\pm$ 0.1			& 27 $\pm$ 1		& 25 $\pm$ 1    \\
       \hline
 \end{tabular}
  \caption[]{The yields of prompt photon background events from MC. The numbers are normalised to the total integrated luminosity. Only statistical uncertainties are included. A dash indicates the yield is $<1$ and therefore negligible.}
  \label{tab:promptBkg}
 \end{table}

\subsubsection{\Zgamma Validation region}

The \Zgamma VR is comprised of the \chll channels with \chee and \chmumu final states. They each have similar object definitions and event selections to the signal region with the exception of the following: number of jets~$\ge~0$, number of $b$-tagged jets = 1 and finally the invariant mass of the two leptons is required to be in a mass window of [60, 100]~\GeV.
With the requirement of $b$-tagged jets = 1, we enter the heavy flavour topology, similar to our signal. 
A second \Zgamma VR is defined with the same cuts above, except there is a requirement to have 0 $b$-tagged jets, i.e., the light flavour topology.

\begin{figure}[!htbp]
\centering
\subfloat[Heavy flavour \chee channel]{
\includegraphics[width=0.44\linewidth]{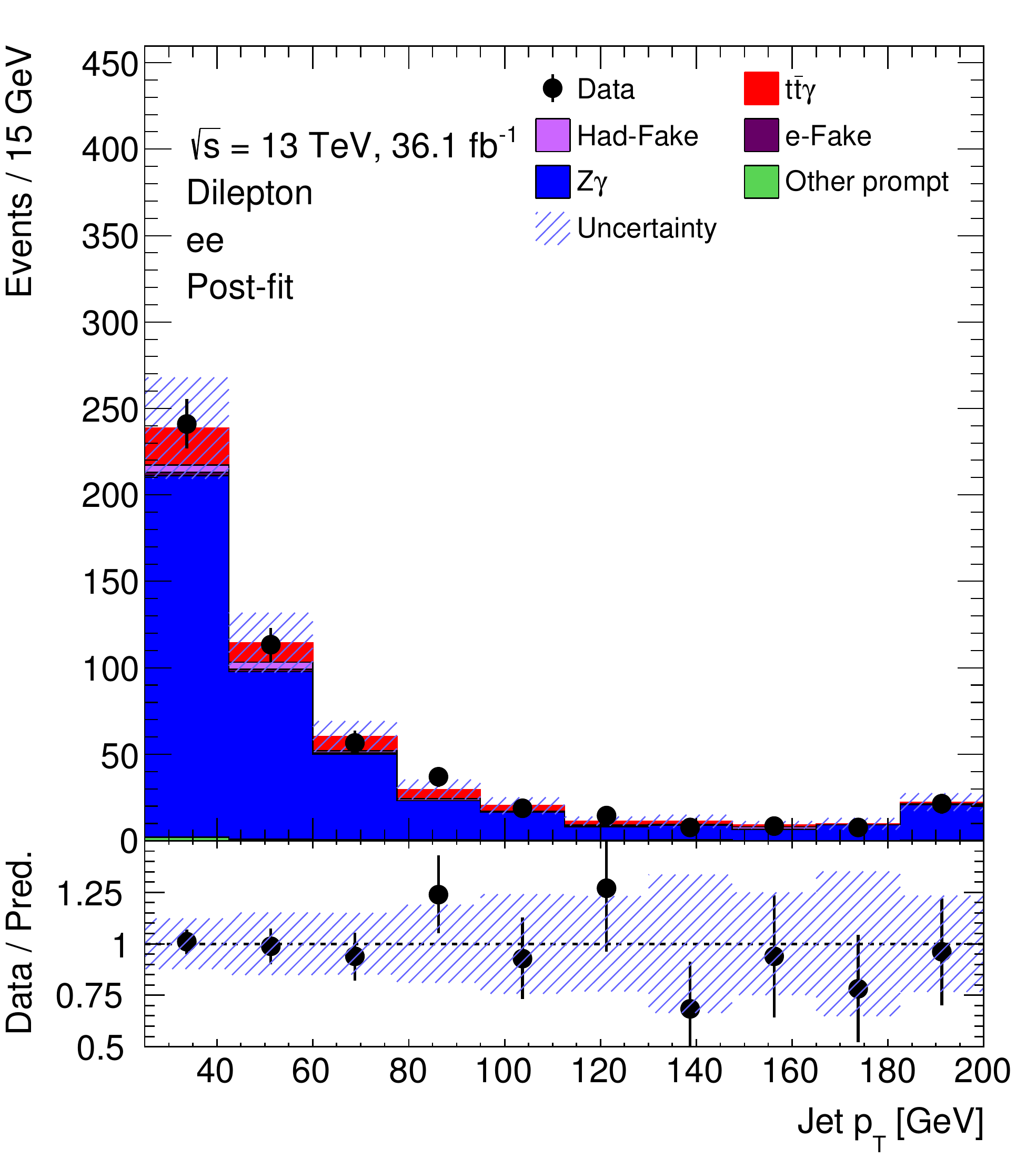}
}
\subfloat[Heavy flavour \chmumu channel]{
\includegraphics[width=0.44\linewidth]{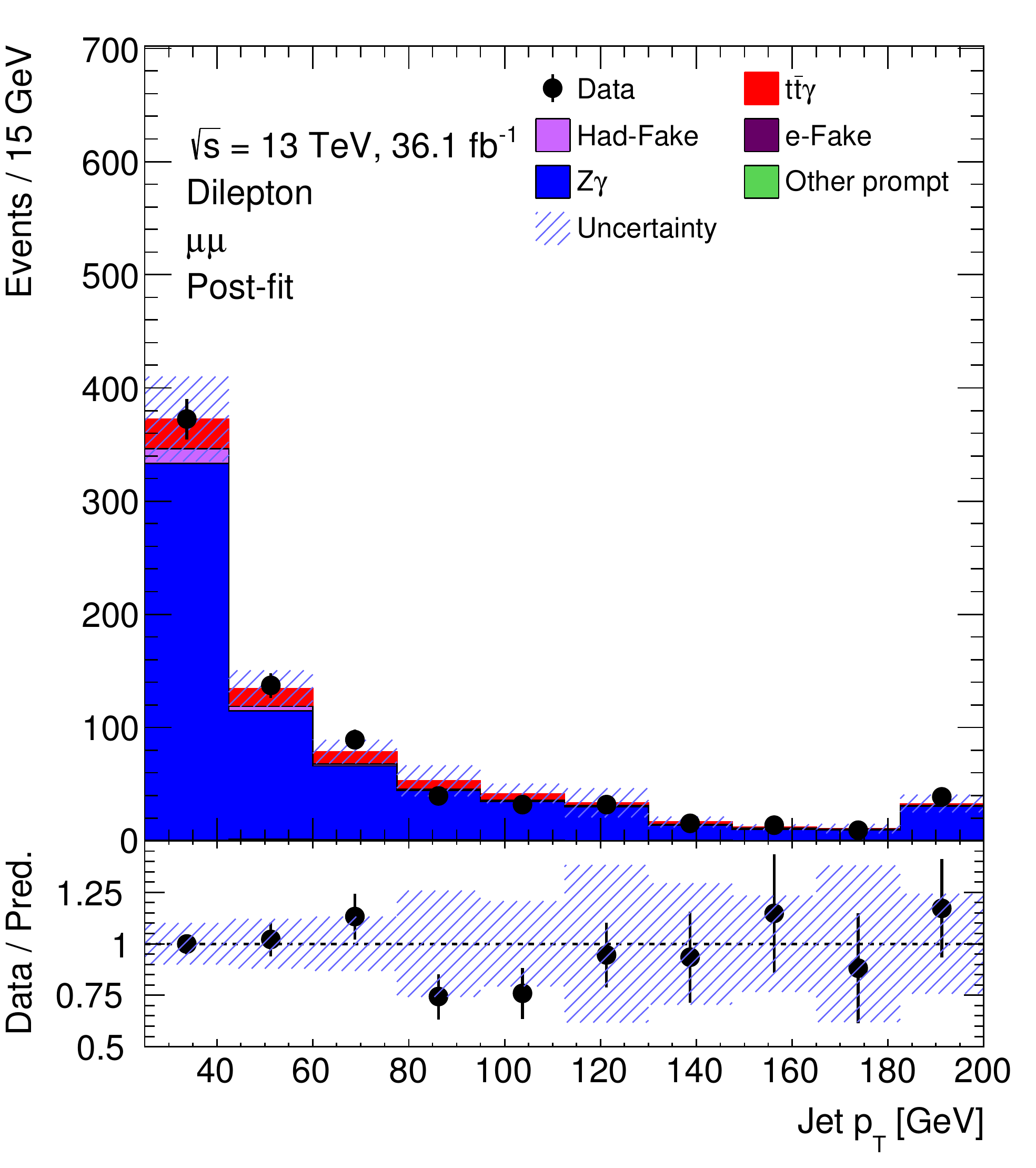}
}

\subfloat[Light flavour \chee channel]{
\includegraphics[width=0.44\linewidth]{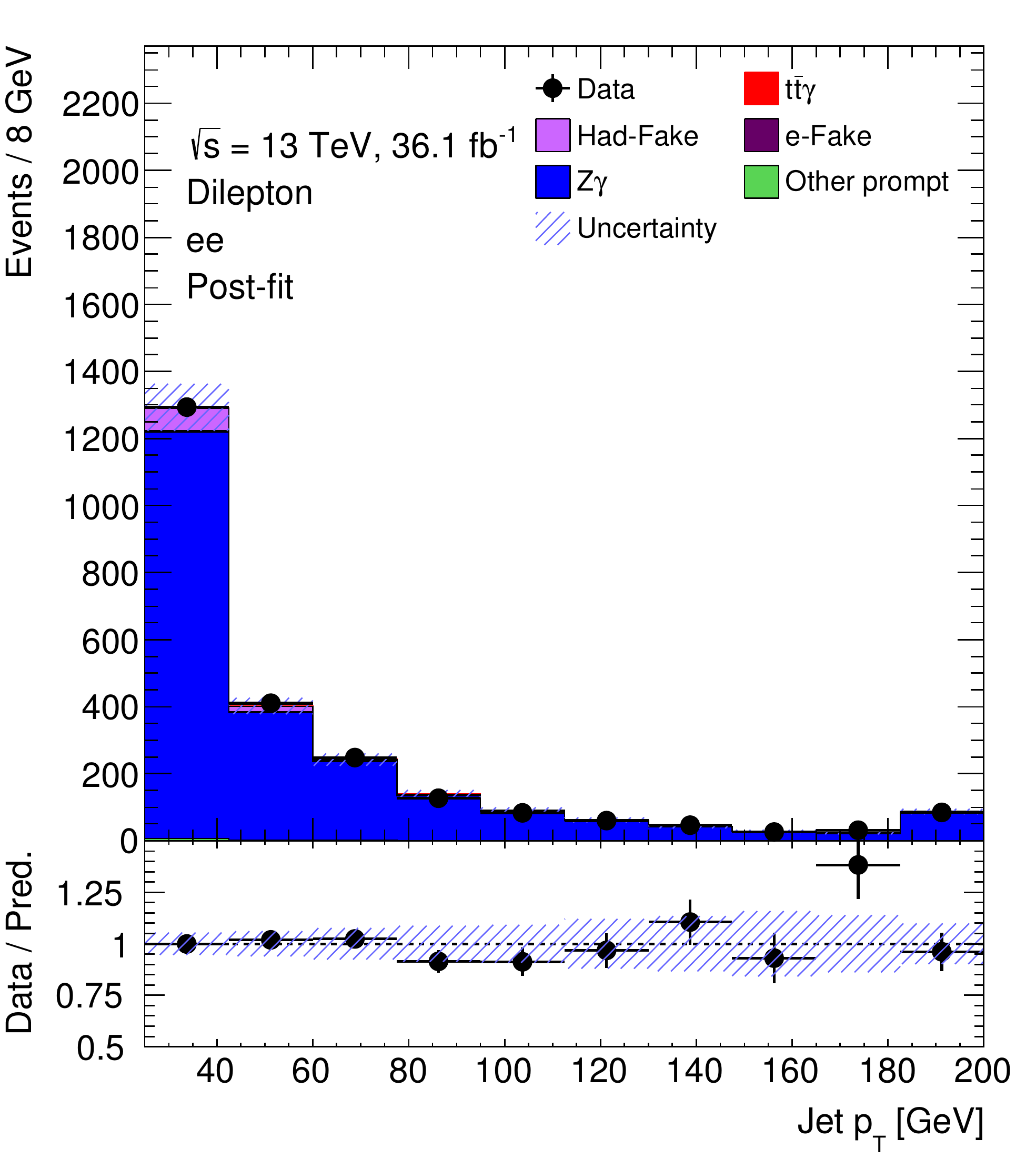}
}
\subfloat[Light flavour \chmumu channel]{
\includegraphics[width=0.44\linewidth]{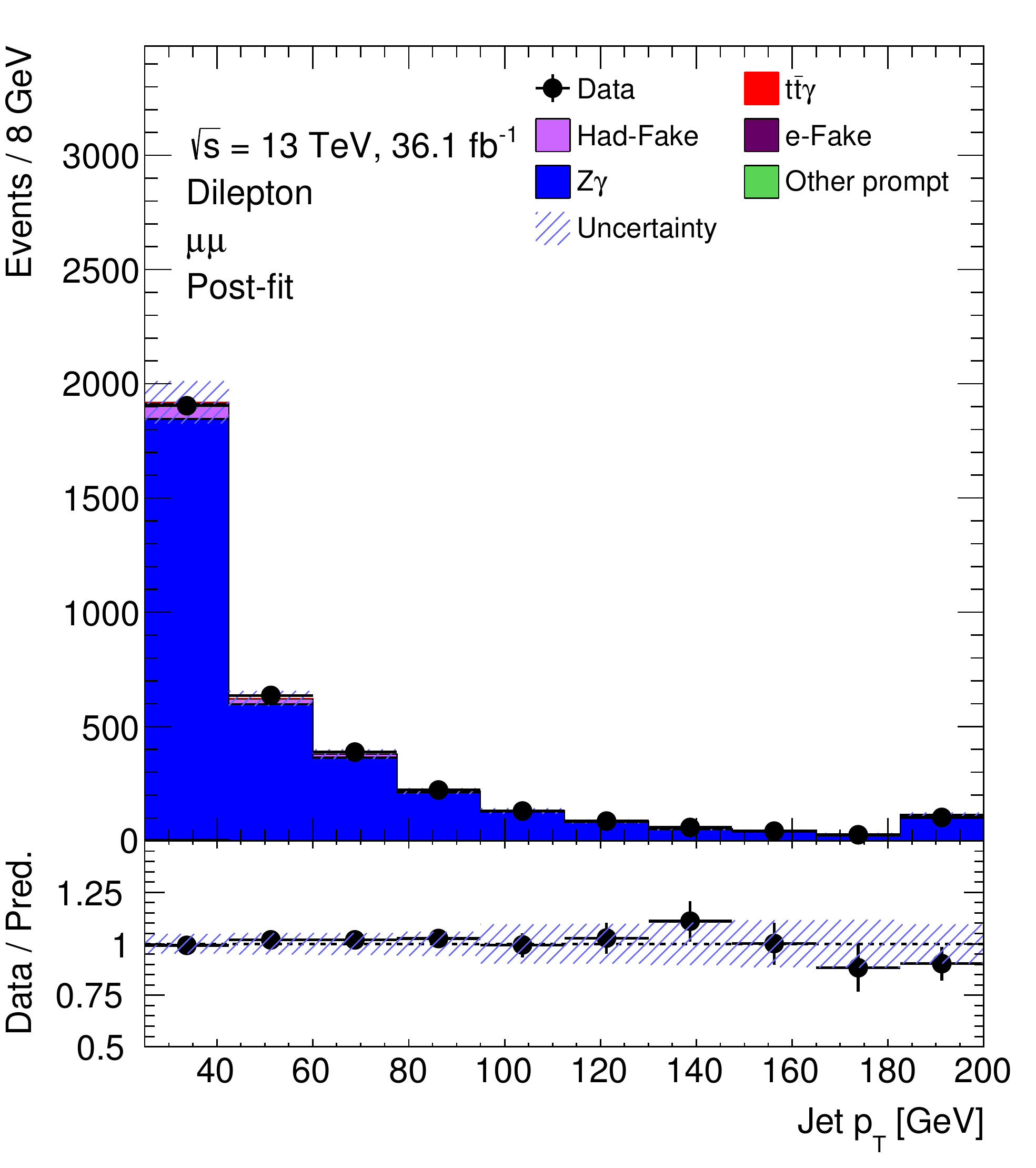}
}

\caption [] {Heavy flavour (top) and light flavour (bottom) post-fit validation regions for the \chll channel. All systematic uncertainties are included.}
\label{fig:prepostlightDL}
\end{figure}

The post-fit histograms for the jet $\pt$ can be seen in Figure~\ref{fig:prepostlightDL}.
The results from the fits are shown in Table~\ref{tab:Zgammamus}. The table shows that there is a 3\% and 10\% overestimation of the \Zgamma MC in the heavy flavour VR for the \chee and \chmumu channels, respectively. The small contribution of \ttgamma could also contribute. However, this result is well within the given uncertainties. Similarly, the light flavour VR overestimates the MC by 19\% and 22\% for the \chee and \chmumu channels, respectively. There is a small contribution from \hfake backgrounds which are modelled by MC and so could contribute to this overestimation. However, these results are within the total uncertainties.

\begin{table}[!htb]
\setlength\extrarowheight{1.5pt}
\begin{center}
\begin{tabular}{|l|c|c|c|c|}
\hline
channel & \multicolumn{2}{c|}{\chee} & \multicolumn{2}{c|}{\chmumu}  \\ \hline
\hline
  & $\mu$ & $\pm$ & $\mu$ & $\pm$  \\ \hline
 b-jets=0	&1.19	& $^{+0.26}_{-0.22}$ & 1.22 & $^{+0.24}_{-0.20}$ \\ \hline
 b-jets=1	&1.03	& $^{+0.35}_{-0.24}$ &  1.10 &$^{+0.29}_{-0.23}$  \\ \hline
\end{tabular}
\caption [] {Fit results for the \Zgamma validation region. Statistical and systematic uncertainties are included.}
\label{tab:Zgammamus}
\end{center}
\end{table}

\FloatBarrier
\subsubsection{\Wgamma Validation region}
\label{sec:wgammavr}
The \Wgamma VR is comprised of the \chljets channels with \chejets and \chmujets final states. They each have similar object definitions and event selections to the signal region with the exception of the following: 
2 $\le$ number of jets $\le$ 3, number of $b$-tagged jets = 1 and \met $>$ 40 GeV. In addition, there is a cut on the ELD distribution (Chapter~\ref{sec:ELD}) of $<$ 0.04 to further reduce signal contamination. For the \chejets channel an invariant mass cut between the photon and the lepton of $m(\gamma,lep)< 80$~\GeV helps reduce the \efake contribution.  
With the requirement of $b$-tagged jets = 1, we enter the heavy flavour regime, similar to our signal. 
A second \Wgamma VR is defined with different cuts to the above. In addition to requiring 0 $b$-tagged jets (i.e., the light flavour regime) we ask for 1 $\le$ number of jets $\le$ 3 and $\Delta R(\gamma,lep)<2.8$. For the \chejets channel we exclude events where the mass between the photon and the lepton falls in a window of [60,100]~\GeV.

\begin{figure}[!htbp]
\centering
\subfloat[Heavy flavour \chejets channel]{
\includegraphics[width=0.44\linewidth]{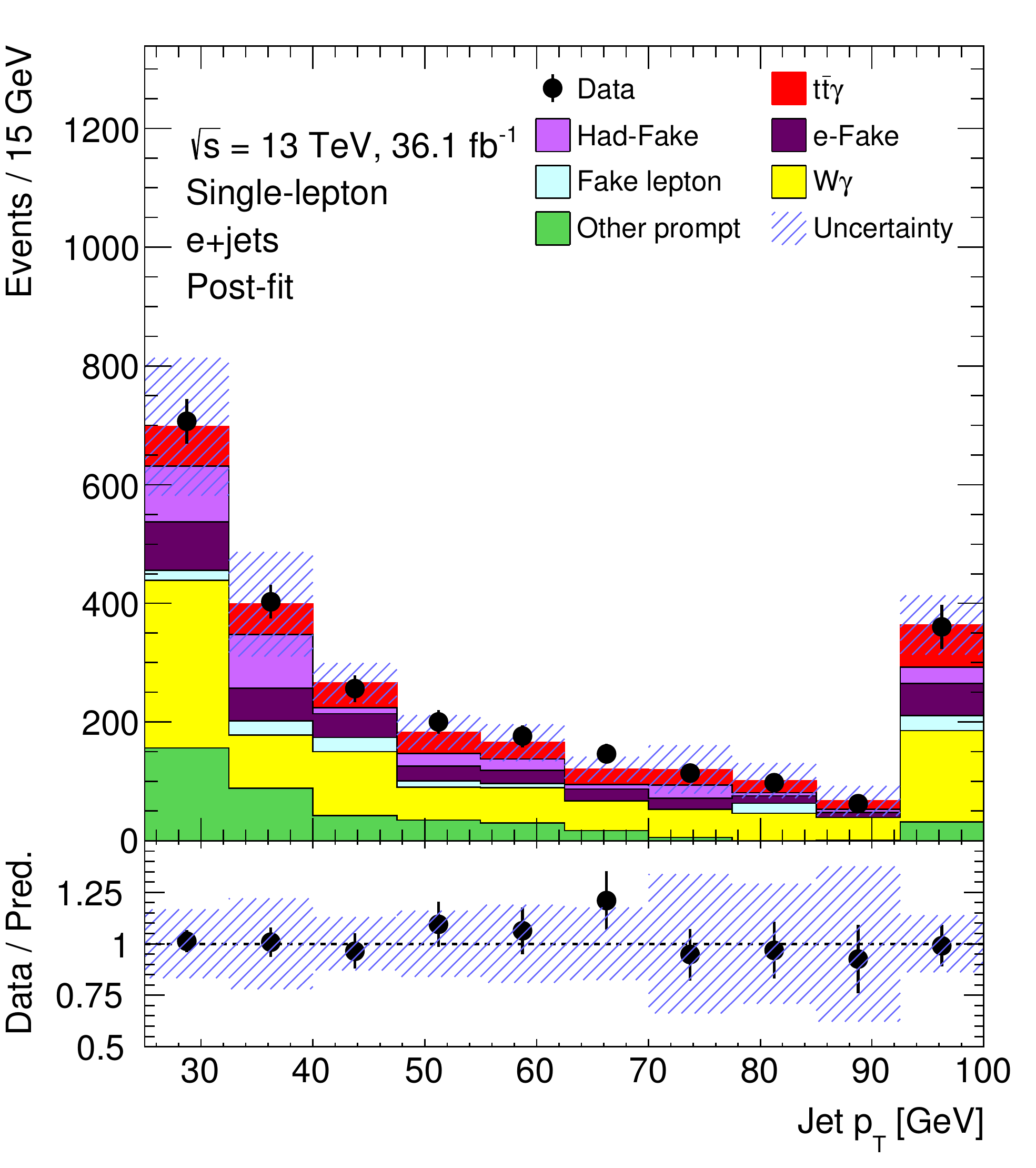}
}
\subfloat[Heavy flavour \chmujets channel]{
\includegraphics[width=0.44\linewidth]{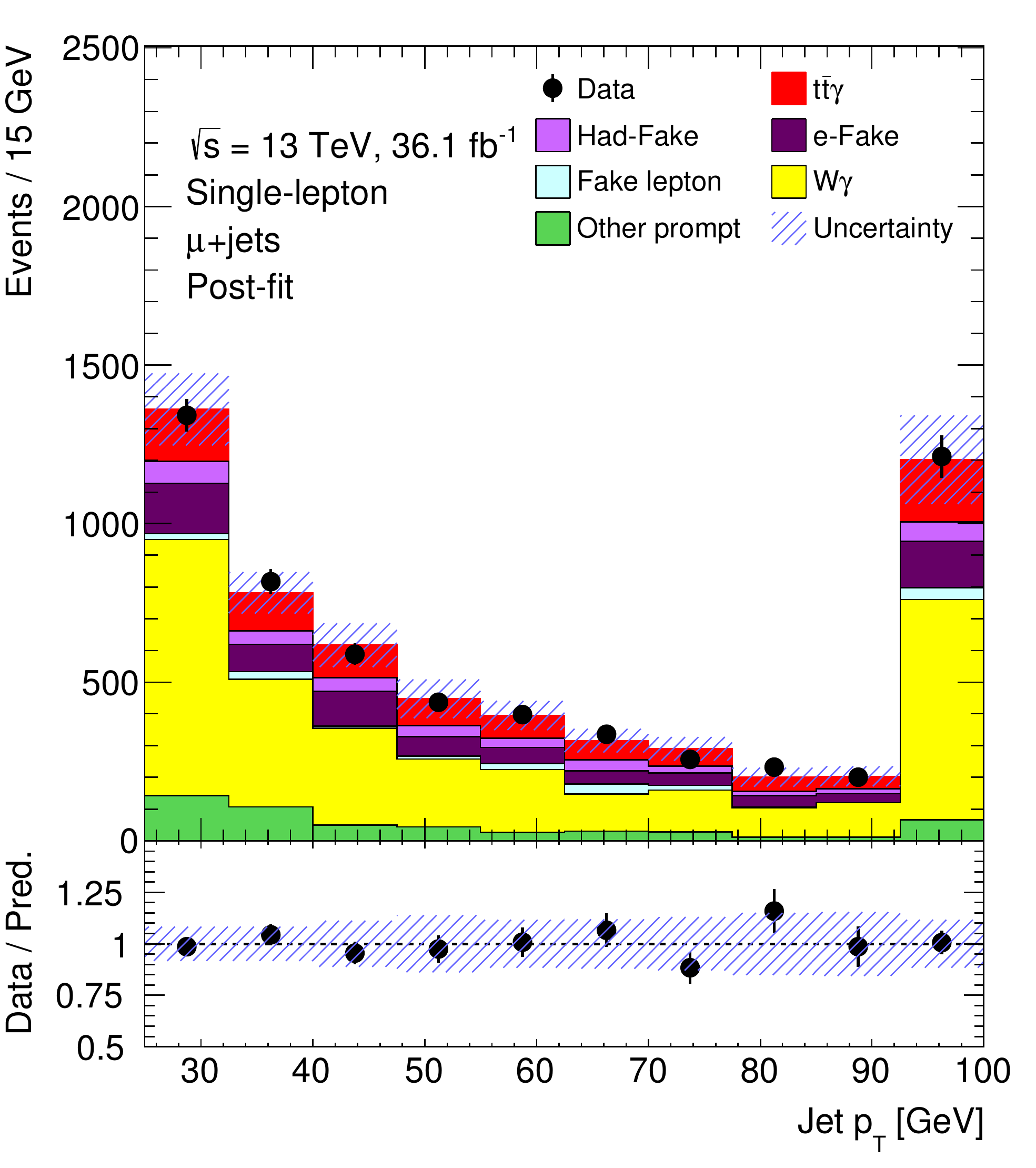}
}

\subfloat[Light flavour \chejets channel]{
\includegraphics[width=0.44\linewidth]{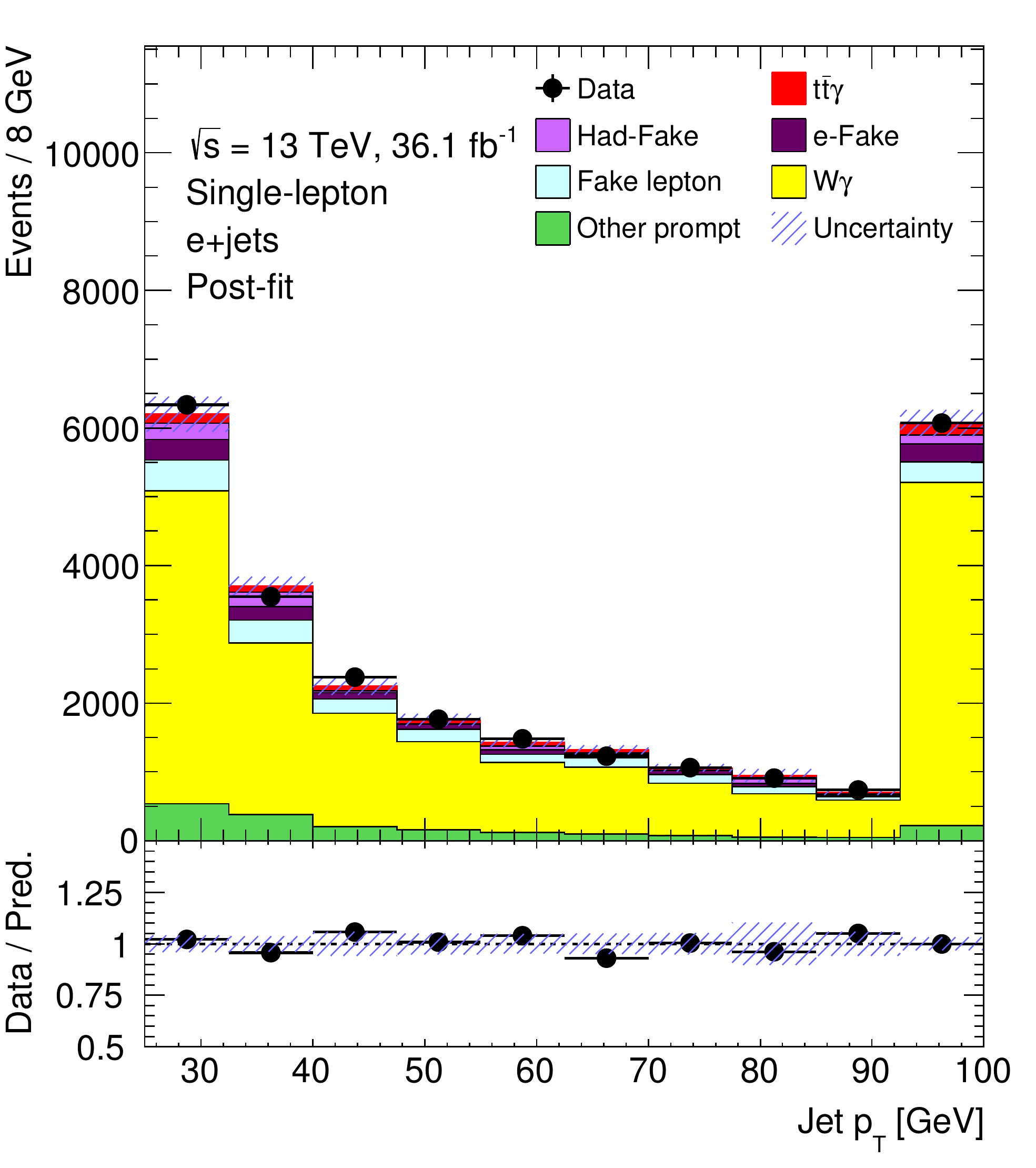}
}
\subfloat[Light flavour \chmujets channel]{
\includegraphics[width=0.44\linewidth]{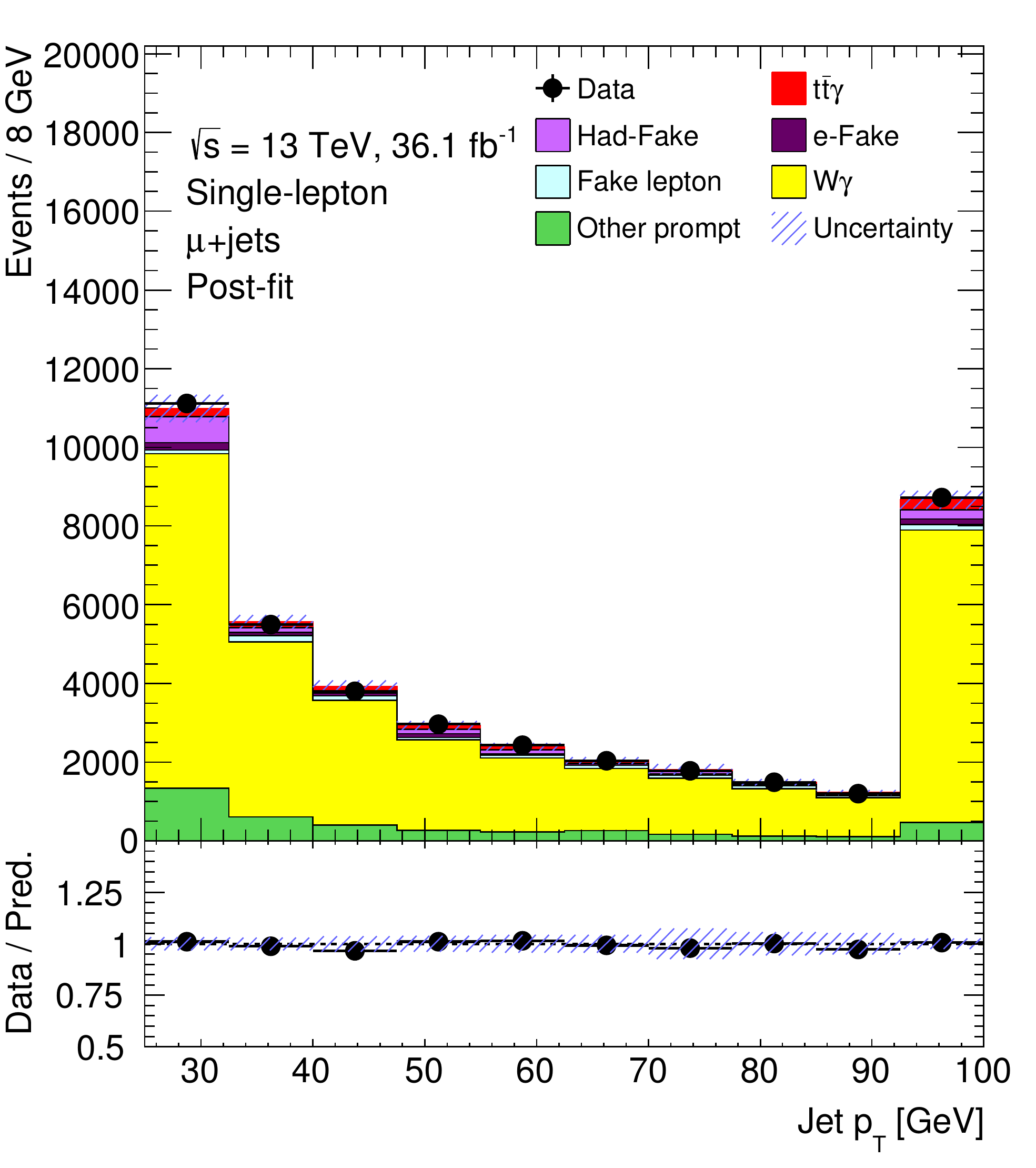}
}

\caption {Heavy flavour (top) and light flavour (bottom) post-fit validation regions for the \chljets channel. All systematic uncertainties are included.}
\label{fig:prepostlightSL}
\end{figure}

The post-fit histograms for the jet $\pt$ can be seen in Figure~\ref{fig:prepostlightSL}.
The results from the fits are shown in Table~\ref{tab:Wgammamus}.
The fit results show that the light flavour VR overestimates the MC contribution between 15\% and 18\%, of which the uncertainties do not fully cover. Similarly for the heavy flavour \chmujets VR, a 26\% over estimation is seen of which uncertainties do not cover. 
Given that none of these regions are completely pure with the \Wgamma process, and no dedicated studies were performed on the backgrounds in this phase space, nothing conclusive can be said. 
For this reason, the \Wgamma background has no constraint on the normalisation in the final fit for \chljets channels; it is a free parameter.

\begin{table}[!htb]
\setlength\extrarowheight{1.5pt}
\begin{center}
\begin{tabular}{|l|c|c|c|c|}
\hline
 channel & \multicolumn{2}{c|}{\chejets} & \multicolumn{2}{c|}{\chmujets}  \\ \hline
 \hline
 
 b-jets=0	&1.15	& $^{+0.05}_{-0.05}$ & 1.18	& $^{+0.06}_{-0.06}$ \\ \hline
 b-jets=1	&1.14	& $^{+0.33}_{-0.32}$ & 1.26	& $^{+0.13}_{-0.14}$ \\
\hline
\end{tabular}
\caption [] {Fit results for the \Wgamma validation region. Statistical and systematics uncertainties are included.}
\label{tab:Wgammamus}
\end{center}
\end{table}

\FloatBarrier

\section{Pre-fit distributions}
\label{sec:preFitDists}

\begin{figure}[!h]
\centering
\includegraphics[width=0.33\linewidth]{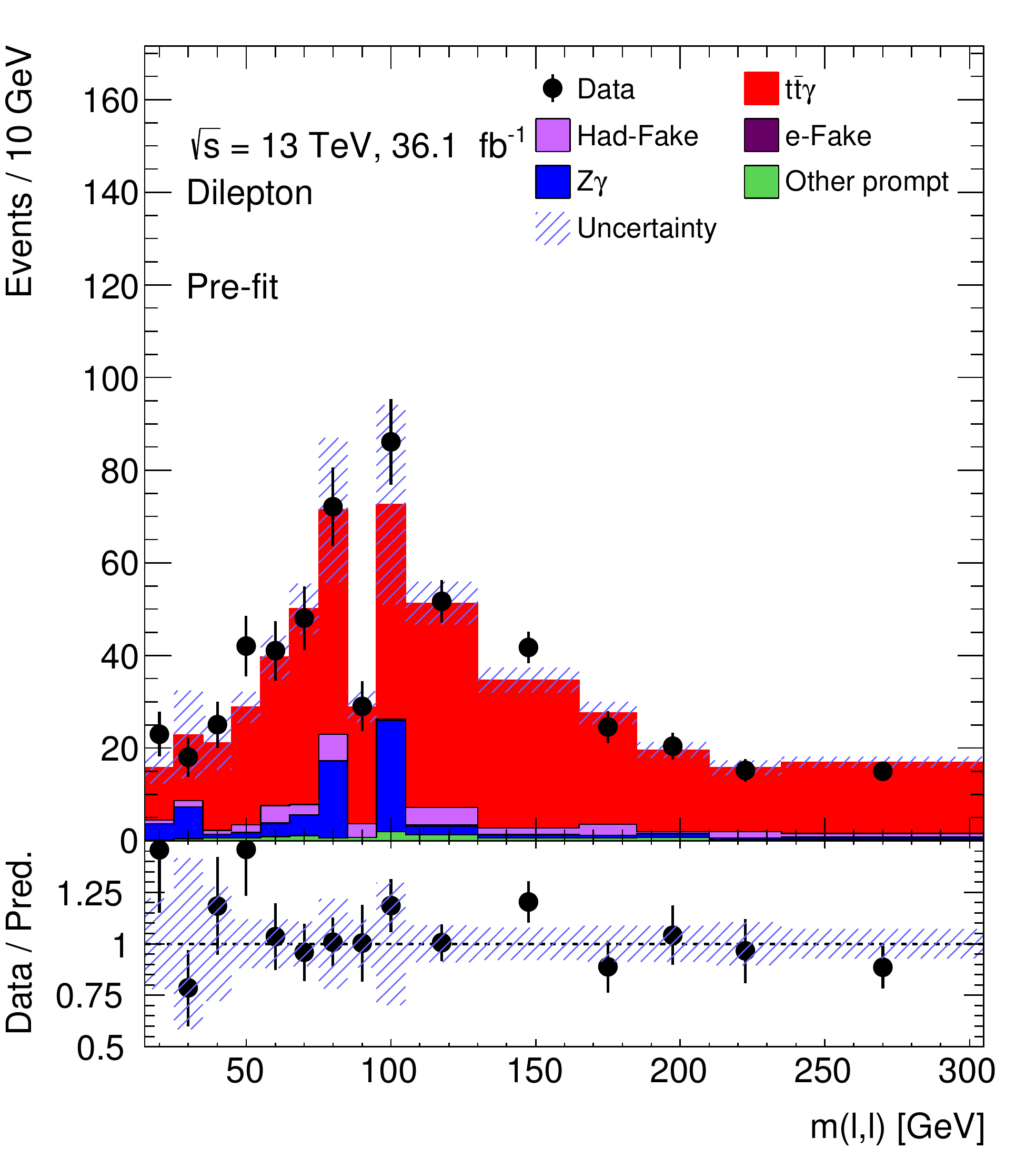}
\hspace{-0.02\linewidth}
\includegraphics[width=0.33\linewidth]{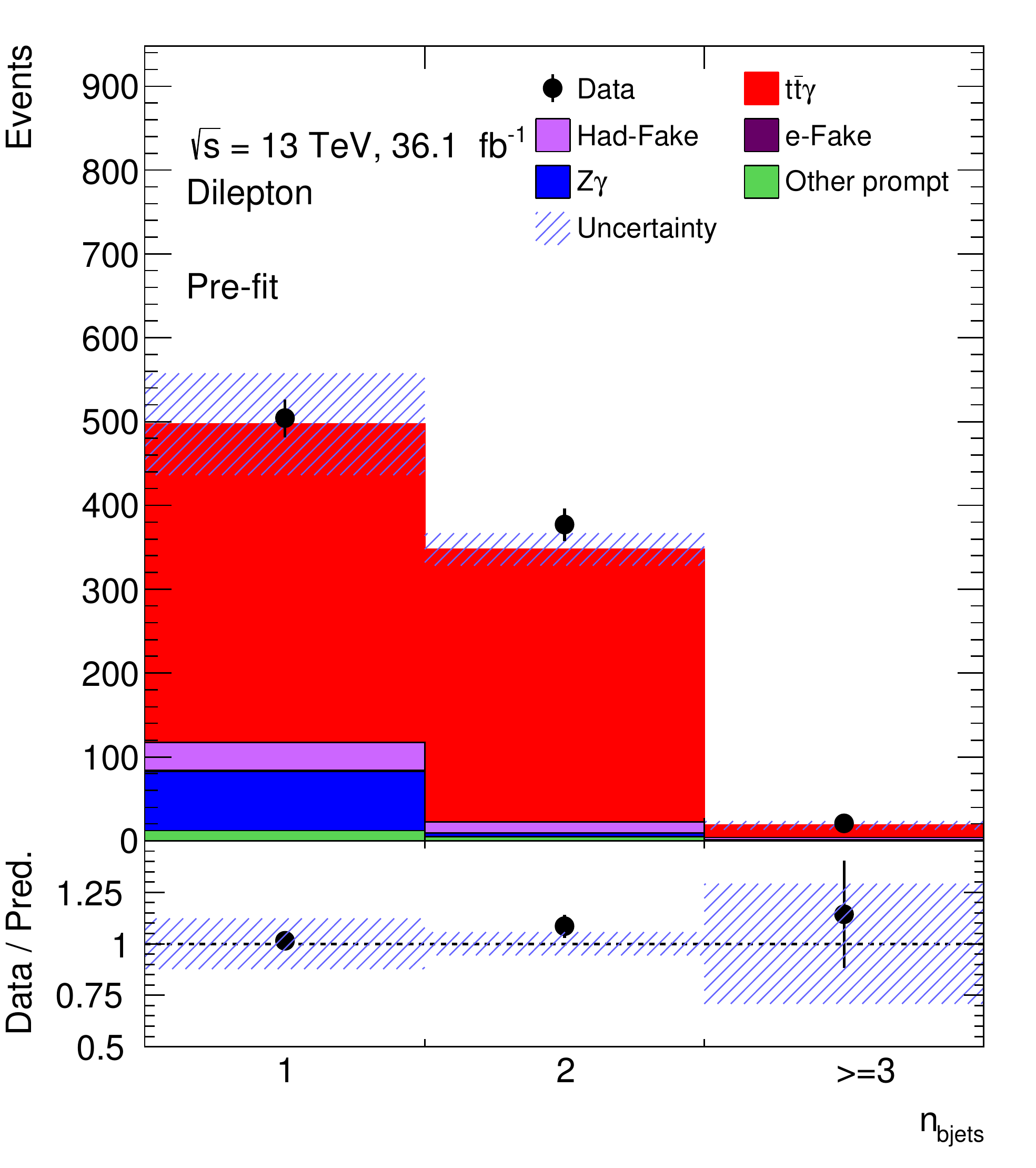}
\hspace{-0.02\linewidth}
\includegraphics[width=0.33\linewidth]{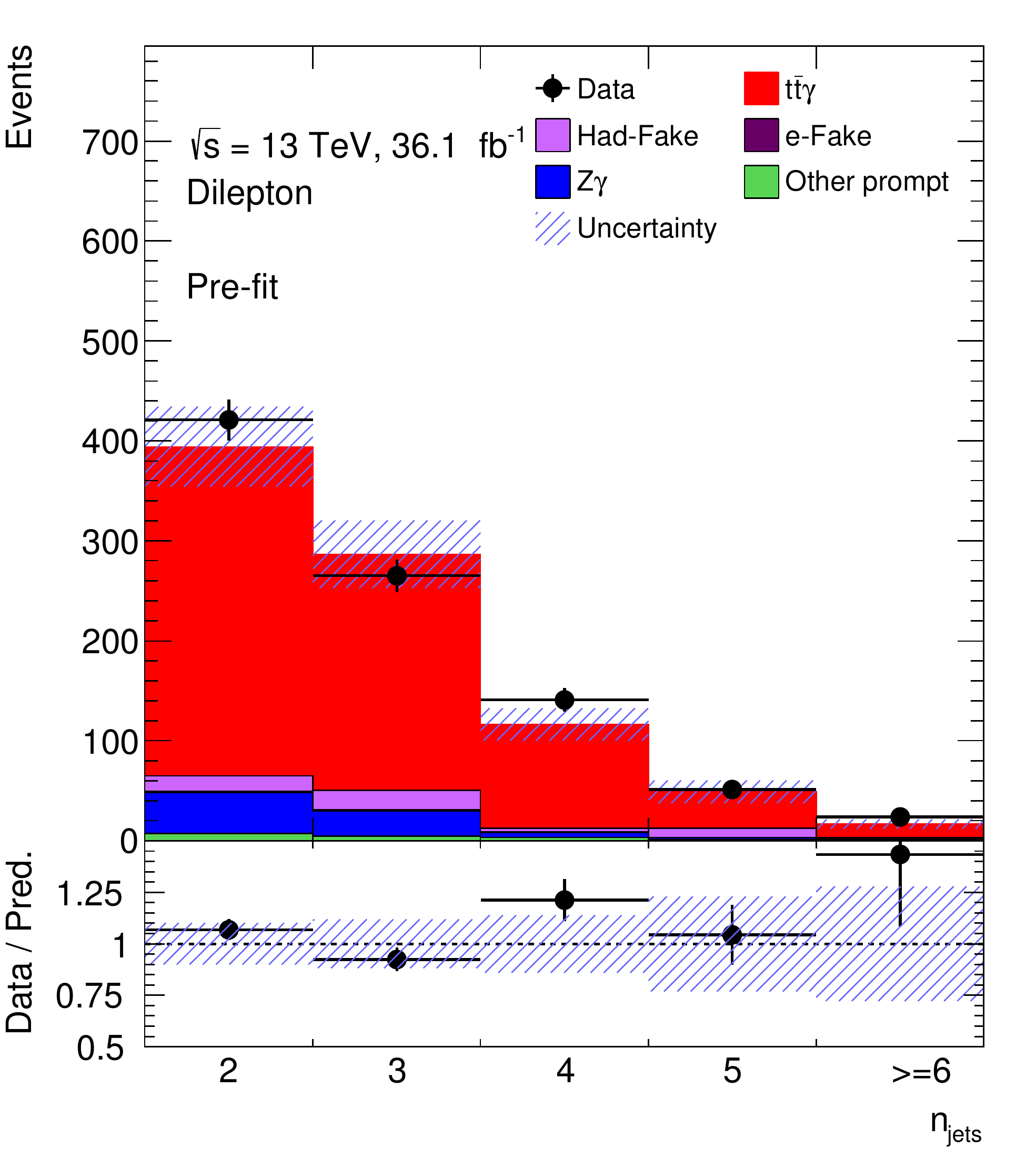}
\hspace{-0.02\linewidth}

\includegraphics[width=0.33\linewidth]{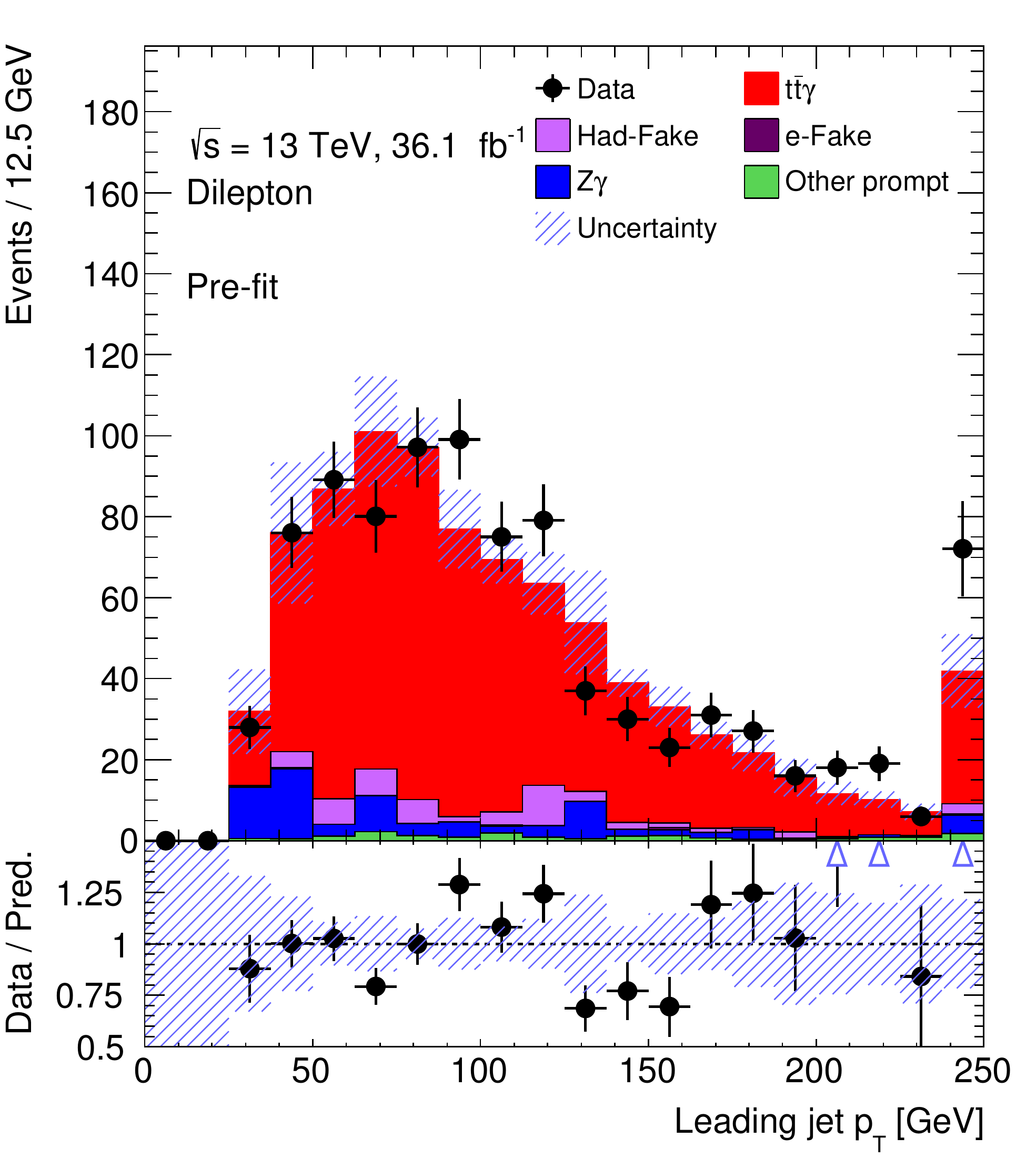}
\hspace{-0.02\linewidth}
\includegraphics[width=0.33\linewidth]{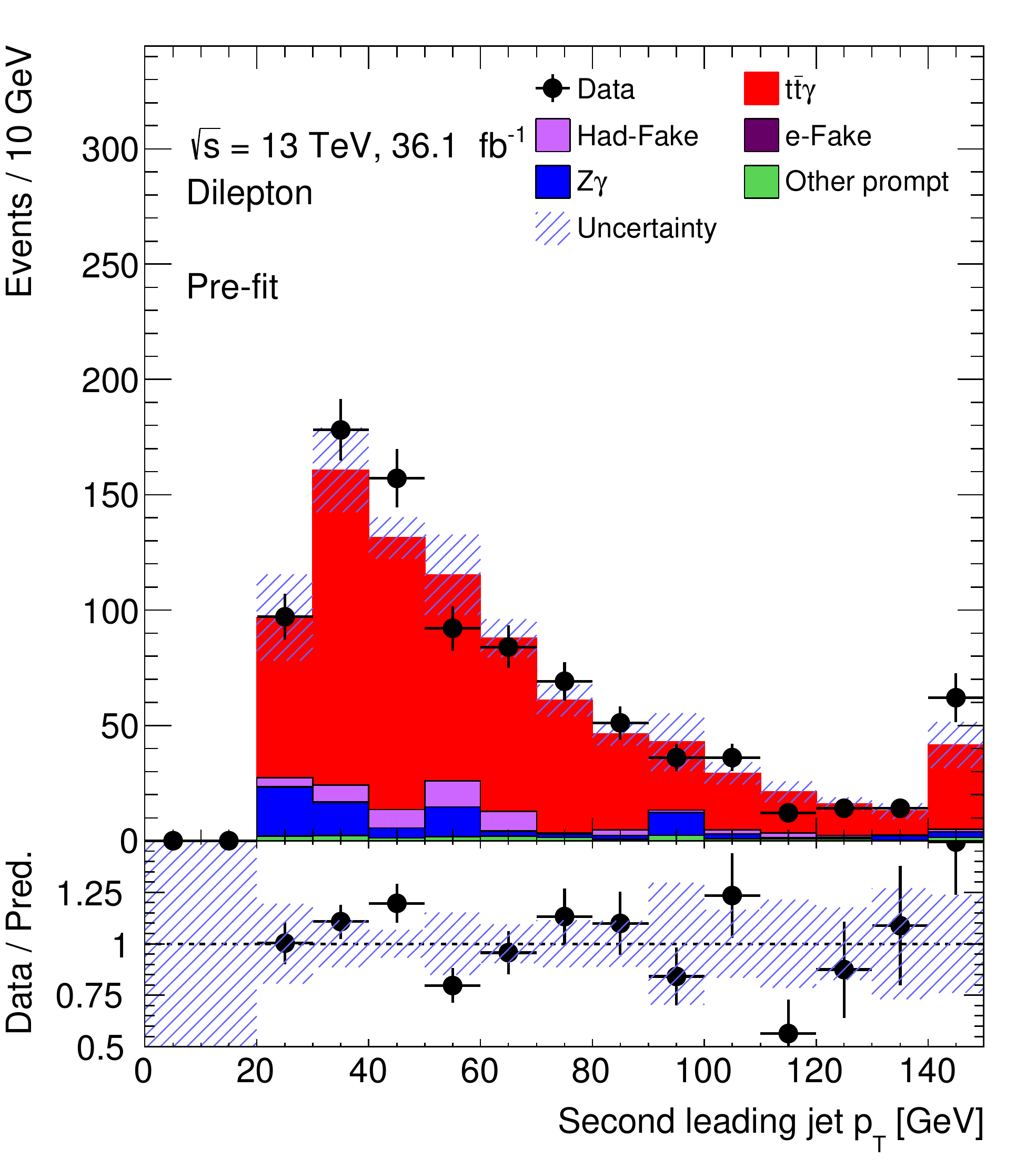}
\hspace{-0.02\linewidth}
\includegraphics[width=0.33\linewidth]{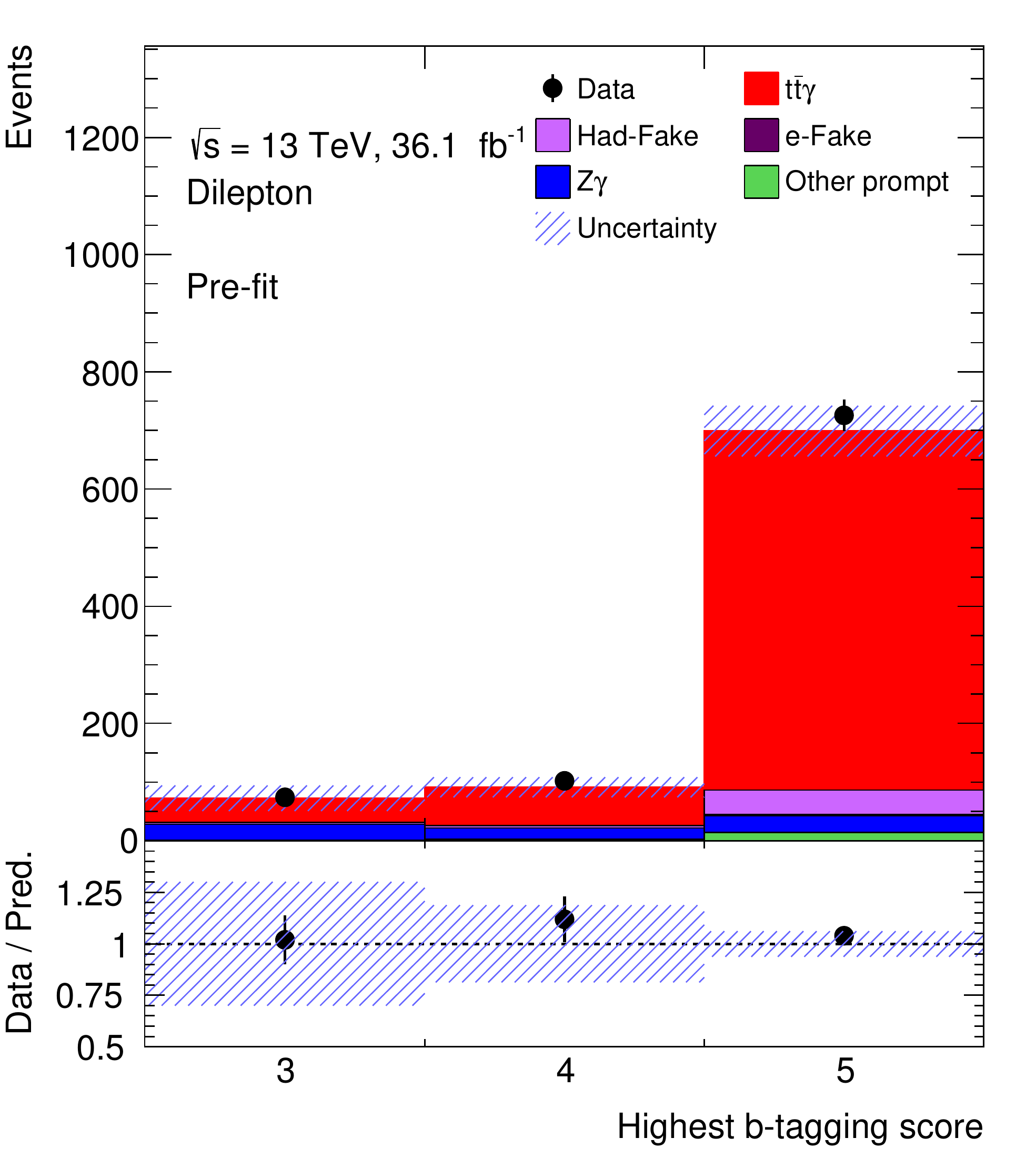}
\hspace{-0.02\linewidth}

\includegraphics[width=0.33\linewidth]{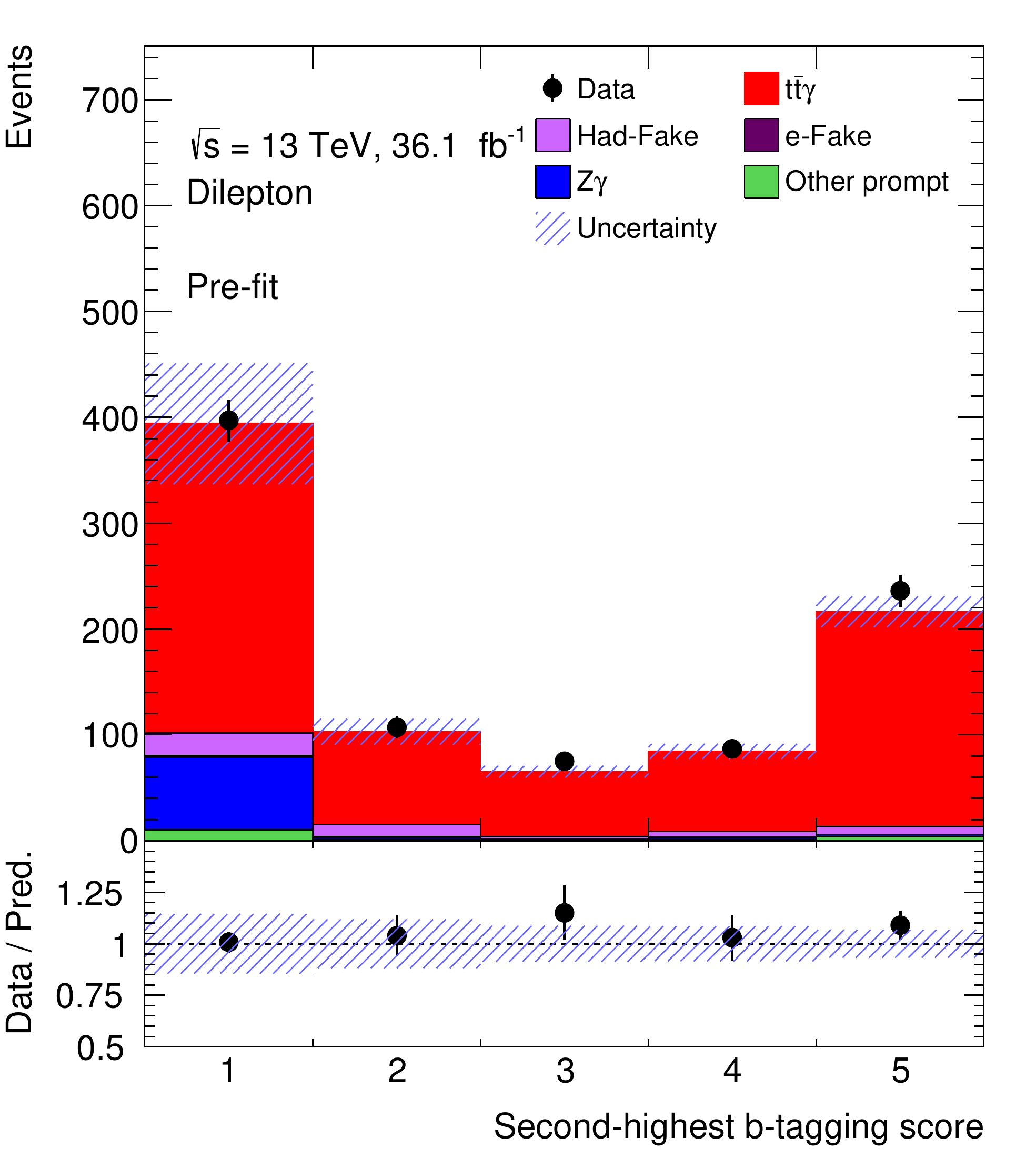}
\hspace{-0.02\linewidth}
\includegraphics[width=0.33\linewidth]{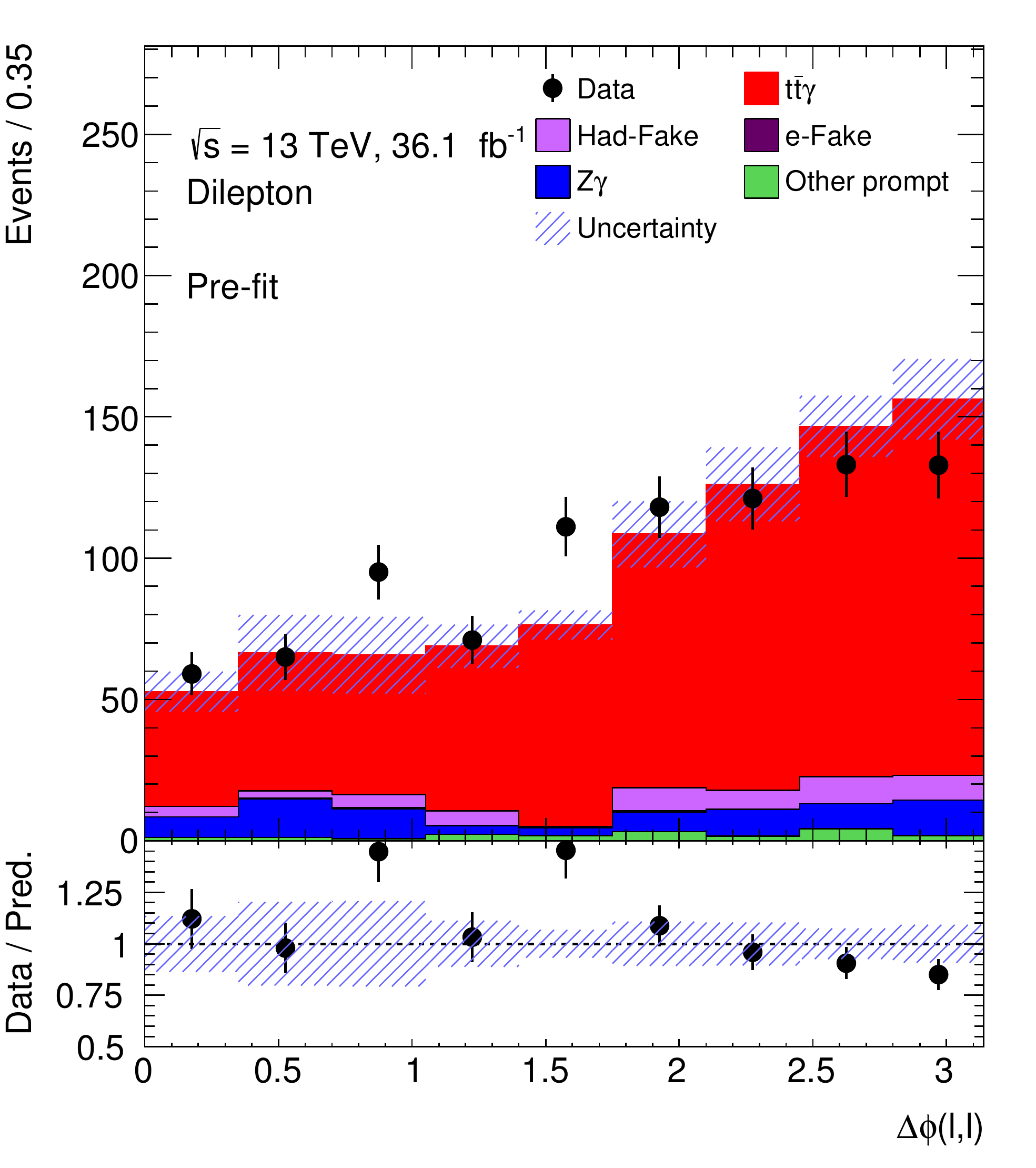}
\hspace{-0.02\linewidth}
\includegraphics[width=0.33\linewidth]{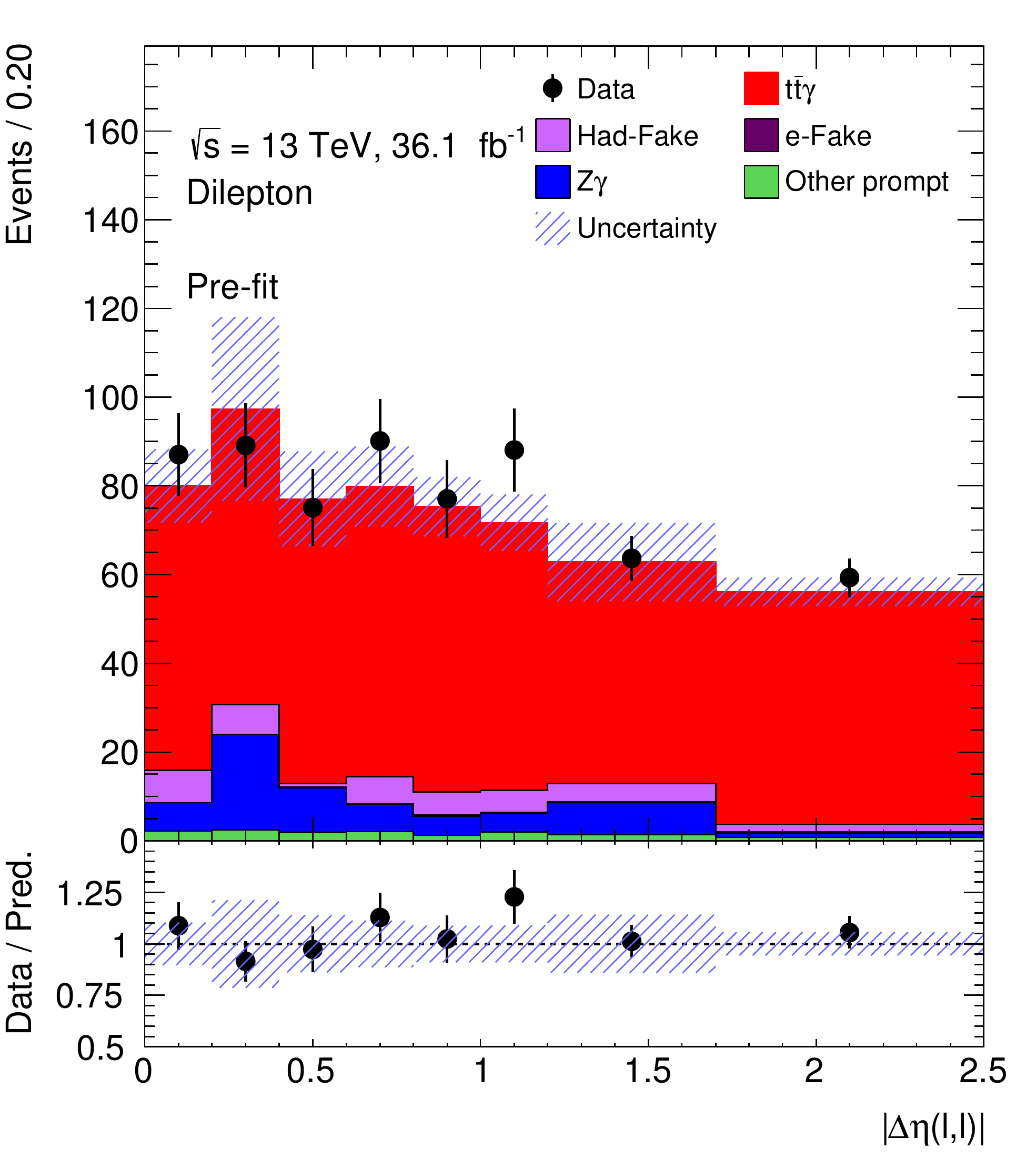}
\hspace{-0.02\linewidth}

\includegraphics[width=0.33\linewidth]{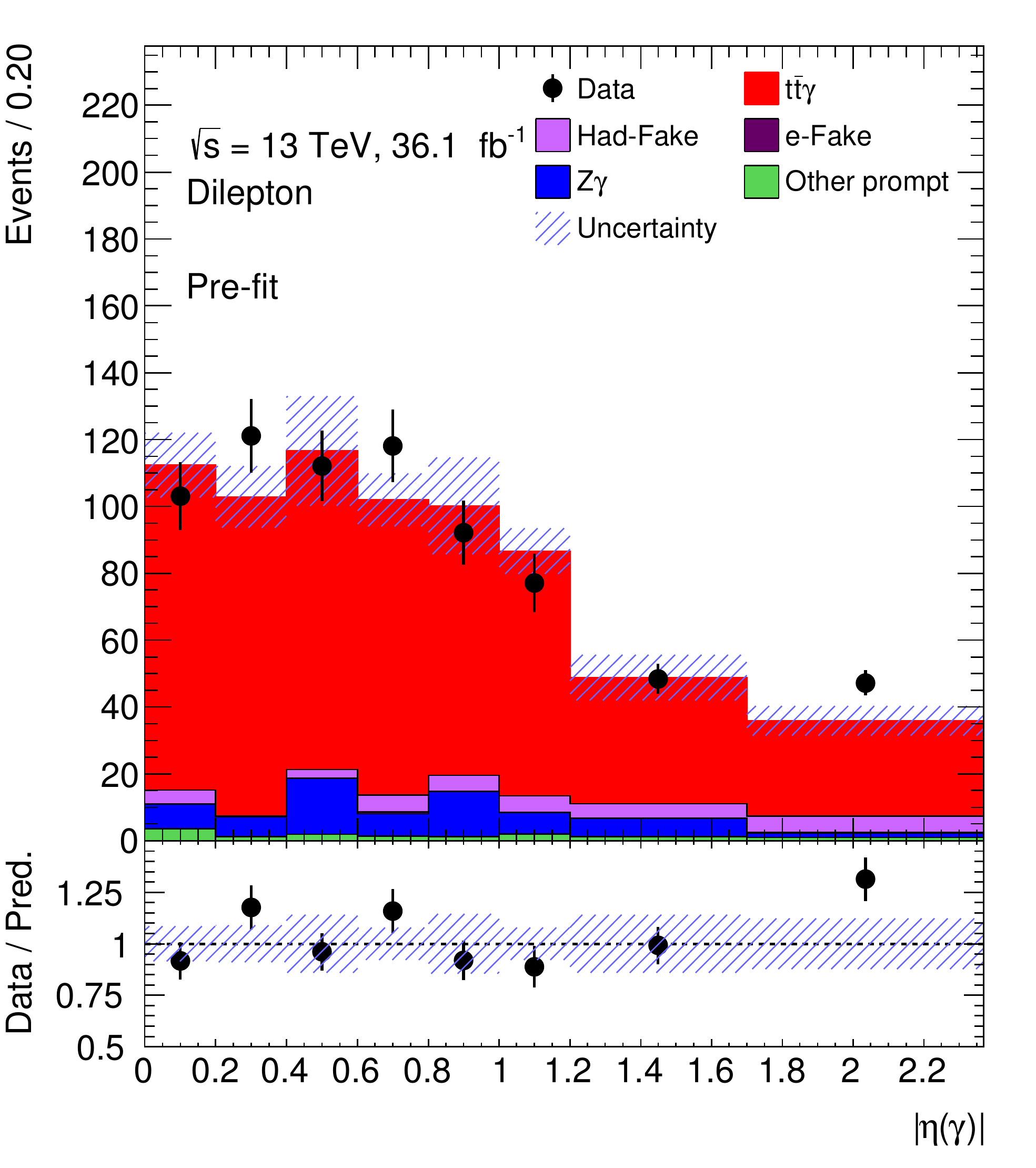}
\hspace{-0.02\linewidth}
\includegraphics[width=0.33\linewidth]{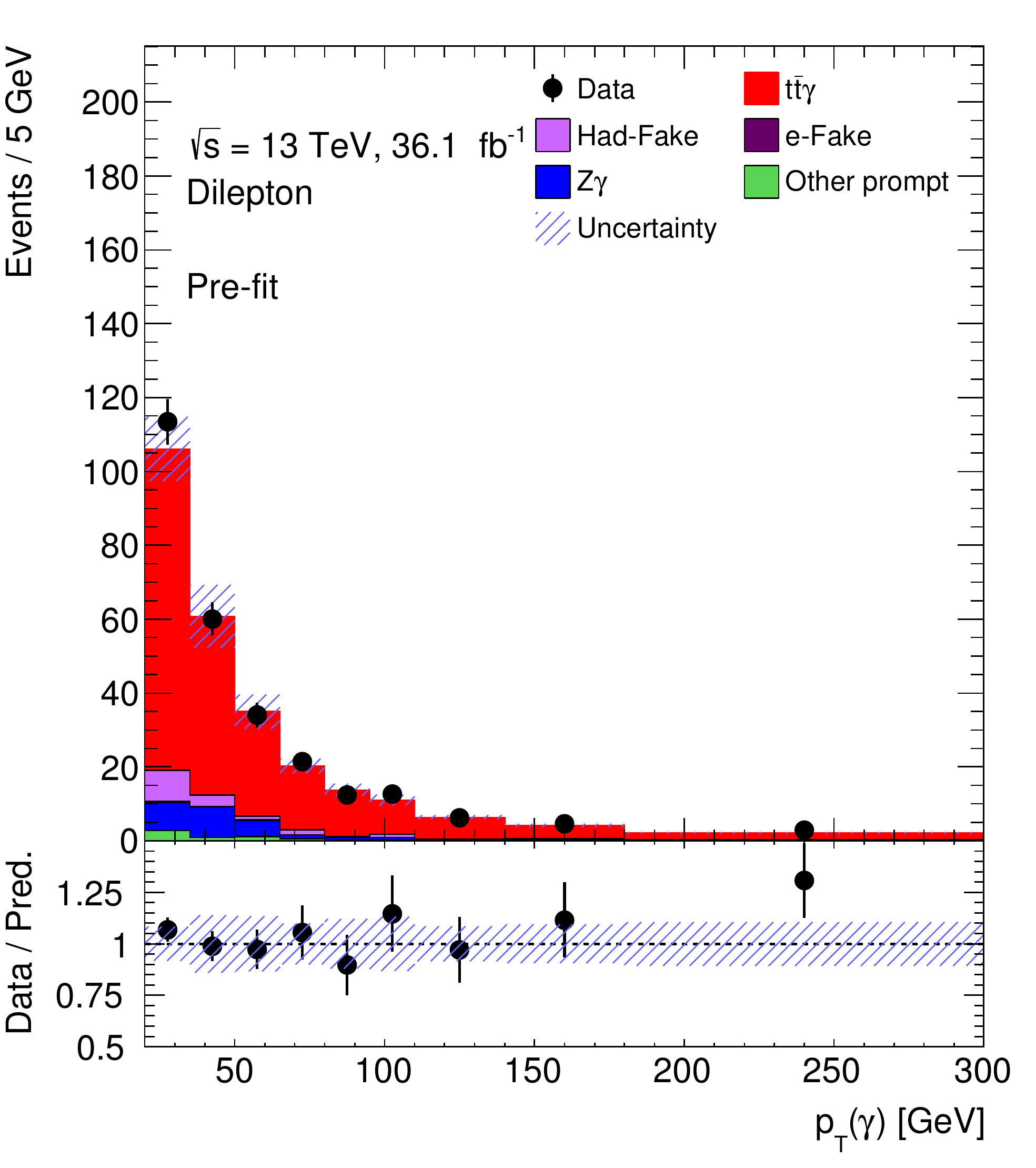}
\hspace{-0.02\linewidth}
\includegraphics[width=0.33\linewidth]{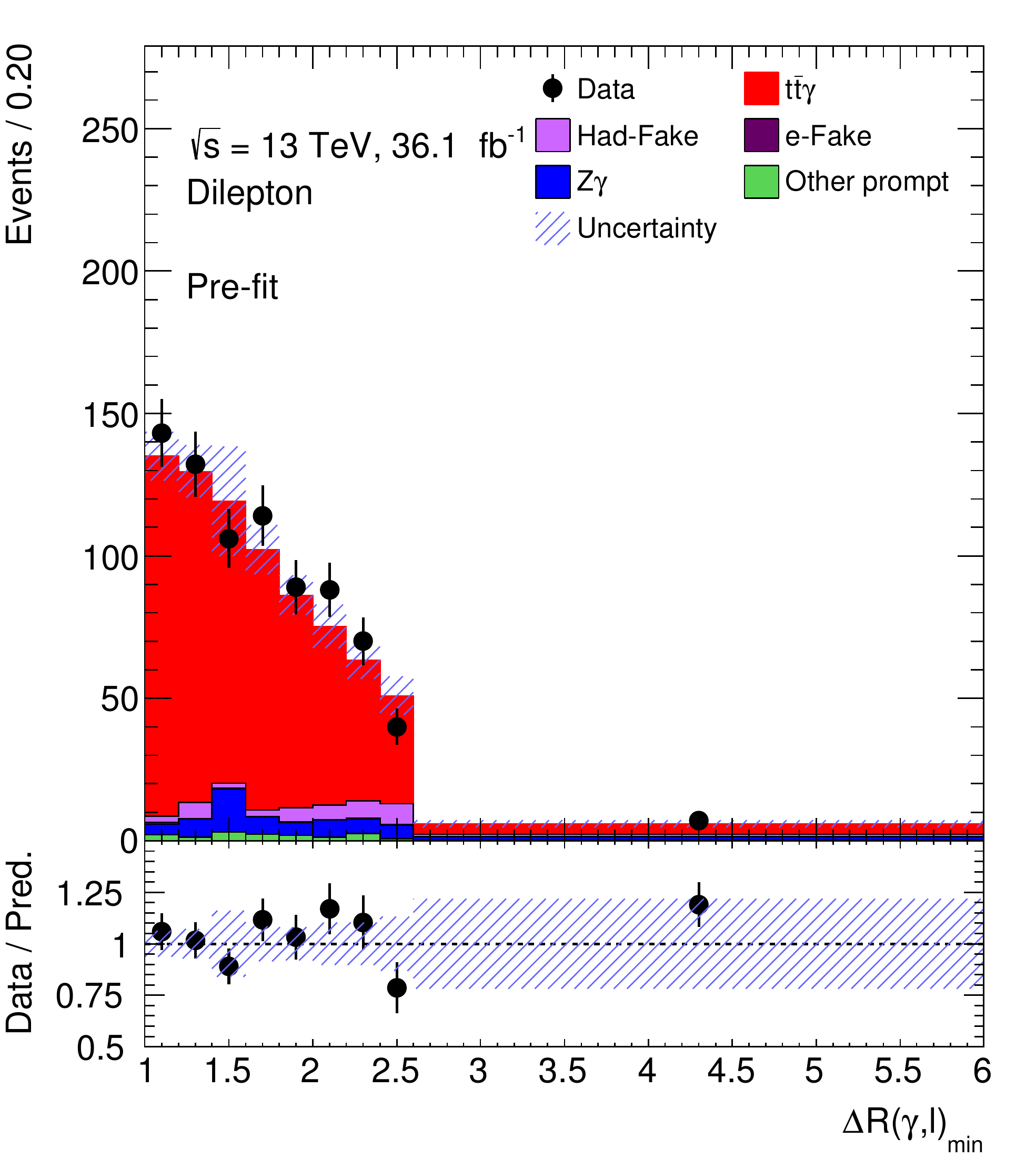}
\hspace{-0.02\linewidth}

\caption [A selection of pre-fit distributions for the \chll channel.] {A selection of pre-fit distributions for the \chll channel. All scale factors and systematic uncertainties are included.}\label{fig:prefitPlotDLSFs}
\end{figure}

This section shows a selection of distributions in the respective SRs comparing data to predicted MC. All appropriate scale factors and scalings from the previous sections have been applied. 
For brevity, only the \chljets and \chll combined channels are shown in Figure~\ref{fig:prefitPlotSLSFs} and ~\ref{fig:prefitPlotDLSFs}, respectively.
All systematic uncertainties (Chapter~\ref{sec:systematics}) have been included and are represented by the blue hashed band (which includes the statistical uncertainty). The sources of these uncertainties will be further discussed in Section~\ref{sec:systematics}.
``Pre-fit" on the distributions indicates this is before any fitting has been performed.
Table~\ref{tab:prefiteventYieldswSFs} shows the yields for the signal, backgrounds, and data.

The variables associated with ``$b$-tagging score" relate to the weights presented in Table~\ref{tab:btaggin_bins} for all jets.
The transverse mass of the $W$-boson (\mwt) is calculated as
\begin{align}
\text{\mwt{}}= \sqrt{2 \times \pt(l) \times \MET \times (1-\text{cos}(\Delta \phi(l,\MET)))}.
\end{align}
The variable \HT is the scalar sum of the transverse momentum for final state jets and leptons.
In the case where a jet does not exist (such as the fifth leading jet \pt), the variable is padded with zeros.

The selected plots shown are also the inputs for the Event-level Discriminator, which will be explained in Chapter~\ref{sec:ELD}. The same phase space is shown for the training of the \chljets NN, while a slightly tighter phase space is shown compared to the training of the \chll NN, as this represents the full signal region.
In all figures, good data/MC can be seen with the projected uncertainties covering any deviations that might arise.

\begin{table}[!htb]
\begin{center}
\scalebox{0.85}{
\begin{tabular}{|c|r|r|}
\hline

 & \chljets & \chll  \\ \hline \hline

\ttgamma & 6488 $\pm$ 420 &  720 $\pm$ 34 \\ \hline  

\hfake & 1435 $\pm$ 290 &  49 $\pm$ 27 \\ \hline  

\efake & 1653 $\pm$ 170 &  2 $\pm$ 1 \\ \hline  

\QCD & 356 $\pm$ 200 &  - \\ \hline

\Wgamma & 1126 $\pm$ 45 & - \\ \hline

\Zgamma & - & 75 $\pm$ 52 \\ \hline  

\Other & 691 $\pm$ 260 &  18 $\pm$ 7 \\ \hline  

Total & 11747 $\pm$ 710 &  863 $\pm$ 78 \\ \hline  

Data & 11662 &  902 \\ \hline 

\end{tabular}
}
\caption{Pre-fit yields for signal and backgrounds for the \chljets and \chll channels. All scale factors and systematic uncertainties are included.}
\label{tab:prefiteventYieldswSFs}
\end{center}
\end{table}

\begin{figure}[!h]
\centering
\includegraphics[width=0.33\linewidth]{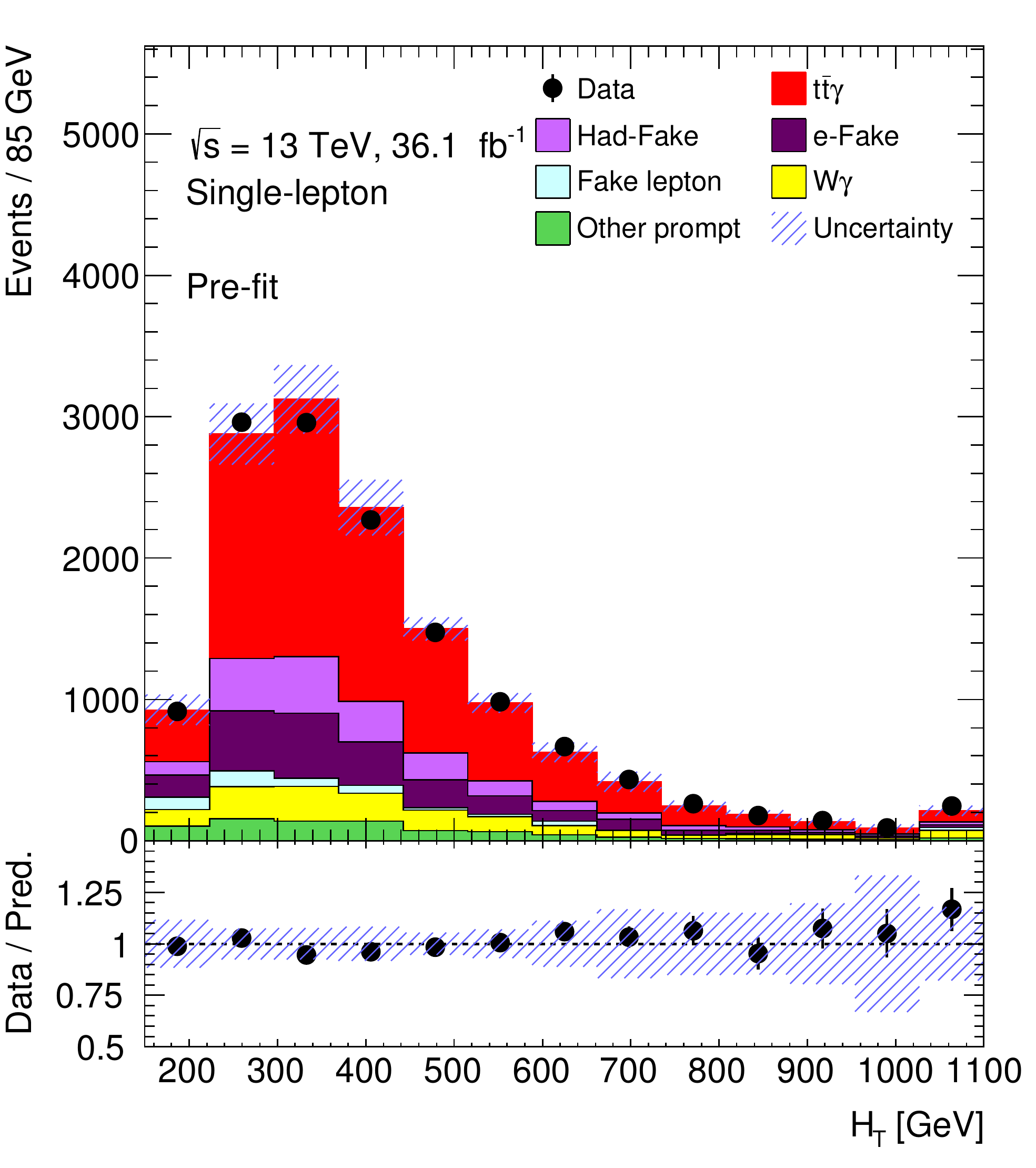}
\hspace{-0.02\linewidth}
\includegraphics[width=0.33\linewidth]{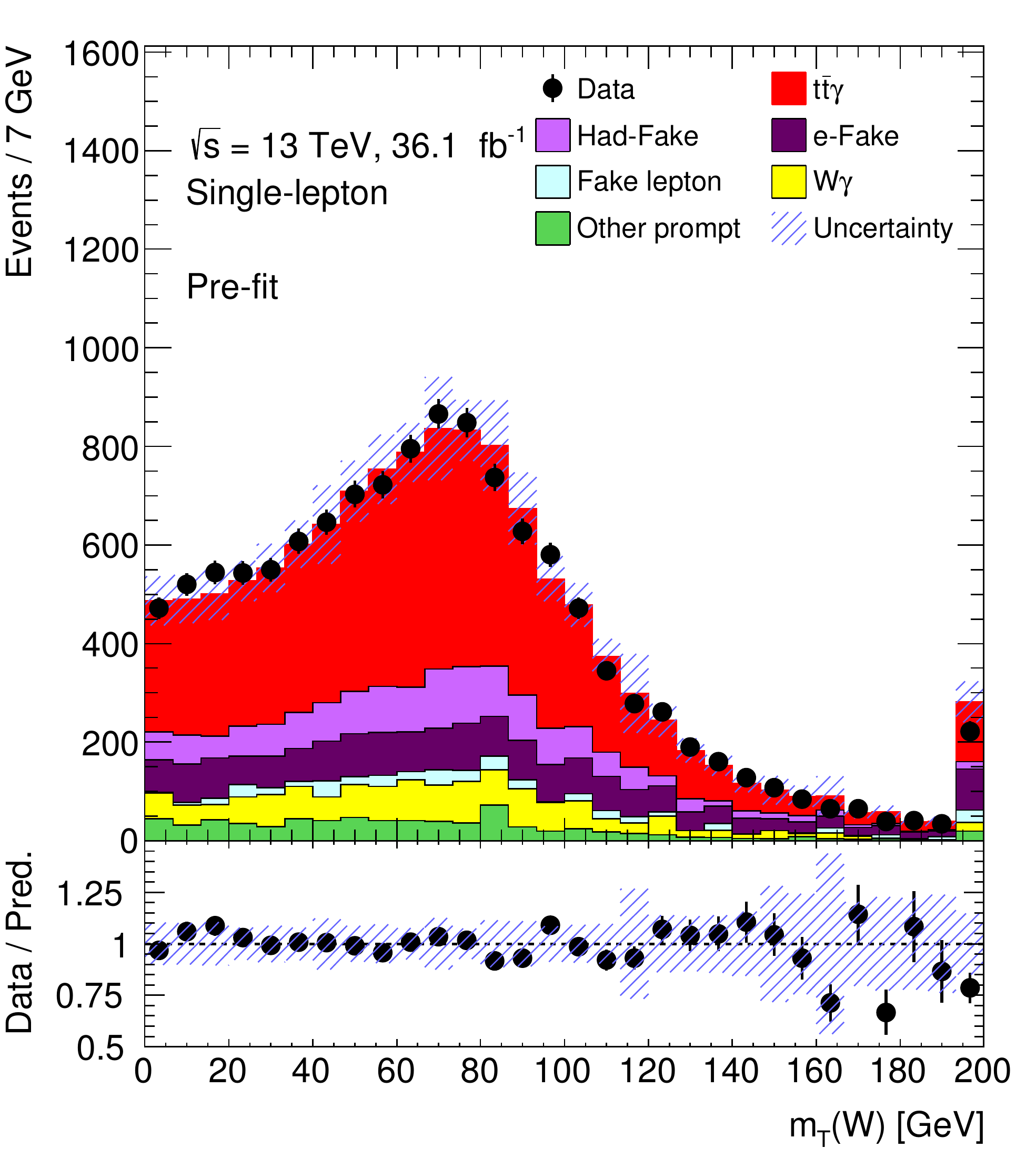}
\hspace{-0.02\linewidth}
\includegraphics[width=0.33\linewidth]{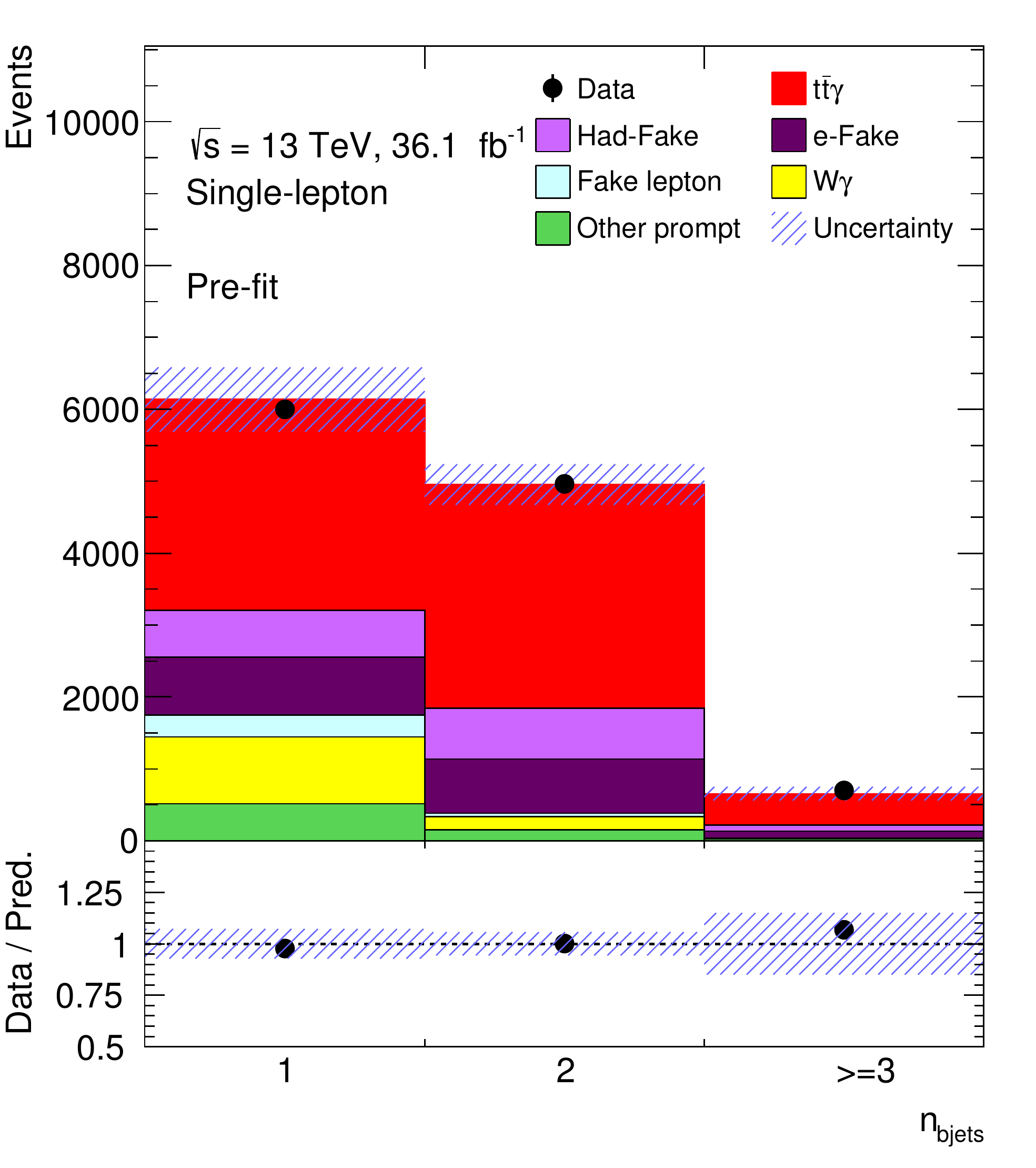}
\hspace{-0.02\linewidth}

\includegraphics[width=0.33\linewidth]{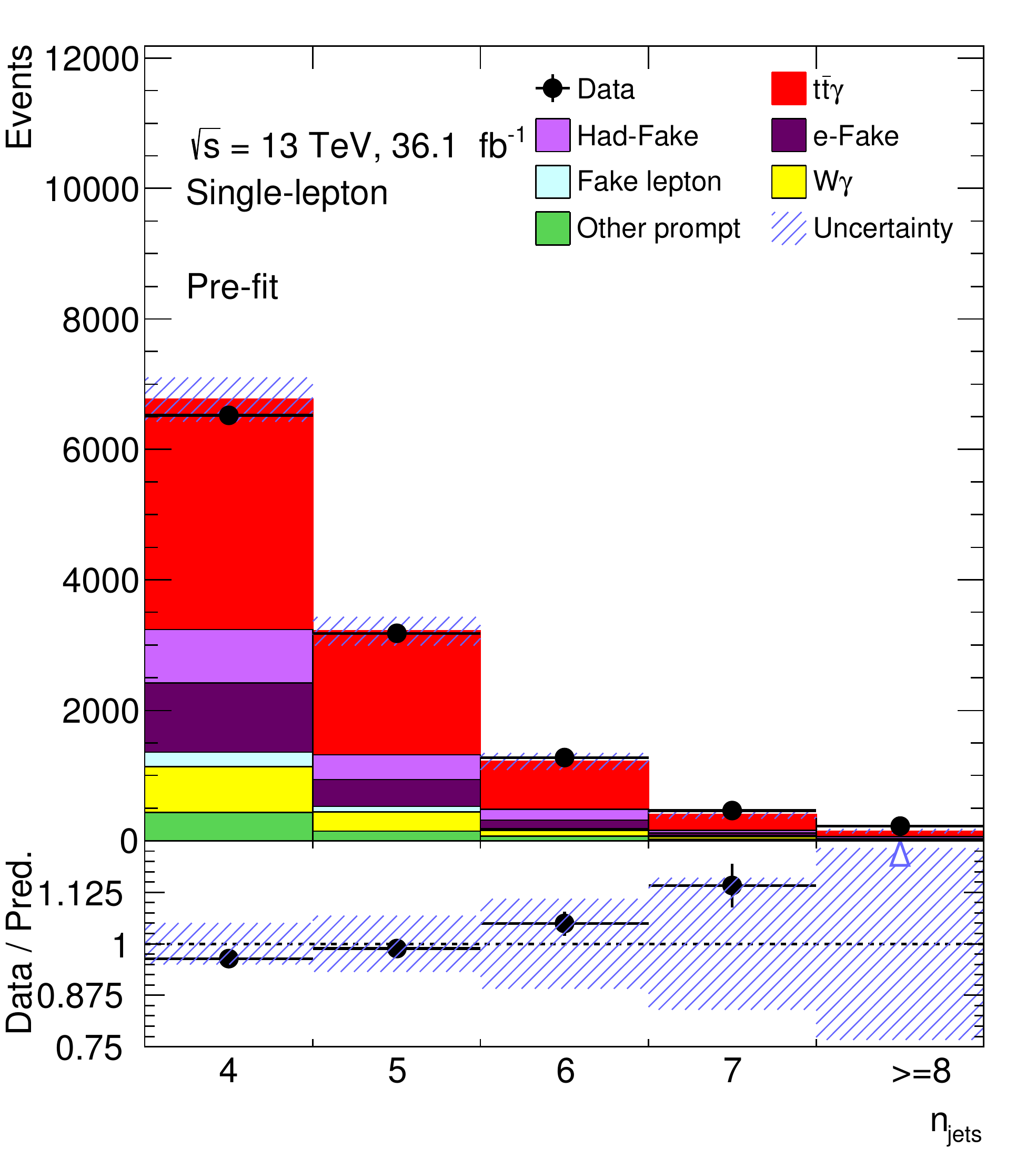}
\hspace{-0.02\linewidth}
\includegraphics[width=0.33\linewidth]{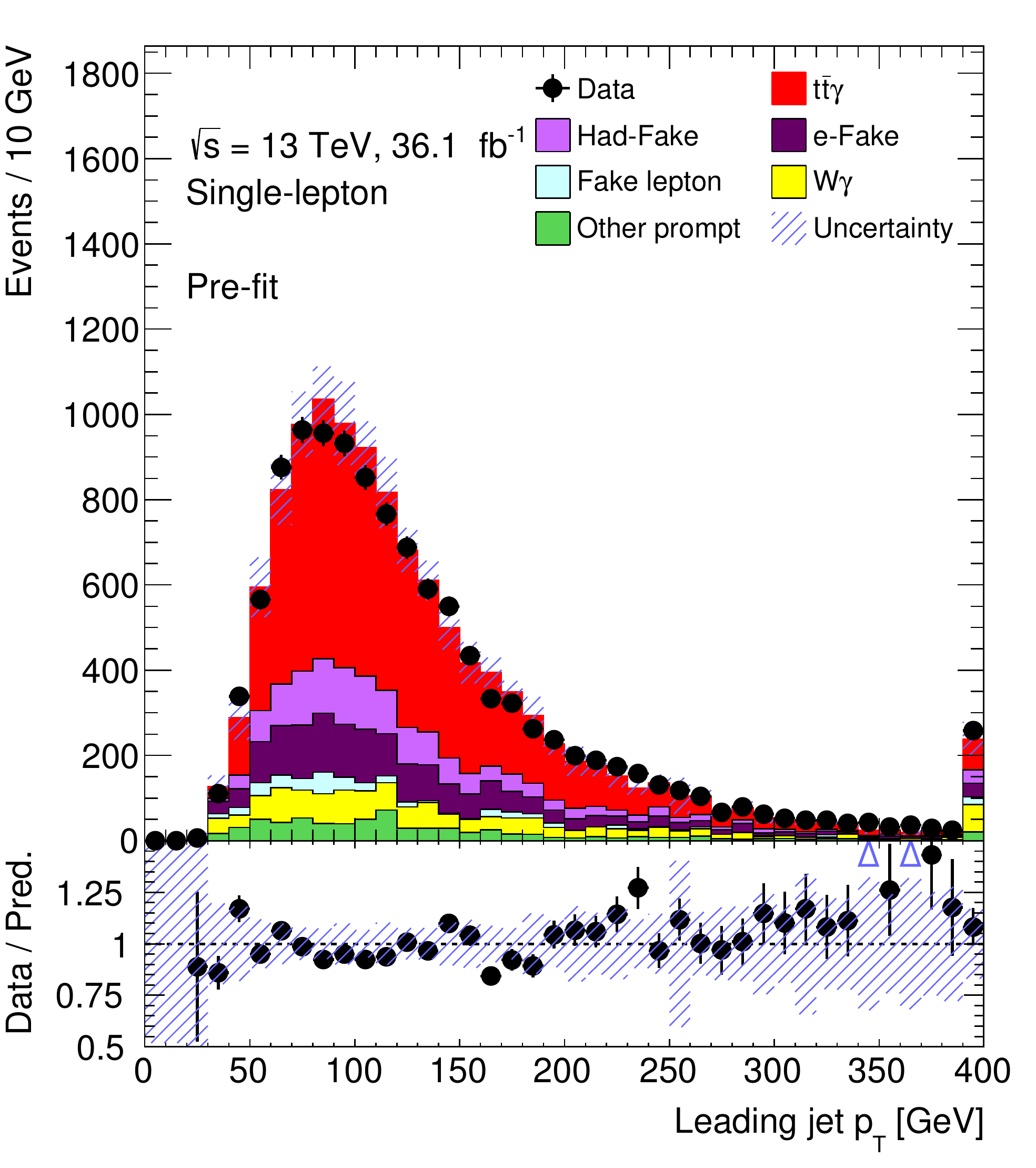}
\hspace{-0.02\linewidth}
\includegraphics[width=0.33\linewidth]{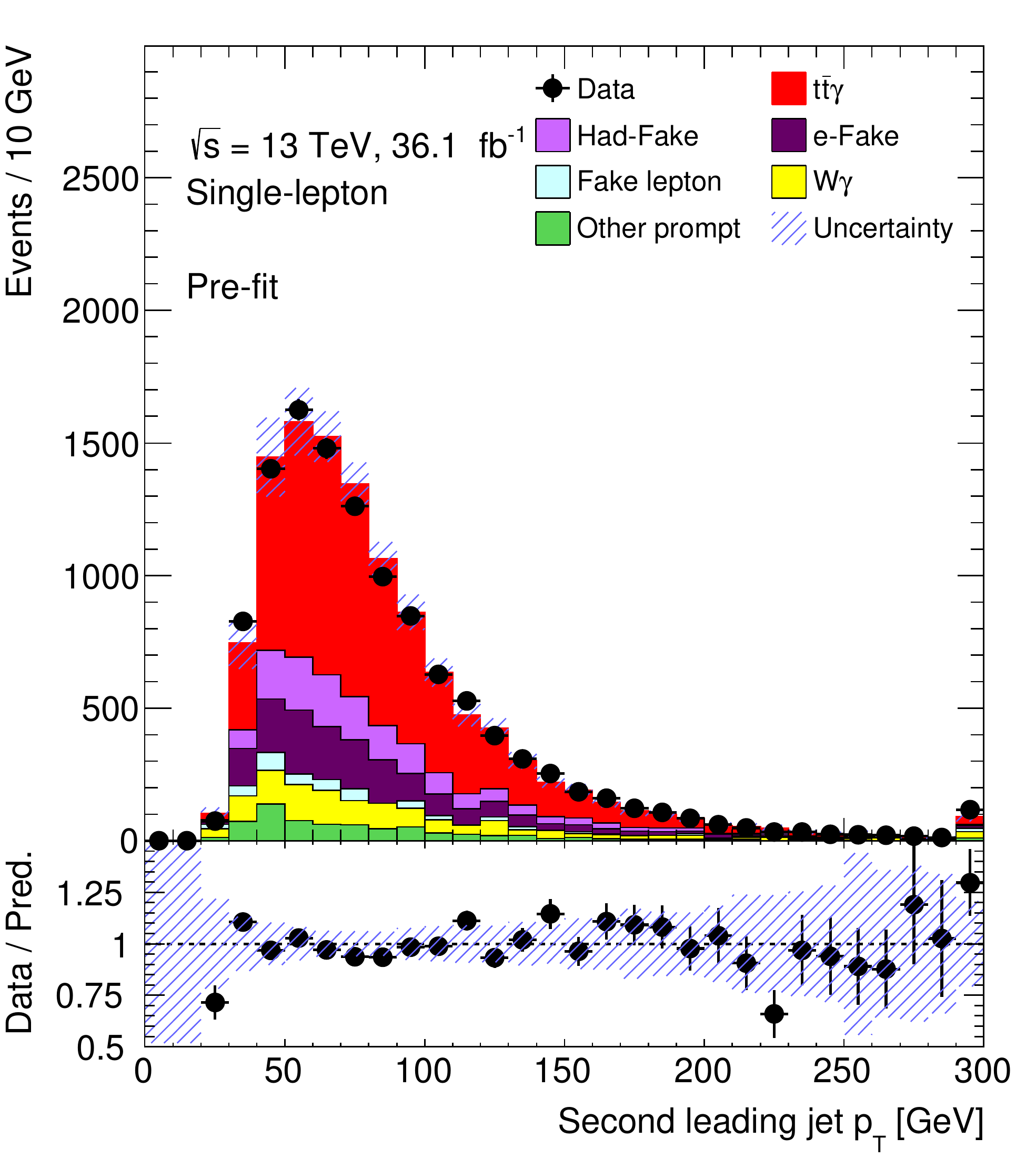}
\hspace{-0.02\linewidth}

\includegraphics[width=0.33\linewidth]{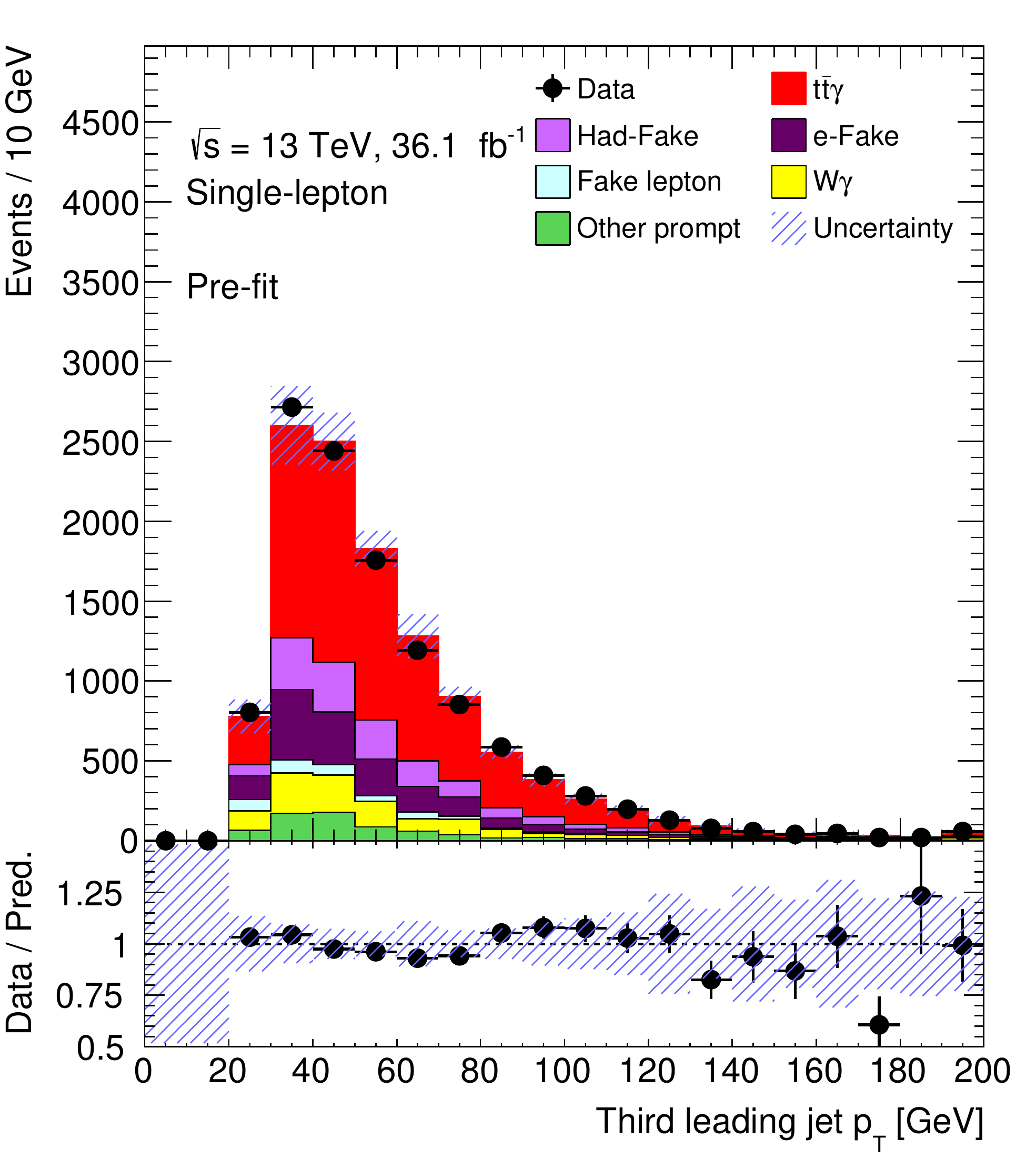}
\hspace{-0.02\linewidth}
\includegraphics[width=0.33\linewidth]{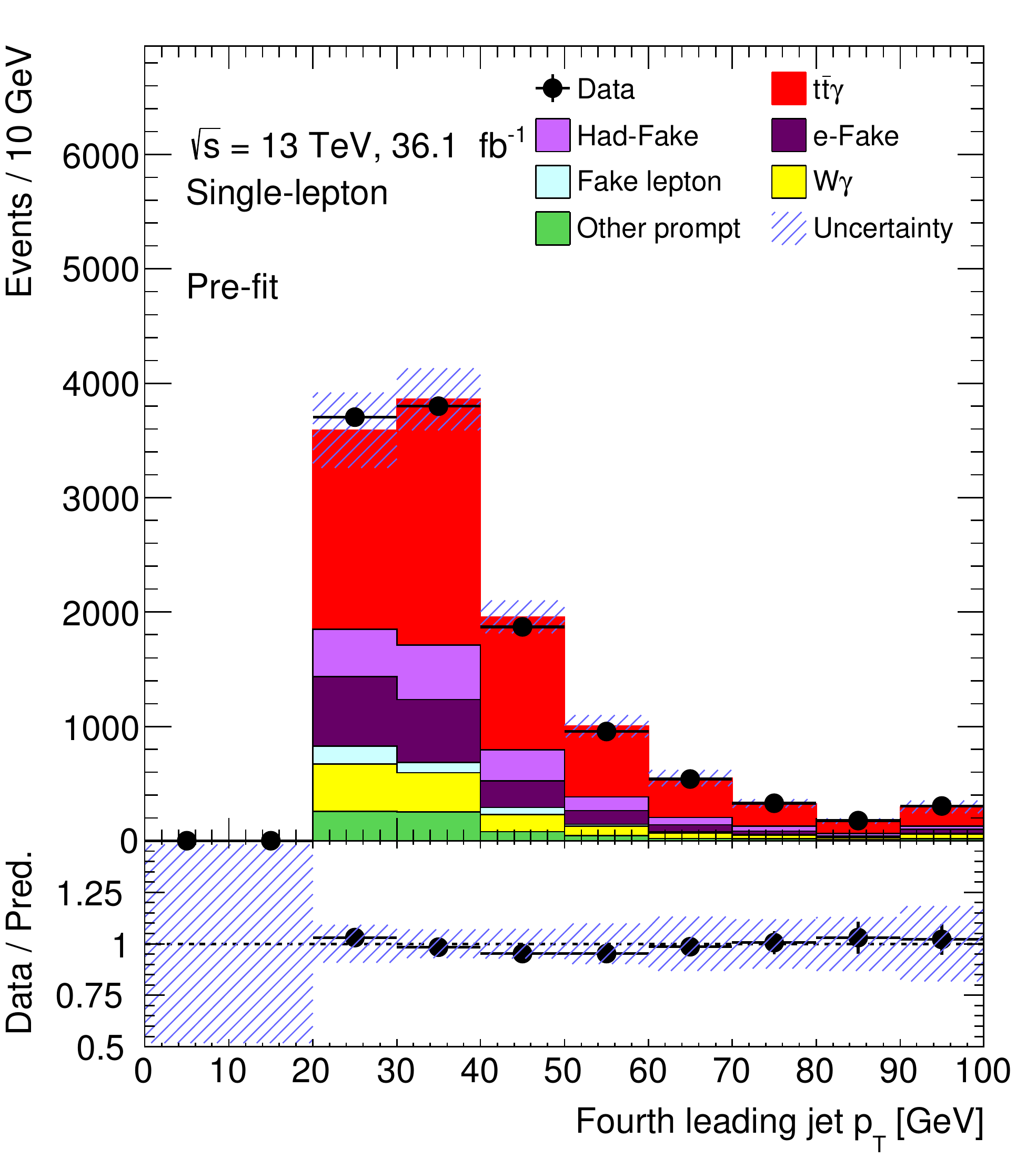}
\hspace{-0.02\linewidth}
\includegraphics[width=0.33\linewidth]{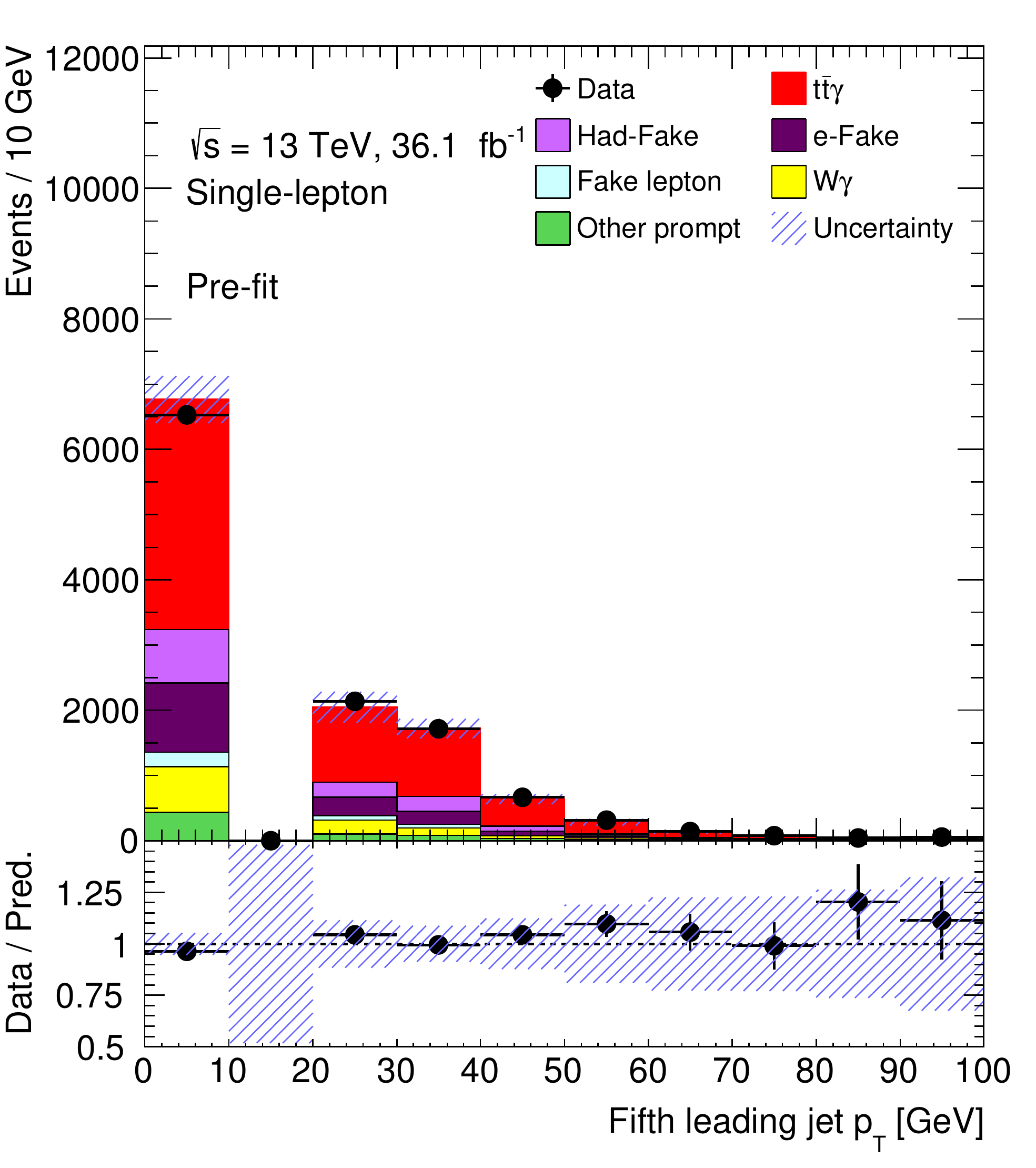}
\hspace{-0.02\linewidth}

\includegraphics[width=0.33\linewidth]{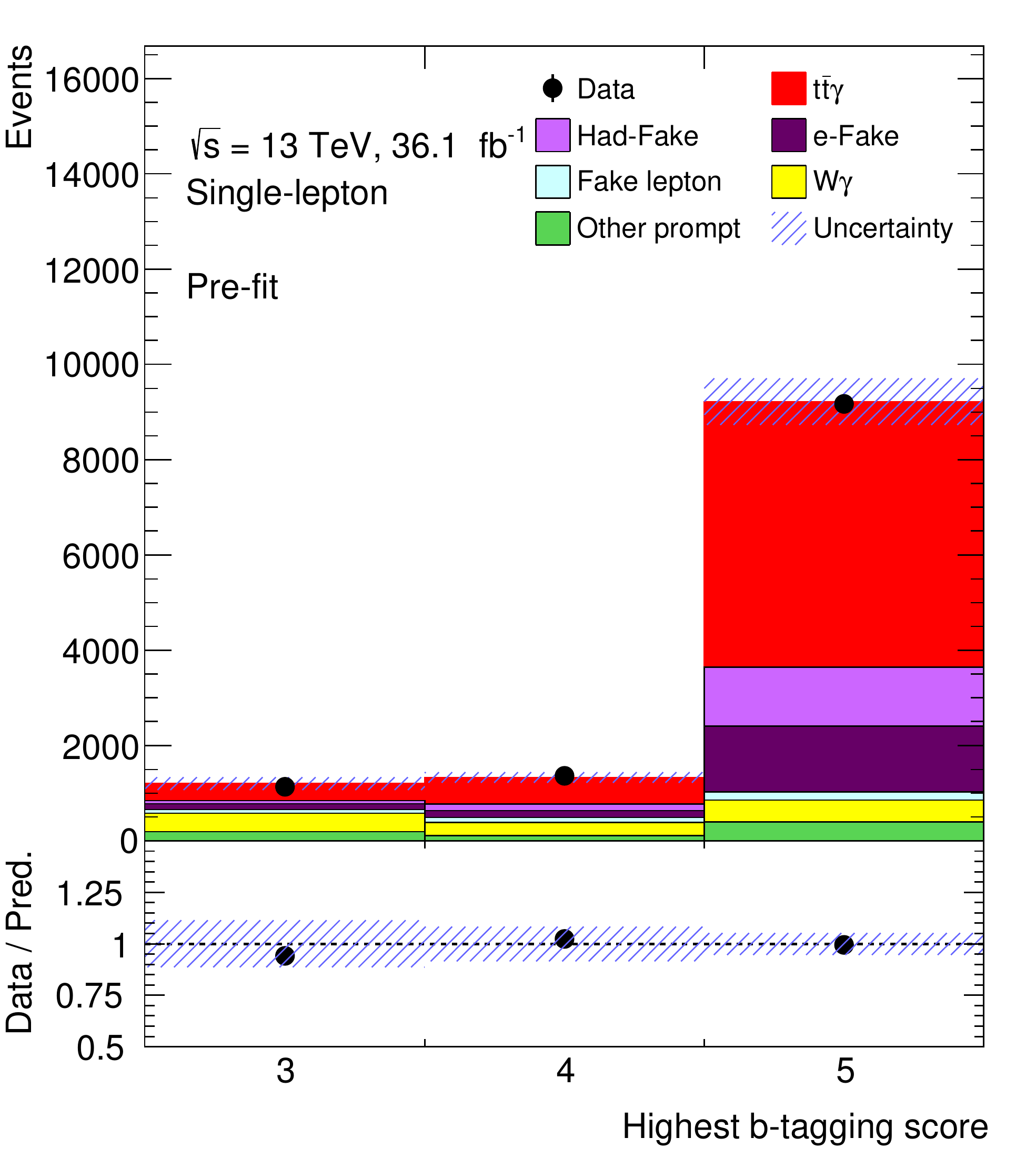}
\hspace{-0.02\linewidth}
\includegraphics[width=0.33\linewidth]{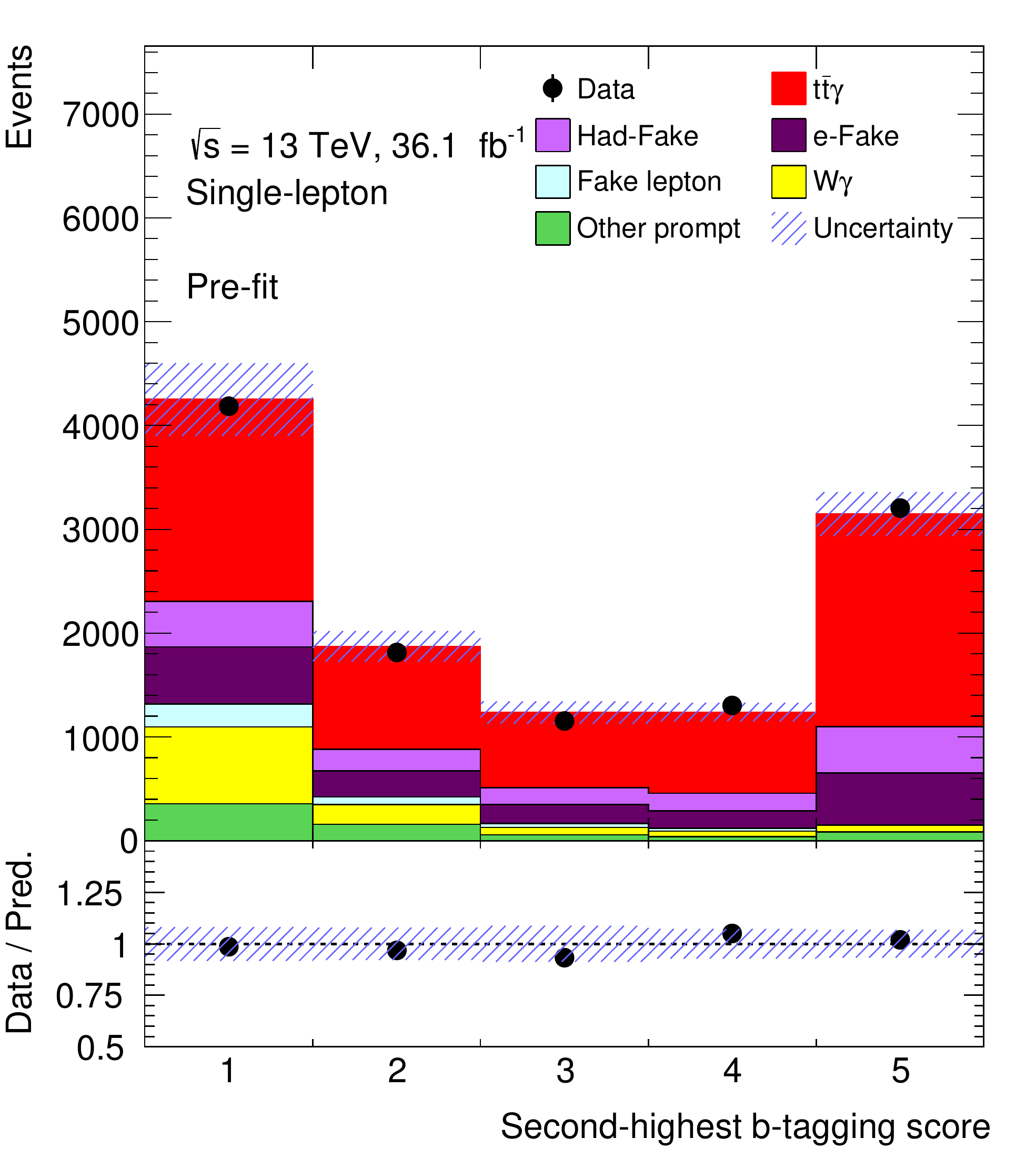}
\hspace{-0.02\linewidth}
\includegraphics[width=0.33\linewidth]{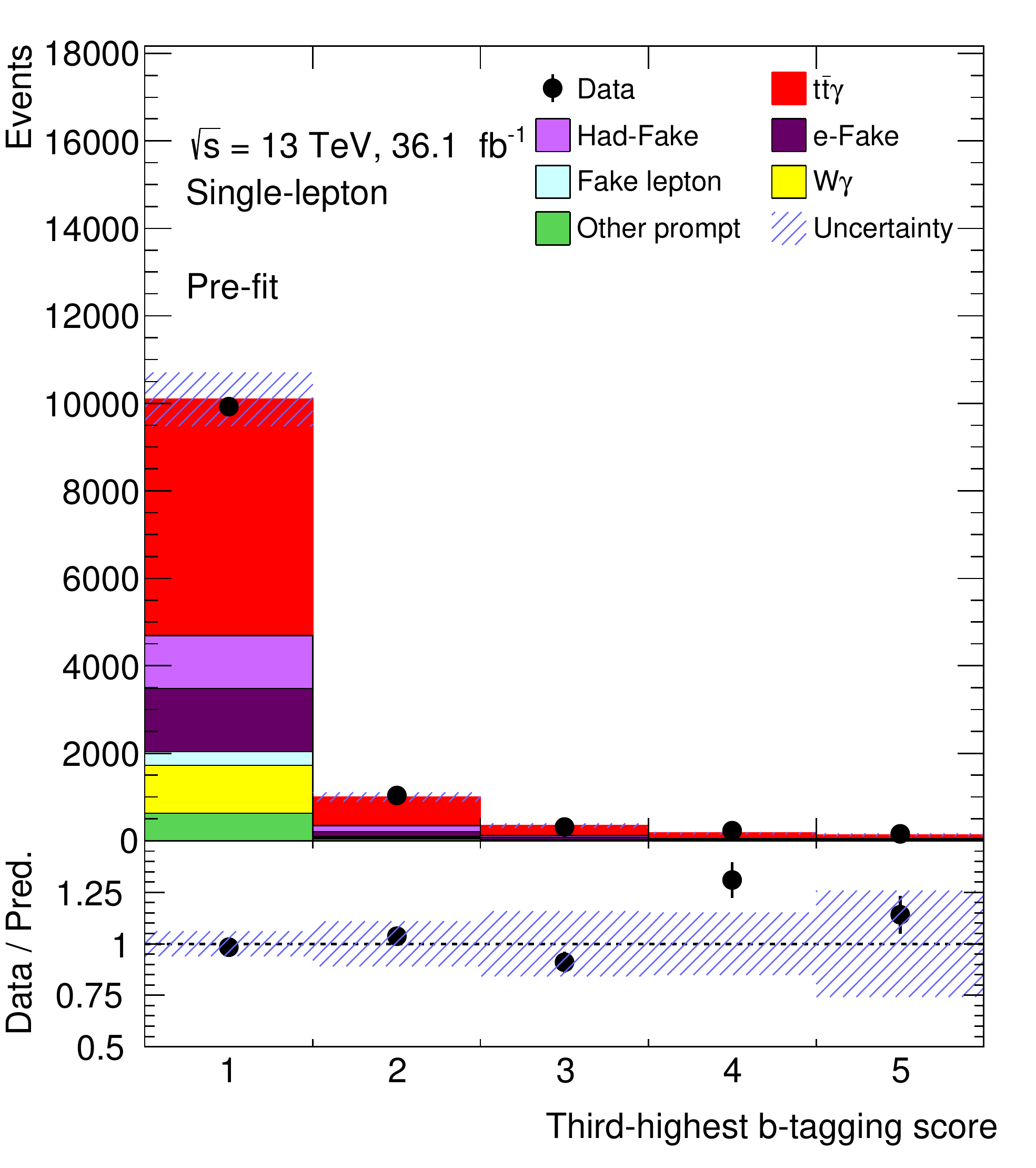}
\hspace{-0.02\linewidth}

\phantomcaption 
\end{figure}

\begin{figure}
\ContinuedFloat
\centering

\includegraphics[width=0.33\linewidth]{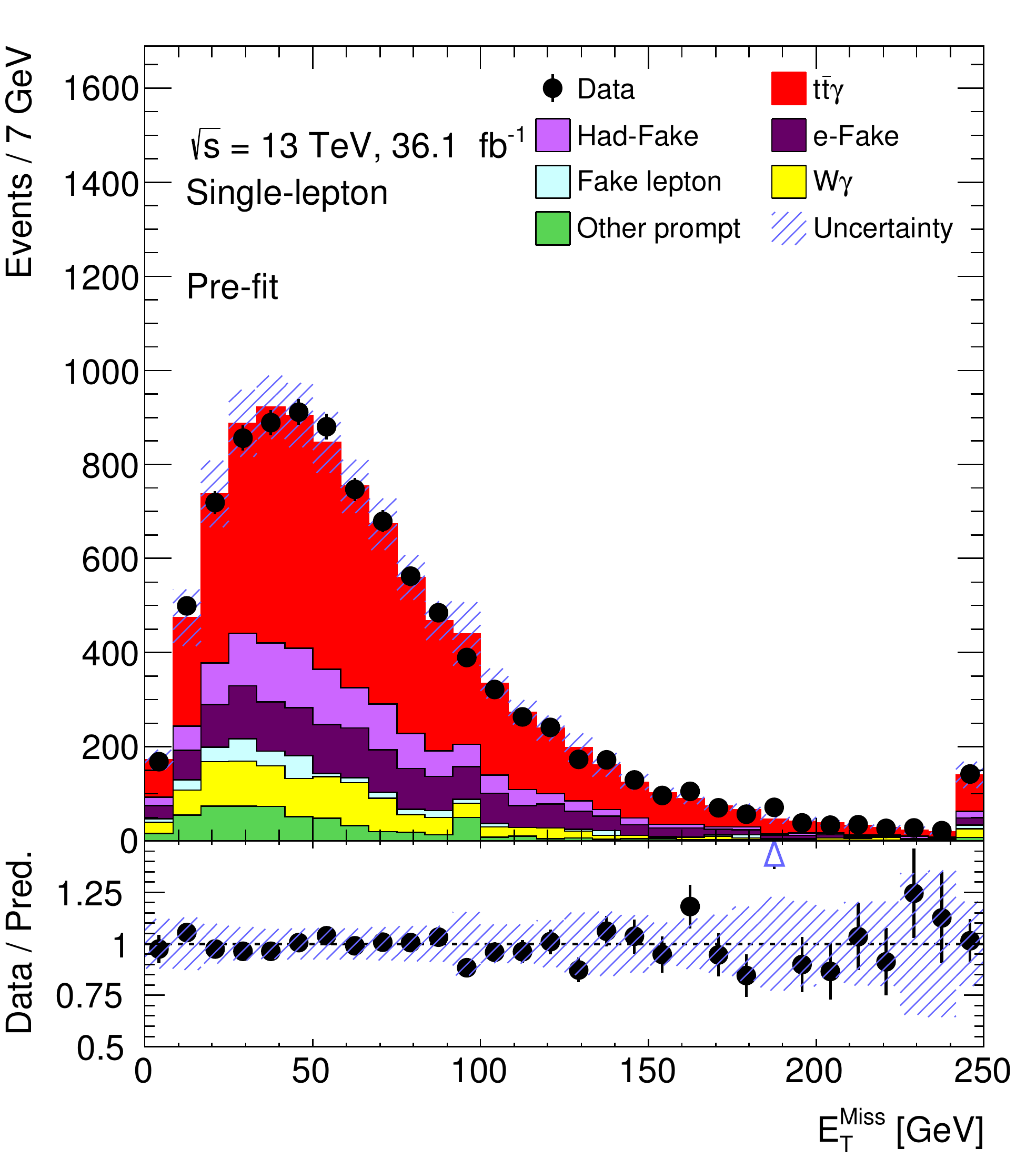}
\hspace{-0.02\linewidth}
\includegraphics[width=0.33\linewidth]{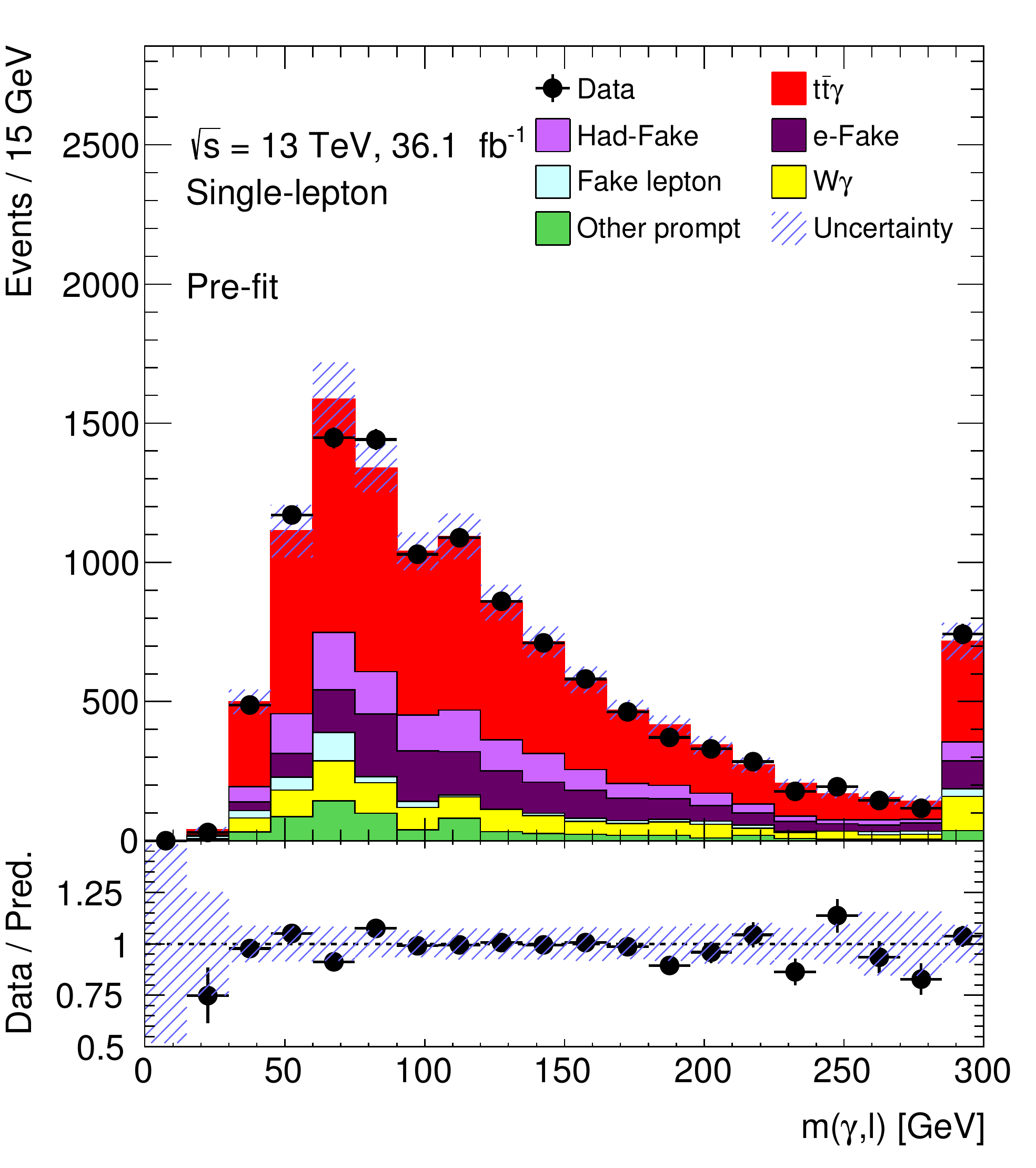}
\hspace{-0.02\linewidth}
\includegraphics[width=0.33\linewidth]{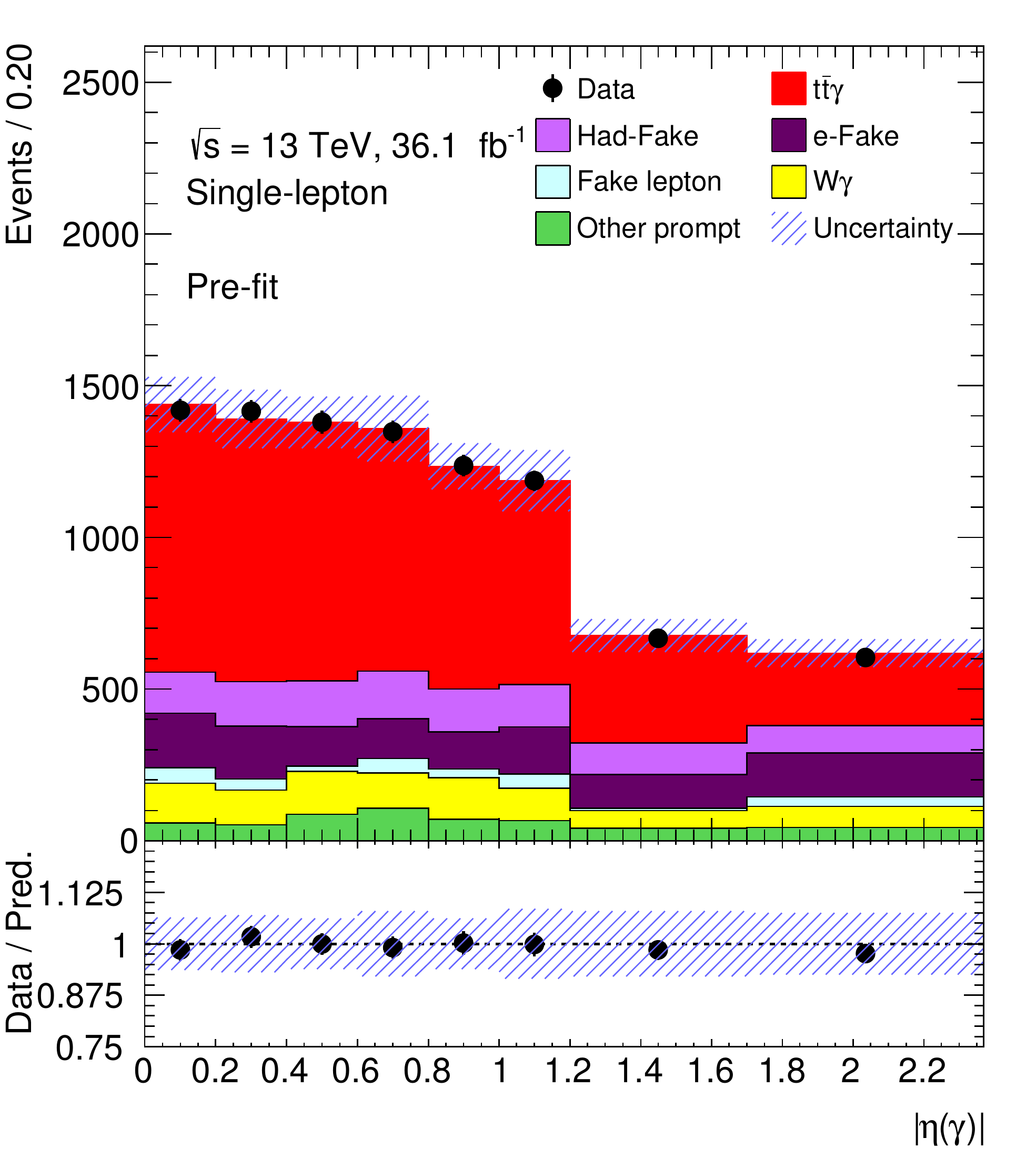}
\hspace{-0.02\linewidth}

\includegraphics[width=0.33\linewidth]{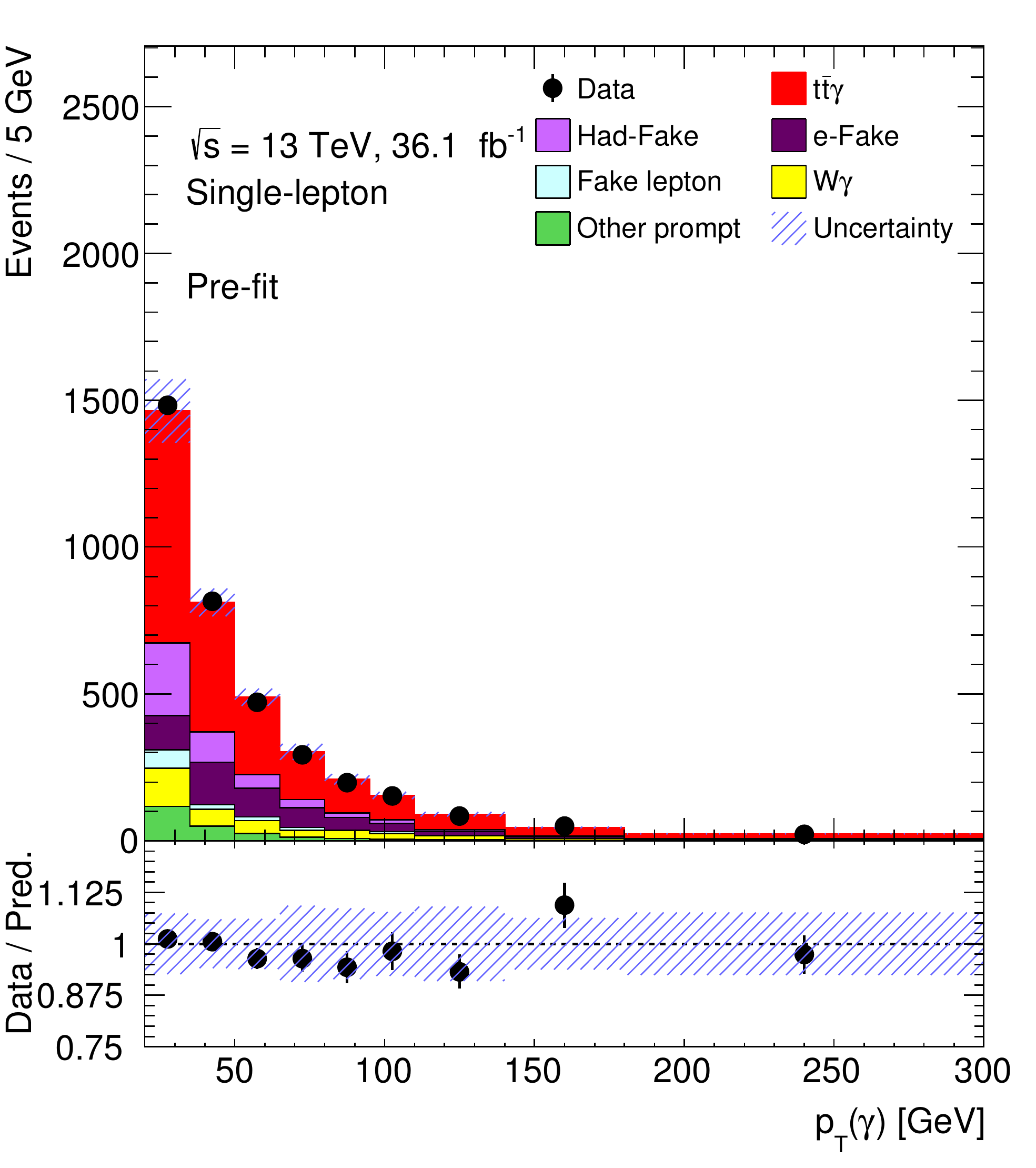}
\hspace{-0.02\linewidth}
\includegraphics[width=0.33\linewidth]{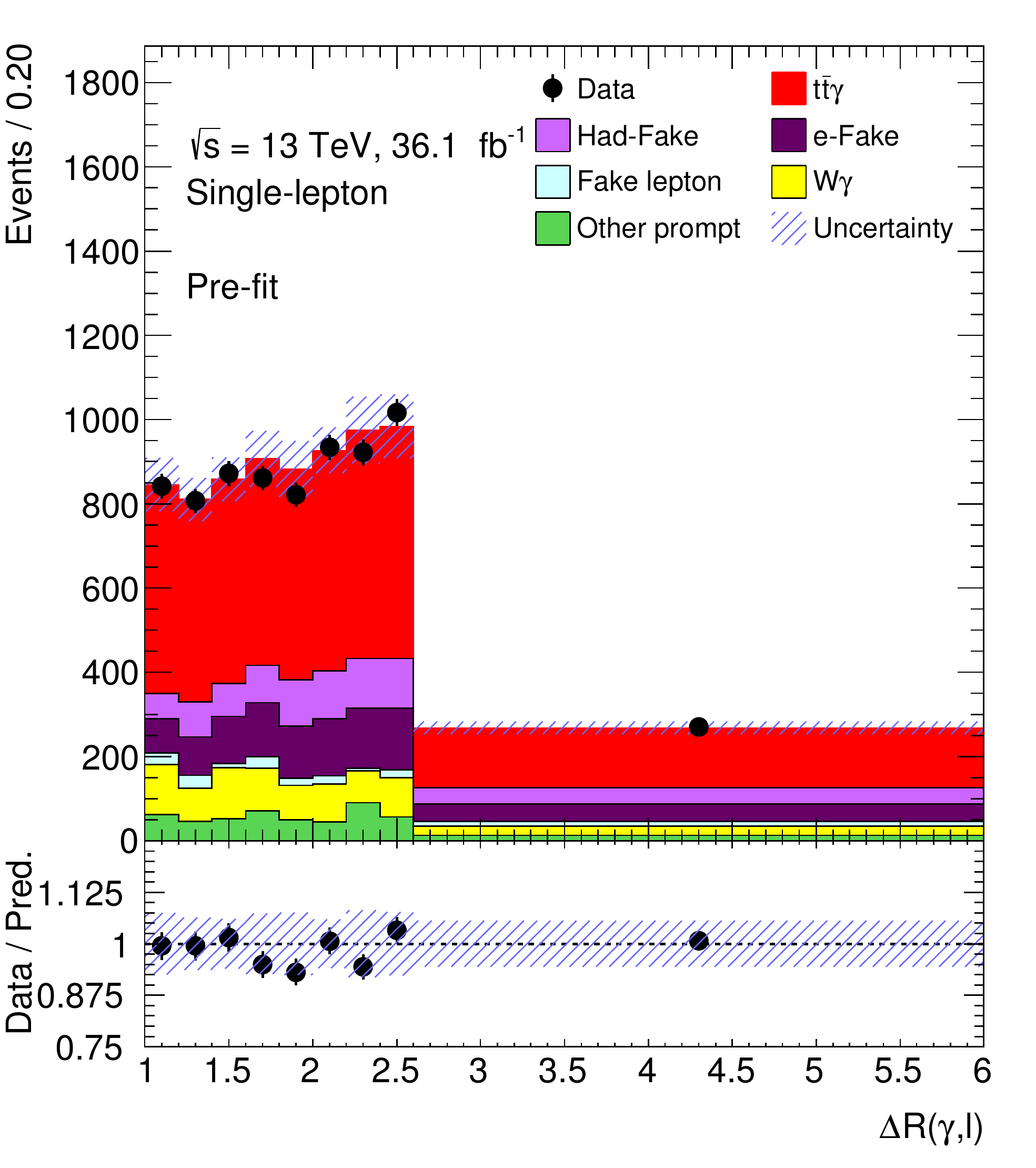}
\hspace{-0.02\linewidth}

\caption [A selection of pre-fit distributions for the \chljets channel.] {A selection of pre-fit distributions for the \chljets channel. All scale factors and systematic uncertainties are included.}\label{fig:prefitPlotSLSFs}
\end{figure}   

\FloatBarrier

\chapter[Object- and event-level neural networks]{Object- and event-level neural networks}
\label{sec:NNs}

An event has an intrinsic topology; the number of jets or leptons, the angular distance between different particles, the \MET, all form the overall defining characteristics of an event for a certain process.
The constituents of an event, the photons, electrons, individual jets etc., have kinematic properties and, in the case of jets, flavours.
The topology of an event as well as the properties of its constituents can be exploited for better signal/background separation. 
NNs are used for this purpose. 

This chapter presents two separate neural networks (NNs) developed in the period of this thesis.
The first (Section~\ref{sec:PPT}) is an analysis independent, object-level NN designed to work on top of the photon reconstruction and identification presented in Chapter~\ref{sec:photonDefs}.
Labelled the Prompt Photon Tagger (PPT), this NN was developed as a G\"ottingen effort with contributions largely from Benedikt V\"olkel during his Master's thesis~\cite{bvoelkel}.
The application and evaluation to \ttgamma events, as well as related studies include significant contributions from myself and Knut Zoch, who focused on crucial systematic uncertainties and scale factor studies.

The second NN developed and presented is the Event-level Discriminator (ELD) for the \ttgamma analysis (Section~\ref{sec:ELD}). The \ELD makes use of the PPT as an input variable in the \chljets channel, as will be explained later.

  
\section[The Prompt Photon Tagger]{The Prompt Photon Tagger}
\label{sec:PPT}

In the previous measurements at 7 and 8~\TeV, the dominant background process was from \hfakes. Thus, a multivariate analysis (MVA) tool in the form of a NN was developed to try mitigate the role this background would play in the final fit. 
Furthermore, since this background is common among many analyses within \ATLAS, the goal was to keep this tool analysis independent. 
To accomplish this the PPT was trained purely on shower shape variables (Chapter~\ref{sec:egamma}) using the same prompt photon and \hfake enriched samples used in photon identification.
Section~\ref{sec:PPTtraining} gives an overview of the procedure to train the \PPT and Section~\ref{sec:PPTEval} evaluates the \PPT's performance on \ttgamma samples. Section~\ref{sec:PPTsysts} gives an overview of how scale factors and systematic uncertainties are derived for this novel tool.

\subsection{Training}\label{sec:PPTtraining}

Training was carried out on two MC datasets provided by the photon identification group within \ATLAS. The first, simulating the QCD-Compton process, contains prompt photons. The second MC dataset contains di-jets events which contains \hfake candidates.
Photon candidates are required to have a $\pt > 25$~GeV as well as $|\eta| < 2.5$. The crack region ($1.37<|\eta|<1.52$) is also excluded. Photons are required to pass \tight identification\footnote{In future versions of the \PPT the \tight requirement will be dropped to cover a larger phase space and cater for a wider range of analyses.}.
Approximately 1 million prompt photons and 200k \hfakes are selected from the MC datasets with 80\% of them used for training and 20\% used for testing purposes.

A subset of shower shape variables was selected based on their separation information from Table~\ref{tab:showershapes}. This is defined as 
\begin{equation}\label{eq:separation}
S=\frac{1}{2}\sum_{i}^{\text{bins}}\frac{(s_{i}-b_{i})^{2}}{s_{i}+b_{i}},
\end{equation}
where for each bin, $i$, $s_{i}$ and $b_{i}$ indicate the number of signal and background events, respectively. 
Once ranked by separation, the top six variables were chosen for the \PPT. These are detailed in Table~\ref{tab:PPTVar} with their definitions explained in Chapter~\ref{sec:photonDefs}.

\begin{table}[!htbp]
	\centering
	\begin{tabular}{cc}
	\hline
	variable & separation \\
	\hline
	\hline
	$f_\mathrm{side}$ & $7.21$~\% \\
	$R_\phi$ & $7.01$~\% \\
	$R_\eta$ & $4.83$~\% \\
	$w_{\eta 1}$ & $4.14$~\% \\
  	$R_\mathrm{had}$ & $3.33$~\% \\
	$w_{\eta 2}$ & $2.01$~\% \\
	\hline
	\end{tabular}
	\caption{The six variables chosen for the \PPT based on separation calculated according to Equation~\ref{eq:separation}.}
	\label{tab:PPTVar}
\end{table}

\begin{figure}[h!]
	\begin{minipage}{0.49\textwidth}
	\subfloat[\label{fig:pptA}Training and test sets]{
		\includegraphics[width=\textwidth]{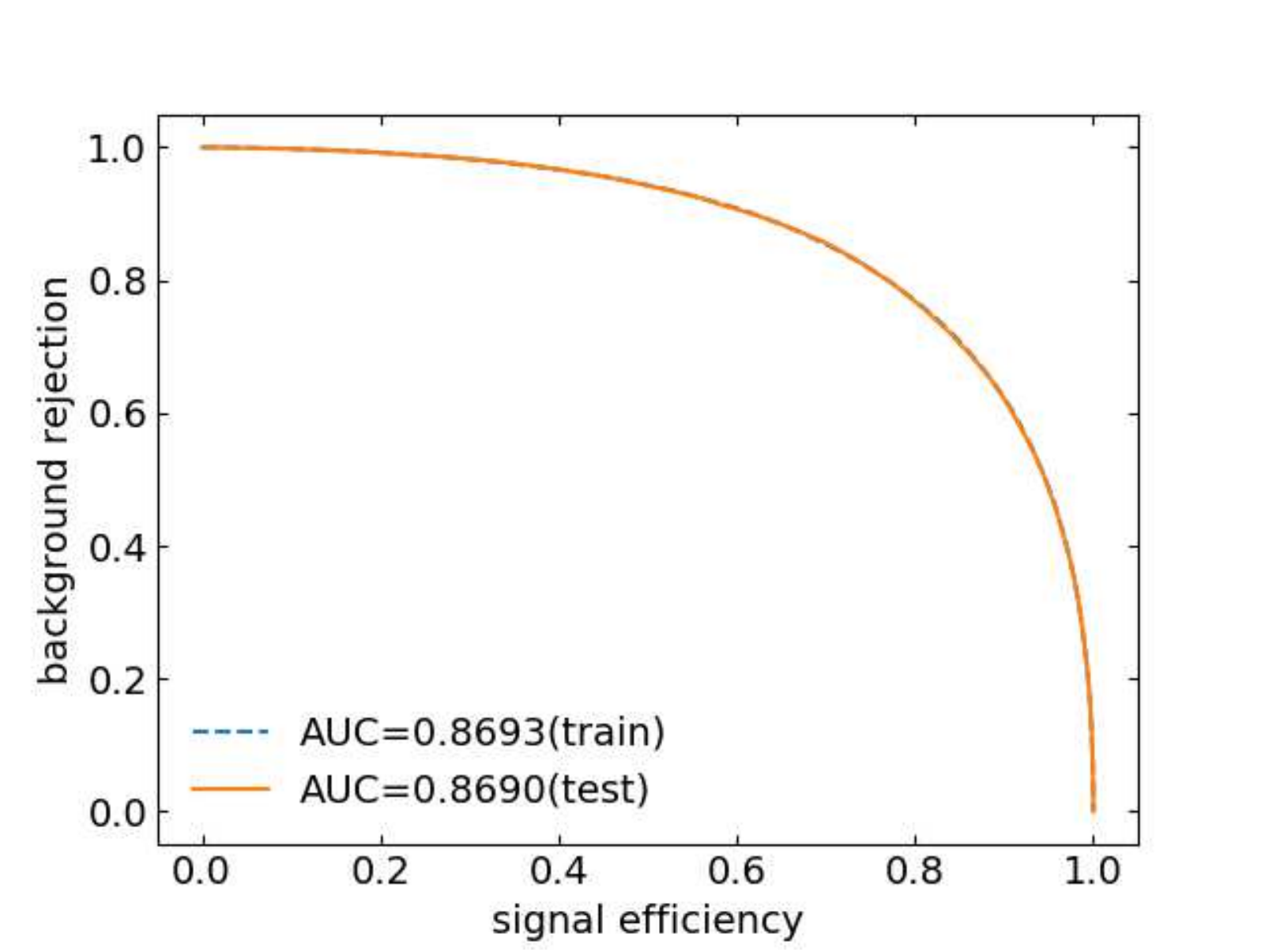}
	}
	\end{minipage}
	\begin{minipage}{0.49\textwidth}
		\subfloat[\label{fig:pptB}5-fold cross validation]{
			\includegraphics[width=\textwidth]{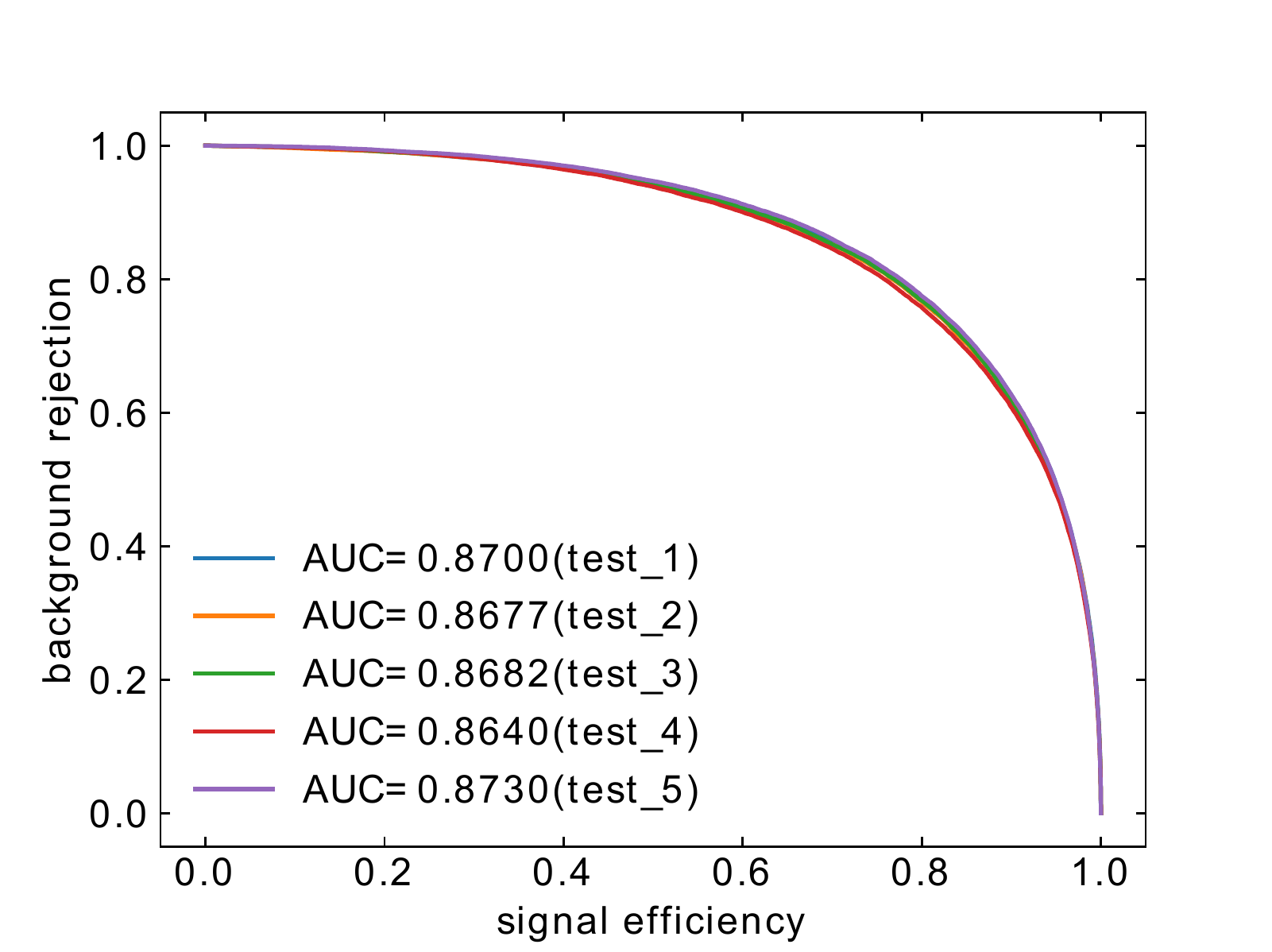}
		}
	\end{minipage}
	\caption{ROC curves of training and test sets as well as those obtained in 5-fold cross validation for the chosen PPT architecture.}
\end{figure}

Grid searches were performed to determine the best \NN architecture. The chosen \PPT consists of five hidden layers.
The first, third and fifth layer contain 64, 40 and 52 neurons, respectively, of which the first uses a ReLu activation function while the other two use softmax activation functions.
The second and fourth layers are batch normalisation layers.
The final output layer comprises one single neuron with a sigmoid activation function making it a binary classifier.
In total, there are 5441 trainable parameters.
The performance of this network is shown in a Receiver Operating Characteristics (ROC) curve in Figure~\ref{fig:pptA}. This shows the background rejection versus the signal acceptance for a given value of the PPT output. Crucially, the testing and training samples have very similar curves, meaning that naively it looks as if the network can generalise to unseen data.
Further stability of the neural network architecture and model can be examined using $k$-fold cross validation. In this case, 5-fold validation was carried out where the datasets were split into five equal samples. Four are used to train while the fifth is used as test set. Each sample has a chance to be the test set and so five ROC curves are drawn. These are shown in Figure~\ref{fig:pptB} for the test samples, where the agreement between each curve can be seen. Thus, the network generalises well to unseen data and is not overtrained.
Further details on the training of the \PPT can be seen in \cite{bvoelkel}.

\FloatBarrier
\subsection{Evaluation}\label{sec:PPTEval}

The PPT was one of the first neural networks in \ATLAS to be trained with modern machine learning libraries (Keras) and be injected into analysis code using \lwtnn. The implications of this are discussed in more detail in Section~\ref{sec:ML}.

\begin{figure}[!htbp]
\centering
\includegraphics[width=0.65\linewidth]{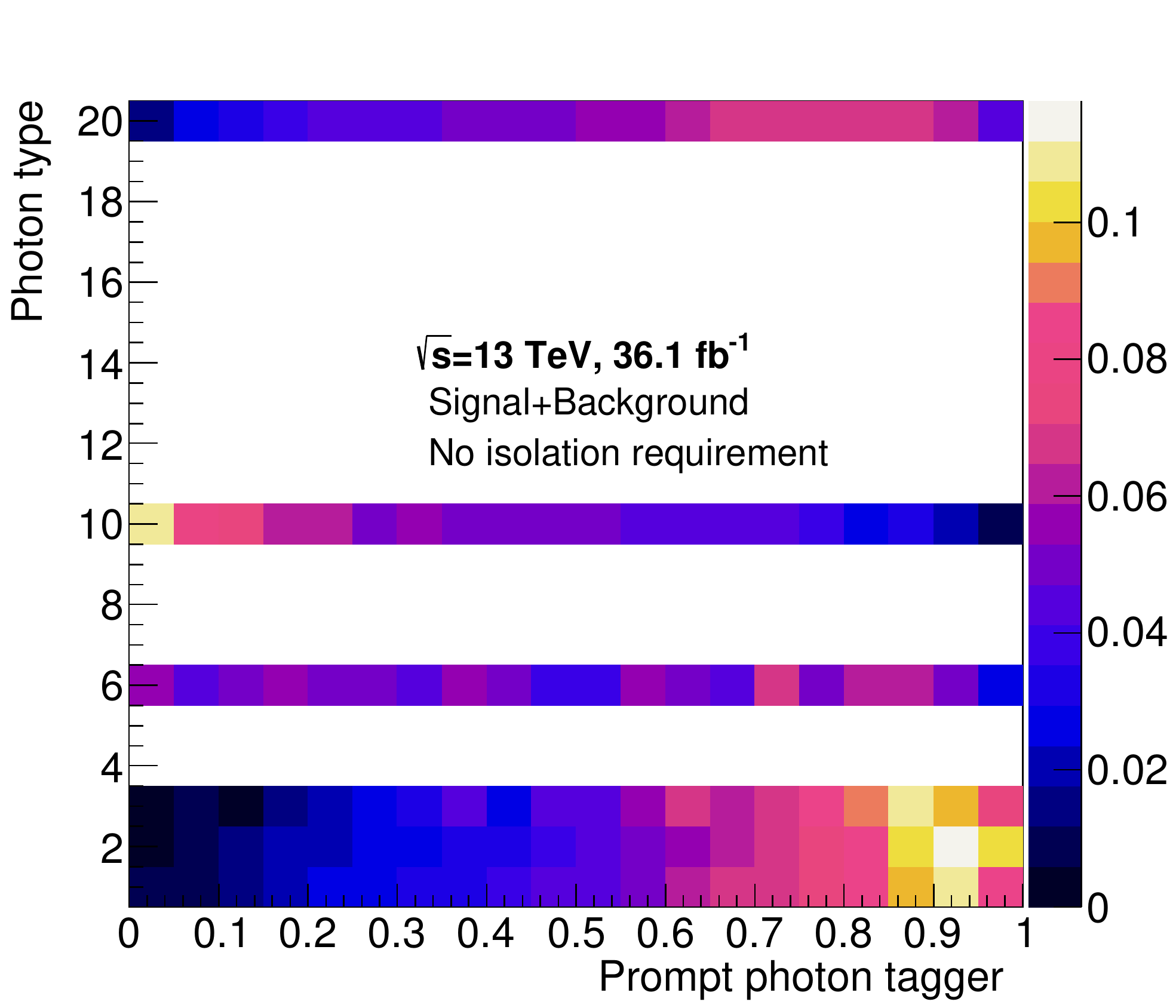}
\includegraphics[width=0.65\linewidth]{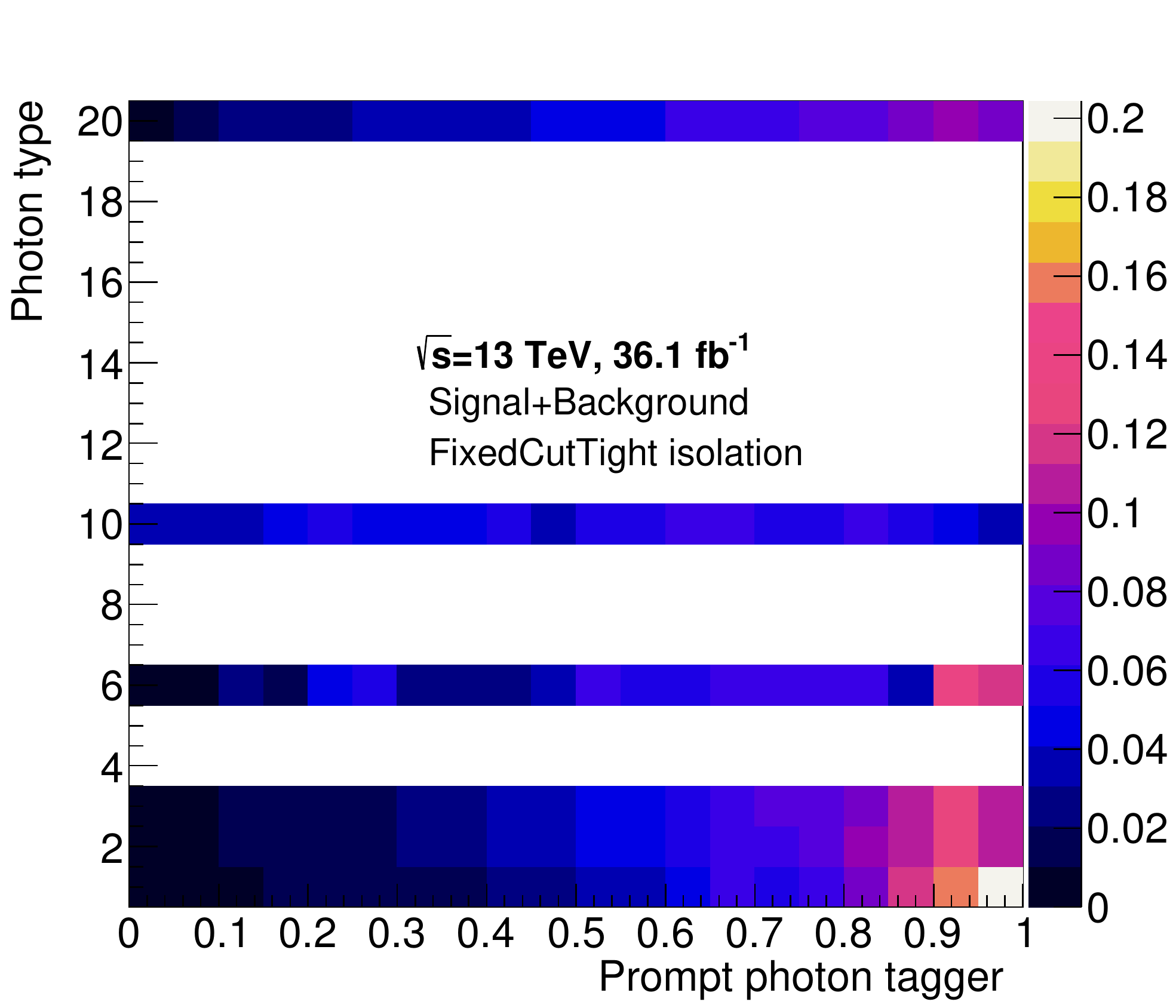}
\caption {\PPT output distributions for different photon truth particle types. The distributions are obtained by applying the \PPT to the \ttgamma analysis MC samples before event-level cuts, but includes object definitions such as the \tight photon requirement. Photon type = 20 corresponds to \efake photons. Photon type = 10 means the photon is classified as a \hfake. Photon type $<$ 10 is a signal-like photon. All rows are normalised to unity individually.}\label{fig:2dPPT}
\end{figure}

We can gauge the performance of the neural network on the \ttgamma MC and data samples (before event-level cuts). Figure~\ref{fig:2dPPT} shows the categorisation for three types of photons; those classified as \efakes (photon type = 20), those as \hfakes (photon type = 10) and those as prompt photons from various mother particles (photon type $<$ 10). 
Prompt photons below 10 are categorised as such since there are ambiguities in the truth records and the exact origin of photons. 
Each row has been normalised separately. 
The first figure shows the behaviour of the \PPT for photons in which there is no requirement on the isolation. For prompt photons there is a clear trend of more events congregating on the right of the distribution. For \hfakes there is a clear trend of being classified to the left. The distribution for \efakes is more or less flat, with a slight trend towards being classified as prompt. This is expected as no information about \efakes is provided in the training of the \PPT.
The second figure shows the response for the same samples, however \FCT isolation is applied to all photons. This significantly reduces the \hfake contribution and the overall phase space. Prompt photons are still classified effectively, however the \hfake distribution is more smeared out.
Interestingly, prompt photons with photon type = 6 have a flat distribution for the figure in which no isolation is required, but are classified more as prompt once \FCT isolation is applied. These prompt photons are classified as ``other", where they are not from ISR, the top quark or the top quark decay products. It is a limitation when trying to directly match prompt truth photons.
This indicates that there is a contribution of hadronic fake photons that are misidentified using the truth classifying scheme that are then removed once including isolation.

\begin{figure}[!htbp]
\centering
\includegraphics[width=0.8\linewidth]{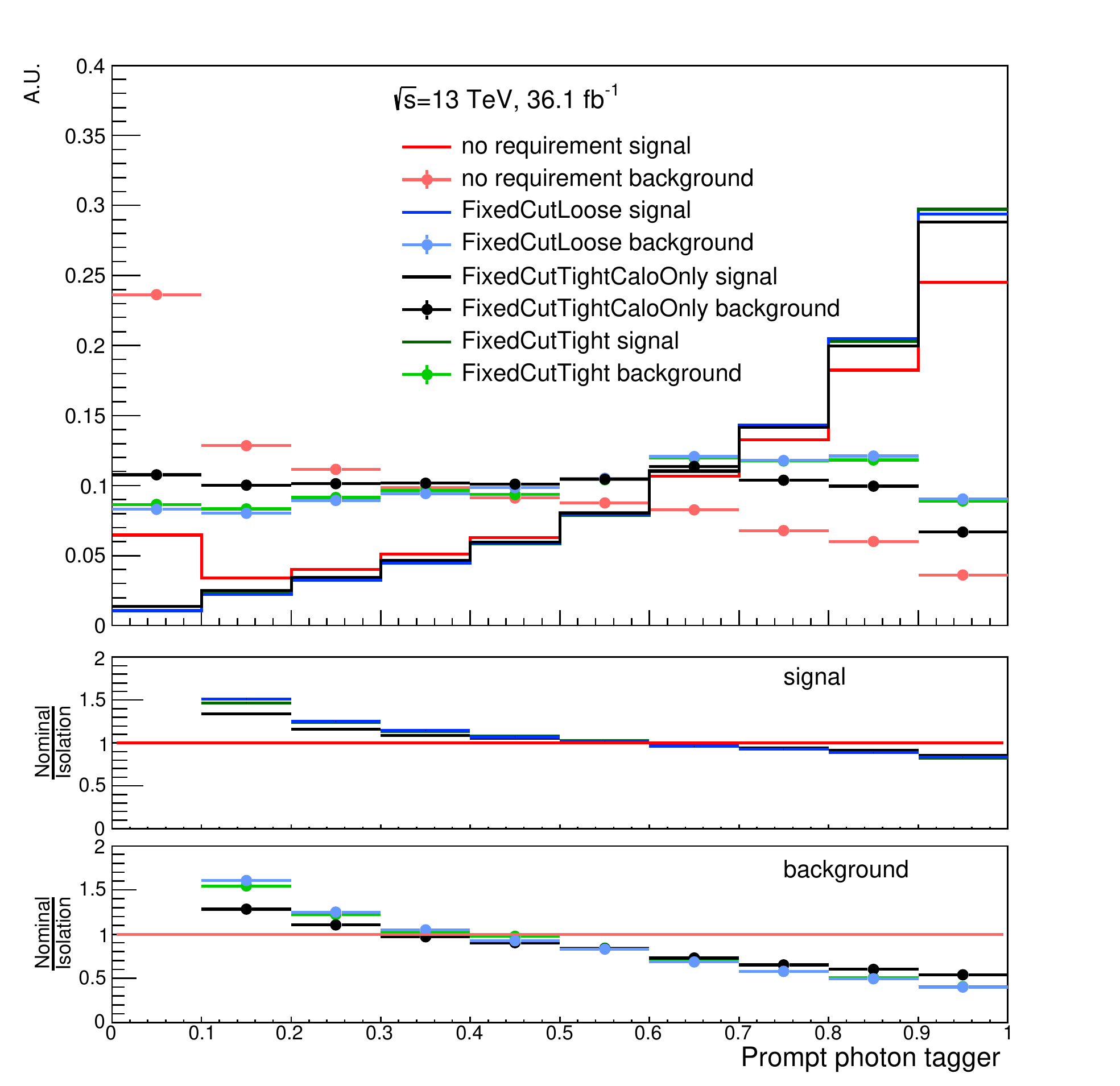}
\caption {\PPT output distributions for increasing isolation cuts. The distributions are obtained by applying the \PPT to the \ttgamma analysis MC samples before event-level cuts, but includes object definitions such as the \tight ID photon requirement.}\label{fig:PPT_isol_dep}
\end{figure}

The correlation between isolated photons and the shape of the \PPT can be further explored as different analyses can have different isolation requirements.
For instance, the photon could be very isolated (as is the case with the \ttgamma analysis) or not isolated at all (such as busy environments involving jets). 
Figure~\ref{fig:PPT_isol_dep} shows the effect of applying increasing ``tightness" of isolation to the output of the \PPT. The shape variation shows that there is a correlation between the \PPT and photon isolation. To account for this in the \ttgamma analysis, a systematic uncertainty is derived and applied to account for the fact that training for the \PPT was performed in a non-isolated phase space. This will be discussed further in the next section.

\subsection{Scale factors and systematic uncertainties}
\label{sec:PPTsysts}

Scale factors to correct for the mis-modelling of the \PPT between MC and data are calculated in a dedicated prompt photon control region. This region corrects for prompt photons in a $Z\to\ll\gamma$ enriched phase space. Exactly one photon and two opposite sign leptons are required, with an invariant mass cut between the lepton pair to be within [60, 100]~\GeV. The 3D scale factors for prompt photons are extracted in bins of \pt, \eta and \PPT output by taking the difference between the MC and the data. 

Closely linked to the scale factors are the systematic uncertainties derived for the \PPT. 
The scale factors are switched on and off and are symmetrised to extract the up and down variations of the \PPT modelling. Only a shape effect is considered and the overall number of events must remain constant.
The prompt photon uncertainties are also applied to the \efake background as the shapes are similar.
For the \hfake background, two further sets of scale factors are derived.
The first is from the \hfake control region C (Chapter~\ref{sec:hfake}). The signal contamination is varied by $\pm$50\% and the maximum variations between data and MC are used to estimate the uncertainty.  This accounts for \tight yet non-isolated photons. The second set of scale factors follows a similar approach in the \hfake control region A. This accounts for isolated yet non-\tight photons.

\begin{figure}[!htbp]
\centering
\subfloat[Prompt photons]{
\includegraphics[width=0.48\linewidth]{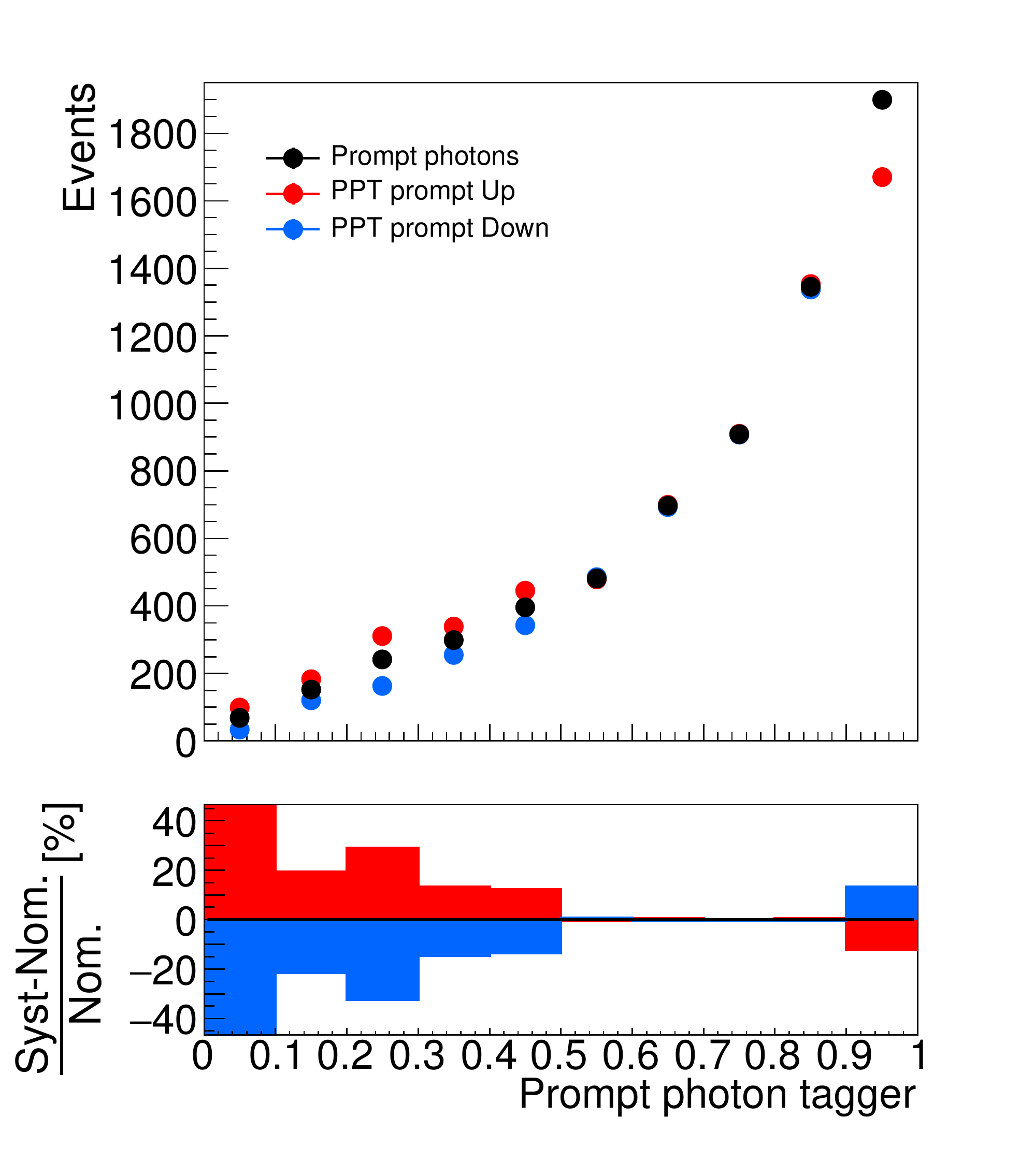}
}
\hspace{-0.036\linewidth}
\subfloat[\efake photons]{
\includegraphics[width=0.48\linewidth]{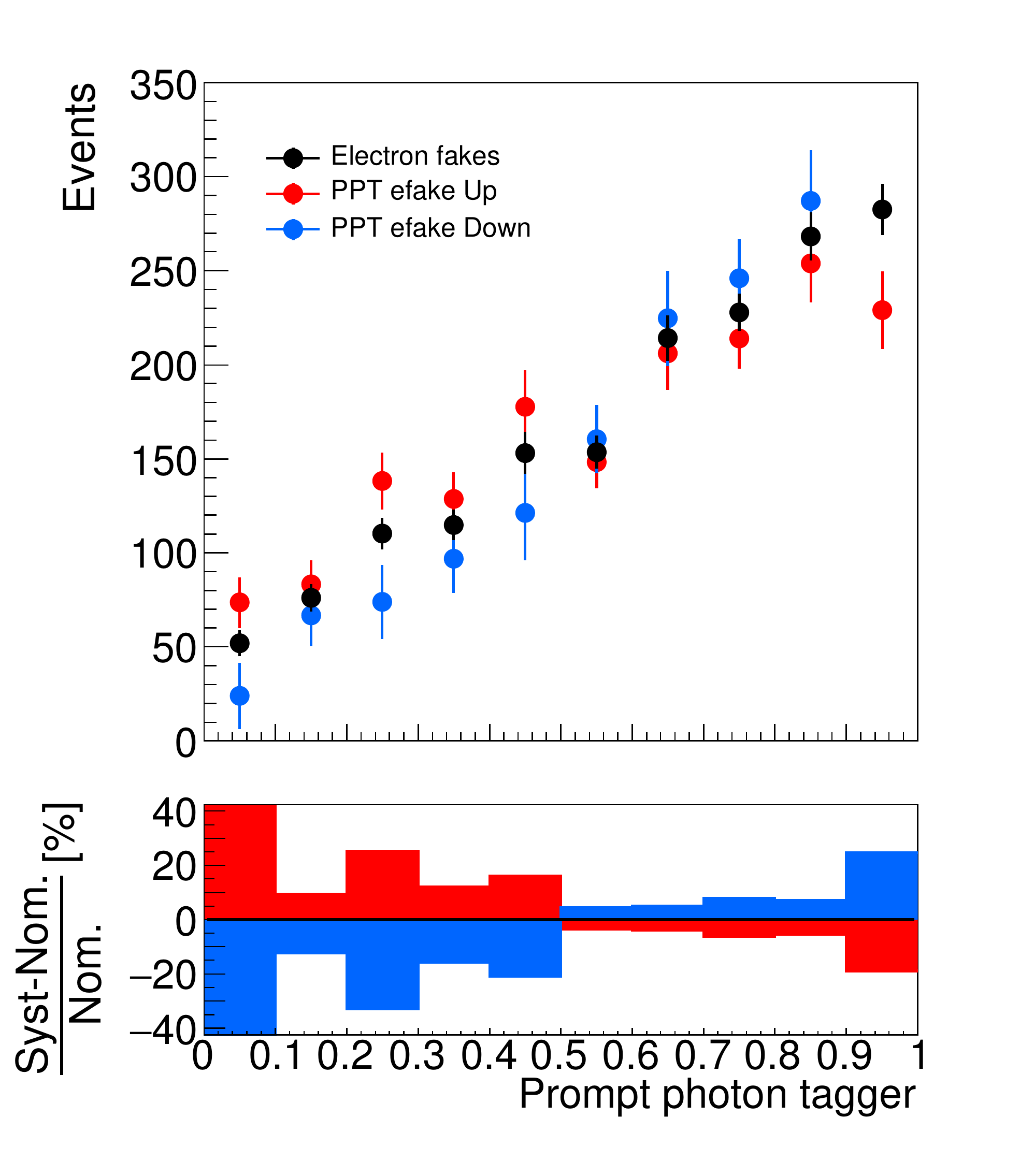}
}
\hspace{-0.036\linewidth}

\subfloat[\hfake photons from CR C]{
\includegraphics[width=0.48\linewidth]{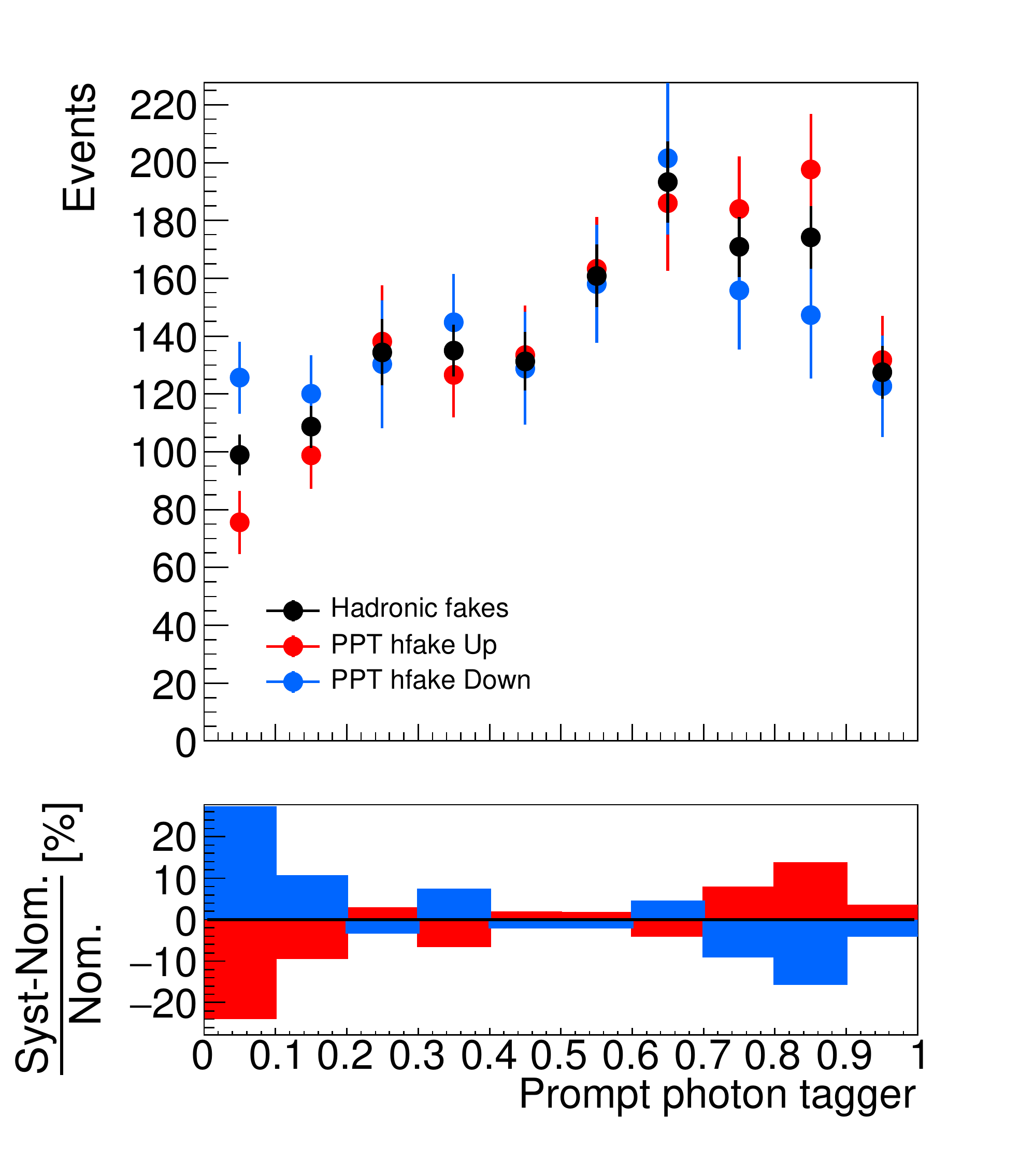}
}
\hspace{-0.036\linewidth}
\subfloat[\hfake photons from CR A]{
\includegraphics[width=0.48\linewidth]{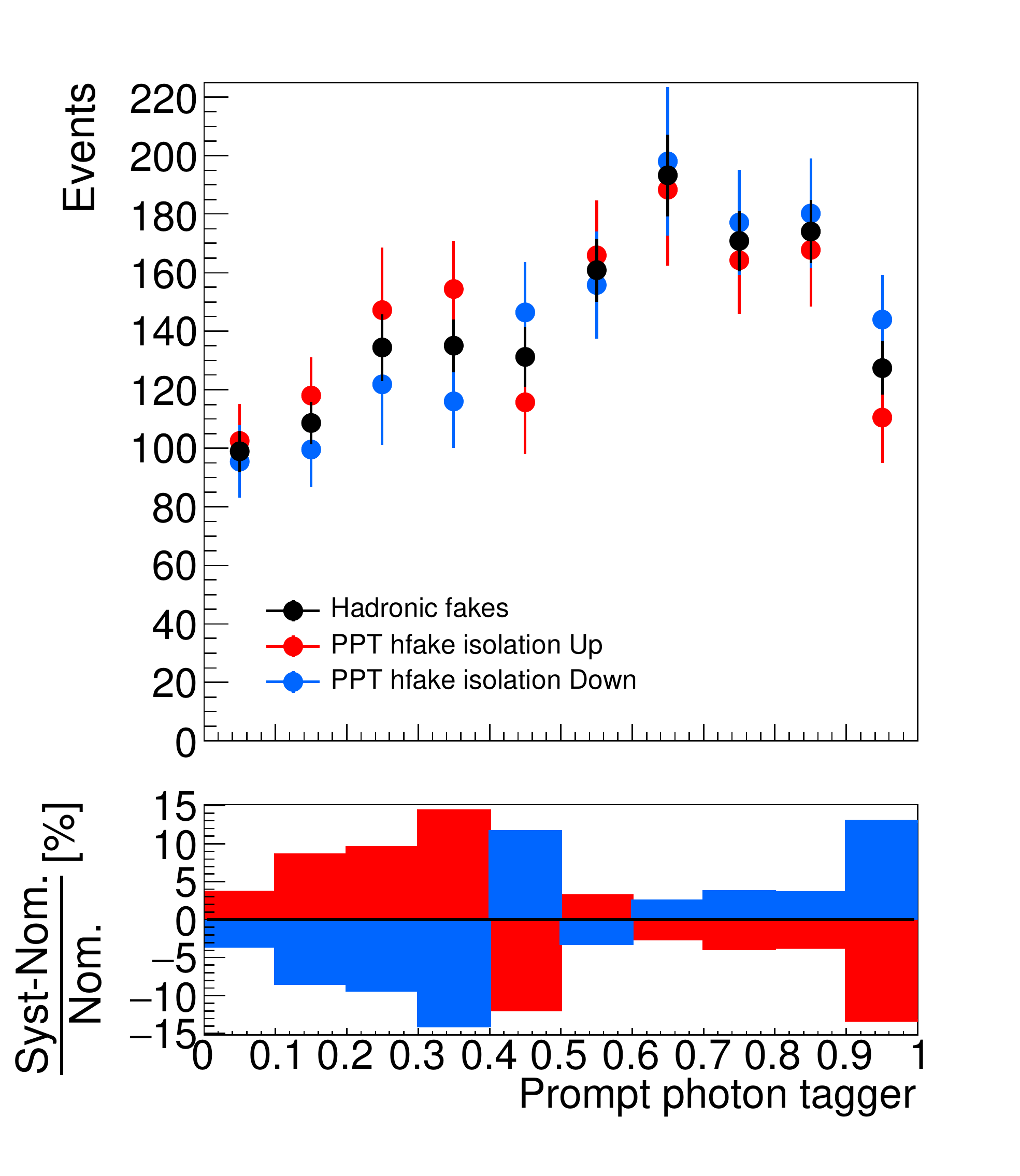}
}
\hspace{-0.036\linewidth}

\caption {The relative sizes of the \PPT scale factors/systematic uncertainties for prompt, \efake and \hfake photons (for the two sources of uncertainties). ``Up" and ``down" labels are arbitrary.}\label{fig:PPTttgammaSFs}
\end{figure}

Overall, the derivation of systematic contributions follows a very conservative approach due to the novelty of the tool and the shower-shape variable mis-modelling.
The effect of turning on and off the scale factors for each process can be seen in Figure~\ref{fig:PPTttgammaSFs}. 
The ratio panel shows the relative contribution the systematic uncertainty has for each process, where the ``up" and ``down" variation labels are arbitrary. For prompt contributions, the extrema of the classifier exhibit large variations. An uncertainty/scale factor of around 15\% is seen in the very prompt rich area of the \PPT, which for signal rich analyses will inflate the final uncertainty.

\begin{figure}[!htbp]
\centering
\subfloat[\label{fig:ttgammaPPTa}]{
\includegraphics[width=0.42\linewidth]{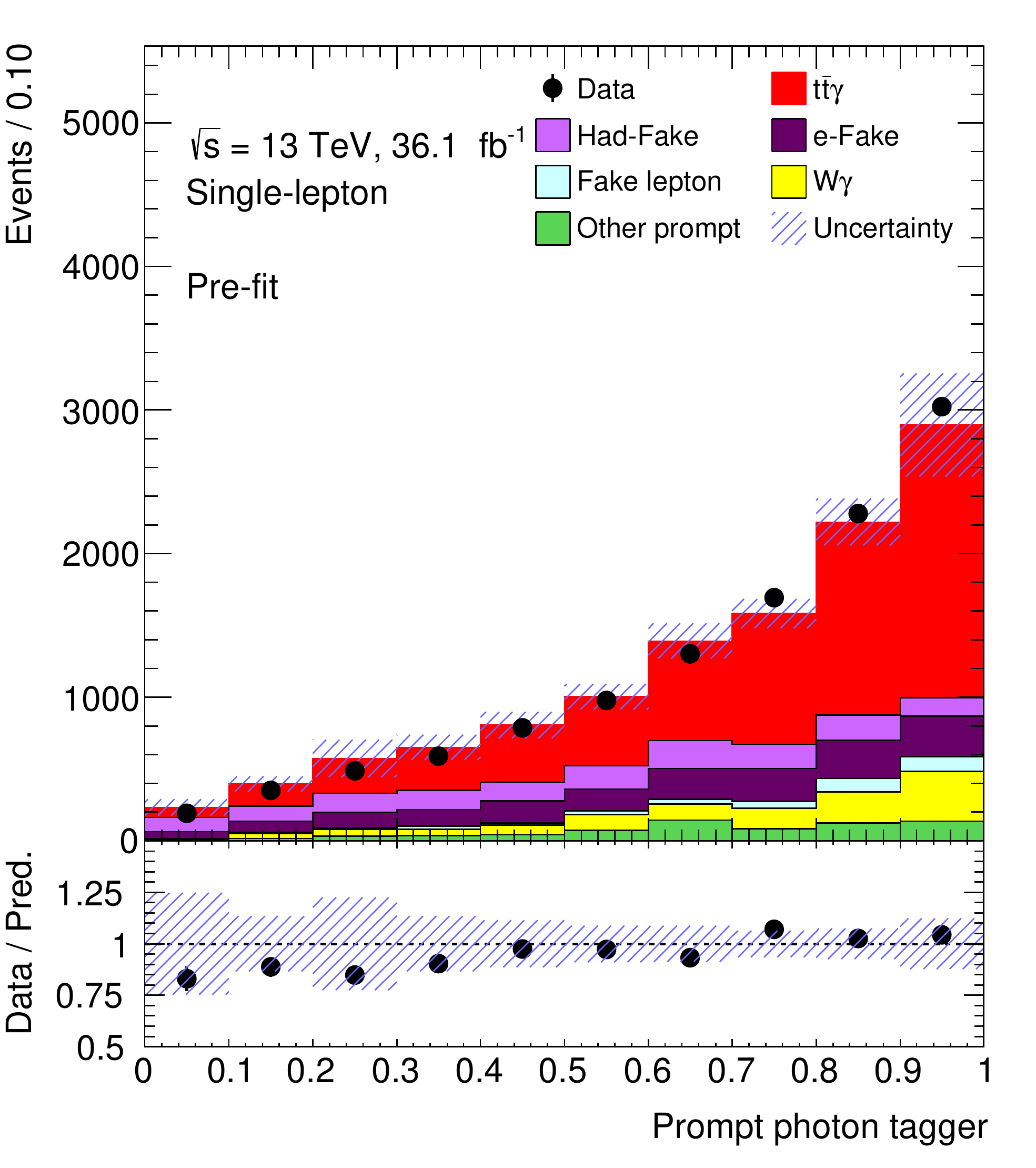}
}
\hspace{-0.028\linewidth}
\subfloat[]{
\includegraphics[width=0.44\linewidth]{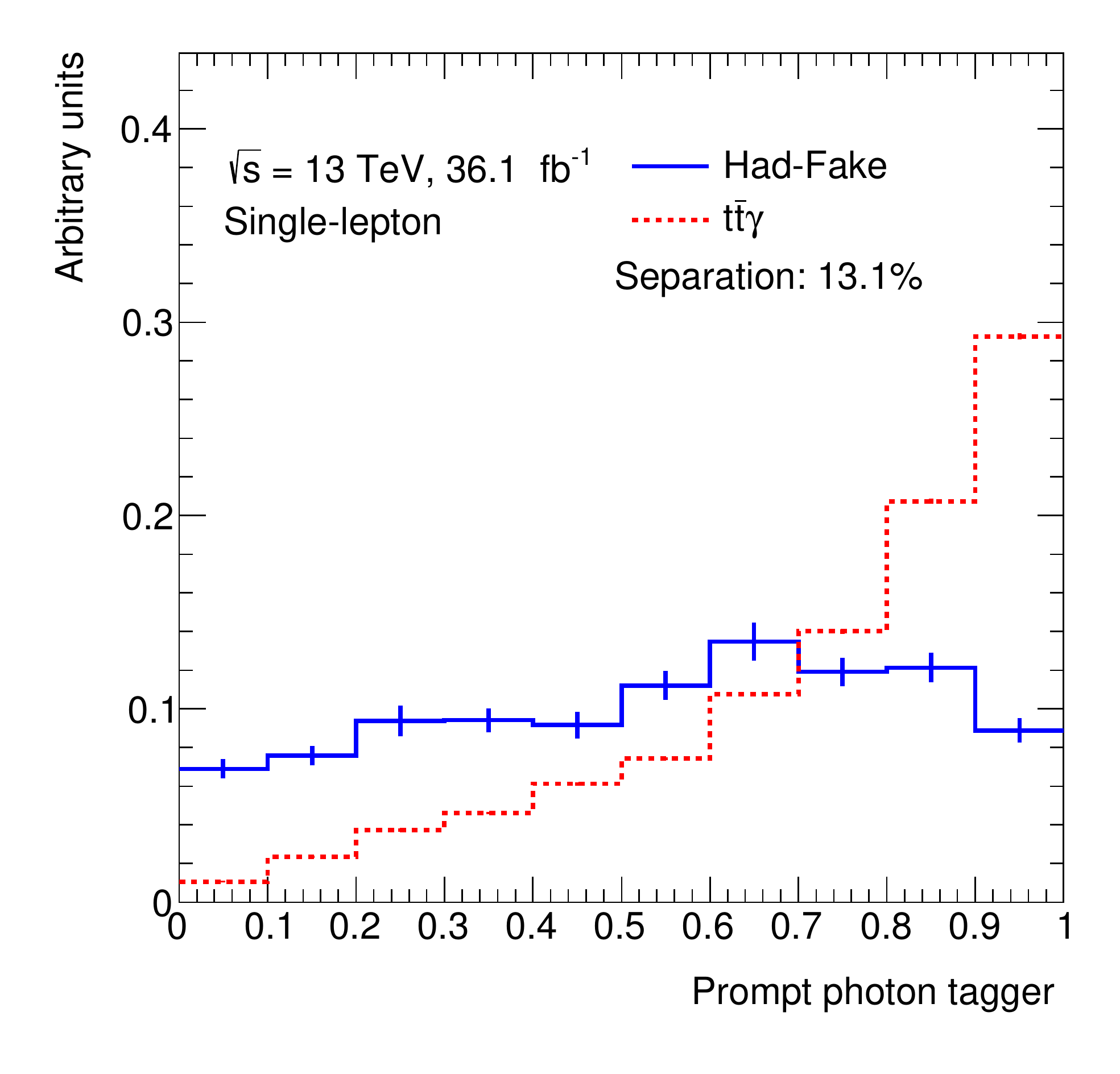}
}
\hspace{-0.028\linewidth}
\caption {a) The \PPT in the \ttgamma analysis after all event selections, scale factors, statistical and systematic uncertainties have been applied. b) The normalised separation between signal and the \hfake background, in which only statistical uncertainties are shown.}\label{fig:PPTttgamma}
\end{figure}

The final \PPT distribution is shown for the \ttgamma analysis in the \chljets channel in Figure~\ref{fig:PPTttgamma}. The full signal region cuts and scale factors from previous sections are applied and total systematic uncertainties are included in the error band. This includes the \PPT scale factors and uncertainties. Also shown is the normalised separation between the signal and the \hfake background contribution, which shows good separation for \tight and \FCT selections. The shape discrepancy seen in Figure~\ref{fig:ttgammaPPTa} arises from the input variables used to train the \PPT, and thus the scale factors. However, this is covered by systematic uncertainties.

\FloatBarrier

\section[The Event-level Discriminator]{The Event-level Discriminator}
\label{sec:ELD}

An Event-level discriminator (\ELD) has been developed separately for the \chljets and \chll channels. Both take the form of NNs.
The \chljets \ELD is trained on events from the \chejets and \chmujets channels while the \chll \ELD is trained on events from the \chee, \chmumu and \chemu channels. The driving factor for this decision is due to events in the respective \chljets and \chll channels having very similar topologies and compositions. While there are some differences in background producing mechanisms\footnote{The \chemu channel has negligible \Zgamma events due to the requirement of one electron and one muon, for which the $Z$-boson can not decay into.
Electron reconstruction efficiencies are generally lower than muon reconstruction efficiencies in the detector, thus, selections relying on muon final state particles will have more events than those with electron final state particles. 
Fake electrons contribute to backgrounds in which final state electrons are required. Thus, the \efake backgrounds will be larger for these channels.
The \QCD background is not only made up of fake leptons, but also leptons which are non-prompt. Thus, due to jets creating and even faking electrons, this background is significantly larger in the \chejets channel compared to the \chmujets channel.}, the negative impact this is meant to have on a NN is small (if any) and is outweighed by the positive impacts: simplicity and an increase in statistics. 
The \ELD exploits topological event information (such as the number of $b$-tagged jets or the \met) and builds a binary classifier. This provides a new variable that helps constrain the signal and background in the maximum likelihood fit, explained in Chapter~\ref{sec:analysisstrategy}.
Section~\ref{sec:varselection} presents the variable selection, Section~\ref{sec:ELDtraining} explains the training methods used. Finally, Section~\ref{sec:ELDapplication} evaluates the performance of the \ELD.

\subsection{Variable selection}
\label{sec:varselection}

Two \ELD{}s were simultaneously trained.
For the \chljets channel \ELD, the full set of selection cuts defined in Section~\ref{sec:eventselections} was applied. 
Looser cuts were used to train the \chll \ELD. This includes dropping the two mass window cuts between the two leptons and the two leptons and the photon, neglecting the \MET cut and not requiring a cut on the number of jets. This allows for more background contamination, enhances statistics and thus makes training more stable. 
Training only in the signal region (or close to the signal region) means the weights reach optimal values in a shorter amount of time. 

\begin{figure}[!htbp]
\centering
\includegraphics[width=0.34\linewidth]{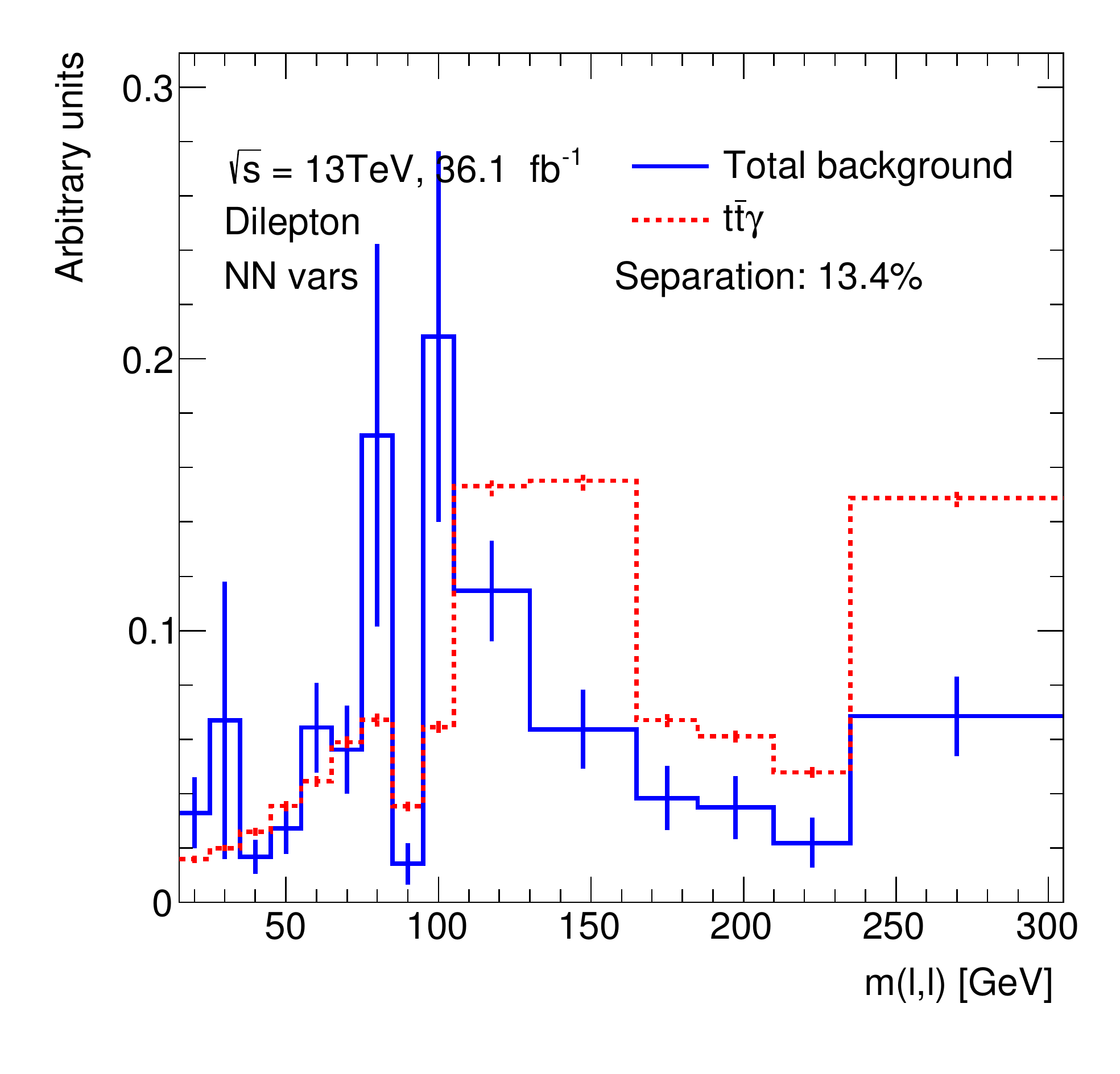}
\hspace{-0.023\linewidth}
\includegraphics[width=0.34\linewidth]{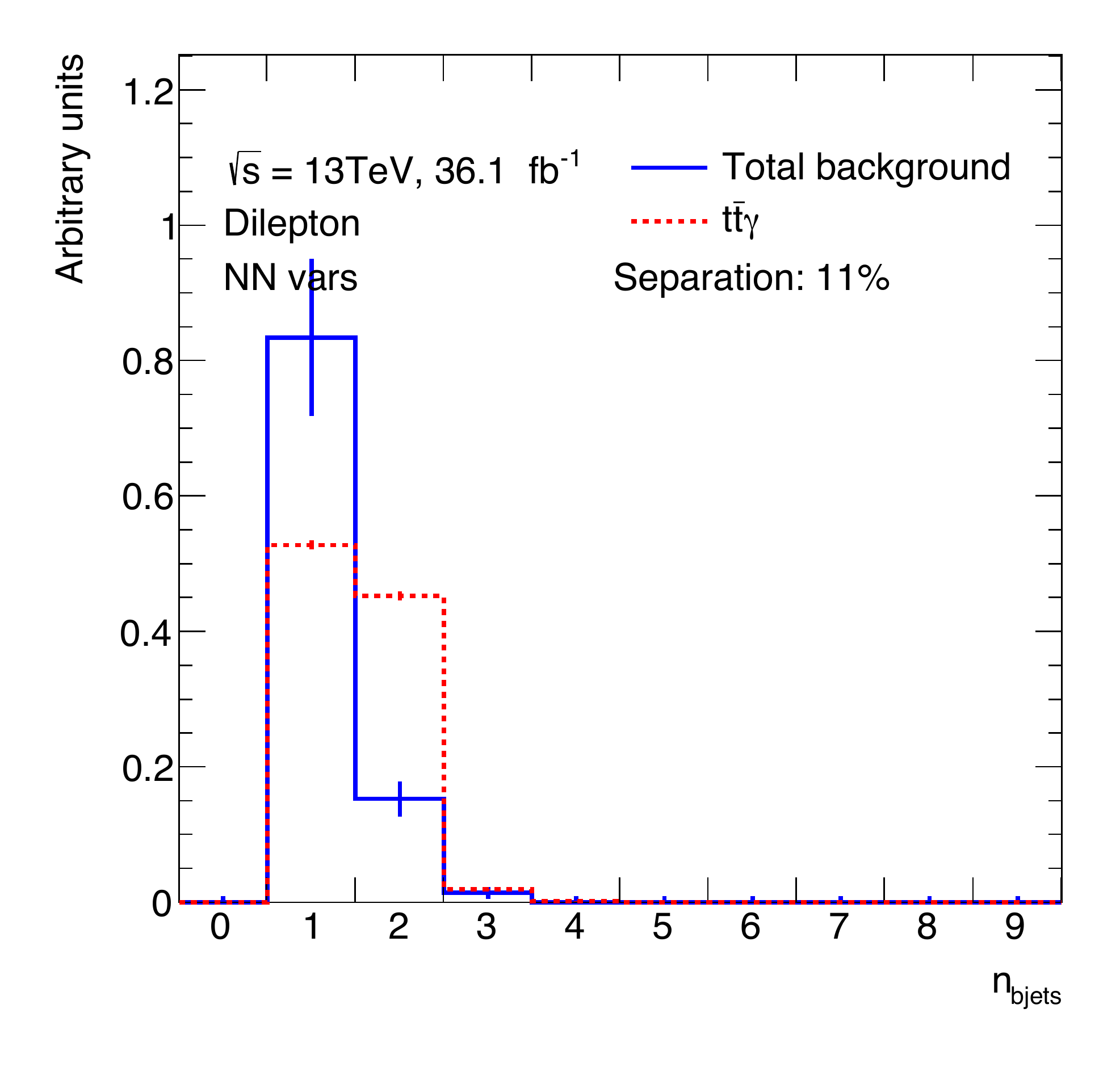}
\hspace{-0.023\linewidth}
\includegraphics[width=0.34\linewidth]{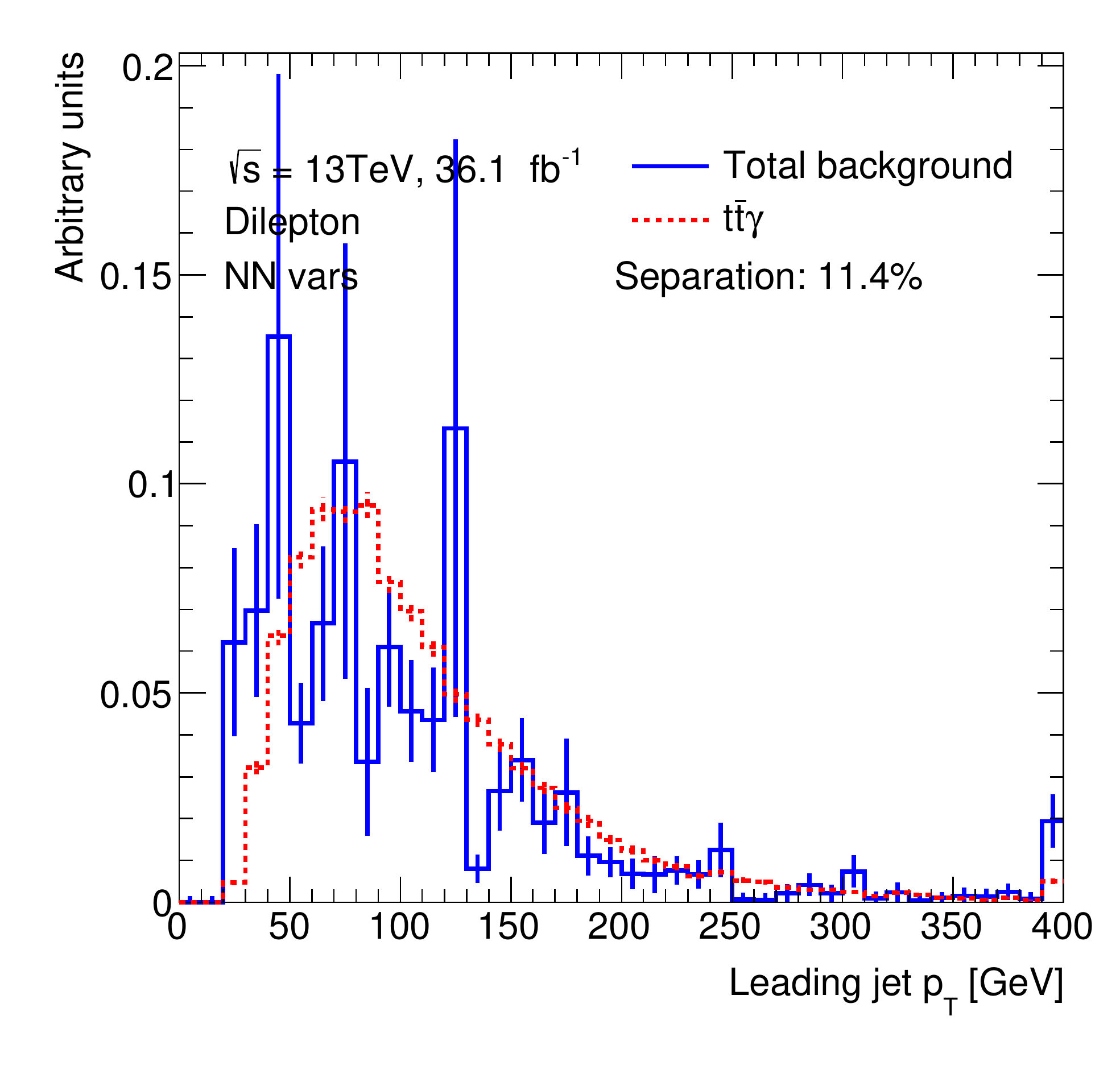}
\hspace{-0.023\linewidth}

\includegraphics[width=0.34\linewidth]{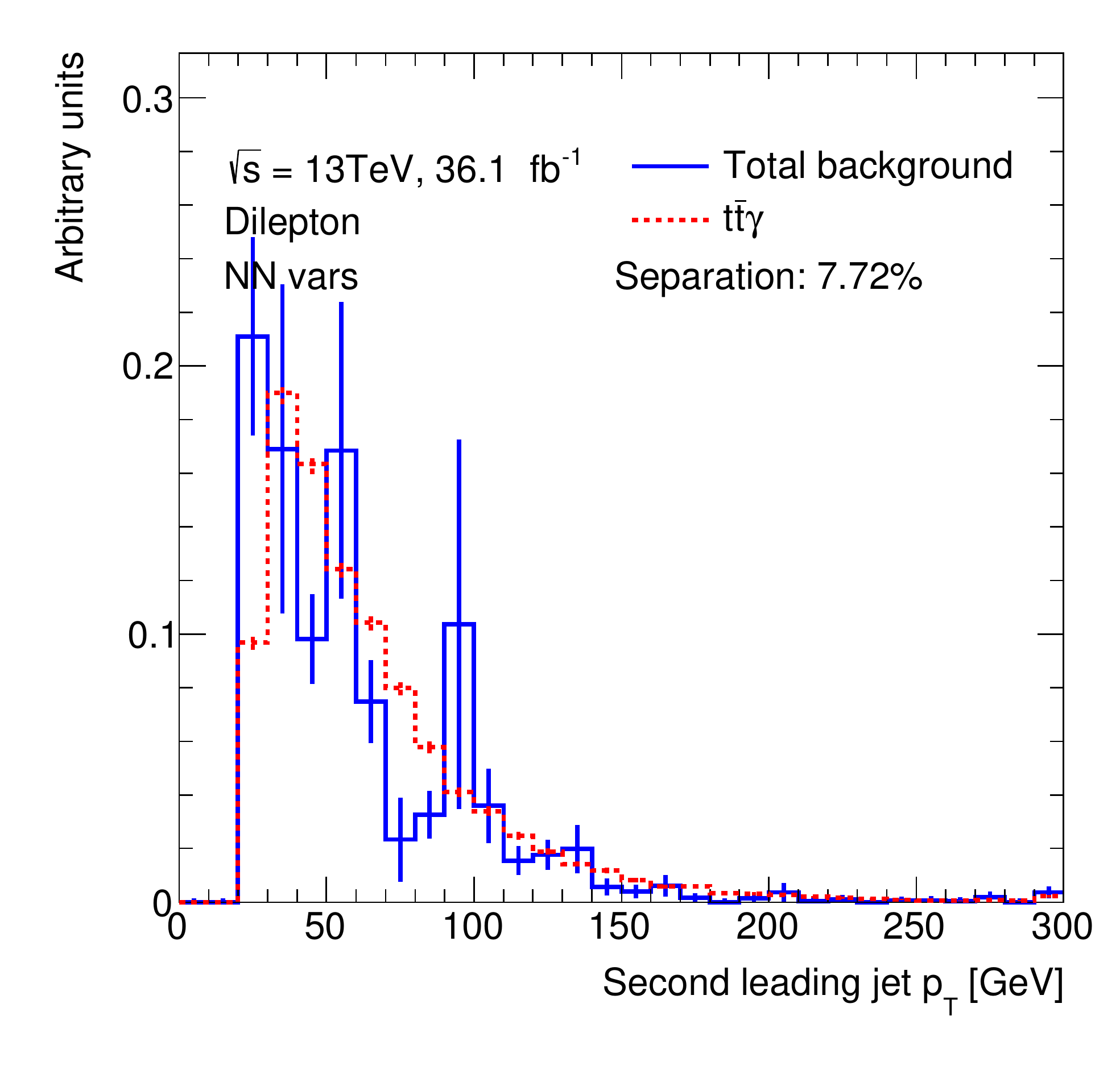}
\hspace{-0.023\linewidth}
\includegraphics[width=0.34\linewidth]{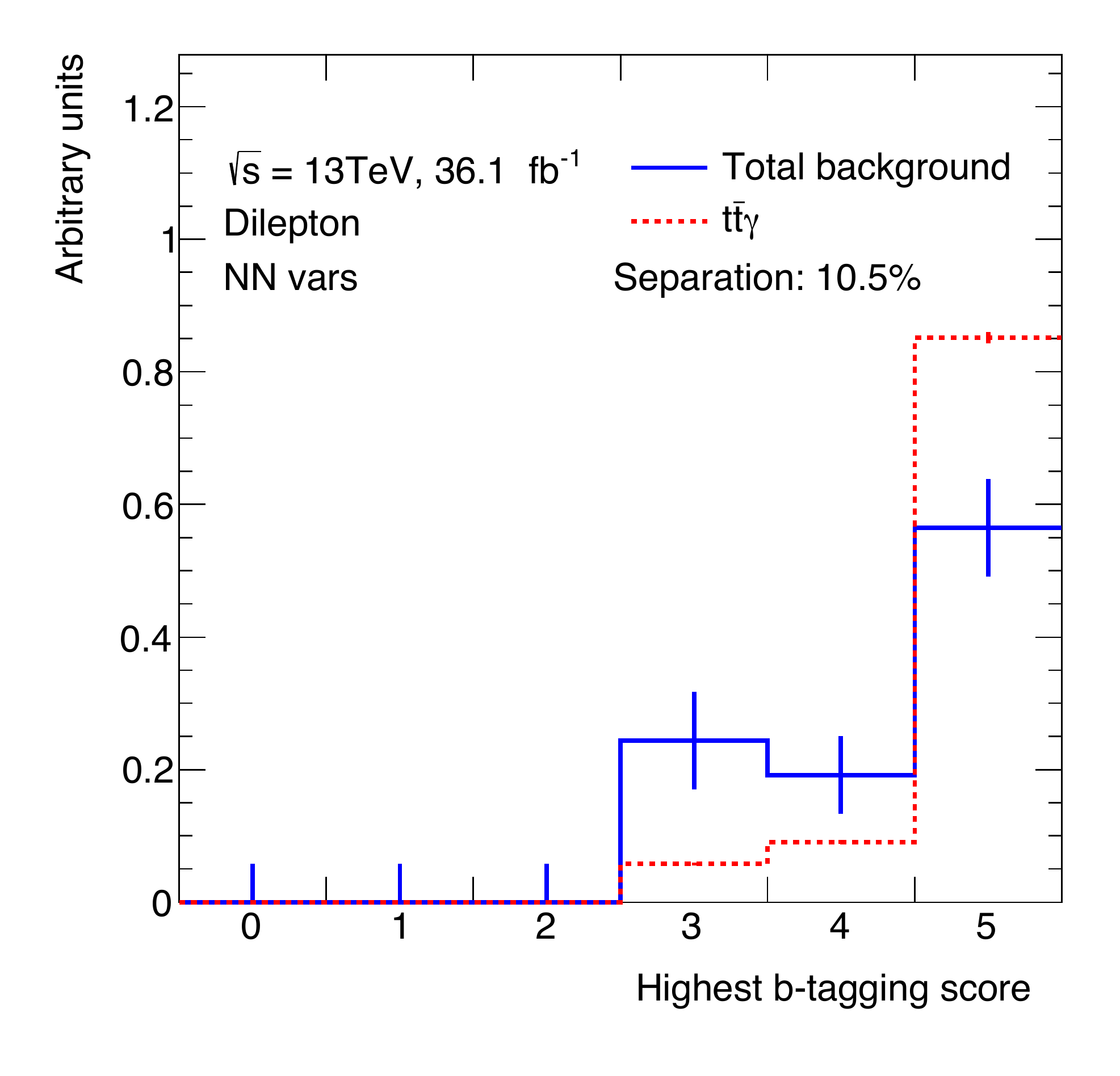}
\hspace{-0.023\linewidth}
\includegraphics[width=0.34\linewidth]{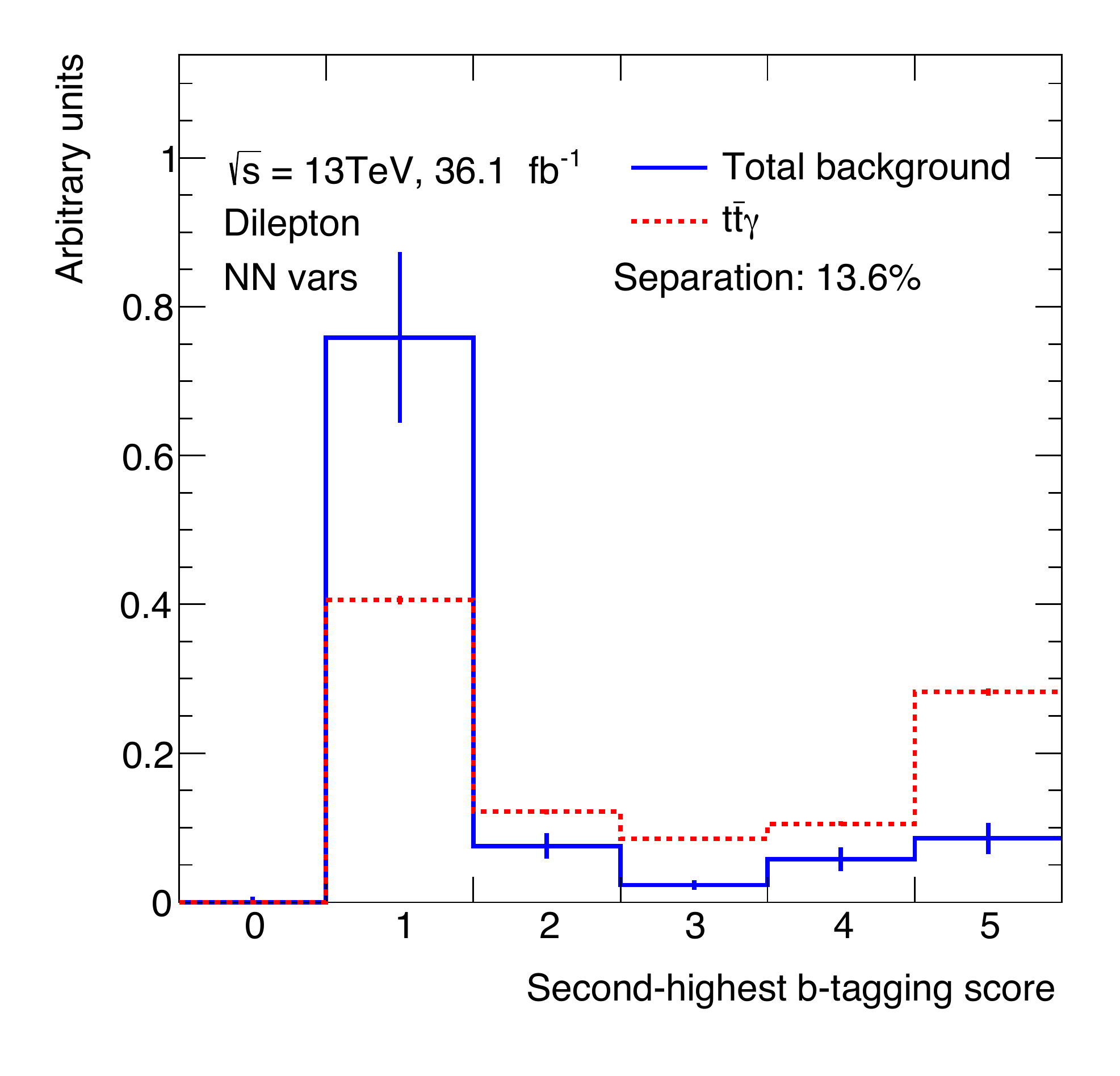}
\hspace{-0.023\linewidth}

\includegraphics[width=0.34\linewidth]{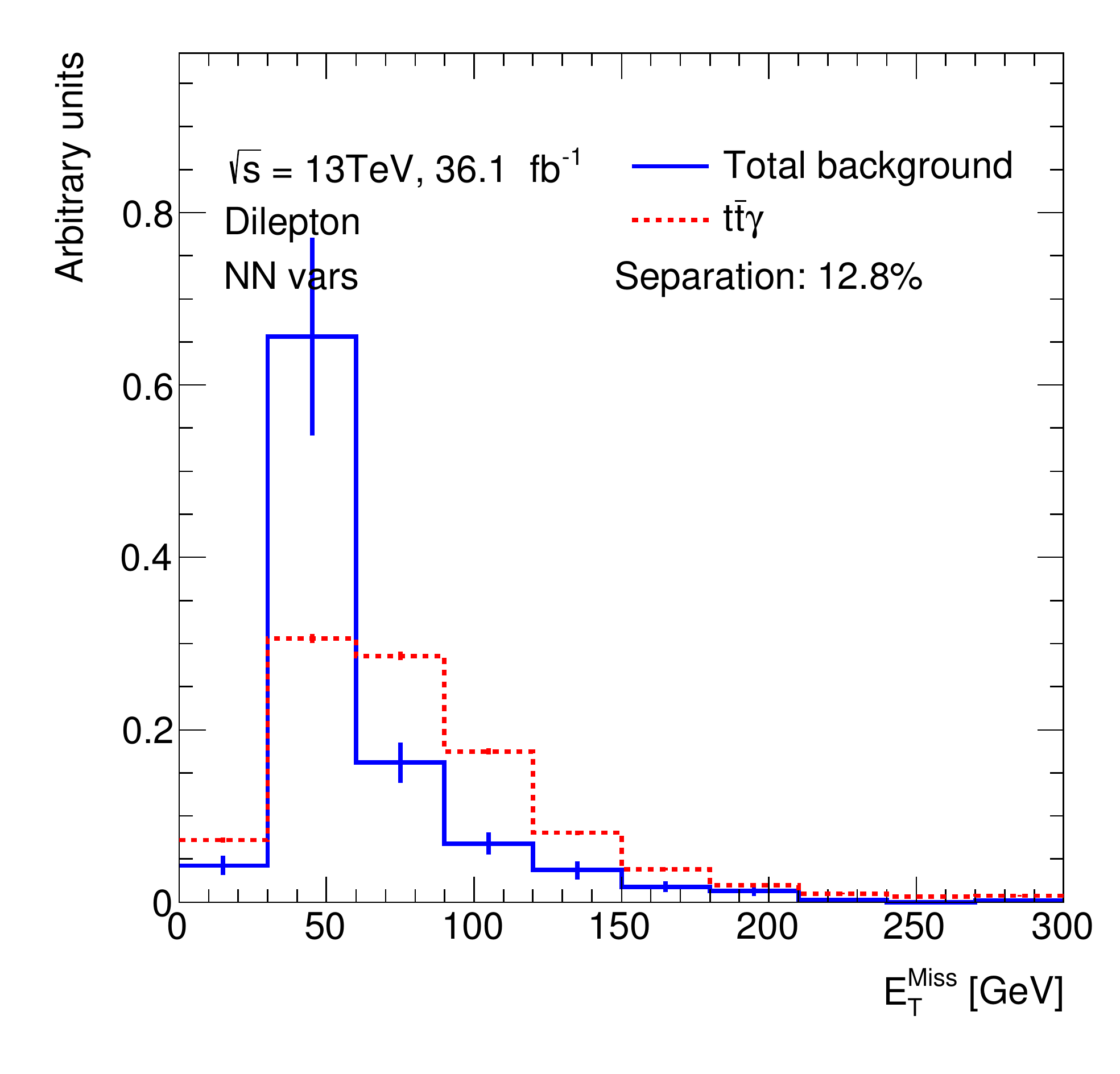}
\caption {Variables used in the \chll ELD showing the separation between the signal and sum of all backgrounds.}
\label{fig:trainingDilepton}
\end{figure}

Variables are used which highlight the different kinematics of the processes considered as signal and background.
Table~\ref{tab:event_MVA_variables_SL} shows a summary of the variables considered for the respective channels. Included is their separation power (Equation~\ref{eq:separation}) of signal versus the sum of all backgrounds. A check-mark indicates the variable is used as an input to the final \ELD\footnote{For the \chll channel, the separation is calculated when including all SR cuts as this is the phase space we are actually interested in.}.
As will be shown in Section~\ref{sec:ELDtraining}, in general for the \chljets \ELD, more variables lead to a better sensitivity. However, the strategy of how differential \xsec measurements would be performed was unclear at this stage. To remove any potential bias, the $\Delta R$ variables were excluded in case they were to be used as variables for the differential \xsecModifyNoun measurements. For the same reason, the \pt and \eta of the photon (not shown in the table) were not considered as inputs to the \ELD.
Thus, a loss of sensitivity in the \ELD{}s was chosen so that the strategy for differential \xsecModifyNoun measurements would not be impaired.
For the \chll \ELD, apart from the differential \xsecModifyNoun argument above, it was found that seven variables was the ideal scenario. This will be shown in Section~\ref{sec:ELDtraining}.
Initially the \PPT was considered for the \chll \ELD as there is a small \hfake background contribution. However, after deliberation over the conservative systematic contributions that would need to be applied, it was decided that any gain would be washed out. Thus, it was dropped for the \chll channel.
The most powerful variables in both channels are due to the number of $b$-tagged jets as well as the pseudo continuous $b$-tagging weights which are described in Chapter~\ref{sec:btagging}. For the \chljets channel this is then followed by the \PPT.
Figure~\ref{fig:prefitPlotSLSFs} and ~\ref{fig:ttgammaPPTa} (the \PPT) show the variables that have been used for training the \chljets \ELD. Figure~\ref{fig:prefitPlotDLSFs} shows a slightly tighter selection (the full signal region) for the variables used in the \chll \ELD.
Figure~\ref{fig:trainingSinglelepton} and \ref{fig:trainingDilepton} show the normalised separation plots for the input variables of the respective \ELD (those with check-marks in Table~\ref{tab:event_MVA_variables_SL}). Only statistical uncertainties are included. 
While some variables show little difference in the normalised signal and background shapes ($m(\gamma,\text{lep})$ or \mwt in the \chljets channel for example), two aspects need to be considered. First, these separation plots are shown for the sum of all backgrounds and might be masking the separation for a given process (such as \QCD). Second, non-linear transformations through a neural network might result in the network learning correlations that are otherwise hard to anticipate, even from seemingly similar shapes of signal and background.
Correlations between input variables and between the final \ELD will be discussed in Section~\ref{sec:ELDapplication}.

\begin{table}[h!]
  \centering
  \begin{tabular}{lcc}
    \hline
    
    \multirow{2}{*}{Variable} &
    \multicolumn{2}{c}{Separation [\%]} \\ 

     & \chljets & \chll \\
    \hline
    \hline
  
   Prompt Photon Tagger (\PPT) & 3.5 \checkmark & 6.8 \\
   \HT & 1.8 \checkmark & 1.3 \\
   number of jets & 0.7 \checkmark & 1.4 \\
   number of $b$-jets & 3.5 \checkmark & 11.0 \checkmark \\
   \MET & 0.8 \checkmark & 12.8 \checkmark  \\
   \mwt & 0.9 \checkmark & - \\
   \pt of first jet & 2.1 \checkmark & 11.4 \checkmark \\
   \pt of second jet & 1.8 \checkmark & 7.7 \checkmark  \\
   \pt of third jet & 1.7 \checkmark & 3.3   \\
   \pt of fourth jet & 1.3 \checkmark &  1.1  \\
   \pt of fifth jet & 0.9 \checkmark &  0.3  \\
   highest $b$-tagging weight & 4.3 \checkmark & 10.5 \checkmark \\
   2$^{\text{nd}}$ highest $b$-tagging weight & 2.8 \checkmark & 13.6 \checkmark  \\
   3$^{\text{rd}}$ highest $b$-tagging weight & 0.8 \checkmark & -  \\
   $m(\ell,\ell)$ & - & 13.4 \checkmark \\
   $m(\gamma,\ell)$ & 0.3 \checkmark & 5.8 \\
   $m(\gamma,\ell,\ell)$ & - & 2.3 \\
   $\Delta R(\gamma,\ell)$ & 0.5 & 13.4  \\
   $\Delta R(\gamma,jet)_{leading}$ & 1.8 & 3.1  \\
   $\Delta R(\gamma,jet)_{subleading}$ & 1.7 & 4.3  \\
    \hline
  \end{tabular}
  \caption{Separation of each variable used in the training of the \chljets and \chll \ELD{}s. A check mark indicates that the variable was used in that \NN. A dash indicates the variable does not apply to that channel.}
  \label{tab:event_MVA_variables_SL}
\end{table}

\begin{figure}[!htbp]
\centering
\includegraphics[width=0.34\linewidth]{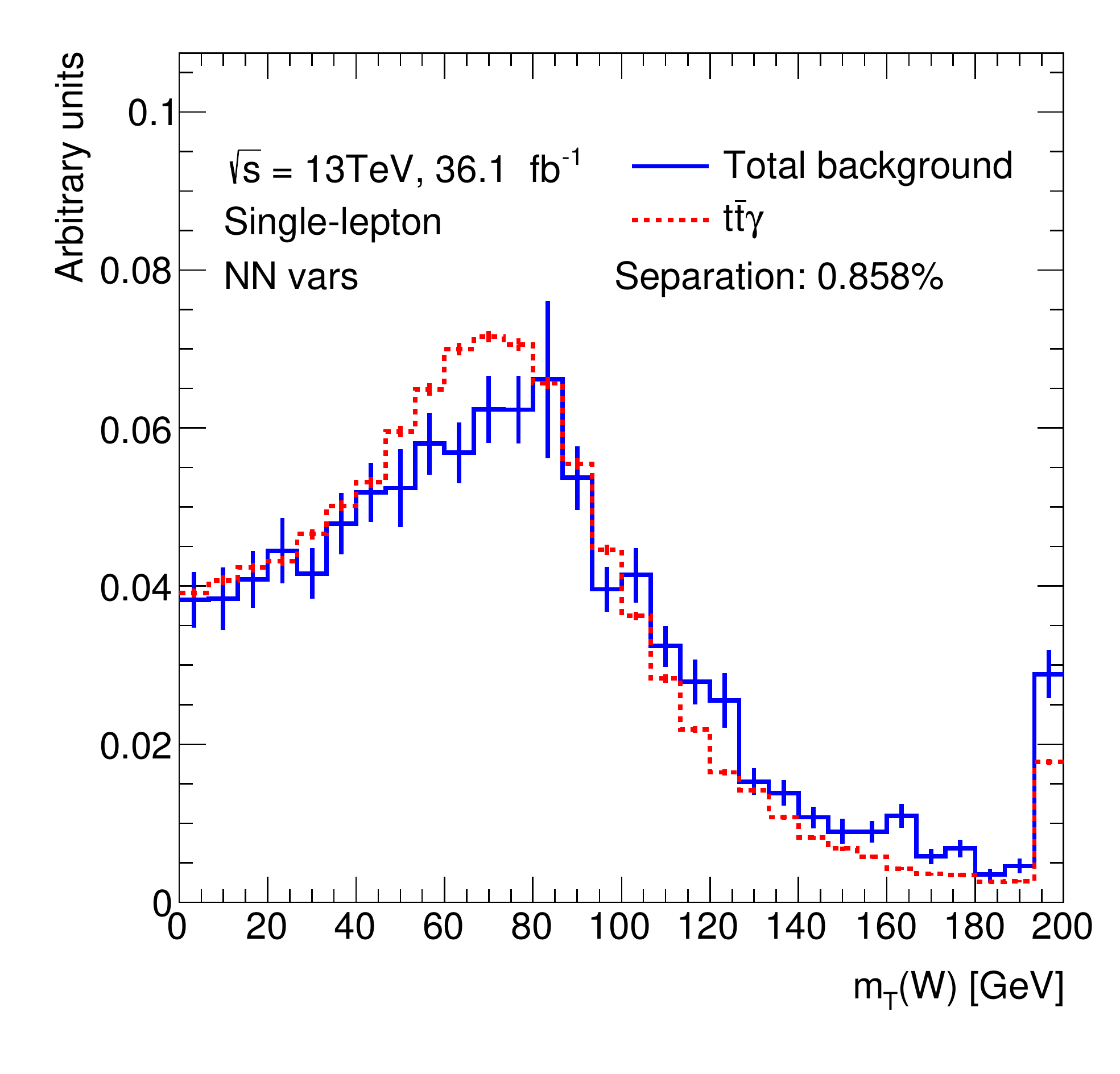}
\hspace{-0.023\linewidth}
\includegraphics[width=0.34\linewidth]{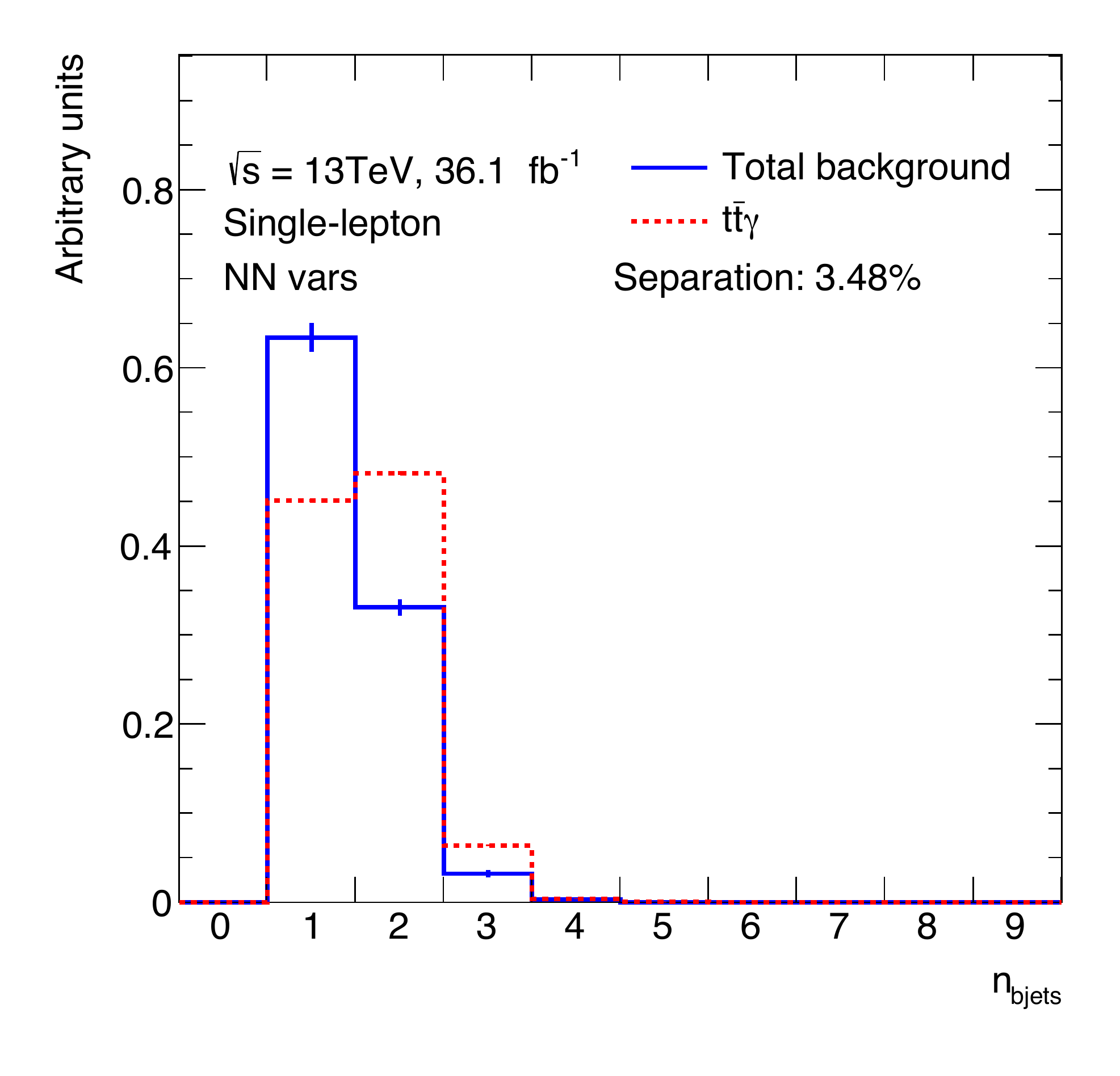}
\hspace{-0.023\linewidth}
\includegraphics[width=0.34\linewidth]{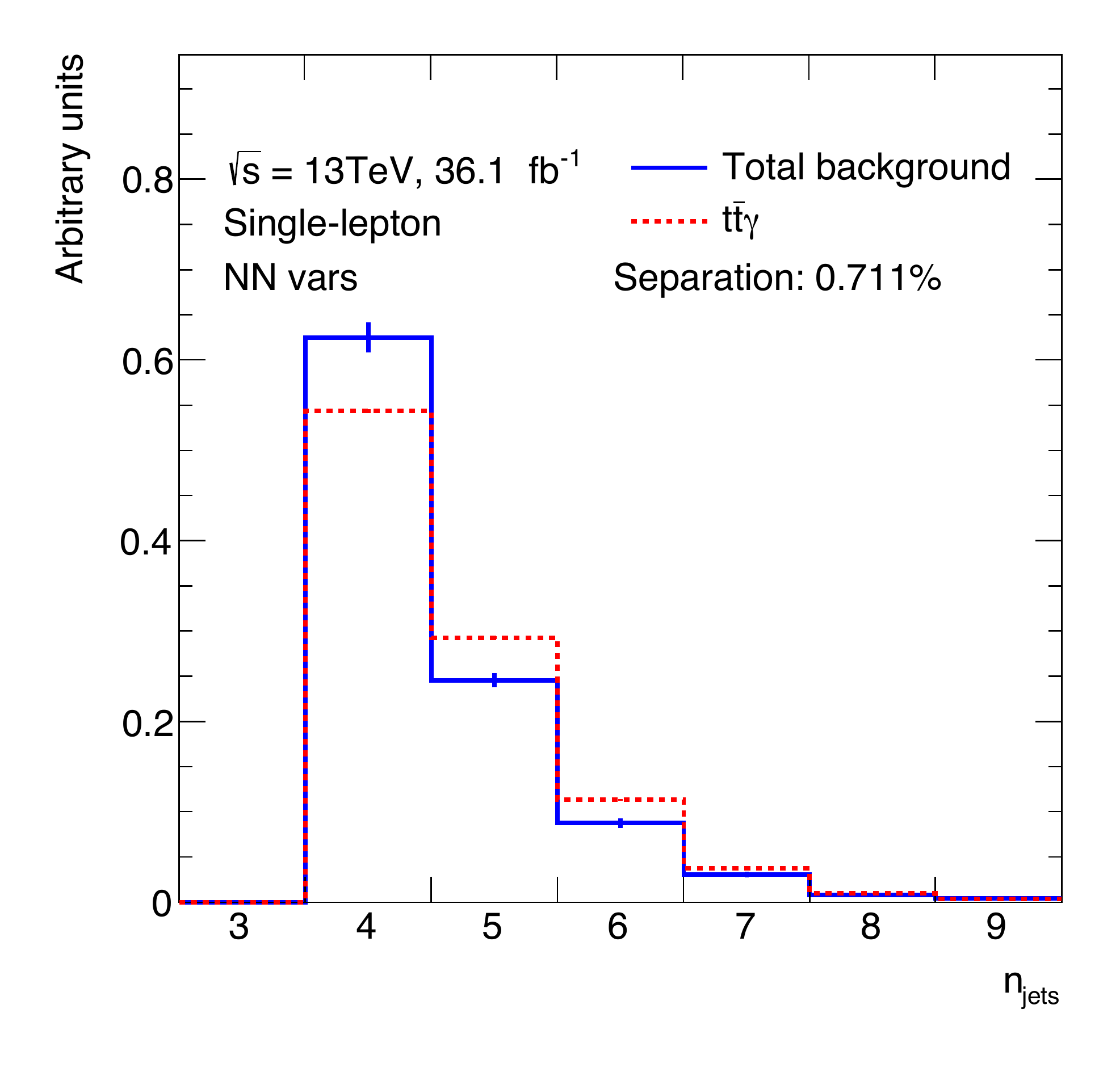}
\hspace{-0.023\linewidth}

\includegraphics[width=0.34\linewidth]{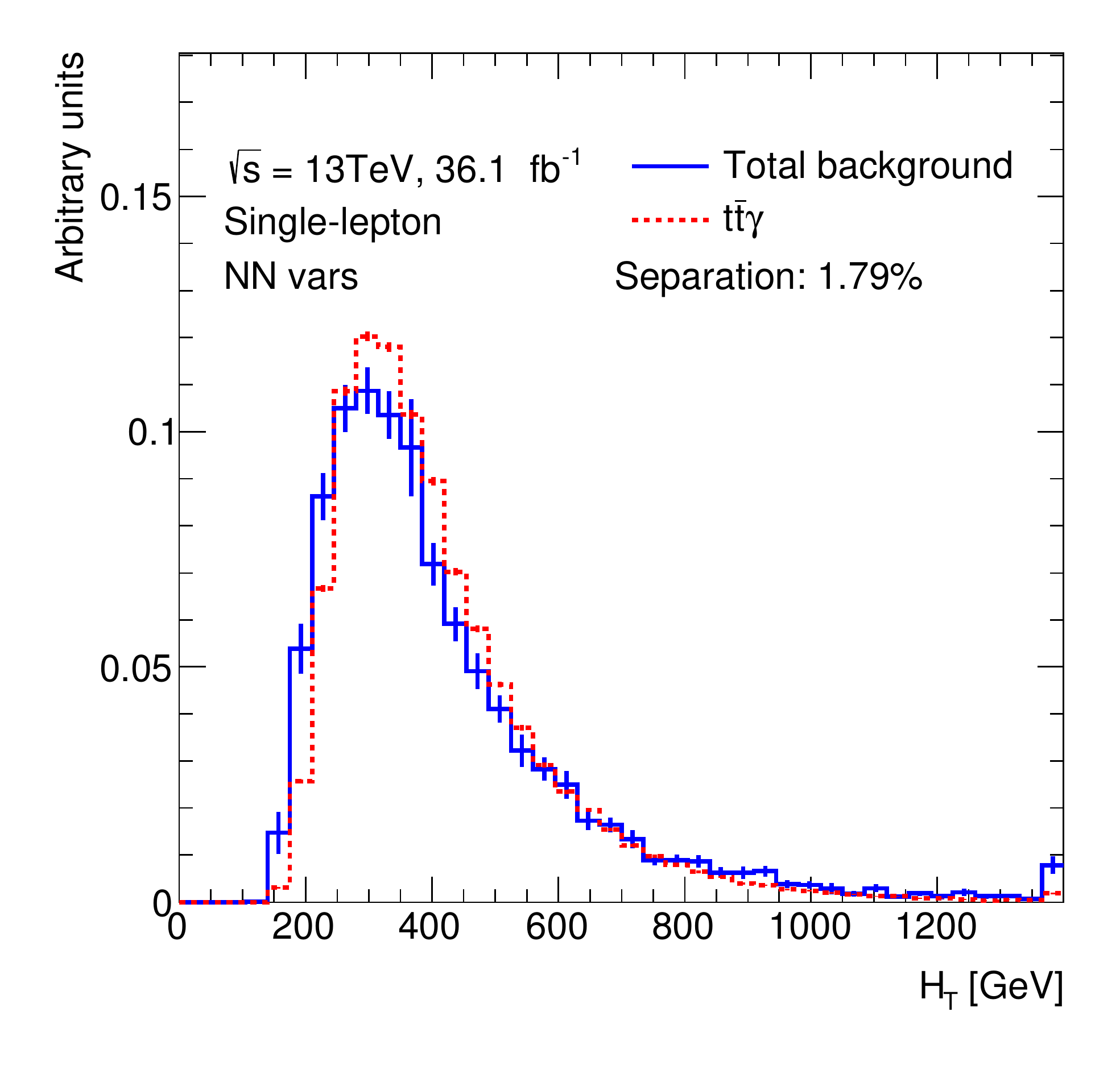}
\hspace{-0.023\linewidth}
\includegraphics[width=0.34\linewidth]{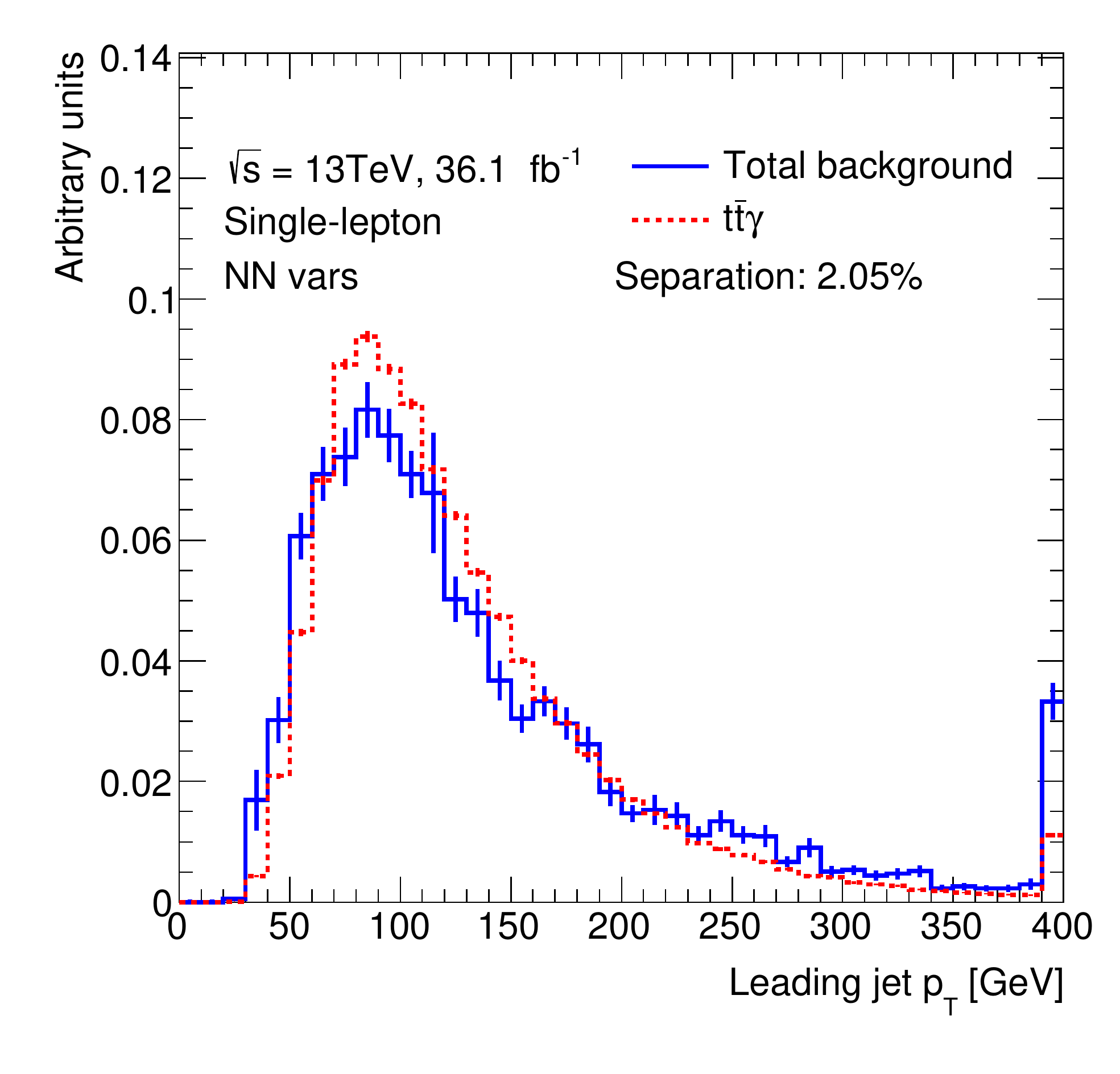}
\hspace{-0.023\linewidth}
\includegraphics[width=0.34\linewidth]{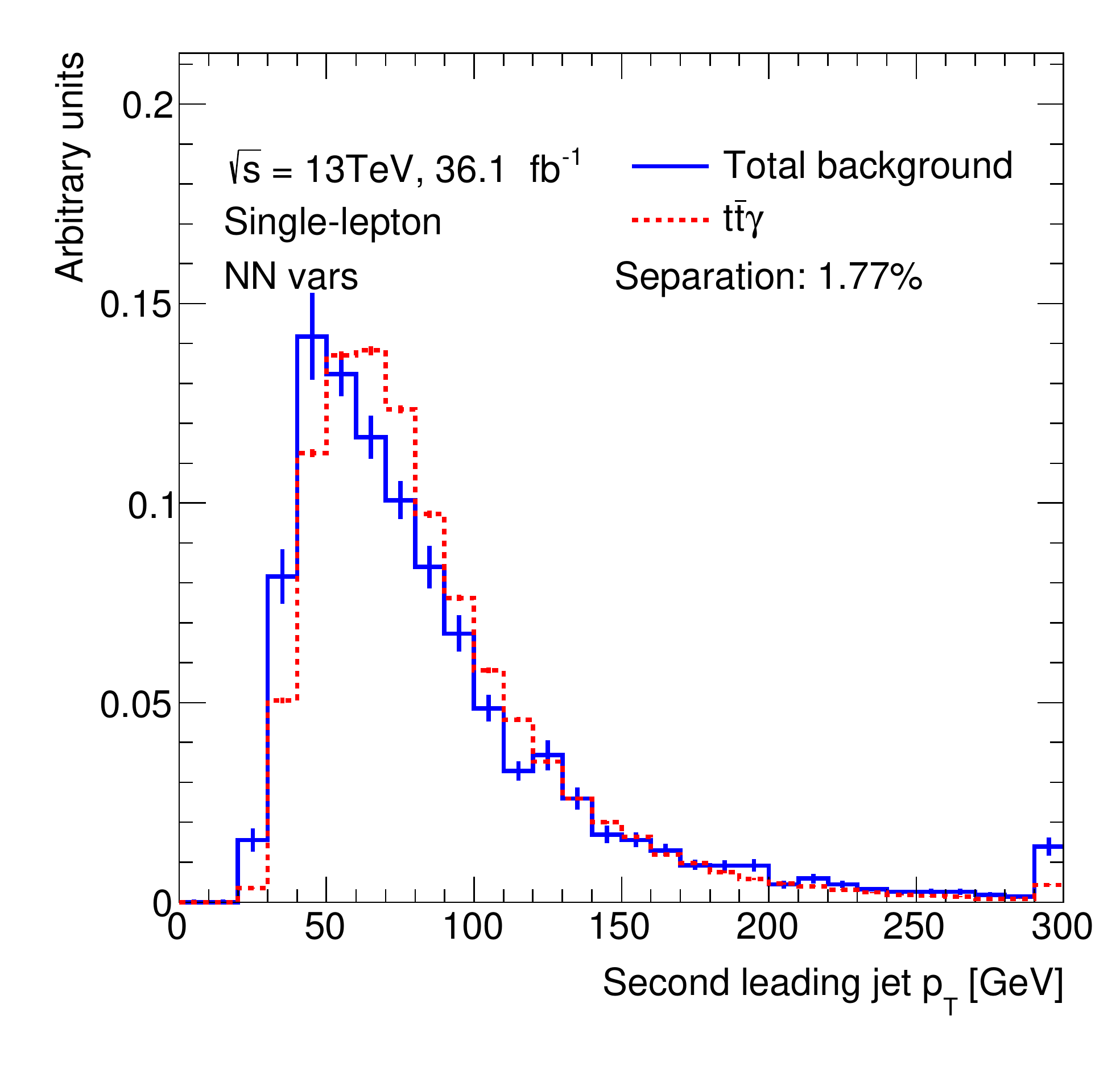}
\hspace{-0.023\linewidth}

\includegraphics[width=0.34\linewidth]{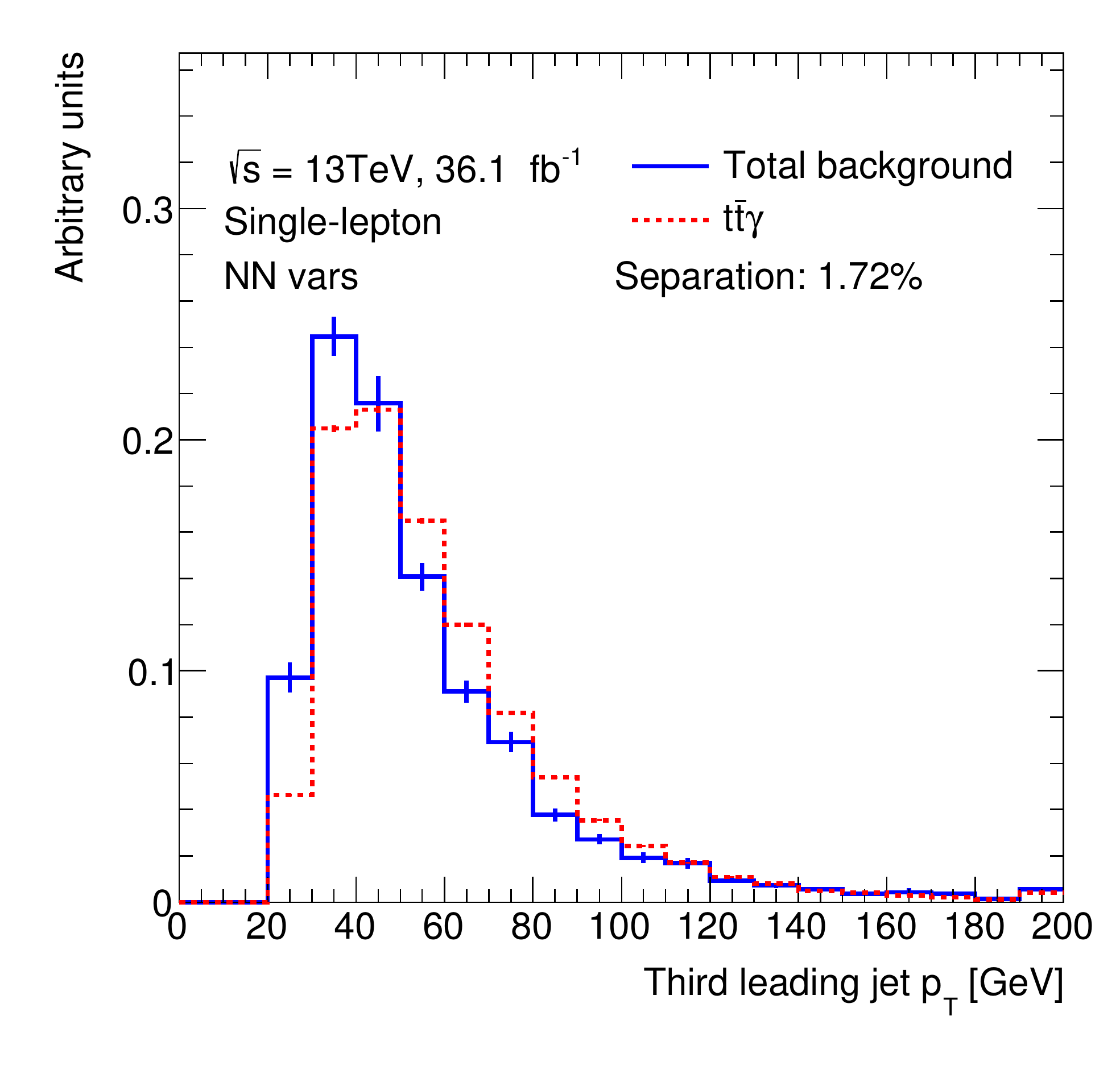}
\hspace{-0.023\linewidth}
\includegraphics[width=0.34\linewidth]{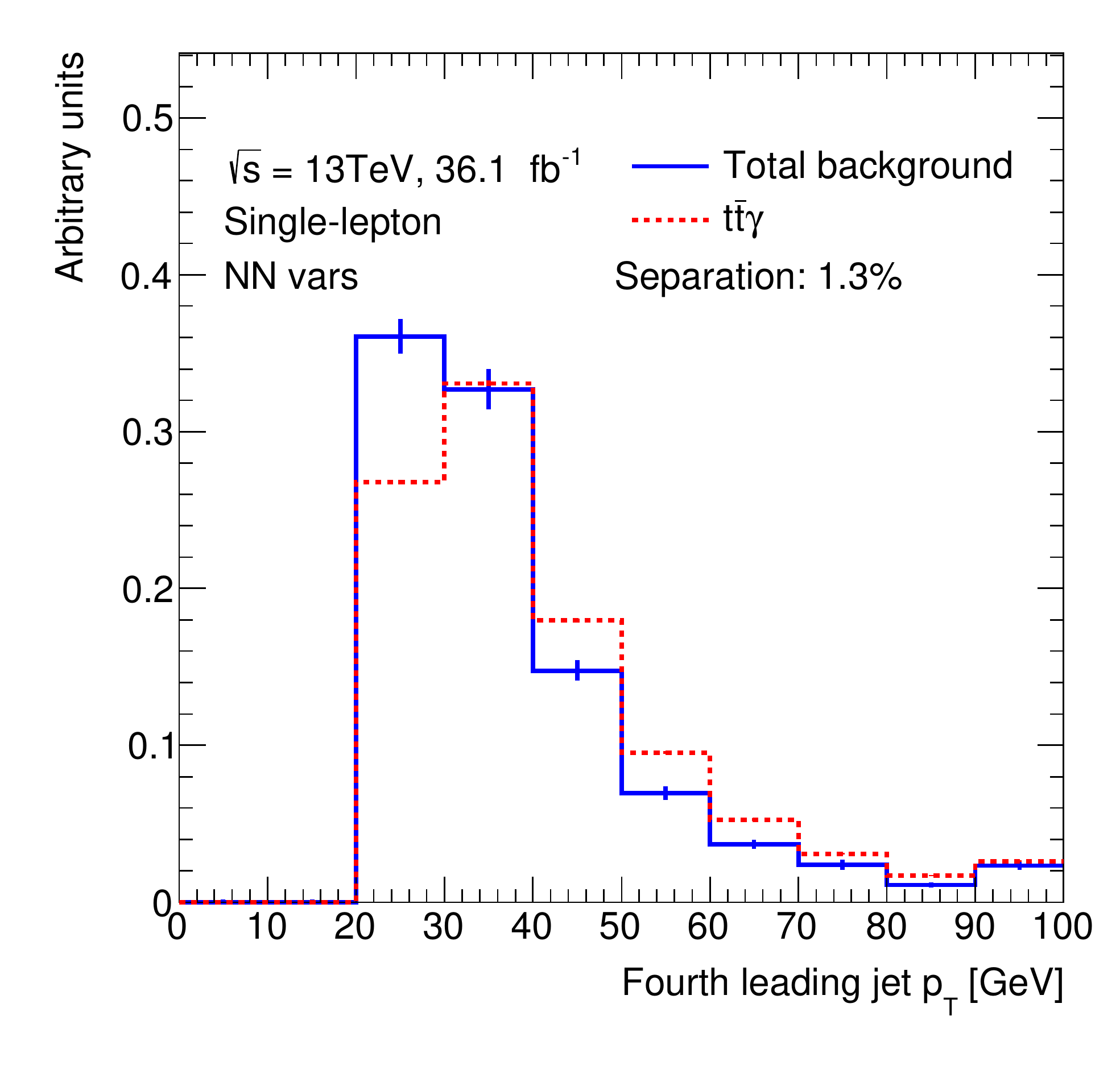}
\hspace{-0.023\linewidth}
\includegraphics[width=0.34\linewidth]{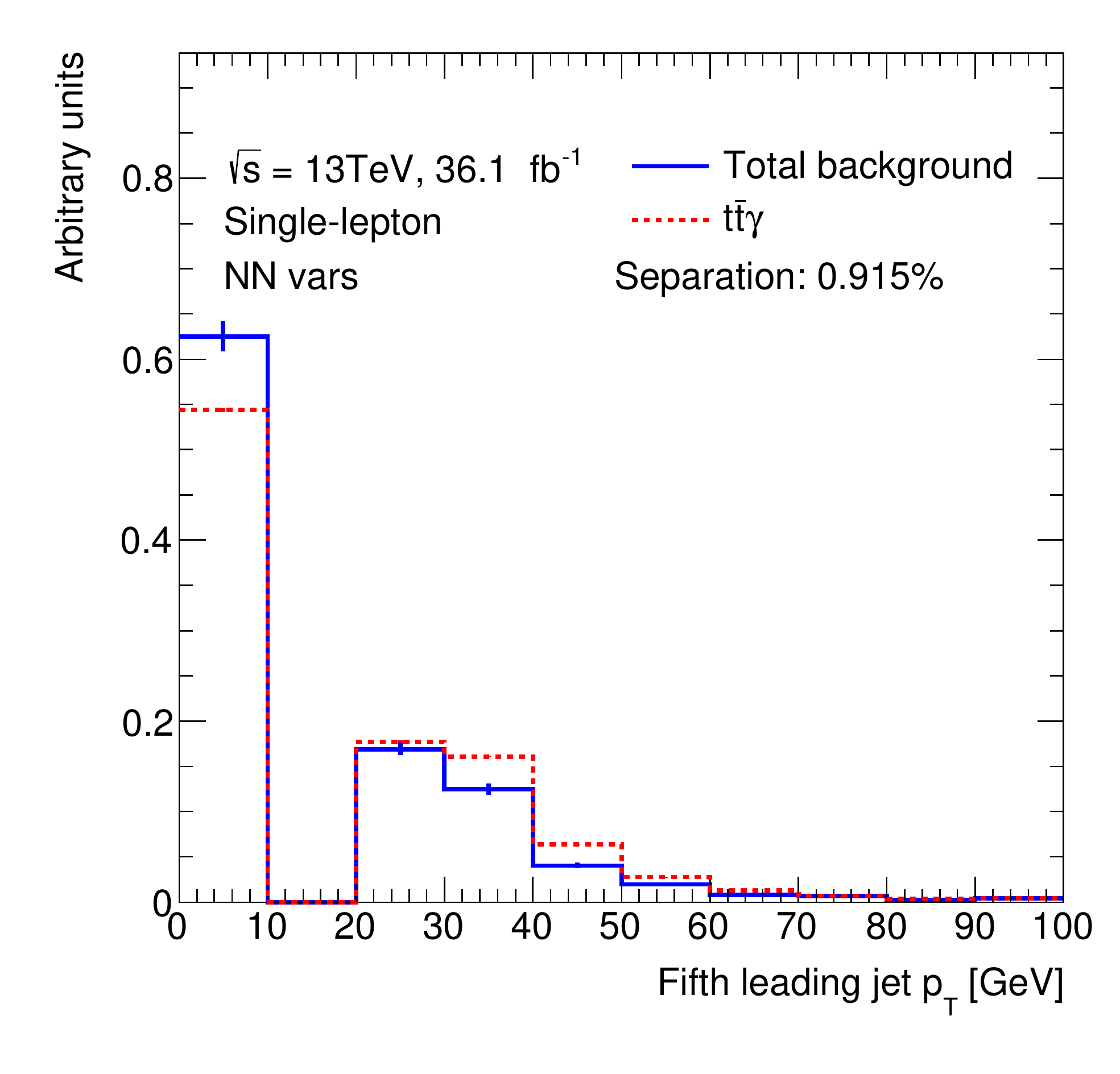}
\hspace{-0.023\linewidth}

\includegraphics[width=0.34\linewidth]{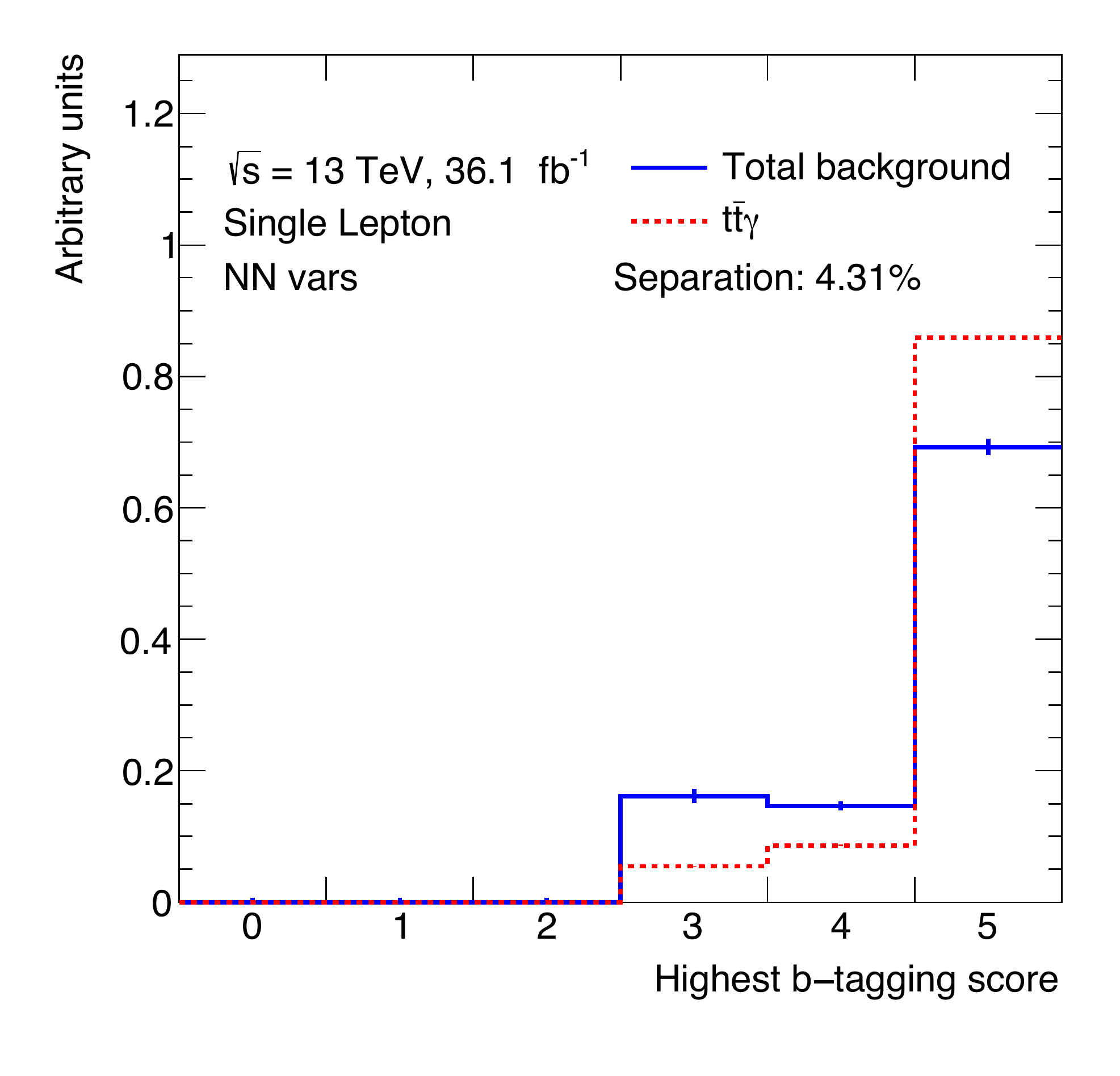}
\hspace{-0.023\linewidth}
\includegraphics[width=0.34\linewidth]{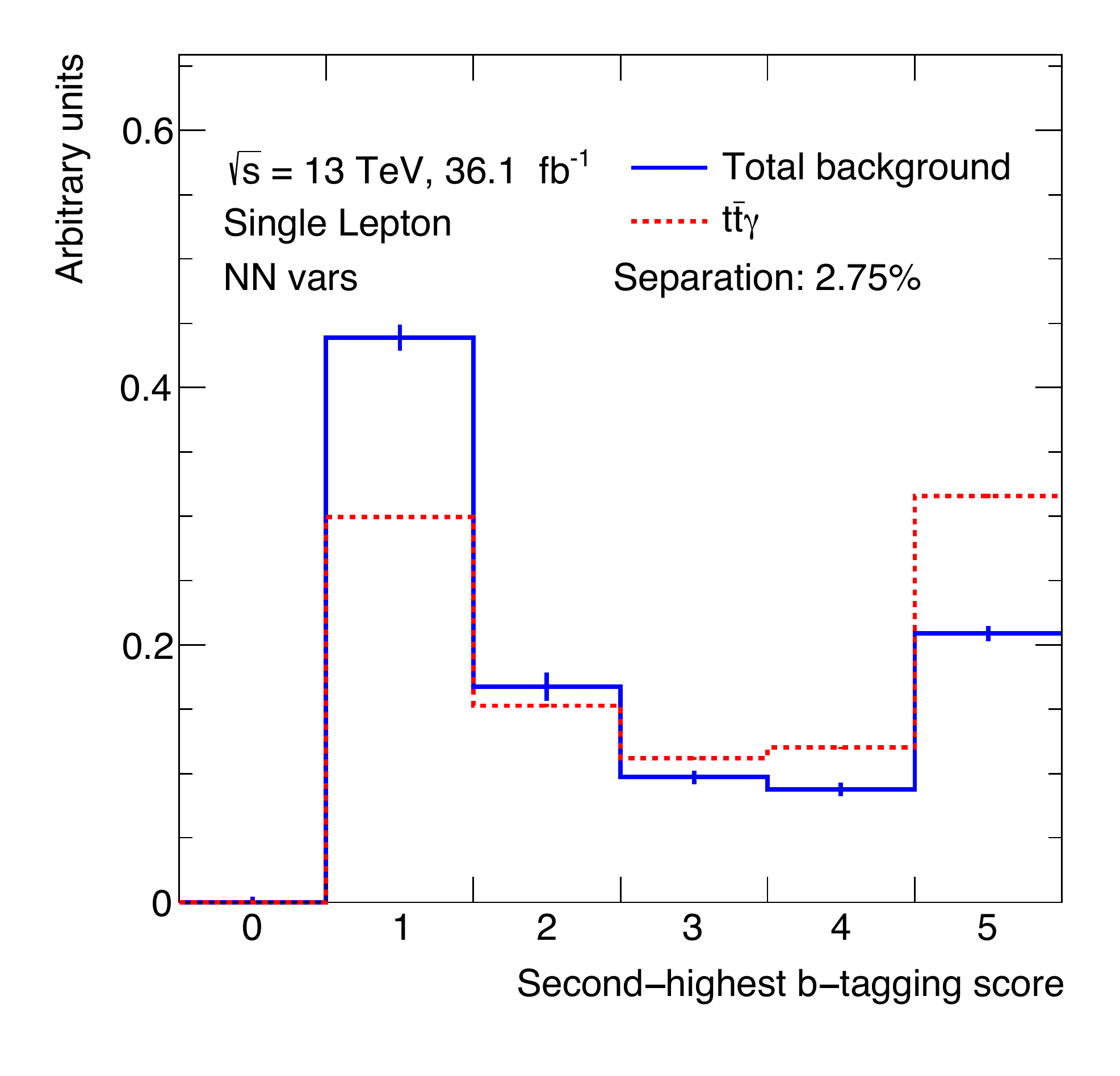}
\hspace{-0.023\linewidth}
\includegraphics[width=0.34\linewidth]{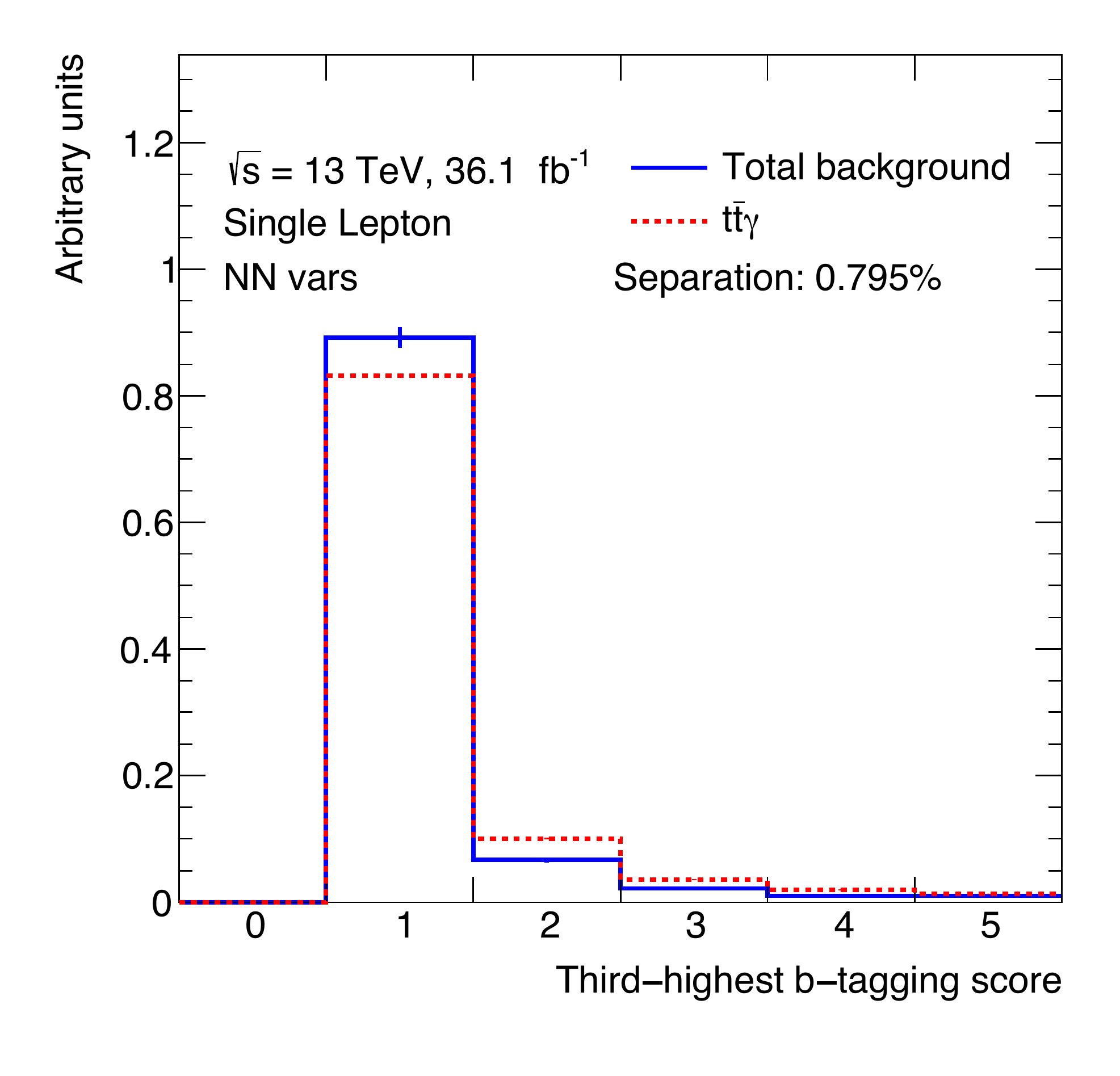}
\hspace{-0.032\linewidth}

\phantomcaption
\end{figure}

\begin{figure}
\ContinuedFloat
\centering

\includegraphics[width=0.34\linewidth]{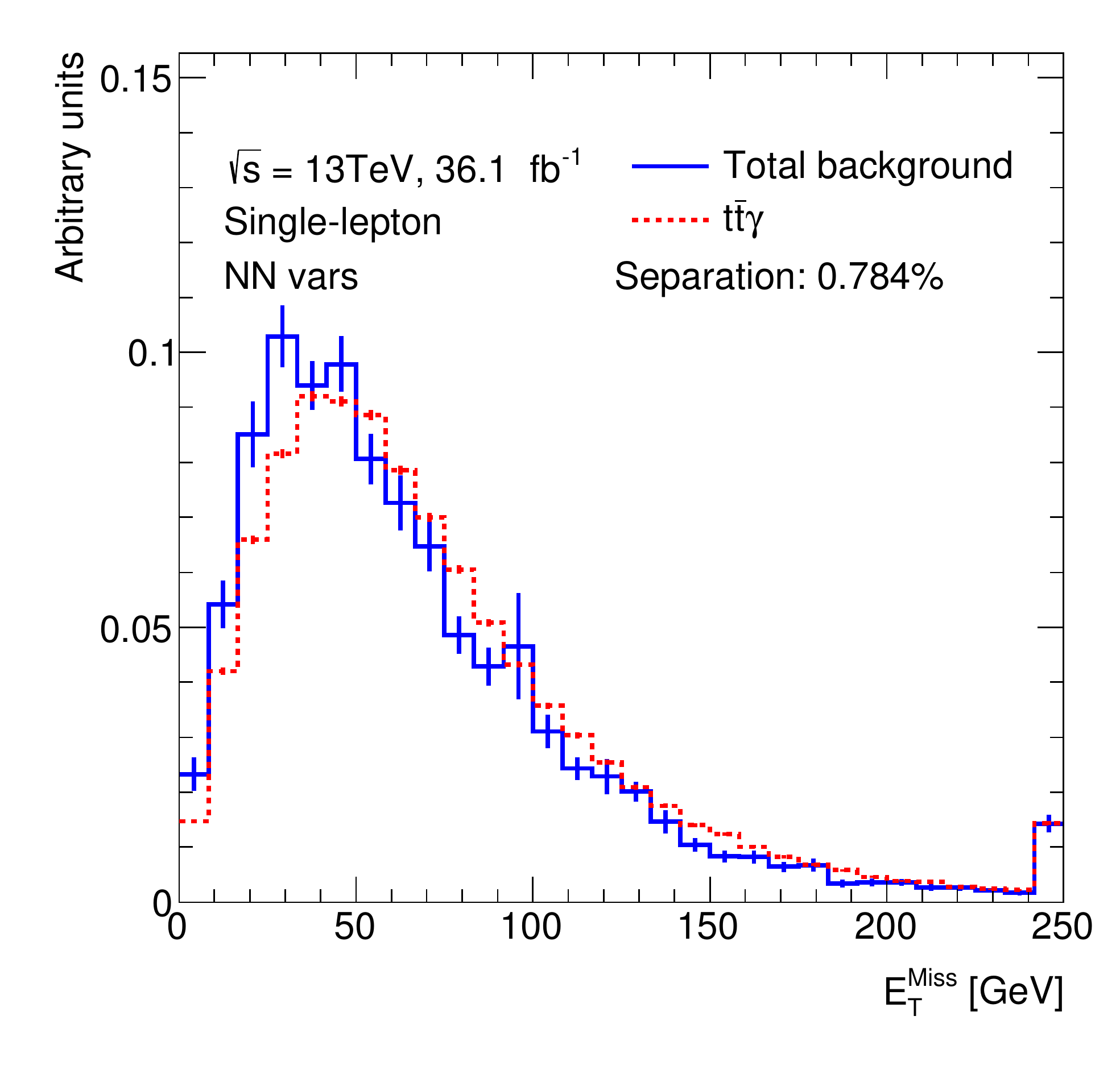}
\hspace{-0.023\linewidth}
\includegraphics[width=0.34\linewidth]{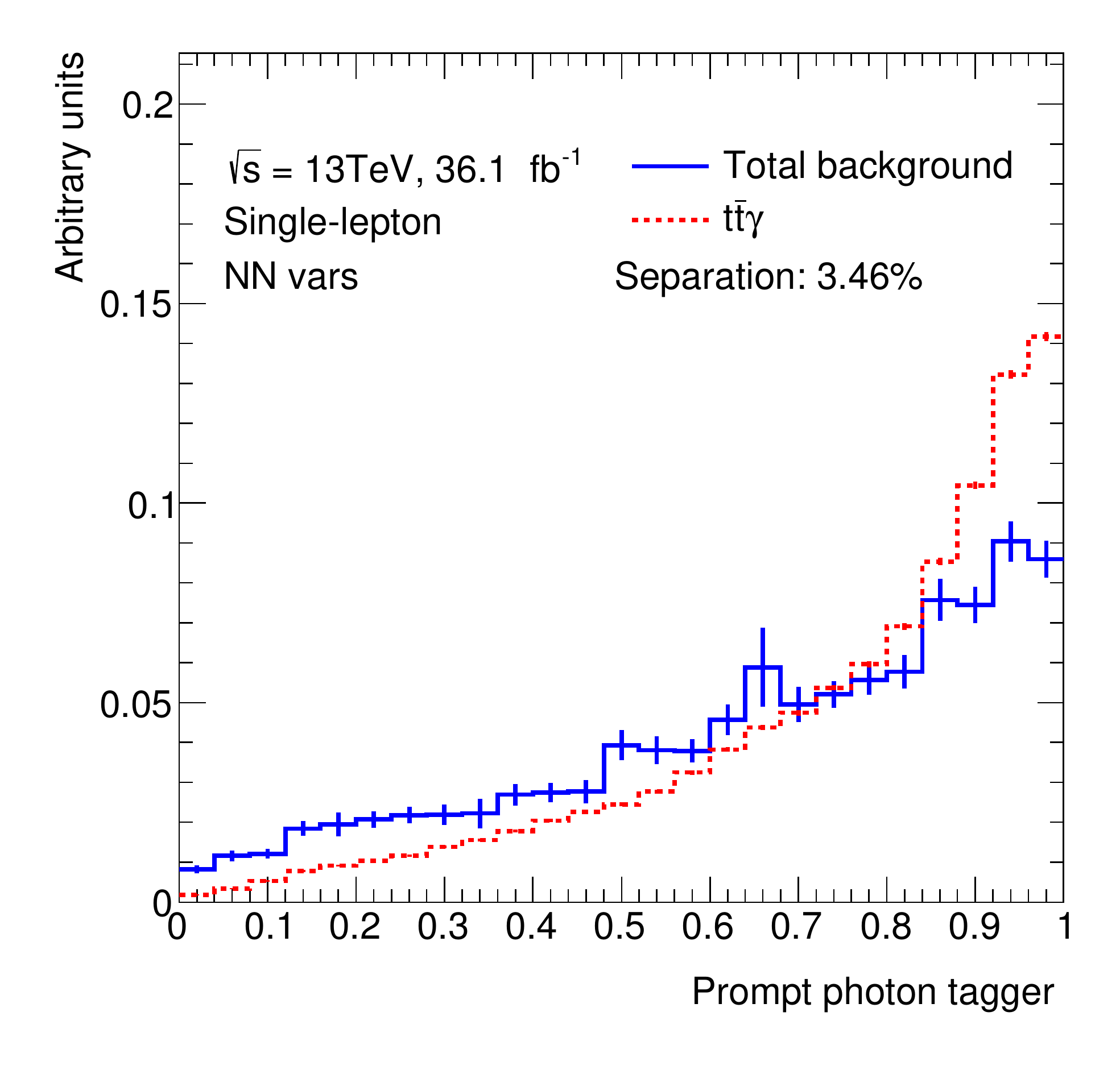}
\hspace{-0.023\linewidth}
\includegraphics[width=0.34\linewidth]{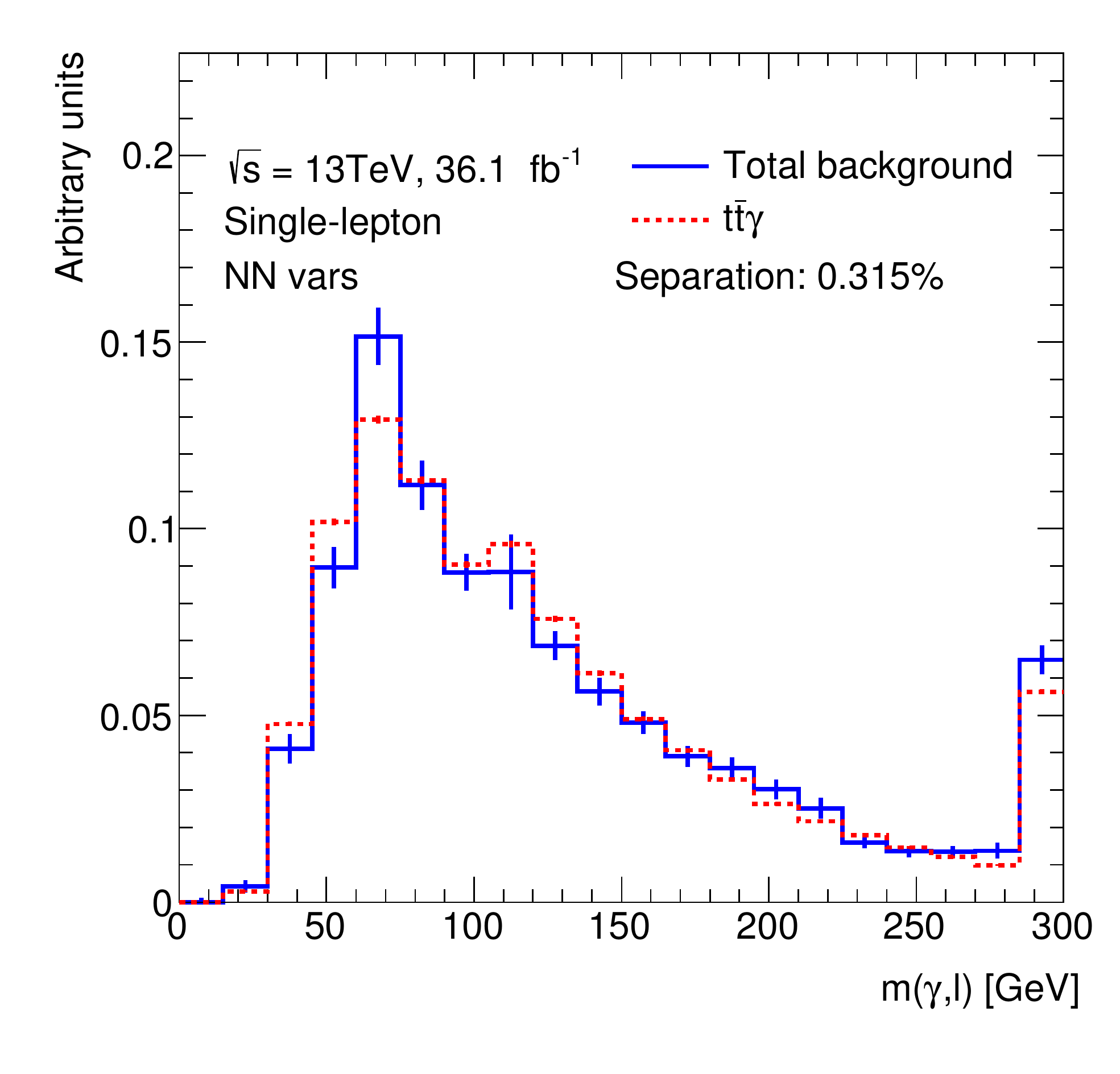}
\hspace{-0.023\linewidth}

\caption {Variables used in the \chljets ELD showing the separation between the signal and sum of all backgrounds.}
\label{fig:trainingSinglelepton}
\end{figure}

\FloatBarrier
\subsection{Training}
\label{sec:ELDtraining}


Training was carried out on the full set of MC samples described in Section~\ref{sec:MCsim} and the data-driven fake lepton samples described in Section~\ref{sec:fakelepton}. 
Each feature (or input variable) is normalised to give a standard deviation of 1 and a mean of 0. Keras is then used to build and train the NNs. 
The final output consists of separate architectural and variable input files.
The variable input file not only contains a map for the variables used in the training, but also the normalisation parameters for each variable. These are essential inputs to \lwtnn when reconstructing a NN inside of \ATLAS code.
Each of the NNs had hyper-parameter searches performed to achieve the best architecture and thus maximum separation. Presented below is a detailed description of the chosen NN architectures.

\subsubsection{Final architecture and hyper-parameters}

The \NN final topology is the same for \chljets and \chll channels: one input dense layer consisting of 30 nodes is followed by the regularisation techniques of a dropout layer then a batch normalisation layer. A dense layer of 20 nodes follows with a final dense output layer consisting of one node with a sigmoid activation function to reduce outputs to values between 0 and 1. All layers' weights are initialised with uniform distributions and input and hidden layers receive a rectified linear unit (ReLU) activation function. 
The dropout layer has a 30\% chance that any node and its weights will randomly be removed from the epoch.

The \chljets{} \NN uses 75\% of the raw input events for training with 25\% used for testing. This equates to 144491 events used for training and 48164 events used for testing. Cumulatively, there are 160679 signal and 31976 background events.
The \chll{} \NN uses 67\% of the raw input events for training with 33\% used for testing. This equates to 20258 events used for training and 10160 events used for testing. Cumulatively, there are 21591 signal and 8827 background events.
The ratio of the split between training and testing is done to ensure that when separate $k$-fold validation is performed the training and test samples can be split consistently into four (three) different tests for the \chljets{} (\chll) channels.
The \chljets{} \NN is trained using a batch size of 150 events while the \chll{} \NN uses 300 events. This describes the number of events used for each gradient update. Optimisation for batch size was performed using a grid search. Event weights are accounted for by scaling the binary cross-entropy loss function during training, with negative event weights set to 0.

\begin{figure}[!htbp]
\includegraphics[width=0.99\linewidth]{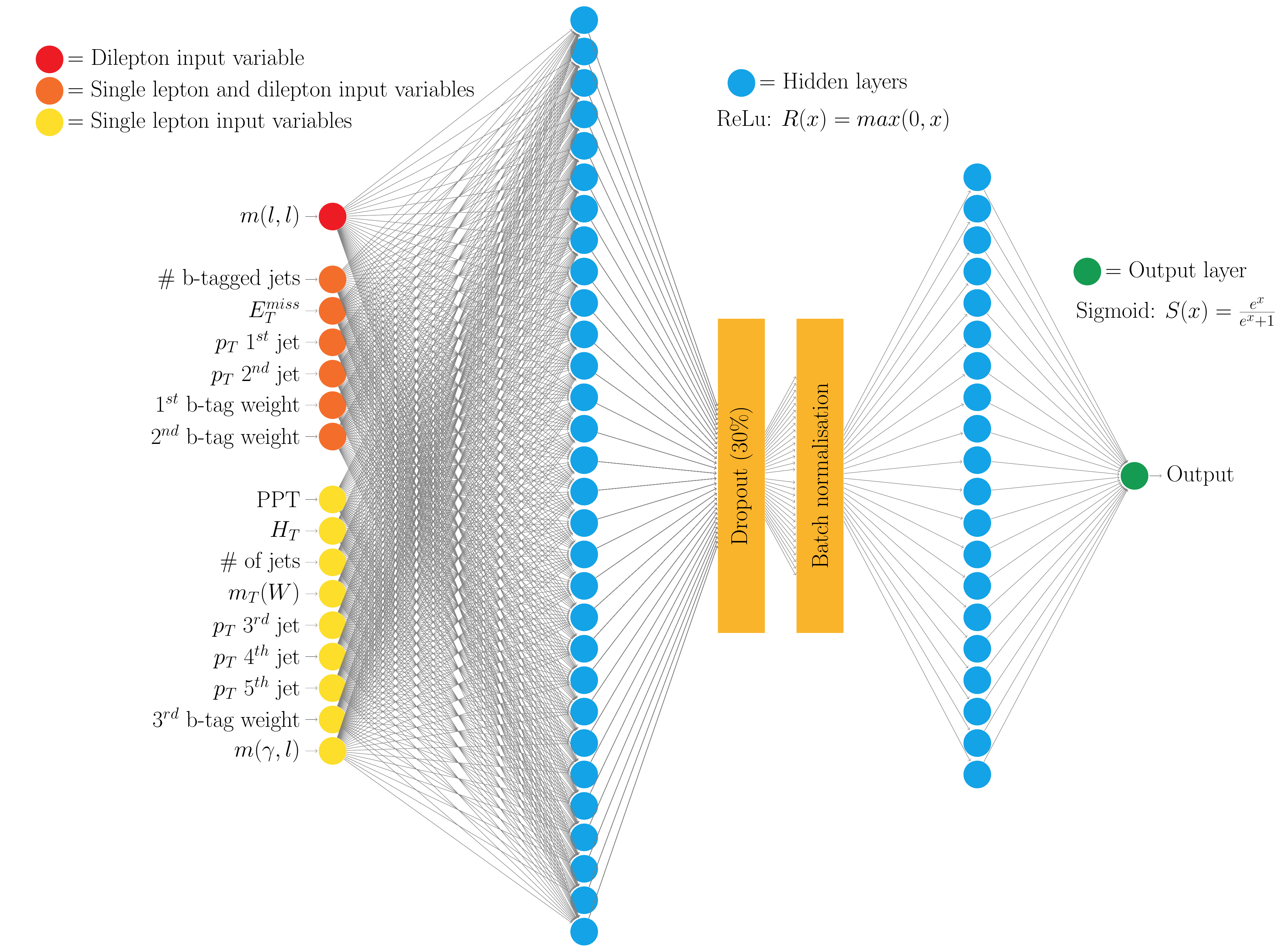}
\caption {Schematic of the \chljets and \chll{} \ELD showing the different input variables and layers.}
\label{fig:tikzELD}
\end{figure}

The previous section introduced the variables considered as inputs for the NNs. Based on their separation a few training scenarios were carried out to find a stable configuration. An NN was trained with three (five) variables ranked by their separation for the \chljets (\chll) channel. The NN was then retrained by adding the next highly ranked variables.
Tables~\ref{tab:SLVariablesOpti} and \ref{tab:DLVariablesOpti} show the area under the curve (AUC) for this process, performed on an increasing number of variables. 
In both cases, the \NN performs the best with the largest number of variables considered. However, for the \chll{} \NN with more than seven variables we see larger variations between multiple trainings and testings and so with the limited statistics it is deemed more unstable with overtraining likely to occur. As a result the network using seven variables was selected.
A schematic of the final \NN{}s is seen in Figure~\ref{fig:tikzELD}, with a summary of the hyper-parameters in Table~\ref{tab:ELDsummary}.

\newcolumntype{L}[1]{>{\raggedright\let\newline\\\arraybackslash\hspace{0pt}}m{#1}}
\begin{table}[!htbp]
  \centering
  \begin{tabular}{cL{6cm}ccc}
    \hline
     Total vars. & Var. names & AUC (train) & AUC (test) & Selected \NN \\
    \hline
    \hline
    5 & highest $b$-tagging~weight, 2$^{\text{nd}}$~highest~$b$-tagging~weight, 3$^{\text{rd}}$~highest~$b$-tagging~weight, \PPT, number of $b$-jets  & 0.6483 & 0.6474 & \\ \hline 
    10 & $+$ \pt~of~first~jet, \HT,  \pt~of~second~jet,  \pt~of~third~jet,  \pt~of~fourth~jet  &  0.6744 & 0.6736 &  \\ \hline 
    14 & $+$ \pt~of~fifth~jet ,$m_{T}(W)$ , \MET, number~of~jets, & 0.6996 & 0.6906 & \\ \hline 
    15 & $+$ $m(\gamma,\ell)$ &  0.7017 & 0.6929 & \checkmark  \\

    \hline
  \end{tabular}
  \caption{Variable optimisation for the \chljets{} \NN. The starting point is the first row. Each subsequent row includes more variables on top of the previous row. Variables are chosen according to separation rank and physics motivation.}
  \label{tab:SLVariablesOpti}
\end{table}

 \newcolumntype{L}[1]{>{\raggedright\let\newline\\\arraybackslash\hspace{0pt}}m{#1}}
 \begin{table}[!htbp]
   \centering
   \begin{tabular}{cL{6cm}ccc}
     \hline
     Total vars. & Var. names & AUC (train) & AUC (test) & Selected \NN \\
     \hline
     \hline
     3 &2$^{\text{nd}}$~highest~$b$-tagging~weight, $m(\ell,\ell)$,\MET & 0.9289 & 0.9233 & \\ \hline 
     7 & $+$ \pt~of~first~jet, number~of~$b$-jets, highest~$b$-tagging~weight,  \pt~of~second~jet  & 0.9313 & 0.9258 & \checkmark \\ \hline 
     8 & $+$ \PPT  & 0.9329 & 0.9261& \\ \hline 
     10 & $+$ \pt~of~third~jet, $m(\gamma,\ell,\ell)$  & 0.9365 & 0.9313 & \\ \hline  
     13 & $+$ \mwt ,number~of~jets, \HT & 0.9428 & 0.9393 &  \\ 
     \hline
   \end{tabular}
     \caption{Variable optimisation for the \chll{} \NN. The starting point is the first row. Each subsequent row includes more variables described on top of the previous row. Variables are chosen according to separation rank and physics motivation.}
   \label{tab:DLVariablesOpti}
 \end{table}

\begin{table}[!htbp]
  \centering
  \begin{tabular}{lcc}
    \hline
     Parameter & \chljets & \chll \\
     \hline
     \hline
     Variables & 15 & 7 \\ 
     \hline
     Raw training events & 144491 (75\%) & 20258 (67\%)\\
     \hline
     Raw test events & 48164 (25\%) &10160 (33\%) \\
     \hline 
     Raw signal events & 160679 (83\%) & 21591 (71\%) \\
     \hline 
     Raw background events & 31976 (17\%) & 8827 (29\%) \\
   \hline
        Batch size & 150 & 300 \\
   \hline
     Optimiser & \multicolumn{2}{c}{Adam} \\
   \hline
     Loss function & \multicolumn{2}{c}{Binary cross entropy} \\
    \hline
     Architecture &  \multicolumn{2}{c}{Dense(30) $\to$ Dropout(30\%) $\to$ BN $\to$ Dense(20) $\to$ Dense(1)}  \\
     \hline
     Activations & \multicolumn{2}{c}{ReLu and Sigmoid} \\
     \hline
    Total trainable parameters & 1181 & 941 \\
    \hline
  \end{tabular}
  \caption{Summary of the parameters used in the training of the \chljets and \chll ELDs. The value between parenthesis for Dense and Dropout indicate the number of nodes and the dropout percentage, respectively. BN is batch normalisation.}
  \label{tab:ELDsummary}
\end{table}

\begin{figure}[!htbp]
\centering
    \subfloat[\chljets]{
      \includegraphics[width=0.51\linewidth]{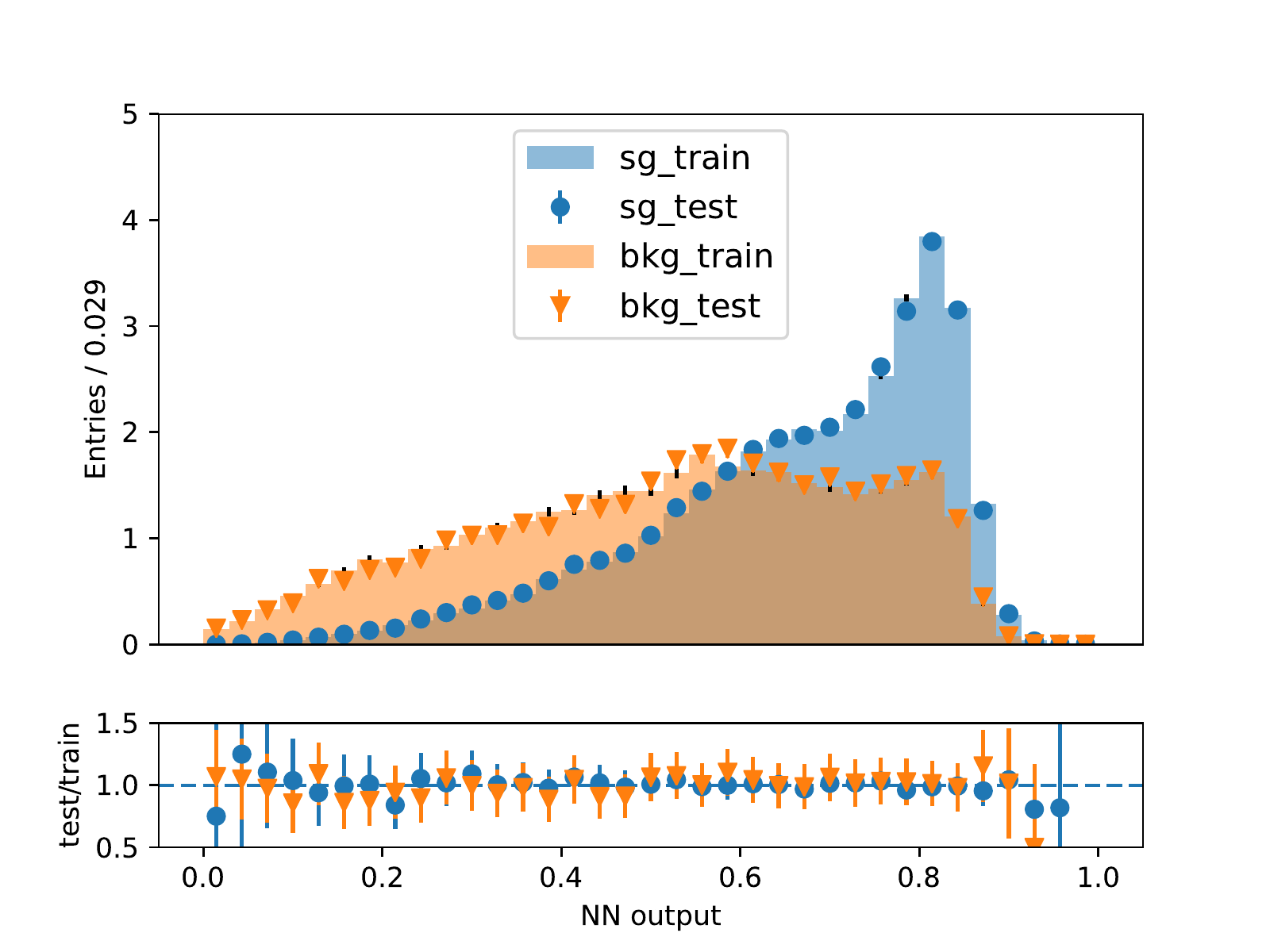}
    }\hspace{-0.072\linewidth}
  \subfloat[\chljets]{
    \includegraphics[width=0.51\linewidth]{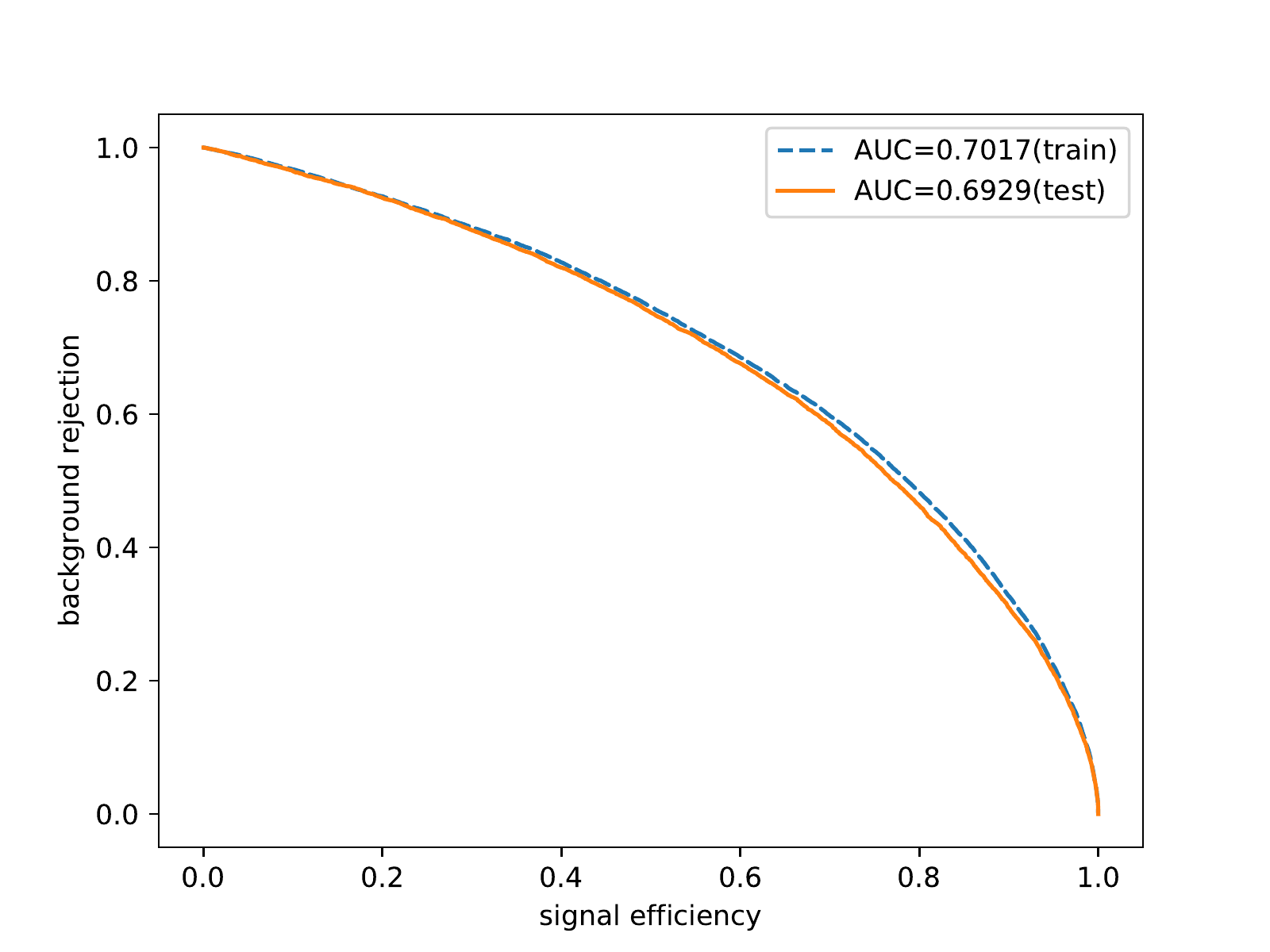}
  }\hspace{-0.072\linewidth}
  
  \subfloat[\chll]{
    \includegraphics[width=0.51\linewidth]{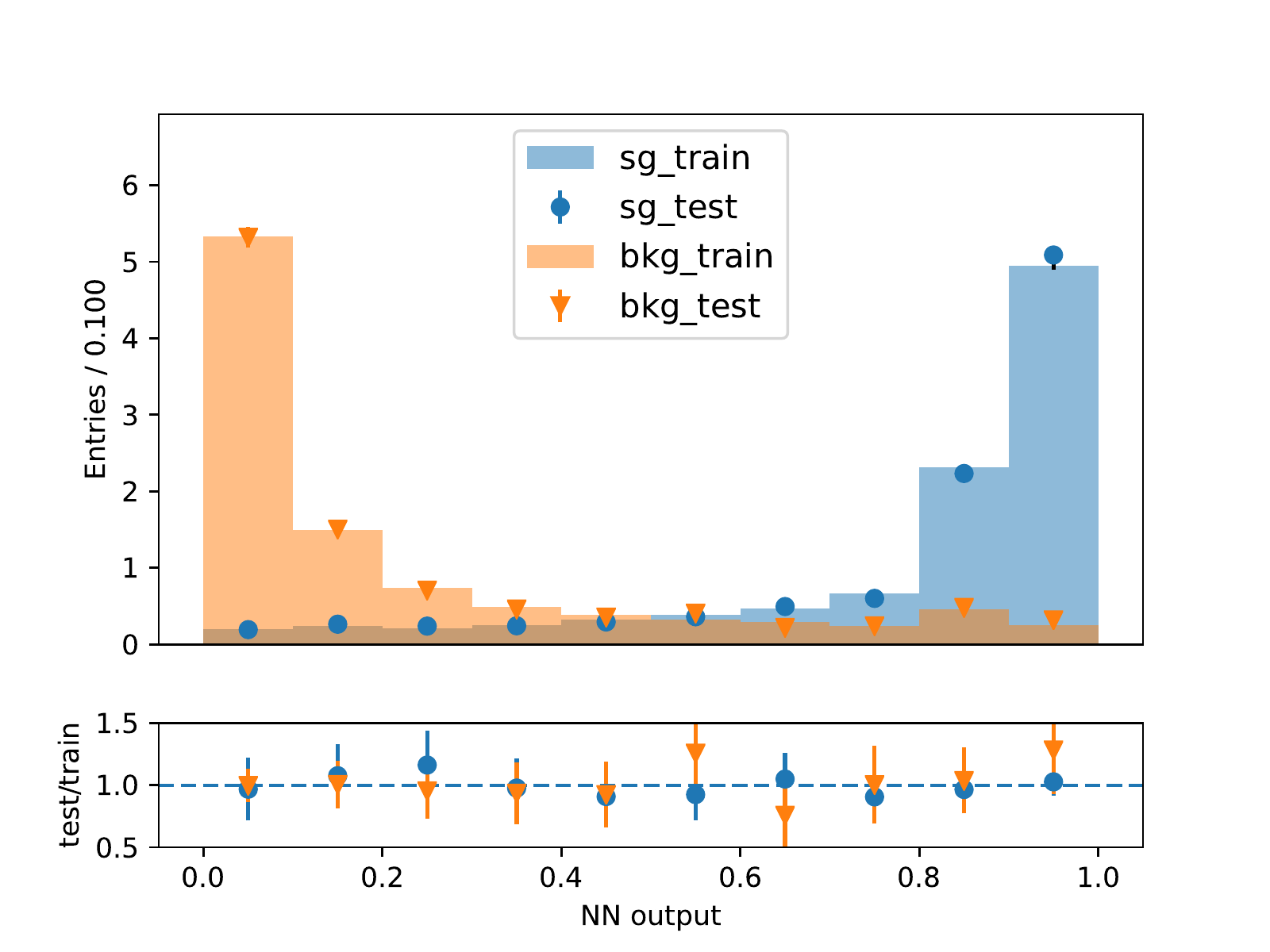}
  }\hspace{-0.072\linewidth}
  \subfloat[\chll]{
    \includegraphics[width=0.51\linewidth]{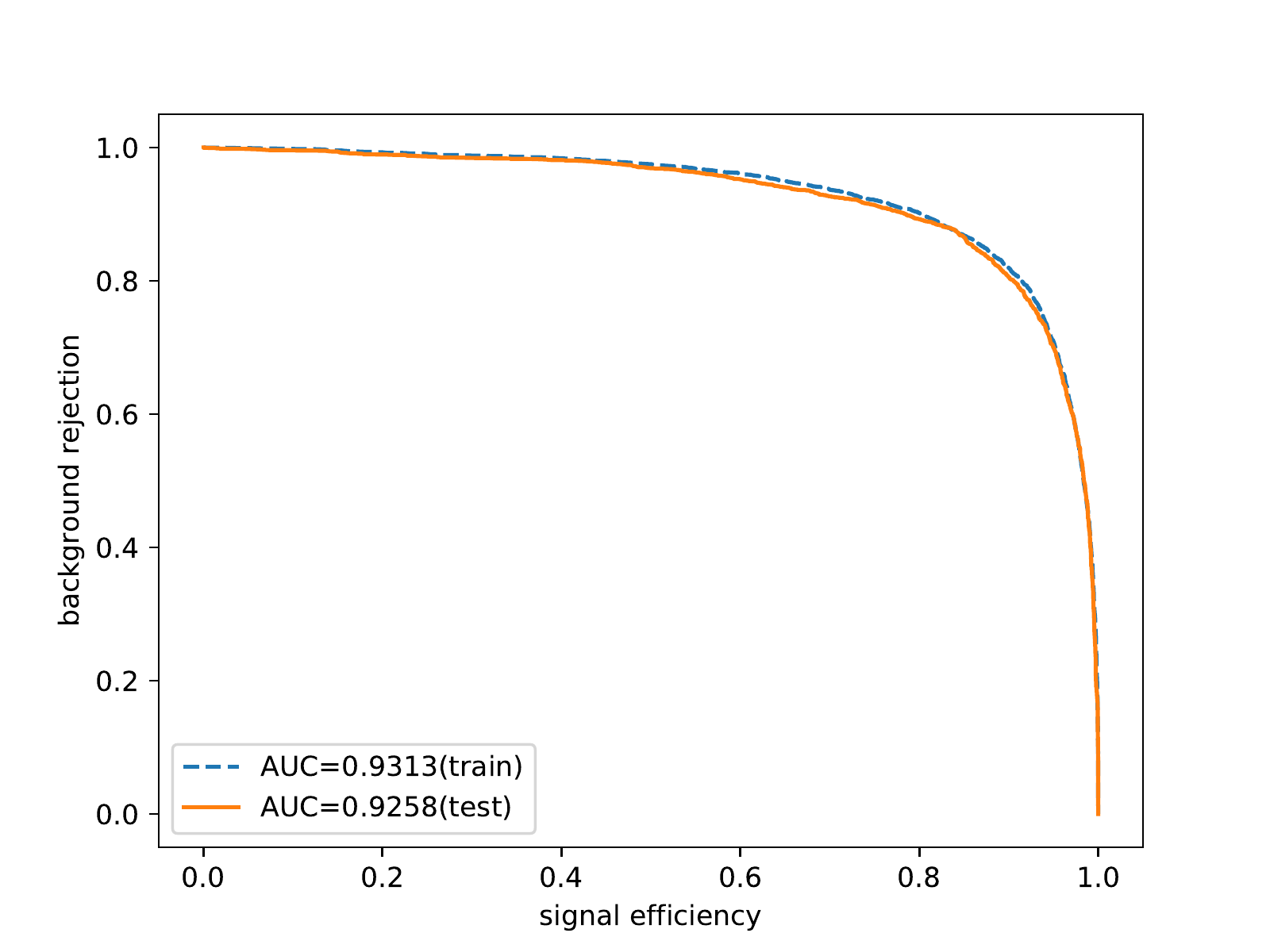}
  }\hspace{-0.072\linewidth}
\caption {Event-level Discriminator training output showing the test and training data agreement for the NN response and the associated ROC curve.}
\label{fig:ELDtraining}
\end{figure}

The performance of the chosen \NN{}s is shown in Figure~\ref{fig:ELDtraining}. The left plots describe the NN's response to training (filled histograms) and test (points) data. As shown in the ratio plots, good agreement can be seen. The right plots show the ROC curve with the respective AUC values for the training and test data. A line going from (0,1) to (1,0) would indicate the classifier is no better than random guesswork. If the classifier were to be 100\% efficient it would be represented by a box with the outermost corner at (1,1). The AUC values for the test and training set are close enough to indicate no overtraining has occurred. 
The \chljets ROC curve represents the actual performance of the \ELD in the signal region. Care must be taken when reading the \chll ROC curve since the training is performed on a larger phase space than the final signal region. 

\begin{figure}[!htbp]
\centering
\subfloat[\chljets]{
\includegraphics[width=0.51\linewidth]{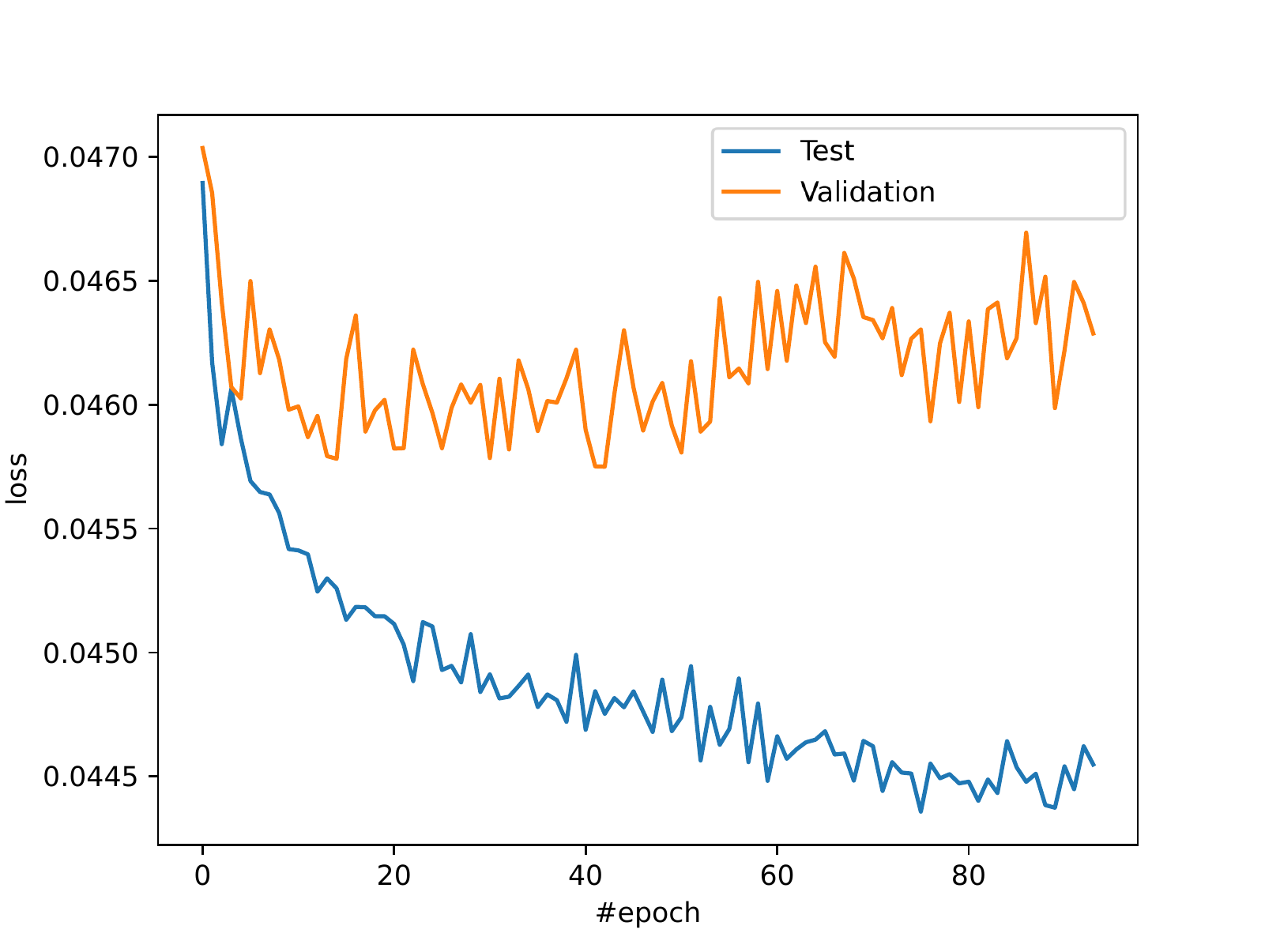}
}\hspace{-0.072\linewidth}
\subfloat[\chll]{
\includegraphics[width=0.51\linewidth]{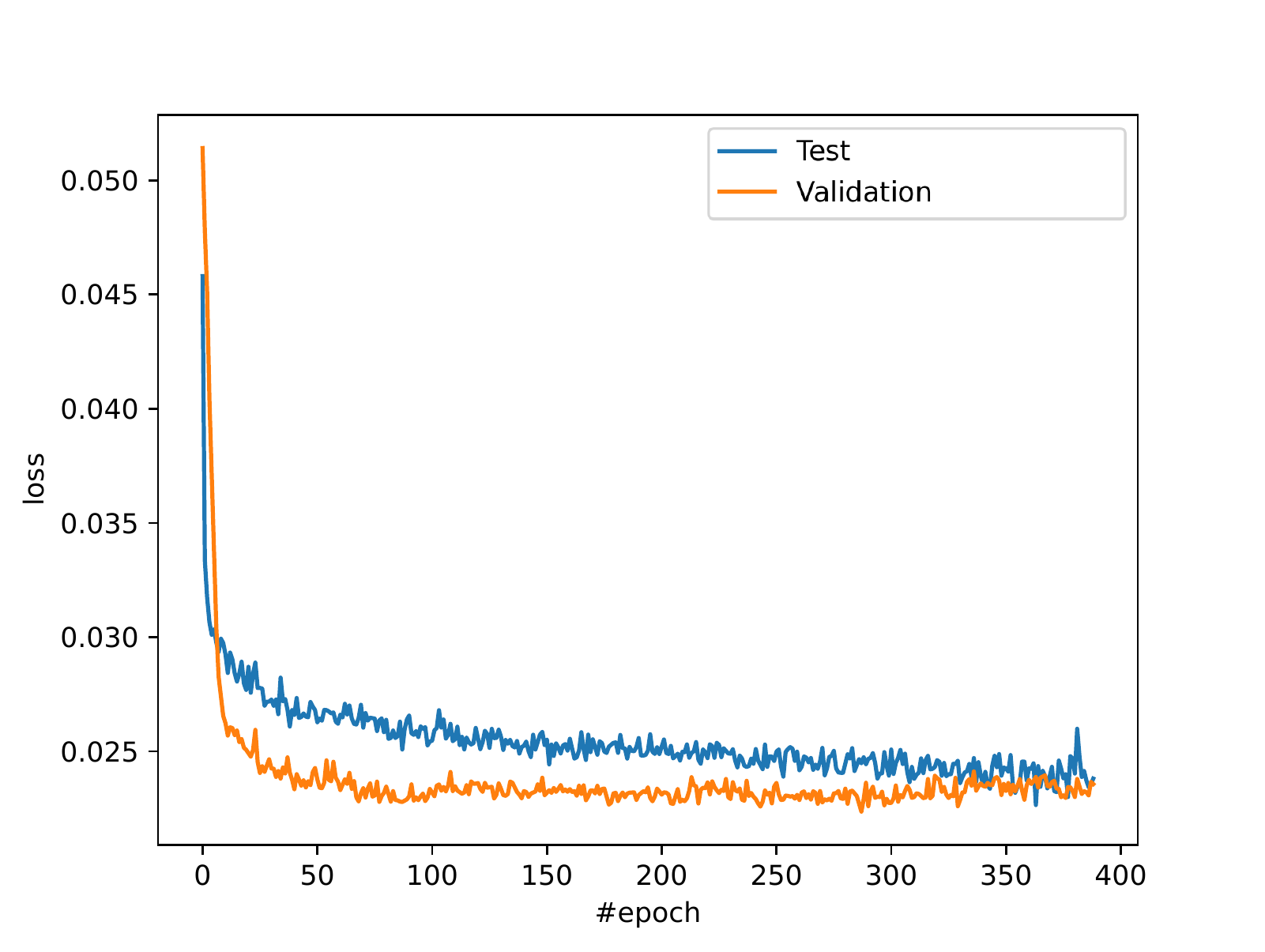}
}\hspace{-0.072\linewidth}
\caption {Test and validation dataset loss curves for the \chljets and \chll \ELD. Training is carried out so that the best model is saved according to the validation loss curve. The model stops training if no improvement is seen in the validation curve for more than 50 epochs. Note the ranges on the $y$-axes.}\label{fig:loss}
\end{figure}

Another important parameter to consider when training a \NN is the loss function, shown for both \NN{}s in Figure~\ref{fig:loss}.
This represents the evolution in the differences between the predicted labels and the actual labels (Equation~\ref{eq:lossfunction}).
Care must be taken when reading the $y$-axis range on these plots. The \chljets is trained on a fairly tight signal region and so from the beginning it must work hard to classify events effectively. Thus, a small improvement in the loss is seen up until the first $\approx 40$ epochs, at which point the validation loss starts to increase; a clear sign of overtraining. A ``patience" parameter is used such that if no improvement is seen for 50 epochs, the training is exited and the best model (at around 40 epochs) is saved. The \chll loss has a sharply falling curve for the first few epochs due to the looser cuts and these events being easily classified as signal or background.

\FloatBarrier

\subsubsection{$K$-fold cross validation}
\begin{figure}[!htbp]
\centering
  \subfloat[\chljets training data]{
     \includegraphics[width=0.51\linewidth]{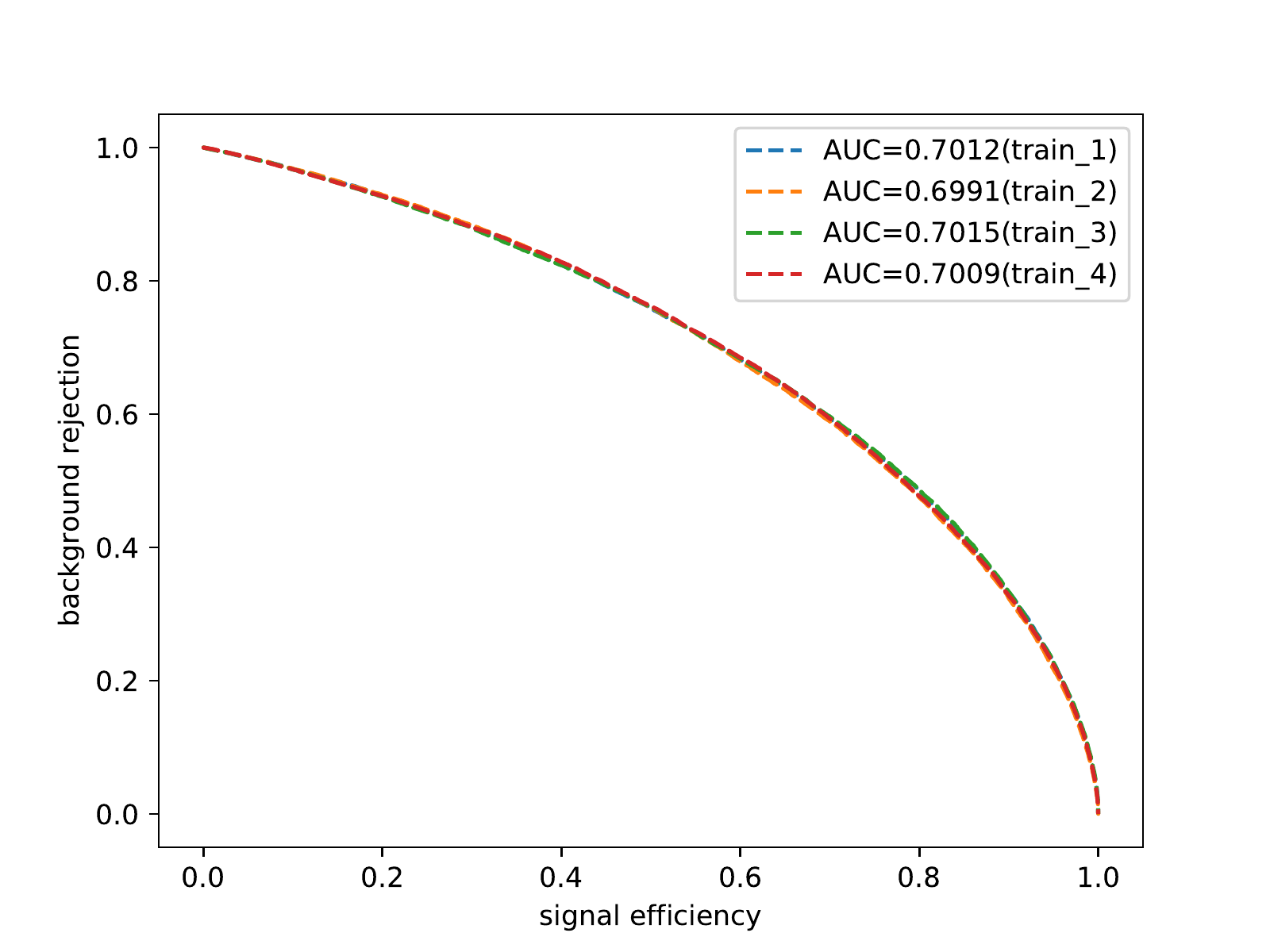}
  }\hspace{-0.072\linewidth}
    \subfloat[\chljets test data]{
     \includegraphics[width=0.51\linewidth]{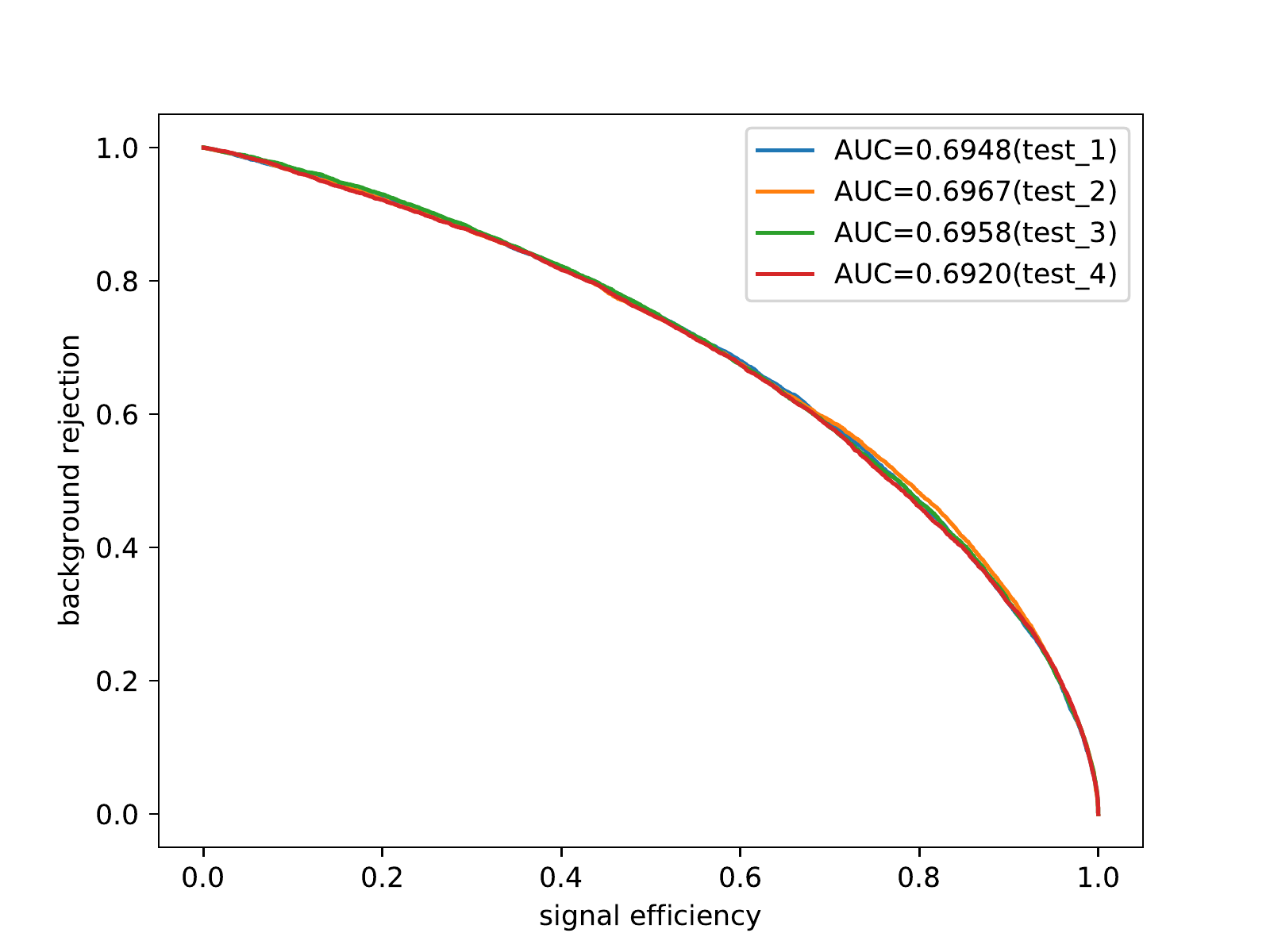}
    }\hspace{-0.072\linewidth}
    
    \subfloat[\chll training data]{
     \includegraphics[width=0.51\linewidth]{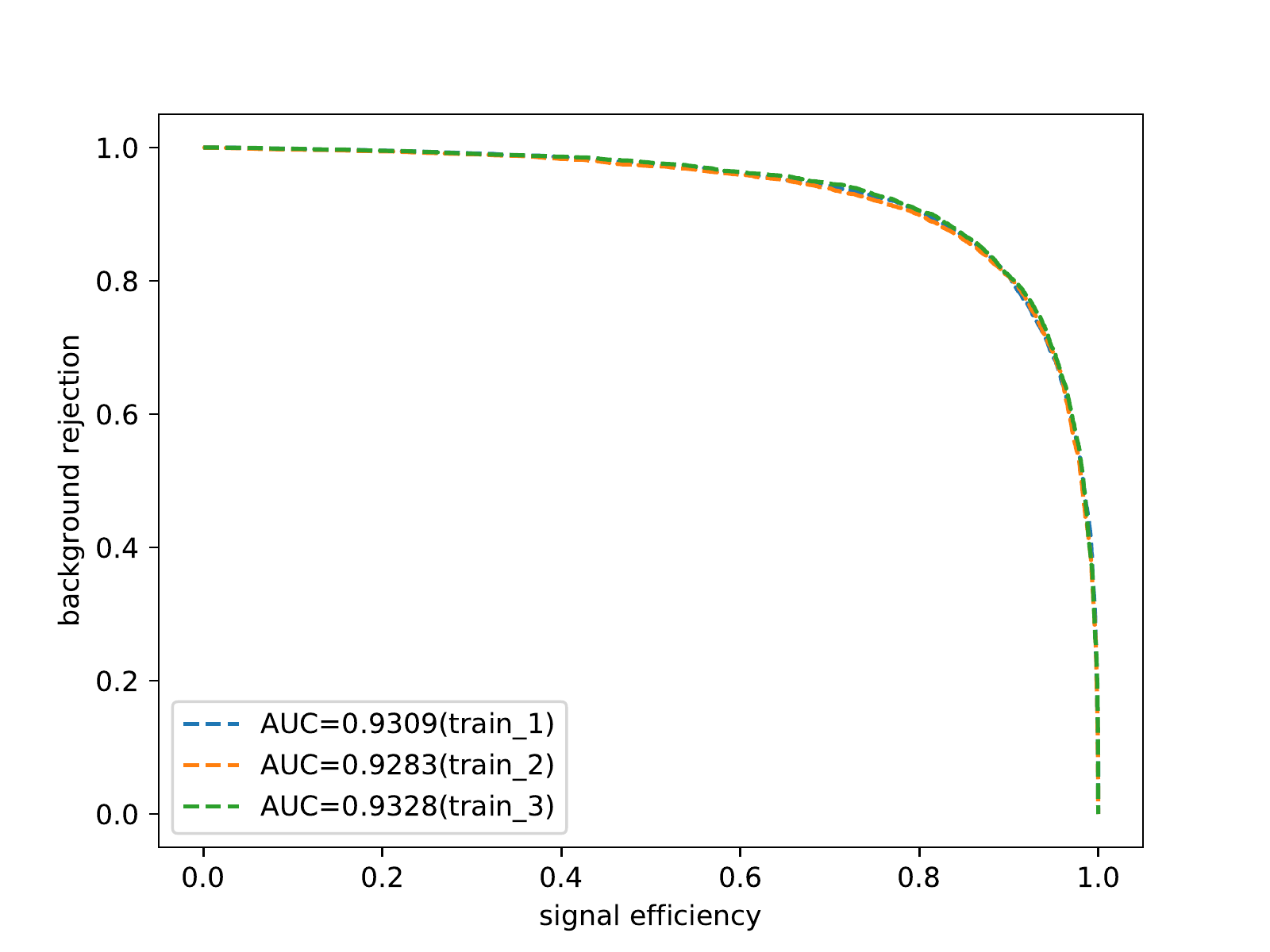}
    }\hspace{-0.072\linewidth}
    \subfloat[\chll test data]{
     \includegraphics[width=0.51\linewidth]{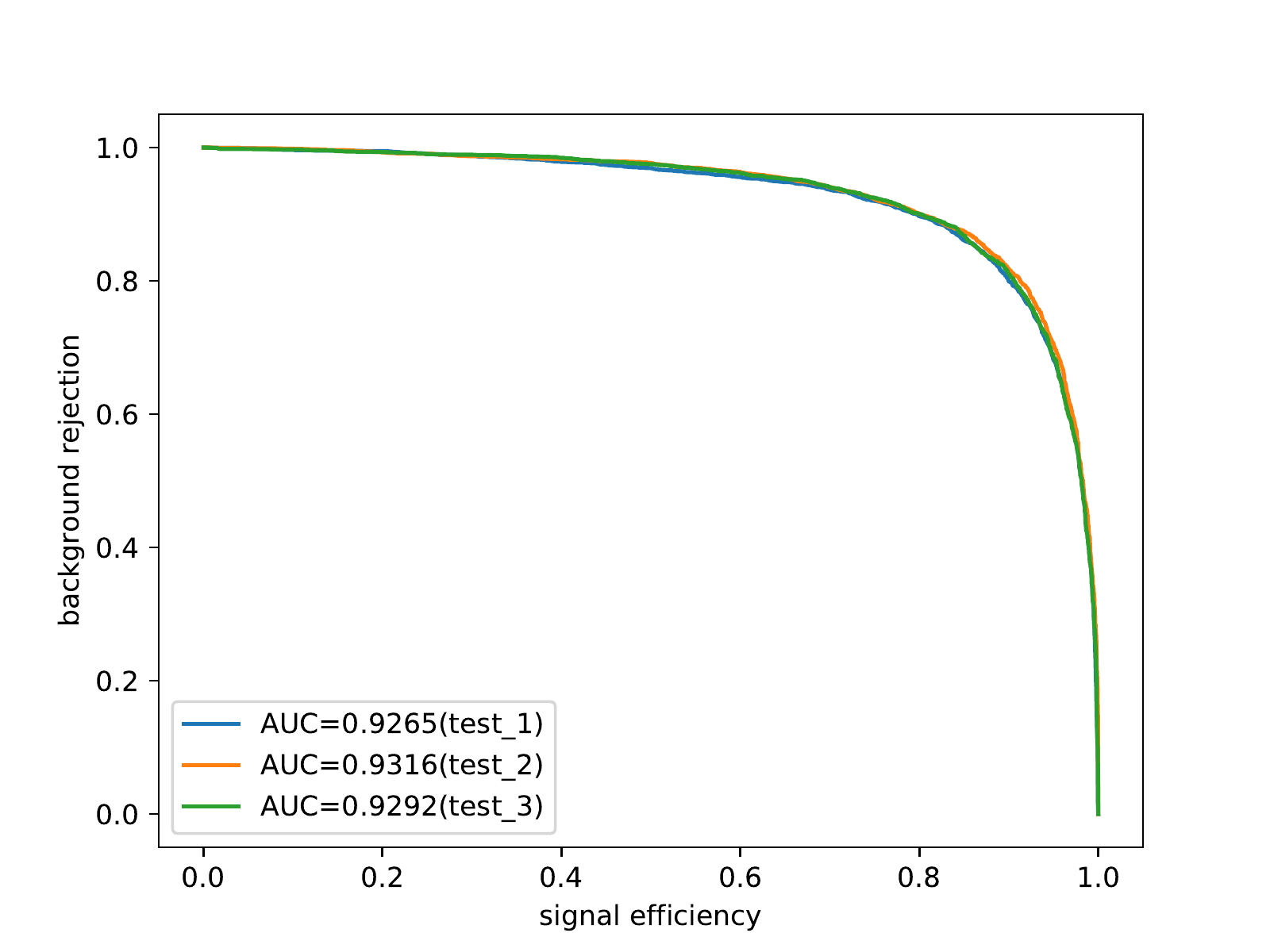}
    }\hspace{-0.072\linewidth}
\caption {The $k$-fold cross validation plots for each \NN. The left column shows the ROC curves for the training dataset, the right column for the test dataset.}
\label{fig:kfold}
\end{figure}

A final check to ensure overtraining has not occurred is to perform $k$-fold cross validation on training data. In this test, the data has been split into four separate batches for the \chljets case and three separate batches for the \chll case. One batch is used as a validation or test set, the other batches are used for training the \NN. Each batch has a turn to be the test set. All ROC curves should be comparable with each other and only deviate slightly from the initial ROC curve (due to statistical fluctuations).
The $k$-fold cross validation plots are shown in Figure~\ref{fig:kfold} for the \chljets and \chll ~{\NN}s. Very little differences in AUC values and ROC curve shapes indicate that no overtraining occurs.

\FloatBarrier
\subsection{Evaluation}
\label{sec:ELDapplication}

\begin{figure}[!htbp]
\centering
\includegraphics[width=0.34\linewidth]{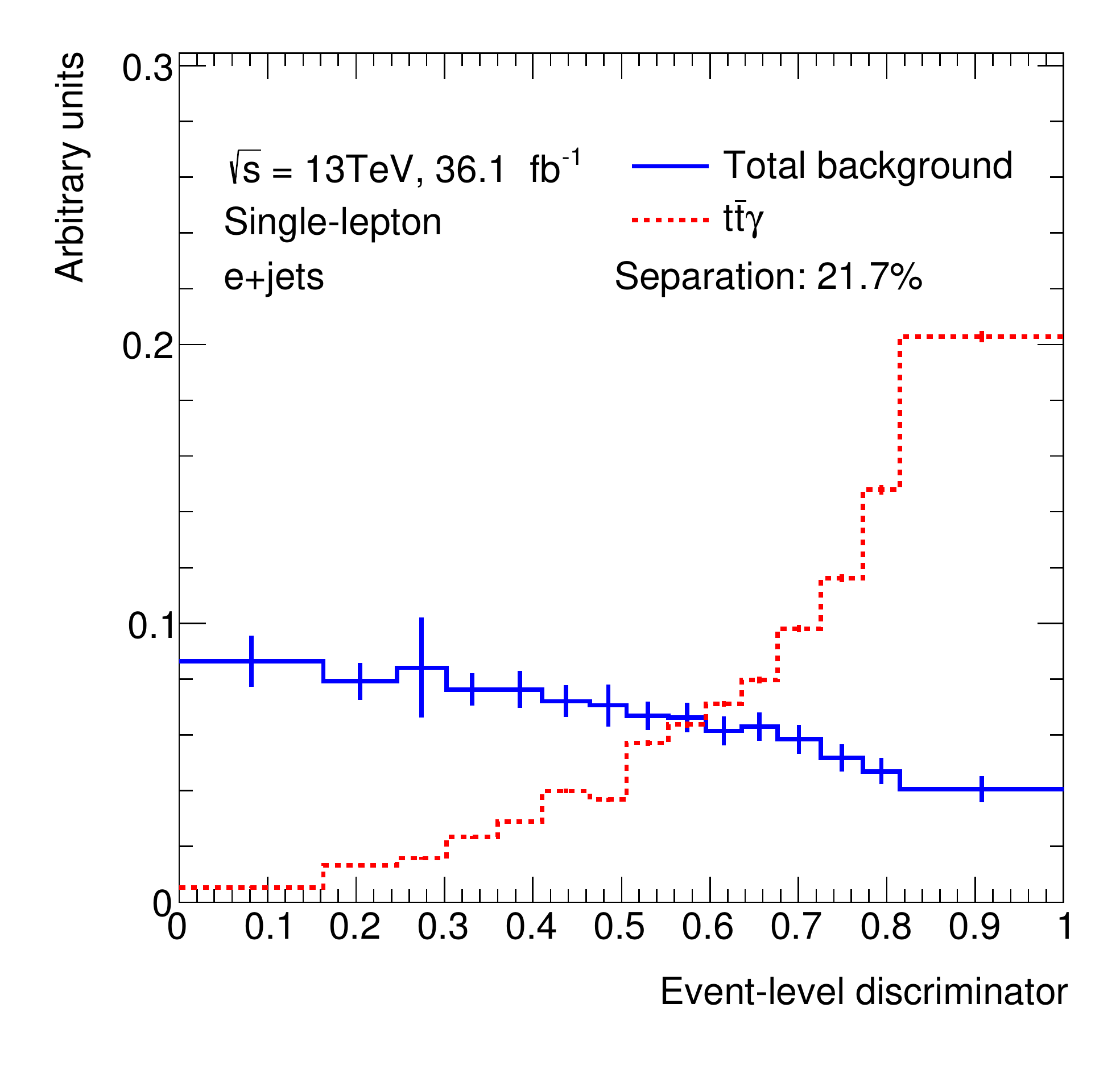}
\hspace{-0.032\linewidth}
\includegraphics[width=0.34\linewidth]{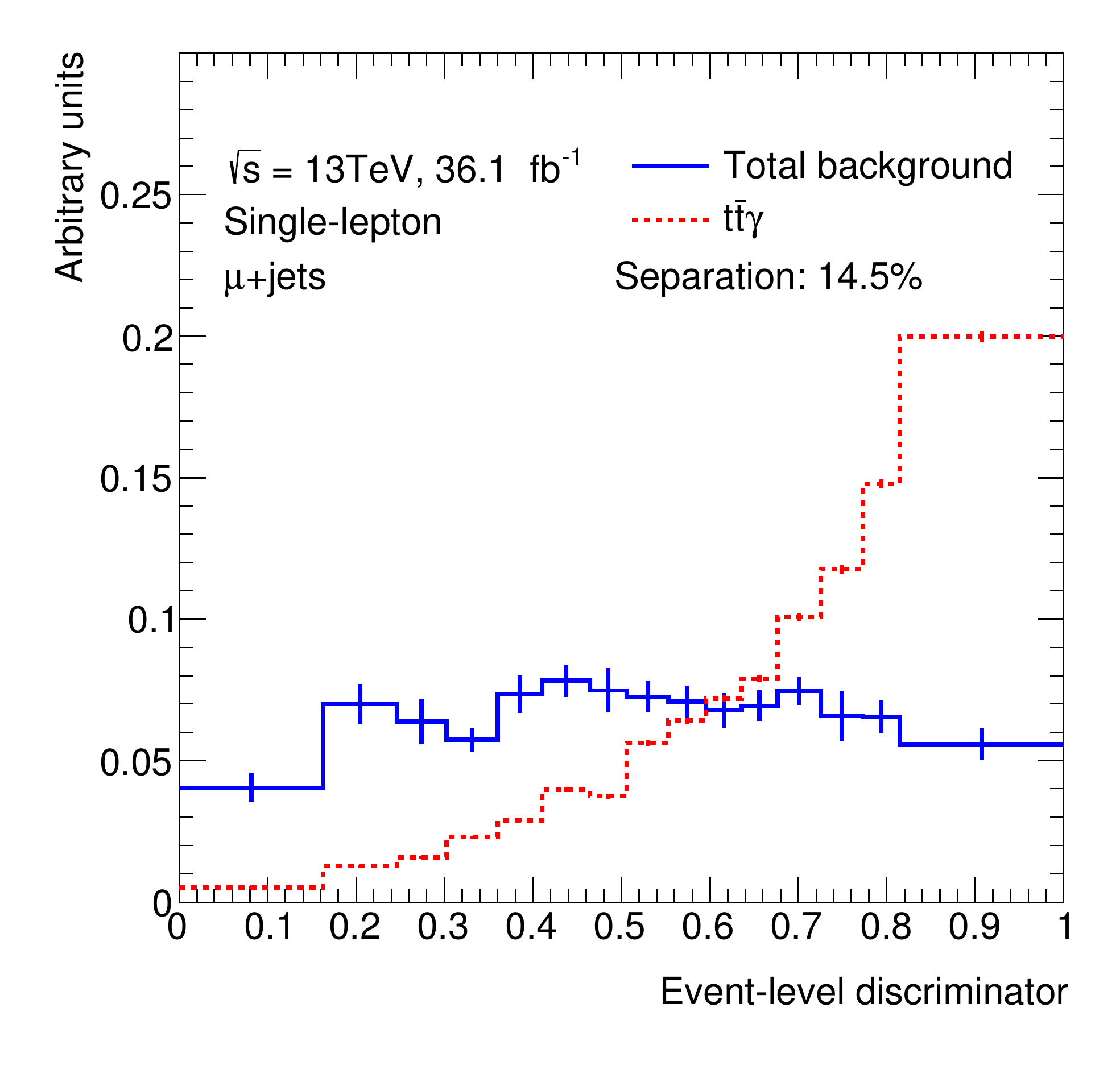}
\caption {Event-level Discriminator variable for the \chljets channels with the corresponding separation plots for all summed backgrounds.}\label{fig:nnoutputsSL}
\end{figure}

\begin{figure}[!htbp]
\centering
\includegraphics[width=0.34\linewidth]{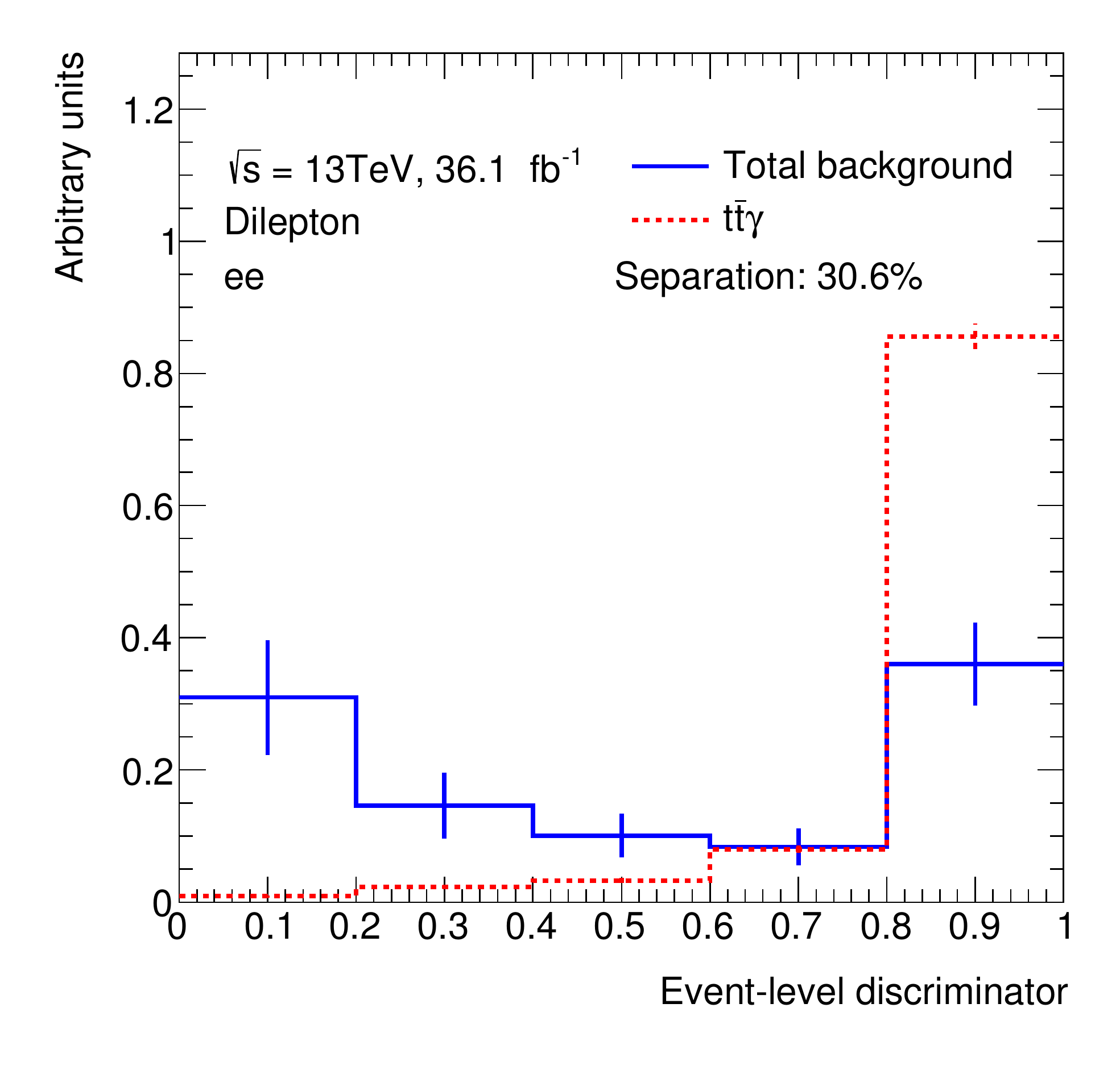} 
\hspace{-0.032\linewidth}
\includegraphics[width=0.34\linewidth]{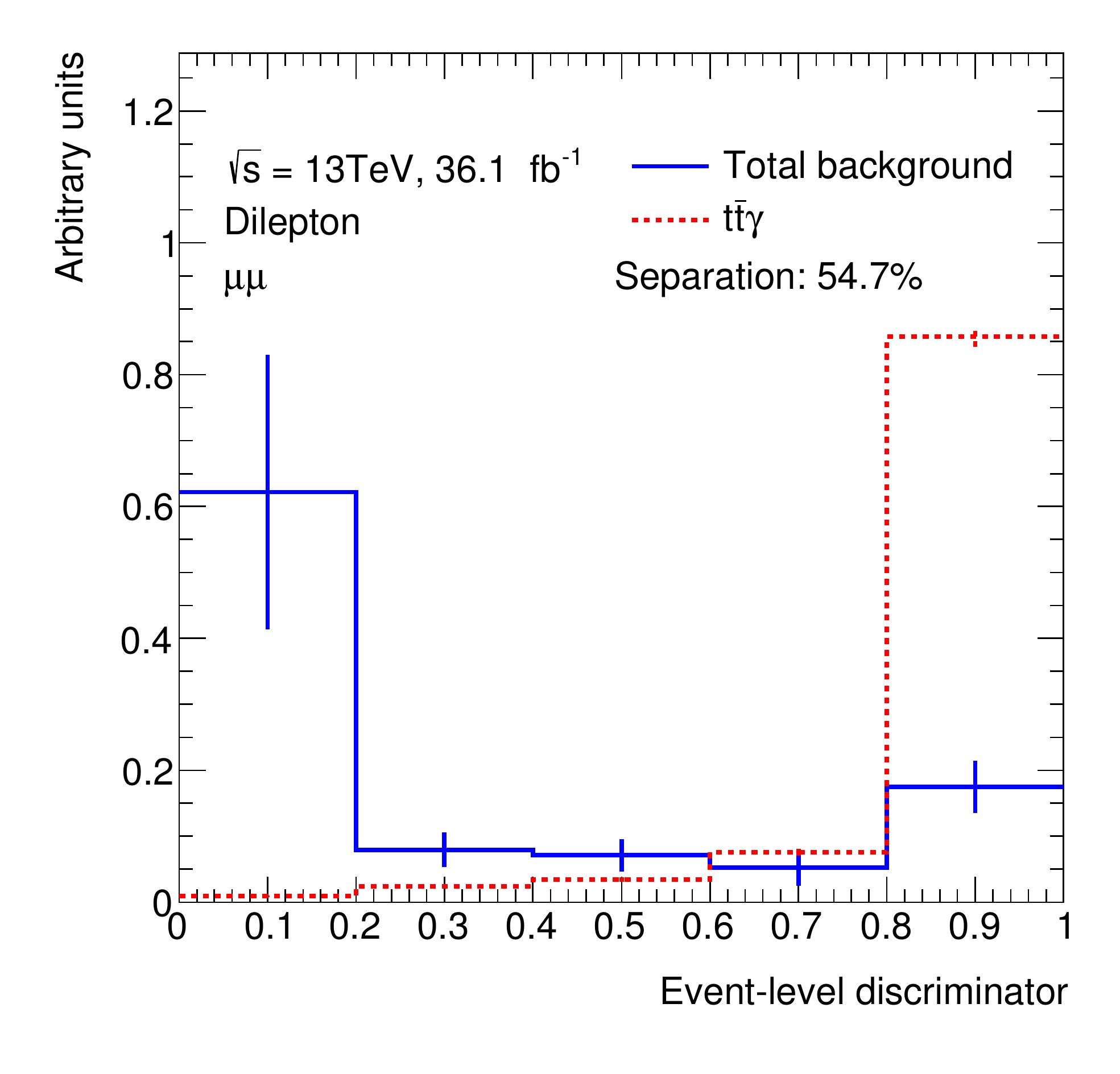} 
\hspace{-0.032\linewidth}
\includegraphics[width=0.34\linewidth]{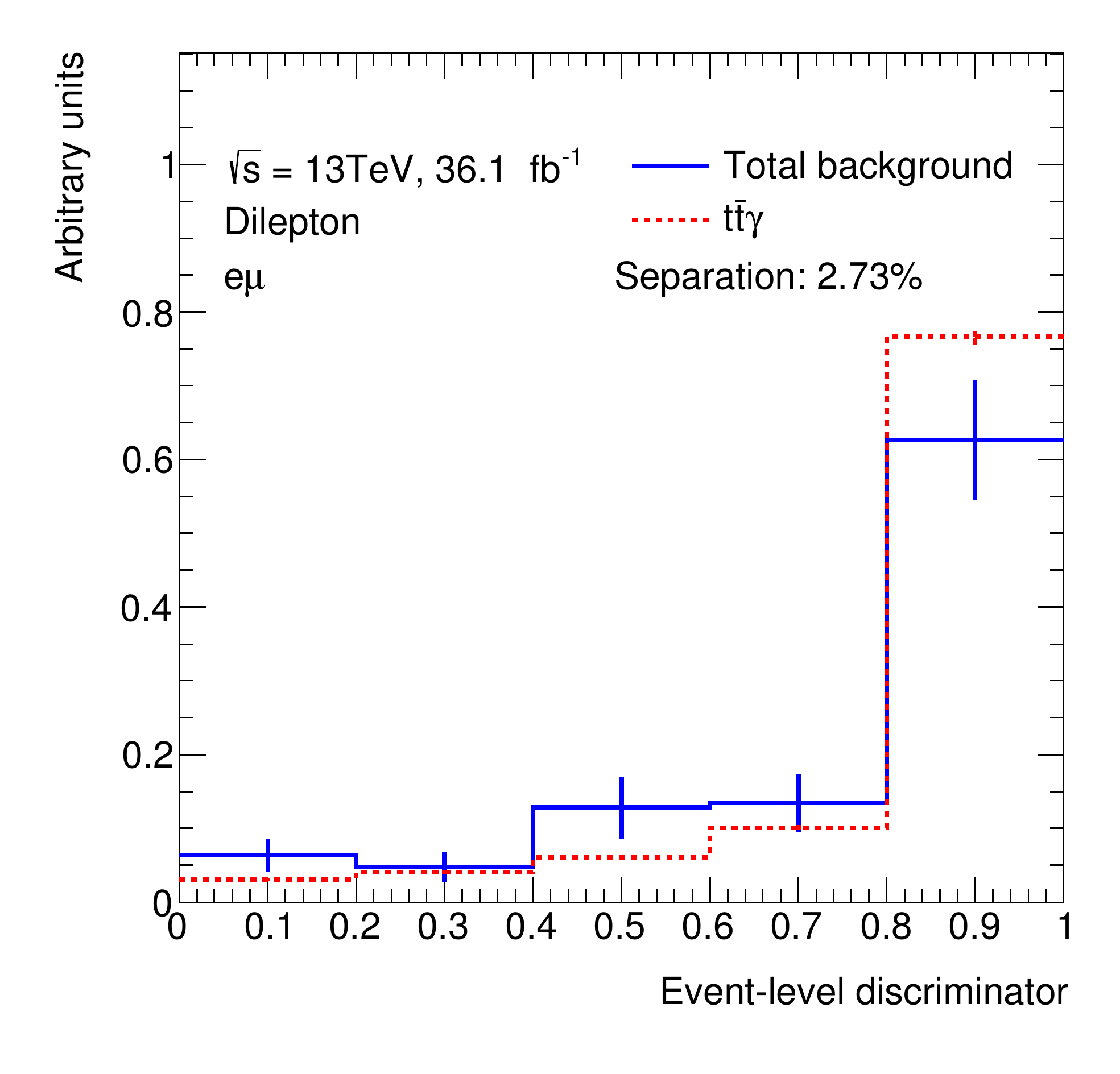} 
\hspace{-0.032\linewidth}
\caption {Event-level Discriminator variable for the \chll channels with the corresponding separation plots for all summed backgrounds.}\label{fig:nnoutputsDilep}
\end{figure}

This section looks into the performance and evaluation of the \ELD for each of the backgrounds. All signal and background are considered with all cuts from Chapter~\ref{sec:eventselections} applied.
The output from the neural networks for each channel is shown in normalised Figures~\ref{fig:nnoutputsSL} and \ref{fig:nnoutputsDilep} for the \chljets and \chll channels, respectively. 
For the \chljets channels the backgrounds are individually compared to the signal in Appendix~\ref{sec:ELDapp}.
Only statistical uncertainties are included. The larger error bars on some of the backgrounds are due to the MC generator weights.
In general, the \ELD performs well in all channels.

For the \chljets channels the \hfake and \efake backgrounds have the smallest separation at 12\% and 14\% for the \chejets channel, and 11\% and 8\% for the \chmujets channel, respectively.
For the \chee and \chmumu channels the main background is from \Zgamma, which is the background peak on the left of the \ELD. A very small \hfake contribution makes up the background peak on the right of the \ELD that sits under the signal. In the case of the \chemu channel, the background peak consists of mainly \hfake events (due to the negligible contribution of \Zgamma events in this channel).  As stacked plots will show later, this is a very small contribution.
These distributions are used to perform a binned maximum likelihood fit, as will be described in Chapter~\ref{sec:analysisstrategy}.


\begin{figure}[!htbp]
\centering
\subfloat[Signal]{
\includegraphics[width=0.46\linewidth]{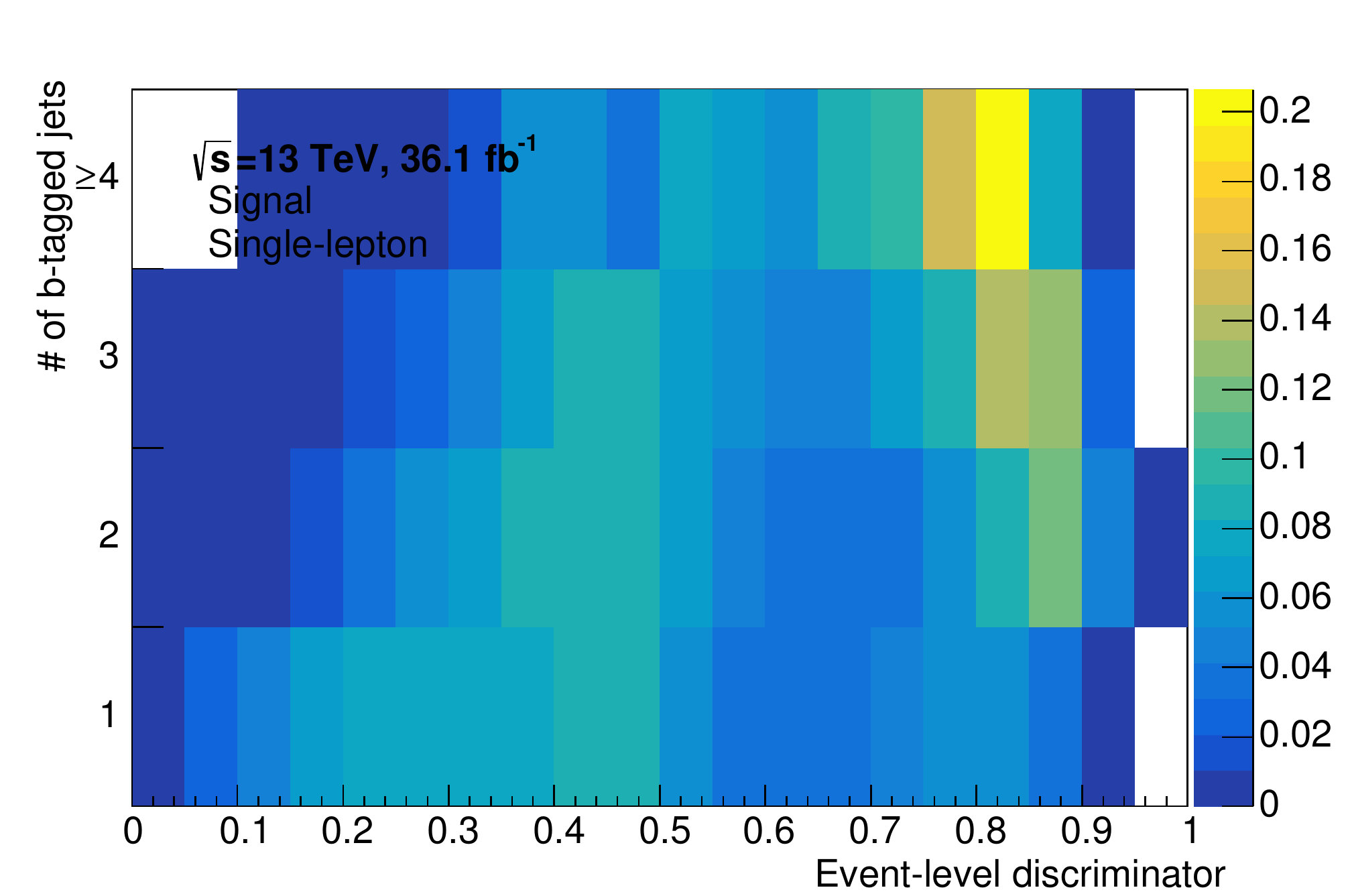}
}
\hspace{-0.03\linewidth}
\subfloat[Background]{
\includegraphics[width=0.46\linewidth]{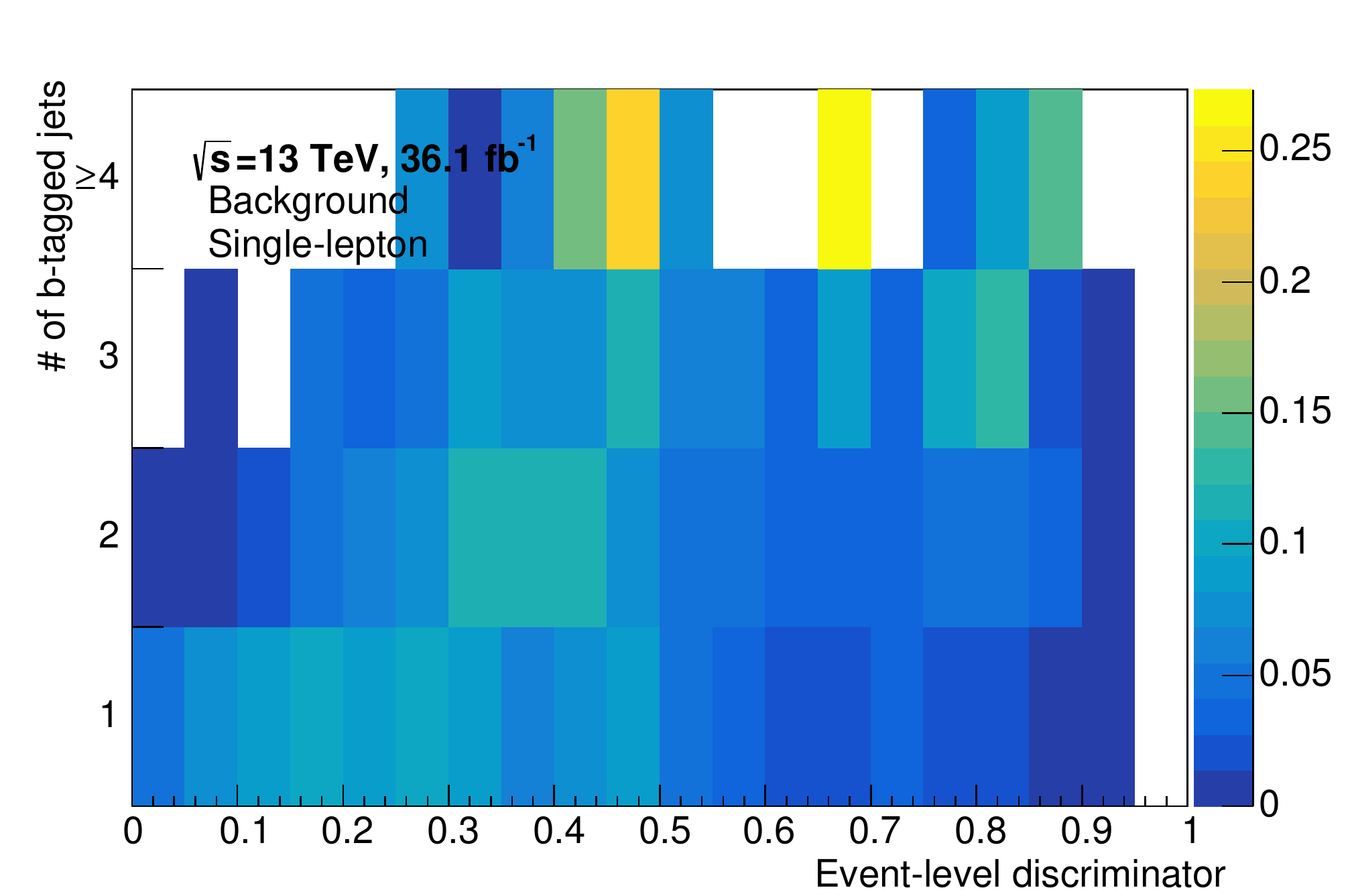}
}
\hspace{-0.03\linewidth}
\caption {The number of $b$-tagged jets versus the \chljets \ELD output for the signal and the background. Each row is normalised separately.}\label{fig:bjetVsELDSL}
\end{figure}

\begin{figure}[!htbp]
\centering
\subfloat[Signal]{
\includegraphics[width=0.46\linewidth]{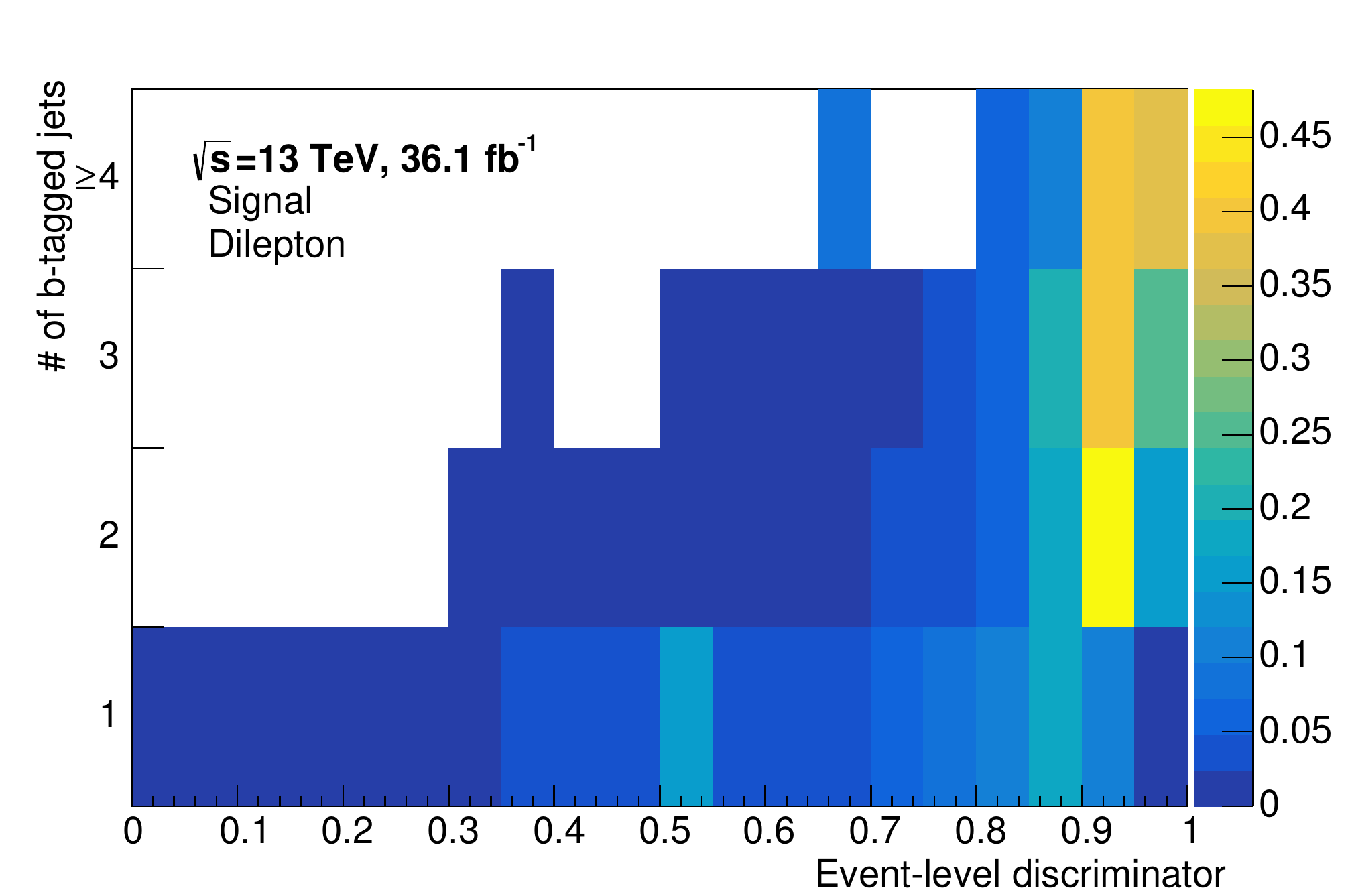}
}
\hspace{-0.03\linewidth}
\subfloat[Background]{
\includegraphics[width=0.46\linewidth]{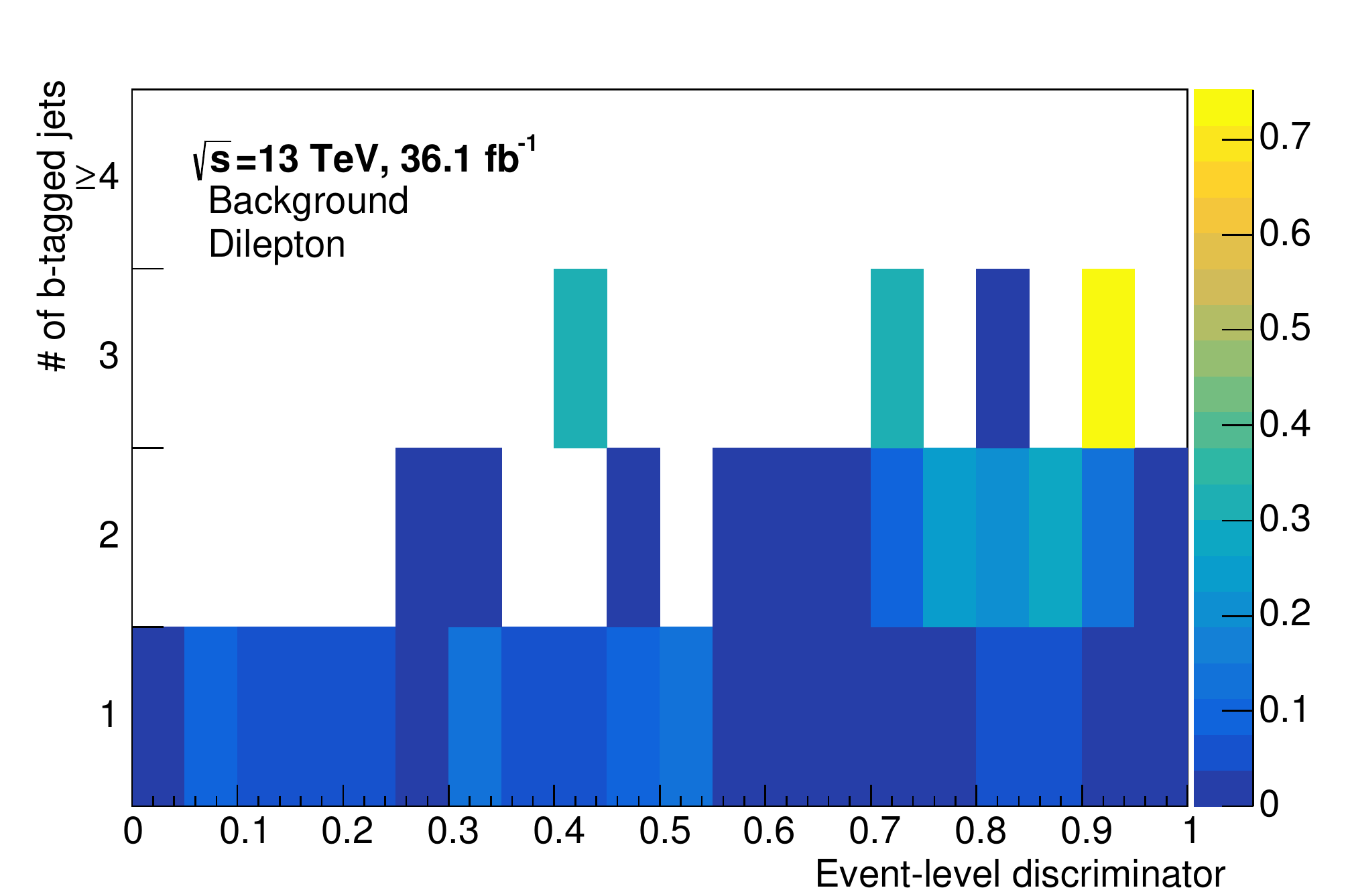}
}
\hspace{-0.03\linewidth}
\caption {The number of $b$-tagged jets versus the \chll \ELD output for the signal and the background. Each row is normalised separately.}\label{fig:bjetVsELDDL}
\end{figure}

Since the $b$-tagging variables play such an import role in the \ELD, we can examine the relationship between the $b$-jet dependence and the \ELD output. Figure~\ref{fig:bjetVsELDSL} and ~\ref{fig:bjetVsELDDL} shows this correlation for the \chljets and \chll channel {\NN}s, respectively, where each row of each plot has been normalised. In general, as more $b$-tagged jets enter into the event, the more signal-like it is. Background processes with a higher number of $b$-tagged jets in the event have an increased chance of being tagged as signal by the \ELD.

\begin{figure}[!htbp]
\centering
\subfloat[Signal]{
    \includegraphics[trim={4cm 0 1cm 0},clip,width=0.46\linewidth]{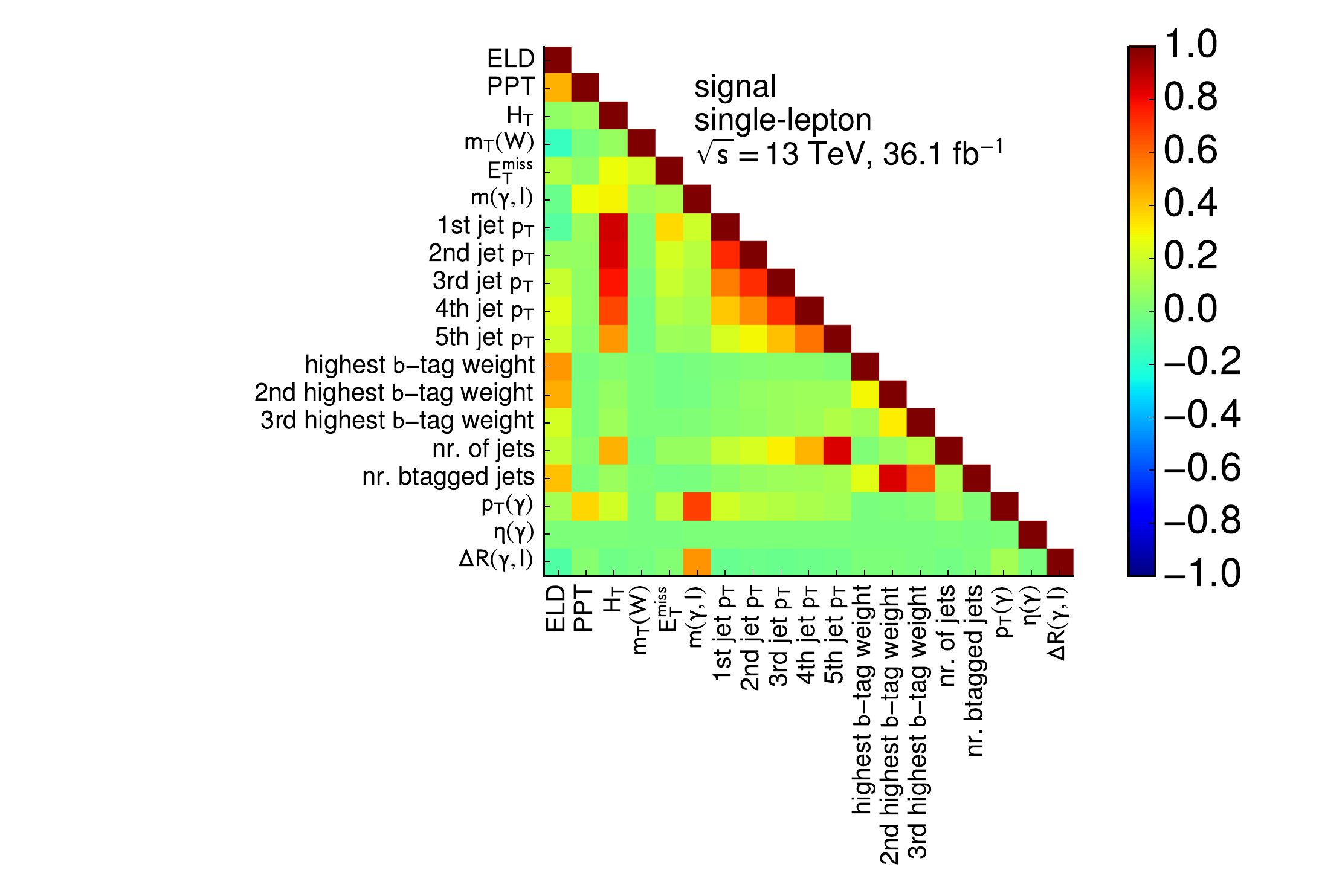}
}\hspace{-0.032\linewidth}
\subfloat[Hadronic fake background]{
    \includegraphics[trim={4cm 0 1cm 0},clip,width=0.46\linewidth]{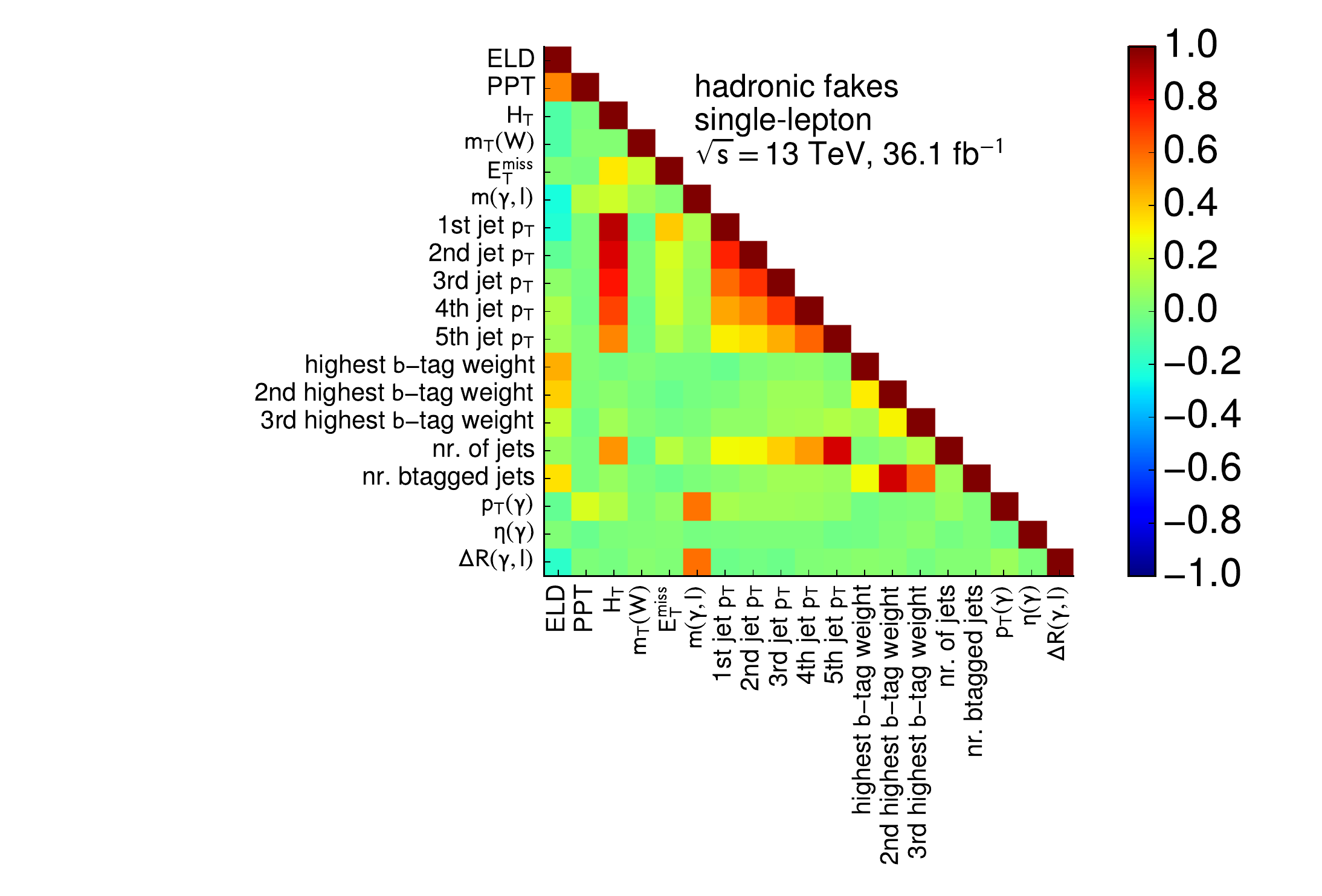}
}\hspace{-0.032\linewidth}

\subfloat[\efake background]{
    \includegraphics[trim={4cm 0 1cm 0},clip,width=0.46\linewidth]{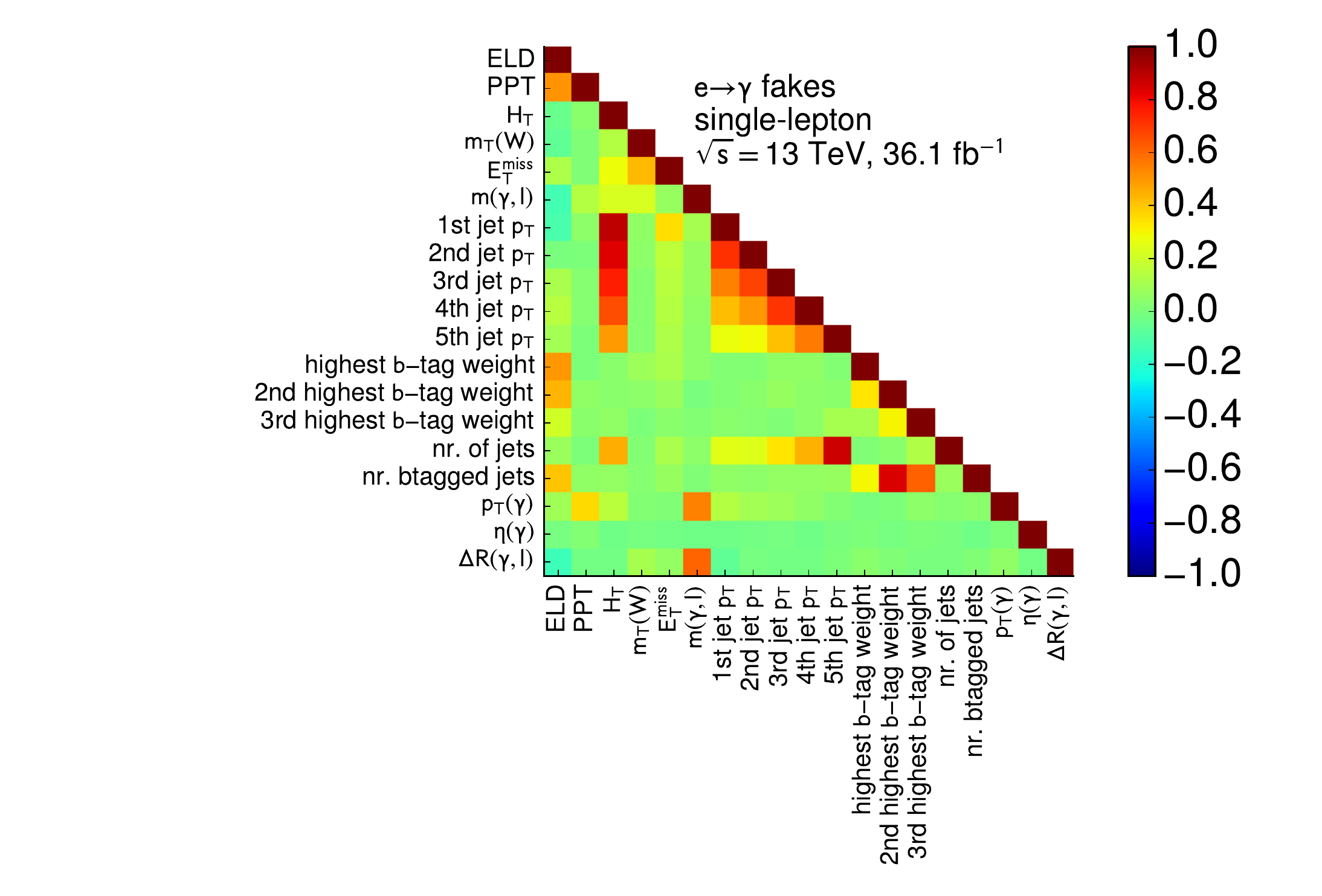}
}\hspace{-0.032\linewidth}
\subfloat[Prompt background]{
    \includegraphics[trim={4cm 0 1cm 0},clip,width=0.46\linewidth]{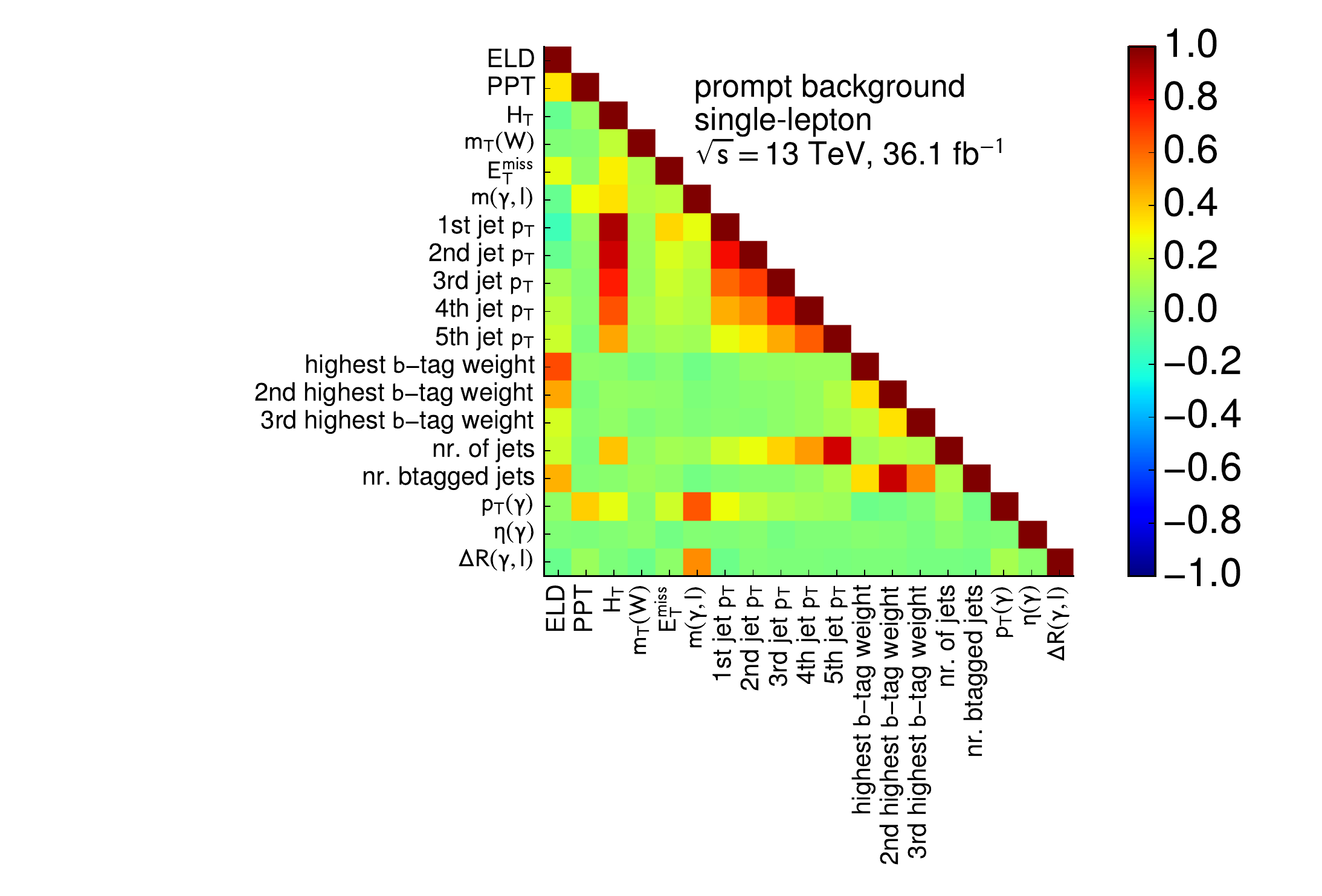}
}\hspace{-0.032\linewidth}

\subfloat[Fake lepton background]{
    \includegraphics[trim={4cm 0 1cm 0},clip,width=0.46\linewidth]{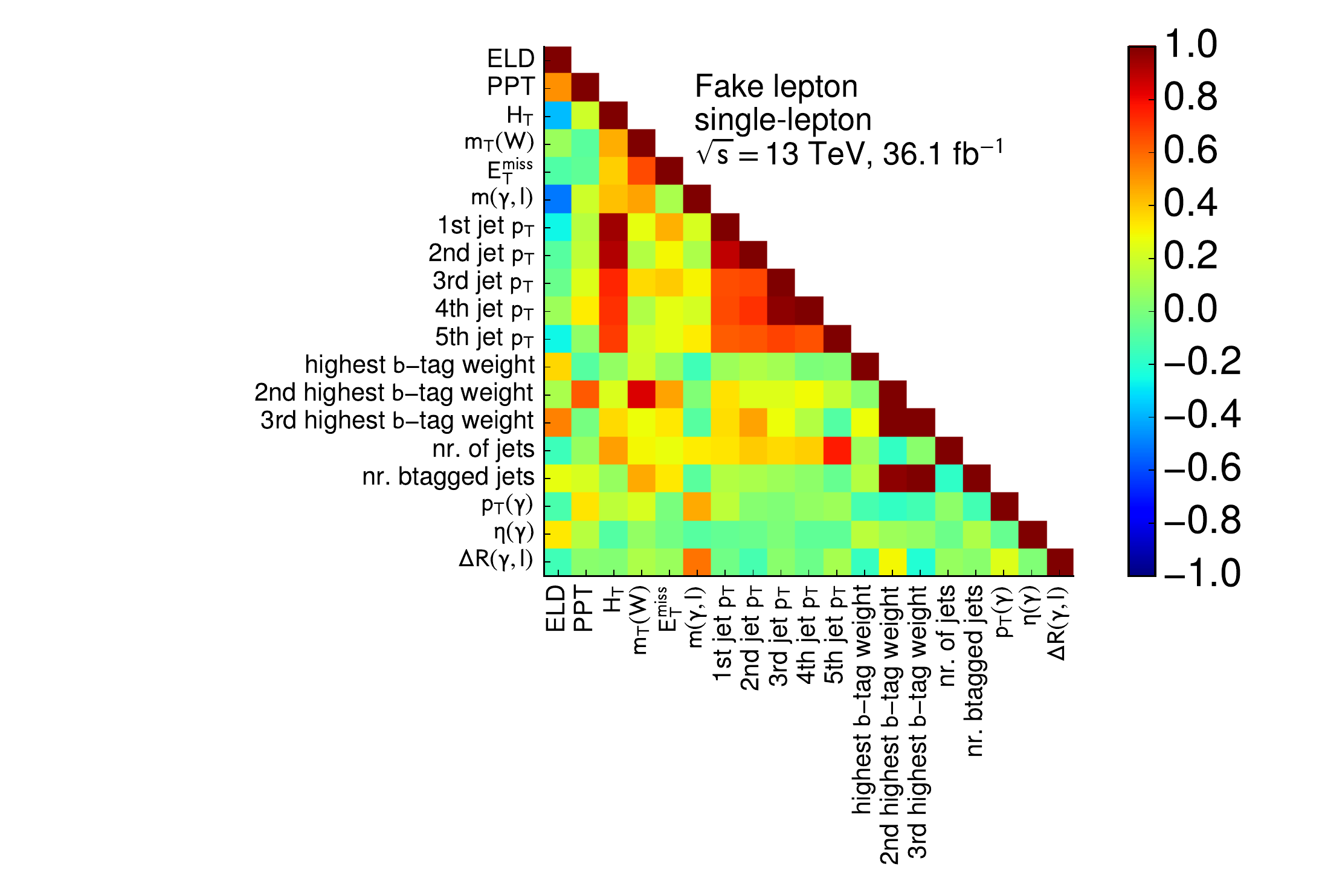}
}\hspace{-0.032\linewidth}
\caption [] {Correlation plots for the signal and separate backgrounds for the \chljets channel comparing the \ELD and training variables. Full event selection and event weights have been applied. The last three variables ($p_{T}(\gamma)$, $\eta(\gamma)$ and $\Delta R(\gamma,l)$) do not enter the training and are used for differential measurements.}\label{fig:eldCorrelationsSL}
\end{figure}

\begin{figure}[!htbp]
\centering
\subfloat[Signal]{
    \includegraphics[trim={4cm 0 1cm 0},clip,width=0.46\linewidth]{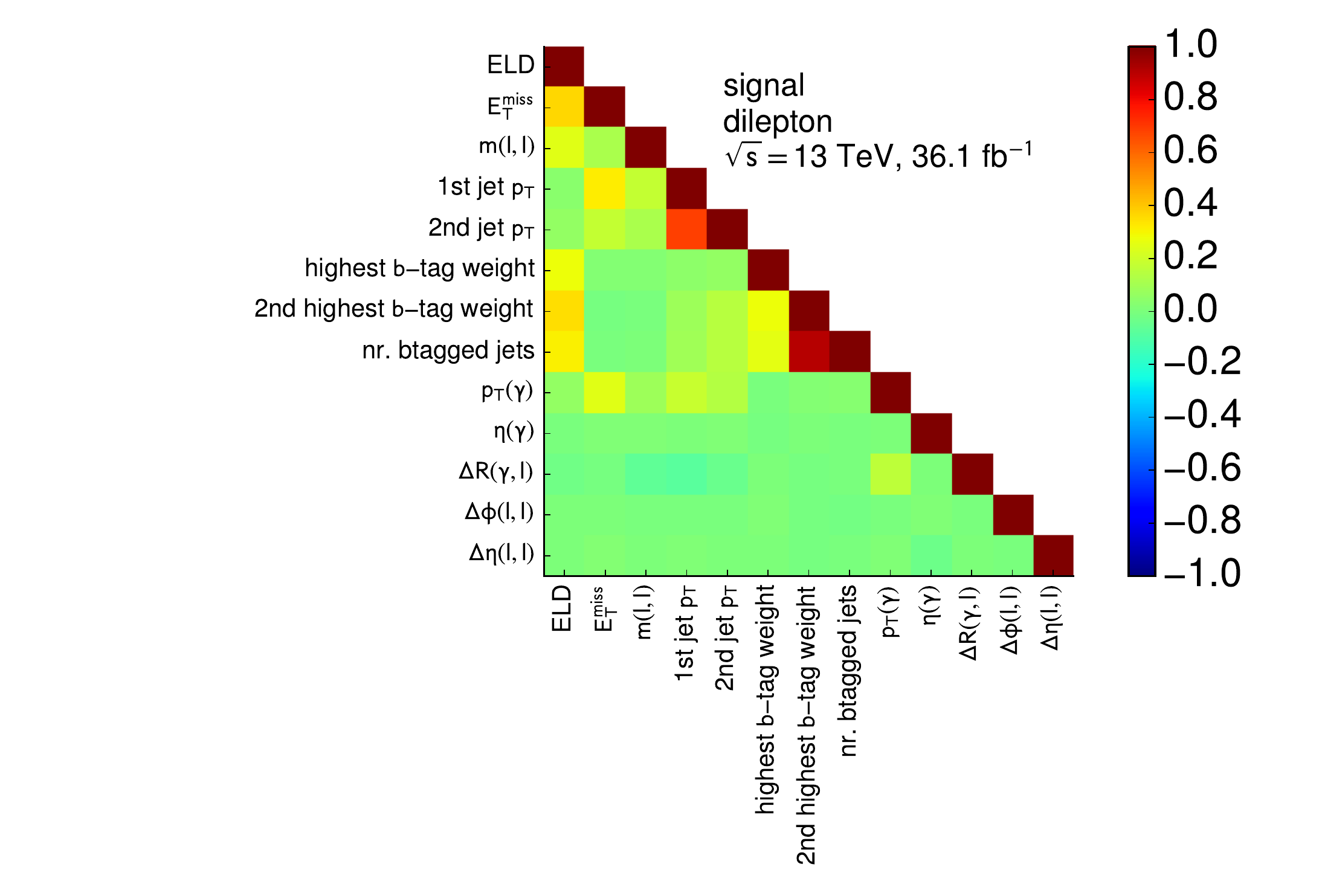}
}\hspace{-0.032\linewidth}
\subfloat[Total background]{
    \includegraphics[trim={4cm 0 1cm 0},clip,width=0.46\linewidth]{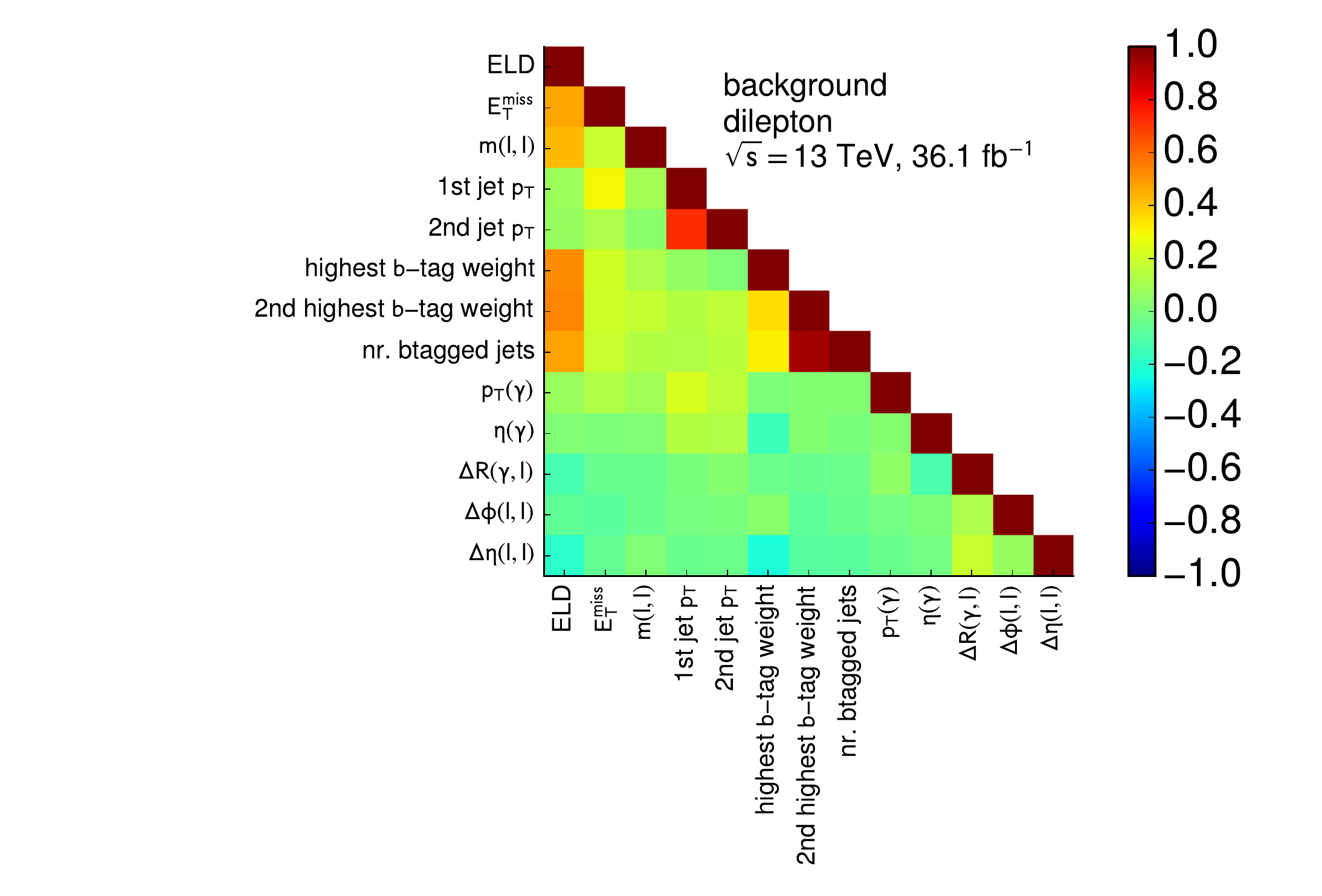}
}\hspace{-0.032\linewidth}
\caption [] {Correlation plots for the signal and sum of backgrounds for the \chll channel comparing the \ELD and training variables. Full event selection and event weights have been applied. The last five variables ($p_{T}(\gamma)$, $\eta(\gamma)$, $\Delta R(\gamma,l)$, $\Delta \phi(l,l)$ and $\Delta \eta(l, l)$) do not enter the training and are used for differential measurements.}\label{fig:eldCorrelationsDL}
\end{figure}

Correlation plots for \ELD output, the training input variables, and the final variables used in the differential measurements (for interest sake) are shown in Figure~\ref{fig:eldCorrelationsSL} and \ref{fig:eldCorrelationsDL} for the \chljets and \chll channels, respectively. 
Full event selections and weights described in the previous chapters have been applied.
The \chljets correlation figures are split into the signal and the different background components. The \chll figures are simply split into signal and the sum of all backgrounds. 

An interesting observation is the correlation between the last three (five) variables used for differential measurements in the \chljets (\chll) \ELD. These variables are not used in the training and so any correlation to the \ELD that is seen has been learned by the \NN. It shows that the two \NN{}s learned very little about photon or lepton related quantities, and thus are potentially missing out on crucial event information. This is not too surprising given that very little photon or lepton kinematic information is provided as input. Future analyses can study the impact of introducing more low-level kinematic variables of the photon and leptons.

The first column essentially shows how much the \ELD learns from each variable. Intuitively, the variables that have the highest separation (as shown in Table~\ref{tab:event_MVA_variables_SL}) should be amongst the most highly correlated to the \ELD, however, this does not always apply. 
Also worthwhile to consider is that the separation values in Table~\ref{tab:event_MVA_variables_SL} show signal against the sum of all backgrounds.
For the \chljets channel, by separating out the different backgrounds we can see that some variables are more important than what Table~\ref{tab:event_MVA_variables_SL} leads us to believe. For instance, the invariant mass of the photon and the lepton have a very high anti-correlation for the \QCD background in the \chljets channel. Similarly, \HT has a high anti-correlation. In separating out the \Other background we see that the \MET of an event plays a fairly important role. This is because the \MET spectrum is softer for events coming from the $V\gamma$, single top, diboson, and \ttV processes.
As expected some of the highest correlations to the \ELD{}s are consistent with respect to the $b$-tagging variables in both signal and backgrounds. 

The second column for the \chljets figures shows the correlations between all the variables and the \PPT. The general trend is that the \PPT is minimally correlated with the rest of the variables. This is expected since it was trained on shower shape variables, which are for the most part analysis independent. An exception occurs for the transverse momentum of the photon, for which shower shape variables are known to have some dependence. Even then, the correlation is never greater than about 30\%.


\chapter{Systematic uncertainties}
\label{sec:systematics}

Many sources of uncertainties need to be considered for any analysis in HEP.
They fall under two classes: statistical and systematic. Statistical uncertainties are related to the amount of data collected or MC available. Thus more data or MC events means lower statistical errors, which is needed for precision measurements. 
Systematic errors are essentially due to limitations when building a detector, or our lack of full understanding for a given physics object/algorithm or detector component. Lack of theoretical knowledge also contributes to systematic uncertainties.
Within \ATLAS some systematic uncertainties are provided as ``recommendations" due to studies carried out by other teams, for example luminosity or Jet Energy Resolution (JER). There are also many systematics that arise from analysis specific techniques, such as those when training a NN or deriving a data-driven background.

Systematic uncertainties in template fits have two components; a shape and an overall normalisation. A shape component can change the shape of the distribution by assigning more or less events in various bins, but the overall integral must remain the same. A normalisation component shifts the integral up or down by some constant amount. A systematic can have either shape, normalisation, or both. A concrete example will be shown in Section~\ref{sec:smoothing}.

This section introduces all the systematics considered for this analysis.
The modelling uncertainties of the signal and backgrounds are shown in Section~\ref{sec:theouncert}. The experimental uncertainties, which are common to both signal and backgrounds are explained in Section~\ref{sec:expsystematics}. 
Finally, a discussion on how systematics can be smoothed, symmetrised, and pruned for the final fit is presented in Section~\ref{sec:smoothing}.

\section{Signal and background modelling uncertainties}\label{sec:theouncert}

In the following systematic sources, references to \hfake and \efake modelling refer to the \ttbar components that make up these backgrounds, since this is the dominant process.


\subsection{QCD renormalisation and factorisation scales}

The QCD renormalisation and factorisation scale uncertainties are applied with a normalisation and shape component to the signal and \hfake processes.
For \efake, \Wgamma and \Zgamma background processes (for the respective channel) only the shape is considered.
The effect of the choice of the renormalisation ($\mu_{r}$) and factorisation ($\mu_{f}$) scales is estimated by varying $\mu_{r}$ and $\mu_{f}$ simultaneously or independently up and down by a factor of two with respect to the nominal sample value, thus resulting in three sets of up-and-down variations.
This implies changing the dependence of the finite order perturbation calculations in QCD.
The final uncertainty on the normalisation is the quadratic sum of the three sets of variations, while for the shape the maximum variation is chosen.

\subsection{Parton Shower}
The parton shower uncertainty is derived and applied to the signal process as well as the \hfake, \efake and where applicable, the \Zgamma backgrounds.

For \ttgamma, the uncertainty is estimated by comparing the nominal sample (\madgraph + \pythia{8}) to the same \madgraph events showered with \herwig{7}.
For \hfake and \efake backgrounds the uncertainty is estimated by comparing the nominal sample (\powhegAll+\pythia{8}) to a \ttbar sample produced by \sherpaAll.
The \Zgamma shape uncertainty in the \chll channels is estimated by comparing the nominal sample (\sherpaAll) to events generated with \madgraph + \pythia{8}.

\subsection{Initial and final state radiation}
The initial and final state QCD radiation systematics (ISR/FSR) are applied to the signal process as well as the \hfake and \efake background processes.
The uncertainty is estimated by comparing the nominal sample to samples produced with variations of the \Afourteen tune.

\subsection{PDF uncertainty}
The PDF systematics uncertainties are only applied to the signal process.
The shape and normalisation aspect is evaluated by the envelope error of the 100 PDF error sets in the \NNPDFLO PDF set, which are stored as weights in the nominal sample.
For the shape component this is inconvenient as it impacts the speed of the fit. Thus, separate studies were done to see the impact on the signal strength and its sensitivity. If this shape affect is negligible, it can be left out of the final fit. 
The impact is calculated by performing 100 fits (a fit for each variation) and computing the standard deviation with the respect to the central value of $\mu$. The errors relative to $\mu$ for each channel are shown in Table~\ref{tab:PDFcontrib}. 
Tests for an ideal pruning value were performed and indicate all of these values are well below the chosen value of 0.7\% (Section~\ref{sec:smoothing}). 

 \begin{table}[h!]
 \centering
 \scalebox{0.80}{
\begin{tabular}{l|c}
\hline
Channel & Relative error on $\mu$ [\%] \\ \hline
\hline
\chejets & 0.066 \\ 
\chmujets & 0.066 \\ 
\chee & 0.137 \\
\chmumu & 0.109  \\ 
\chemu & 0.027  \\ \hline
\chljets & 0.066 \\ 
\chll & 0.200  \\ \hline
 \end{tabular}
 }
  \caption{The relative uncertainty on $\mu$ for the 100 PDF shape uncertainty fits.}
 \label{tab:PDFcontrib}
 \end{table}

\subsection{Other background modelling uncertainties}

Further sources of systematic uncertainties for the \hfake and \efake backgrounds have already been presented in Chapter~\ref{sec:efake} and \ref{sec:hfake}. These result from the tag-and-probe and ABCD methods.

For non-floating MC backgrounds without a modelling systematic error (single top, diboson and \ttV backgrounds) a flat uncertainty of 50\% is placed on the cross section. As a cross-check, this uncertainty is doubled and leads to negligible changes. This is shown in  Appendix~\ref{sec:promptSystCheck}. 
In the \chljets channels the \Zgamma background is also assigned a 50\% normalisation uncertainty, similarly the same is done for the \Wgamma background in the \chll channels.

The estimation for the \QCD background systematics is presented in Section~\ref{sec:fakelepton}. An envelope of the different parameterisations used in the matrix method is calculated, with the outlying parameterisations taken as the up/down variations. This source of uncertainty has both shape and normalisation components. This uncertainty is large and so no other contributions are considered.

\section{Experimental uncertainties}\label{sec:expsystematics}

A large source of experimental systematic uncertainties arise from algorithms that simulate and reconstruct leptons, photons, jets and \met in the \ATLAS detector. Different pileup profiles also need to be simulated and therefore have an associated error.
Another source of error associated  with the LHC and detectors is the integrated luminosity (used in the normalisation of the MC).
The above mentioned sources can change the overall shape and normalisation of the MC, and thus this needs to be taken into account.
For data-driven corrected backgrounds the normalisations are already corrected but the shapes of the distributions could still be biased. Thus, only shape components are considered. Similarly for when the \Wgamma background is a free parameter in the final fit, the overall normalisation is adjusted. Thus, only shape systematics are considered for the \Wgamma background in the \chljets channels.

\subsubsection{Leptons and photons} 
Leptons (electrons and muons) and photons are corrected with \ET/\pt and \eta data-driven scale factors from identification and isolation efficiency measurements discussed in Chapter~\ref{sec:egamma} and \ref{sec:muons}. 
Corresponding efficiencies are derived, for which a few examples include: varying the amount of material in front of the calorimeters, studying the effects of different generators, varying the background contamination or background subtraction in the fits, varying the fraction of converted photons in data and simulation (for photon related systematics) and varying the isolation cone sizes~\cite{ATL-PHYS-PUB-2016-014,ATLAS-CONF-2016-024,PERF-2015-10}.

\subsubsection{Prompt photon tagger}

The systematic uncertainties associated with the \PPT are discussed in Chapter~\ref{sec:PPTsysts}. They are derived separately for prompt and \hfake photons in the form of scale factors, which are then turned on and off.
One source is assigned to the \efake background (due to similarity in shape to the prompt photons), a source is assigned to each of the prompt contributions (both signal and background and correlated in the fit), and two sources enter for the \hfake background. All \PPT systematic variations have only a shape component.

\subsubsection{Jets}

Systematic contributions assigned to jets include: pileup, flavour composition (quarks and gluons), $\eta$-calibrations, calorimeter responses to different jet flavours, punch-through corrections and single-particles (for high \pt uncertainties). These are derived from simulation, test beam data and \emph{in-situ} measurements~\cite{PERF-2011-03,PERF-2011-05,PERF-2012-01,PERF-2016-04}.
An important contribution to the JES uncertainty scheme is due to pileup corrections. Specifically, the uncertainty of the per-event \pt density modelling in the $\eta$-$\phi$ plane for MC simulations. This is termed \emph{Pileup RhoTopology}.
The JVT systematic contribution is derived by varying the cut on the multivariate output.

\subsubsection{$b$-tagging}
The numerous uncertainties on $b$-tagging come from separate data-driven methods applied to different jet flavours: $b$-jets, $c$-jets and $light$-jets, which have 30, 15 and 80 NPs, respectively~\cite{PERF-2012-04}.

\subsubsection{Missing Transverse Energy}
The \met is calculated from all objects in an event according to Equation~\ref{eq:met}. Thus, the associated uncertainties of each object are propagated through. Additionally, the scale and resolution of the soft terms are considered.

\subsubsection{Pileup}
Events are re-weighted in MC to match the number of interactions per bunch crossing in data (Chapter~\ref{sec:data}). Systematic uncertainties for pileup are evaluated by scaling these distributions up and down.

\subsubsection{Luminosity}
The integrated luminosity of the 2015/2016 data at 13~\TeV is measured to a precision of 2.1\%.
This uncertainty is used to normalise the MC signal and background samples, and as such the uncertainty is applied to these MC samples. The derivation of this uncertainty has been derived using $x$-$y$ beam separation scans using a similar methodology described in~\cite{DAPR-2013-01}.

Table~\ref{tab:systOverview} details all the systematics and the number of components considered for this analysis.

 \begin{table}[!htbp]
 \centering
 \scalebox{0.9}{
\begin{tabular}{lccp{5cm}}
\toprule
 Systematic name & Type & Components  & Notes \\ \hline
 \hline

\textbf{Signal modelling} & & \\ 
ISR/FSR & SN & 1  \\
\hdashline
Parton Shower & SN & 1  \\
\hdashline
QCD Scales & SN & 1 \\ 
\hdashline
PDF& SN & 1 \\
\hline
\hline

\textbf{Background modelling} & & \\

ISR/FSR & SN*, S** & 2 & *(\hfake), **(\efake) \\
\hdashline
Parton Shower & SN*, S** & 3 & *(\hfake), **(\efake, \Zgamma) \\
\hdashline
QCD Scales & S & 4 & \hfake, \efake, \Wgamma, \Zgamma \\ 
\hdashline

\xsec normalisation & N & 4 & 50\% for VV, ST, \ttV, \Zgamma,  \Wgamma in \chll \\
\hdashline

\QCD & SN &1 \\ 
\hdashline
\efake & SN & 11 \\

\hdashline
\hfake & SN& 25 &  \\ 
\hline
\hline

\textbf{Prompt Photon Tagger} & & \\
Prompt photons & S & 1 & Split into \ttgamma, \Wgamma and \Other contributions and correlated in fit \\ 
\hdashline
\efake & S &1 \\ 
\hdashline
\hfake & S &1 \\ 
\hdashline
\hfake isolation & S &1 \\ 
\hline
\hline

\textbf{Object reconstruction} & &  &All, except data-corrected backgrounds receive only S \\
Electrons (trigger, reco, ID, isolation) & SN & 5 \\
\hdashline
Egamma (resolution, scale) & SN & 2\\
\hdashline
Muons (trigger, reco, ID, isolation) & SN & 15 \\
\hdashline
\MET (resolution, scale) & SN & 3 \\ 
\hdashline
Photons (efficiency, isolation) & SN & 3 \\ 
\hdashline

Jet energy scale & SN & 21 \\
\hdashline

Jet energy resolution &SN& 1 \\
\hdashline
Jet vertex tagger & SN & 1 \\
\hdashline

b-tagging efficiency  & SN & 30 \\ 
\hdashline
c-tagging efficiency  & SN & 15 \\ 
\hdashline
Light-jet tagging efficiency  & SN & 80 \\ 
\hdashline
b-tagging extrapolation  & SN & 1 \\ 
\hline
\hline

\textbf{Miscellaneous} & & \\
Luminosity & N & 1 & All except data-corrected backgrounds   \\ 
\hdashline
Pileup re-weighting & SN & 1  & All, except data-corrected backgrounds receive only S\\
\bottomrule
 \end{tabular}
 }
  \caption{Complete list of systematics considered. A ``N" indicates that only normalisation has been considered as a systematic. A ``S" indicates that only the shape is considered as a systematic. ``SN" indicates that both shape and normalisation have been considered as a systematic.}
 \label{tab:systOverview}
 \end{table}

\section{Smoothing, symmetrisation and pruning}\label{sec:smoothing}

\emph{Smoothing} and \emph{symmetrisation} are methods in which statistical fluctuations in various systematic sources are minimised. In the case of symmetrisation the systematic is centred around a mean value. \emph{Pruning} occurs when a systematic contribution is below some threshold in the fit and removed. This serves to remove a potential cause of instabilities and speeds up the fit.

\subsubsection{Smoothing and symmetrisation}
\emph{Two-sided} symmetrisation is performed when an up and a down variation is provided for any given systematic. The difference between the two variations is calculated then divided by the mean of the variations. This value is then taken as positive (for up) and negative (for down). This is described by Equation~\ref{eq:symmetry}.

 \begin{equation}\label{eq:symmetry}
 \text{Variation up/down} = \pm \left\lvert \frac{\text{up}-\text{down}}{(\text{up}+\text{down})/2}\right\rvert
 \end{equation}

Thus, the $\pm1\sigma$ variation is centred around the nominal value. It is important to notice that if the up/down variations are symmetric then applying this procedure has no effect.
\emph{One-sided} symmetrisation is when only an up or down variation is provided. An example is the \PPT systematics. In this case the variation is simply mirrored to reflect the supplementary variation.
In general, experimental systematic sources are symmetrised, while signal and background modelling contributions are not.

Smoothing algorithms average statistics across bins. This prevents large statistical spikes in many of the systematic uncertainties that are expected to give small contributions. In certain cases such as for signal, \ttbar and \Zgamma modelling, this option is turned off to allow for maximum shape variations\footnote{For the \chll channels, the \ttbar modelling is smoothed due to very small background contributions and large statistical fluctuations.}.
The smoothing algorithms make use of bin and neighbouring bin information such as integrals, statistical uncertainties and derivative sign changes. The smoothing does not change overall normalisations.

\begin{figure}[!htbp]
\centering
\subfloat[\label{fig:redblueExampleA} Shape and normalisation]{
\includegraphics[width=0.49\linewidth]{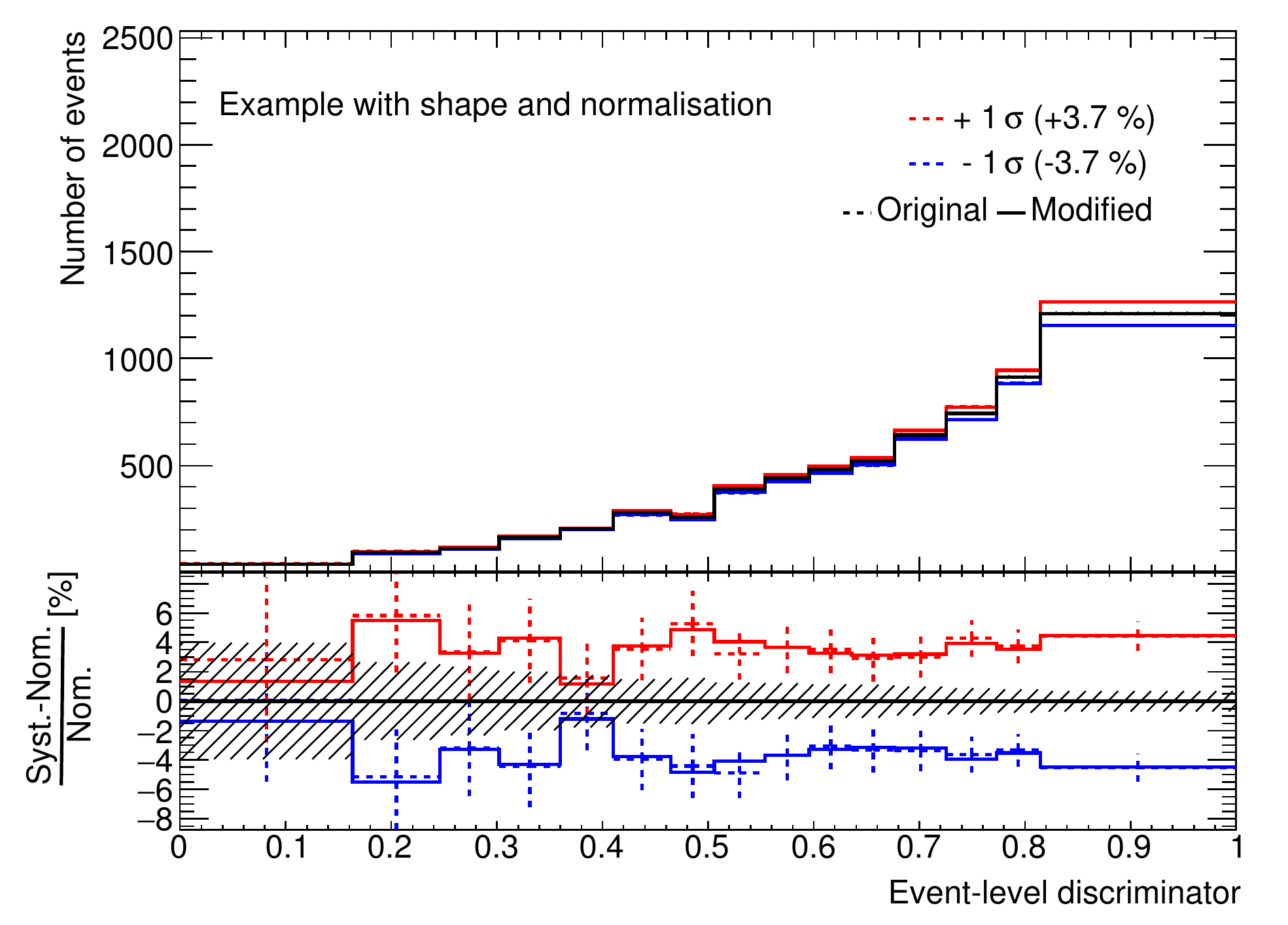}
}\hspace{-0.038\linewidth}
\subfloat[\label{fig:redblueExampleB}Shape only]{
\includegraphics[width=0.49\linewidth]{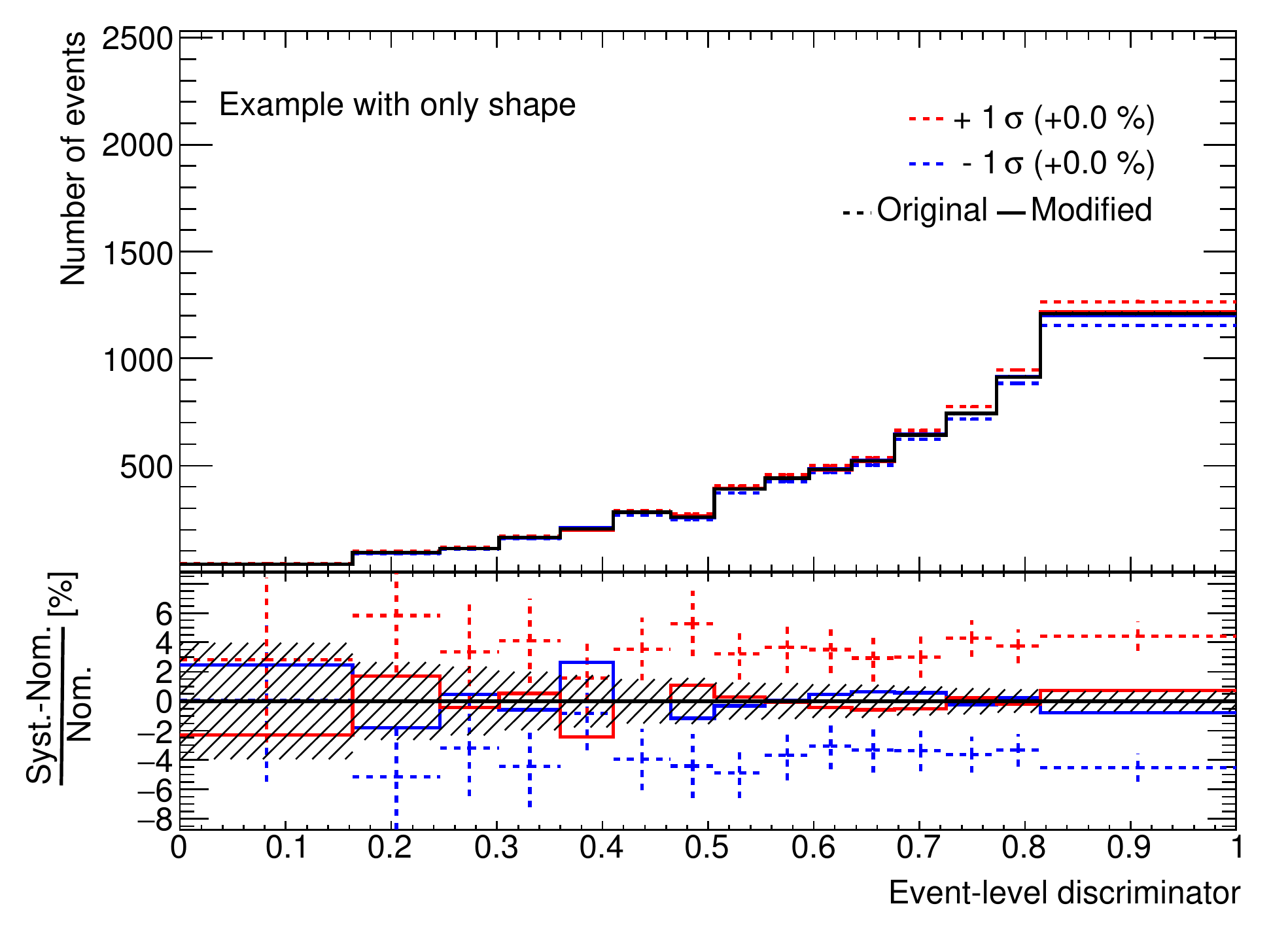}
}

\subfloat[\label{fig:redblueExampleC}Normalisation only]{
\includegraphics[width=0.49\linewidth]{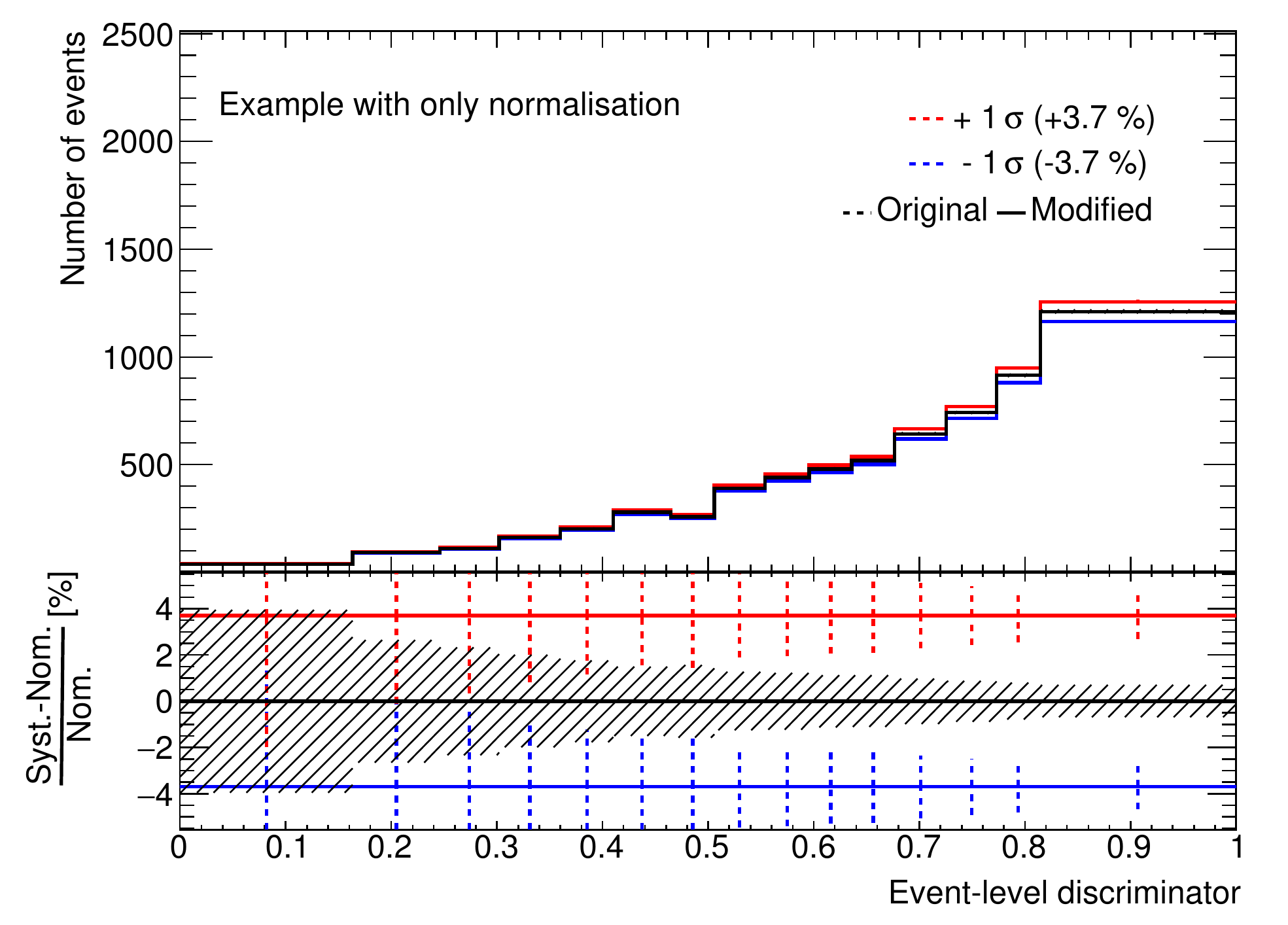}
}
\caption [] {An arbitrary example showing the effects of smoothing and symmetrisation for the same systematic source applied to the same physics process for systematic uncertainties that have contributions from shape and normalisation, shape only, and normalisation only.
The dotted line represents the uncertainty before smoothing and symmetrisation has been applied. The solid line represents the after effect and the nuisance parameter that will enter the fit.}
\label{fig:redblueExample}
\end{figure}

An arbitrary nuisance parameter is shown for three toy examples in Figure~\ref{fig:redblueExample} as an example of smoothing and symmetrisation.
Dotted lines show the original nominal and up/down variations before any smoothing or symmetrisation has been performed. The solid red/blue lines represent the systematic effect that will enter the fit after the symmetrisation and smoothing has been performed. If a systematic uncertainty is not symmetrised and is not smoothed, or if these operations have very little effect, the respective dotted and solid lines will be overlaid.
The hatched black line is the statistical error for the nominal signal or background sample.
Figure~\ref{fig:redblueExampleA} shows an arbitrary, non-negligible systematic uncertainty applied to some arbitrary signal or background process. The overall normalisation contribution is shown in the parenthesis on the top right (for both the up and down variations). 
The ratio plot shows the shape and the normalisation contribution. It is important to note that the shape contribution is the relative difference of one bin to the bordering bins. One can not have a shape-only systematic contribution if there is only a single bin.
Figure~\ref{fig:redblueExampleB} shows this same systematic uncertainty applied to the same MC, but with the normalisation effect removed. Thus, this systematic uncertainty only has a shape aspect. This is shown in the ratio plot by looking at bin height and derivative changes, with respect to the previous bin.
Figure~\ref{fig:redblueExampleC} shows this same systematic uncertainty where only an overall normalisation effect of 3.7\% has been applied. An example of such a systematic uncertainty would be the luminosity or the cross-section normalisation for the \Other backgrounds.

These plots are crucial in diagnosing problematic NPs in the final fit and so will be shown frequently in Chapter~\ref{sec:results}.

\subsubsection{Pruning}

The pruning of systematic uncertainties is done for the normalisation and shape separately. The $\pm 1 \sigma$ variation is calculated for each NP and an initial fit performed. For either shape or normalisation, if the effect on the uncertainty is less than the given threshold, the component is removed from further fits.
To determine the threshold a range of scenarios were tested. The chosen value of 0.7\% sees little difference between all systematic uncertainties used in the fit and only those that are greater than this threshold.
The NPs that pass this threshold are shown in Figure~\ref{fig:pruningSL} for the \chljets channels and Figure~\ref{fig:pruningDL} for the \chll channels.
Differences between the \chljets, \chejets and \chmujets (\chll, \chee, \chmumu and \chemu) channels mainly arise from different efficiencies for electrons and muons. Small differences in background compositions will also result in very slight pruning disparities.
``Not Present" indicates that either the NP does not apply for that process (for example ``\ttgamma Parton shower" for the \hfake background) or that the NP has been pruned.

\begin{figure}[!htbp]
\centering
\includegraphics[trim={0 45cm 0 0},clip,width=0.98\linewidth]{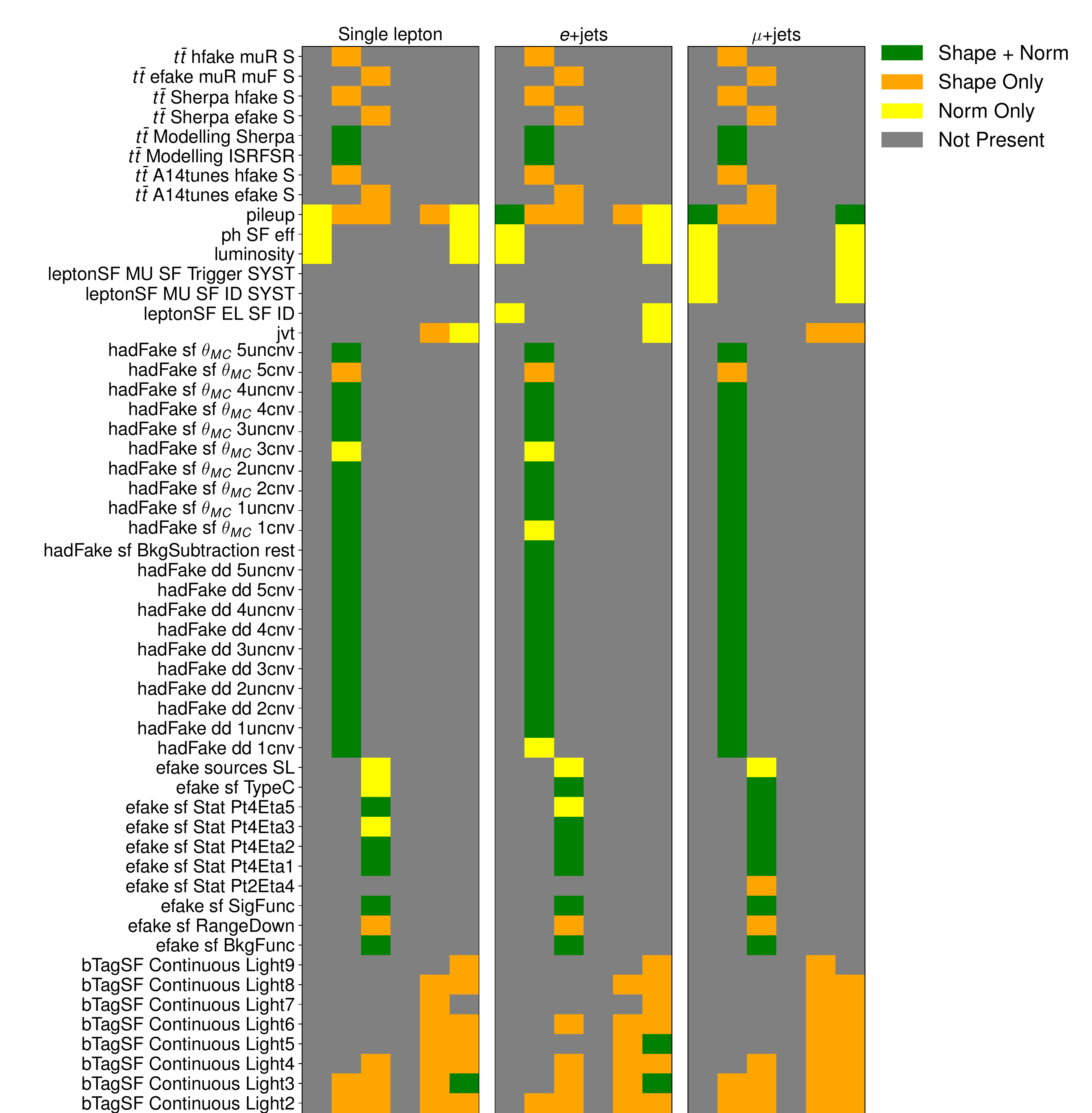}
\includegraphics[trim={0 0cm 0 80.3cm},clip,width=0.98\linewidth]{./figures/results/pruning/singlelepton}
\phantomcaption
\end{figure}
\begin{figure}
\ContinuedFloat
\centering
\includegraphics[trim={0 0 0 38.9cm},clip,width=0.98\linewidth]{./figures/results/pruning/singlelepton}
\caption [] {Schematic showing the systematic sources that survive the pruning procedure for each process. From left to right the \chljets, \chejets and \chmujets channels are shown. Only systematic uncertainties that pass the pruning for at least one process are included. Grey either indicates the systematic was pruned, or it does not apply for the process.}
\label{fig:pruningSL}
\end{figure}

\begin{figure}[!htbp]
\centering
\includegraphics[trim={0 51.5cm 0 0},clip,width=0.98\linewidth]{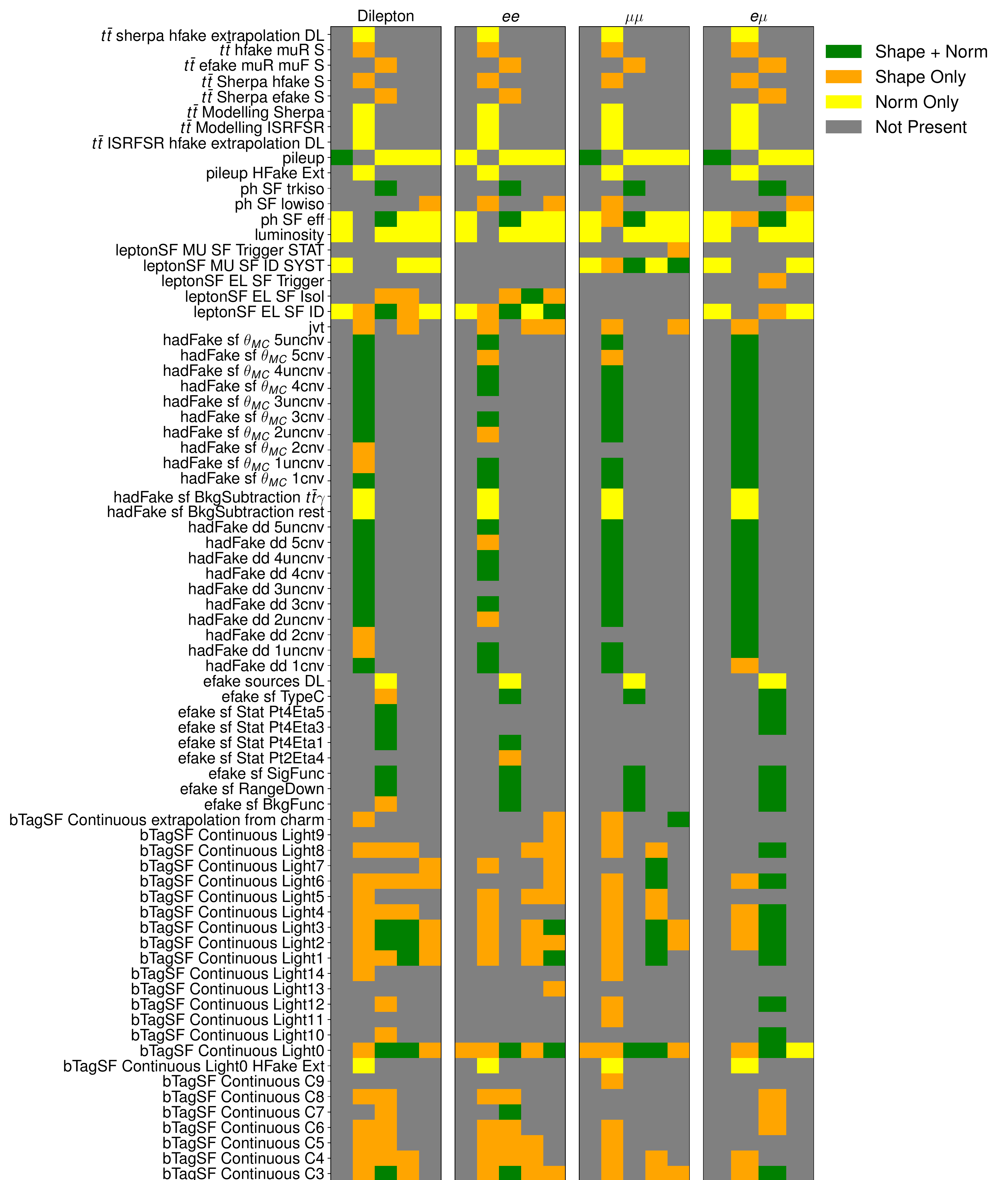}
\includegraphics[trim={0 0cm 0 95.6cm},clip,width=0.98\linewidth]{./figures/results/pruning/dilepton}
\phantomcaption
\end{figure}
\begin{figure}
\ContinuedFloat
\centering
\includegraphics[trim={0 0 0 47.6cm},clip,width=0.98\linewidth]{./figures/results/pruning/dilepton}
\caption [] {Schematic showing the systematic sources that survive the pruning procedure for each process. From left to right the \chll, \chee, \chmumu and \chemu channels are shown. Only systematic uncertainties that pass the pruning for at least one process are included. Grey either indicates the systematic was pruned, or it does not apply for the process.}
\label{fig:pruningDL}
\end{figure}

\FloatBarrier


\FloatBarrier
\chapter{Results}
\label{sec:results}

The section presents the culmination of the material introduced and studies performed in the previous chapters. Results directly relating to the final measurements, as well as the final measurements themselves are presented here.
Section~\ref{sec:asimovresults} documents studies and tests done using pseudo-datasets. This includes binning optimisations for the \ELD as well as preliminary nuisance parameter checks.
Section~\ref{sec:finalresults} presents the final fit-to-data results along with further nuisance parameter discussions.
Lastly, Section~\ref{sec:fidcrosssections} presents the final fiducial \xsecModifyNoun results.

\section{Asimov fits}
\label{sec:asimovresults}

Asimov\footnote{The name ``Asimov" originates from Isaac Asimov's short story, \emph{Franchise}~\cite{asimov1989franchise}. In it, the single most representative member of the population is elected to vote on behalf of everyone.} datasets are pseudo datasets that essentially imply our signal and background shapes and normalisations are completely understood, thus pseudo data can be built as signal+background for each bin. A fit using this dataset (an \emph{Asimov fit}) will by design yield a signal strength of one with some uncertainty. Asimov fits can be used to check the expected sensitivity\footnote{The sensitivity of a model is a measure of how large the (relative or absolute) uncertainties are.} of a model as well as how the NPs behave.

\subsection{\ELD binning optimisation}
\label{sec:binoptis}

The choice of binning for the final discriminating variable used in the maximum likelihood fit plays an important role in constraining signal and backgrounds. 
A bin that contains only a single background process will help constrain the NPs that apply to that background for the rest of the distribution. 
Three different Asimov fit scenarios are tested on the \chljets and \chll channels to determine the best bin sizes for the \ELD. All systematic uncertainties are included in these scenarios.
The first scenario uses a uniform binning distribution with ten and six bins for the \chljets and \chll channels, respectively. Another scenario uses an automatic binning algorithm. The algorithm scans the distribution and merges bins from the right until a certain fraction of signal and background remains in the bin. For each bin, a threshold for this merging is defined as 
\begin{equation}
Z  = z_b \frac{n_b}{N_b} + z_s \frac{n_s}{N_s} ,
\end{equation} 
where $n_s$ and $n_b$ are the signal and background content, respectively. $N_s$ and $N_b$ are the total number of signal and background events in the full \ELD distribution. This function takes two free parameters, $z_s$ and $z_b$, which define the maximum allowed fraction of signal and background events in each bin for $z_s+z_b = \text{number of bins}$. The algorithm stops iterating when $Z>1$.

The results for each of the scenarios are shown in Table~\ref{tab:binOptiTab}.
For the \chljets channel small differences are seen between each scenario. However, the AutoBin 2 binning algorithm consisting of 15 bins yields the highest sensitivity and so will be used. For simplicity, this binning is also applied to the individual \chejets and \chmujets channels. The final binning is $[0,0.16,0.25,0.30,0.36,0.41,0.46,0.51,0.55,0.60,0.64,0.68,0.73,0.77,0.81,1]$.
For the \chll channel, similarly, small differences are seen. The AutoBin 2 algorithm performs slightly better, however with fewer events as is seen in the \chee, \chmumu and \chemu channels it seems prudent to limit the number of bins. For simplicity sake, the uniform binning with six bins is chosen for all \chll (combined and separate) channels.

\begin{table}[htbp]
        \centering
        \begin{tabular}{l|l|c|c}
        \hline
        Channel & Binning & ``+" error & ``-" error \\
        \hline
        \hline
        \multirow{3}{*}{\chljets} & Uniform & 0.0934 & 0.0915  \\
         & AutoBin 1 $(z_s = 5, z_b=5)$ & 0.0923  & 0.0897 \\
         &  \textbf{AutoBin 2} $(z_s = 12, z_b=3)$& \textbf{0.0911} & \textbf{0.0888} \\
         \hline
        \multirow{3}{*}{\chll} &  \textbf{Uniform} & \textbf{0.0678} & \textbf{0.0629} \\
         & AutoBin 1 $(z_s = 5, z_b=5)$ & 0.0663 & 0.0626 \\
        &  AutoBin 2 $(z_s = 12, z_b=3)$ & 0.0652 & 0.0606 \\
         \hline
        \end{tabular}
        \caption{The different Asimov fit scenarios for the \chljets and \chll channels. All systematics are included in the fit. The error represents the absolute up/down error of $\mu$. The chosen binning is highlighted in bold text.}
        \label{tab:binOptiTab}
\end{table}

\subsection{Asimov fit cross checks}
\label{sec:asimovcrosschecks}

NPs enter the fit with a Gaussian prior constraint with a mean of 0 and a width (root mean square) associated with the normalisation uncertainty of 1$\sigma$. If an NP should be \emph{constrained} in the fit, the posterior width will be narrower than the prior. If an NP should be \emph{pulled} in a fit, the central value of the Gaussian distribution will no longer be 0 (thus having a normalisation different to the prediction). 
In an Asimov fit all NPs are kept constant at 0 and the constraining power of each is checked. Significant constraints need to be explored as it could mean we do not understand our data (or our NP) as well as we should.

\begin{figure}[!htbp]
\centering
\includegraphics[trim={0 30cm 0 0},clip,width=0.49\linewidth]{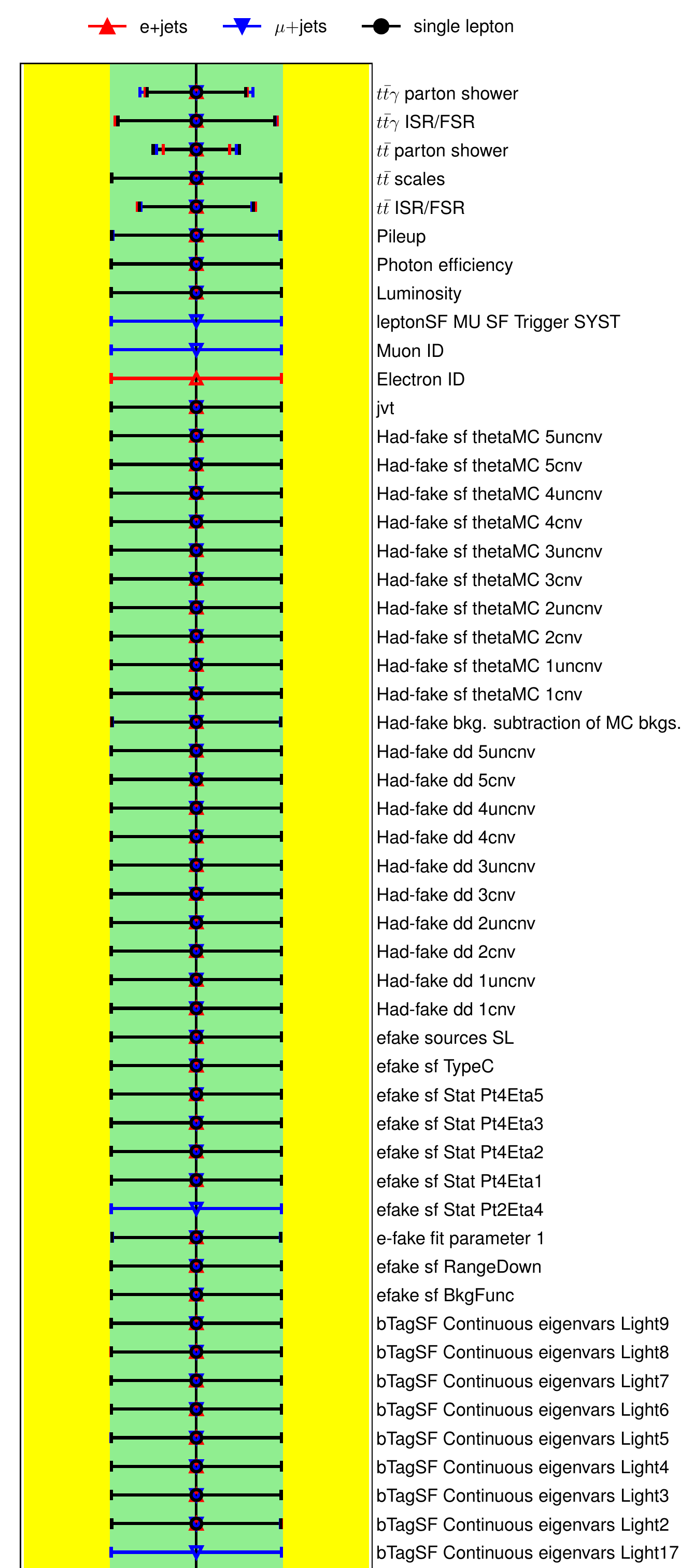}
\includegraphics[trim={0 0 0 30.7cm},clip,width=0.49\linewidth]{./figures/results/nps/asimov/pullPlot_SLchannels}
\caption [] {Asimov post-fit pull plots for all nuisance parameters in the \chljets channels. The ``Signal strength" and ``$W\gamma$ floating NP" both have expectation values held fixed at one. ``Had-fake" and ``efake" NPs relate to the data-driven methods and the uncertainties that arise from estimating these backgrounds. The ``bTagSF" NPs are associated with pseudo-continuous $b$-tagging for different flavour jets.}
\label{fig:AsimovpostFitBrazilSLchannels}
\end{figure}

\begin{figure}[!htbp]
\centering
\includegraphics[trim={0 30.2cm 0 0},clip,width=0.49\linewidth]{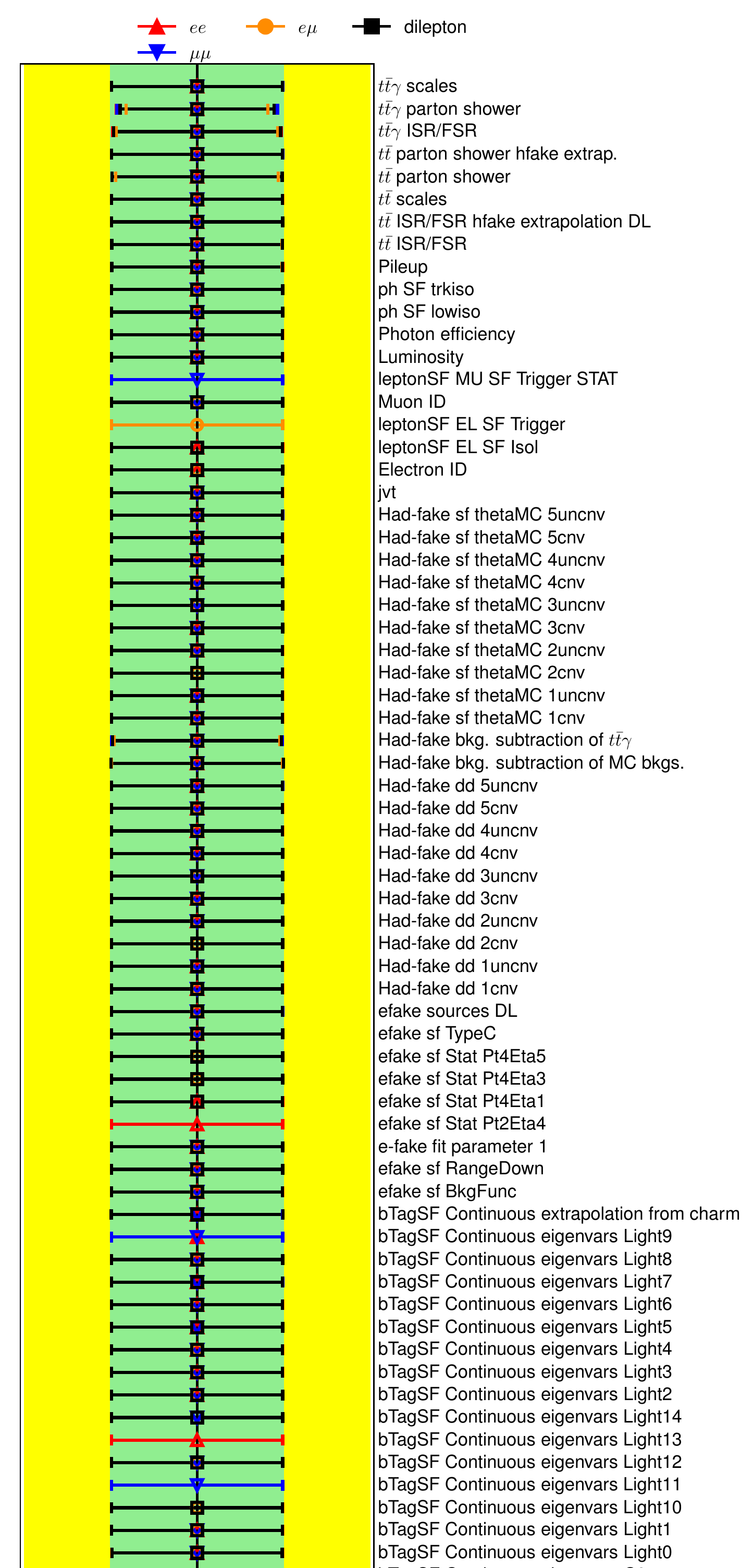}
\includegraphics[trim={0 0 0 30.5cm},clip,width=0.49\linewidth]{./figures/results/nps/asimov/pullPlot_DLchannels}
\caption [] {Asimov post-fit pull plots for all nuisance parameters in the \chll channels. The ``Signal strength" has an expectation value held fixed at one. ``Had-fake" and ``efake" NPs relate to the data-driven methods and the uncertainties that arise from estimating these backgrounds. The ``bTagSF" NPs are associated with pseudo-continuous $b$-tagging for different flavour jets.}
\label{fig:AsimovpostFitBrazilDLchannels}
\end{figure}

The posteriors for all NPs and channels are shown in Figure~\ref{fig:AsimovpostFitBrazilSLchannels} and \ref{fig:AsimovpostFitBrazilDLchannels} in \emph{pull-plots} for the individual and combined \chljets and \chll channels, respectively\footnote{The Asimov pull-plot for the 5-channel inclusive fit is not shown as any observed constraints would need to be traced back to the individual channels.}. Also included is the signal strength point-of-interest (POI) and in the case of the \chljets channels, the floating \Wgamma normalisation (which in an Asimov fit is held constant at one).
From Section~\ref{sec:smoothing}, a pruning threshold of 0.7\% is applied to all systematics in each channel. One NP might be over the threshold for one channel but fail it for another. Thus, the pull-plots could have a ``missing" contribution for a given channel. 
For those systematics that are constrained, further discussions on the priors are presented to ensure that no problems exist.

For the \chljets channels there are four constrained NPs that warrant closer checks, of which three are related to \ttbar modelling. This is not too surprising since dedicated scale factors are derived for the nominal sample and not the systematic variation samples. 

\begin{figure}[!htbp]
\centering
\subfloat[Shape]{
\includegraphics[width=0.48\linewidth]{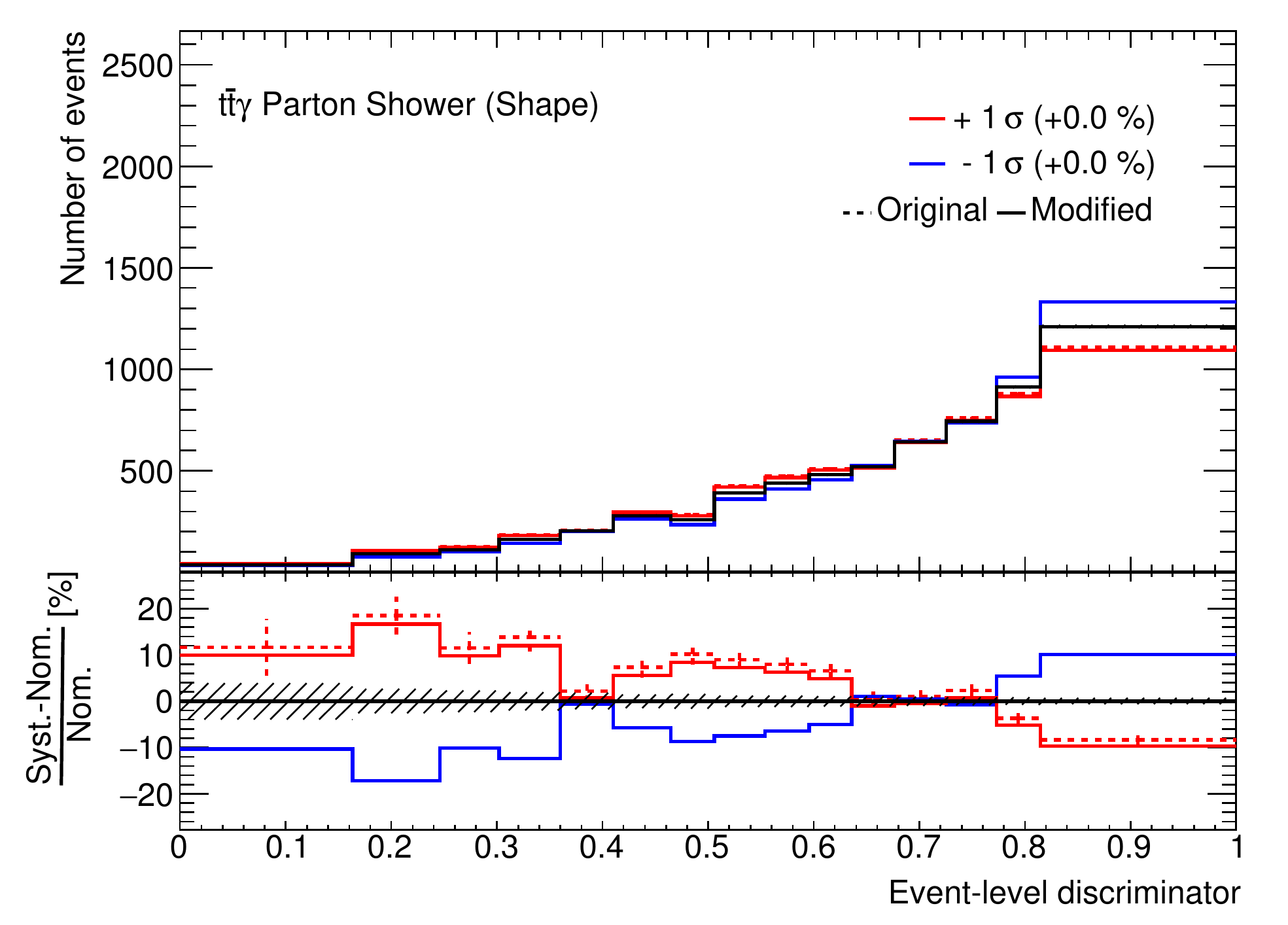}
}\hspace{-0.036\linewidth}
\subfloat[Normalisation]{
\includegraphics[width=0.48\linewidth]{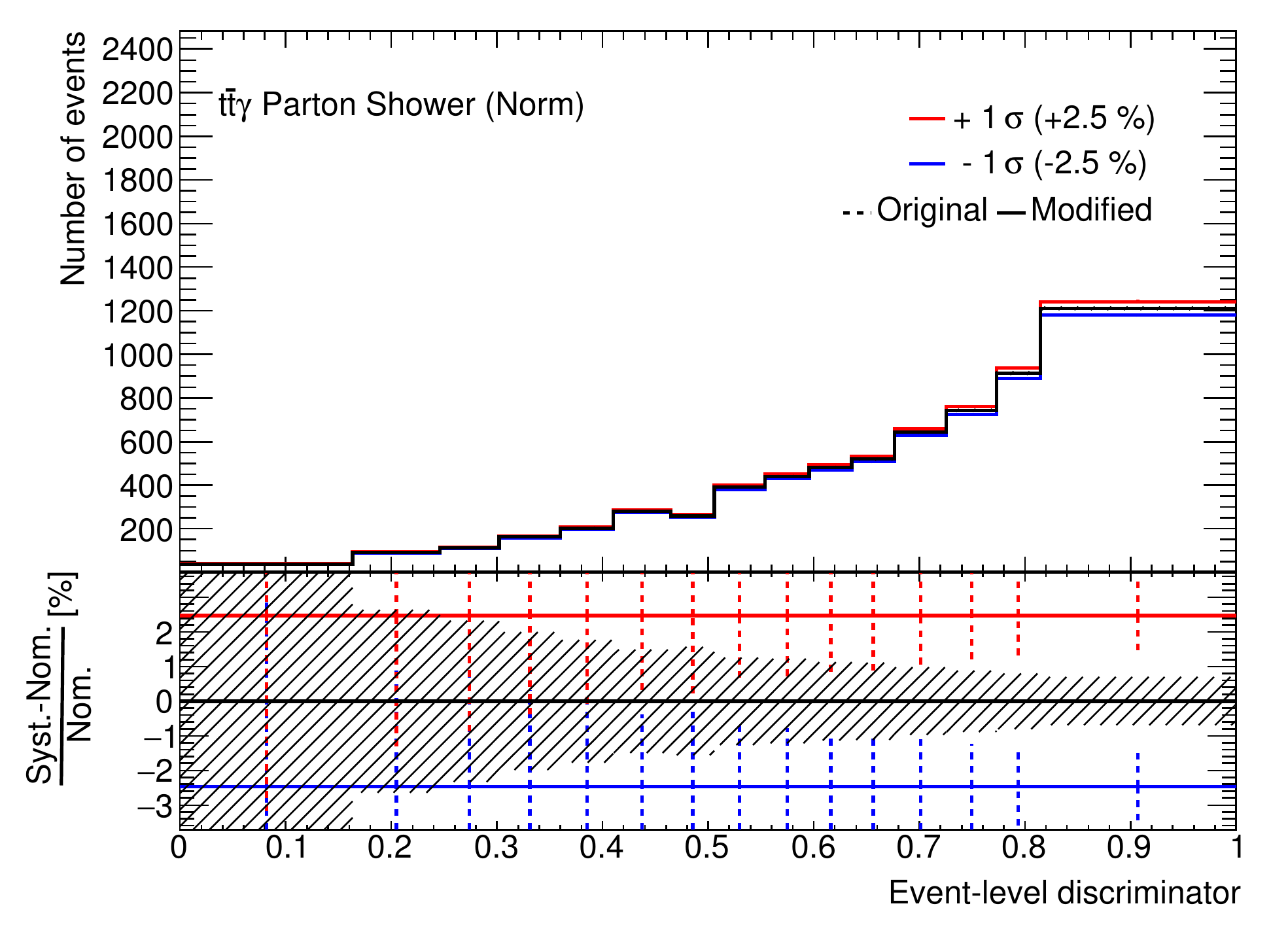}
}
\caption {The signal parton shower systematic for the \chljets channel.}
\label{fig:signalpshower}
\end{figure}

\begin{figure}[!htbp]
\centering
\subfloat[\efake]{
\includegraphics[width=0.48\linewidth]{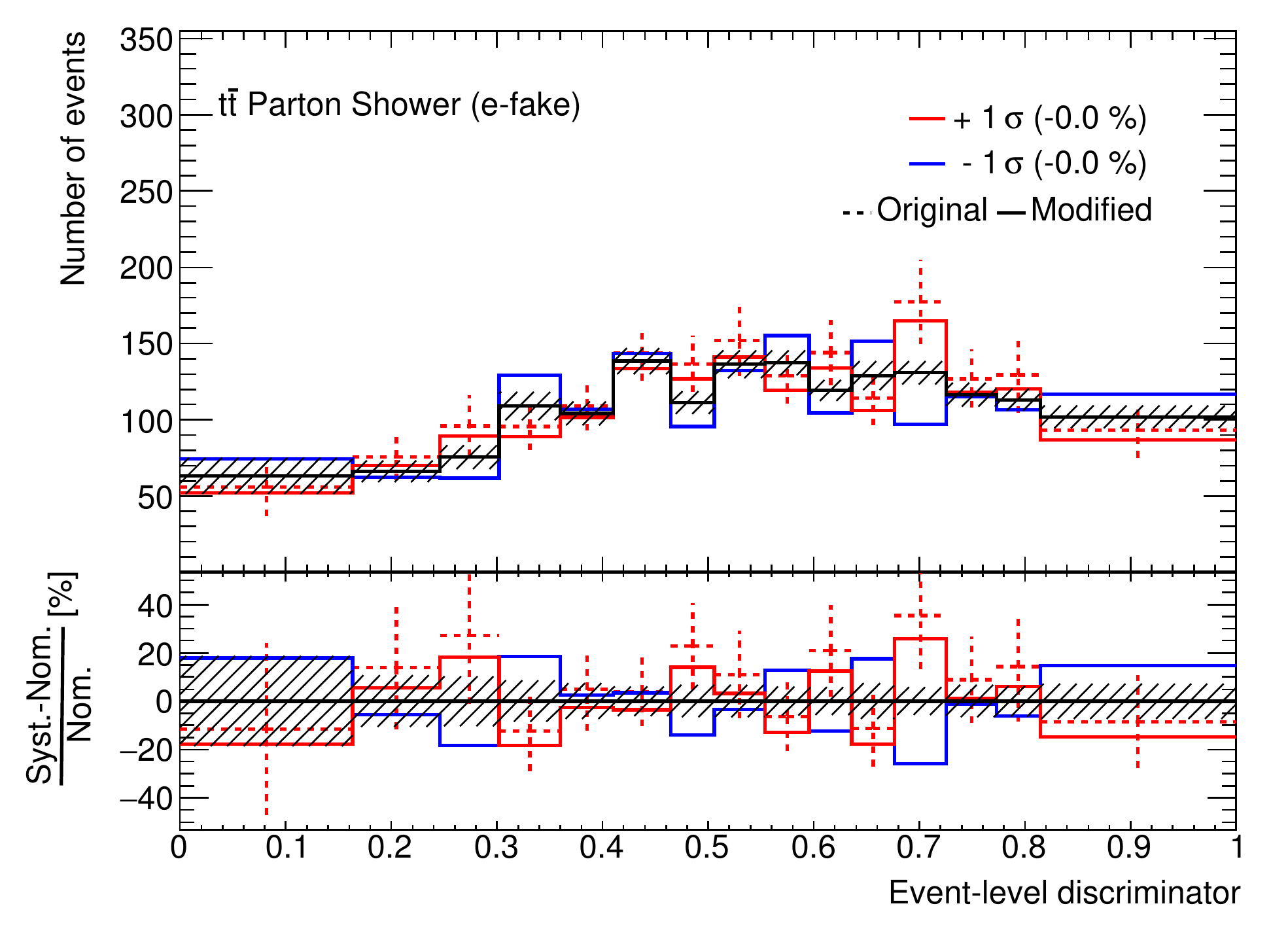}
}\hspace{-0.036\linewidth}
\subfloat[\hfake]{
\includegraphics[width=0.48\linewidth]{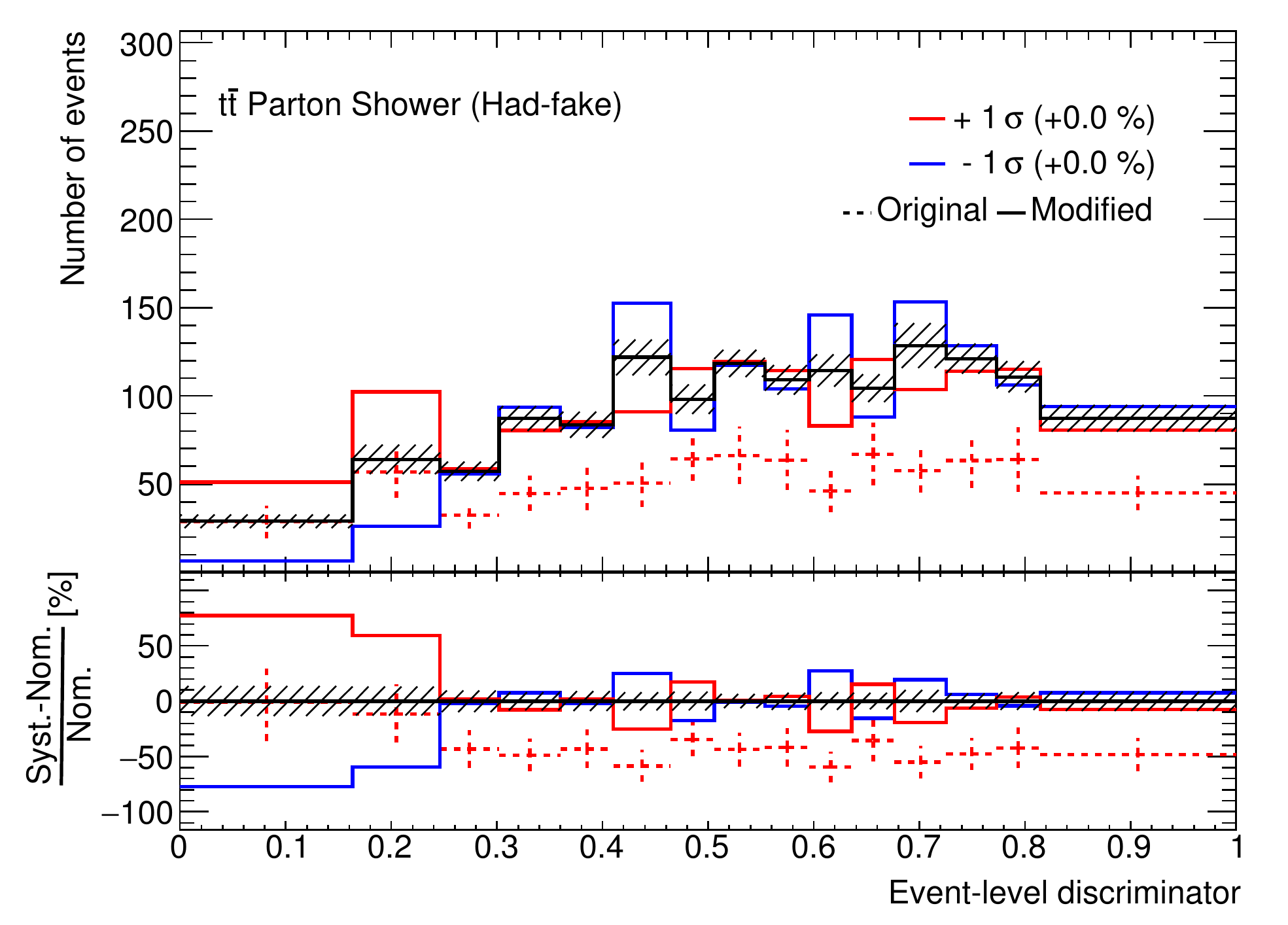}
}
\caption {The shape only component of the \ttbar parton shower NP for the \hfake and \efake background processes in the \chljets channel.}
\label{fig:ttbarpartonshower}
\end{figure}

\begin{figure}[!htbp]
\centering
\subfloat[\efake]{
\includegraphics[width=0.48\linewidth]{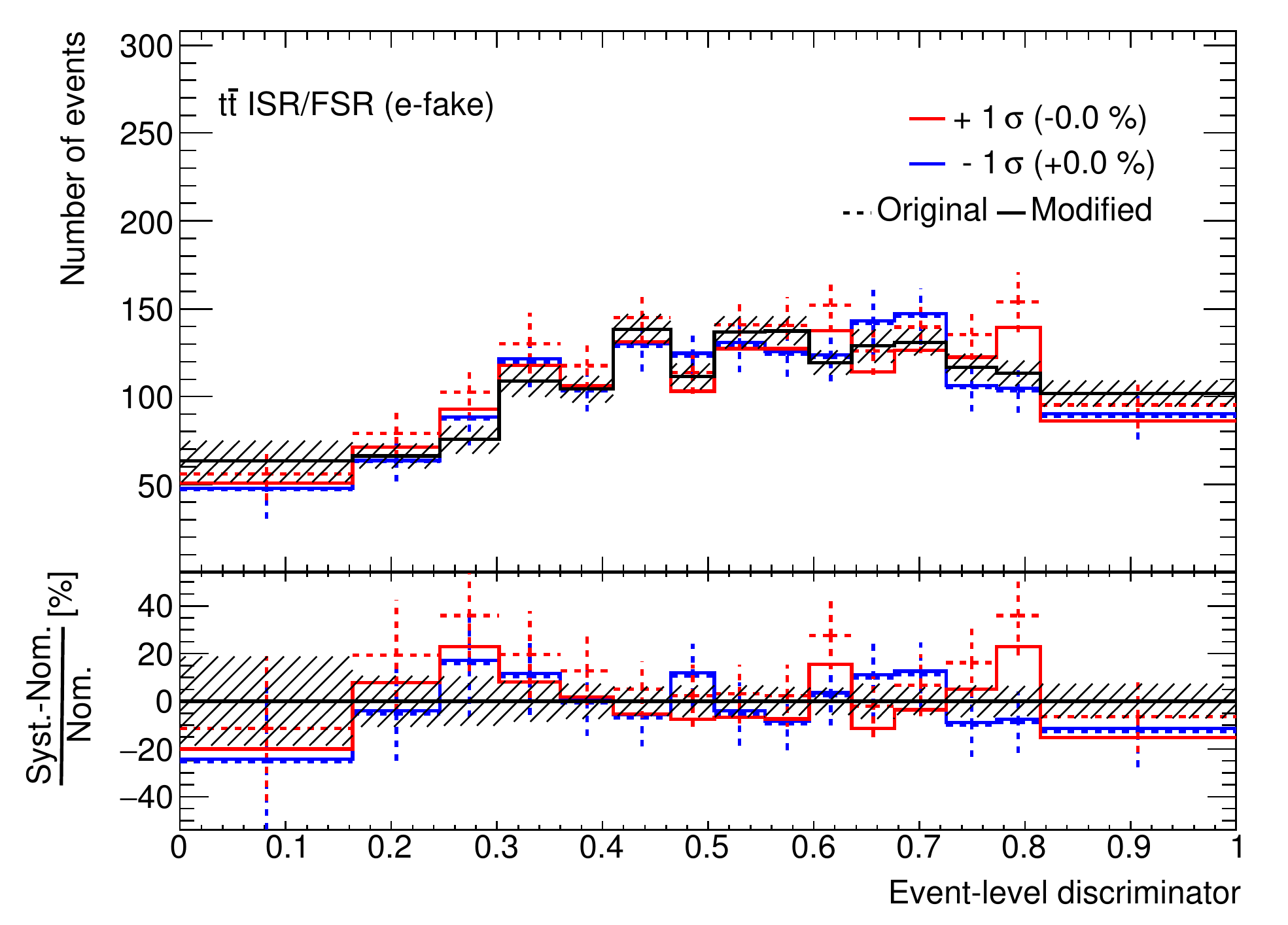}
}\hspace{-0.036\linewidth}
\subfloat[\hfake]{
\includegraphics[width=0.48\linewidth]{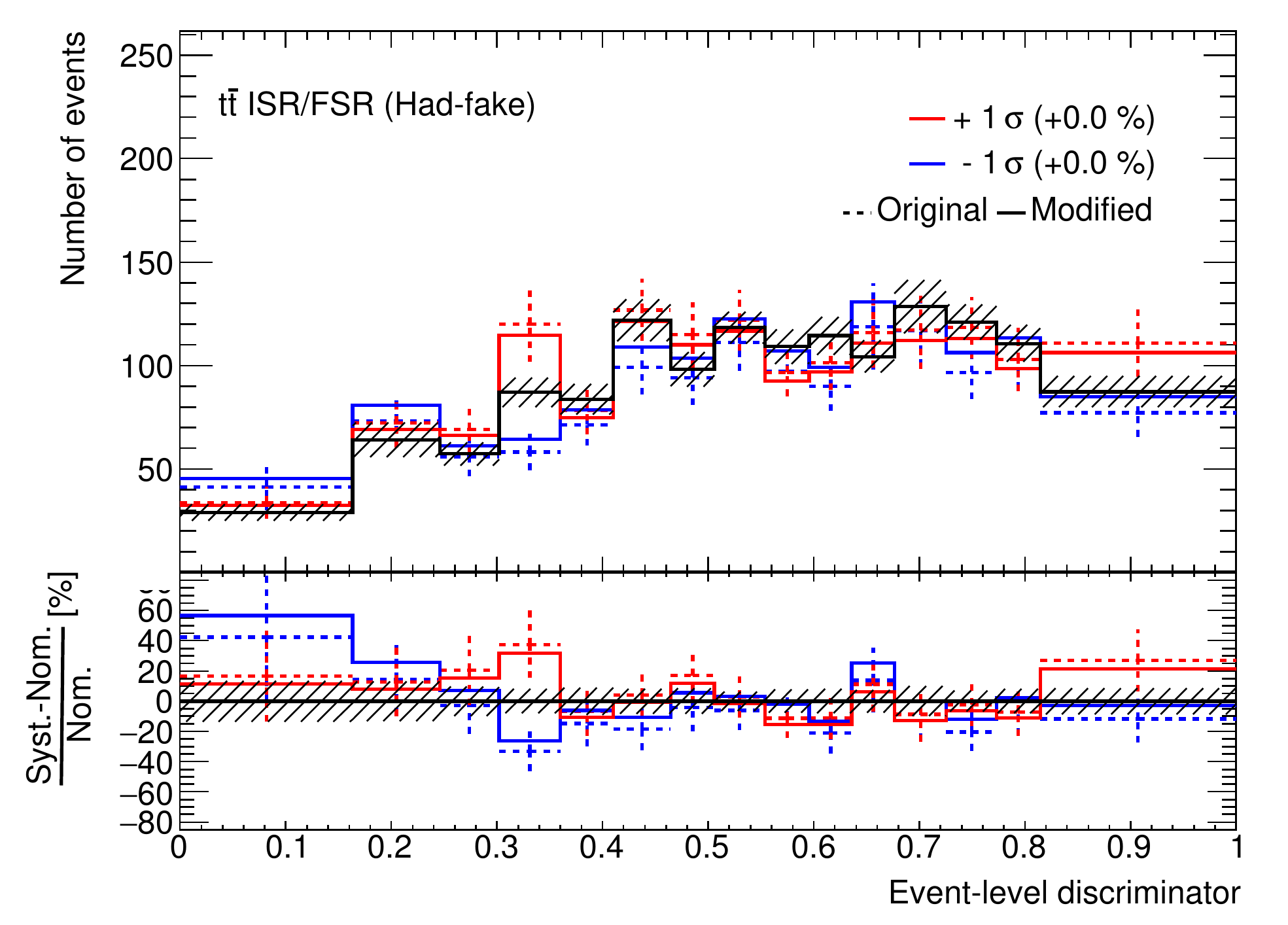}
}
\caption {The shape only component of the \ttbar ISR/FSR NP for the \hfake and \efake background processes in the \chljets channel.}
\label{fig:ttbar_ISRFSR}
\end{figure}

\begin{figure}[!htbp]
\centering
\includegraphics[width=0.50\linewidth]{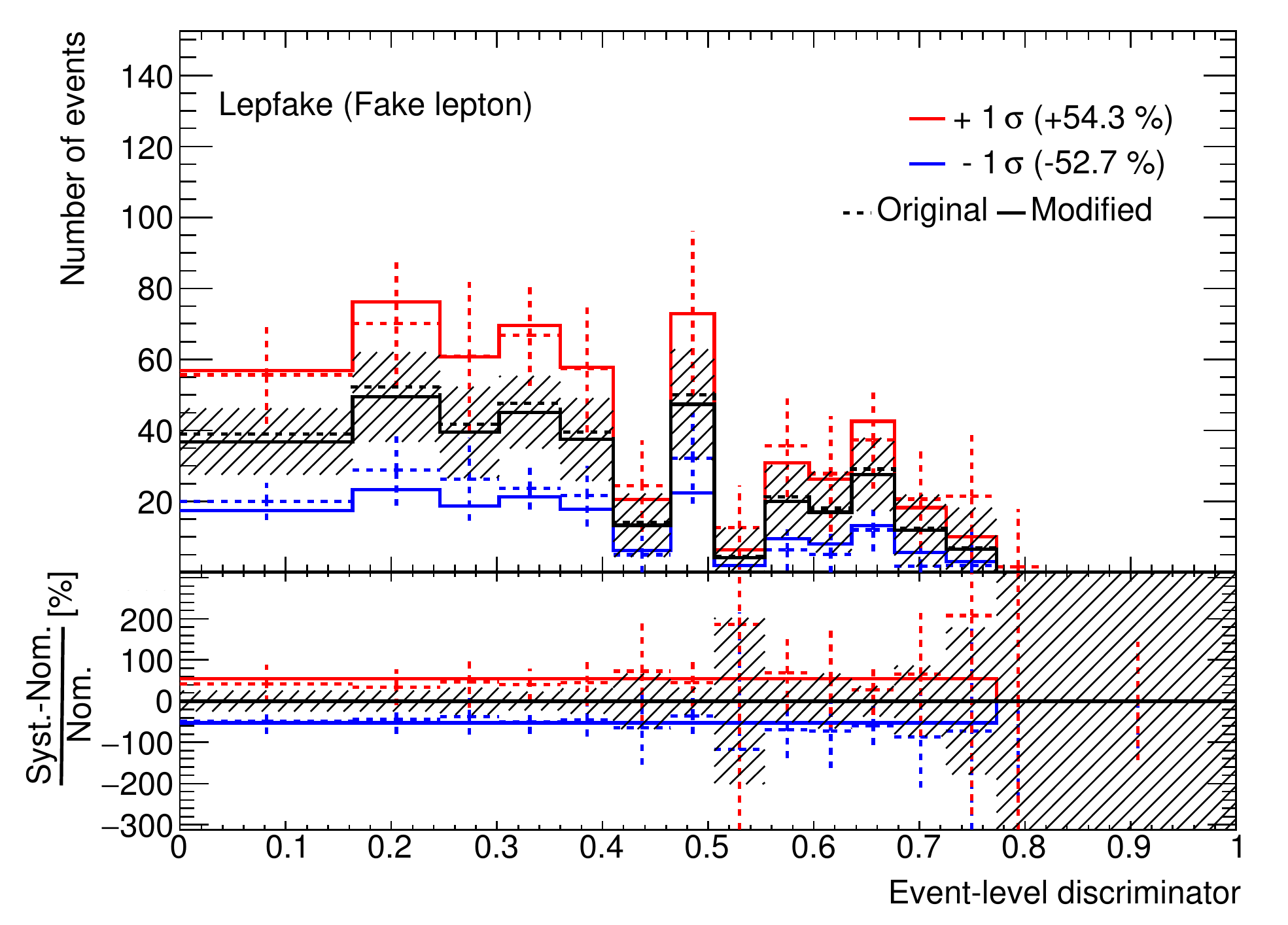}
\caption {The \QCD background NP for the \chljets channel.}
\label{fig:QCDsysts}
\end{figure}

\begin{itemize}
\item \ttgamma parton shower: In all \chljets channels this systematic uncertainty is constrained. The shape component of the NP that enters each fit is shown in Figure~\ref{fig:signalpshower} for the \chljets channel (these figures are similar for the \chejets and \chmujets channels). Large up/down variations from the \herwigall sample can be seen. Given that this NP is correlated to the signal (and thus the signal strength), it is not surprising there is a potential for it to be constrained in the fit.

\item \ttbar parton shower: This NP is constrained in all \chljets channels. The shape-only contributions (due to being corrected with data-driven scale factors) for the \hfake and \efake backgrounds are shown in Figure~\ref{fig:ttbarpartonshower}. In the low end of the \ELD distribution large shape discrepancies are seen.

\item \ttbar ISR/FSR: This NP is constrained in all \chljets channels, with the un-symmetrised, shape-only contributions for the \hfake and \efake backgrounds shown in Figure~\ref{fig:ttbar_ISRFSR}. Large shape discrepancies are seen.

\item Fake lepton background: The up/down variations for the \QCD background are from the extremes of the parameterisation ``envelope" used in the data driven method (Chapter~\ref{sec:fakelepton}). Thus, one expects large relative differences to the nominal distribution as shown in Figure~\ref{fig:QCDsysts}. Given that the overall contribution of the \QCD background is small and spread out due to the \ELD, the effect that this NP has on the final result is small.

\end{itemize}

\begin{figure}[!htbp]
\centering
\includegraphics[width=0.50\linewidth]{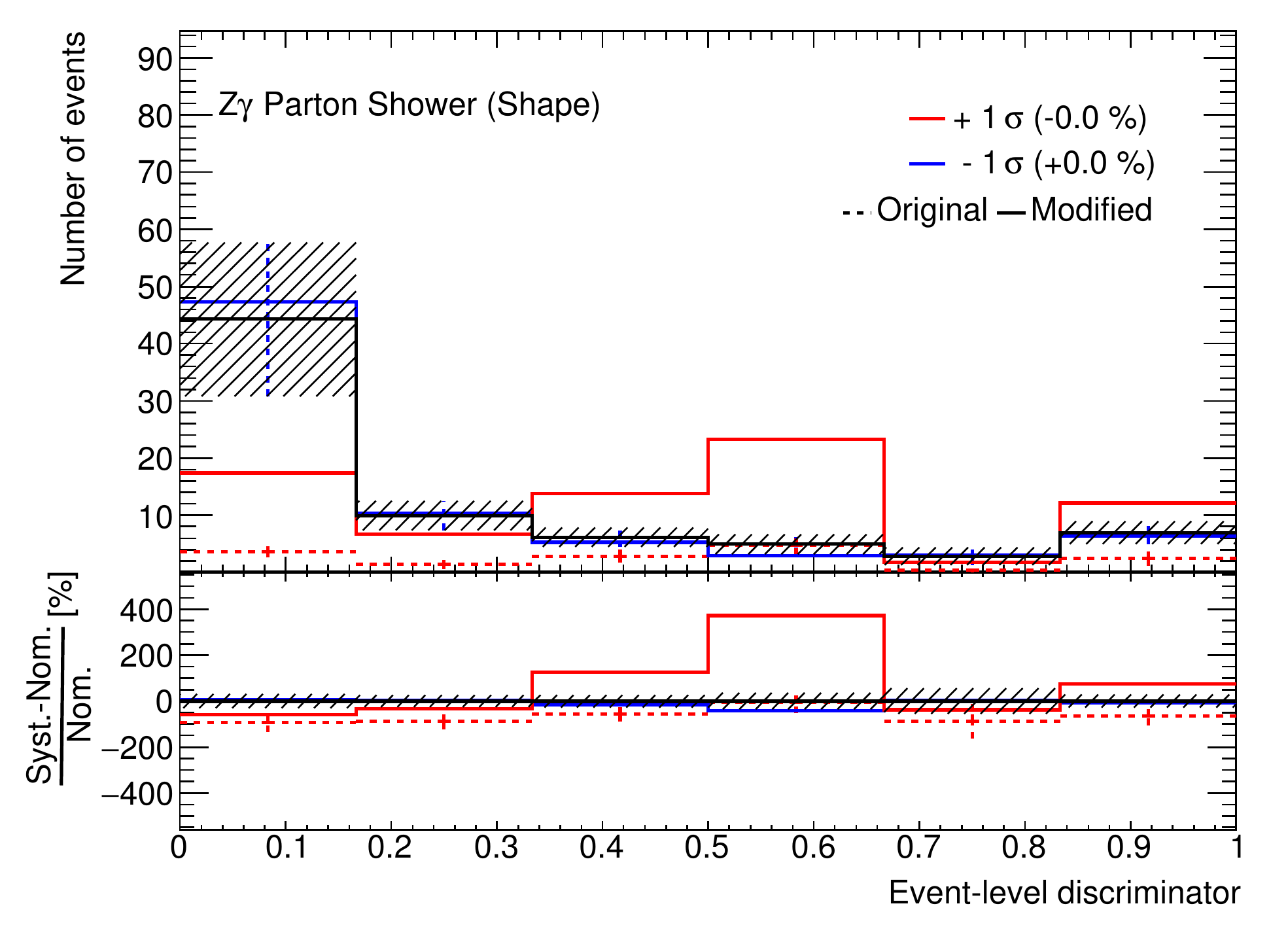}
\caption {The \Zgamma parton shower NP for the \chll channel.}
\label{fig:Zgammapartonshower}
\end{figure}

For the \chll channels the main constrained NPs arise because of the modelling of the \Zgamma background. 
This includes the shape due to the parton shower and the conservative normalisation of 50\% placed on the \xsec. 
The \Zgamma parton shower shape can be seen in Figure~\ref{fig:Zgammapartonshower}, which shows clear differences to the nominal sample. The outcome of this mis-modelling is that uncertainties in the \chee and \chmumu channels will be larger and thus more conservative.

In most channels the JER is also seen to be slightly constrained. This is simply due to the size of the NP. This and the other few constraints observed in the Asimov fits will be further discussed in the upcoming section with fits to real data.


\FloatBarrier
\section{Full fit results}
\label{sec:finalresults}

Final fits to data are presented in this section for all channels.
The signal strength measurements are presented in Section~\ref{sec:sigstrength}, followed by post-fit plots, yields, and dominant uncertainties in Section~\ref{sec:plotsyieldsuncerts}. This is followed by a discussion on the nuisance parameters in Section~\ref{sec:NPdiscuss}.  

\subsection{Signal strength}
\label{sec:sigstrength}

\begin{figure}[!htbp]
\centering
\includegraphics[width=0.8\textwidth]{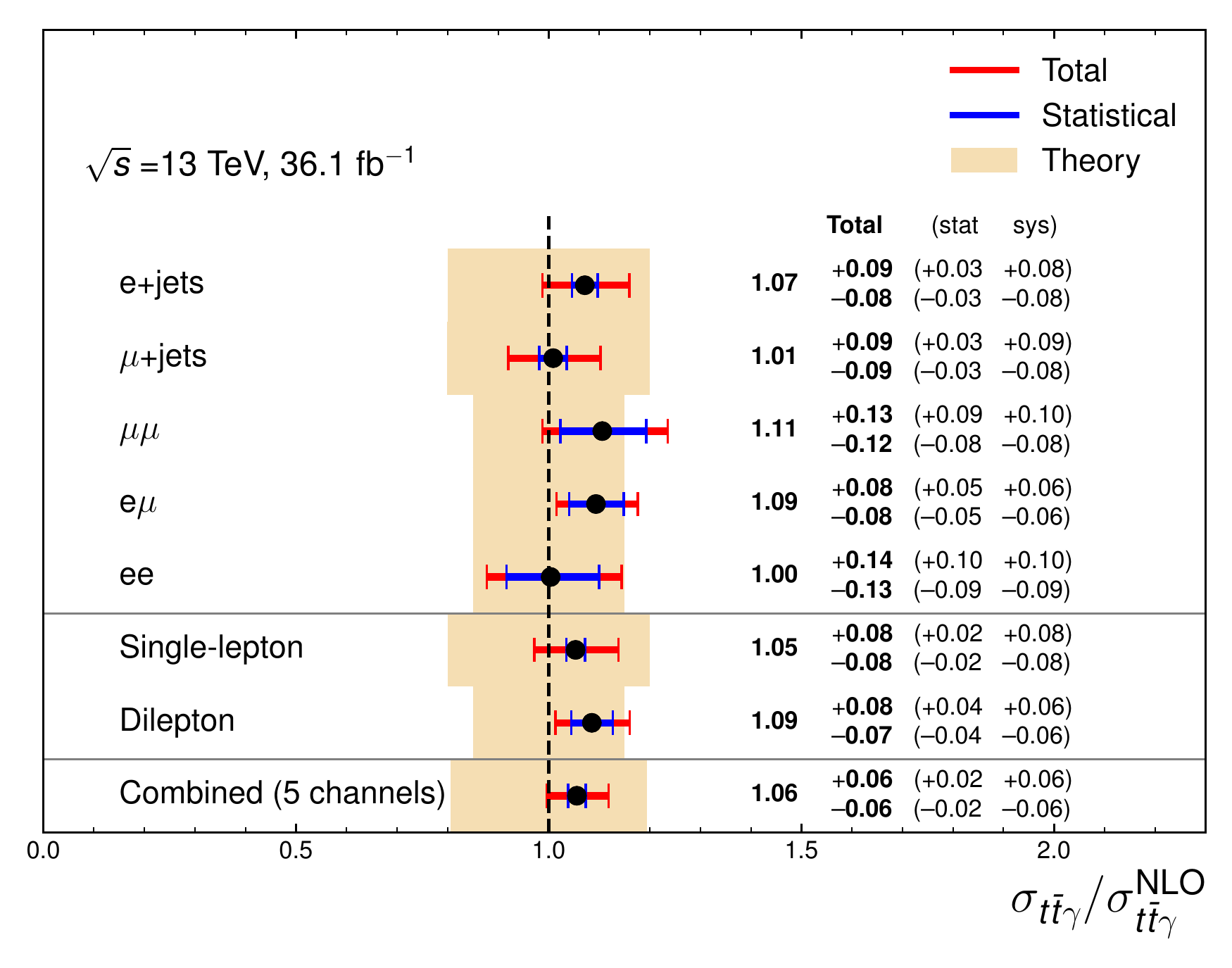}
\caption {The observed signal strength for individual, \chljets, \chll and combined channels. The NLO SM prediction is represented by the dashed vertical line.  The theoretical uncertainty for each fit is represented by the shaded region.}
\label{fig:crossSectionMu}
\end{figure}

The signal strength ($\mu$) for each fit is shown in Figure~\ref{fig:crossSectionMu} along with the total uncertainty split into statistical and systematic components. 
The vertical dotted line at one represents the SM prediction of the signal strength. The shaded region represents the theoretical uncertainty derived from Equation~\ref{eq:nlo}, which includes the uncertainty on the $k$-factor calculation.

All \chljets measurements indicate that we are dominated by systematic uncertainties. For \chll channels, while the systematic uncertainty is dominant, much can still be gained from more data. 
There also seems to be a clear trend of signal strengths greater than one. Given that all measurements agree within the theory prediction uncertainties, nothing conclusive can be said. Future measurements need to put high priority on reducing the theoretical uncertainty.
For completeness, this same figure but with the addition of all fit scenarios detailed in Chapter~\ref{sec:fitscenarious} is shown in Appendix~\ref{sec:resultsappendix}, where negligible differences are seen.

\clearpage

For the \chljets channels the \Wgamma background is a free parameter in the fit and so has a corresponding normalisation.
These are: 
\begin{align*}
W\gamma \text{SF}_{e+\text{jets}} &= 0.70 \pm 0.40\  \text{(tot.)}, \\ 
W\gamma \text{SF}_{\mu+\text{jets}} &= 0.94 \pm 0.32\  \text{(tot.)}, \\ 
W\gamma \text{SF}_{\text{SL}} &= 0.80  \pm 0.34\ \text{(tot.)},\\
W\gamma \text{SF}_{\text{5-chan}} &= 0.86 \pm 0.26\ \text{(tot.)}.
\end{align*}
These normalisations agree within uncertainties with the expectation value of one, as well as the validation region fit results in Chapter~\ref{sec:wgammavr}.

\FloatBarrier
\subsection{Distributions, yields and uncertainties}
\label{sec:plotsyieldsuncerts}

\begin{figure}[!htbp]
\centering
\subfloat[Pre-fit \chljets]{
\includegraphics[width=0.45\linewidth]{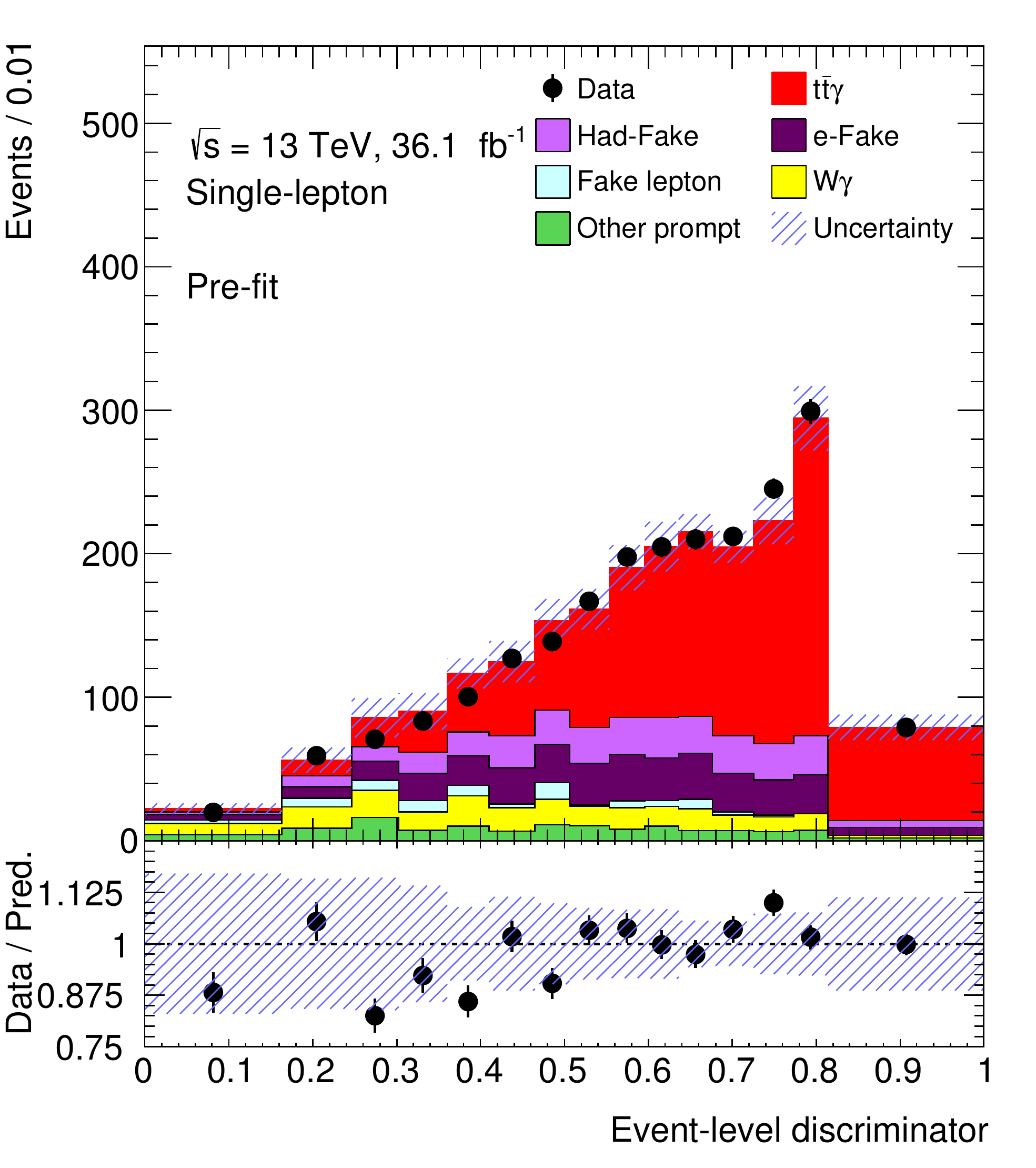}
}\hspace{-0.038\linewidth}
\subfloat[Post-fit \chljets]{
\includegraphics[width=0.45\linewidth]{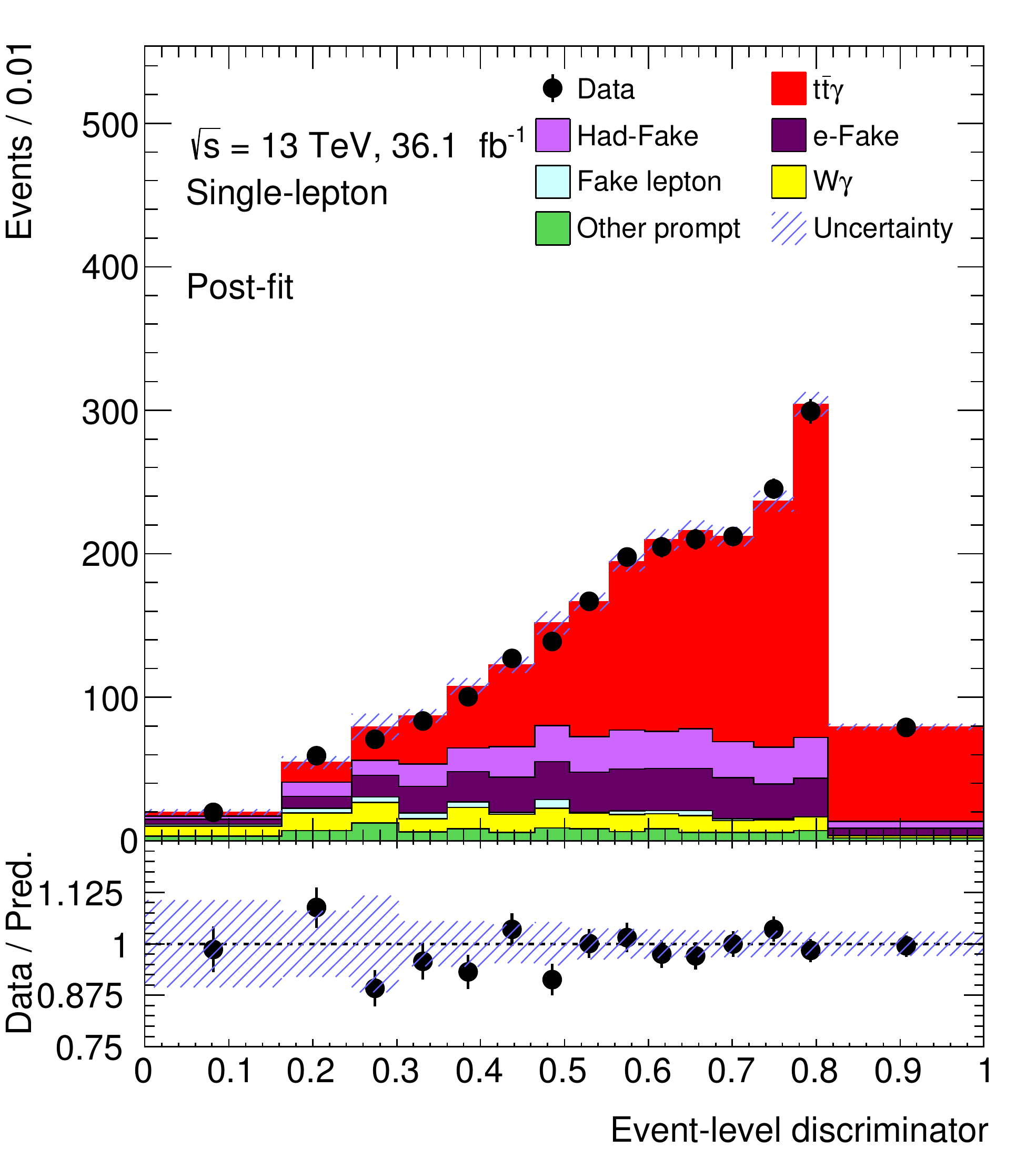}
}
\caption {Pre- and post-fit plots for the \chljets channel where the \ELD distribution is used as the discriminating variable in the fit.}
\label{fig:postFitELDSL}
\end{figure}

The pre-fit and post-fit \ELD distributions for the \chljets channel are shown in Figure~\ref{fig:postFitELDSL}, while the distributions for the \chll channel are shown in Figure~\ref{fig:postFitELDDL}.
Individual channel distributions are shown in Appendix~\ref{sec:resultsappendix}.
The post-fit yields for all channels are shown in Table~\ref{tab:postFitYields}, which includes all uncertainties.
In all bins the ratio of predicted MC and measured data agree within uncertainties.
By looking at the uncertainty bands for pre- and post-fit distributions, the reduction of the size is a clear indication that the fitting mechanism is able to constrain NPs.

\begin{figure}[!htbp]
\centering

\includegraphics[width=0.33\linewidth]{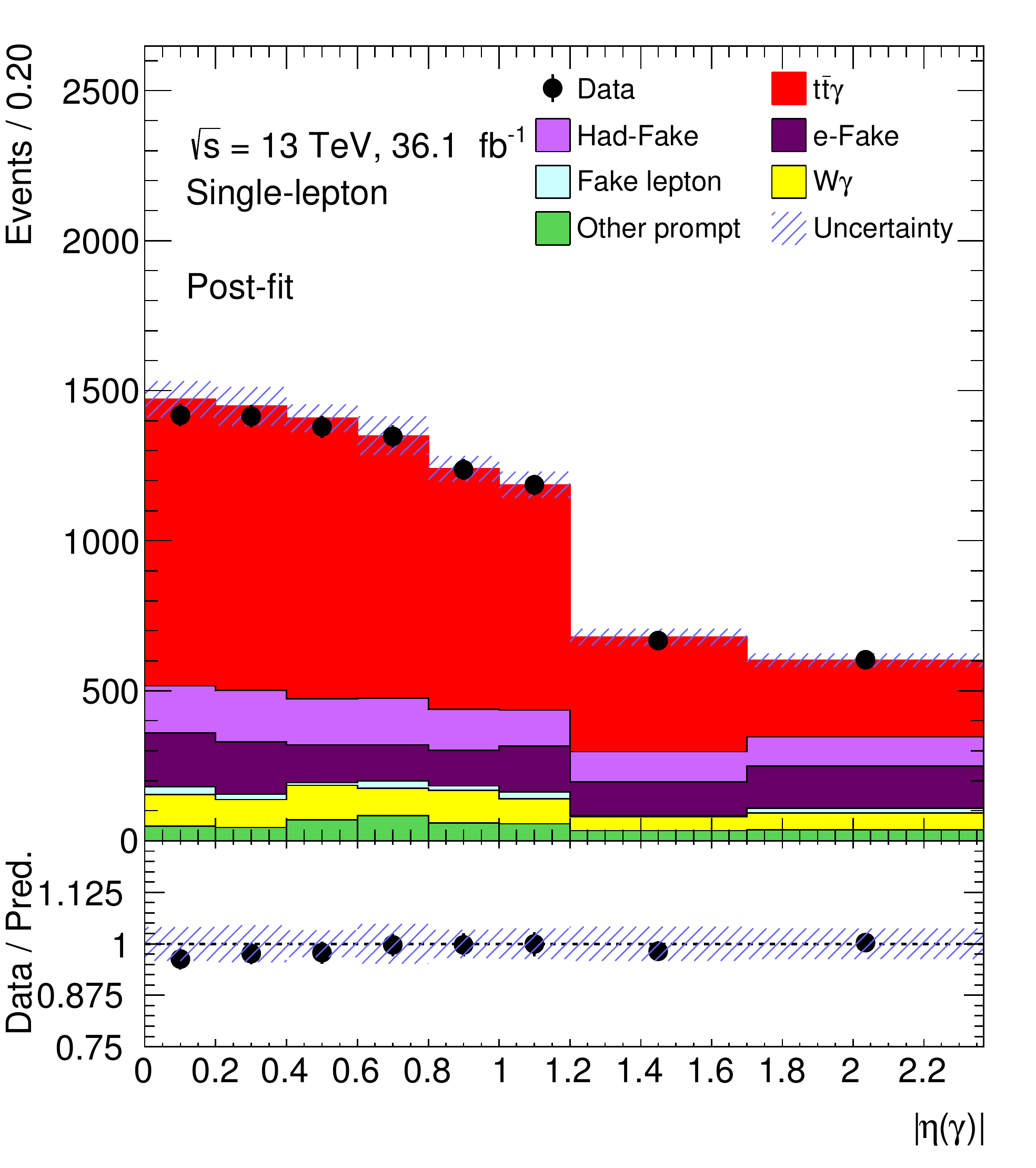}
\hspace{-0.023\linewidth}
\includegraphics[width=0.33\linewidth]{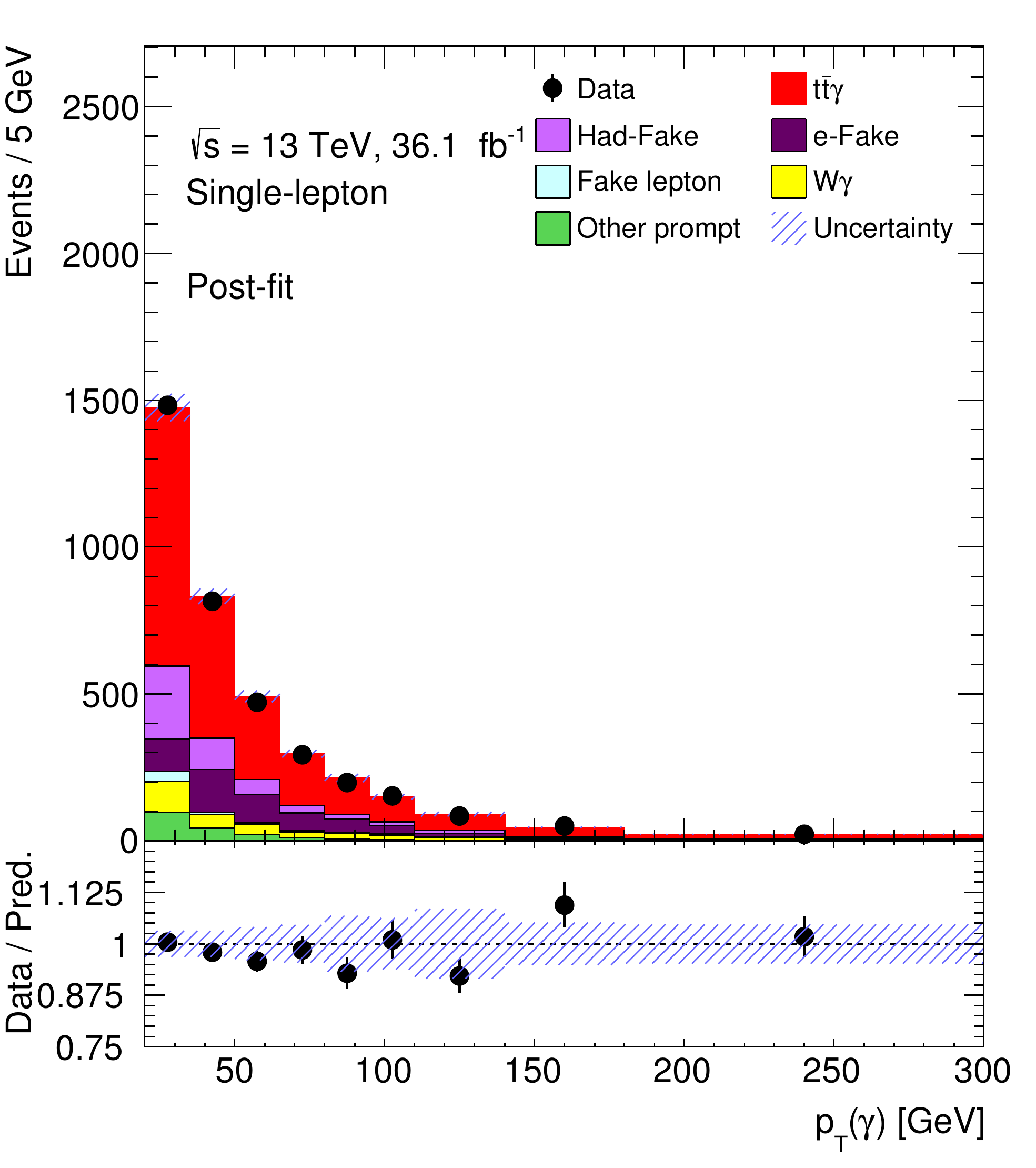}
\hspace{-0.023\linewidth}
\includegraphics[width=0.33\linewidth]{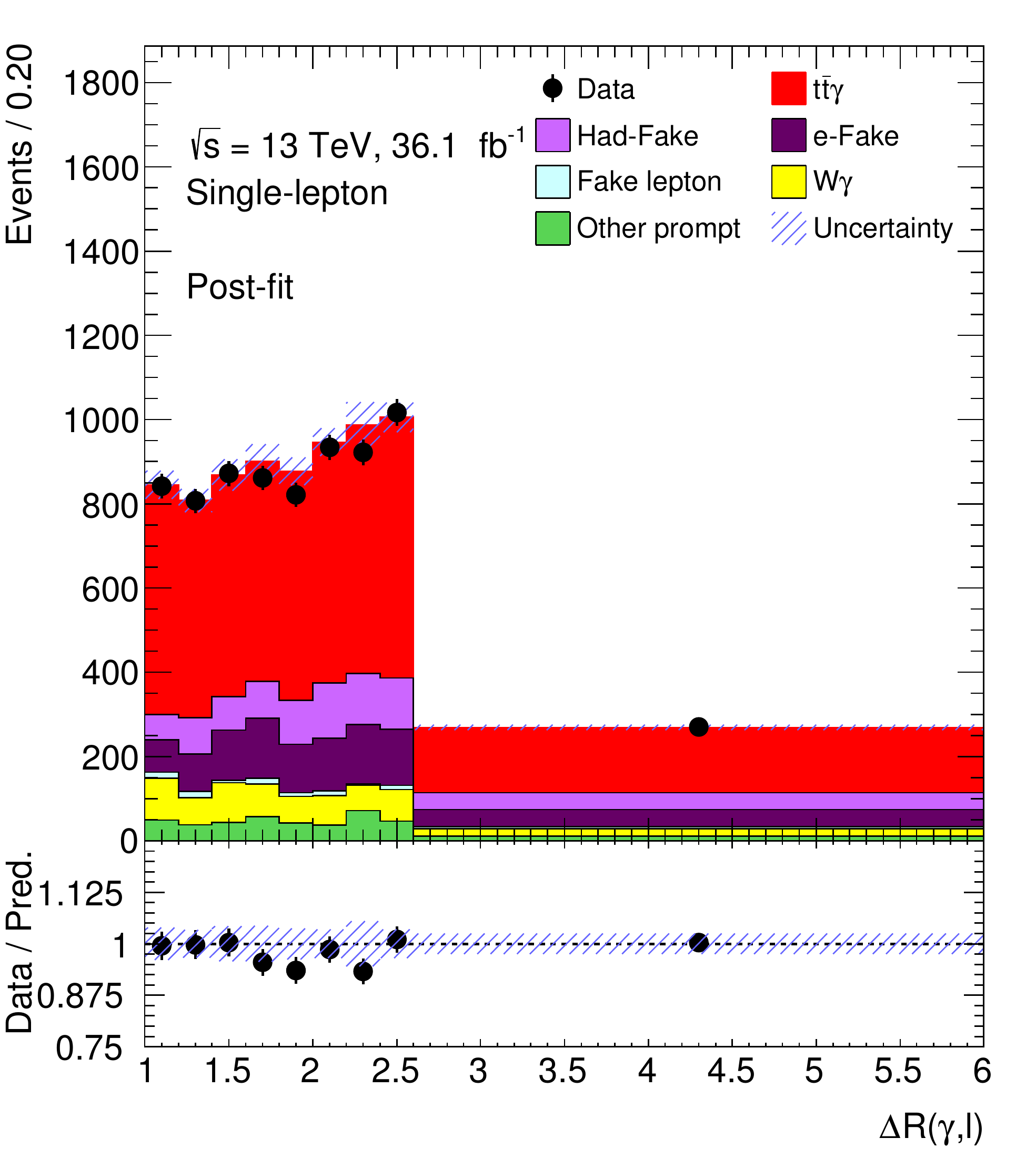}
\caption {Post-fit distributions for the \chljets channel where the \chljets \ELD distribution is used to extract uncertainties and signal strength parameters.}
\label{fig:postFitSL}
\end{figure}

\begin{figure}[!htbp]
\centering
\includegraphics[width=0.33\linewidth]{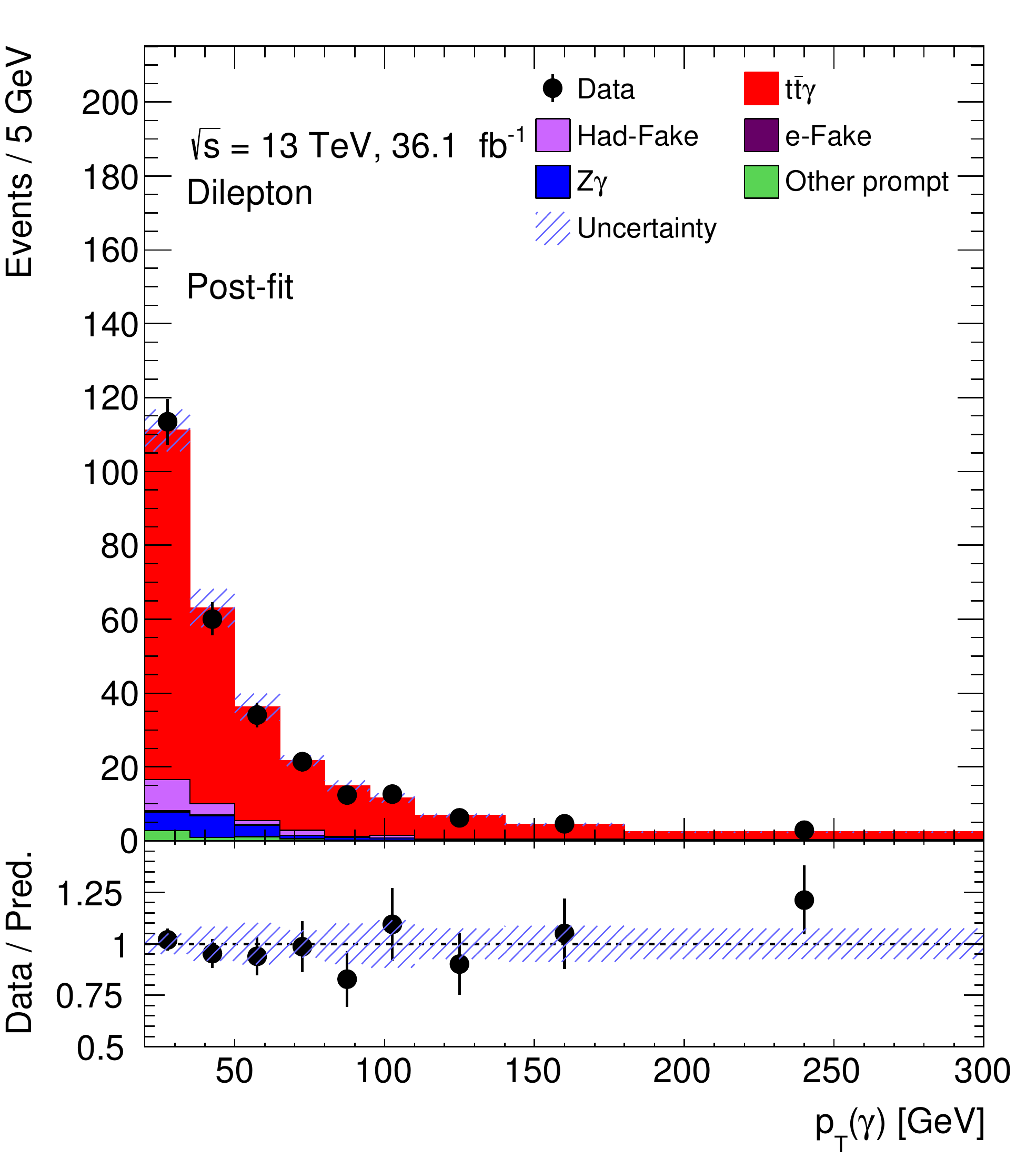}
\hspace{-0.023\linewidth}
\includegraphics[width=0.33\linewidth]{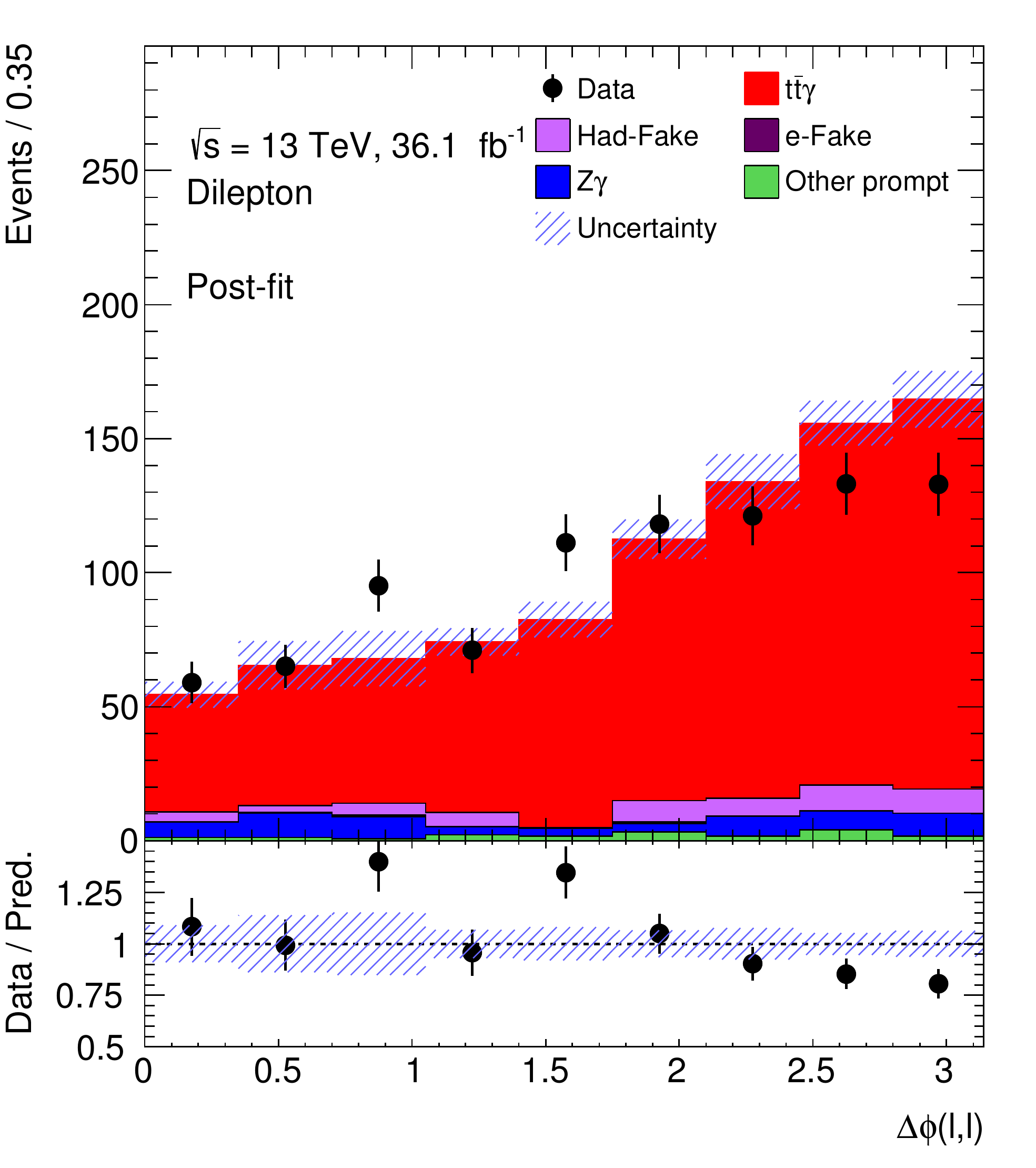}
\hspace{-0.023\linewidth}
\includegraphics[width=0.33\linewidth]{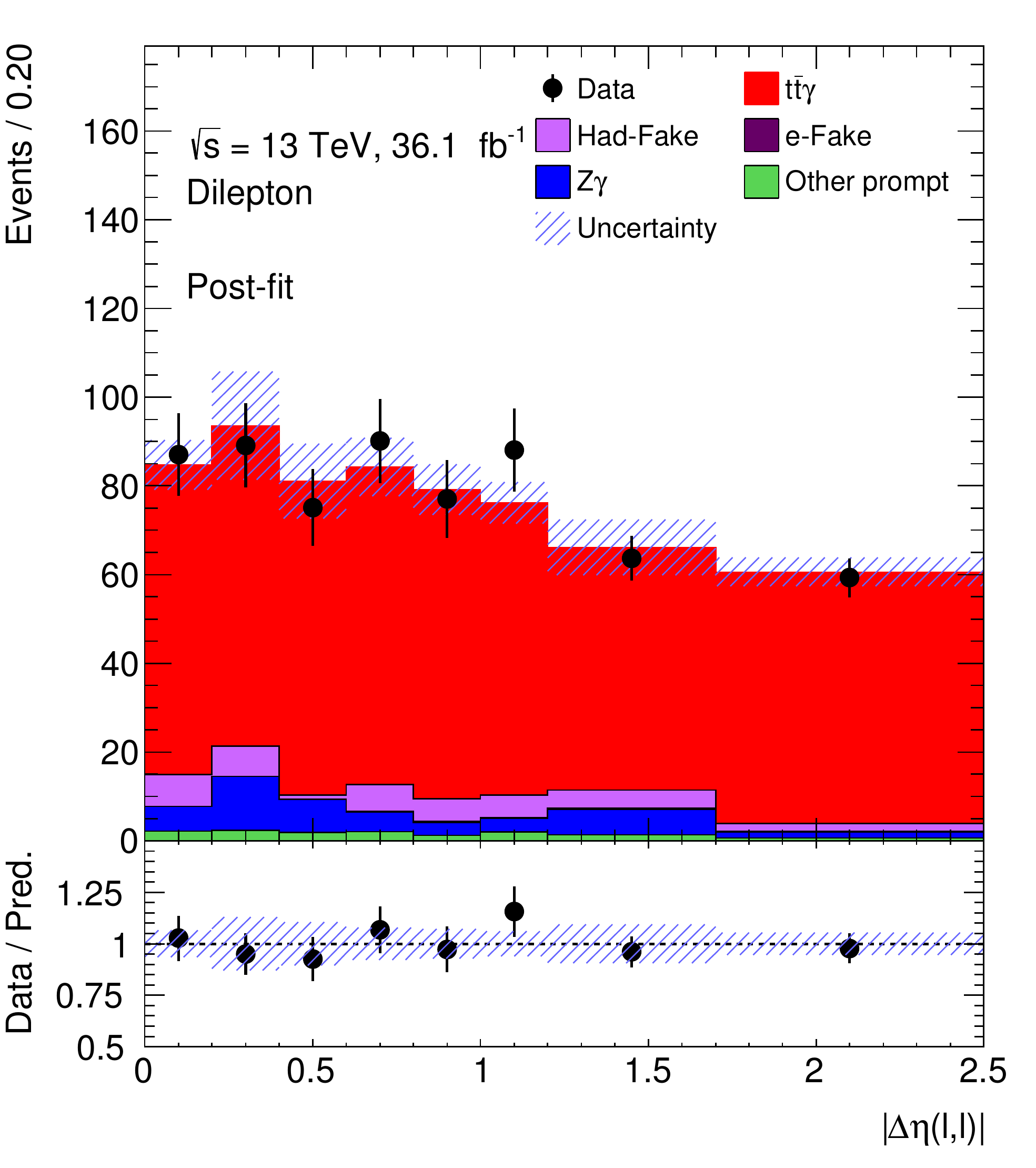}
\caption {Post-fit distributions for the \chll channel where the \chll \ELD distribution is used to extract uncertainties and signal strength parameters.}
\label{fig:postFitDL}
\end{figure}

The extracted signal strengths and NP results from the \ELD{}s can be applied to other distributions in the respective channels\footnote{Additionally, the post-fit results from the 5-channel combined fit can be applied to distributions of each channel, the advantage being that the total uncertainties will be smaller. This has been done and uncertainties still cover most of the deviations seen (the exception being the $\Delta \phi (l,l)$ distribution). This is not presented in this thesis due to the quantity of figures.}.
For the \chljets and \chll channels, a selection of distributions are shown in Figure~\ref{fig:postFitSL} and Figure~\ref{fig:postFitDL}. For the \chljets distributions the prediction and data agree. Similarly, this is the case for all but one of the \chll variables. The $\Delta \phi (l,l)$ distribution is sensitive to the \ttbar spin correlation and represents the azimuthal opening between the two leptons. 
This distribution translates to observing a higher spin correlation than that predicted by the SM.
This is consistent with findings found in~\cite{ATLAS:2018rgl} for the \chemu channel in \ttbar events\footnote{Since the \ttgamma events are a subset of the \ttbar events, this is not a new finding, but rather corroborates what the other analysis measured.}. 
For further differential \ttgamma measurements of this variable one can refer to~\cite{ATLAS-CONF-2018-048}.

\begin{figure}[!htbp]
\centering

\subfloat[Pre-fit \chll]{
\includegraphics[width=0.45\linewidth]{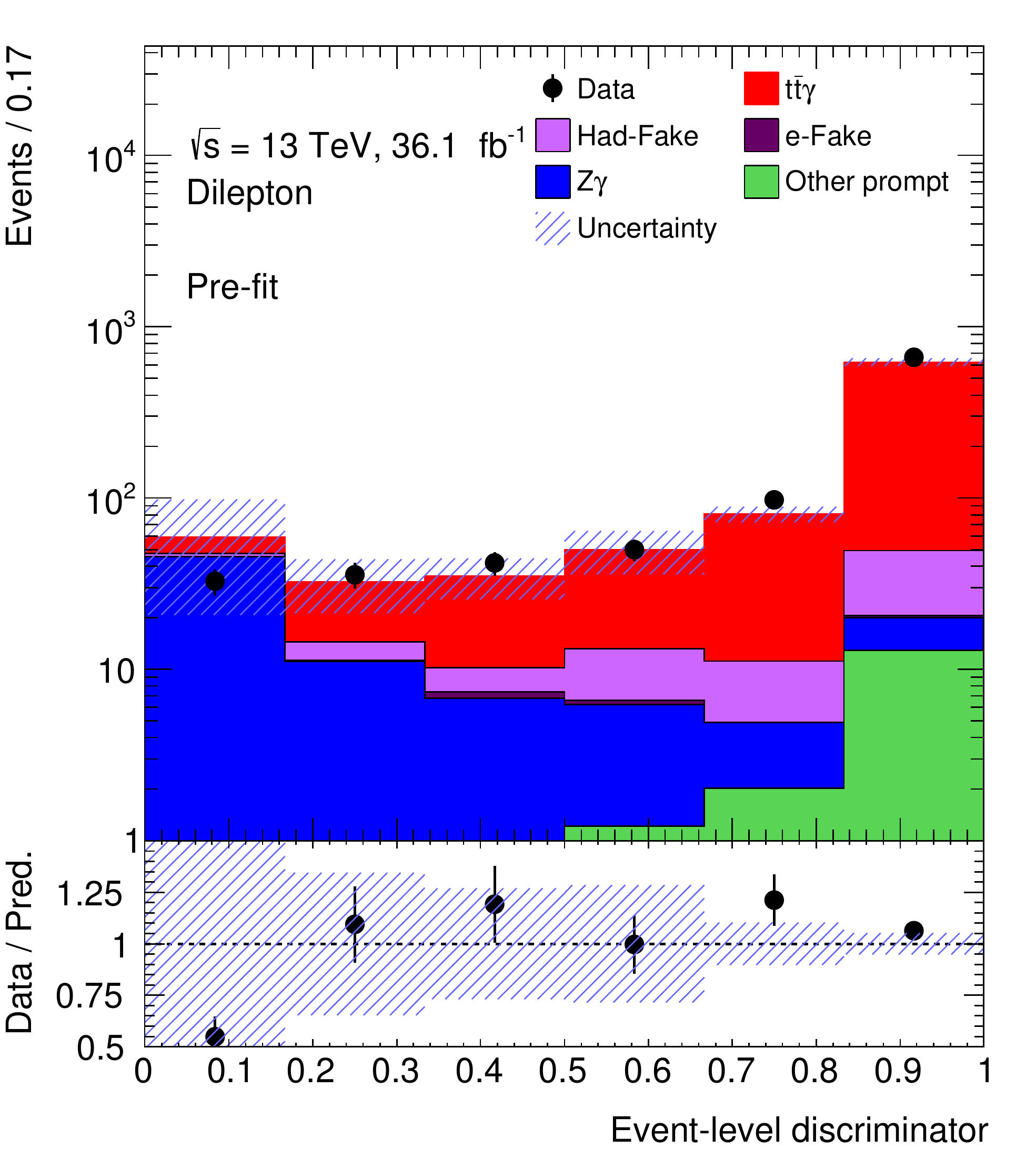}
}\hspace{-0.032\linewidth}
\subfloat[Post-fit \chll]{
\includegraphics[width=0.45\linewidth]{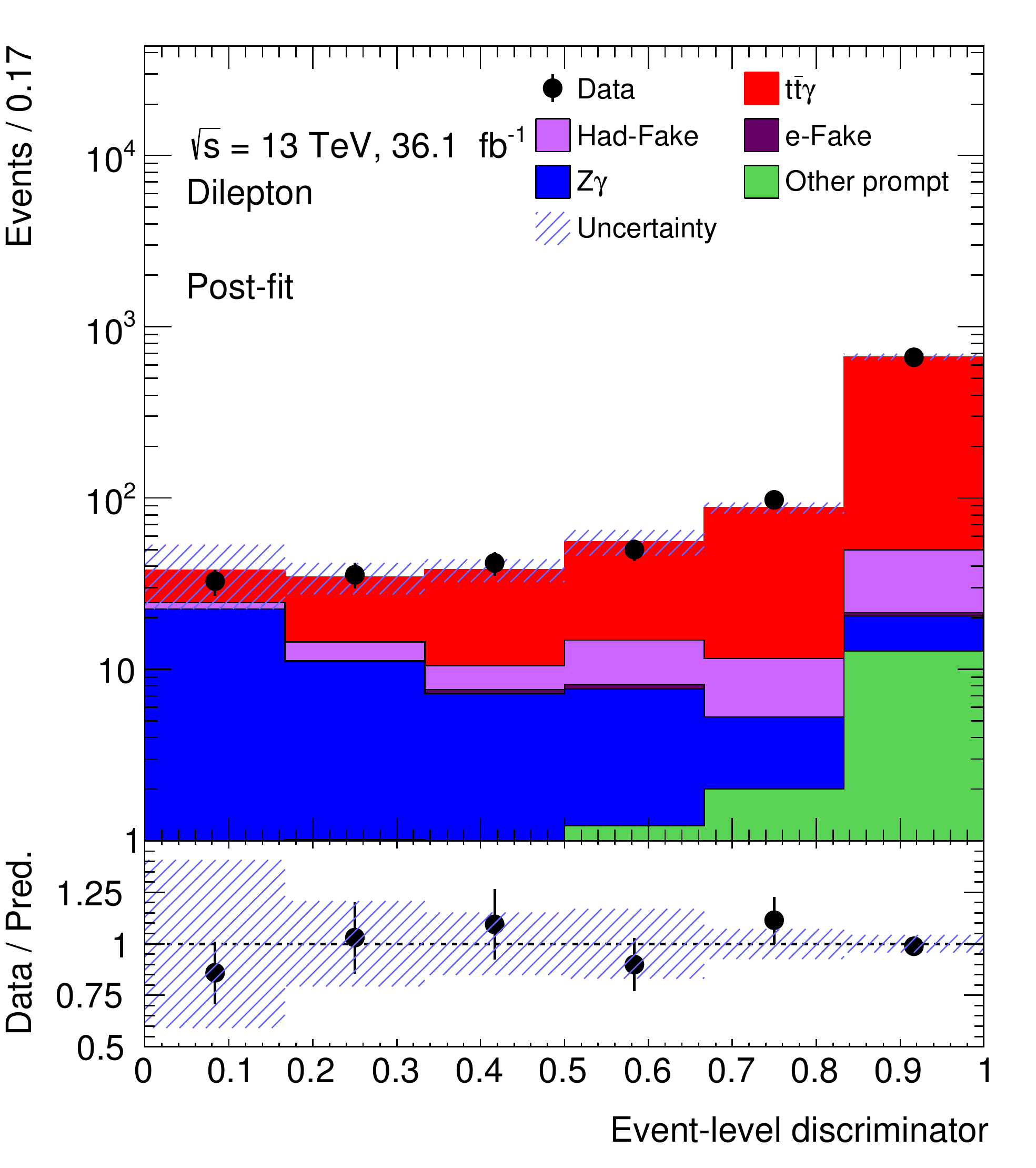}
}
\caption {Pre- and post-fit plots for the \chll channel where the \ELD distribution is used as the discriminating variable in the fit.}
\label{fig:postFitELDDL}
\end{figure}


\begin{table}[!htbp]
\begin{center}
\scalebox{0.85}{
\begin{tabular}{|c||r|r||r|r|r|r|r|}
\hline

 & \chljets & \chll  & \chejets & \chmujets & \chee & \chmumu & \chemu  \\ \hline \hline

\ttgamma & 7044 $\pm$ 350 &  782 $\pm$ 44 & 3548 $\pm$ 190 &  3326 $\pm$ 220 &  147 $\pm$ 17 &  192 $\pm$ 19 &  440 $\pm$ 27 \\ \hline  

\hfake & 1473 $\pm$ 180 &  49 $\pm$ 26 & 734.0 $\pm$ 83 &  736 $\pm$ 110 &  13 $\pm$ 7 &  8 $\pm$ 7 &  25 $\pm$ 13 \\ \hline  

\efake & 1621 $\pm$ 160 &  2 $\pm$ 1 & 939 $\pm$ 95 &  723 $\pm$ 74 &  1 $\pm$ 1 &  0 $\pm$ 1 &  1 $\pm$ 1 \\ \hline  

\QCD & 186 $\pm$ 68 &  - & 171 $\pm$ 58 &  36 $\pm$ 30 &  - & - & - \\ \hline

\Wgamma & 896 $\pm$ 370 &  - & 397 $\pm$ 210 &  545 $\pm$ 180 &  - & - & - \\ \hline

\Zgamma & - & 55 $\pm$ 29 & - & - & 29 $\pm$ 17 &  31 $\pm$ 20 & - \\ \hline

\Other  & 569 $\pm$ 180 &  18 $\pm$ 7 & 312 $\pm$ 110 &  304 $\pm$ 99 &  5 $\pm$ 2 &  4 $\pm$ 2 &  8 $\pm$ 3 \\ \hline  
\hline
Total & 11787 $\pm$ 180 &  906 $\pm$ 38 & 6102 $\pm$ 140 &  5669 $\pm$ 110 &  195 $\pm$ 19 &  235 $\pm$ 24 &  474 $\pm$ 24 \\ \hline  
\hline
Data  &11662 &  902 & 6002 &  5660 &  196 &  233 &  473 \\ \hline 

\hline

\end{tabular}
}
\caption{Post-fit yields for all \chljets and \chll channels. All uncertainties are included.}
\label{tab:postFitYields}
\end{center}
\end{table}

\FloatBarrier

All uncertainties that enter the fit can be grouped into appropriate categories and their impact on the fit explored. Table~\ref{tab:systSummary} shows grouped systematics and their relative contributions to the uncertainty on the signal strength, $\Delta \mu/\mu$. To calculate $\Delta \mu$, all NPs in a category are held fixed while the rest are free to float. 
The resulting uncertainty is then subtracted in quadrature from the initial total uncertainty.
Thus, what is shown in the table is a relative uncertainty given in percentage for an ``up" and ``down" variation.
The same table for the individual channels is shown in Appendix~\ref{sec:resultsappendix}.

Some systematics of interest on Table~\ref{tab:systSummary} are shaded in grey. We can see that the largest contributions come from jets, background modelling and in the case of \chljets channels, the PPT. For the \chll channels the signal modelling also plays a large role which is no surprise given how much signal there is. These groups can be further dissected. 

\begin{figure}[!htbp]
\centering
\includegraphics[width=0.70\textwidth]{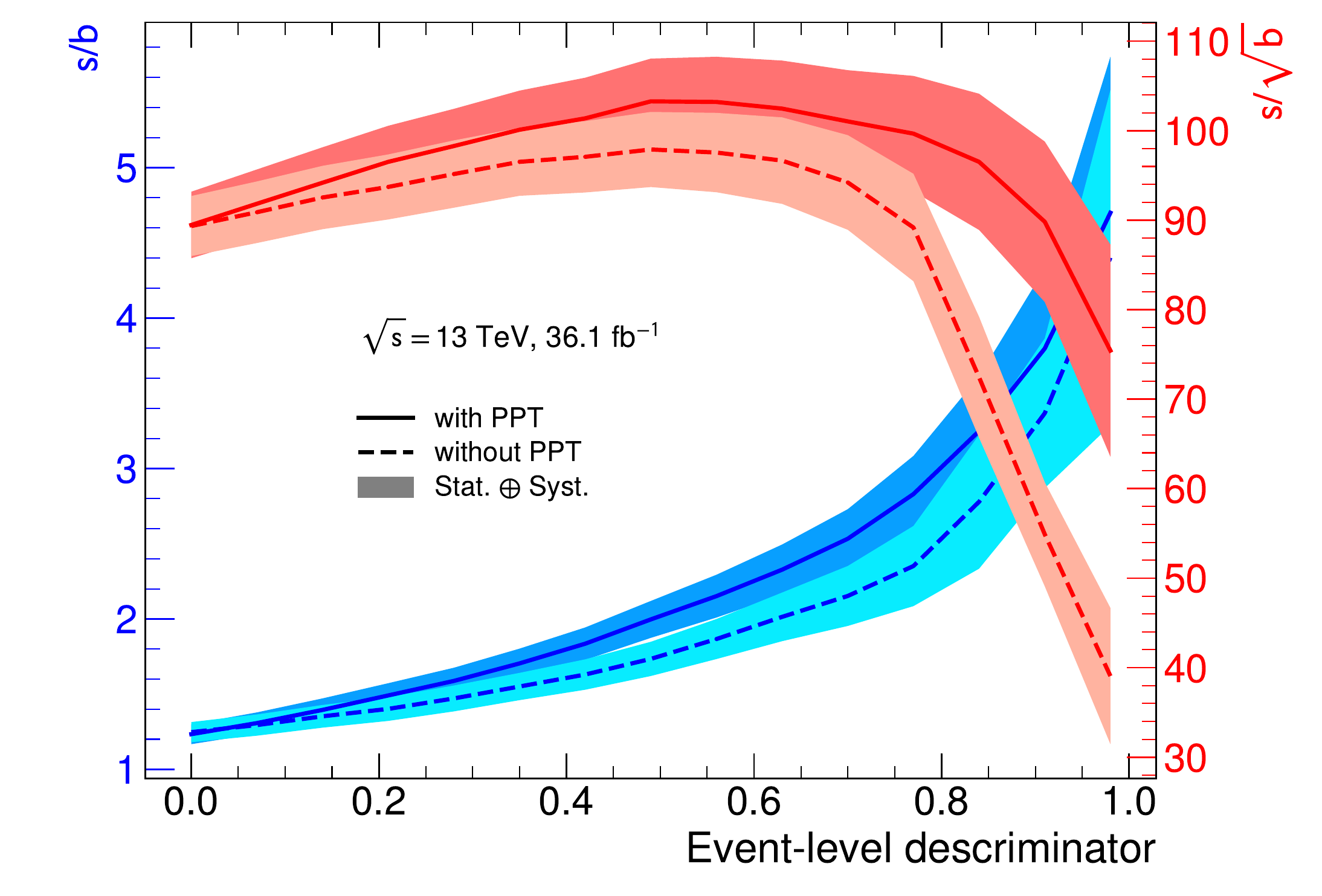}
\caption {Curves showing $s/b$ and $s/\sqrt{b}$ for different cuts on the \ELD distributions trained with and without the \PPT. 
The point at zero on the $x$-axis represents how this current analysis was conducted. Total uncertainties are included in the shaded bands.}
\label{fig:pptsoverbcurves}
\end{figure}

\begin{itemize}
\item Jets: Table~\ref{tab:systSummaryJets} further breaks down the jet contributions. The Jet Pileup RhoTopology component~\cite{ATL-PHYS-PUB-2015-015} is the largest contribution in all cases. This has already been studied within \ATLAS and future analyses will make use of a reduced Jet Pileup RhoTopology systematic uncertainty.
 
 \item Background modelling: Table~\ref{tab:systSummaryBkgs} categorises the background modelling according to the three major backgrounds (\efake, \hfake and \Other) where contributions largely come from analysis specific techniques used in Chapter~\ref{sec:ttgammaprocess}. 
Included separately is the \ttbar modelling, which is essentially theoretical uncertainties that are more difficult to reduce. 
For the \chljets channels a significant contribution of uncertainty in background modelling comes from the \Other category. This is because it includes all cross-section normalisation uncertainties for the \Other backgrounds, and also the uncertainty on the floating \Wgamma background, which is the largest contribution. The \efake modelling is one of the dominant systematics from the \chljets channels, which is not surprising since it is the largest background. 
For the \chll channel the dominant systematic component comes from the \hfake background. This analysis took  a generally conservative approach when deriving the \hfake systematics, even more so when extrapolating to the \chll channels. For future analyses this component can certainly be reduced using different techniques.
 
 \item \PPT: Table~\ref{tab:PPTsysts} breaks down the systematic contributions of the \PPT for the \chljets channels. As expected, the dominant component comes from the signal and background sources that have the most statistics. The \PPT's systematic uncertainties need to be revised as the current approach is too conservative.
 An additional study was done to assess the impact the \PPT has on the \ELD and the potential gain it can bring for future analyses. The \ELD was retrained in the identical fashion with identical parameters. The only difference being that the \PPT variable was removed. As expected, the ROC curve values are lower, and the discrimination for \hfake events disappears. The $s/b$ and $s/\sqrt{b}$ figures of merit ($s$=signal, $b$=sum of all backgrounds) are shown in Figure~\ref{fig:pptsoverbcurves}. From left to right, at each bin on the respective \ELD distribution a cut is made and the two figures of merit calculated using the remaining events. Thus, the point at zero reflects how this analysis was conducted. It can be seen that our total uncertainty is slightly larger due to the inclusion of the \PPT. However, the figure also shows the increase in sensitivity the \PPT can bring should we cut on the \ELD distribution in future analyses. 
 
 \end{itemize}
 
\begin{figure}[!htbp]
\centering
\includegraphics[trim={1.9cm 1.3cm 0.3cm 0.3cm},clip,width=0.65\textwidth]{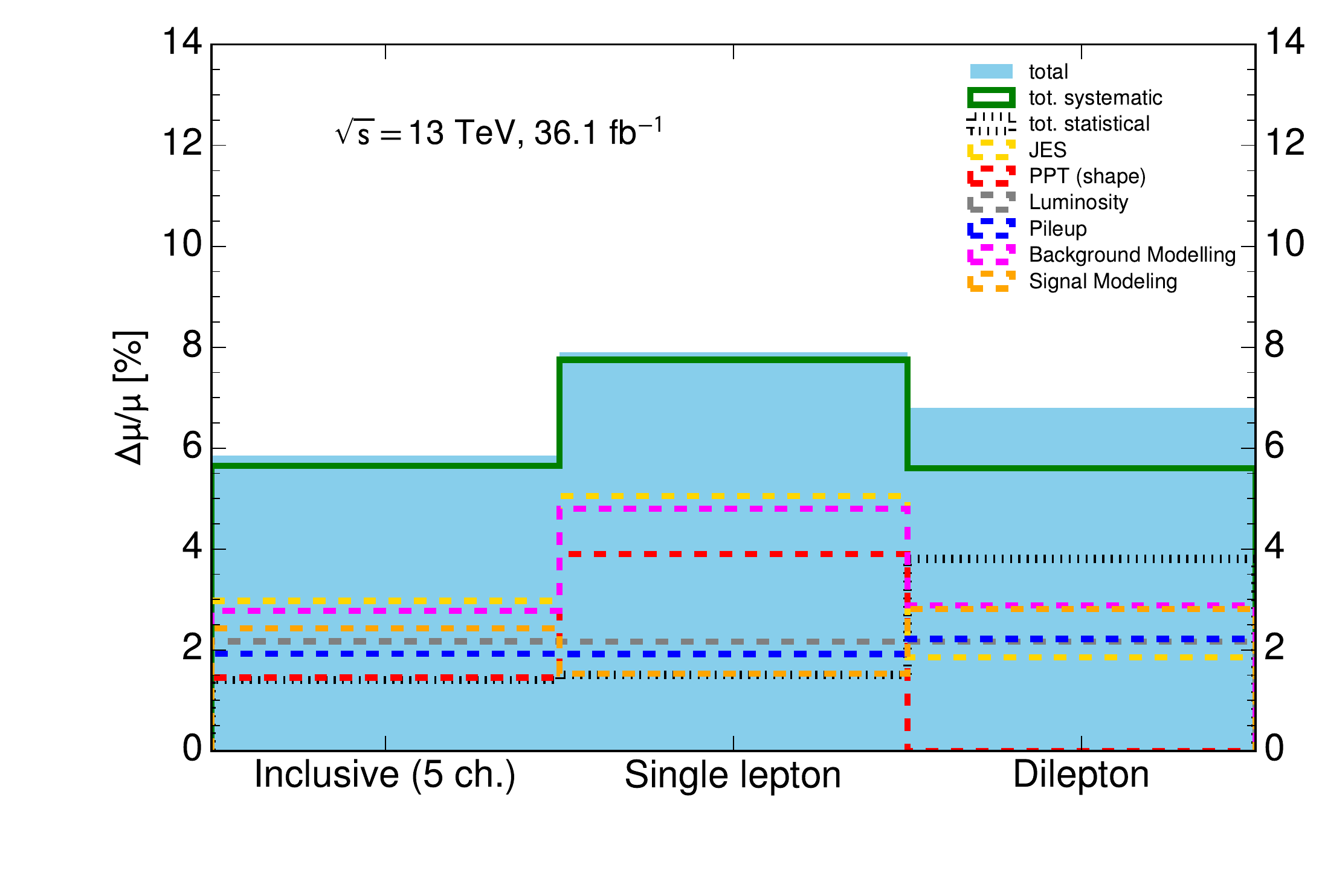}
\caption {The relative contribution of uncertainties shown in groups. A fit is performed where all NPs in a group are kept fixed while the rest are left floating. The resulting uncertainty is subtracted in quadrature from the original uncertainty to obtain $\Delta \mu$.}
\label{fig:systsOverview}
\end{figure}

 A summary of the above discussion is shown in Figure~\ref{fig:systsOverview} for all fits. The relative uncertainty ($\Delta \mu/\mu$) of a few selected groups is shown along with the total, systematic and statistical contributions. This plot gives a better feeling for the size of the uncertainties discussed above.
The next section discusses the stability (and thus the reliability of the results) of the fit by examining the individual NPs in more detail.

\begin{table}[!htbp]
\centering 
\scalebox{0.85}{
\begin{tabular}{|l|rr||rr|rr|}
 \hline
&  \multicolumn{2}{c||}{Inclusive}  &  \multicolumn{2}{c|}{Single-lepton}  &  \multicolumn{2}{c|}{Dilepton}  \\
\hline
\hline
& + [\%] & - [\%] & + [\%] & - [\%] & + [\%] & - [\%]  \\
\cline{2-3}
\cline{4-5}
\cline{5-7}

Signal Modelling & 2.50 & 2.36 & 1.59 & 1.47 & 2.87 & 2.75\\
\rowcolor{Gray!40}Jets & 3.14 & 2.83 & 5.40 & 4.80 & 2.02 & 1.89\\
Luminosity & 2.28 & 2.06 & 2.32 & 2.01 & 2.29 & 2.04\\
Pileup & 1.99 & 1.86 & 2.04 & 1.80 & 2.32 & 2.12\\
Photon Efficiencies & 1.06 & 0.97 & 1.08 & 0.95 & 1.07 & 0.96\\
$b$-Tagging & 0.41 & 0.40 & 0.76 & 0.86 & 0.36 & 0.40\\
\rowcolor{Gray!40}Background modelling & 2.81 & 2.74 & 4.80 & 4.80 & 2.91 & 2.86\\
Leptons & 0.96 & 0.88 & 0.26 & 0.27 & 1.30 & 1.20\\
\rowcolor{Gray!40}Prompt photon tagger (shape) & 1.45 & 1.45 & 3.80 & 4.00 &  - & -\\
E$\gamma$ (Resolution and scale) & 0.06 & 0.08 & 0.02 & 0.03 & 0.16 & 0.20\\
Template Statistics & 1.50 & 1.38 & 1.89 & 1.78 & 1.65 & 1.56\\
\hline
Total systematic & 5.8 & 5.5 & 7.9 & 7.6 & 5.8 & 5.4\\
Total statistical & 1.4 & 1.4 & 1.5 & 1.5 & 3.8 & 3.8\\
\hline
Total & 6.0 & 5.7 & 8.1 & 7.7 & 7.0 & 6.6\\
\hline
\end{tabular}
}
\caption{Relative effects on $\mu$ due to the grouped systematic sources. Fits are performed with nuisance parameters in each group held constant with the rest floating. This new uncertainty is subtracted in quadrature from the total uncertainty to obtain $\Delta \mu$.}
\label{tab:systSummary} 
\end{table}
\begin{table}[!htbp]
\centering 
\scalebox{0.85}{
\begin{tabular}{|l|rr||rr|rr|}
 \hline
Jets &  \multicolumn{2}{c||}{Inclusive}  &  \multicolumn{2}{c|}{Single-lepton}  &  \multicolumn{2}{c|}{Dilepton}  \\
\hline
\hline
& + [\%] & - [\%] & + [\%] & - [\%] & + [\%] & - [\%]  \\
\cline{2-3}
\cline{4-5}
\cline{5-7}
Jet Pileup (RhoTopology) & 2.31 & 2.10 & 3.80 & 3.38 & 1.66 & 1.52\\
Jet Flavour Composition & 0.05 & 0.05 & 0.13 & 0.14 & 0.10 & 0.10\\
Jet EtaInterCalibration & 0.13 & 0.12 & 1.10 & 1.01 & 0.21 & 0.18\\
Jet Effective NP & 1.79 & 1.61 & 2.91 & 2.53 & 0.97 & 0.89\\
Jet Flavour Resp. & 0.41 & 0.39 & 1.94 & 1.79 & 0.02 & 0.00\\
\hline
\end{tabular}
}
\caption{Breakdown of the effects that the larger jet systematic components have on $\mu$.}
\label{tab:systSummaryJets} 
\end{table}
\begin{table}[!htbp]
\centering 
\scalebox{0.85}{
\begin{tabular}{|l|rr||rr|rr|}
 \hline
Background modelling &  \multicolumn{2}{c||}{Inclusive}  &  \multicolumn{2}{c|}{Single-lepton}  &  \multicolumn{2}{c|}{Dilepton}  \\
\hline
\hline
& + [\%] & - [\%] & + [\%] & - [\%] & + [\%] & - [\%]  \\
\cline{2-3}
\cline{4-5}
\cline{5-7}
\efake modelling & 1.23 & 1.19 & 2.51 & 2.54 & 0.05 & 0.05\\
\ttbar modelling & 1.10 & 1.06 & 2.02 & 2.05 & 0.98 & 1.13\\
\hfake modelling & 1.93 & 1.87 & 1.41 & 1.42 & 2.13 & 2.17\\
\Other bkg modelling & 1.57 & 1.55 & 3.55 & 3.70 & 1.26 & 1.25\\
\hline
\end{tabular}
}
\caption{Breakdown of the effects that various background modelling components have on $\mu$. The \Wgamma floating normalisation uncertainty is included in ``Other prompt bkg modelling".}
\label{tab:systSummaryBkgs} 
\end{table}
\begin{table}[!htbp]
\centering 
\scalebox{0.85}{
\begin{tabular}{|l|rr|} 
\hline 
PPT &  \multicolumn{2}{c|}{Single-lepton}  \\
\hline 
\hline
& + [\%] & - [\%]  \\
\cline{2-3}

Prompt sources& 3.17 & 3.30\\
\efake sources& 1.74 & 1.92 \\
\hfake sources & 0.60 & 0.60\\
\hfake isolation & 0.25 & 0.23 \\

\hline
\end{tabular} 
}
\caption{Breakdown of the effects that the \PPT components have on $\mu$.}
\label{tab:PPTsysts} 
\end{table}

\FloatBarrier
\subsection{Fit cross checks}
\label{sec:NPdiscuss}

\begin{figure}[!htbp]
\centering
\includegraphics[trim={0 30cm 0 0},clip,width=0.49\linewidth]{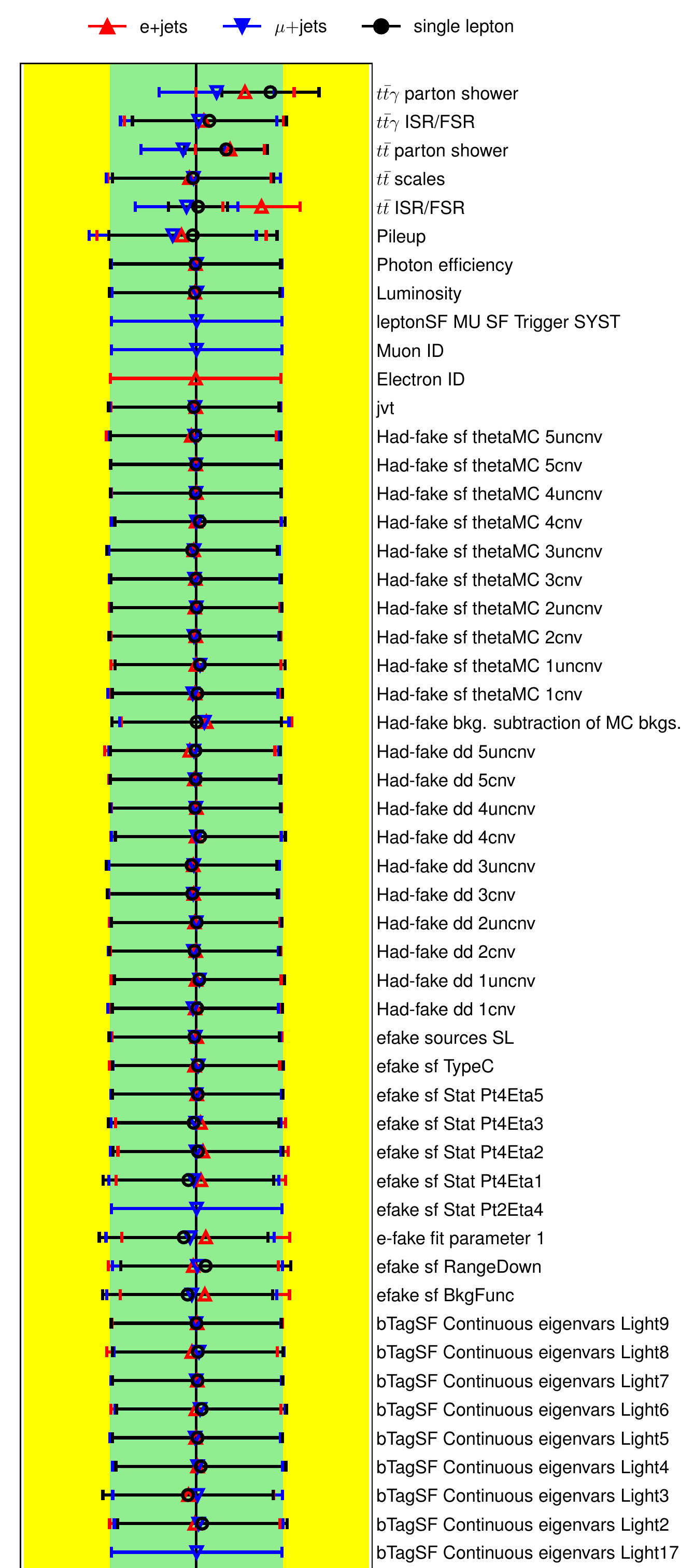}
\includegraphics[trim={0 0 0 30.9cm},clip,width=0.49\linewidth]{./figures/results/nps/pullPlot_SLchannels}
\caption [] {Post-fit pull plots for all nuisance parameters in the \chljets channels. Each set of points constitutes posteriors for a different fit. The ``Signal strength" and ``$W\gamma$ floating NP" both have expectation values of one. ``Had-fake" and ``efake" NPs relate to the data-driven methods and the uncertainties that arise from estimating these backgrounds. The ``bTagSF" NPs are associated with pseudo-continuous $b$-tagging for different flavour jets.}
\label{fig:postFitBrazilSLchannels}
\end{figure}

\begin{figure}[!htbp]
\centering
\includegraphics[trim={0 30.2cm 0 0},clip,width=0.49\linewidth]{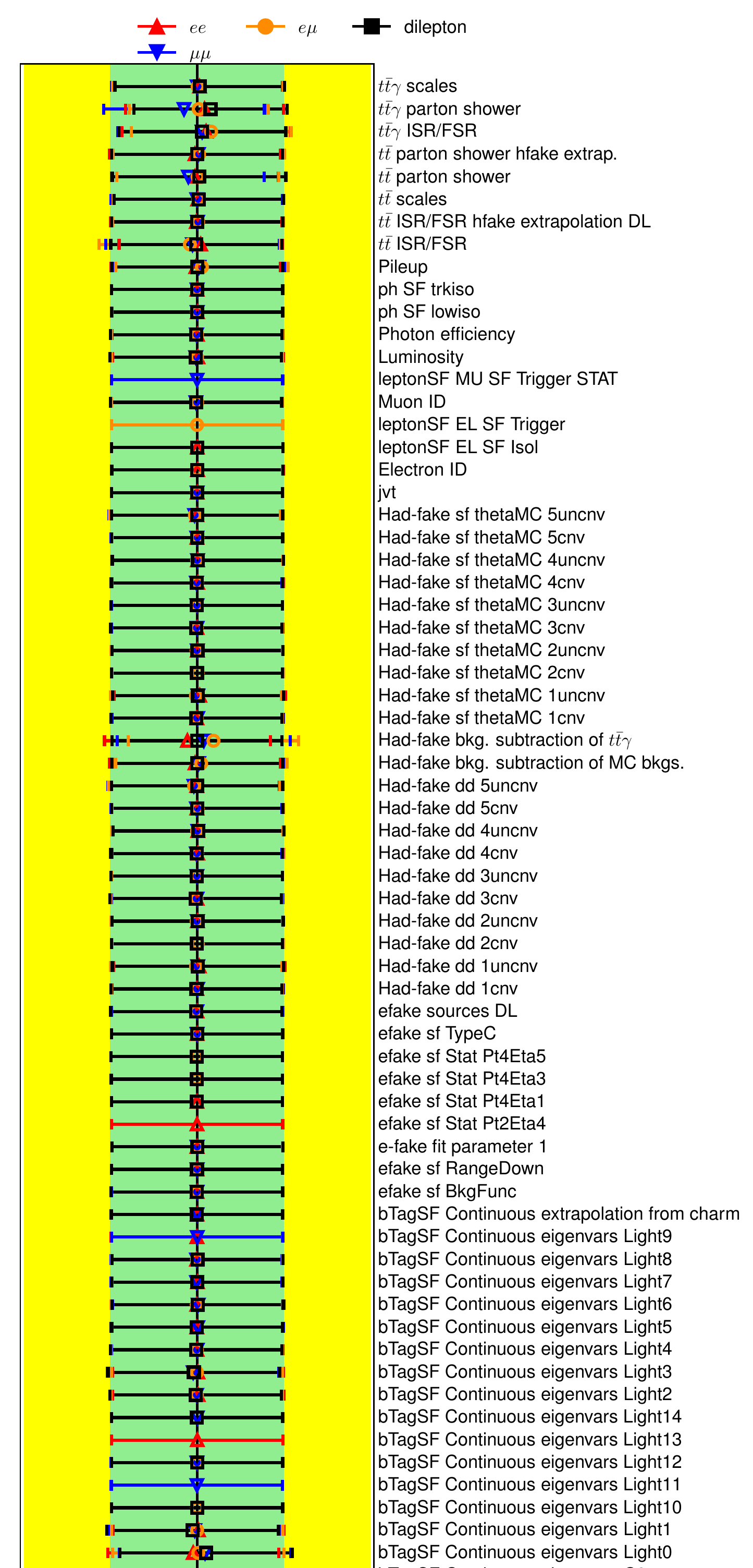}
\includegraphics[trim={0 0 0 30.5cm},clip,width=0.49\linewidth]{./figures/results/nps/pullPlot_DLchannels}
\caption [] {Post-fit pull plots for all nuisance parameters in the \chll channels. Each set of points constitutes posteriors for a different fit. The ``Signal strength" has an expectation value of one. ``Had-fake" and ``efake" NPs relate to the data-driven methods and the uncertainties that arise from estimating these backgrounds. The ``bTagSF" NPs are associated with pseudo-continuous $b$-tagging for different flavour jets.}
\label{fig:postFitBrazilDLchannels}
\end{figure}

\begin{figure}[!htbp]
\centering
\includegraphics[trim={0 30.1cm 0 0},clip,width=0.49\linewidth]{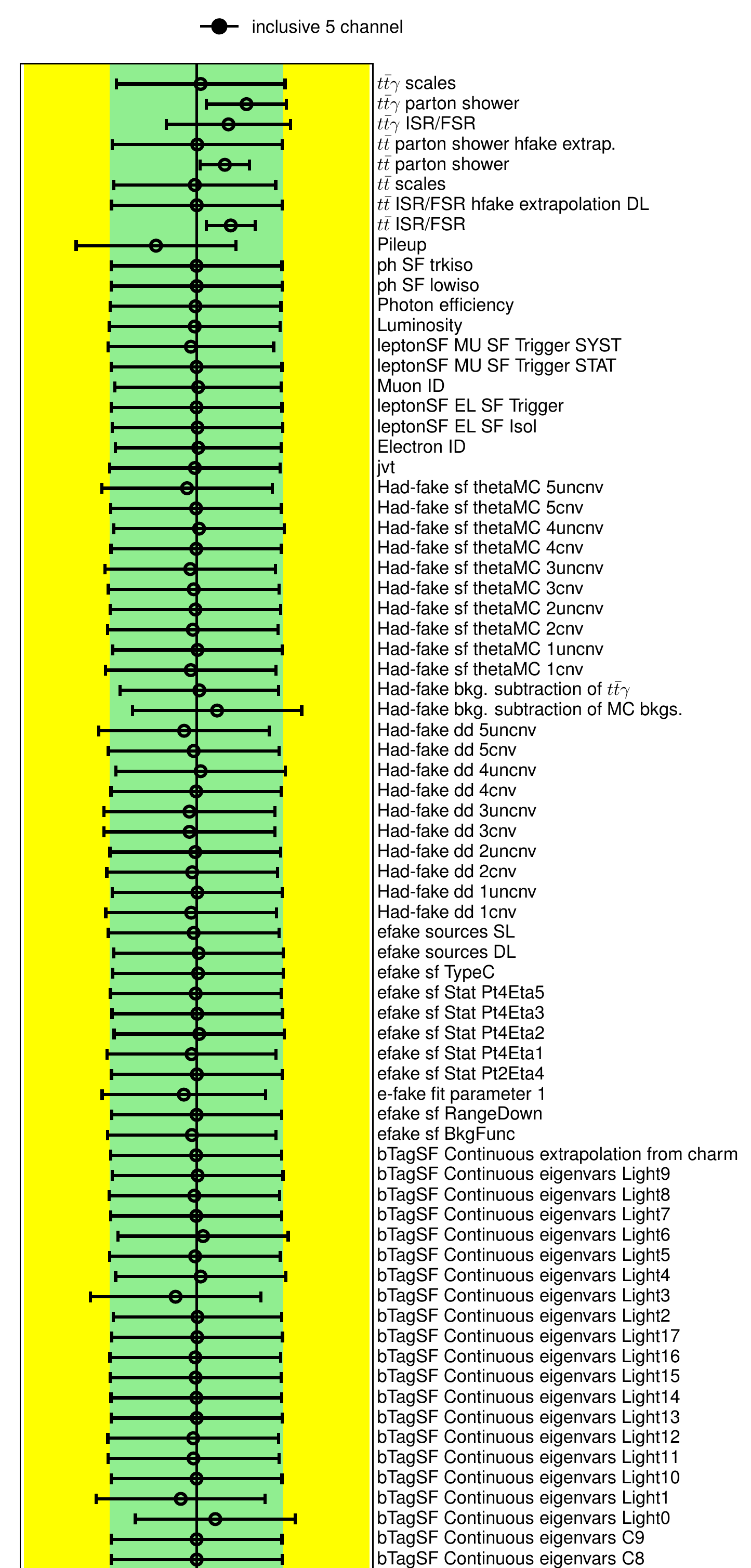}
\includegraphics[trim={0 0 0 30.7cm},clip,width=0.49\linewidth]{./figures/results/nps/pullPlot_Inclusivechannels}
\caption [] {Post-fit pull plots for all nuisance parameters in the 5-channel inclusive fit. The ``Signal strength" and ``$W\gamma$ floating NP" both have expectation values of one. ``Had-fake" and ``efake" NPs relate to the data-driven methods and the uncertainties that arise from estimating these backgrounds. The ``bTagSF" NPs are associated with pseudo-continuous $b$-tagging for different flavour jets.}
\label{fig:postFitBrazilInclusivechannels}
\end{figure}

This section presents cross checks similar to Section~\ref{sec:asimovcrosschecks} in order to ensure the fits are reliable as well as gain a deeper insight into the NPs. The major difference compared to the Asimov studies is that in a full fit the NPs and POIs are allowed to float. 
Figure~\ref{fig:postFitBrazilSLchannels} shows the posterior NPs in separate \chljets channels (overlaid into one figure). Similarly, Figure~\ref{fig:postFitBrazilDLchannels} shows the same for all the \chll channels.
It is important to note that for the two above mentioned plots each of the overlaid set of posterior NPs makes up the first seven fits in Figure~\ref{fig:crossSectionMu}. The posterior NPs for the final fit on Figure~\ref{fig:crossSectionMu} (5-channel combined) are shown in Figure~\ref{fig:postFitBrazilInclusivechannels}.

The fits in general are stable with only a few pulls and constraints that need to be further checked.
For the \chljets channels the main pulls/constraints are for: \ttgamma parton shower, \ttbar parton shower, \ttbar ISR/FSR, \Zgamma normalisation, \QCD and JER.
For the \chll channels the main pulls/constraints are for: \Zgamma factorisation and renormalisation scales, \Zgamma parton shower, \Zgamma normalisation and JER.
For the most part, pulls are seen for the same NPs that are constrained in the Asimov fit.
First, it is useful to see how much of a role these NPs actually play in the final fit. Figure~\ref{fig:rankingPlots} shows ranking plots for the top ten systematic uncertainties of the \chljets, \chll and 5-channel inclusive fit. The individual channels (\chejets, \chmujets, \chee, \chmumu and \chemu) are not shown due to being very similar. 
The top axis shows absolute difference and applies to the pre-fit and post-fit impact on $\mu$. Here, similar to tables presented earlier the NP in question is held fixed while the fit is performed with the remaining NPs floating. This is then subtracted in quadrature from the original uncertainty.
The bottom axis applies to the black points and shows the constraints in the same way as the previous pull plots.

For the \chljets rankings the Jet Pileup RhoTopology NP contributes the largest uncertainty. It has already been mentioned that dedicated groups within \ATLAS have reduced this for future analyses. The \PPT NPs have both prompt and \efake sources within the top 5, evidence that the approach to deriving the systematics for PPT can be revisited.
The \chll channel ranking shows that the overall systematic contributions are already quite small. 
The highly ranked systematics have little dependence on the techniques used to analyse this channel (with the exception of one \hfake derived NP in the top ten). Thus, it is a more straightforward channel to work with, which offers a very clean signal (largely due to the \chemu channel). With 36~fb$^{-1}$ of data collected, the \chll (\chemu) channel can already be considered the ``golden channel".
For the 5-channel inclusive measurement common systematics between the five channels (or subsets of the five channels) are treated as fully correlated. This allows for larger possible constraining power. 
Table~\ref{tab:rankedsyst} summarises the pulled or constrained NPs from the \chljets, \chll and 5-channel combined fits and their ranked impact on the signal strength. A dash indicates the NP lies outside the largest 40 contributions. The main NPs that are constrained are not ranked particularly high, and so do not impact the fit as much. 

\begin{figure}[!htbp]
\centering

\subfloat[\label{fig:efakecombined}\ttbar parton shower (\efake)]{
\includegraphics[width=0.48\linewidth]{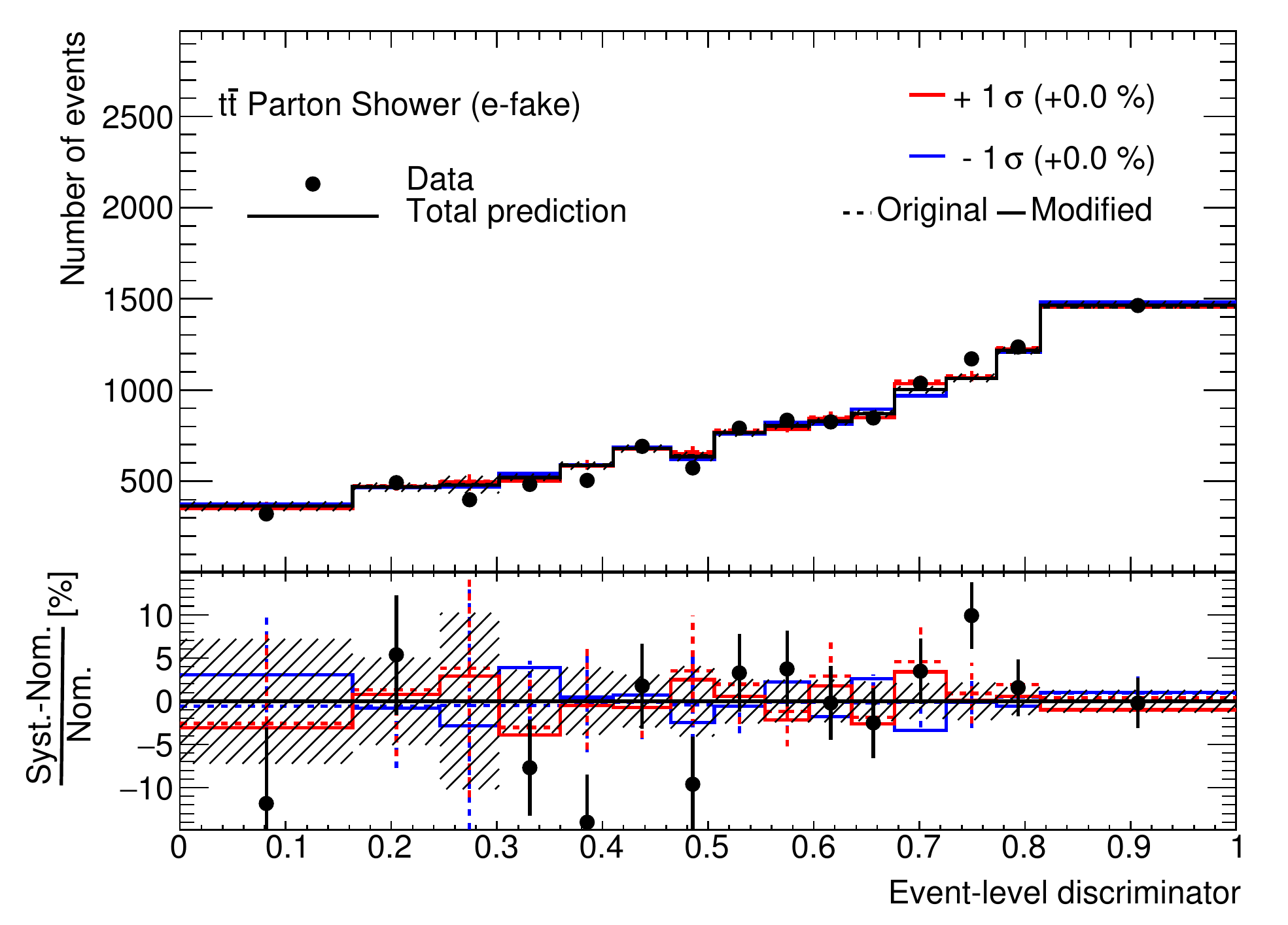}
}\hspace{-0.036\linewidth}
\subfloat[\label{fig:hfakecombined}\ttbar parton shower (\hfake)]{
\includegraphics[width=0.48\linewidth]{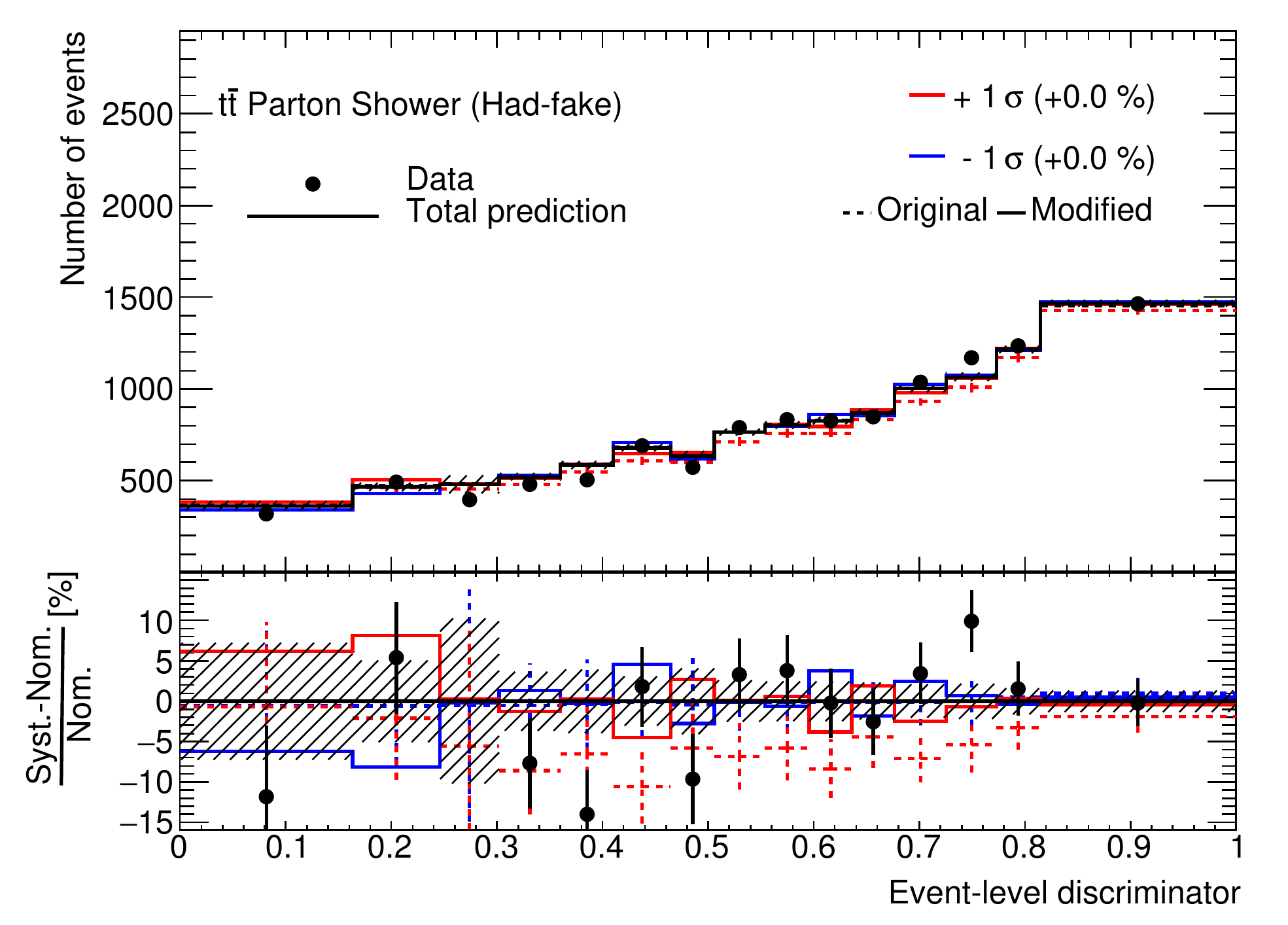}
}

\subfloat[\label{fig:ttgammapartonshowercombined}\ttgamma parton shower (shape)]{
\includegraphics[width=0.48\linewidth]{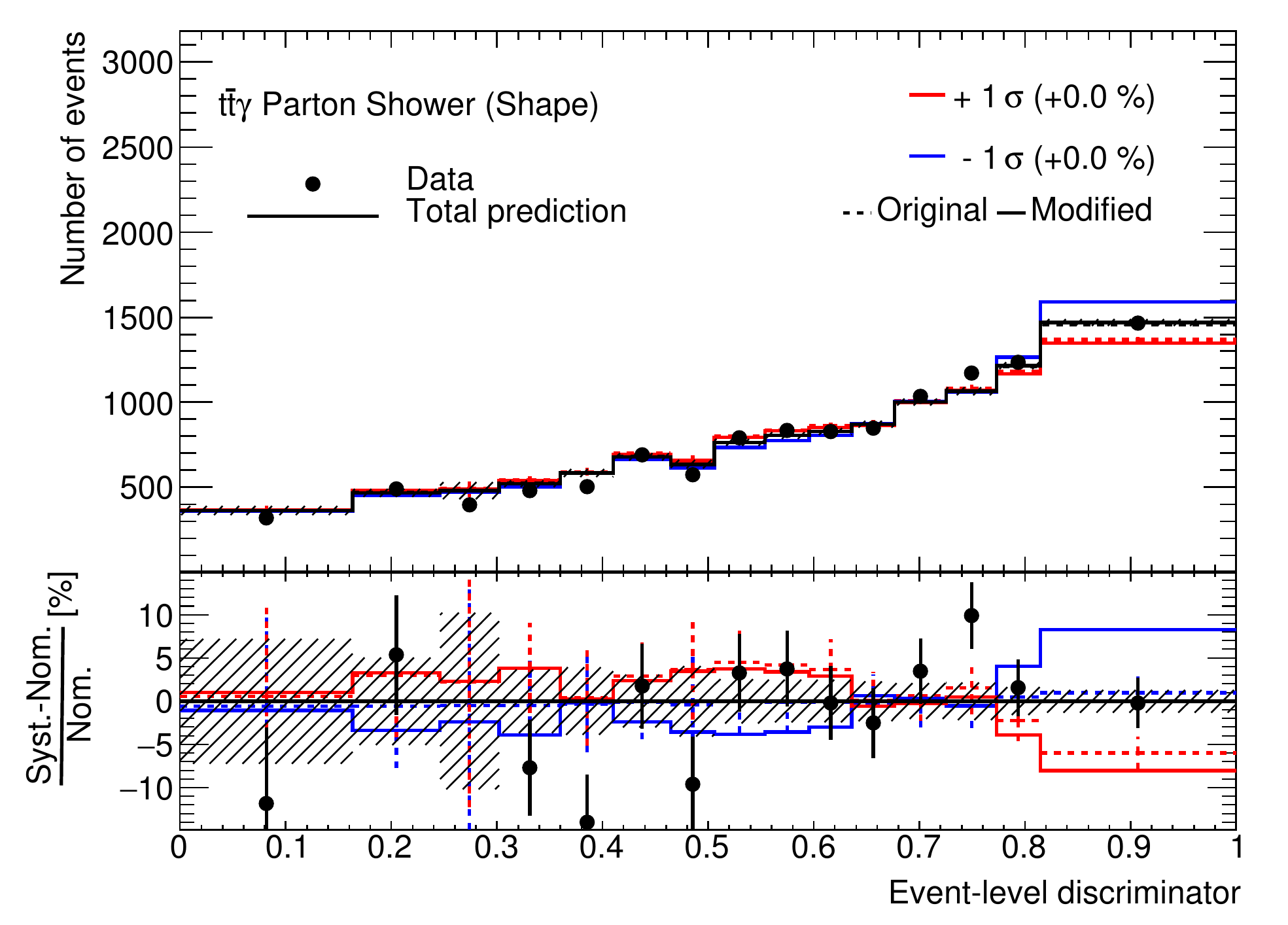}
}\hspace{-0.036\linewidth}
\subfloat[\label{fig:ttgammapartonshowercombinednorm}\ttgamma parton shower (norm)]{
\includegraphics[width=0.48\linewidth]{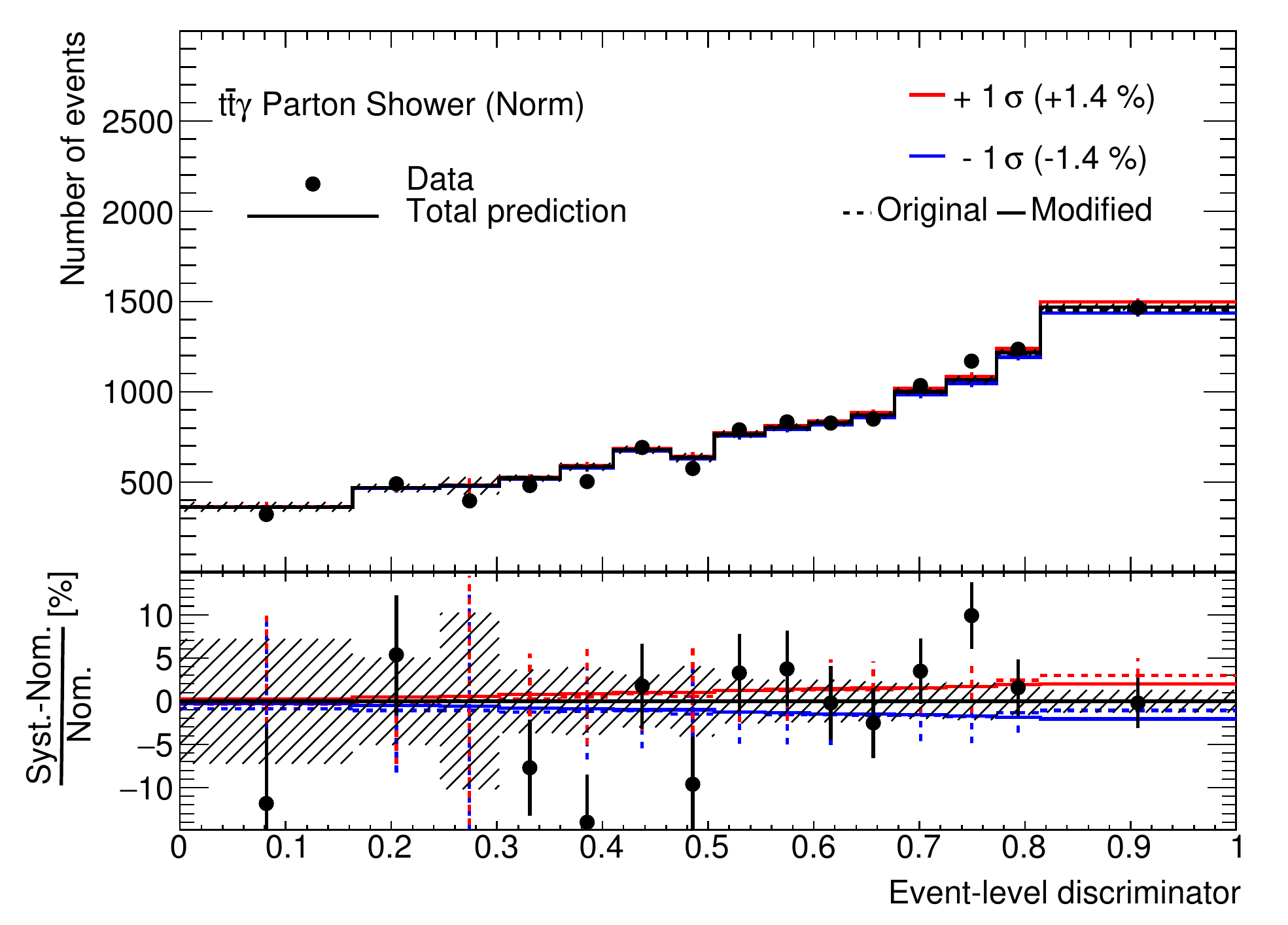}
}

\subfloat[\label{fig:rhotopcombined}Pileup RhoTopology]{
\includegraphics[width=0.48\linewidth]{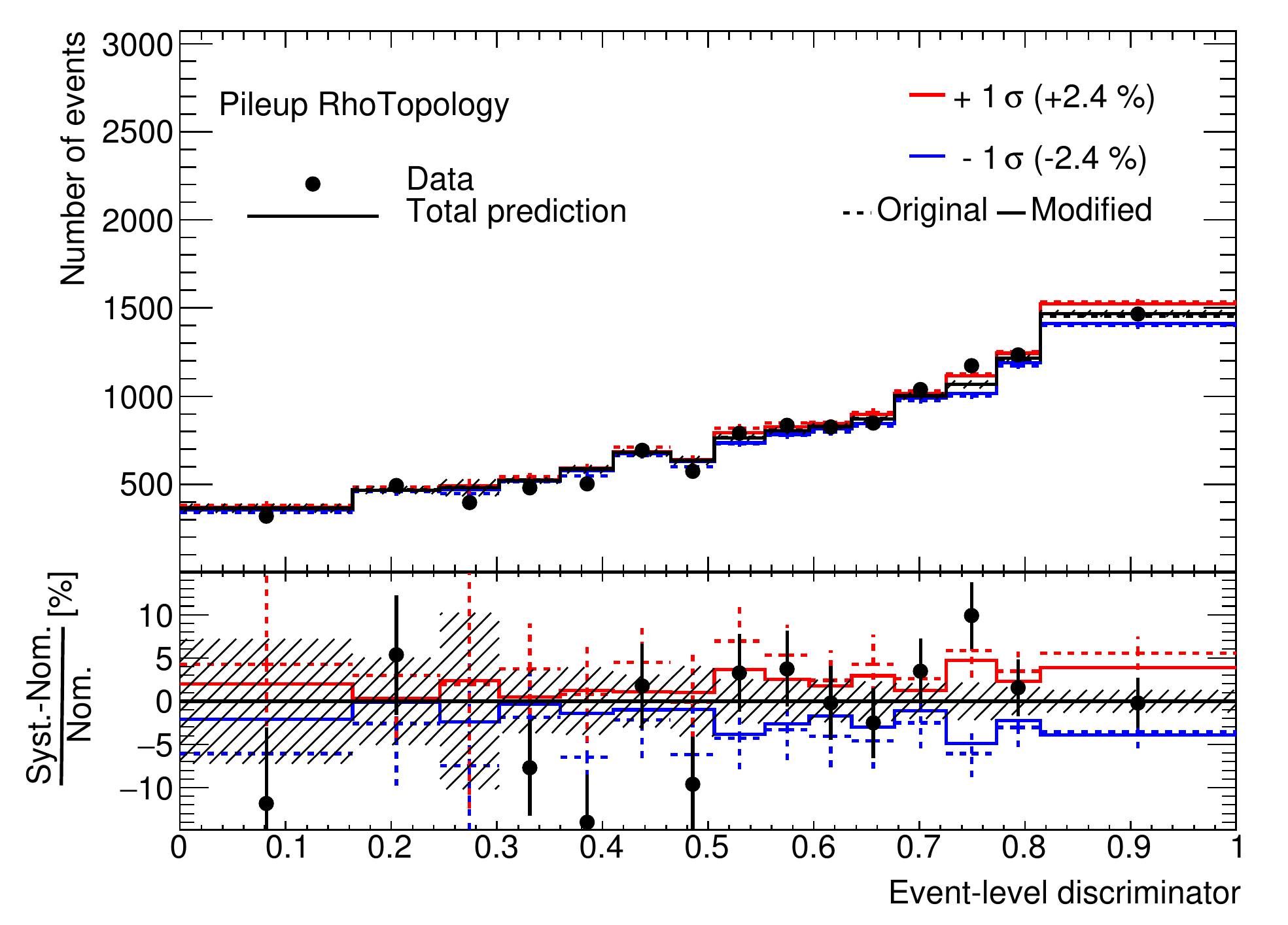}
}\hspace{-0.036\linewidth}
\subfloat[\label{fig:jercombined}Jet energy resolution]{
\includegraphics[width=0.48\linewidth]{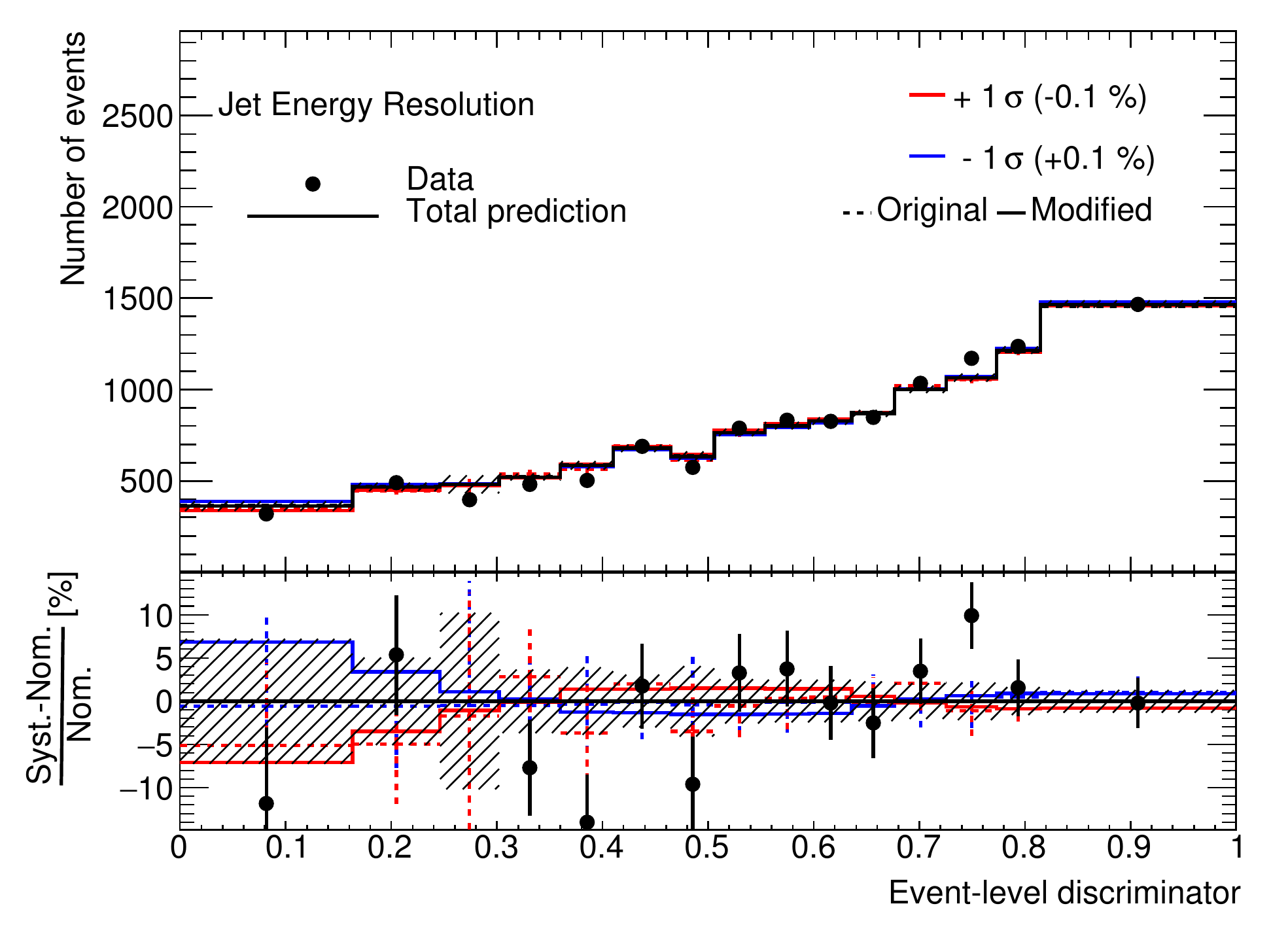}
}

\caption {A selection of NPs that are either pulled or constrained, or have large contributions in the final \chljets fit.}
\label{fig:redblucombinedSL}
\end{figure}

\begin{figure}[!htbp]
\centering
\subfloat[\label{fig:zgammacombinedps}\Zgamma parton shower]{
\includegraphics[width=0.48\linewidth]{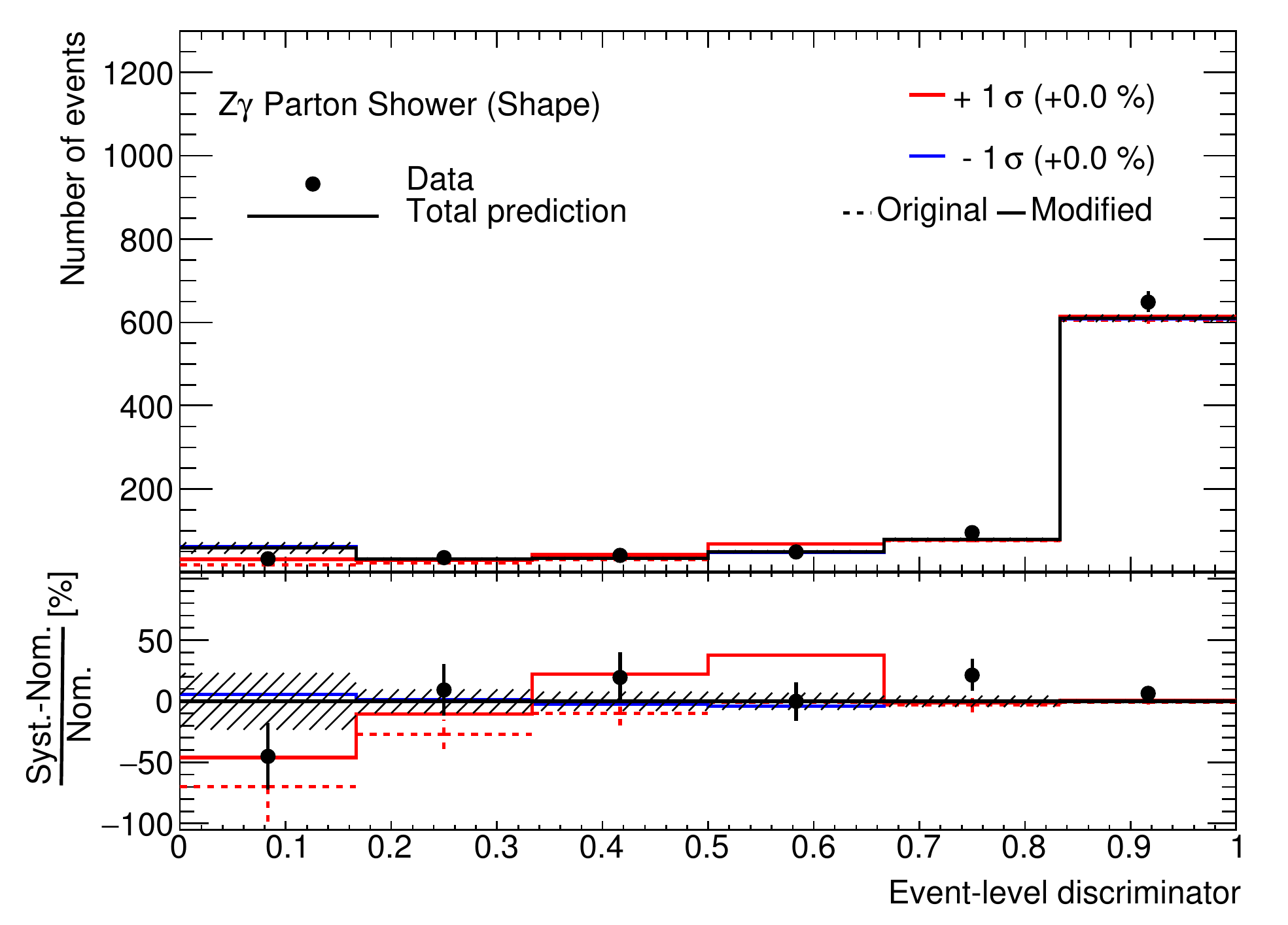}
}\hspace{-0.036\linewidth}
\subfloat[\label{fig:jercombinedDL}Jet energy resolution]{
\includegraphics[width=0.48\linewidth]{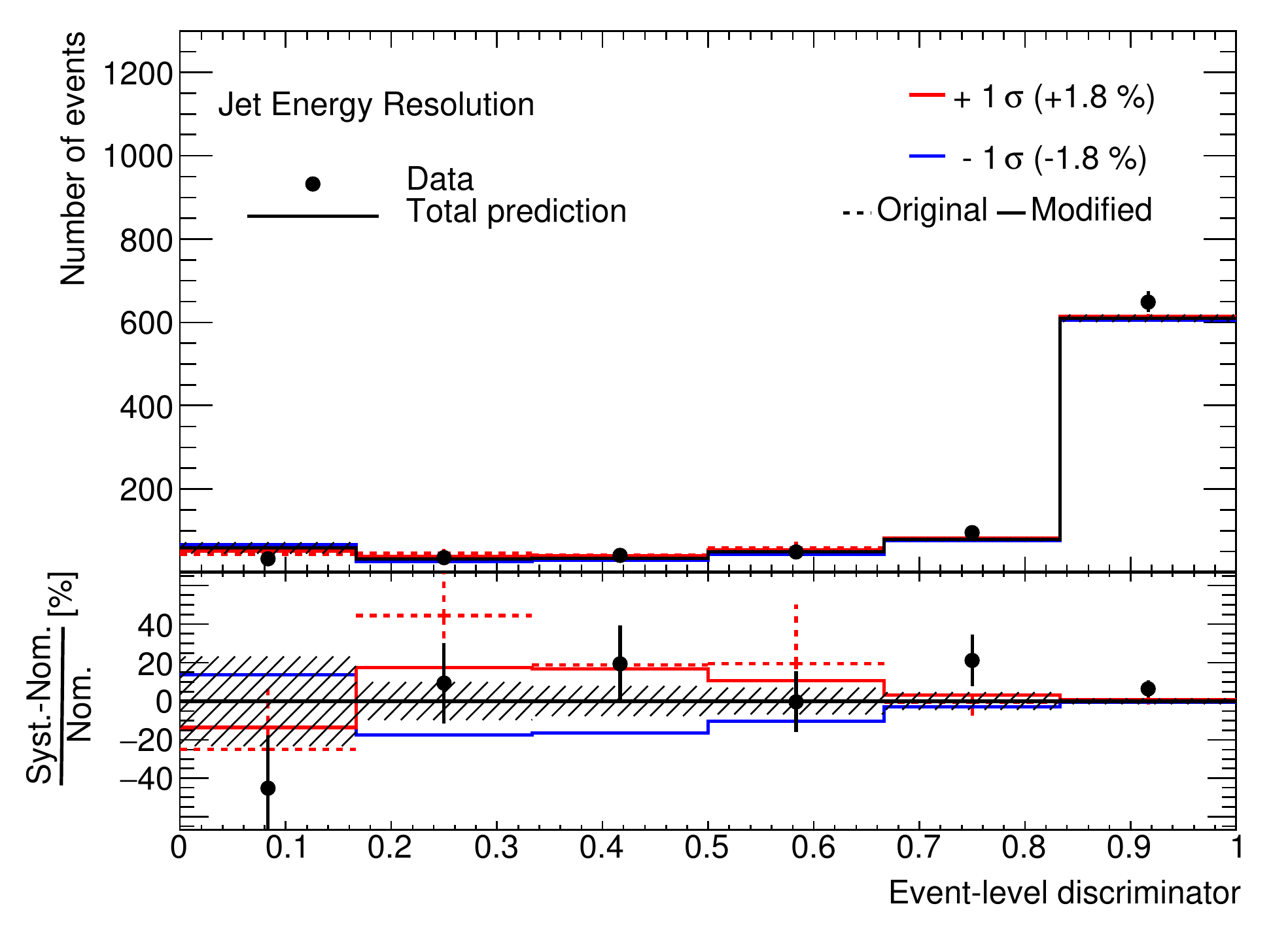}
}

\caption {A selection of NPs that are either pulled or constrained, or have large contributions in the final \chll fit.}
\label{fig:redblucombinedDL}
\end{figure}

A selection of NPs is presented based on whether or not they are pulled in the fit as well as how high they are ranked, and is shown in Figures~\ref{fig:redblucombinedSL} and \ref{fig:redblucombinedDL} for the \chljets and \chll channels, respectively. These figures are different to the previous NP figures in that they show the systematic contribution compared to the full signal + background prediction, as well as data. This enables us to see why certain NPs are pulled, and in which bin(s) the mis-modelling originates from.
For the \chljets channel there are a few aspects to notice:

\begin{itemize}
\item Figure~\ref{fig:efakecombined} and \ref{fig:hfakecombined} show the two main contributions of parton shower modelling to the \efake and \hfake background. More sources are included (and correlated) with these NPs, which arise from the data-driven methods and the effect varying the parton shower has when deriving the scale factors. However, since the pulls are in opposite directions for the \chejets and \chmujets channels, and well within $1\sigma$, this is no cause for concern.

\item For the \ttgamma parton shower the largest shape differences can be seen in the most signal-rich bin (Figure~\ref{fig:ttgammapartonshowercombined}). This can account for the constraint. While in the third bin from the right a fluctuation in data, also in a signal-rich bin, will cause a small pull towards $+1\sigma$. This is compounded by the normalisation part of the parton shower (Figure~\ref{fig:ttgammapartonshowercombinednorm}), where the third bin from the right also prefers the $+1\sigma$ variation.

\item Figure~\ref{fig:rhotopcombined} shows the highest ranked NP for the \chljets channel, which does not show any major pulls or constraints.

\item For JER (Figure~\ref{fig:jercombined}), the third bin from the right includes a data fluctuation. In this bin the ``up" and ``down" variations (which are arbitrary) flip, and so the $-1\sigma$ variation is preferred. This is indicated by the pull towards $-1\sigma$ on the pull plots.
\end{itemize}

For the \chll channel two NPs are shown. 

\begin{itemize}
\item The \Zgamma parton shower NP (Figure~\ref{fig:zgammacombinedps}) is heavily pulled and constrained in the \chll fits. This is due to the mis-modelling between the nominal \sherpaAll and systematic variation \madgraph samples. The data prefers the shape provided by the systematic variation. The overall impact this and other \Zgamma related NPs have on the fit are small, as can be seen from Table~\ref{tab:rankedsyst}.

\item  The JER (Figure~\ref{fig:jercombinedDL}) has a mild preference for the $+1\sigma$ variation as can be seen in the signal-rich (and generally event-rich) right side of the \ELD.

\end{itemize}

\begin{figure}[!htbp]
\centering
\subfloat[\chljets]{
\includegraphics[width=0.78\linewidth]{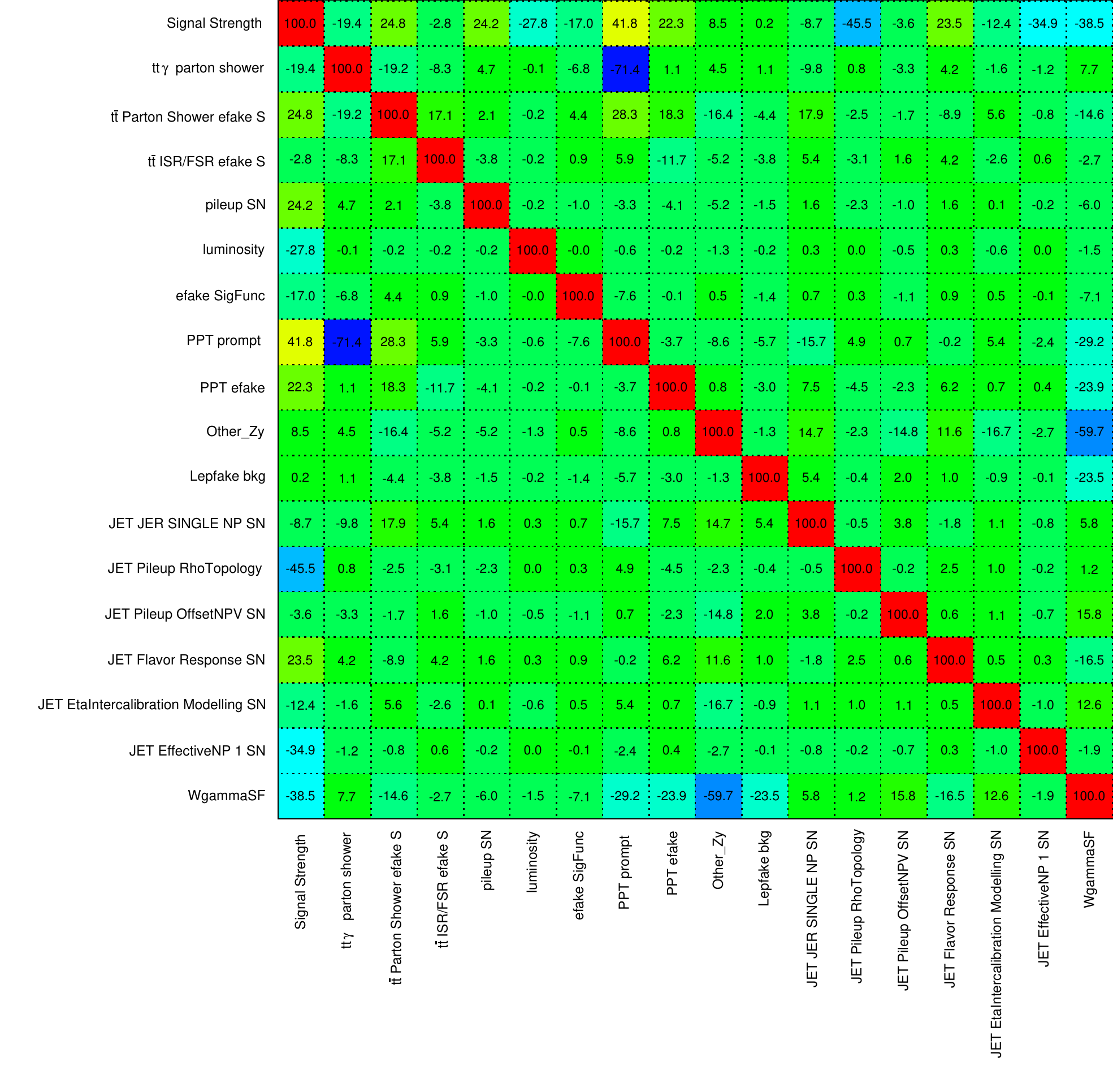}
}\hspace{-0.02\linewidth}

\subfloat[\chll]{
\includegraphics[width=0.64\linewidth]{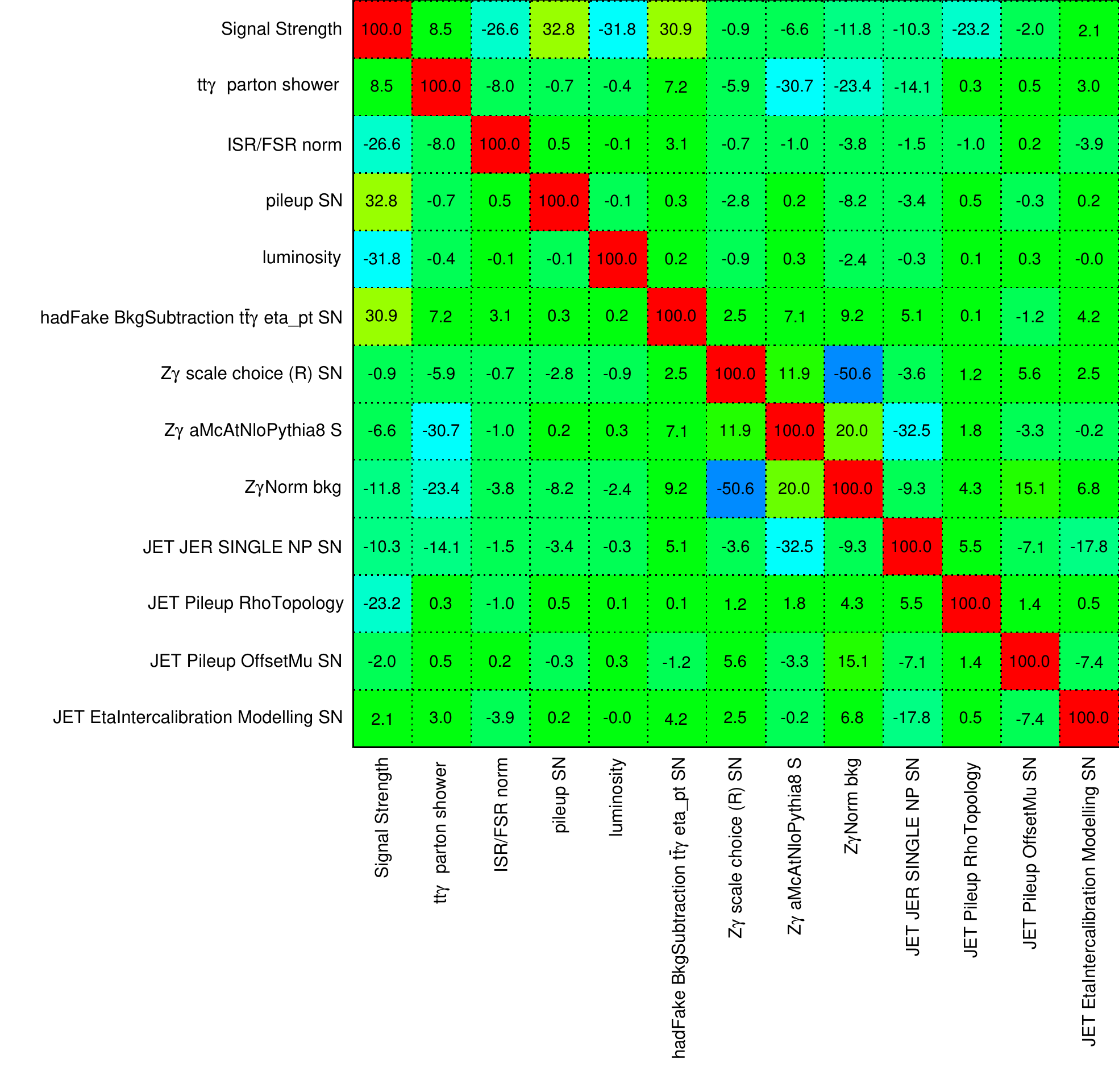}
}\hspace{-0.02\linewidth}

\caption {Correlations for the \chljets and \chll channel for each nuisance parameter and the signal strength. Only NPs with at least one correlation above 15\% are shown.}
\label{fig:correlations}
\end{figure}

Finally, Figure~\ref{fig:correlations} shows the correlations from the maximum likelihood fit for the \chljets and \chll channels. Only NPs that have at least one correlation above 15\% are included. 
For the \chljets channel, large correlations between \PPT prompt, Jet Pileup RhoTopology and the signal strength are unsurprising given their very conservative nature and thus their impact on the measurement. These are large NPs that predominantly affect the signal. For the \chll channel there are no alarming or surprisingly large correlations.
Thus, taking into account all studies presented above, we can conclude that the fit behaves as expected and is stable.

\begin{table}[htbp]
\centering
\begin{tabular}{|c|l|c|}
\hline
& Nuisance Parameter & Rank \\ \hline
\hline
\multirow{ 6}{*}{\chljets} & \ttbar Parton shower & 6 \\ 
& \ttgamma Parton shower & 9 \\ 
& JER & 17   \\ 
& \ttbar ISR/FSR & 18    \\ 
& \Zgamma Normalisation & 23    \\ 
& Lepfake & -   \\  \hline

\multirow{ 4}{*}{\chll} & \Zgamma Normalisation & 12    \\ 
& JER & 14  \\ 
& \Zgamma Parton shower & 28  \\ 
& \Zgamma F+R scales & -   \\ \hline

\multirow{ 9}{*}{5-channel combined} & Pileup & 3 \\
& \ttgamma ISR/FSR & 5 \\
& PPT prompt photons & 7 \\
& \ttbar Parton shower & 17  \\ 
& \ttgamma Parton shower & 18 \\ 
& \ttbar ISR/FSR & 19 \\
& Lepfake & - \\
& \Zgamma Normalisation & -    \\ 
& \Zgamma Parton shower & -  \\ 

\hline

\end{tabular}
\caption{The most pulled/constrained NPs in the \chljets and \chll fits and their ranking. A dash indicates the NP is not within the top 40 entries.}
\label{tab:rankedsyst}
\end{table}

\FloatBarrier

\begin{figure}[!htbp]
\centering
\subfloat[\chljets]{
\includegraphics[width=0.50\linewidth]{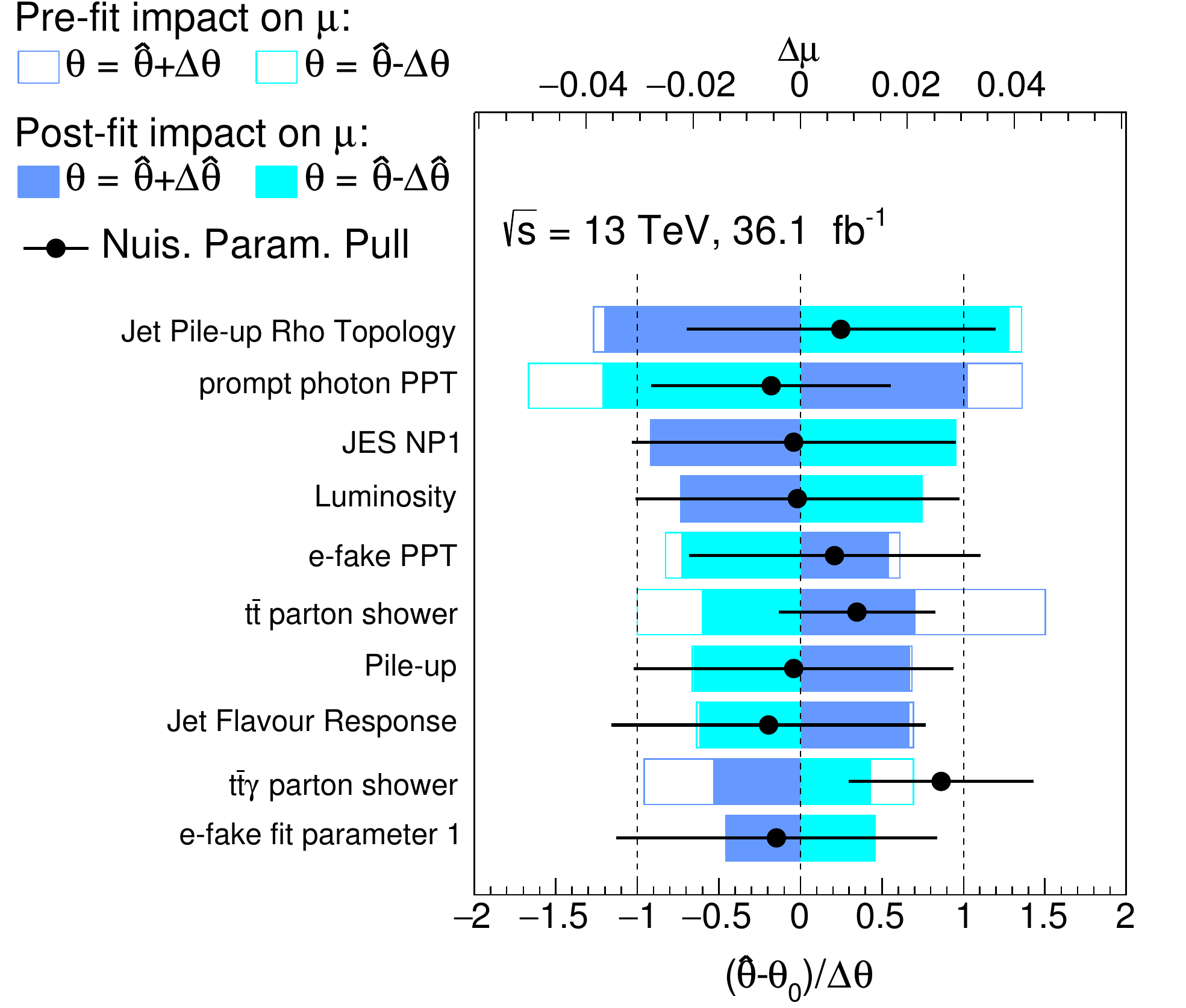}
}\hspace{-0.043\linewidth}
\subfloat[\chll]{
\includegraphics[width=0.50\linewidth]{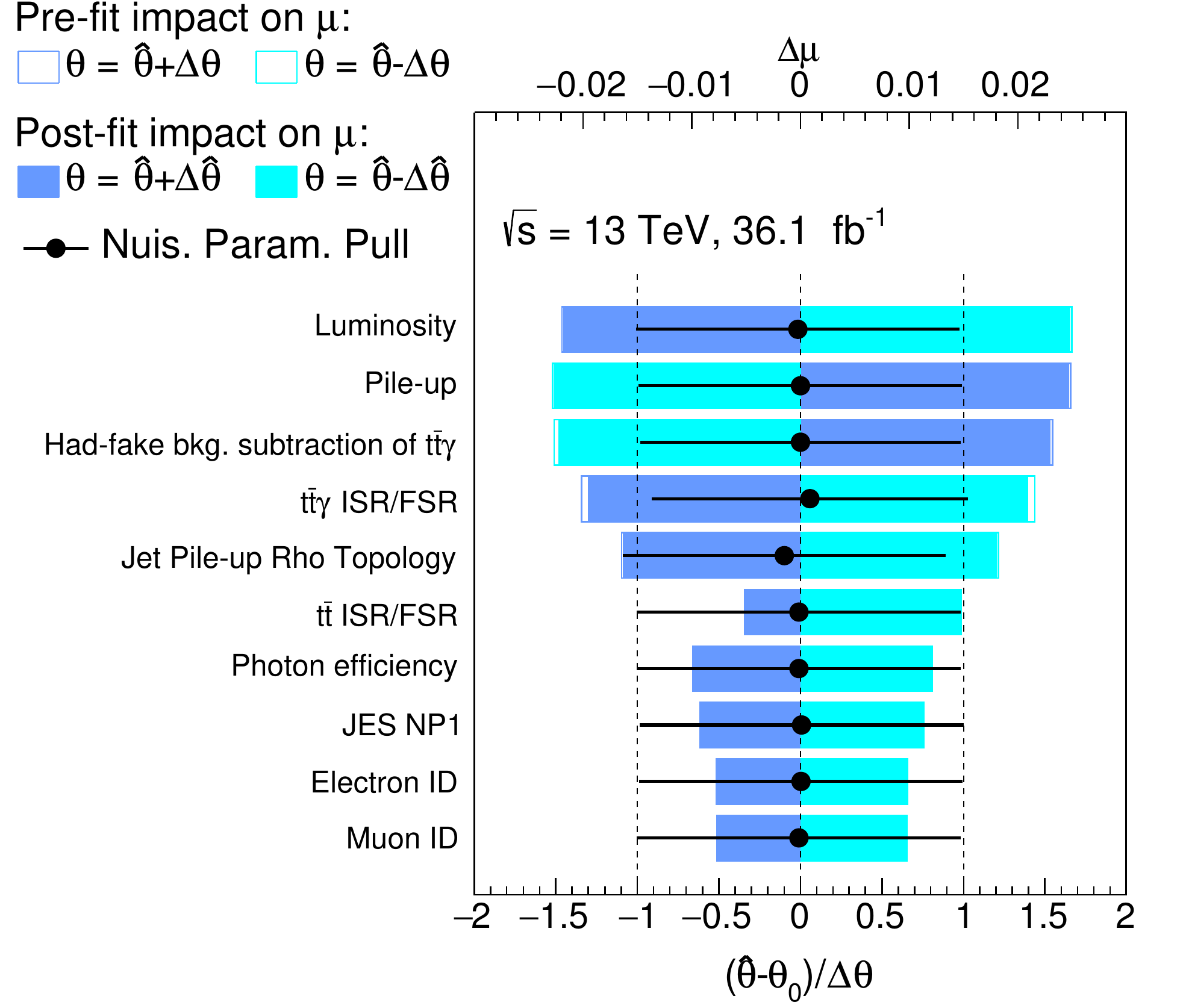}
}
\phantomcaption
\end{figure}

\begin{figure}
\centering
\ContinuedFloat
\subfloat[5-channel inclusive]{
\includegraphics[width=0.50\linewidth]{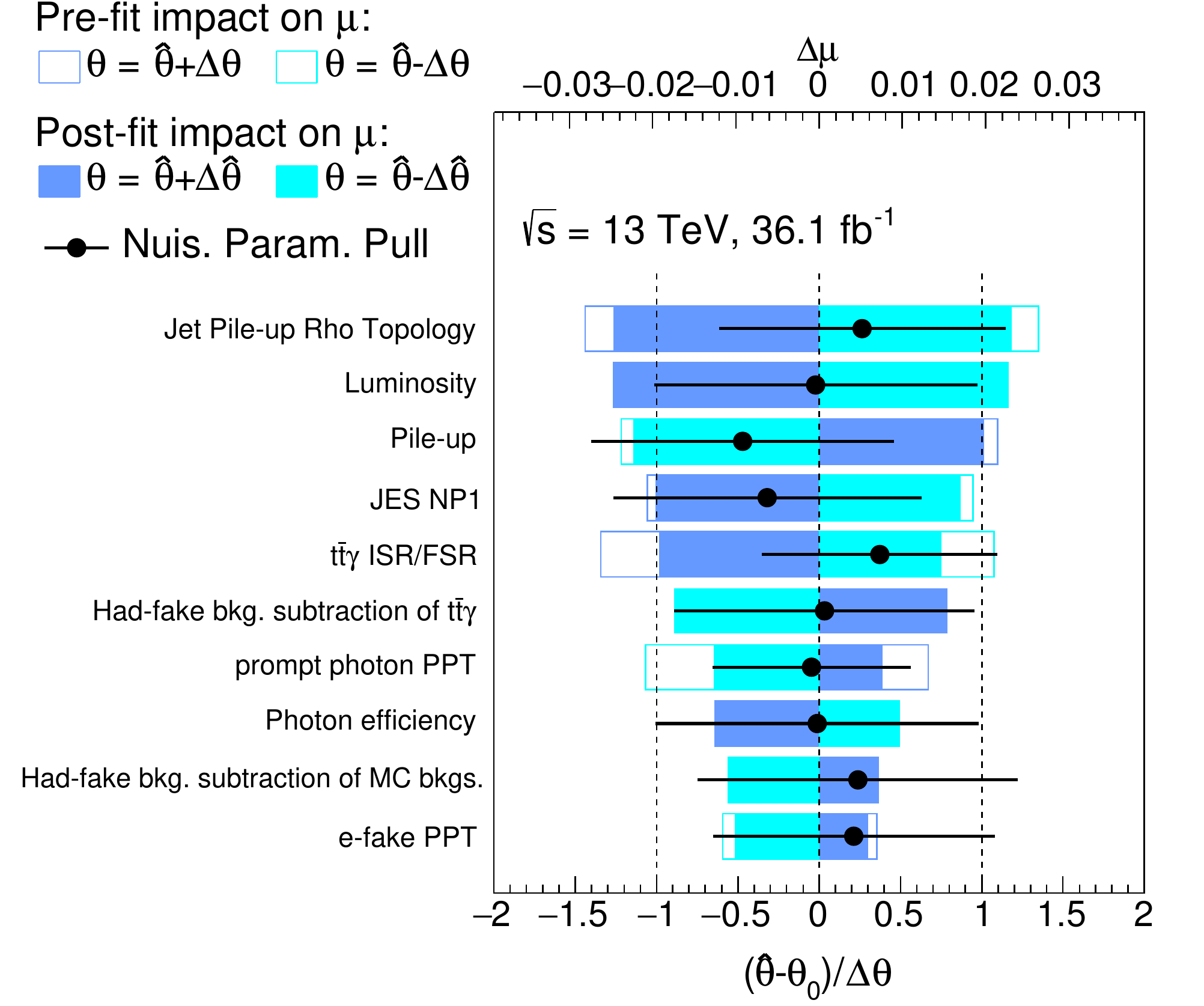}
}
\caption {Largest ten NPs ranked according to their impact on the \chljets, \chll and 5-channel inclusive fits.}
\label{fig:rankingPlots}
\end{figure}

\FloatBarrier
\section{Fiducial cross sections}
\label{sec:fidcrosssections}

The fiducial \xsec{}s can be calculated using the theoretical calculations from Chapter~\ref{sec:fiducial} and the results in Section~\ref{sec:finalresults}. These are summarised in Table~\ref{tab:fidcrosSections}, which also show theoretical predictions. All measurements agree with the predictions within uncertainty.

\begin{table}[h]
\centering 
\begin{tabular}{|l|r|rr|rr|r|}
 \hline
& $\sigma_{\text{fid}}$ [fb]  &  \multicolumn{2}{c|}{Statistical}  &  \multicolumn{2}{c|}{Systematic} & \multicolumn{1}{c|}{Theory [fb]}   \\
\hline
\hline
& & + [fb] & - [fb] & + [fb] & - [fb] & \\
\cline{3-4}
\cline{5-6}

\chejets & 265  & 6 & 6 &  21 & 20 & 247 $\pm$ 49 \\
\chmujets &  250  & 7 & 7 &  22 & 21 & 248 $\pm$ 50 \\
\chee &  16  & 2 & 1 &  2 & 1 & 16 $\pm$ 2 \\
\chmumu &18  & 1 & 1 &  2 & 1 & 16 $\pm$ 2  \\
\chemu &  34  & 2 & 2 &  2 & 2 & 31 $\pm$ 5 \\
\hline
\chljets & 521  & 9 & 9 &  41 & 39 & 495 $\pm$ 99 \\
\chll & 69  & 3 & 3 &  4 & 4 &  63 $\pm$ 9\\ 
\hline
Inclusive (5 channels)  &  589  & 10 & 10 &  34 & 32 & 558 $\pm$ 110 \\
\hline
\end{tabular}
\caption{Fiducial \xsecModifyNoun summary for all channels with uncertainty split into statistical and systematic components. The theoretical prediction is included in the last column.}
\label{tab:fidcrosSections} 
\end{table}

\chapter{Conclusions}
\label{sec:conclusions}

This thesis presents fiducial \xsecModifyNoun measurements of the $pp\to t\bar{t}\gamma$ process using $36.1~\fb^{-1}$ of data collected with the \ATLAS detector at $\sqrt{s} = 13$~\TeV. The data were collected during the LHC running periods of 2015 and 2016.
In total, eight cross sections are measured including the individual \chljets and \chll channels (\chejets, \chmujets, \chee, \chmumu and \chemu), the merged \chljets and \chll channels, and a 5-channel inclusive \xsec, where no distinction between leptons and jets from tau decays are made.
This thesis presents the first measurements of \chll, and single- and dilepton inclusive cross sections.

The strategy for this analysis deviated significantly from the previous results at 7 and 8~\TeV, both in \ATLAS and CMS. Previously, the discriminating variable was the track isolation of the photon. This analysis places a cut on this variable, which removes a significant number of background events. Then, two separate neural network algorithms trained using Keras and applied using \lwtnn are introduced to further separate signal from background. These tools present a paradigm shift for how machine learning is typically carried out at the LHC.
The first neural network, called Prompt Photon Tagger, makes use of energy deposits in the calorimeter to help discriminate prompt photons from those coming from hadron decays or from hadrons misidentified as photons. The second network (the Event-level Discriminator) makes use of a range of variables related to the kinematics and topology of an event. This includes information about $b$-tagging, \MET, number of jets etc. In the case of the \chljets channels the Prompt Photon Tagger serves as a powerful input to the Event-level Discriminator.

The various backgrounds fall into four main classes; \efake{}s, \hfake{}s, prompt photons and \QCD{}s.
The \efake background is dominant in the \chljets channels and consists of events where electrons have been misidentified as photons. The main contribution comes from the \ttbar dileptonic decays in the \chee and \chemu channels. This background is negligible in the \chll channels. The estimation of this background follows a data-driven approach in which a fake photon enriched control region is created and compared to another control region of similar phase space but with the photon required to be an electron. Thus, the fake photon can be probed, efficiencies calculated and correction factors derived.

Hadrons, or photons from hadron decays that are misidentified as prompt photons, form the \hfake background.
This background is mainly seen in the \chljets channels, but also a small contribution exists in the \chll channels. This background is estimated following a data-driven approach where three orthogonal control regions to the signal region are defined. Isolation and identification algorithms in these control regions are reversed. Correction factors to the fake photon contribution can then be derived.

The \QCD background is predominantly found in the \chljets channels with a negligible contribution in the \chll channels. It is estimated using a purely data-driven approach in which looser criteria on the leptons are required, thus the sample is assumed to contain mainly fake leptons. Weights can be derived and are applied to data.

Prompt photons that are not from top quarks contribute towards a significant portion of the total background for the \chljets and \chll channels. In the \chljets channel the dominant process is from \Wgamma, while in the \chll channels this contribution is from \Zgamma. Small contributions arise from single top, diboson and \ttV processes.
The estimation of these backgrounds is based on MC prediction.

A profile maximum likelihood fit is performed on the \ELD distribution to extract the signal strength, and thus the final cross sections. 
In the \chljets channels the \Wgamma background enters as a free-floating parameter.
Dedicated $k$-factors are used in the respective channels to scale the leading-order predictions to next-to-leading order.

A summary of all measured \xsec{}s is shown in Table~\ref{tab:fidcrosSectionsConclusions}. This includes the five individual channels, the two combined channels, and the inclusive measurement. Theoretical \xsec{}s are also included.
All measurements agree with theoretical predictions.

\begin{table}[!ht]
\centering 
\scalebox{1}{
\begin{tabular}{|l|c|c|}
 \hline
 \rule{0pt}{13pt} &  \multicolumn{2}{c|}{$\sigma^{\text{\ttgamma}}_{\text{fid}}$ [fb]} \\ \hline
Channel & Measured $\pm$(stat.) $\pm$(syst.) & Theory $\pm$(total) \\ \hline
\chejets & $265 \pm 6 \pm 21$   &  $247 \pm 49$ \\ 
\chmujets & $250 \pm 7 \pm 22 $   &  $248 \pm 50$ \\ 
\chee & $16 \pm 2 \pm 2 $   &  $16 \pm 2$ \\ 
\chmumu & $18 \pm 1 \pm 2 $   &  $16 \pm 2$ \\ 
\chemu & $34 \pm 2 \pm 2 $   &  $31 \pm 5$ \\ \hline

Single-lepton & $521 \pm 9 \pm 41 $   &  $495 \pm 99$ \\ 
Dilepton & $69 \pm 3 \pm 4 $   &  $63 \pm 9$ \\ \hline

Inclusive (5 channels) & $589 \pm 10 \pm 34 $   &  $558 \pm 110$ \\ 
\hline
\end{tabular}
}
\caption{Fiducial \xsecModifyNoun summary for all channels, as well as theoretical predictions.}
\label{tab:fidcrosSectionsConclusions} 
\end{table}

For the \chljets channels the largest systematic contributions come from the modelling of jets, background estimation techniques and the \PPT. For the \chll channel more statistics will help increase the sensitivity, as well as reducing signal and background modelling systematics. To reduce background modelling one could focus on just the \chemu channel, which is dominated by signal with very little (to almost negligible) background contributions.

Further studies on the \ttgamma process should focus on reducing the theoretical uncertainties associated with the $k$-factor, developing techniques to reduce the contamination from photons being radiated from the top quark decay products, and EFT interpretations. 
The relationship between the \xsec and the $t\gamma$ coupling ($Q_{t}$) is expected to be quadratic. Thus, one method to determine the $t\gamma$ coupling could be achieved by performing template fits with various hypotheses of $Q_{t}$.
This will further help to improve our understanding of the electromagnetic coupling to the top quark.

\cleardoublepage
\phantomsection
\bibliographystyle{Style/atlas}
\bibliography{Bib/atlas,Bib/misch,Bib/ConfNotes,Bib/MLpapers,Bib/PubNotes}

\cleardoublepage

\printindex
\cleardoublepage

\addappheadtotoc
\appendix
\appendixpage
\noappendicestocpagenum

\chapter{Porting the \ATLAS software stack to the ARM architecture}
\label{sec:arm}

By using fewer instructions on a silicon chip, fewer transistors are needed, and thus the power consumption of a CPU is reduced.
This makes such chips ideal for portable devices as the battery life is extended. 
Around 60\% of mobile devices use the ARM (Advanced RISC (Reduced Instruction Set Computing) Machine) architecture. This includes smartphones, tablets, wearables and e-readers, with the percentage much higher for just smartphones. As of 2016, more than 86~billion ARM based chips have been shipped\footnote{Taken from \url{https://www.arm.com}.}.
A large part of the appeal is that ARM does not produce any of their own CPUs and so has very few overhead costs. Instead, they sell their intellectual property. This gives companies more freedom to design their own CPUs.

Interest in ARM at the LHC began around 2013/14 with feasibility studies and the potential payoffs. 
Many standard benchmark studies were performed, centred around the 32-bit embedded ARM development boards. Experiments such as \ATLAS, CMS, and LHCb also started work on porting their software stacks to the 32-bit ARM architecture~\cite{Abdurachmanov:2013bxa,Abdurachmanov:2013kla,Abdurachmanov:2014mea,Cox:2015yza,Smith:2015xza,Smith:2015eta}.
However, it was found that while it is indeed very power conservative, ARM 32-bit did not have the available memory to make HEP computing feasible.

In 2016 ARM entered the server market and introduced their 64-bit architecture\footnote{The bottleneck is no longer CPU clock cycles, but rather the memory bandwidth. The 32-bit architectures limit the amount of random access memory (RAM).}. 
The latest results from porting and benchmarking a subset of the \ATLAS software stack to the ARM 64-bit architecture (Aarch64) can be found in~\cite{Smith:2017afo} and is summarised in this appendix.
It should be noted that the fast paced environment of computing will see these types of benchmark results continuously changing as each architecture vie for the better efficiency. However, the message is the same; Aarch64 presents a feasible alternative to Intel x86.

\section{Hardware}

In early 2016 CERN openlab installed a cluster of Aarch64 evaluation prototype servers, which will be referred to as (Aarch64\_Proto). Each server is comprised of a single-socket, ARM 64-bit system-on-a-chip, with 32 Cortex-A57 cores. In total, each server has 128 GB RAM connected with four fast memory channels. 
Another type of ARM server is also maintained, called HP Moonshot.
Two types of Intel servers (Intel Atom and Intel Xeon) are also available. 
The Intel Atom was the company's initial response to ARM and is fairly old, however it still provides interesting results. The features of each server are described in Table~\ref{tab:setups}.

\begin{table}[!hbtp]
\begin{center}
\scalebox{0.7}{
\begin{tabular}{ p{2.2cm} p{3.2cm} p{2.5cm}p{2.8cm} p{3.5cm} p{1.9cm} p{1.9cm} } 
\hline
\\[-1.5ex]
Name & Processor & Cores & RAM & Cache & Fabrication (Release) & OS
\\ [0.5ex] 
\hline 
\hline
\\[-1ex]
 \raggedright{HP Moonshot} & \raggedright{X-Gene, 2.4 GHz} &  \raggedright{8 Armv8} &  \raggedright{64 GiB DDR3 (1600 MHz)} &\raggedright{32 KiB L1/core, 256 KiB L2/core pair, 8 MiB L3}&\raggedright{40 nm (2014)} &Ubuntu 14.04 
\\[1ex]
 \raggedright{Aarch64\_Proto} & \raggedright{-, 2.1 GHz} &  \raggedright{32 Cortex-A57} &  \raggedright{128 GiB DDR3 (1866 MHz)} & \raggedright{32 KiB L1, 1 MiB L2} & \raggedright{16 nm (-)}&Ubuntu 14.04 
\\[1ex]
 \raggedright{Intel Atom} & \raggedright{Intel Atom Processor C2750, 2.4GHz} &  \raggedright{8} &  \raggedright{32 GiB DDR3 (1600 MHz)} & \raggedright{24 KiB L1d, 32 KiB L1i, 1 MiB L2 }&\raggedright{22 nm (2013)}&Fedora 21
\\[1ex]
\raggedright{Intel} & \raggedright{ Intel Xeon CPU E5-4650, 2.70 GHz} &  \raggedright{32} &  \raggedright{512 GiB DDR3 (1600 MHz)} &
\raggedright{32 KiB L1(d)(i)/core, 256 KiB L2/core, 20 MiB L3} & \raggedright{32 nm (2012)} &Scientific Linux CERN 6 
 \\[1ex]
\hline
\end{tabular}
}
\end{center}
\caption{\label{tab:setups} The hardware for Intel and Aarch64 servers~\cite{Smith:2017afo}.}

\end{table}

\subsection{AthSimulation}

The \ATLAS codebase (Athena) consists of around 2400 packages and 6.5 million lines of code. Due to its size and complexity, porting to alternative architectures is difficult. Thus, a project called AthSimulation was chosen, which consists of a subset of packages from a full Athena release. AthSimulation is capable of carrying out CPU intensive MC simulations needed for the experiment. At around 350 packages, the porting process becomes significantly easier and faster. 
Typical non-\ATLAS specific packages include ROOT, \textsc{Geant4} and \textsc{Gaudi}~\cite{Barrand:2001ny}. \ATLAS specific simulation code forms the peak of the pyramid of dependencies.
In general, compilation options needed to be added for Aarch64, Intel specific compiler options removed and various build configurations made more general.

To enable multiple builds on various servers a continuous integration tool, Jenkins, was used.
AthSimulation was compiled on both Aarch64 servers, while the Intel servers made use of the equivalent pre-existing versions built within \ATLAS.
The benchmark consists of simulating 100 \ttbar events on the respective servers.
Simulation took between 3.5 minutes (Intel Xeon) to just under 12 minutes (Intel Atom) per event depending on the architecture. 

\subsubsection{Validation}
\begin{figure}[!hbtp]
\centering
\subfloat[]{
\includegraphics[width=0.46\textwidth]{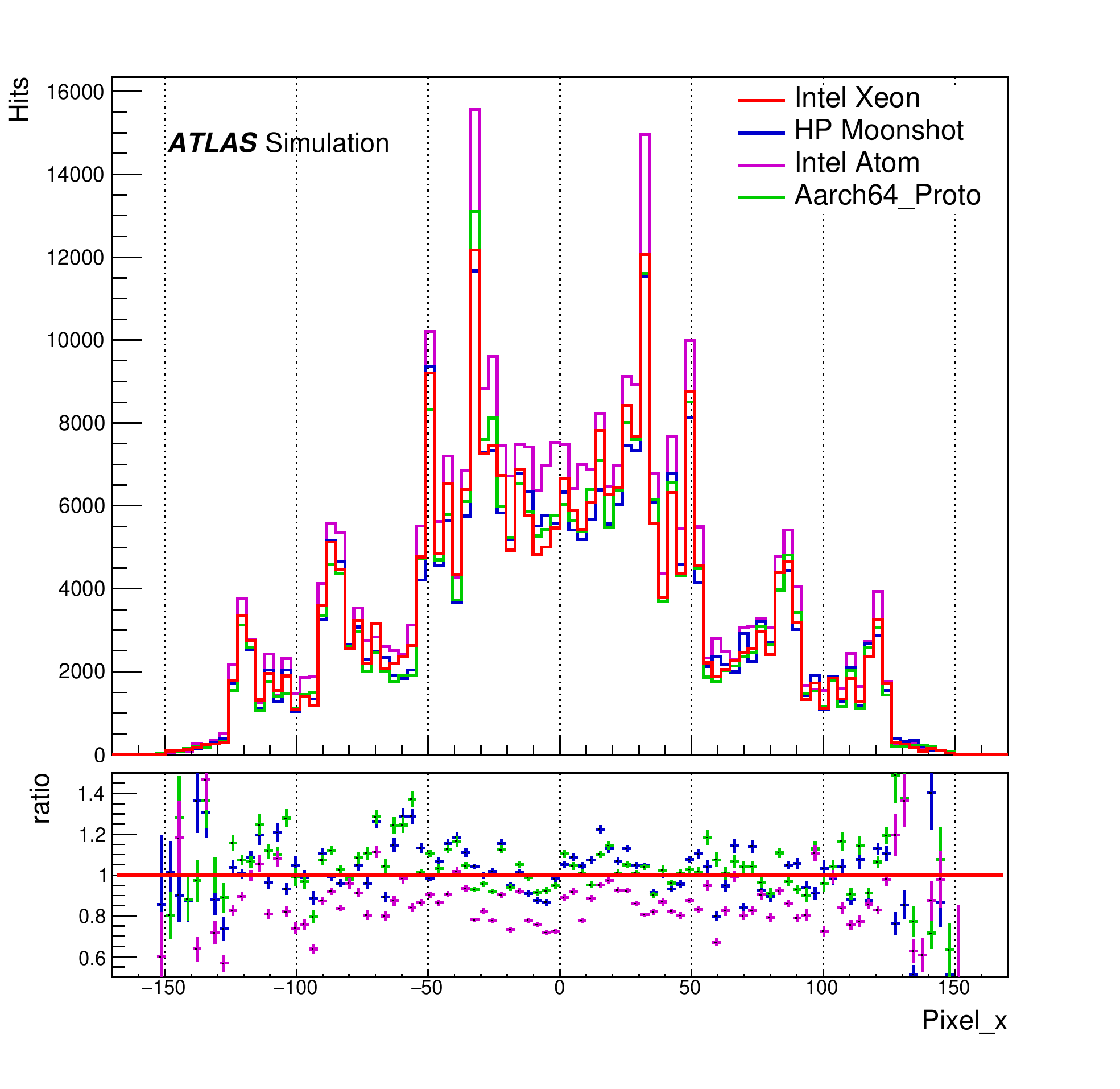}
\label{fig:pixelx}
}\hspace{-0.045\linewidth}
\subfloat[]{
\includegraphics[width=0.46\textwidth]{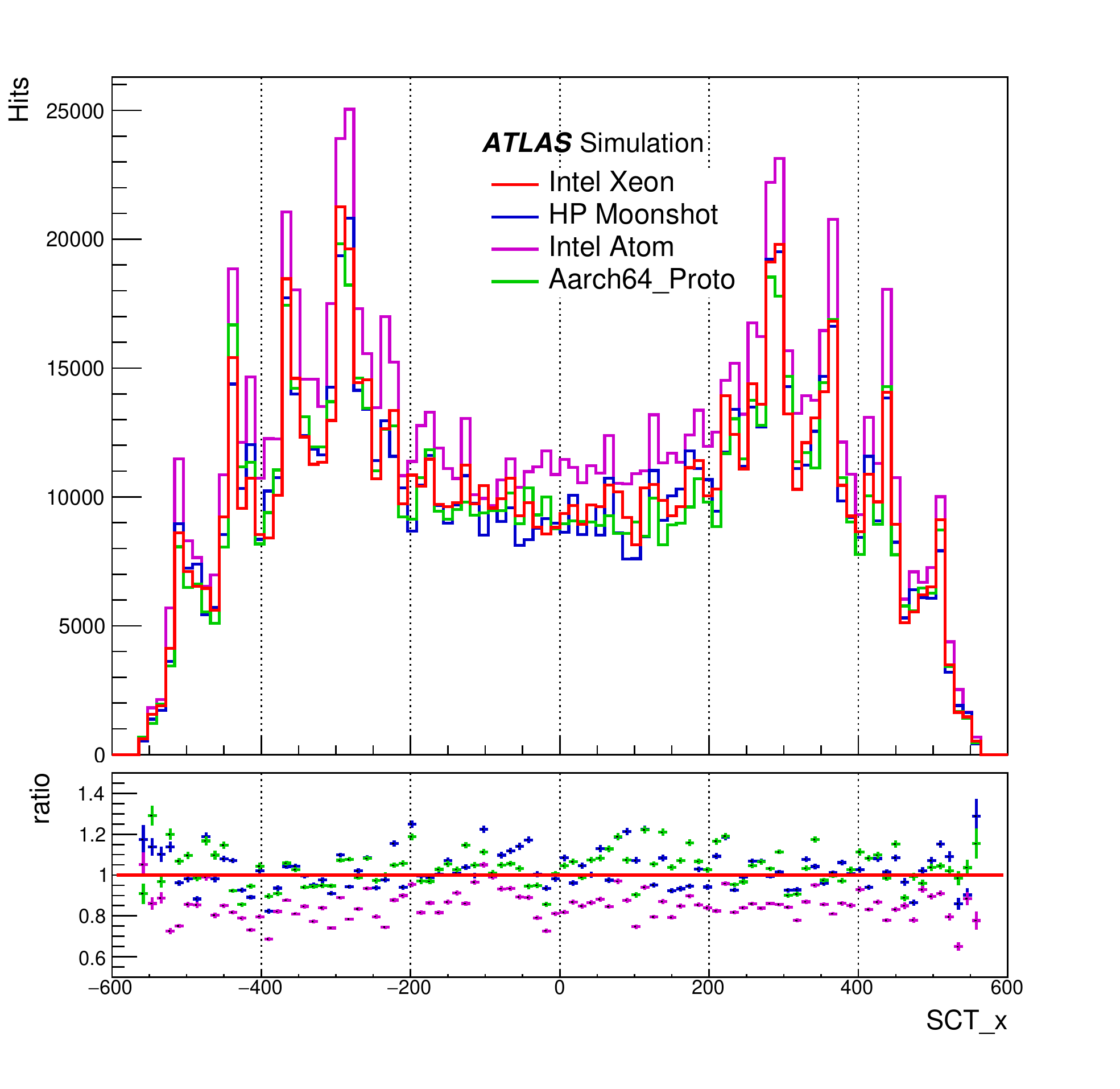}
\label{fig:sctx}
}
\caption{Results showing the hits in the a) pixel and b) SCT detectors in \ATLAS. The ratio between the three servers and the Intel Xeon is shown in the ratio plots. The Intel Xeon is taken as the ``accepted'' distribution.~\cite{Smith:2017afo}}
\label{fig:validation}
\end{figure}

Validating the results from different architectures was made a high priority since this had not been checked before.
Due to the nature of MC simulation, numerical identity is not expected. Some reasons can include and are not limited to random numbers being generated in an architecturally specific way, as well as the way floating point numbers are handled by each compiler. However, overall trends are expected to be similar.
Figure~\ref{fig:validation} shows the hits in the pixel and SCT detectors for different architectures compared to the Intel Xeon. When compared to the Intel Xeon, the ARM and Intel Atom servers give similar distribution shapes, but around 10-15\% less hits on ARM and 15-20\% more hits on Intel Atom. Further studies with larger simulated datasets needs to be done to understand why this occurs.

\begin{figure}[!hbtp] 
\centering
\subfloat[]{%
\includegraphics[width=0.65\textwidth]{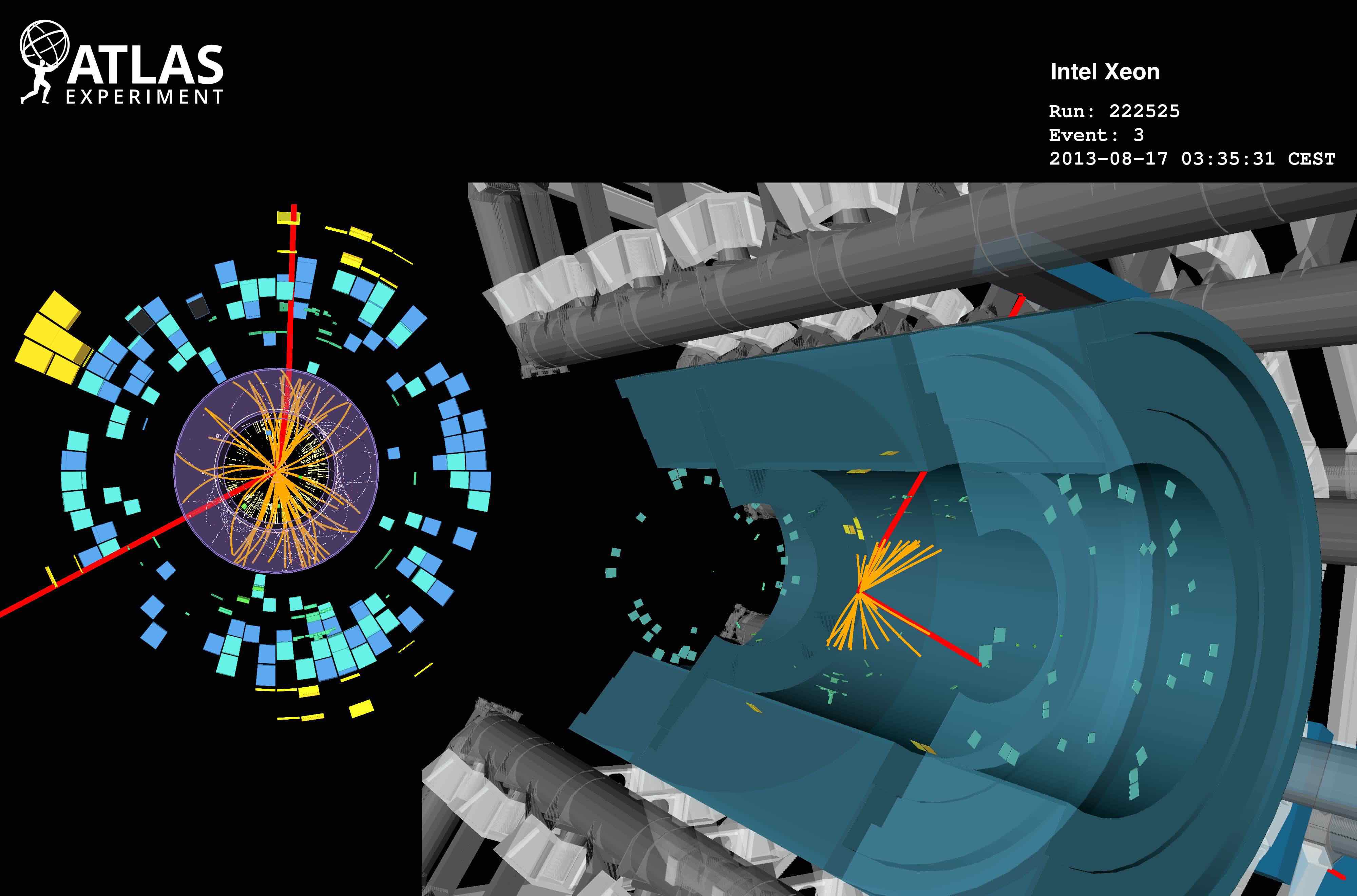}%
\label{fig:vp1Xeon}%
}

\subfloat[]{%
\includegraphics[width=0.65\textwidth]{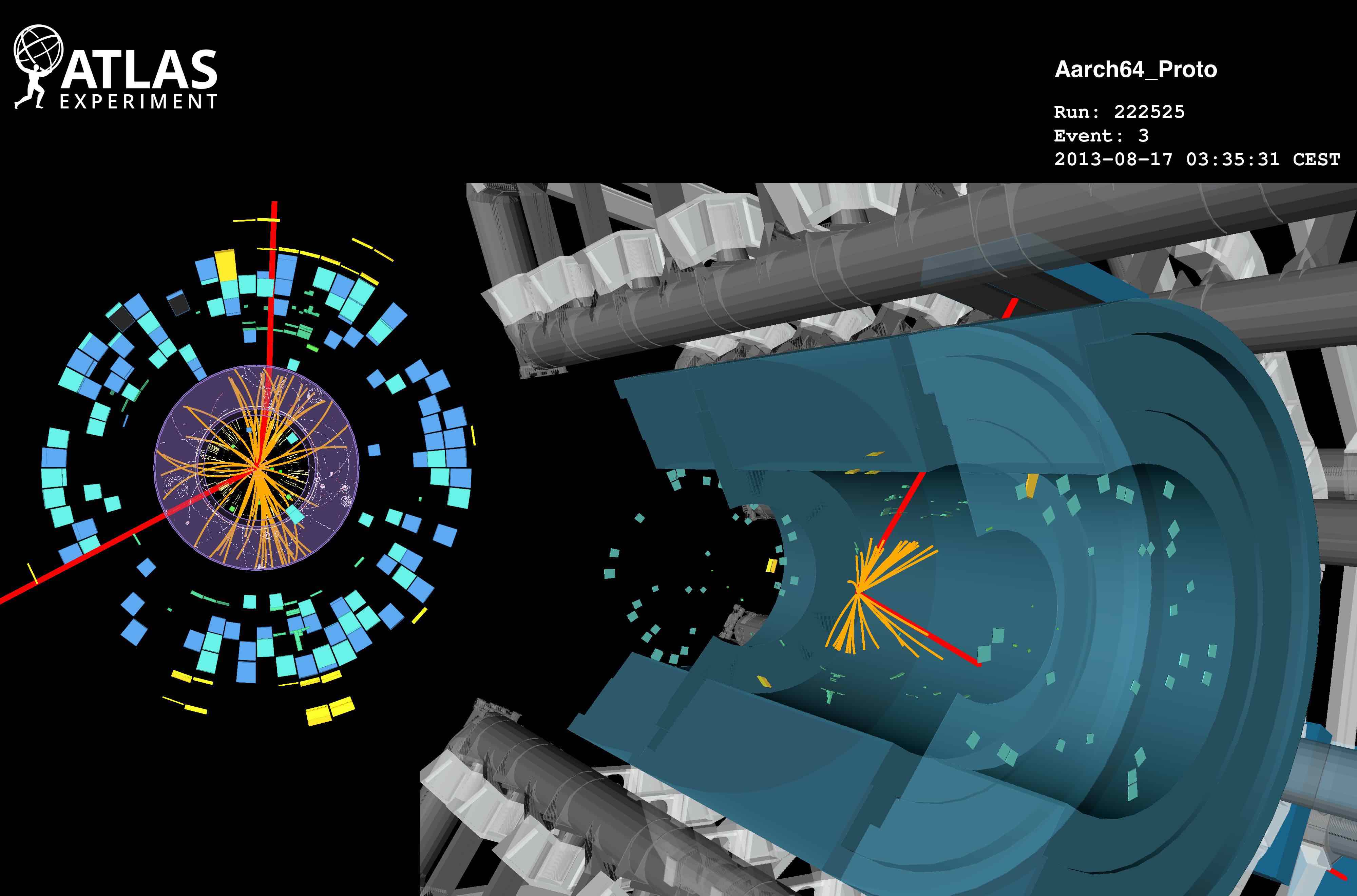}%
\label{fig:vp1Proto}%
}
\caption{Event displays for the same $t\bar{t}$ event simulated on the a) Intel Xeon and b) Aarch64\_Proto server. Minor differences can be seen in the calorimeter energy deposits and tracks~\cite{Smith:2017afo}.}
\label{fig:EvDisplays}
\end{figure}

Figure~\ref{fig:EvDisplays} shows a selected \ttbar event simulated on Intel Xeon and Aarch64\_Proto. There are subtle differences in energy deposits due to a different number of hits and also a very minor difference in tracks. However, the general topology of the event matches well.

\subsection{Power measurements}
\begin{figure}[!htpb]
\centering
\includegraphics[width=.9\textwidth]{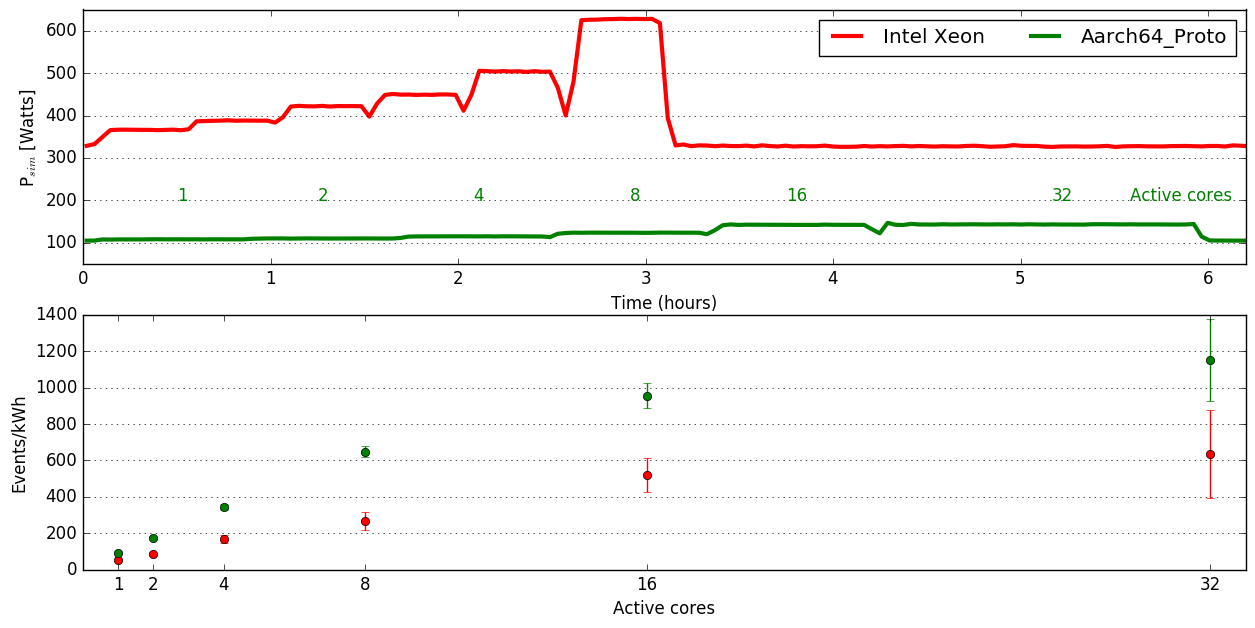}
\caption{Power measurements for the benchmark running on an increasing number of cores. Top: total time and power for all tests. Bottom: Events/kWh calculated for each test~\cite{Smith:2017afo}.}
\label{fig:pow2}
\end{figure}

To take memory bandwidth into account while performing power measurements, the benchmark was repeated on the Aarch64\_Proto and the Intel Xeon server with multiple jobs running. In this benchmark, eight \ttbar events are simulated on an increasing number (2,4,8,16,32) of cores.
Figure~\ref{fig:pow2} shows the results. The top plot shows the net power usage during each test, while the bottom image shows the events/kWh for each test.
Taking time and power usage into account, this shows that for these CPUs, the Aarch64 server is more efficient.

\subsection{Outlook for ARM in HEP}

It is clear that ARM servers have improved dramatically over the past few years. Their 64-bit architecture is now competitive with the traditional Intel machines. 
Studies like these show academia and industry alike that there is no need to be reliant on a single computing architecture. Various reasons ranging from cost-effectiveness to geopolitical issues could result in server farms being made available that might not necessarily be the standard Intel architecture. 
CERN needs to be ready to utilise whatever computing power is available at the most reasonable price, especially with the ever increasing computing demands.
As an experiment \ATLAS needs to be flexible in how they can deploy their software. 
Steps have been undertaken to make the \ATLAS software more independent of operating system and CPU architecture.


\chapter[Support material for the \texorpdfstring{\ttgamma}{ttgamma} process]{Support material for the \ttgamma process}
\label{sec:analysisstrategyappendix}

\section[Validation plots for \texorpdfstring{\ttbar}{ttbar} selections]{Validation plots for \ttbar selections}
\label{sec:validation_plots}

\begin{figure}[!htbp]
\centering
\subfloat[\chejets]{
\includegraphics[width=0.3\linewidth]{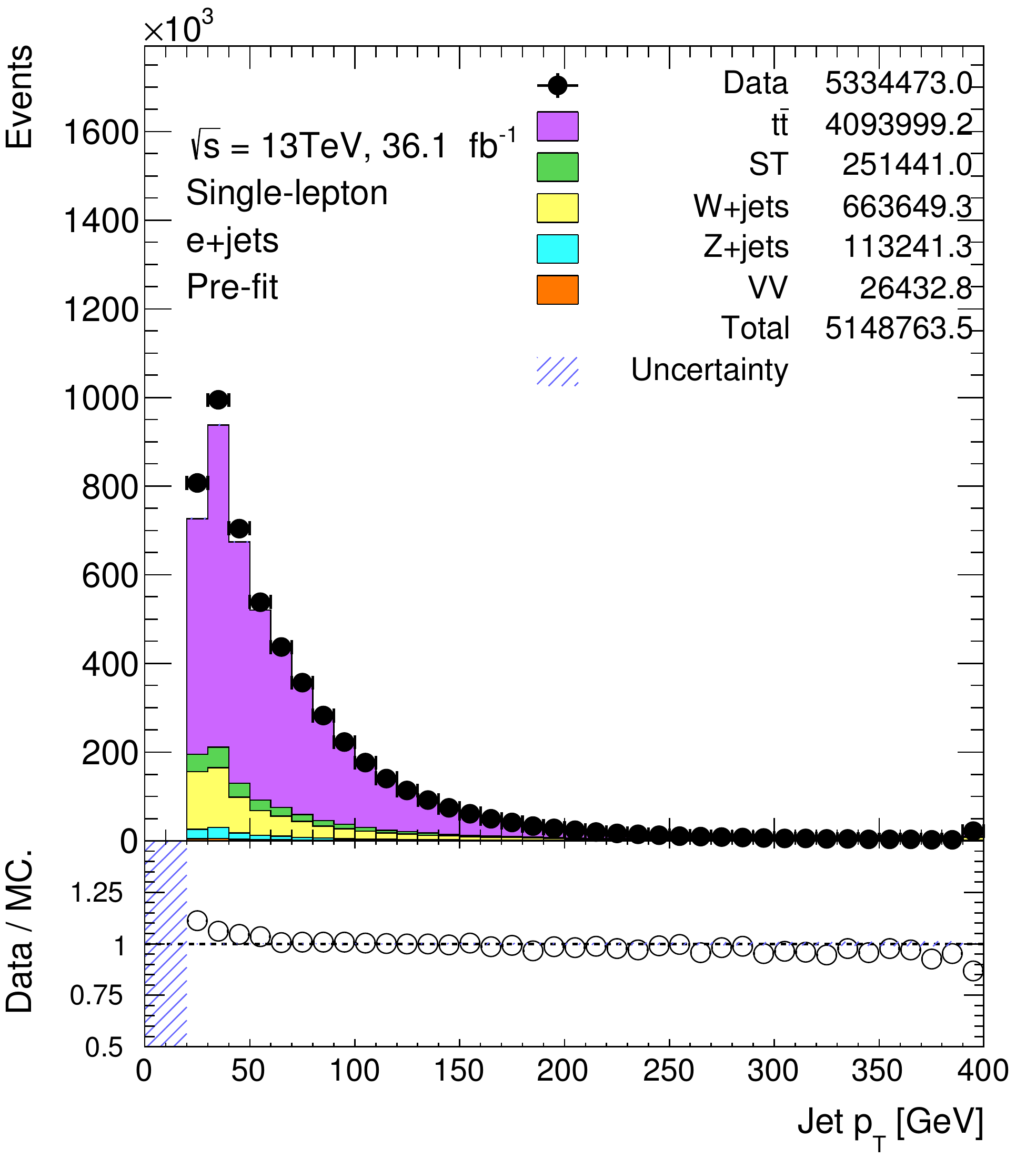}
}
\subfloat[\chmujets]{
\includegraphics[width=0.3\linewidth]{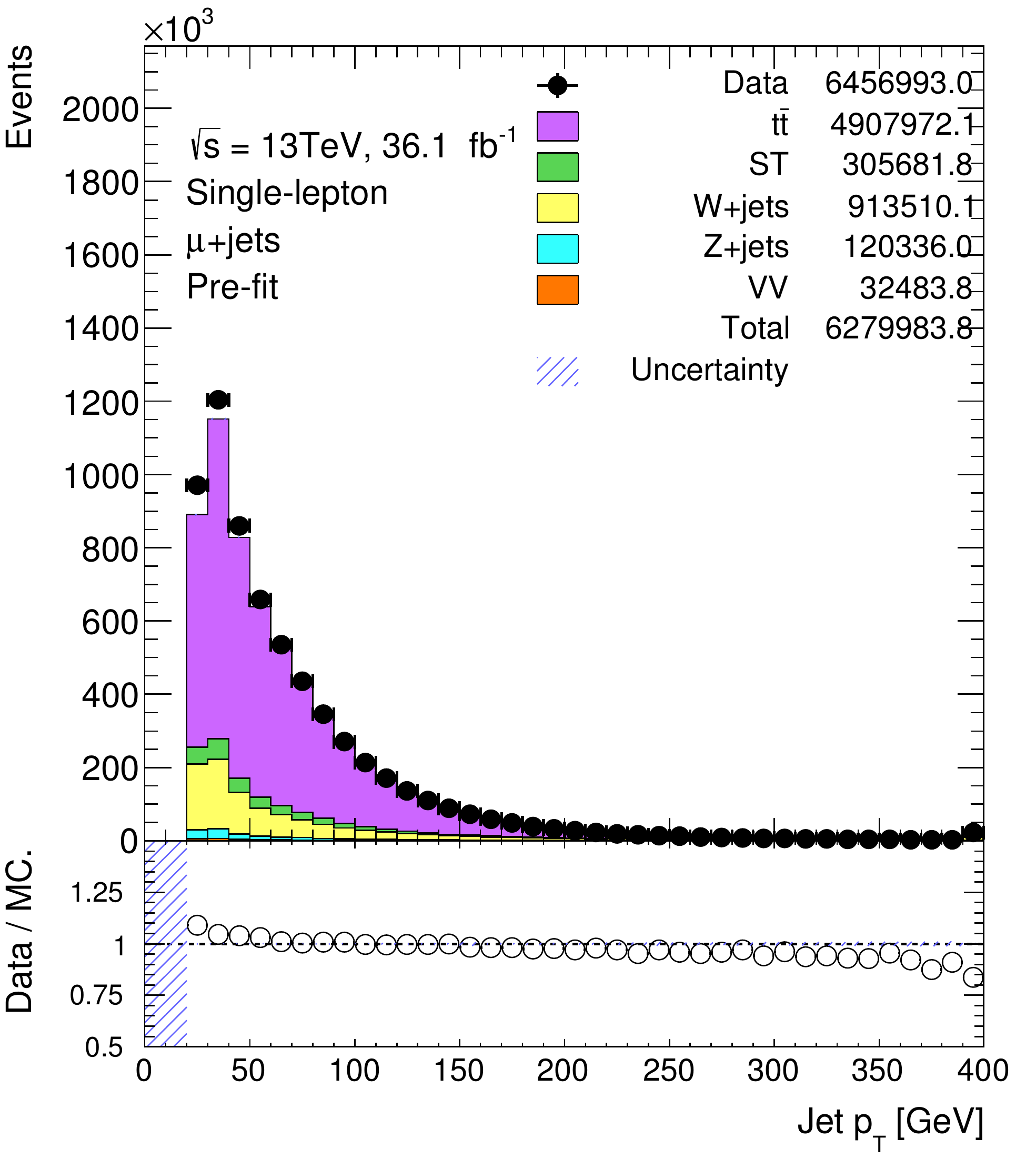}
}

\subfloat[\chee]{
\includegraphics[width=0.3\linewidth]{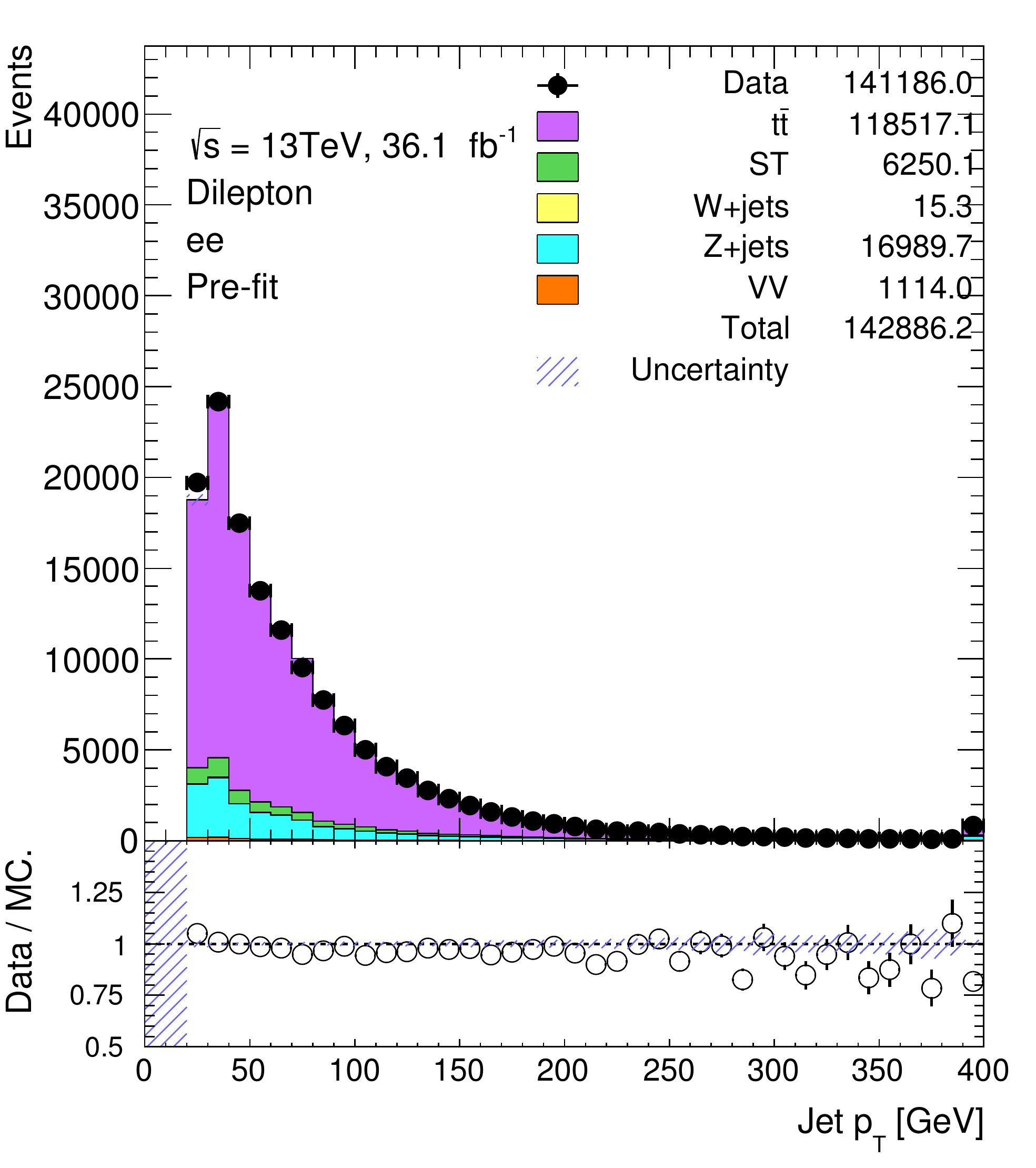}
}
\subfloat[\chemu]{
\includegraphics[width=0.3\linewidth]{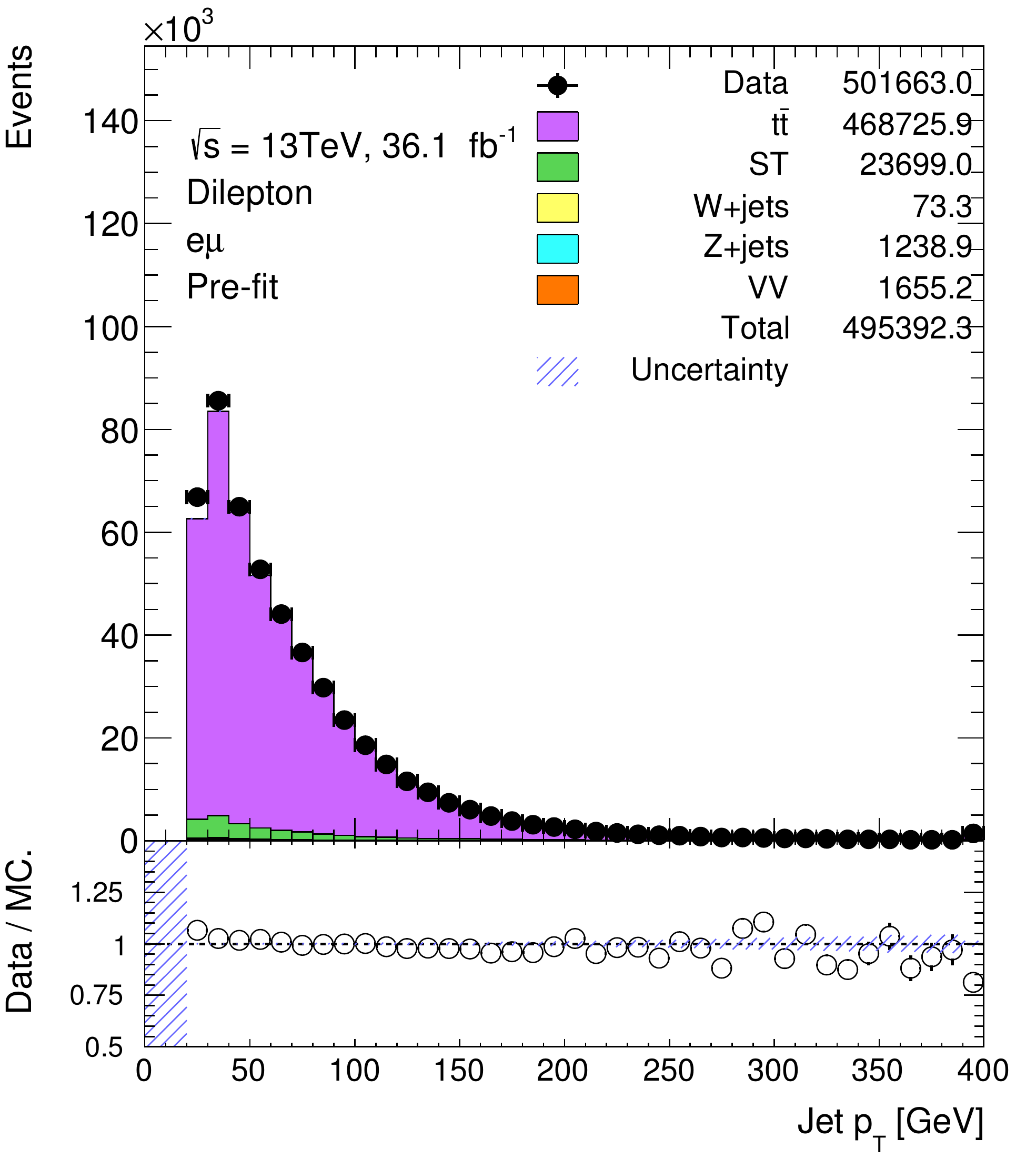}
}
\subfloat[\chmumu]{
\includegraphics[width=0.3\linewidth]{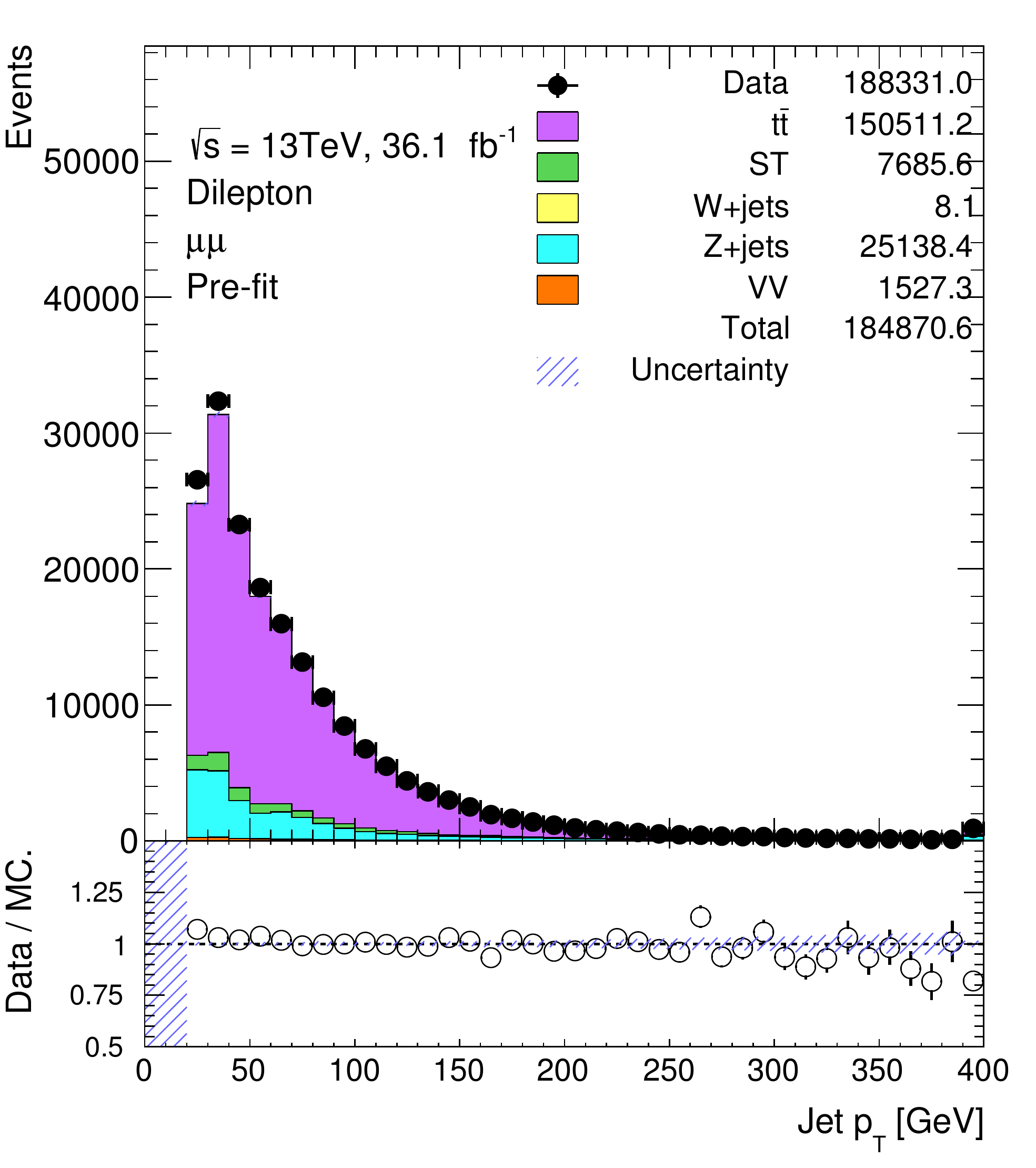}
}

\caption [Validation plots for event selections before the requirement of a photon] {Validation plots of the jet \pt for event selections before the requirement of a photon. Only statistical uncertainties are included. Not included is the \QCD background. All processes shown are MC.}
\label{fig:ttbarvalidation}
\end{figure}  

\chapter{Object- and event-level neural network support material}
\label{sec:results_appendix}

\section{Event-level Descriminator}
\label{sec:ELDapp}
\begin{figure}[!htbp]
\centering
\includegraphics[width=0.34\linewidth]{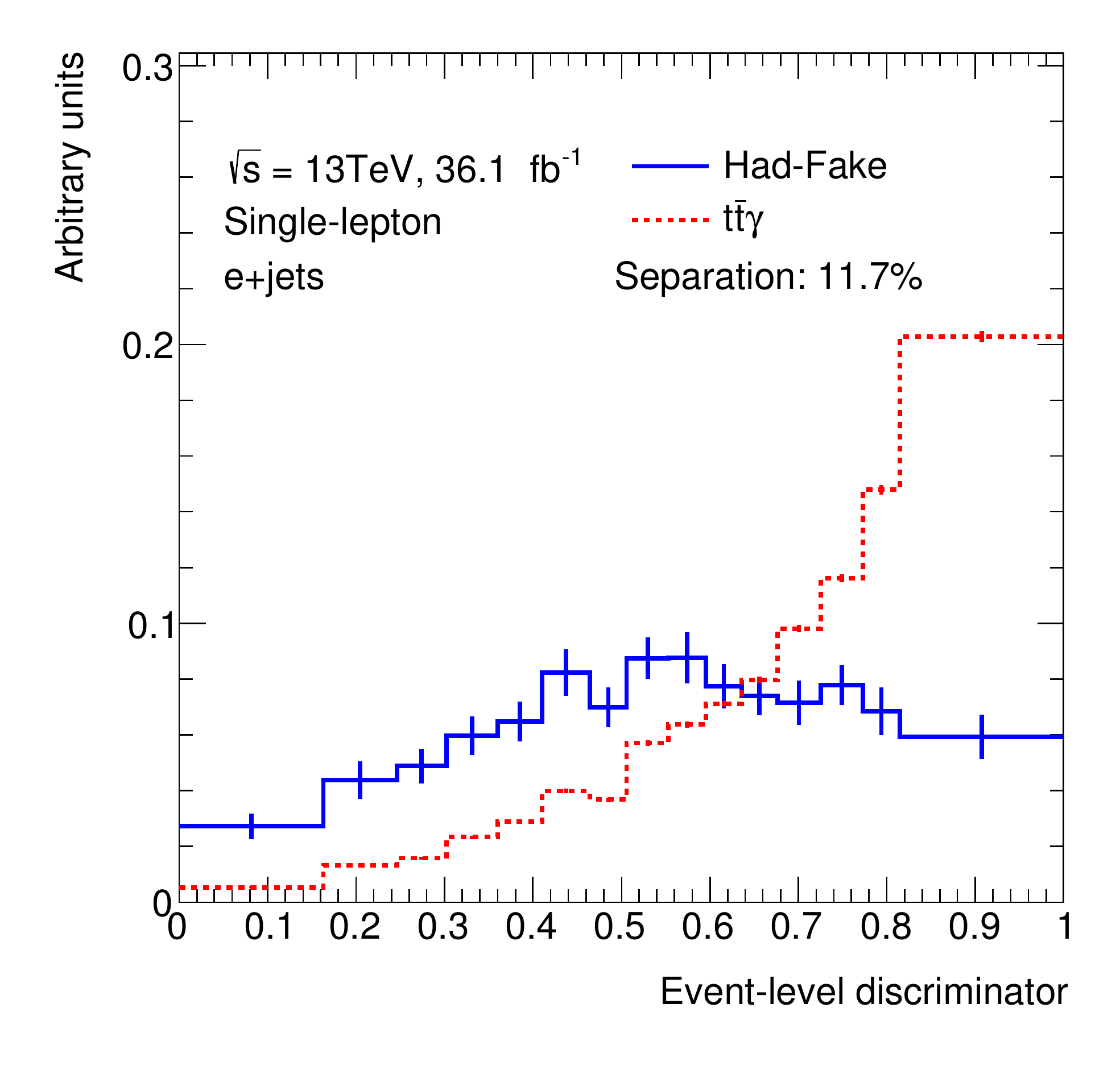}
\hspace{-0.032\linewidth}
\includegraphics[width=0.34\linewidth]{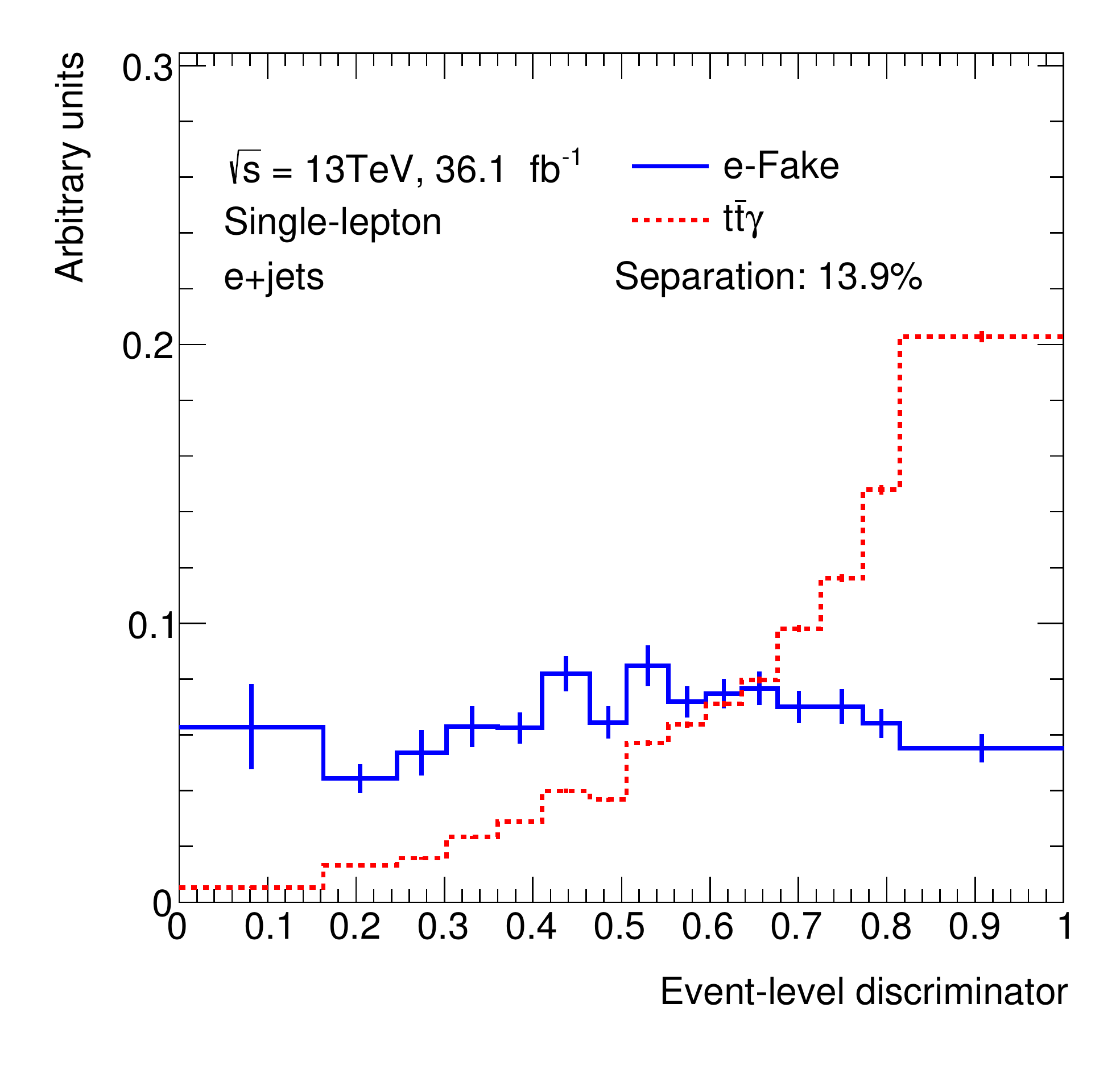}
\hspace{-0.032\linewidth}

\includegraphics[width=0.34\linewidth]{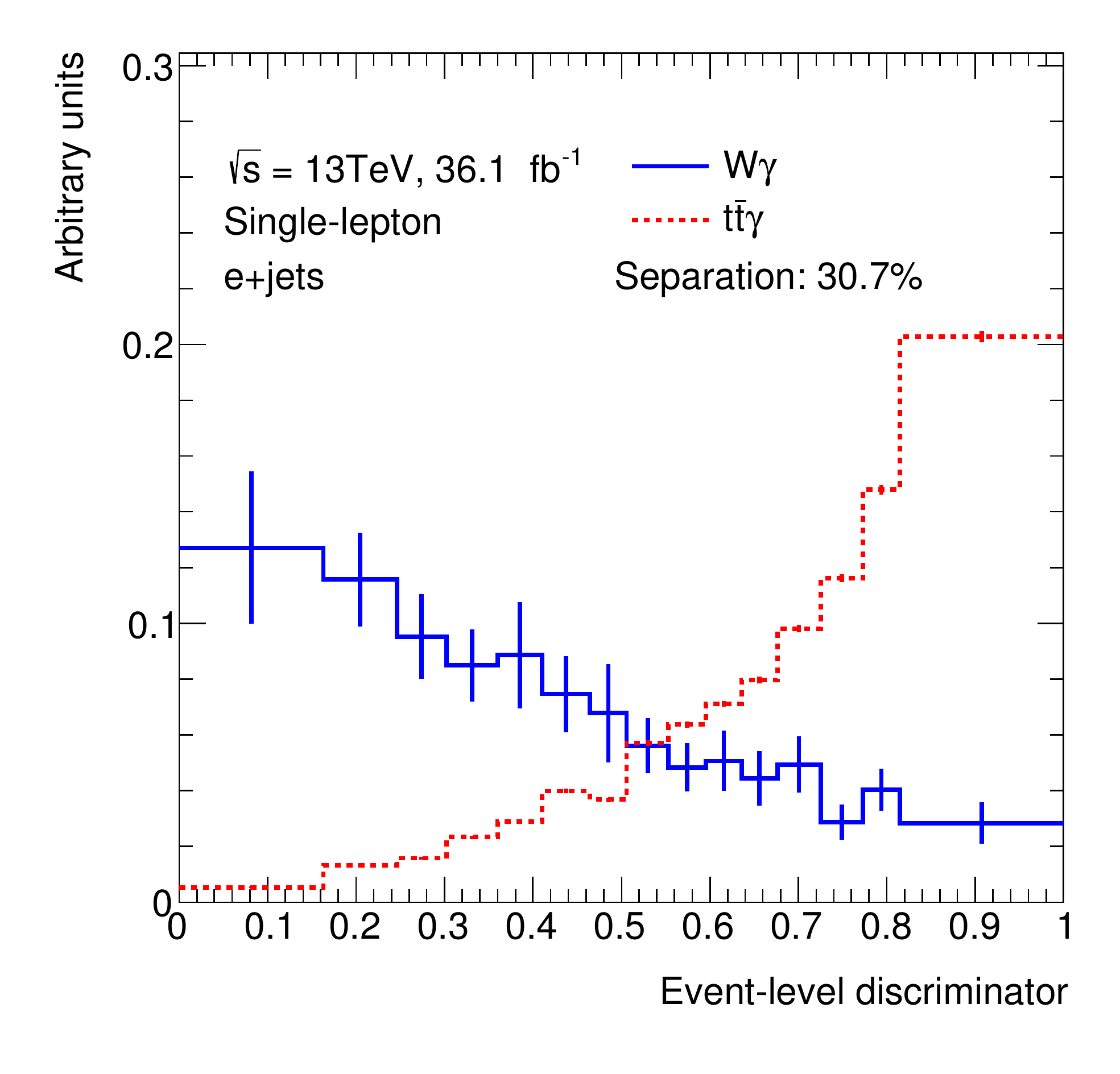}
\hspace{-0.032\linewidth}
\includegraphics[width=0.34\linewidth]{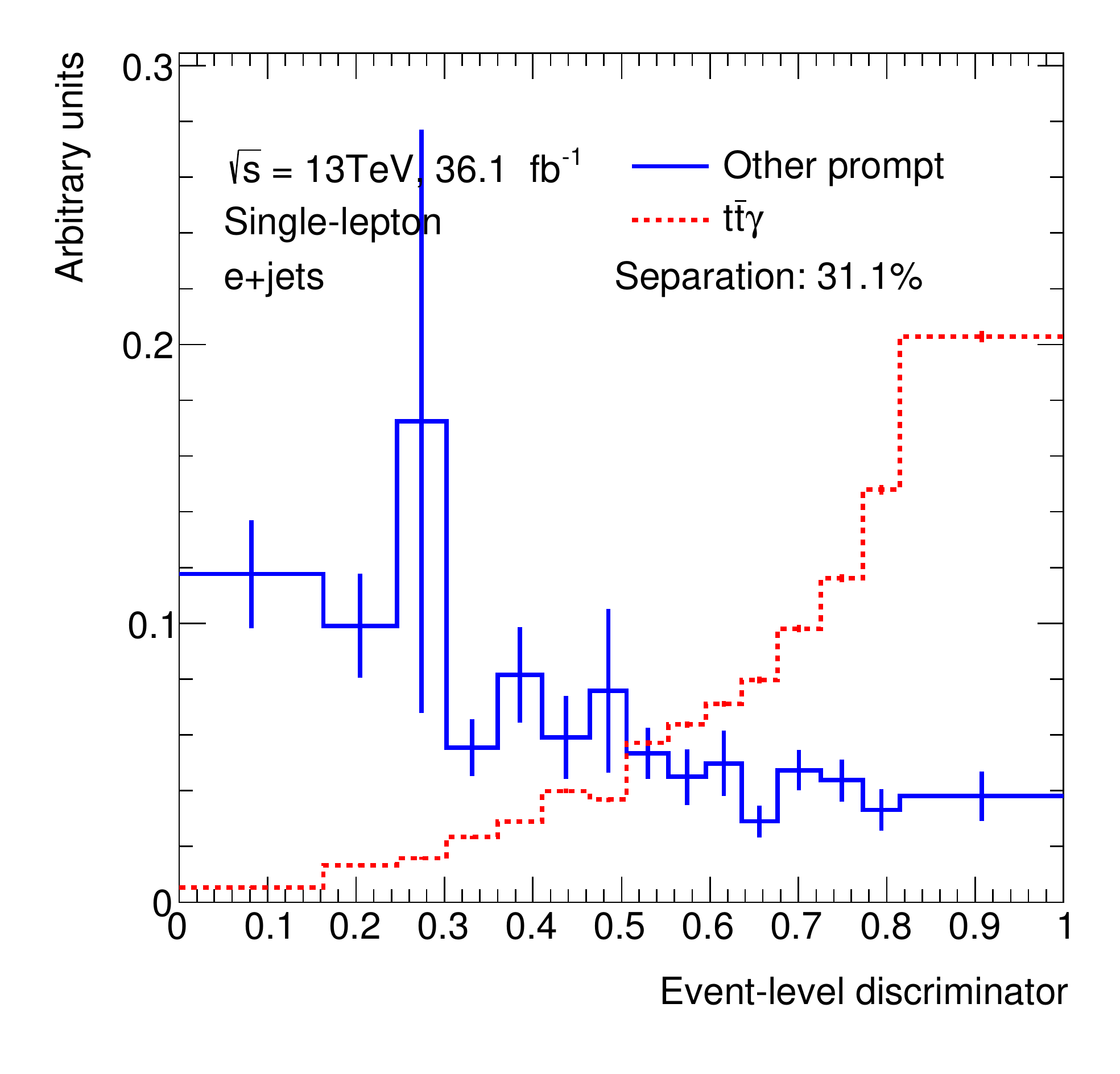}
\hspace{-0.032\linewidth}
\includegraphics[width=0.34\linewidth]{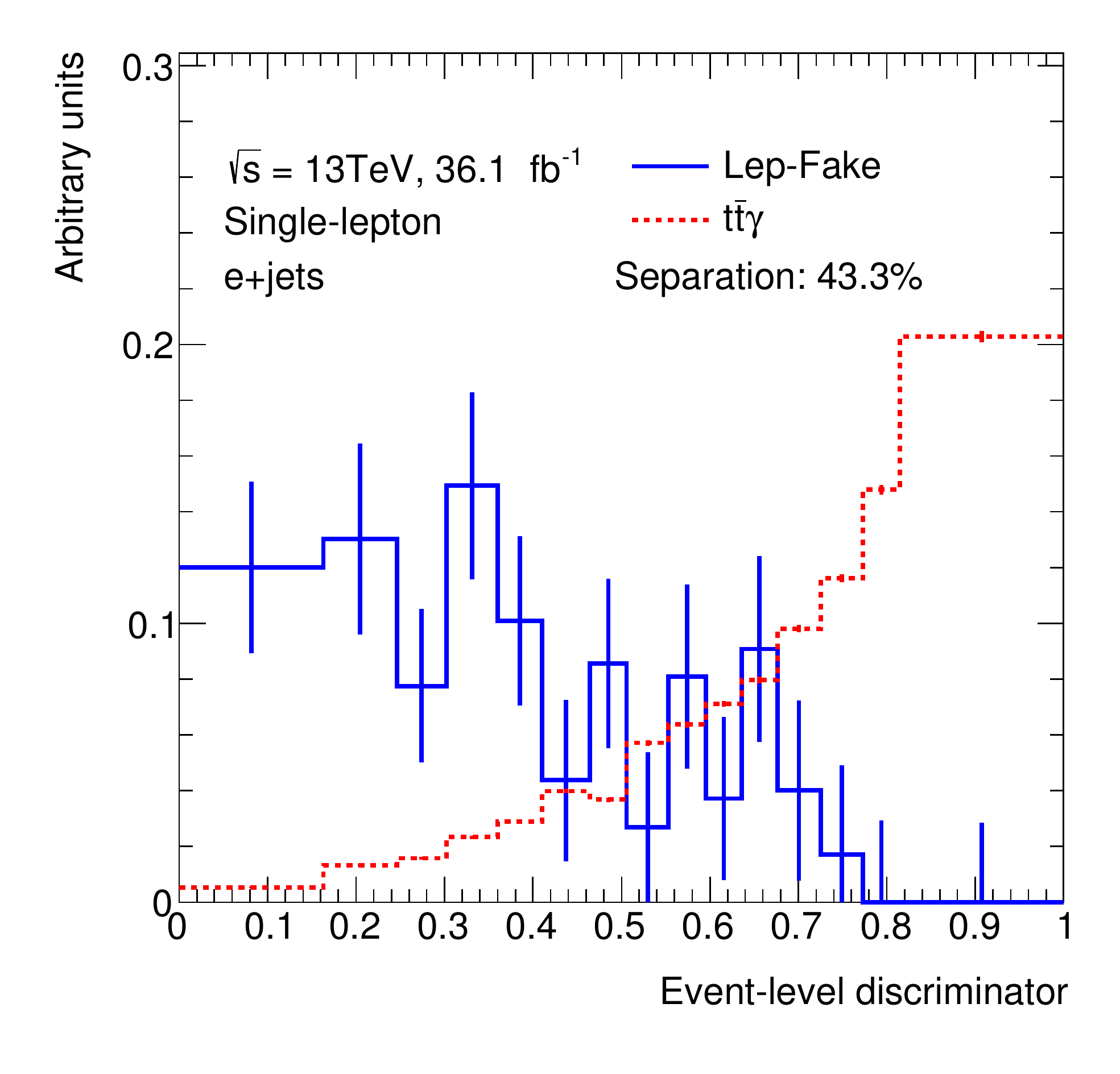}
\hspace{-0.032\linewidth}
\caption {Event-level Discriminator variable for the \chejets channel with separation plots shown for signal and the different background components.}\label{fig:nnoutputsEjets}
\end{figure}   

\begin{figure}[!htbp]
\centering
\includegraphics[width=0.34\linewidth]{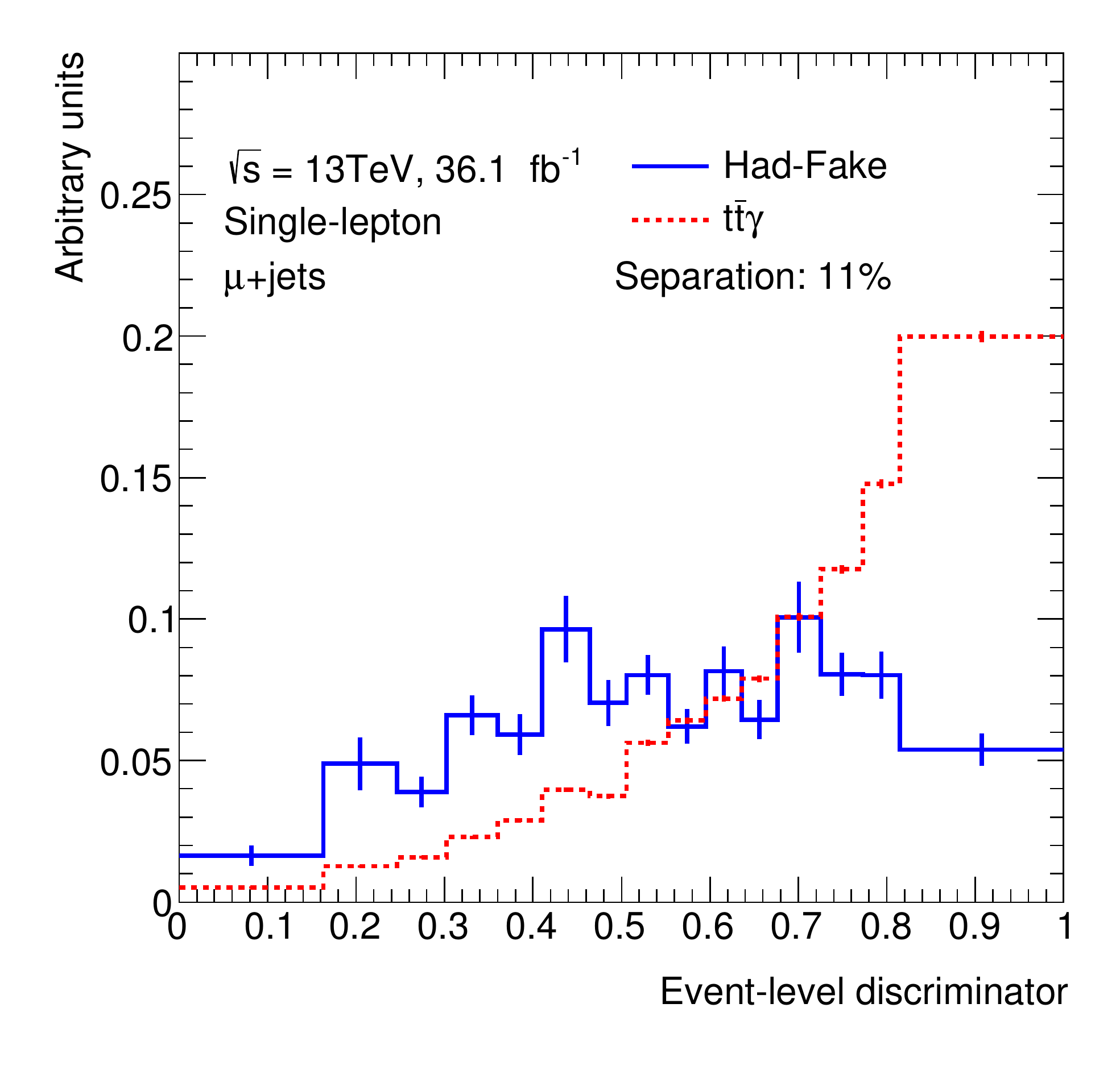}
\hspace{-0.032\linewidth}
\includegraphics[width=0.34\linewidth]{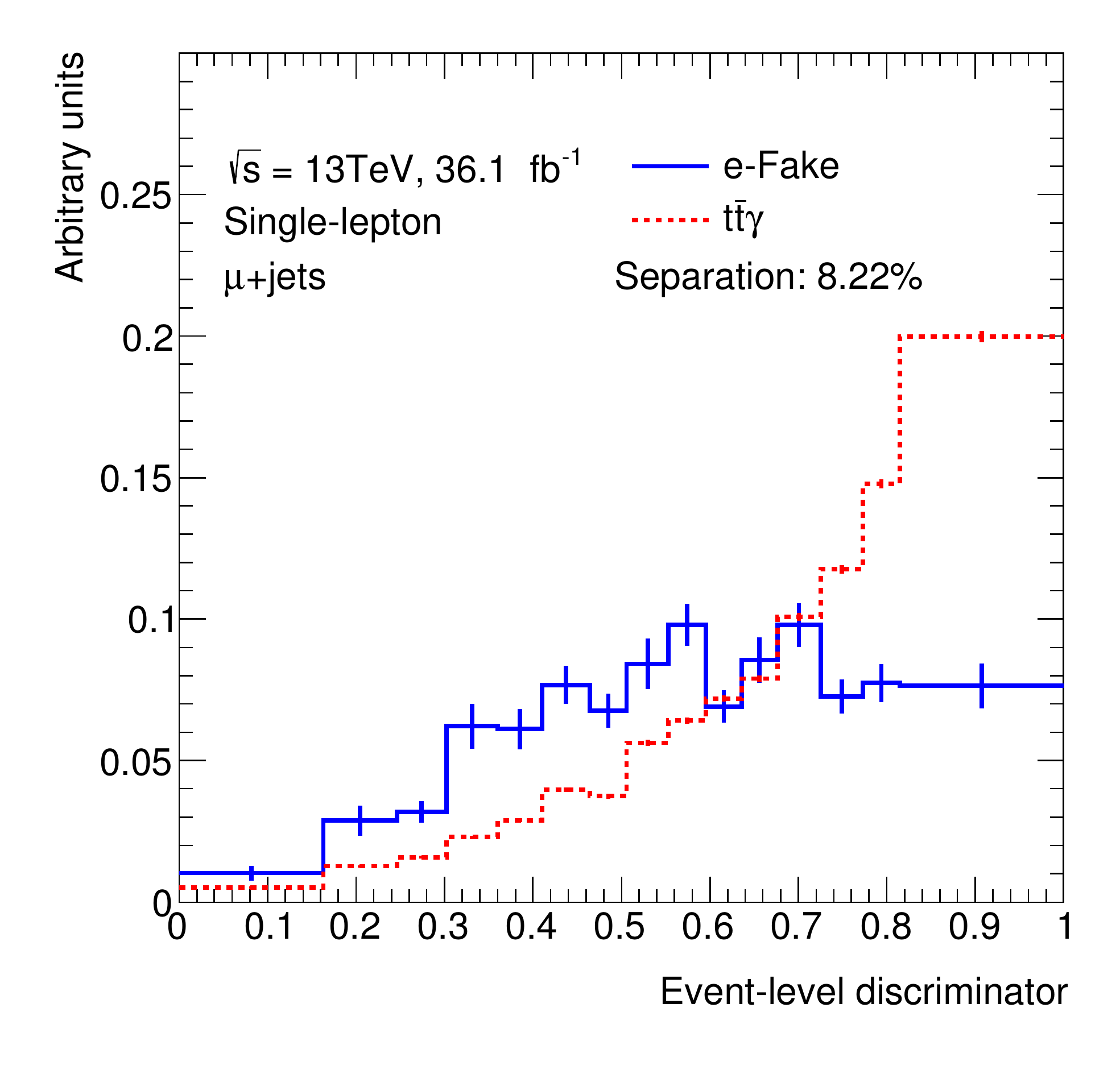}
\hspace{-0.032\linewidth}

\includegraphics[width=0.34\linewidth]{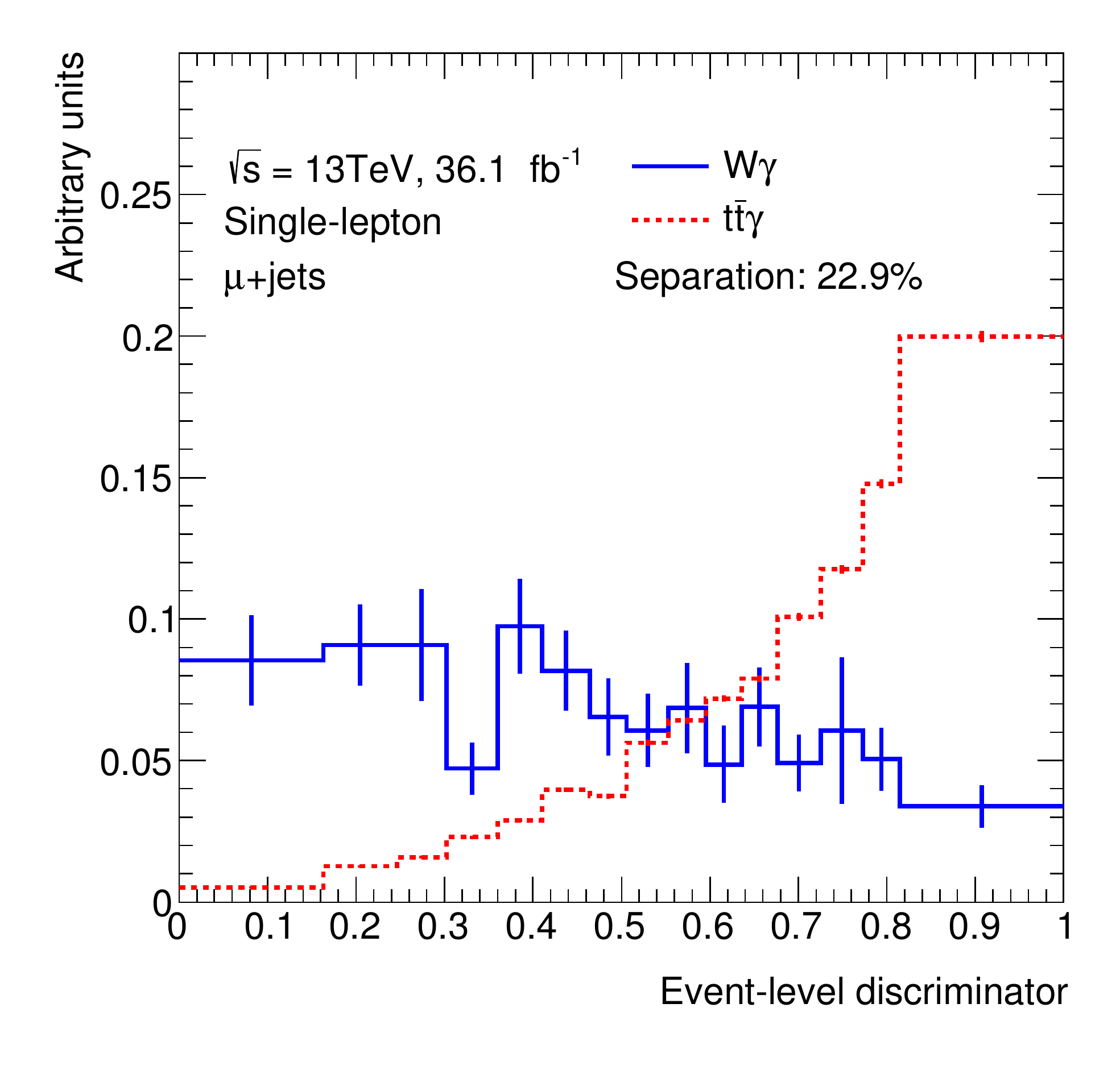}
\hspace{-0.032\linewidth}
\includegraphics[width=0.34\linewidth]{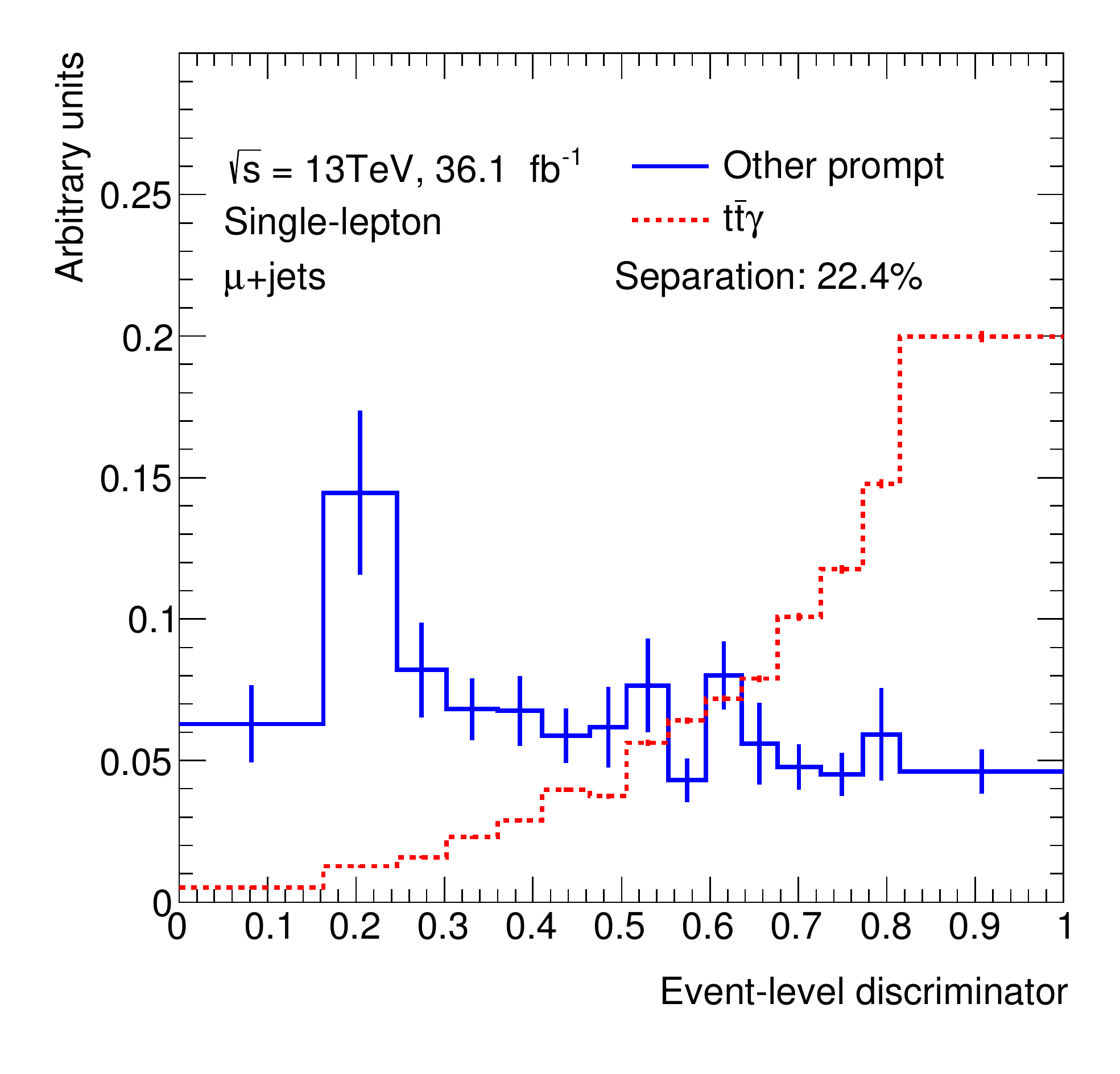}
\hspace{-0.032\linewidth}
\includegraphics[width=0.34\linewidth]{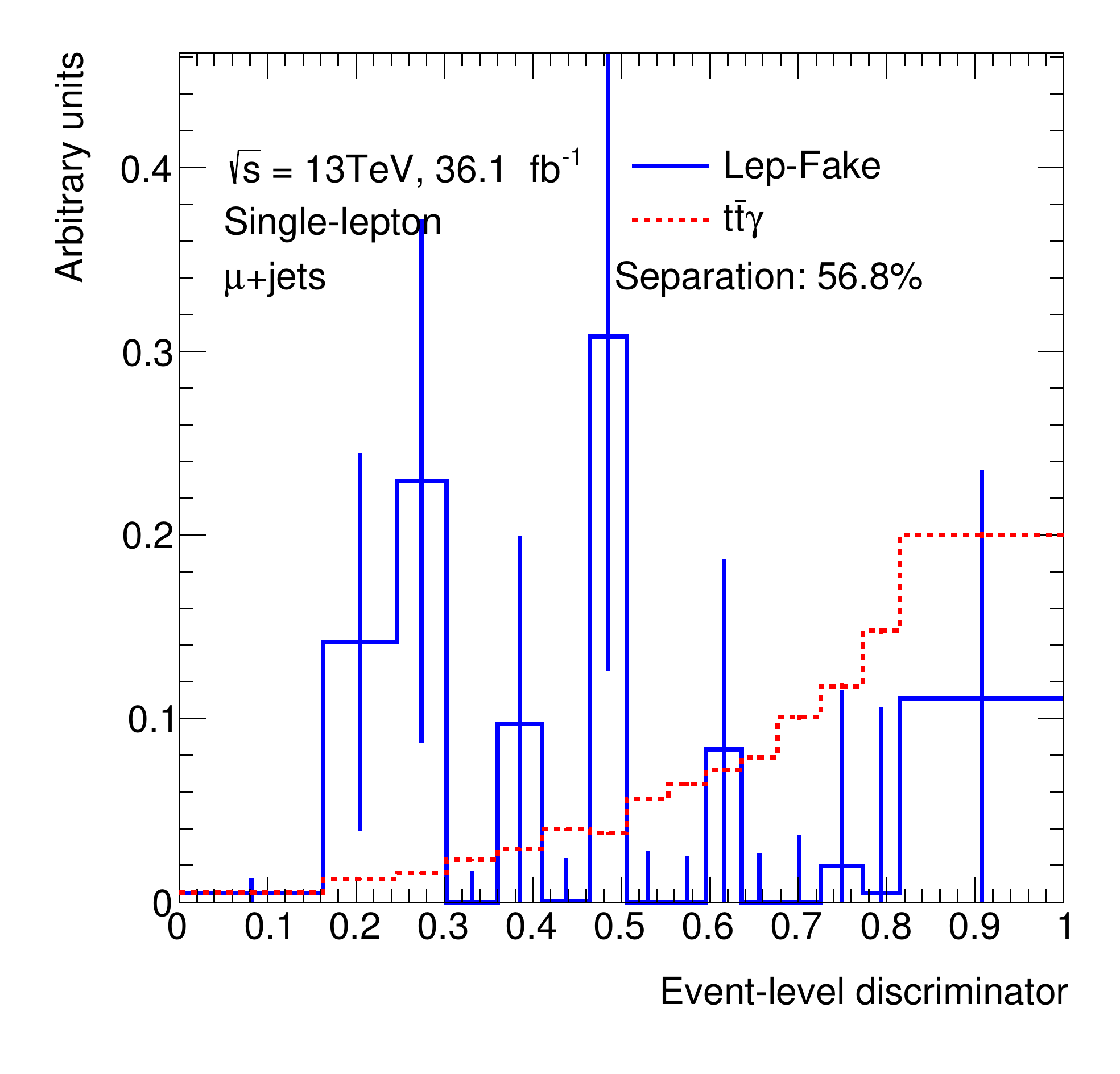}
\hspace{-0.032\linewidth}
\caption {Event-level Discriminator variable for the \chmujets channel with separation plots shown for signal and the different background components.}\label{fig:nnoutputsMujets}
\end{figure}   

\chapter{Systematic uncertainties support material}
\label{sec:systematics_appendix}

\section{Cross-section closure test for small MC backgrounds}\label{sec:promptSystCheck}
There is currently a 50\% normalisation uncertainty applied to each of the single top, diboson and \ttV backgrounds
This section checks the impact the systematic uncertainty has for these prompt backgrounds, grouped as \Other. For the \chljets and \chll channel two Asimov fits are performed with all systematic sources included. In the first fit a 50\% uncertainty is assigned to each of these three backgrounds. In the second fit the systematic uncertainty is doubled, i.e. a 100\% uncertainty is assigned. The results for the fits are shown in Table~\ref{tab:otherbkgAsimovFits}. Doubling the uncertainty on the three backgrounds leads to an absolute increase in the error of $\approx 0.2\%$ for both the \chljets and \chll channels. This impact is negligible and means these backgrounds do not play an important role in the fit. For this reason, the normalisation uncertainty is left at 50\%.  

\begin{table}[htbp]
        \centering
        \begin{tabular}{|l|l|c|c|}
        \hline
        Channel & Normalisation uncertainty & ``+" error & ``-" error \\
        \hline
        \hline
        \multirow{2}{*}{\chljets} & 50\% & 0.0929 & 0.0914  \\
         &  100\%& 0.0945 & 0.0927 \\
         \hline
        \multirow{2}{*}{\chll} & 50\% &  0.0678 & 0.0629 \\
        & 100\% & 0.0697 & 0.0647 \\
         \hline
        \end{tabular}
        \caption{Two different Asimov fit scenarios for the \chljets and \chll channels where the normalisation uncertainty on the \Other backgrounds has been varied. All systematics are included in the fit. The error represents the absolute up/down error of $\mu$. }
        \label{tab:otherbkgAsimovFits}
\end{table}

\FloatBarrier

\chapter{Results section support material}
\label{sec:resultsappendix}

\section{Final fits}

\begin{figure}[!htbp]
\centering
\includegraphics[width=0.8\textwidth]{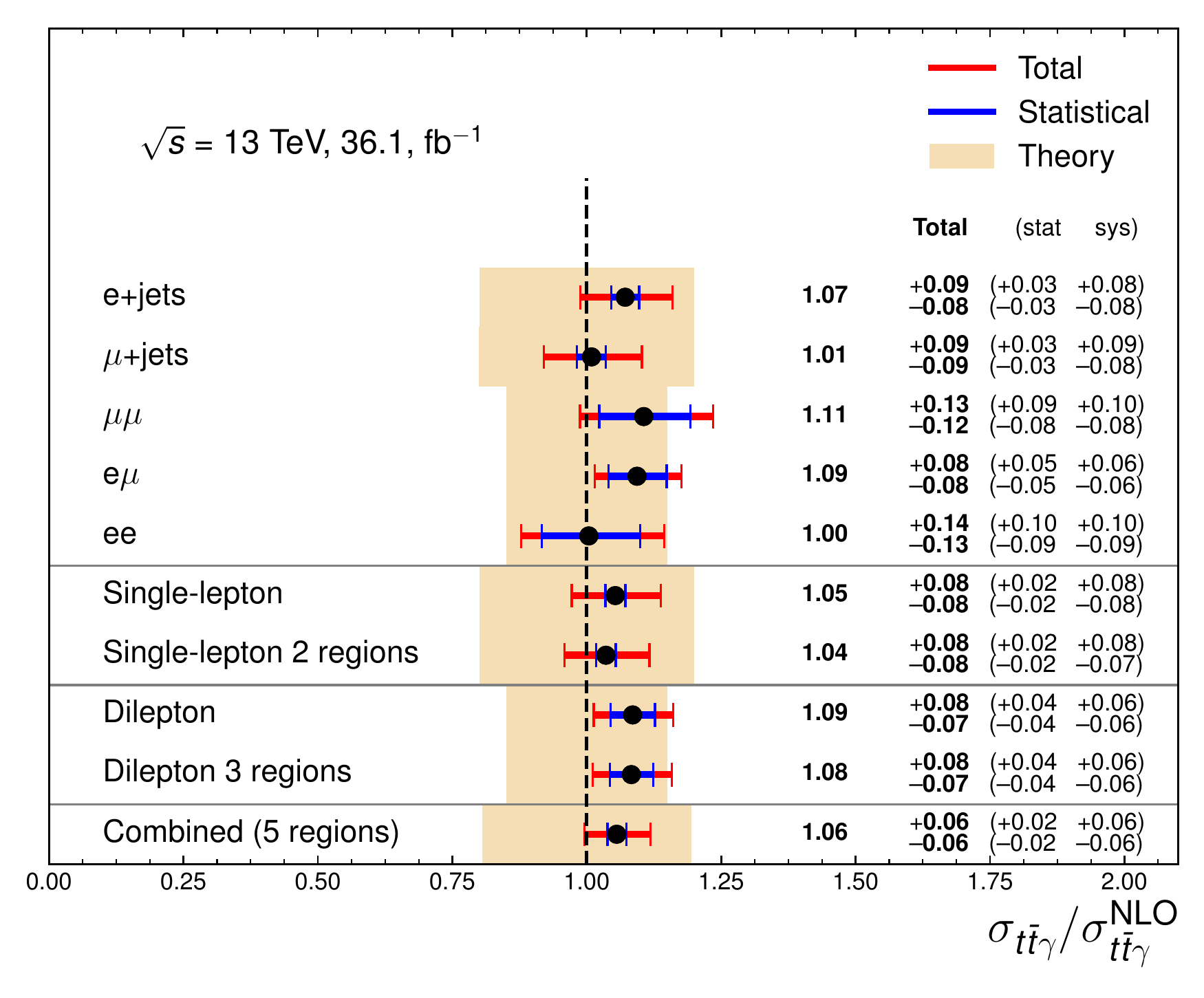}
\caption {The observed signal strength for individual, merged and combined \chljets and \chll, and combined channels. The NLO SM prediction is represented by the dashed vertical line.  The theoretical uncertainty for each fit is represented by the shaded region.}
\label{fig:crossSectionMu_all}
\end{figure}

\begin{figure}[!htbp]
\centering
\subfloat[Pre-fit \chejets]{
\includegraphics[width=0.39\linewidth]{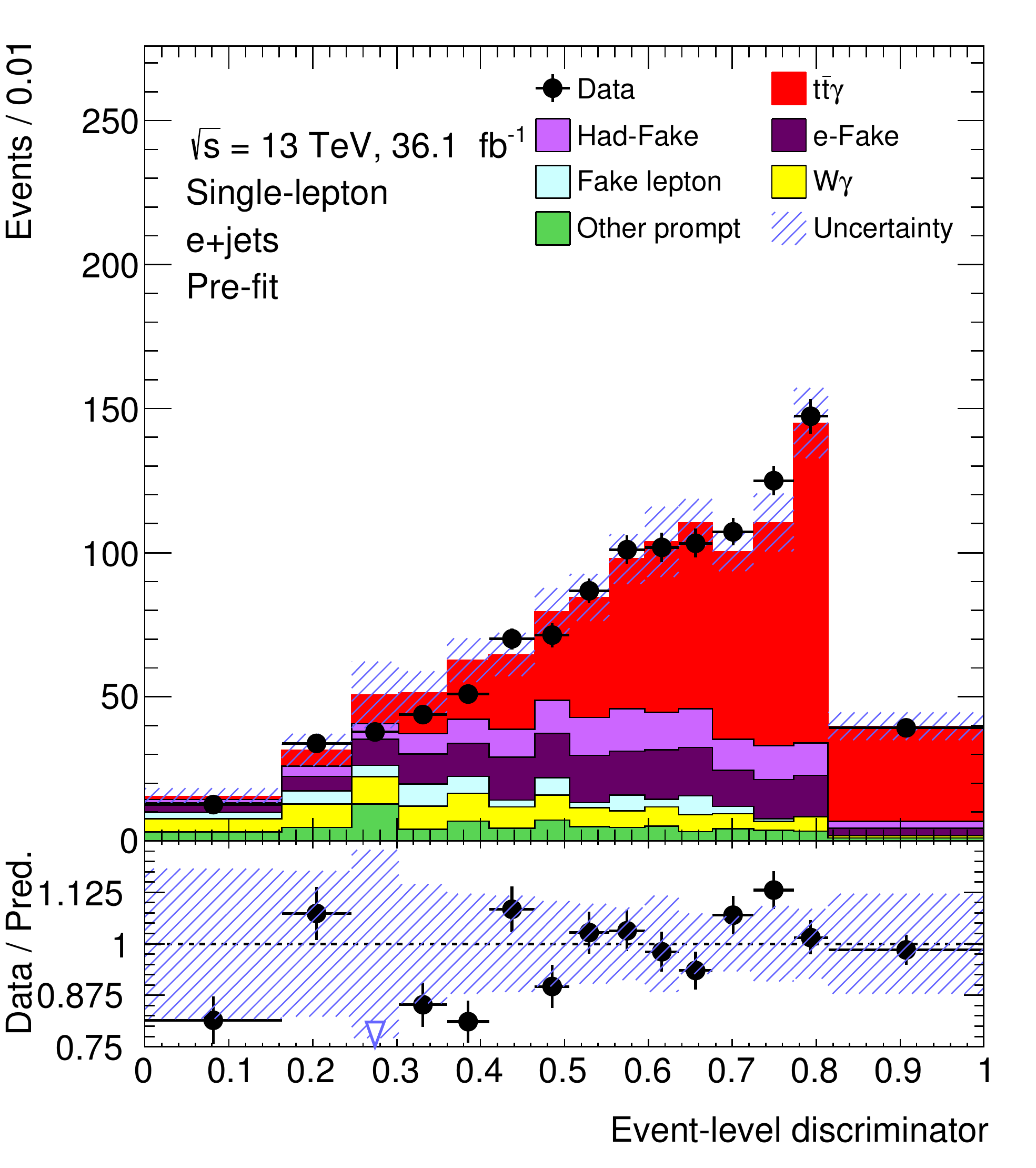}
}\hspace{-0.038\linewidth}
\subfloat[Post-fit \chejets]{
\includegraphics[width=0.39\linewidth]{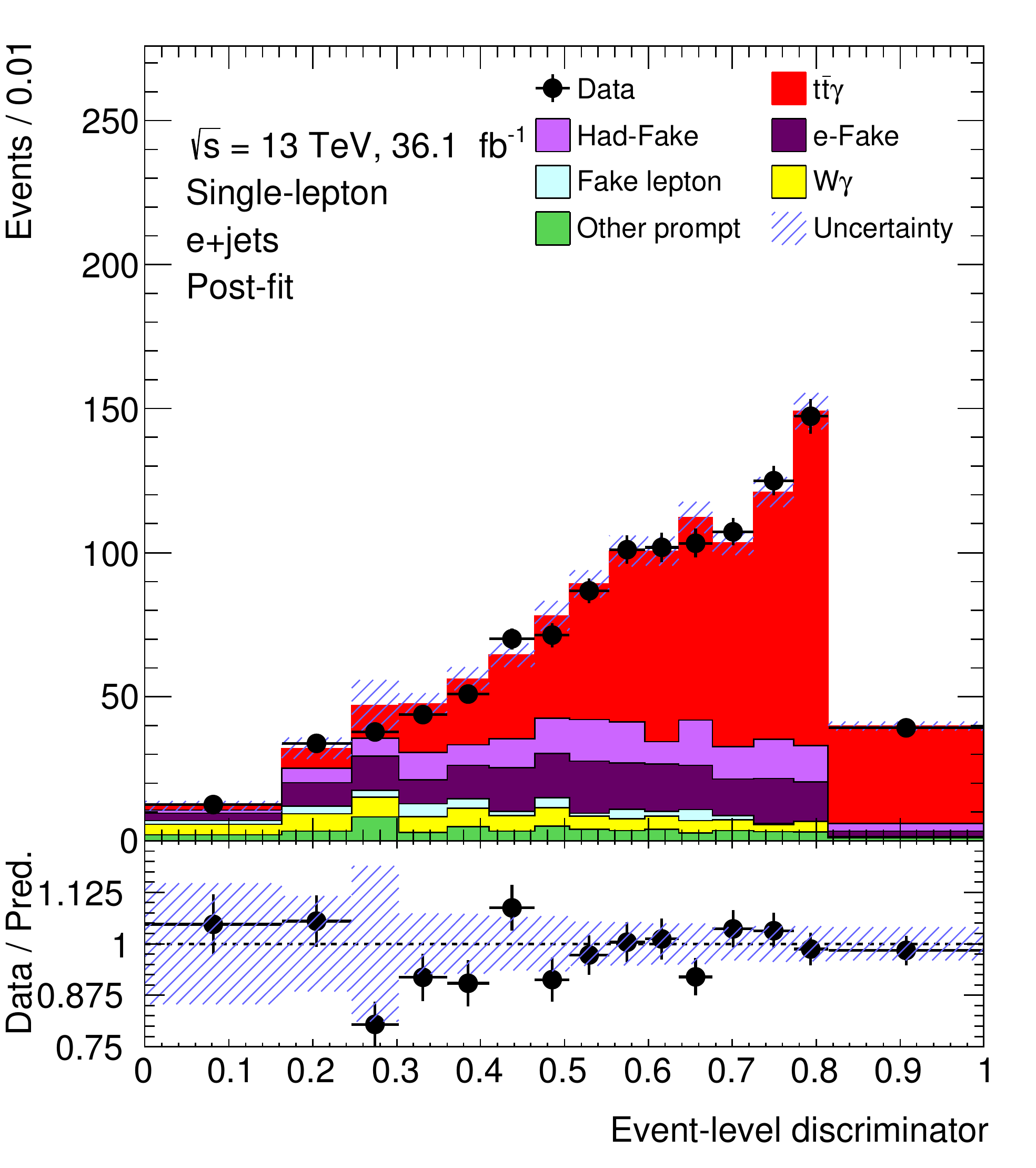}
}\hspace{-0.038\linewidth}

\subfloat[Pre-fit \chmujets]{
\includegraphics[width=0.39\linewidth]{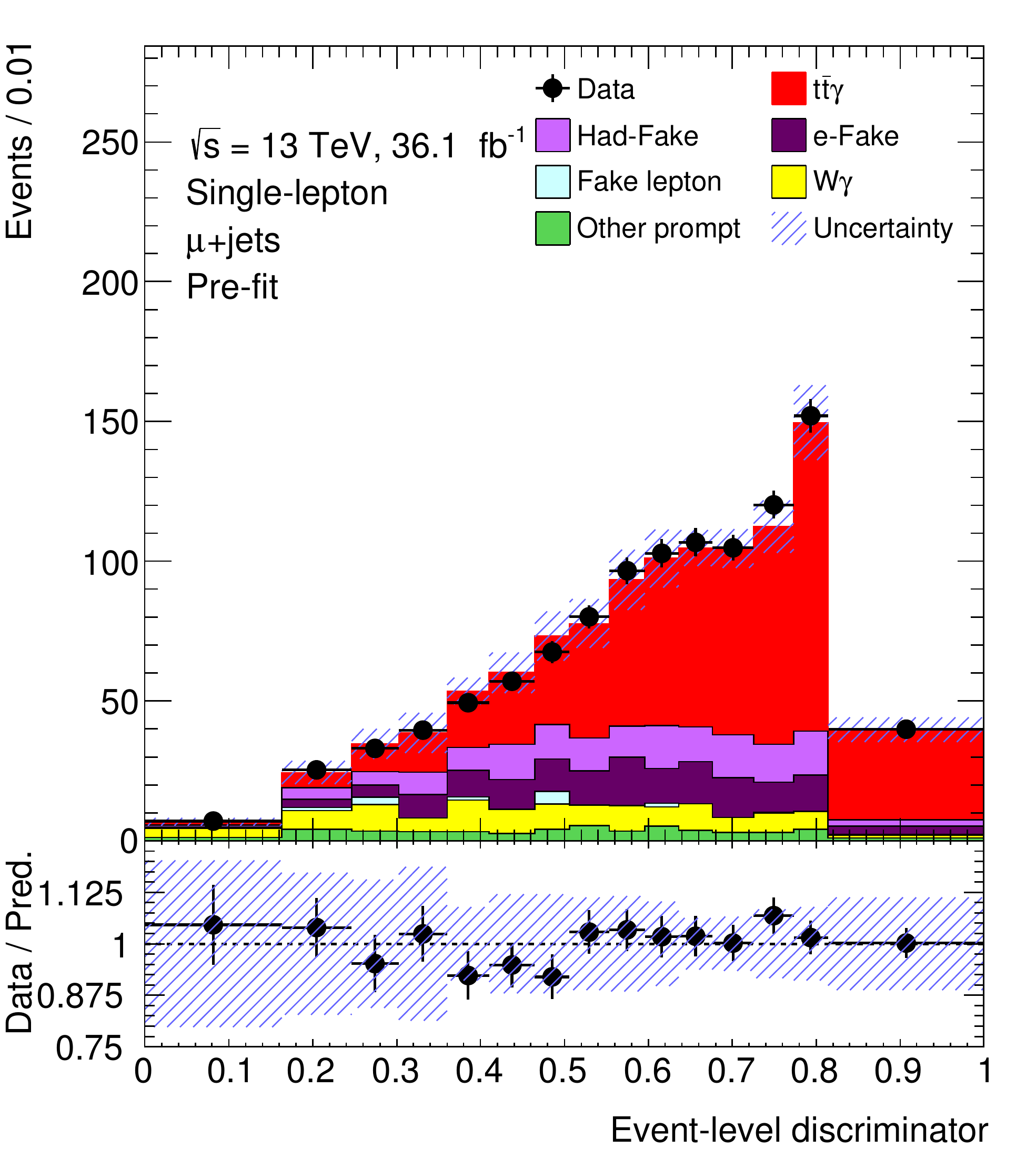}
}\hspace{-0.038\linewidth}
\subfloat[Post-fit \chmujets]{
\includegraphics[width=0.39\linewidth]{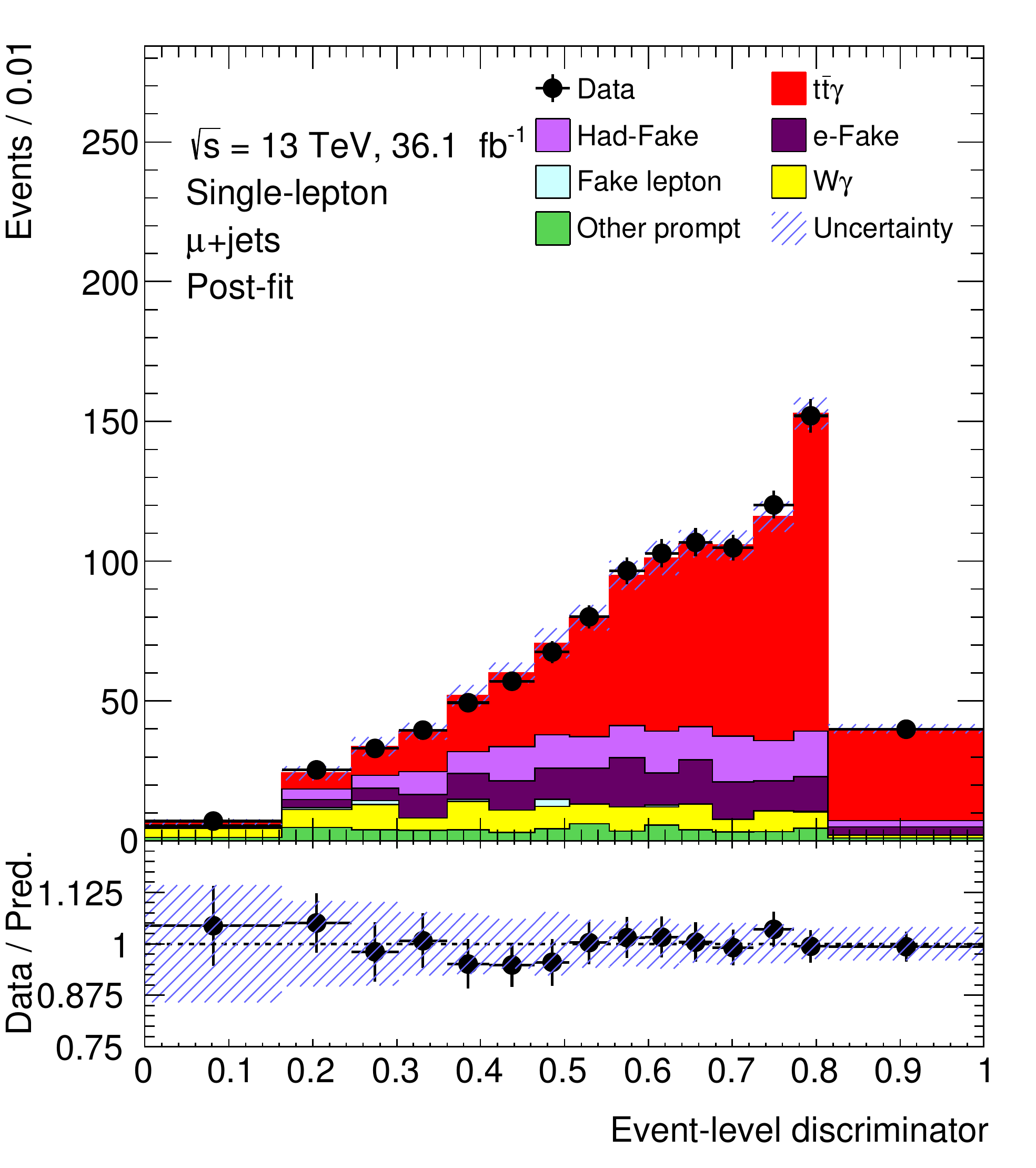}
}\hspace{-0.038\linewidth}

\caption {Pre- and post-fit plots for the individual \chljets channels where the respective \ELD distribution is used as the discriminating variable in the fit.}
\label{fig:postFitELDSLIndividual}
\end{figure}

\begin{figure}[!htbp]
\centering
\subfloat[Pre-fit \chee]{
\includegraphics[width=0.39\linewidth]{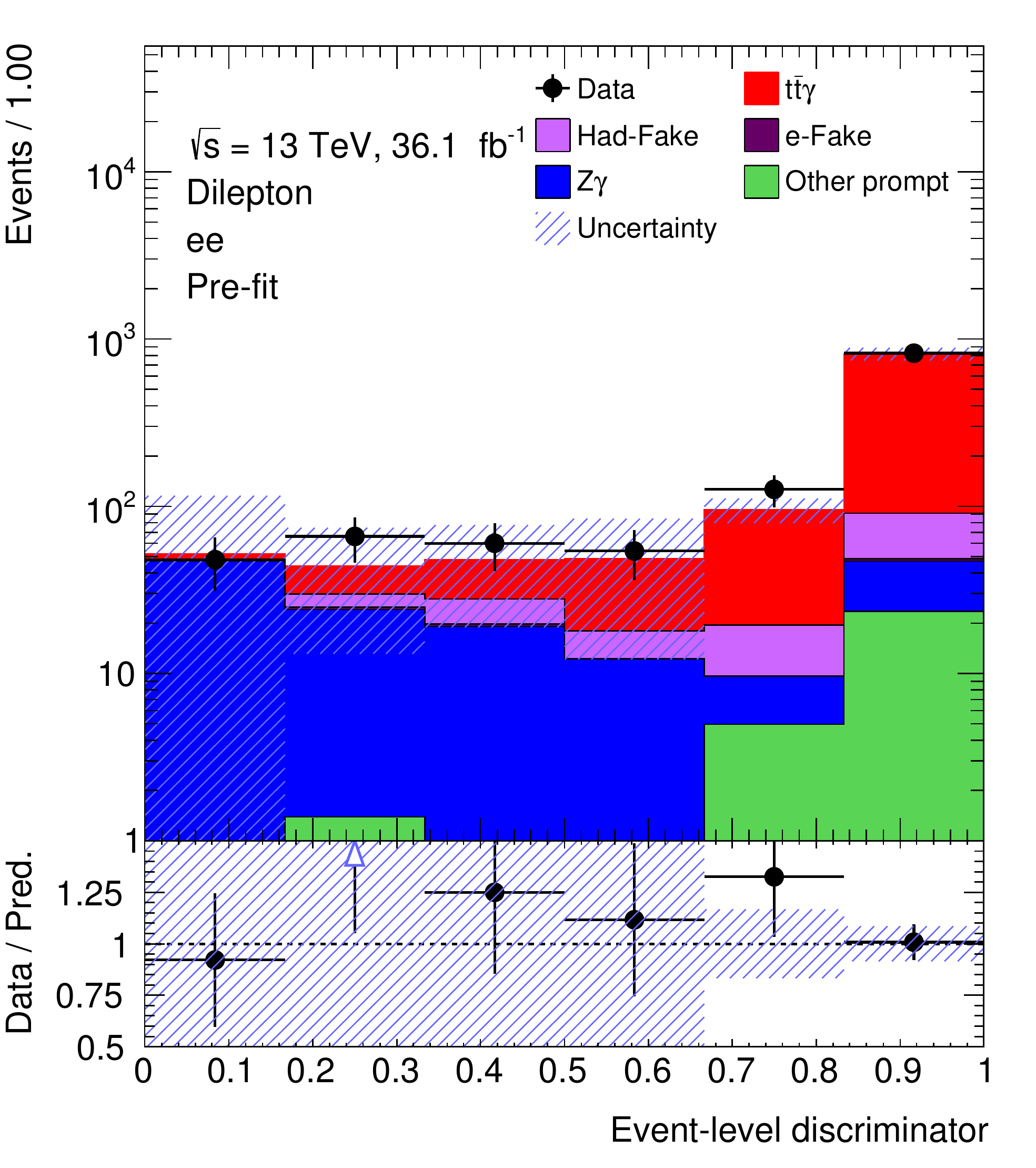}
}\hspace{-0.032\linewidth}
\subfloat[Post-fit \chee]{
\includegraphics[width=0.39\linewidth]{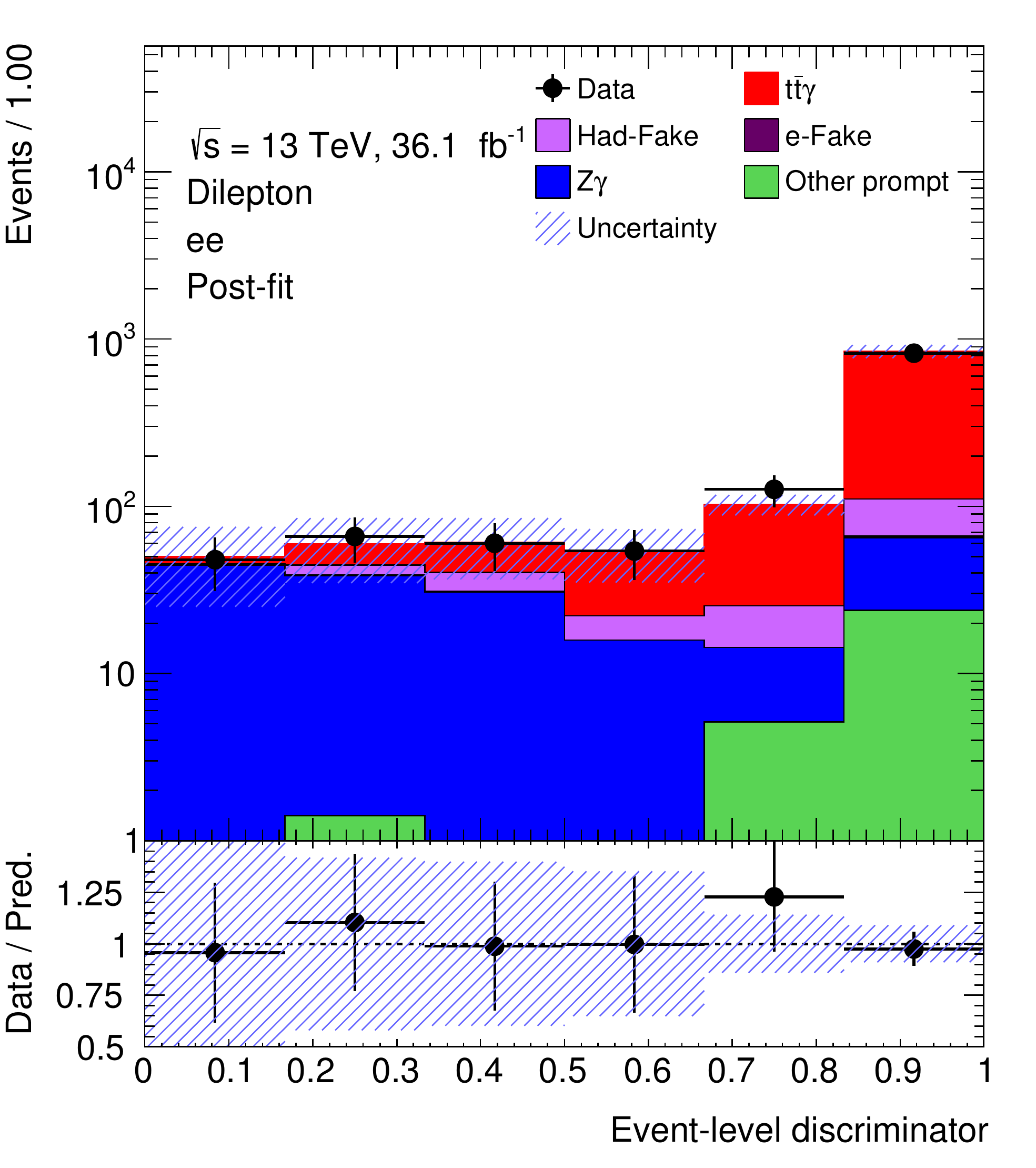}
}\hspace{-0.032\linewidth}

\subfloat[Pre-fit \chmumu]{
\includegraphics[width=0.39\linewidth]{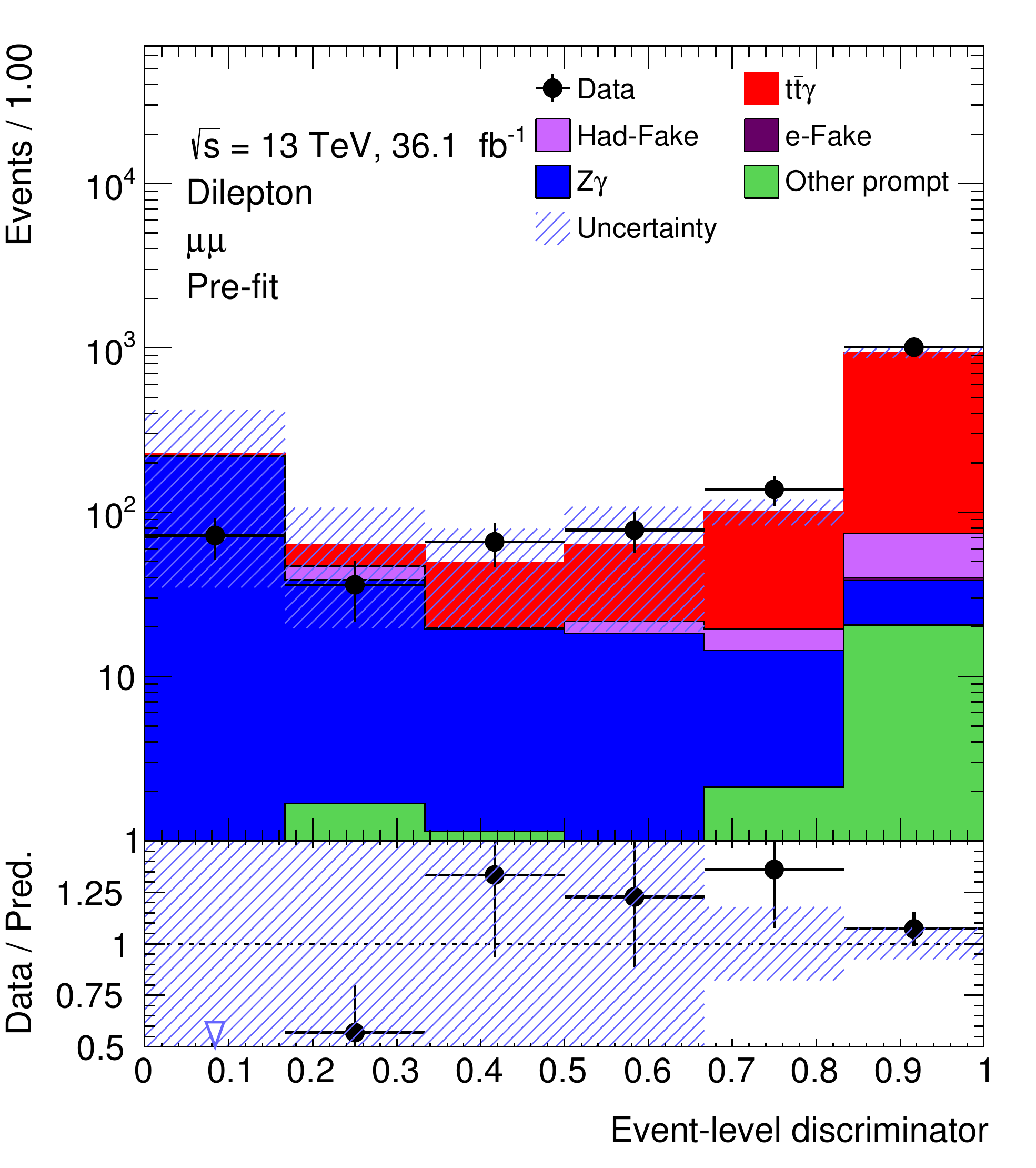}
}\hspace{-0.032\linewidth}
\subfloat[Post-fit \chmumu]{
\includegraphics[width=0.39\linewidth]{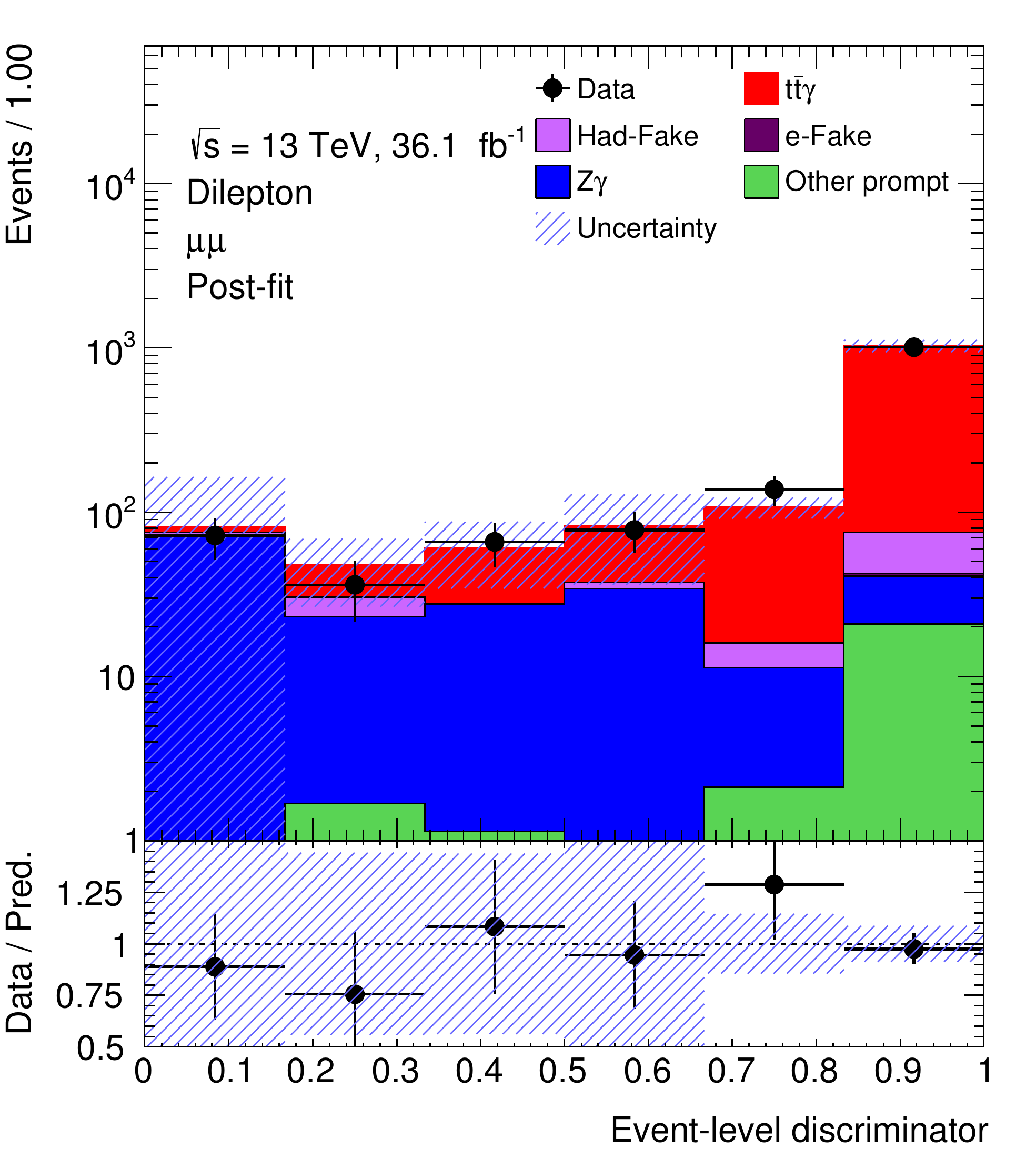}
}\hspace{-0.032\linewidth}

\subfloat[Pre-fit \chemu]{
\includegraphics[width=0.39\linewidth]{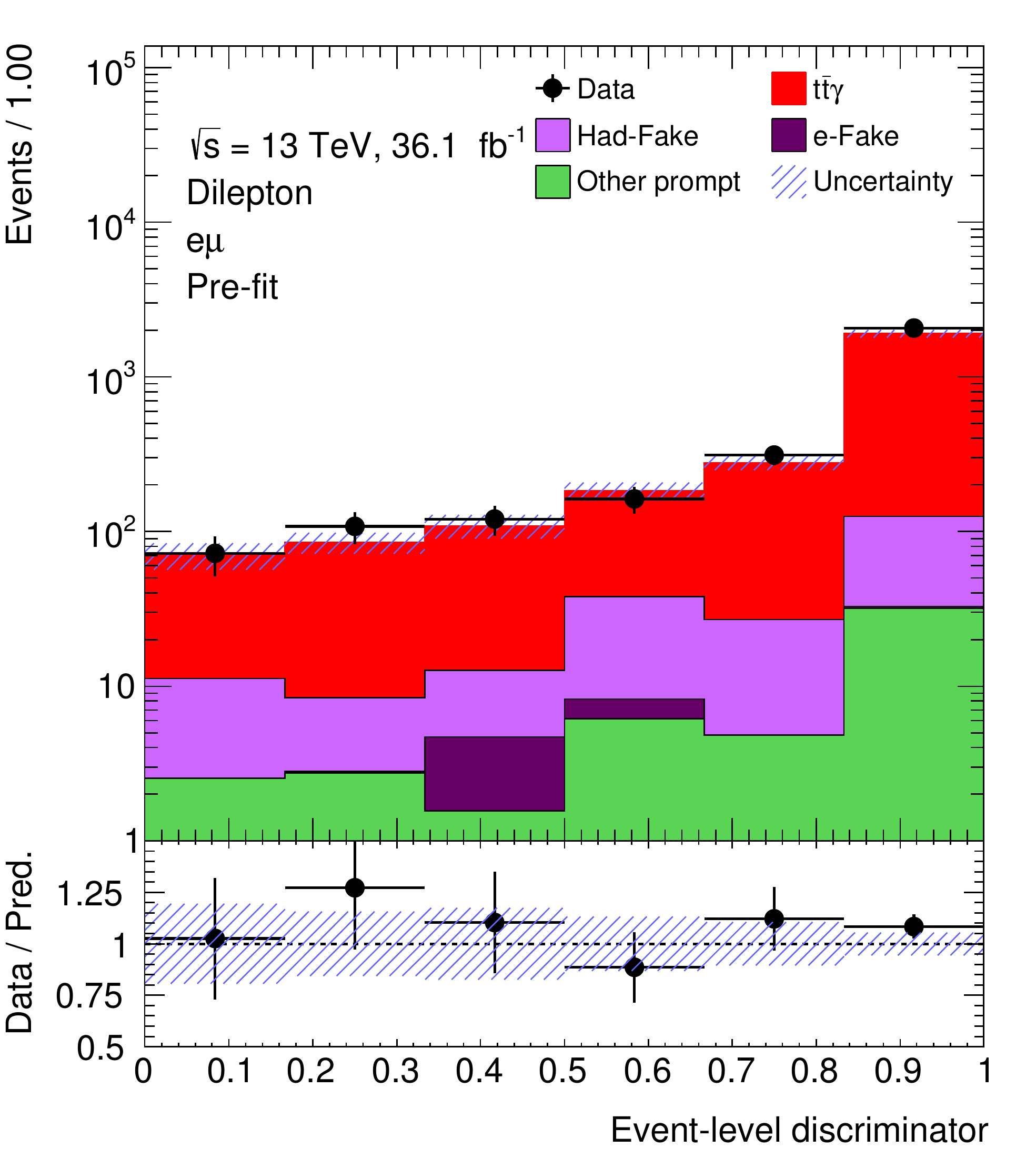}
}\hspace{-0.032\linewidth}
\subfloat[Post-fit \chemu]{
\includegraphics[width=0.39\linewidth]{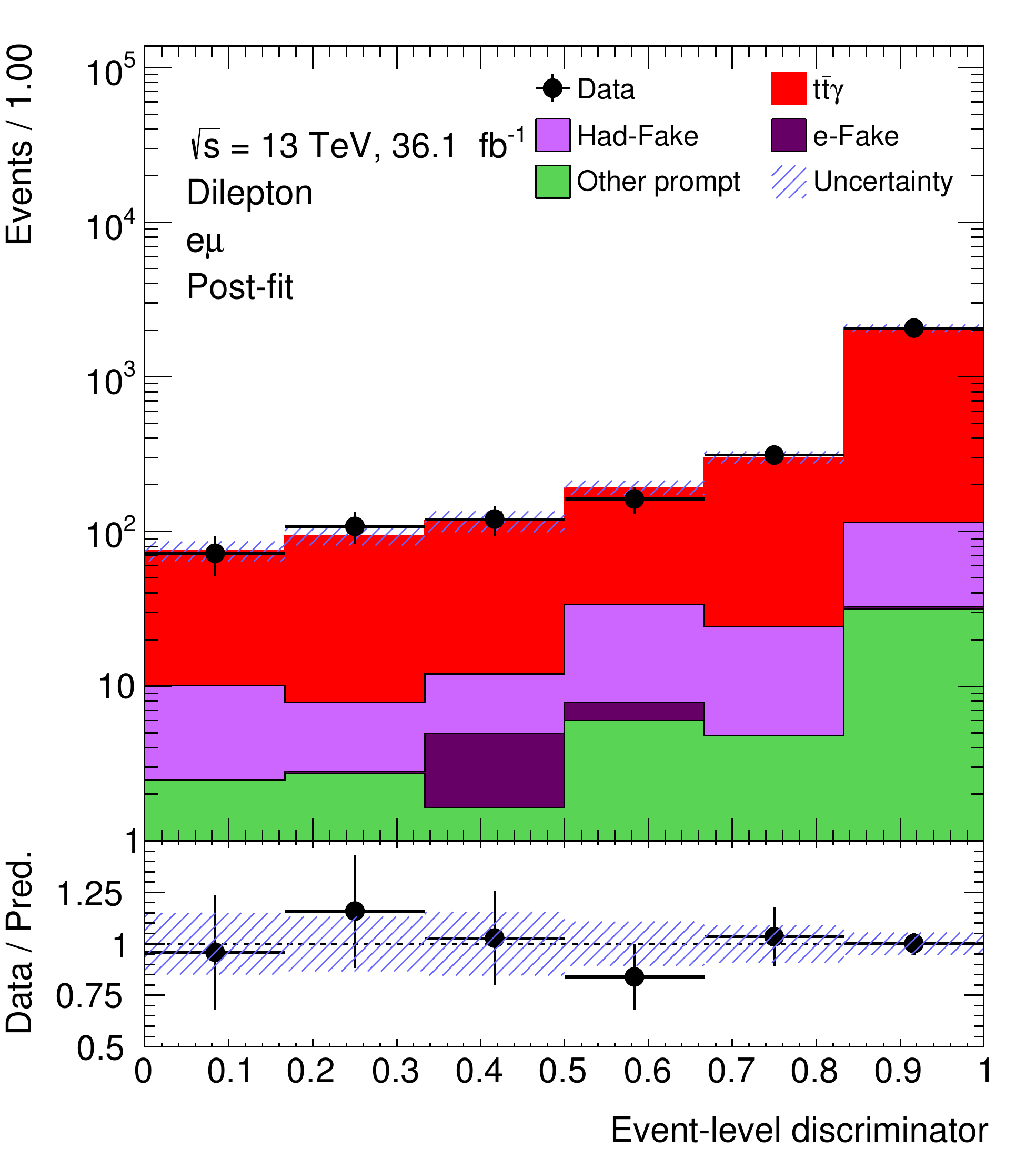}
}\hspace{-0.032\linewidth}

\caption {Pre- and post-fit plots for the individual \chll channels where the respective \ELD distribution is used as the discriminating variable in the fit.}
\label{fig:postFitELDDLIndividual}
\end{figure}

\begin{table}[!htbp]
\begin{center}
\scalebox{0.85}{
\begin{tabular}{|l|rr|rr|rr|rr|rr|}
 \hline
&  \multicolumn{2}{c|}{\chejets}  &  \multicolumn{2}{c|}{\chmujets}  &  \multicolumn{2}{c|}{\chee} &  \multicolumn{2}{c|}{\chmumu} &  \multicolumn{2}{c|}{\chemu} \\
\hline
\hline
& + [\%] & - [\%] & + [\%] & - [\%] & + [\%] & - [\%] & + [\%] & - [\%] & + [\%] & - [\%]  \\
\cline{2-11}

Signal Modelling & 1.36 & 1.21 & 0.50 & 0.61 & 7.40 & 5.80 & 3.20 & 2.88 & 3.60 & 3.53\\
Jets & 5.30 & 4.60 & 5.60 & 5.00 & 3.34 & 2.96 & 2.91 & 2.66 & 1.61 & 1.45\\
Luminosity & 2.33 & 2.03 & 2.35 & 2.01 & 2.47 & 1.99 & 2.35 & 2.01 & 2.28 & 2.02\\
Pileup & 2.35 & 2.08 & 2.37 & 2.21 & 2.34 & 1.96 & 4.30 & 3.70 & 1.38 & 1.33\\
Photon Efficiencies & 1.09 & 0.95 & 1.10 & 0.94 & 1.16 & 0.93 & 1.11 & 0.97 & 1.08 & 0.95\\
$b$-Tagging & 0.86 & 0.97 & 0.80 & 0.84 & 0.58 & 0.57 & 0.27 & 0.31 & 0.14 & 0.16\\
Background modelling & 4.50 & 4.50 & 6.00 & 5.90 & 3.53 & 3.41 & 3.04 & 3.24 & 2.98 & 3.08\\
Leptons & 0.95 & 0.87 & 1.64 & 1.52 & 2.16 & 1.81 & 1.78 & 1.56 & 1.18 & 1.04\\
Prompt photon tagger (shape) & 3.26 & 3.42 & 4.20 & 4.30 & - & - & - & - & - & -\\
E$\gamma$ (Resolution and scale) & - & - & 0.06 & 0.05 & 0.35 & 0.31 & 0.08 & 0.10 & 0.08 & 0.12\\
Template Statistics & 2.40 & 2.28 & 2.84 & 2.66 & 3.80 & 3.25 & 4.10 & 3.70 & 1.76 & 1.61\\
\hline
Total systematic & 7.9 & 7.5 & 9.0 & 8.6 & 10.5 & 8.8 & 8.5 & 7.7 & 5.6 & 5.3\\
Total statistical & 2.1 & 2.1 & 2.2 & 2.2 & 9.4 & 8.9 & 7.9 & 7.5 & 5.0 & 4.9\\
\hline
Total & 8.2 & 7.8 & 9.3 & 8.8 & 14.0 & 12.5 & 11.6 & 10.8 & 7.5 & 7.2\\
\hline
\end{tabular}
}
\caption{Relative effects on $\mu$ due to the grouped systematic sources. Fits are performed with nuisance parameters in each group held constant with the rest floating. This new uncertainty is subtracted in quadrature from the total uncertainty to obtain $\Delta \mu$.}
\label{tab:systSummary_individual} 
\end{center}
\end{table}

\begin{figure}[!htbp]
\centering
\includegraphics[trim={1.9cm 1.3cm 0.3cm 0.3cm},clip,width=0.65\textwidth]{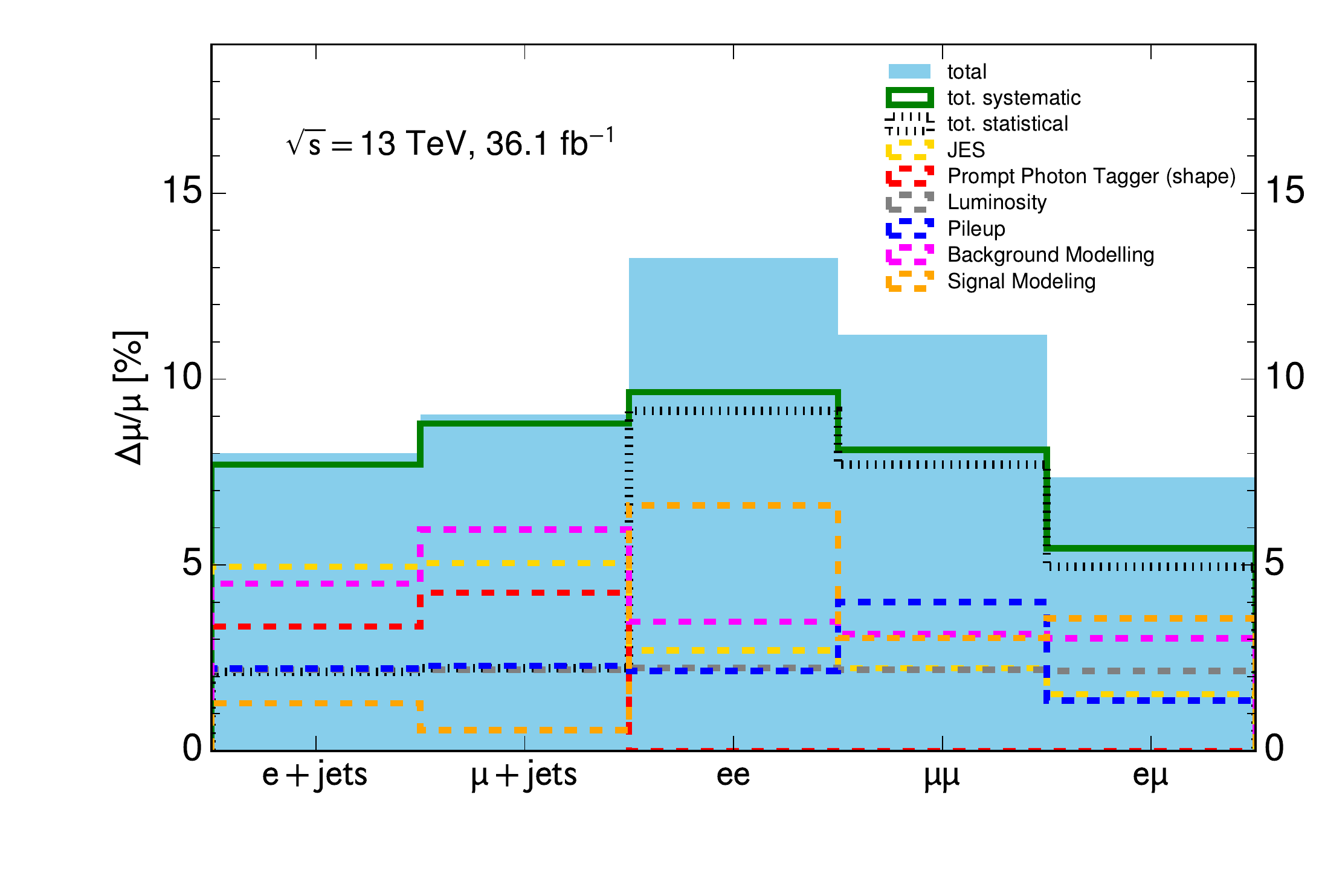}
\caption {The relative contribution of uncertainties shown in groups. A fit is performed where all NPs in a group are kept fixed while the rest are left floating.The resulting uncertainty is subtracted in quadrature from the original uncertainty to obtain $\Delta \mu$.}
\label{fig:systsOverview_5channels}
\end{figure}

\end{document}